# PRINCIPLES OF A SECOND QUANTUM MECHANICS

# *QM2*

## ROOTED IN FACTUALITY
## AND CONSTRUCTED BOTTOM-UP*

**10 JUILLET 2019**

**Mioara MUGUR-SCHÄCHTER**
http://www.mugur-schachter.net/

# AKNOWLEDGEMENT

This work would not have been possible without the life-long support of Sully Schächter.

## GRATITUDE

I am grateful to my sons François and Vincent for their constant support.
I express my deep gratitude to those who have encouraged me.

My early and basic views owe much to my very numerous exchanges with
Daniel Evrard throughout the years 1970-1994.

I feel particularly indebted to my friends
Jean-Paul Baquiast, Henri Boulouet, Jean-Marie Fessler, Geneviève Rivoire.
They have constantly encouraged me throughout long periods of troubled
time.

## DEDICATION

This work is dedicated to

Louis de Broglie

whose deep unconventional work has founded Quantum Mechanics
and 90 years later permits to re-found it.

# TABLE OF CONTENTS

## GÉNÉRAL ABSTRACT

*This work is **not** an interpretation of the Hilbert-Dirac quantum mechanics* $QM_{HD}$ . It exposes the principles of a new representation of microstates, *a second quantum mechanics* denoted *QM2*. This representation is rooted directly into the a-conceptual physical reality wherefrom it is constructed bottom-up – by in physical-operational-conceptual terms, in uninterrupted relation with the physical factuality and with the constraints imposed by the involved cognitive situation.

First – independently of any mathematical formalism – is constructed in qualitative but general and *formalized* terms, a framework for reference-and-imbedding of *any* acceptable theory of physical micro-entities that – radically – cannot be perceived directly by human observers-conceivers. The construction is achieved under exclusively the *physical*-operational, methodological and conceptual constraints entailed by the cognitive situation that is involved. The obtained representation is called *infra-(quantum mechanics)* and is denoted *IQM*.

By reference to *IQM* is then worked out a preliminary critical examination of $QM_{HD}$. Thereby it appears that:

***(a)*** $QM_{HD}$ is devoid of any general formal representation of the *individual* physical entities and operations that – quite essentially – it does involve: *The whole level of individual conceptualization of the microstates is lacking inside* $QM_{HD}$, *massively*. Inside $QM_{HD}$ are endowed with clear but directly mathematical definitions – flawed by lacunae, approximations or a priori idealizations – exclusively abstract statistics that express results of ghostly and *fallaciously* specified acts of measurement performed on ghostly sketched out physical entities.

***(b)*** Contrary to the orthodox general interdiction of any model of a microstate and to the unanimous assertion of absence of such a model inside $QM_{HD}$, it is found that the mathematical formalism of $QM_{HD}$ ***does*** involve a model of microstate, and in a quite fundamental role, namely de Broglie's model of a wave with a "corpuscular-*like*" singularity in its amplitude. But inside $QM_{HD}$ both this fact and its meaning remain implicit. So their consequences are not systematically recognized and made use of. This entails a catastrophic hole in the very definability of the acts of measurement: *no general and explicit rules are formulated for **coding** the directly observable effect of an act of quantum-measurement – just marks on registering devices – in meaningful terms, of a definite value of the measured quantity*.

> From *(a)* and *(b)* it follows that the nowadays Hilbert-Dirac Quantum-Mechanics $QM_{HD}$ *is simply devoid of an acceptable representation of the quantum measurements***.**

The lacunae mentioned above are then compensated via a radically constructive bottom-up approach that starts from local zeros of knowledge on, specifically, the physical microstate to be studied. It starts there by a conceptual-*physical* operation of extraction from the physical factuality, of specimens of the microstate to be studied. These specimens carry in them the factual germs of any possible individual knowledge on a given sort of microstate.

*This* is what entails the emergence of a Second Quantum Mechanics *QM2* that is rooted in factuality and is constructed bottom-up.

The elaboration of *QM2* begins by a new representation *of the quantum measurements*. The measurements on *un*-bound microstates devoid of a quantum-potential are treated in the first place. The elaborated treatment is then extended to also the *un*-bound interference-microstates, by recourse to the de Broglie-Bohm guiding law.

The case of bound states does not raise specific questions of principle).

The predictive probability measures on results of quantum measurements are constructed by a factual-conceptual procedure – directly via measurements, in a way *identical to that which one is **obliged** to make use of for **verifying** the predictions* – and they are expressed in Hilbert-space terms on the basis of Gleason's theorem.

The factual output of this procedure parallels the mathematical output of the Schrödinger equation of the problem when this equation can be written and solved. When the equation involves gross idealizations or/and approximations, this entails control upon tits mathematical output. While when the equation of the problem cannot be solved or cannot be defined, it can be radically replaced by the conceptual-factual procedure defined in this work. This is a notable improvement consonant with the progresses realized in nanotechnology and informatics.

Around the core constituted by the new representation of the quantum measurements is then structured a very synthetic global outline of a fully intelligible Second Quantum Mechanics *QM2*. This theory emerges as a synthesis between *IQM*, $QM_{HD}$ and the de Broglie-Bohm approach. Its basic descriptional structure is represented by *a "factual-probability-tree" rooted in the physical factuality and endowed with a two-level statistical-probabilistic crown*.

> Thereby *QM2* – via the particular case of microstates – illustrates a strongly deepened and extended, new and ***general*** *"Theory of Factual Probabilities" belonging to the Mathematical Physics*, where all the involved probability measures tied with *one* studied physical entity can be collectively expressed by a Hilbert-vector and are genetically inter-connected by meta-probabilistic correlations. Dirac's bra-ket calculus permits to express formally certain *semantic* contents involved by a factual-probability-tree.

This explains the universal characters perceived by many authors in the Hilbert-Dirac formulation of Quantum Mechanics, while *distinguishing* clearly between the features mentioned above – that are ***purely*** *probabilistic* – and on the other hand the concept of "microstate" to which these features do just apply in particular.

***The GENESES of IQM and QM2 indicate a path toward a METHODOLOGICAL sort of unification of Physic***s.

# GENERAL INTRODUCTION

« The book will, therefore, draw a limit to thinking, or rather, not to thinking, but to the expression of thoughts; for, in order to draw a limit to thinking we should have to be able to think both sides of this limit (we should therefore have to be able to think what cannot be thought). »

Wittgenstein, Preface of the Tractatus

The first attempts at a representation of microscopic physical entities started in terms of usual 'objects' endowed with delimited spatial volumes. Therefrom classical models and ways of reasoning were more and more deeply lowered into the domain of small space-time dimensions. This process however has come to a clear crisis around 1900: The connections with classical physics ceased being compatible with the experimentally established facts. To meet this Bohr and Plank introduced *non*-classical ad hoc "principles". But thereby the intelligibility dissolved.

And then, de Broglie’s 'corpuscular-wave' *model* fractured the evolution: *It changed the* ***origin*** *on the vertical that connects knowledge of macroscopic physical entities to knowledge concerning microscopic entities*. Indeed de Broglie’s model is placed just upon the extreme frontier between the microscopic, still *a-conceptual* physical reality, and the realm of the already conceptualized. And therefrom it tended to push the process of conceptualization upward toward the previously conceptualized in classical terms, and to connect to this, intelligibly.

> So the *direction* of the actions of construction of knowledge along the vertical of conceptualization was *reversed*; and this changed also the *nature* of these actions and the semantic contents that were successively involved.

Instead of continuing to try to guess top-down starting from the classical level and advancing 'downwards' into the realm of microscopic space-time dimensions via *mental extrapolating procedures* that were unconsciously trussed up into inertial strings developed since millennia inside the classical knowledge and thinking, there timidly began to emerge a new, fluctuating tendency to construct representations bottom-up by a sort of conceptual climb in the dark guided by operational-observational-formal constraints and requirements.

The direction of constructive thought – *wher*e it begins and *how* it acts in order to reach a definite representational purpose – is quite determinant; the *succession* of the acts of conceptualization is tied with specific constraints and perennial consequences of these. So the mentioned inversion of the direction of the process of conceptualization of the microstates involved quite fundamental changes in the nature of the constraints to be overcome; and these in their turn induced obscure but strong mental confrontations between ancestral habits of thought and new unavoidable conceptual and physical-operational procedures that still lacked definite and stable contours, but of which the necessity was strikingly sensed though feebly understood.

The *method* of constructing scientific representations of physical reality was undergoing mutation.

The mathematical representations of Schrödinger and their results – directly initiated by de Broglie's model – and on the other hand Heisenberg's algorithms that were founded on different principles but for bound states offered equivalent results, led to impressing first successes. These, for a while, neutralized the conceptual disquietudes.

Meanwhile Bohr – strongly aware of the radically new characters of the emerging theory of which however the source and nature withstood identification even inside his very perceptive mind – tried to protect these characters from premature restrictions via a preventive *interdiction of any model of a microsystem*. He founded this interdiction upon the assertion of a general philosophical requirement of a strictly 'positivistic' attitude in science, consisting of the acceptance of – *exclusively* – purely operational procedures, free of *any* ontological assumption.

***But this was an impossible requirement***.

Indeed when the entity to be studied is quintessentially un-observable and is unknown, if strictly no model is assigned to it any criteria are lacking for deciding *what* ***sort*** *of operatio*n deserves being considered to be a "measurement-interaction" upon *"that"* entity for estimating the value, for it, of a *given* qualifying quantity; and also for deciding *what* such value is entailed by the observable marks obtained by each one given measurement-interaction of a chosen sort; one cannot even know in advance where in space-time the entity to be studied "is", nor what extension and contours the space-time support of this entity possesses. Nothing insures even that such classical delimiting notions possess meaning with respect to what, a priori, is called 'microsystem' and 'state of a microsystem'. No specifically adequate *language* has been constructed as yet, nor criteria for constructing such a language. No intuitive basis whatever is offered for *beginning* to construct the desired knowledge.

While when one wants to enter upon a bottom-up process of conceptualization of physical entities – as de Broglie conceived to attempt – the perspective of a whole implicit order of *constructability* opens up like a ladder from the *as yet never conceptualized before* toward the sky of classical knowledge. This ladder has to be constructed-and-climbed step by step. And if this is attempted in a purely formal-algorithmic way, void of any explanation, the procedure cannot but seem arbitrary from the point of view of meaning. The result *cannot* be endowed with intelligibility.

And precisely this happened indeed.

In a certain very warped way Bohr's interdiction of any model of a microstate protected indeed the development of the emerging Schrödinger-Heisenberg mathematical representation, and later its mutation into the nowadays Hilbert-Dirac reformulation. But on the other hand this interdiction led to hidden violations of certain laws of human thought that – remarkably – *do* irrepressibly work inside the human constructive processes of conceptualization. And this entailed non-intelligibility of the achieved formalism. Moreover, it nourished a hidden inner vice. Namely, de Broglie's 'corpuscular-wave' model that had triggered Schrödinger's contributions, though rejected by Bohr's positivistic philosophical diktat, *remained quite essentially involved in the quantum mechanical formalism as well as in the current language that accompanied its manipulations*. But it remained there in an only minored way, hidden in mathematical forms and superficially utilized words. It subsisted devoid of an explicitly formulated semantic status. In consequence of this – up to this very day – *de Broglie's model keeps acting most fundamentally inside the formalism without being offered to overt control and optimization*. This sort of imbalance leads always to unpredictable problems.

In the case considered here this circumstance led to the occultation, inside the quantum mechanical formalism, of also many other features, factual, operational and conceptual, that *do act* irrepressibly but without being mutually distinguished, named, and genuinely dominated from a semantic point of view.

> The most massive such occultation is that of the radical difference of nature and role between *individual* representations and *statistical* ones. This occultation emerged already inside Louis de Broglie's Thesis as well as in Schrödinger's prolongation of it.

Therefore since 90 years our representation of microstates nourishes endless waves of questionings that systematically are pulverized by a paradoxical negative dike of *absence of criteria* for defining the exact *meaning* and the adequacy of this or that mathematical representation. In this conceptual climate the mathematical representations proliferated densely and they still do so, while in their core there subsists a deleterious semantic magma. Mathematics can *carry* physical meanings and they can control the ways of expressing and of connecting them correctly. But mathematics alone cannot generate all the physical meanings that they are asked to transport. These have to be established otherwise and then charged in the mathematical vehicles.

There is an urgent need *to overtly construct and organize the* ***meanings*** *from our representation of the microscopic physical entities*. We are not yet robots. We are still essentially organic human beings that need to understand in order to optimize with depth, generality and precision, in the full light of our own sort of rationality. Heaps and powders of purely algorithmic, 'technical' ad hoc solutions, amorphous with respect to rationality, do not yet fully satisfy everybody. The singularity may be near, but it is not here as yet.

> And what lacks – dramatically – for organizing meaning, is *a structure of insertion-and-reference* constructed independently of the quantum mechanical formalism and ***outside it***, that offer a clear and thorough understanding of the specific non-classical requirements of a human process of construction of a satisfactory representation of *radically non-perceptible microscopic entities*.

Only this could permit an explicit, exhaustive and coherent specification of the way in which a mathematical representation of micro-phenomena can be brought to signify in a controlled and adequate way. In the first part of this work I construct such a structure.

In the *first* part of this work I construct such a structure of insertion-and-reference. It is the very first one of this kind and it might open the way toward many others of the same type but tied with other representational aims[2].

In the *second* part, by reference to this structure, are identified the main lacunae that vitiate the nowadays quantum mechanics. Moreover some of these are immediately compensated locally. The global result is a cleansed and pre-organized ground for a new construction.

In the *third* part are outlined the main contours of *a second quantum mechanics* baptized in advance *QM2*, that is rooted in the physical factuality, is constructed bottom-up, and emerges freed of interpretation problems and fully intelligible, via a step by step explicit identification of the incorporated semantic contents and a constant control of the semantic-syntactic consistency[3]. And this entails an unexpected and major reward:

> *QM2* permits to discern a natural and – a posteriori – a rather obvious path toward the unification between nowadays microphysics and Einstein's theories of Relativity.

---

[2] For Maxwell's classical electromagnetism, because "fields" are not directly perceptible, a fully new syntax of specifically adequate field-descriptors has been independently created before the formulation of the theory itself.

[3] For the sake of effectiveness throughout the whole following work all the involved descriptional elements are posited finite, so discrete.

The approach practised in this work is not usual. So it will surprise, and many readers might feel rejected. I am deeply aware of this. But I think that a less singular way toward a fully intelligible representation of the modern microphysics cannot be constructed.

This way was not conceivable in the first phase of the attempt at a fundamental theory of the microstates. But now the time for it to burgeon and bear fruit has arrived.

# PART I

## INFRA - QUANTUM MECHANICS (*IQM*)

**ABSTRACT OF PART I**

A qualitative but *formalized* representation of the general characteristics of *any* physical theory of the microstates is developed quite independently of the quantum mechanical formalism and outside it, under exclusively the [*physically operational*-conceptual-methodological] constraints entailed by the requirement of a scientific character (consensual, predictive, and verifiable) for the descriptions of entities that – radically – cannot be perceived directly by human conceivers-observers. This representation is called infra-(quantum mechanics) and is denoted *IQM*. The specific purpose of *IQM* is to permit to overcome the ties that immobilize the minds inside a phase of the representation of the microstates that historically has been unavoidable but that nowadays subsists by inertia.

# INTRODUCTION TO INFRA-(QUANTUM MECHANICS) *IQM*

A human being who wants to construct knowledge concerning states of microsystems – let us call this 'microstates' – makes use of physical entities to which he associates this denomination, of instruments and operations, and he introduces representational *aims* and corresponding *methods* of acting and thinking. Thereby the human observer introduces severe constraints that structure the process of construction of knowledge. It is not possible to preserve the process from such constraints. They are precisely what 'forms' it. Nor is it possible to eliminate a posteriori the effects of theses constraints from the constructed knowledge, these are essentially incorporated to the achieved form to which they have led. *Any* piece of knowledge is a construction and this construction remains irrepressibly relative to its whole genesis. So, if the observer-designer wants to stay in control of the knowledge that he has generated, to be able to understand and to freely optimize it – he has to be thoroughly aware of the conceptual-operational-methodological weft of this knowledge.

> In what follows – quite independently of the mathematical formalism of quantum mechanics – is elaborated a structure of the necessary and sufficient features of ***a GENERAL procedure, a general METHOD for creating "scientific knowledge" on the microstates**, **so on physical entities that are radically non-perceivable directly by the human biological sensorial apparatuses***.

In this Part I of the present work we do ***not*** yet research a definite *mathematical* representation. In our view and in terms of the language practised throughout this work[4] a mathematical representation tied with knowledge on physical entities – *alone*, in the absence of an explicitly defined structure of conceptual-*physical* operations for generating the factual data tied with that mathematical representation – is ***not*** "scientific knowledge on physical entities", it is not what we call a "scientific theory belonging to the Mathematical *Physics*". It is just a proposed expression mentally tied with the above-mentioned structure of operations and factual data.

So the data and the operational structure that generates them – a conceptual, physical, methodological operational structure – come *first*. And in the case of microstates it is particularly important to state this from the outset, for the following reasons.

[4] Cf. the glossary from the end of the Part 1 of this work.

With respect to the processes of construction of the classical conceptualization, the procedural structure researched here involves a deliberate change of the ***origin*** on the vertical of conceptualization, of the processes of construction of consensual knowledge: Deliberately, these processes will be started from *beneath* the extreme limit between [what has been drawn before inside the volume of the already initiated actions of conceptualization] and [what we imagine to still remain non-singularized inside the strictly a-conceptual universal physical substance]. Thereby *the **order** of conceptual constructability is inversed* and this entails a fundamental change in also the ***content*** of the classical concept indicated by the historically introduced word 'microstate':

> This content transmutes into that of ***a factually defined concept***, because at the newly chosen origin purely verbal definitions in the classical sense cannot be realized any mor*e*. And it will appear that ***this***, the factual character of the modified definition of a microstate, is what entails the emergence of the famous 'problem' of the "*essentially* probabilistic character" of the modern microphysics.

Nevertheless I have kept in use, unchanged, the same word – 'microstate' – because, paradoxically, it brings into maximal evidence the genesis of the pivotal feature called "essentially probabilistic character". Indeed it will appear that the word 'microstate' introduces a key-connection, and so comparability, with the classical top-down historical progression of the scientific conceptualization toward microscopic space-time dimensions, such as it has been achieved inside the first, classical phases of the initial formulations of molecular and atomic physics. Thereby this word entails intelligibility. And so the basic misunderstandings that since a century plague Physics become perceptible and they dissolve into the outline of a new global framework for generating scientific knowledge on factually defined microstates.

The approach proposed in the first part of this work is structured in qualitative terms, but in *explicit, **formalized***[5] ***and finite, effective terms***. The result is called in advance *Infra-(Quantum Mechanics)* and is denoted *IQM*.

I would like to convey to the reader from the start what follows.

*Nothing* – throughout the construction elaborated below – is conceived as an assertion of "intrinsic *truth*". The whole construction consists of just a succession of conceptual-(***physically operational***)-***methodological*** steps each one of which is imposed with necessity by:
- The ***global aim*** to construct a guiding structure for the elaboration of a satisfactory scientific representation of physical entities that – radically – *cannot be directly perceived*.
- The ***local aim*** of the considered step.
- The corresponding ***cognitive situation***.

[5] We employ the word 'formalized' in the sense that the posits are explicitly stated; all the specific basic terms are endowed with explicit and finite definitions; and the elements introduced in this explicit way are constructed as general and syntactically related void loci for receiving in them particular unspecified semantic data. Furthermore, once posited or constructed, the elements are explicitly connected in full agreement with the current syllogistic.

In order to instil maximal intelligibility, each step from the mentioned structure is explicitly referred to our classical thought-and-languages that have emerged and settled inside our minds by interactions with entities that *are* perceived. But on the other hand each constructive methodological step achieved in this work transgresses our classical forms of thought by definite features commanded by the radical novelty of the cognitive situation entailed by the global aim specified above. Each such transgression is patiently explicated, fixed by definitions, named and denoted.

> *IQM* is the global procedural whole that is obtained when the constructive methodological steps indicated above are put together under the constraint of logical coherence. *IQM* is an explicitly and strictly defined procedural, operational reference-and-hosting-structure specifically appropriated for assisting the elaboration of ***any*** scientific theory of *factually* defined micro-entities.

I hold that in the absence of such a structure it simply is not possible to construct a fully intelligible scientific theory of the microstates belonging to the Mathematical Physics: I hold that unavoidably a theory of the microstates has to be a theory of the factually defined microstates and that such a theory has to obey the procedural structure called here *IQM*.

*IQM* is a strict application of a general method for constructing human scientific knowledge, namely the Method of Relativized Conceptualization, *MRC* (MMS [2002A], [2002B], [2006]). This method offers the general framework for constructing all the "infra-disciplines" of any scientific theory of physical entities; and the Infra Quantum Mechanics *IQM* is the very first such infra-discipline that has been actually achieved. In the parts II-VIII of this work it will appear how this achievement leads to a Second Quantum Mechanics *QM2* that is fully intelligible as well as to a *physical* theory of probabilities.

The genesis of the second Quantum Mechanics developed by use of the Infra-Quantum Mechanics that is constructed accordingly to the general Method of Relativized Conceptualization, draws into evidence a *fundamental methodological unity between* ***all*** *the domains of the modern physics; but it also illustrates that* ***any given sort of cognitive situation entails specific*** ***representational features***. And this is what calls forth on the horizon the first contours of a Unified Modern Physics where the theory of gravitation and the microphysics are inserted into a same basic methodological scheme, while the unavoidable relativities to the operational cognitive actions imposed by two very different cognitive situations, entail two *very different final conceptual-mathematical representations*. It thereby becomes clear that the constantly increasing distance, in modern Physics, between a direct sensorial perceptibility of what is represented, and the constructed mathematical representation, *obliges to relativize with method and rigor* **the whole way of constructing scientific, consensual knowledge**, *if one wants to achieve a fully intelligible mathematical representation*. Einstein's theories of "Relativity" are very insufficient for reaching the aim of general intelligibility of all the theories from the Mathematical Physics. The final representation of a given domain of physical facts is fully intelligible only if the *whole* semantic structure carried by it stays exposed, controllable and open to optimization inside each constructive step; and this, inside each phase of the process of conceptualization, requires explicit and exhaustive relativization to *each* sort of descriptive element that is brought into play.

This fact, implicitly, burgeons already into attention everywhere inside the sciences of matter. By the Method of Relativized Conceptualization *MRC* and the Infra-(Quantum Mechanics) *IQM* I have tried to take this ubiquitous emergence into account and, for the particular case of microstates, to construct an expression of this essential fact, as explicit, exhaustive and coherent as I have been able to achieve.

# PROLOGUE

What follows is the original French text of an extract from the volume "*Einstein 1879-1955 (6-9 juin 1979), colloque du centenaire, Collège de France*, Editions du Centre National de la Recherche Scientifique". It is useful here for reminding of the 1979-state-of-mind concerning the fundamental problems in Quantum Mechanics: it is striking to what a degree *this same state of mind still persists today*.

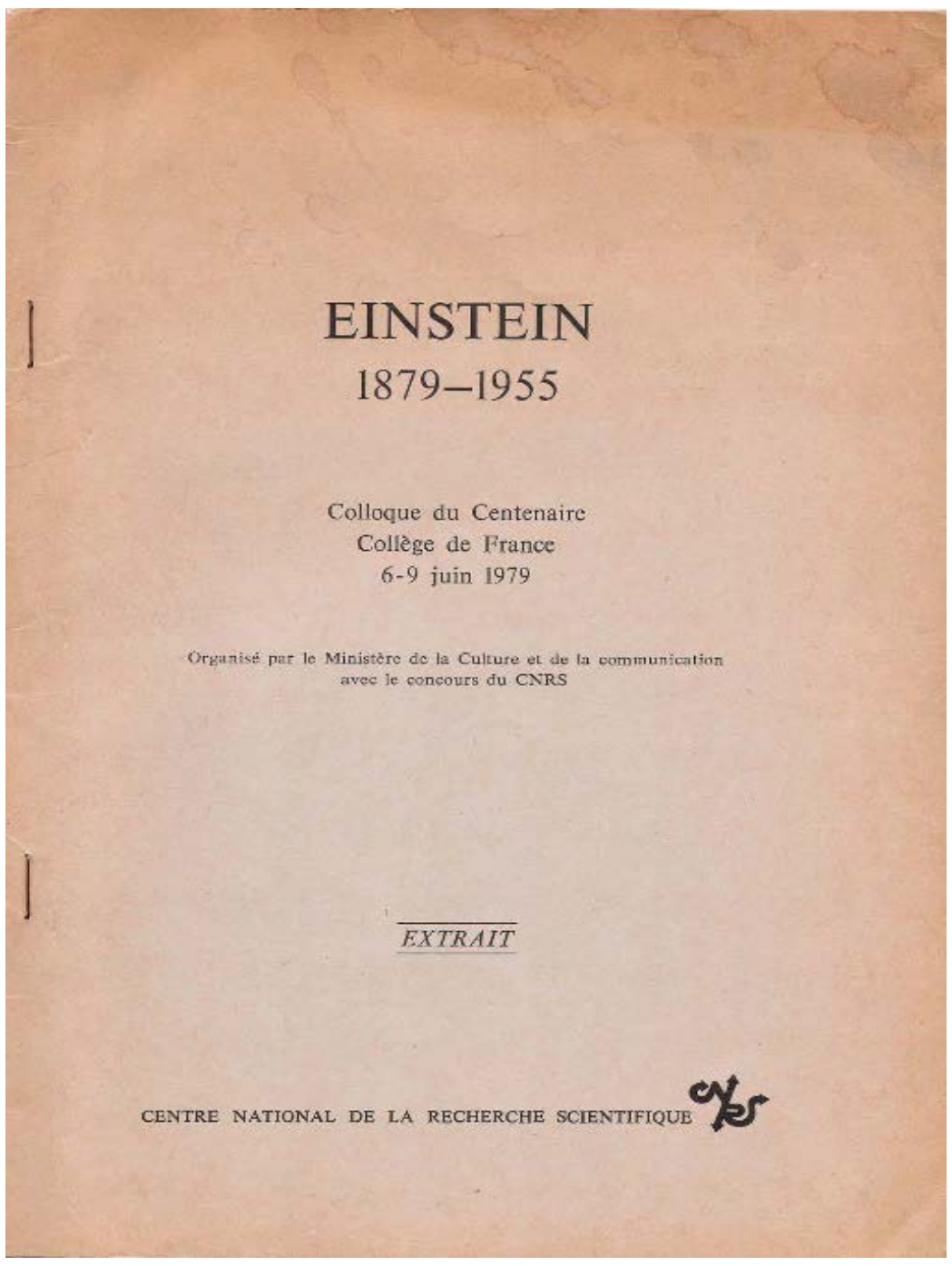

**(EXTRAIT)**

# REFLEXION SUR LE PROBLEME DE LOCALITE

**M. Mugur–Schächter**

**UNIVERSITE DE REIMS**
**B.P 347 51062 REMS CEDEX**

## But

Depuis huit ans ce que l'on appelle le problème de localité retient de plus en plus l'attention. Des théoriciens, des expérimentateurs, des penseurs pluridisciplinaires investissent des efforts importants pour élucider ce problème. Les aspects techniques – mathématiques et expérimentaux – ont été déjà examinés dans un grand nombre de travaux et ils sont bien connus de ceux qui font à ce sujet des recherches spécialisées. Mais la configuration conceptuelle qui est en jeu me paraît avoir des contours beaucoup moins définis. Le but de l'exposé qui suit est d'examiner cette configuration conceptuelle. J'essaierai de procéder à cet examen d'une manière aussi simple et frappante que possible, presque affichistique, à l'aide de schémas et de tableaux. Ces moyens me paraissent être les plus adéquats pour donner le maximum de relief aux insuffisances que je perçois dans la définition même du problème de localité.

## Bref rappel

***Le paradoxe EPR (I935)***. Le problème de localité est soulevé par un théorème bien connu de J. Bell (1) qui se rattache à un raisonnement formulé en 1935 par Einstein, Podolsky et Rosen (2). Ce raisonnement, connu sous la dénomination de "paradoxe EPR", et été construit pour démontrer que le formalisme de la Mécanique Quantique ne fournit pas une description complète des microsystèmes individuels. Les hypothèses qui constituent la base de départ du paradoxe EPR sont indiquées dans le tableau suivant (où des notations abrégées leur sont associées):

| | |
|---|---|
| Toutes les prévisions de la Mécanique Quantique sont vraies. | $\forall$ MQ |
| La Mécanique Quantique fournit une description complète des microsystèmes. | C (MQ) |
| La réalité physique existe indépendamment de l'observation. Elle est "déterministe" et locale (ou"séparable")* | $\exists$(r.d.l.) |

Le "paradoxe EPR" consiste dans la démonstration du fait que les hypothèses énumérées ne sont pas compatibles.

L'interprétation proposée par Einstein, Podolsky et Rosen, de cette démonstration, a été la suivante:

Les prévisions du formalisme quantique se montrent correctes. Il n'existe donc aucune base pour abandonner l'hypothèse ∀MQ.

L'hypothèse ∃(r.d.l.) exprime un credo métaphysique que l'on est libre d'accepter ou de rejeter. Mais *si* on l'accepte, alors il faut l'adjoindre aux prévisions de la Mécanique Quantique. En ce cas la démonstration de l'incompatibilité du système d'hypothèses [∀MQ + C(MQ) + ∃(r.d.l.)] oblige à abandonner hypothèse de complétude C(MQ). En d'autres termes cette démonstration oblige alors à accepter la possibilité d'une théorie déterministe et locale (TDL) des microphénomènes, où le formalisme quantique sera complété par des éléments descriptifs additionnels, des paramètres cachés (par rapport au formalisme quantique) déterministes et locaux (p.c.d.l.) qui permettent d'accomplir une description complète des microsystèmes individuels. Cette description complète fournie par TDL doit être compatible avec la Mécanique Quantique – en vertu de l'hypothèse ∀MQ – et avec la Relativité, en vertu de l'hypothèse ∃(r.d.l.) qui se trouve intégrée dans la théorie de la relativité. Cette structure d'idées peut être représentée par le schéma suivant:

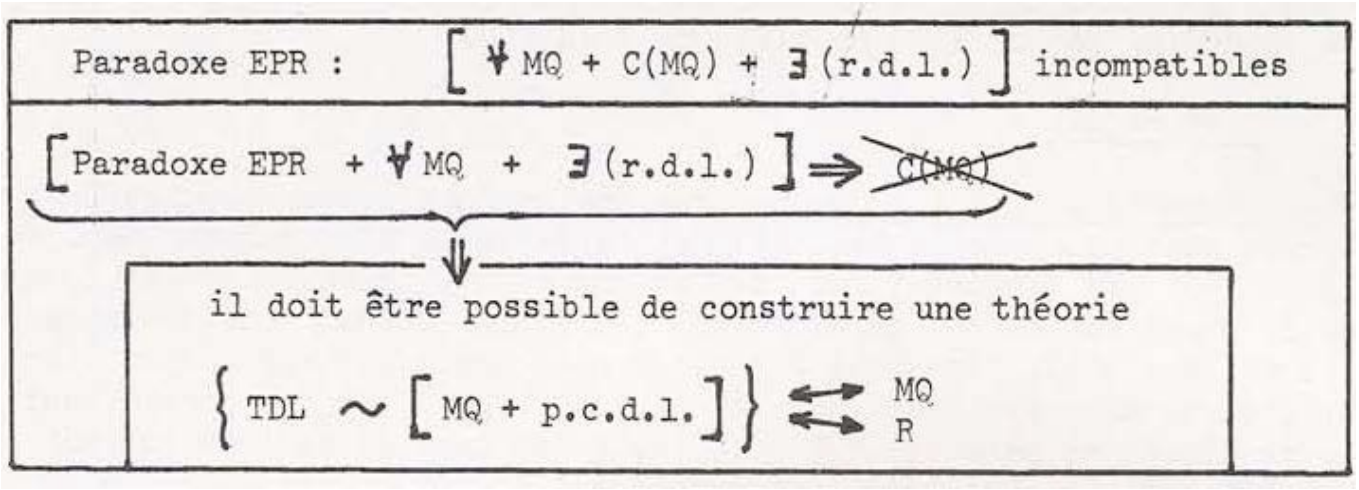

***Les réactions pendant 30 ans.*** Les réactions ont été diverses. Pourtant la note dominante a été nettement celle du positivisme: l'hypothèse "réaliste" ∃(r.d.l.) est dépourvue de toute signification opérationnelle. Elle est donc essentiellement métaphysique, extérieure à la démarche scientifique. L'incompatibilité dénommée "paradoxe EPR" n'existe que par rapport à cette hypothèse non scientifique, et donc elle ne constitue pas un problème scientifique. Pour la science il s'agit là d'un faux problème.

***Le théorème de J. Bell (1964).*** Trente années plus tard J.Bell a démontré un théorème qui semble contredire l'interprétation associée par Einstein Einstein, Podolsky et Rosen à leur propre démonstration. La conclusion du théorème de Bell peut s'énoncer ainsi (ou de manières équivalentes): il n'est pas possible, à l'aide de paramètres cachés déterministes et locaux, d'obtenir dans tous les cas les mêmes prévisions que la Mécanique Quantique ; en certains cas, de tels paramètres conduisent à d'autres prévisions. Si alors on veut rétablir l'accord avec les prévisions de la Mécanique Quantique, il faut supprimer le caractère local des paramètres cachés introduits, ce qui contredira l'hypothèse ∃(r.d.l.), que la théorie de la Relativité incorpore. Par conséquent la théorie déterministe TDL compatible à la fois avec la Mécanique Quantique et la Relativité, dont Einstein Podolsky et Rosen ont cru avoir établi la possibilité, est en fait impossible.

La démonstration repose sur la production d'un exemple. On considère deux système $S_1$ et $S_2$ à spins non nuls et corrélés, créés par la désintégration d'un système initial S de spin nul. On envisage des mesures de spin sur $S_1$ selon trois directions a, b, c, à l'aide d'un appareil $A_1$, et des mesures de spin sur $S_2$ selon ces mêmes directions, à l'aide d'un appareil $A_2$ qui peut se trouver à une distance arbitrairement grande de $A_1$. *L'hypothèse ∃(r.d.l.) est ensuite formalisée*: des paramètres cachés sont introduits et ils sont soumis à des conditions telles qu'elles fournissent une traduction mathématique des qualifications de "déterministes" et "locaux". Ainsi la conceptualisation introduite auparavant au niveau d'une sémantique claire, mais qualitative, est élevée jusqu'à un niveau sémantique syntaxisé. Un tel pas est souvent important, car il peut permettre des déductions mathématiques à conclusions quantitatives. Et en effet Bell a démontré que l'hypothèse ∃(r.d.l.) ainsi formalisée entraîne nécessairement une certaine inégalité concernant les corrélations statistiques entre les résultats de mesures de spin enregistrés sur les appareils $A_1$ et $A_2$. Or, cette inégalité n'est *pas* satisfaite par les corrélations statistiques prévues par la Mécanique Quantique. On pourrait retrouver les corrélations quantiques en supprimant la condition qui traduit mathématiquement le caractère "local" des paramètres cachés introduits, c'est-à-dire en renonçant à une partie de l'hypothèse ∃(r.d.l.). On exprime ceci en disant que, dans la circonstance considérée, "la Mécanique Quantique est non-locale" ou "implique des effets non-locaux" qui la rendent incompatible avec ∃(r.d.l.). Schématiquement, on peut résumer l'apport de Bell ainsi (en notant $(p.c.d.l.)_B$ les paramètres cachés soumis aux conditions de Bell).

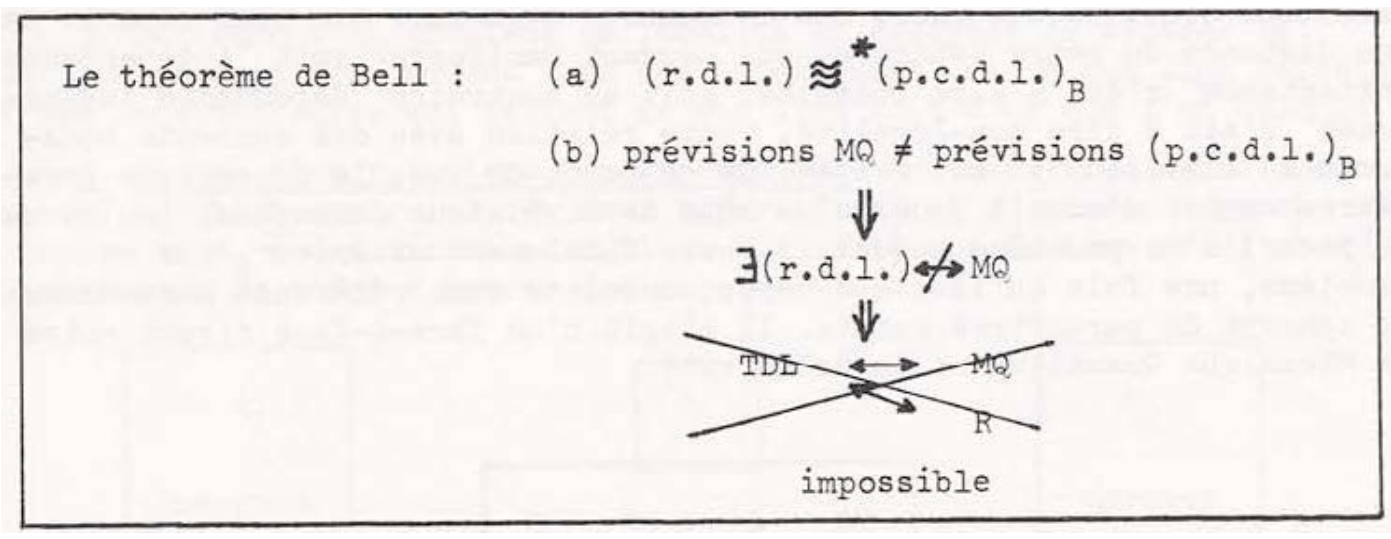

Comme les statistiques dont il s'agit sont observables, il est en principe possible d'établir expérimentalement si les faits physiques correspondent aux prévisions de la Mécanique Quantique ou à celles entraînées par les paramètres cachés déterministes et locaux au sens de Bell. C'est l'un des traits les plus forts du théorème de Bell.

Si l'expérience infirmait la Mécanique Quantique, la situation conceptuelle créée paraîtrait claire. On devrait admettre la possibilité d'une théorie déterministe et locale des microphénomènes, mais différente de celle envisagée par Einstein, Podolsky et Rosen, car elle n'obéirait pas à l'exigence d'identité prévisionnelle avec la Mécanique Quantique, pour tous les cas.

Mais un certain nombre d'expériences de vérification a déjà été fait et il se trouve que les résultats obtenus à ce jour – bien qu'ils ne tranchent pas encore définitivement – étayent fortement la supposition que la prévision de la Mécanique Quantique s'impose comme correcte.

Il s'agit donc de comprendre la situation conceptuelle qui semble s'établir et que l'on dénomme "problème de localité".

**Interprétations**

Le problème de localité est ressenti diversement. Je distinguerai en gros trois interprétations, en omettant ou en bousculent beaucoup de nuances.

***I- Interprétations de refus***. Un certain de nombre de physiciens semble considérer cette fois encore qu'il s'agit d'un problème métaphysique qui n'existe que par rapport au concept non opérationnel de paramètre cachés, mais qui se dissout dès qu'on refuse ce concept. D'autres physiciens considèrent que le problème n'existe parce qu'il est faussement posé (3).

***2- Interprétation minimale***. Selon d'autres physiciens (4), (5), (6), (7), etc…., le problème satisfait cette fois aux normes positivistes les plus draconiennes, parce qu'il conduit à des testes expérimentaux. Toutefois, ils refusent de conceptualiser au-delà de ce que ces tests mettent en jeu. Ils ne prennent en considération strictement que des corrélations statistiques entre des évènements de mesure qui sont séparés par une distance du genre espace et qui peuvent manifester soit "indépendance instantanée" c'est-à-dire localité, soit au contraire "dépendance instantanée" c'est-à-dire non-localité. Toute relation avec des concepts sous-jacents "explicatifs" est évitée. De ce point de vue, le concept de paramètres cachés n'aurait qu'un rôle de révélateur conceptuel (ou de catalyseur) d'un problème auquel il reste finalement extérieur. Car ce problème, une fois qu'il a été perçu, subsiste sans référence nécessaire au concept de paramètres cachés. Il s'agit d'un face à face direct entre la Mécanique Quantique et de Relativité.

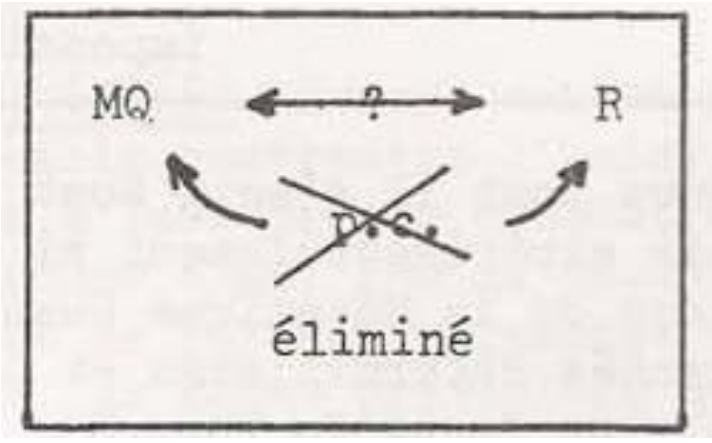

*3- **L'interprétation épistémologique***. Il existe enfin une tendance (8) à connecter le problème de localité à notre conceptualisation la plus courante de la réalité, qui postule l'existence d'objets isolés possédant des propriétés intrinsèques et permanentes. La violation des inégalités de Bell serait incompatible avec ces suppositions. Il s'agirait donc en dernière essence d'un face-à-face entre la Mécanique Quantique et – à travers le concept de paramètres cachés et à travers la Relativité – des postulats épistémologiques fondamentaux.

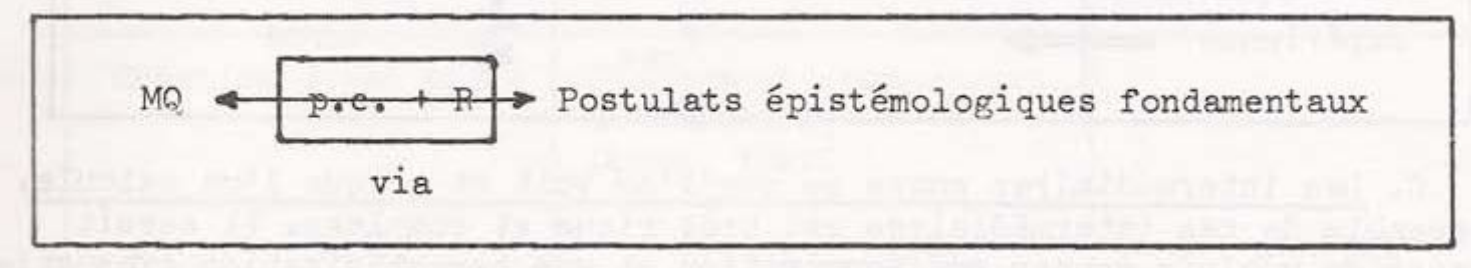

Je n'examinerai pas l'interprétation de refus, car elle ne peut conduire à aucun élément nouveau.

Quant aux deux face-à-face impliqués par les deux autres interprétations, aucun d'eux ne me semble s'imposer dans la phase actuelle du débat. Seule une question ressort clairement:

Qu'est ce qui est en jeu – au juste – dans le problème de localité?

L'examen qui suit montrera que, pour fixer une réponse, les conceptualisations existantes et les tests sur l'inégalité de Bell ne peuvent pas suffire. Inévitablement d'autres conceptualisations encore, et les tests correspondants, devront être abordés. Sinon, aucune conclusion définitive ne pourra être tirée, même si l'inégalité de Bell est clairement violée.

**Le problème de localité et le terrain conceptuel sous-jacent**

Reconsidérons le problème de localité en essayant de séparer ce que l'on perçoit directement lors des expériences, de ce que l'on calcule, et des intermédiaires qui relient ce que l'on voit à ce que l'on calcule.

***A. Ce qu'on voit lors des expériences***. On voit (tous les détails mis à part) un objet central $A_o$ et deux appareils $A_I$ et $A_2$ placées à gauche et à droite de $A_o$ à des distances égales. Sur certaines parties de $A_I$ et $A_2$ apparaissent de temps à autres des marques visibles.

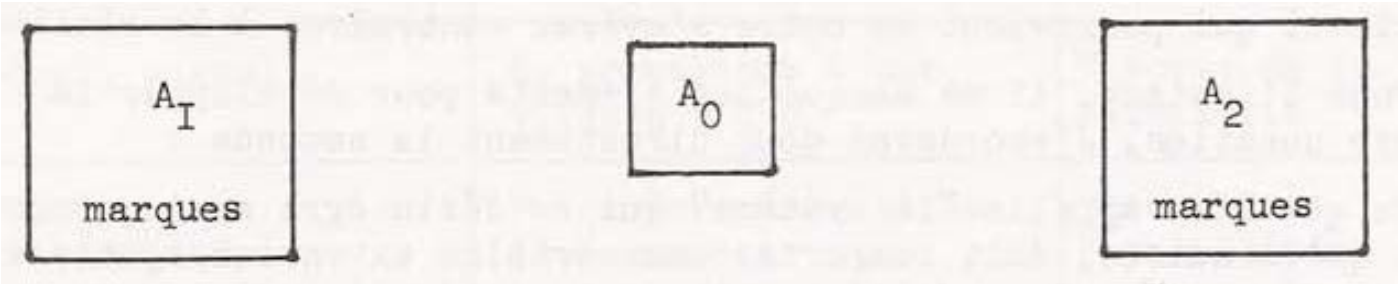

***B. Ce qu'on calcule***. On calcule des corrélations statistiques en employant trois sortes de distributions de probabilités conduisant à trois fonctions de corrélation, une fonction $F_{(TDL)B}$ caractéristique d'une théorie déterministe locale au sens de Bell, une fonction $F_{MQ}$ obéissant aux algorithmes de la Mécanique Quantique, et une fonction $F_{obs}$ correspondant aux statistiques observées. L'inégalité de Bell distingue $F_{(TDL)B}$ de $F_{MQ}$. L'expérience doit montrer si la réalité observée reproduit $F_{MQ}$ ou $F_{(TDL)}$.

| | |
|---|---|
| | $F_{(TDL)_B}$ , $F_{MQ}$ , $F_{obs}$ |
| Inégalité de Bell $\Longrightarrow$ | $F_{(TDL)_B} \leqslant$ borne , $F_{MQ} >$ borne |
| Expérience $\Longrightarrow$ | $F_{obs} \overset{?}{\gtrless}$ borne |

***C. Les intermédiaires entre ce qu'on voit et ce qu'on calcule***. L'ensemble de ces intermédiaires est très riche et complexe. Il serait insensé de vouloir donner une énumération et une caractérisation déterministes et locaux de Bell violent la pudeur sémantique dictée par le positivisme. Alors autant aller jusqu'au bout et avouer l'ensemble des questions sémantiques liées aux interprétations 2 et 3 du problème de localité telles que je les ai distinguées plus haut.

Je commence par l'interprétation minimale. Je perçois deux questions.

En premier lieu, les contenus sémantiques assignés aux qualificatifs "déterministes" et "locaux", tels qu'ils sont impliqués par la modélisation mathématisée de Bell, permettent-ils la représentation la plus générale concevable d'un processus d'observation d'un "microétat" à l'aide d'un "appareil" macroscopique?

En second lieu, en supposant que la modélisation de Bell d'un processus d'observation n'introduit vraiment aucune restriction non nécessaire, quelle sorte de non-localité, exactement, la violation des inégalités de Bell démontrerait-elle? La non-localité que la théorie de la Relativité interdit clairement, ou bien des prolongements spontanés et encore flous de celle-ci qui pourraient en outre s'avérer contraire à la réalité?

Pour l'instant, il me manque les éléments pour développer la première question. J'aborderai donc directement la seconde:

Ce qu'on appelle "le système" qui se désintègre en $A_o$, pour autant qu'il existe, doit comporter une certaine extension spatiale non nulle de départ $\Delta x_s(t_o)\neq 0$ (ce qui peuple ce domaine d'espace, est-ce un "objet" ou un "processus", ou les deux à la fois? les définitions même manquant pour répondre). Ce qu'on désigne par les termes "désintégration" ou "création d'une paire $S_1$ et $S_2$", comment le concevoir? Les mots indiquent dans le substrat conceptuel l'hypothèse d'un processus, d'une entité réelle en cours de changement. Pour exister, ce processus doit se produire quelque part et il doit durer, il doit occuper un certain domaine non nul d'espace-temps $\Delta s_c(t).\Delta t_c \neq 0$ (l'indice c: création) à l'intérieur duquel "le système de départ S" existe encore mais change, cependant que $S_1$ et $S_2$ n'existent pas encore mais se forment.

| MOTS | CONCEPTS | ORGANISATIONS SYNTAXIQUES |
|---|---|---|
| 1 système | Macroobjet, objet | Logique des classes d'objets et des prédicats |
| Création d'une paire | Changement, processus, SUCCESSIONS, DUREE, TEMPS | ? |
| 2 systèmes corrélés | Objets ⇄ corrélés / isolés | |
| Appareil | Macroobjet capable de réagir d'une manière significative avec un microsystème | |
| Spin | Propriété d'observation | |
| Mesure | Processus d'observation<br>Evénement de connexion entre microobjets et macroobjets | |
| Paramètres cachés | Propriétés intrinsèques et permanentes (d'objets ou de processus) | |
| Déterministe | Prévisible?<br>A partir de quoi? | |
| Local, signal | Se propageant à une vitesse $v < c$ | Théorie de la Relativité |
| Statistiques Probabilités | Phénomènes aléatoires (événements, processus, objets) | Théorie des Probabilités |
| Prévisions quantiques | Vecteurs d'états, algorithmes quantiques | Mécanique Quantique |
| Influence à distance | Changement transporté à vitesse $v < c$ | ? |

Dans l'écriture qui désigne ce domaine d'espace-temps, le facteur de durée $\otimes t_c = t_{12,\,o} - t_o$ s'étend – par définition – d'une certaine "valeur initiale de temps" $t_o$ où le changement de création commence, jusqu'à une "valeur finale de temps" $t_f \equiv t_{12,\,o}$ à partir de laquelle "la paire $S_1$, $S_2$ de systèmes corrélés" commence à exister (des objets? des processus eux aussi? les deux à la fois?). Quand au facteur d'extension spatiale $\Delta s_c$ (t), il semble obligatoire de concevoir, puisqu'il s'agit d'un processus, qu'il change en fonction de la "valeur de temps" t, avec ($t_o < t < t_f$ ), en restant toutefois métastablement connexe tant que $t < t_f$ (c'est-à-dire tant que S subsiste encore et que $S_I$ et $S_2$ ne sont pas encore créés). Pour tout $t > t_f$, toutefois, ce domaine spatial devrait être devenu non connexe *via* une scission plus ou moins "catastrophique" conduisant à cette nouvelle forme de stabilité à laquelle on rattache l'expression "la paire $S_I$, $S_2$ de deux systèmes corrélés".

Je m'arrête un instant et je regarde ce que je viens d'écrire. Quel mélange de "nécessités" et d'arbitraire, de signes et de mots qui ont l'air de pointer vers un désigné précisé et sous lesquels pourtant on ne trouve que des images floues et mouvantes accrochées à ces mots et ces signes de manière non séparée. J'écris entre guillemets "valeur de temps", par exemple, parce qu'à chaque fois que je réfléchis au degré d'inexploration où se trouvent encore les concepts de durée et de temps et leur relation, je ressens une réticence à écrire quoi que ce_soit en dehors d'un algorithme qui fixe une règle du jeu. Le paramétrage de la propriété fondamentale de durée à l'aide de la variable de temps t, telle que cette paramétrisation est pratiquée dans les théories existantes et même dans la Relativité, est encore certainement très simplificatrice et souvent falsificatrice, rigidifiante, mécanisante en quelque sorte. Les changements ne sont pas toujours des déplacements d'entités stables intérieurement. Pour pouvoir rendre compte pleinement de l'entière diversité des types et des intensités de changements, il faudrait une sorte de grandeur vectorielle, un *champ* de temps processuel défini en chaque point de l'espace abstrait encadré par l'axe de durée et par les axes des changements envisagés.

Mais un tel temps se transformerait-il selon Lorentz? Quel rôle joue la vitesse d'un "signal" lumineux face aux vitesses de propagation "d'influences" (?) dans un tel espace processuel? Qu'est ce que la Relativité impose véritablement aux processus *quelconques* et qu'est-ce qu'elle laisse en blanc? Lorsqu'il s'agit de processus très "intenses" localement, "catastrophiques", comme l'est probablement la "création d'une paire", que devient "le temps"?

En théorie relativiste générale de la gravitation, par exemple, un gradient non nul du champ de gravitation est lié à une impossibilité de définir un temps unique, pour les observateurs d'un même référentiel, si ces observateurs sont spatialement distants l'un de l'autre. Quand à l'invariance de la vitesse *de la lumière* elle-même (et non la vitesse d'autres sortes "d'influences") lorsqu'on passe d'un référentiel à un autre, elle n'est postulée que localement, car il n'existe aucune définition uniforme des distances et des temps dans des champs gravitationnels variables (9) (espace-temps courbes). Comment savoir quelle sorte de "courbure" locale de l'espace-temps est produite (ou non) par un processus – essentiellement variable – de création d'une paire?

Enfin, la Relativité n'introduit aucune quantification au sens de la Mécanique Quantique, sa description est continue. Lorsqu'on écrit [vitesse = distance/temps], le temps est un paramètre continu.

Si ensuite on se demande comment on trouve la valeur de t, on s'aperçoit qu'elle est de la forme $NT_H$ où N est un entier et $T_H$ une période "d'horloge" (supposée constante !) ce qui ramène au discret. En macroscopie cela peut être négligeable aussi bien sur le plan du principe que sur le plan numérique. Mais lorsqu'on considère des processus quantiques et relativement très brefs, quel est le degré de signifiance d'une condition comme

$$v = \frac{\text{distance}}{\text{temps}} = \frac{\text{distance}}{NT_H} = \text{const?}$$

Quelle horloge faut-il choisir, avec quel $T_H$, et comment par ailleurs s'assurer que lorsqu'on écrit $\otimes t = 10^{-x}$, on fait plus qu'un calcul vide de sens?

On comprend, devant de telles questions, les prudences positivistes et les normes qui conseillent de se maintenir dans la zone salubre de l'opérationnellement défini et du syntaxisé, où la pensée circule sur des voies tracées et consolidées. Au dehors, on s'enfonce dans une véritable boue sémantique. Pourtant ce n'est que là, dans cette boue, et lorsqu'on force le regard à discerner les formes mouvantes, que l'on peut percevoir du nouveau. De toute façon le problème de localité nous y force: c'est un problème très fondamental où tout comportement inertiel, inanalysé ou approximatif, conduit inéluctablement à l'arrêt de la capacité de raisonnement, ou à des problèmes et perspectives illusoires. On ne peut pas cette fois suivre un chemin parce qu'il est construit. On est obligé de choisir et de suivre la direction qui convient.

Je reviens donc sur la création d'une paire corrélée $S_1$, $S_2$. J'imagine ce processus comme ayant des analogies avec la formation de gouttes. (Ceci peut être faux, mais ce n'est pas *a priori* impossible, et *on n'a besoin que d'un exemple de possibilité*). Je dessine donc ainsi la projection spatiale (en deux dimensions) du domaine d'espace- temps $\otimes s_c(t).\otimes t_c$, $t_o < t < t_f$, pour 4 époques:

* $t = t_o$;

* $t_o < t < t_f$

* $t_o < t < t_f^-$ (c'est-à-dire immédiatement *avant* $t_f$) ;

et $t = t_f^+$ (c'est-à-dire immédiatement *après* $t_f$)

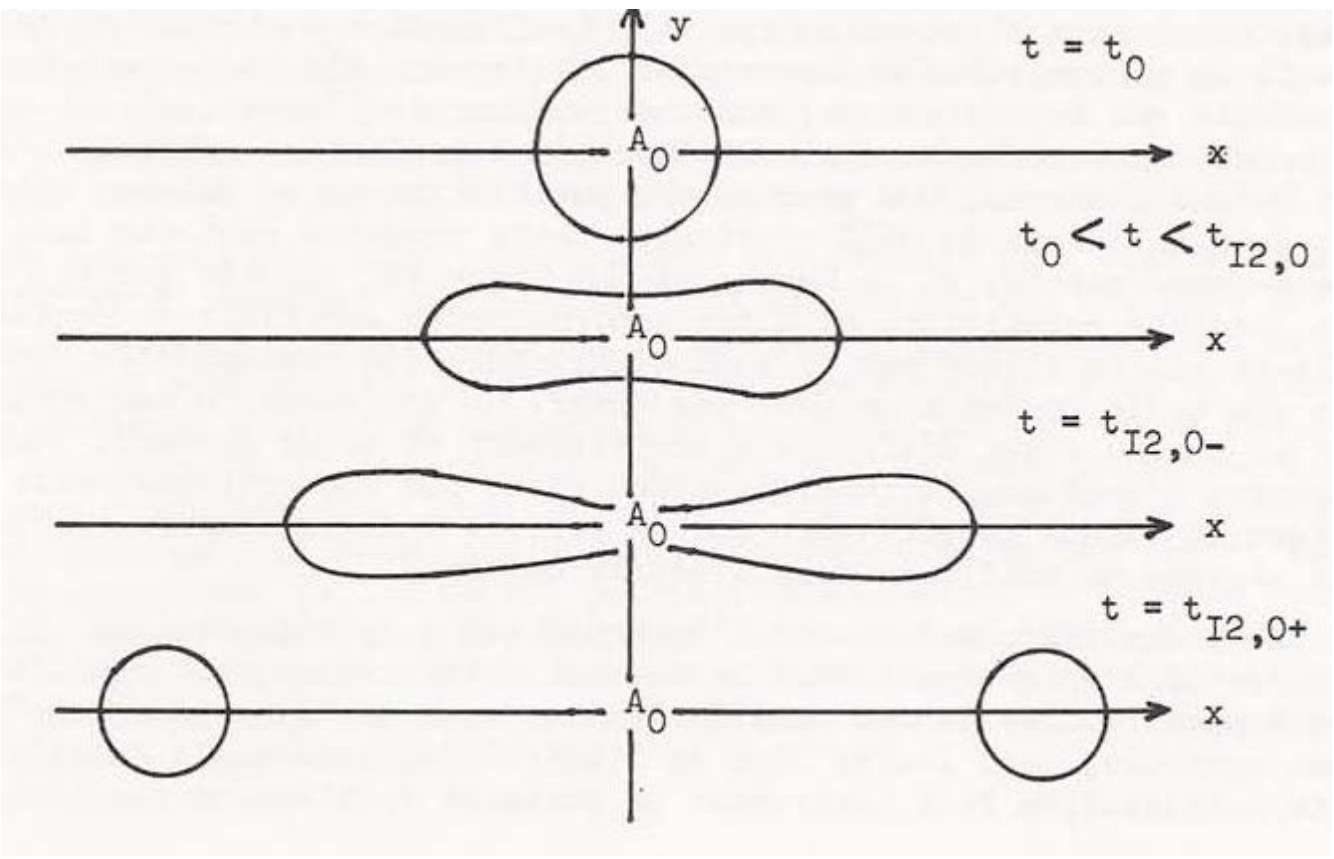

Supposons maintenant que la distance $d_{12}$ entre les appareils $A_1$ et $A_2$ est plus petite que la projection maximale de $\otimes s_c(t)$ sur l'axe x correspondant à $t = t_f^-$.

Les appareils $A_1$ et $A_2$ seront donc atteints non pas par "$S_1$" et "$S_2$" respectivement, mais par "S en cours de désintégration", qui *peut* néanmoins déclencher des enregistrements sur $A_1$ et $A_2$. Supposons encore que la durée des évènements de mesure se trouve être telle par rapport à $d_{12}$, que la distance d'espace-temps entre les événements de mesure soit du genre espace. Enfin, supposons que, en dépit de tout cela, les évènements de mesure ne soient pas indépendants au sens de Bell, c'est-à-dire supposons qu'un changement de $A_2$ puisse agir à une vitesse $v > C$ sur le résultat de l'un des enregistrements de $A_2$. Les statistiques de résultats d'enregistrements sur $A_1$ et $A_2$ seront alors "non localement corrélées" et l'inégalité de Bell sera violée. Mais serait-il en ce cas justifié de conclure qu'on a démontré une contradiction avec la théorie de la Relativité? La théorie de la Relativité ne statue *que* sur des "signaux" (quelle est exactement la définition?) se propageant "dans le vide". Elle ne statue rien du tout concernant la transmission "d'influences" (définition?) à travers un "système" (objet? processus?). En particulier, elle n'impose rien du tout concernant "l'ordre temporel" (?) ("causal" ou "non causal") (?) d'événements placés à des endroits spatiaux différents d'"un même système". L'exemple imaginé – un modèle de "création d'une paire" – n'appartient tout simplement pas au domaine de faits que la Relativité décrit. Aucune théorie constituée ne le décrit encore. Pourtant cet exemple, quelles que soient ses inadéquations face à la réalité inconnue, caractérise certainement d'une manière en essence acceptable ce qui mérite la dénomination de processus de création d'une paire: un tel processus doit occuper un domaine non nul d'espace-temps, dont la projection spatiale, connexe au départ, évolue, devenant non connexe.

Cet exemple de possibilité me semble suffire comme base pour la conclusion suivante: les tests destinés à vérifier l'inégalité de Bell, même s'ils violaient définitivement l'inégalité, ne pourront jamais établir à eux seuls que le principe einsteinien de localité a été enfreint. Pour préciser ce qui est en jeu, la modélisation de Bell et le test correspondant devront être associés à d'autres modélisations et à d'autres tests, concernant l'extension d'espace-temps des évènements qui interviennent, non observables ("création") et observables (mesures). La minimalité de l'interprétation minimale n'est en fait qu'une prudence, une peur encore positiviste de se laisser entraîner trop loin en dehors du déjà construit. Cette prudence cantonne dans un face-à-face indécis, où la Mécanique Quantique est opposée *indistinctement* à la localité relativiste et à des prolongements inertiels et confus de celle-ci qui ne s'insèrent en aucune structuration conceptuelle constituée. Mais une telle prudence ne peut pas durer. Un processus de conceptualisation en chaîne s'est déclenché subrepticement et aucun obstacle factice ne pourra l'arrêter. Cette affirmation n'est pas une critique, elle désigne la valeur la plus sûre que je perçois dans la démarche de Bell, et elle exprime ma confiance dans l'esprit humain.

Je considère maintenant l'interprétation épistémologique. Celle-ci s'avance déjà précisément dans le sens de cette inéluctable modélisation supplémentaire. Les termes considérés sont ceux de "1 système" et "2 systèmes corrélés mais isolés l'un de l'autre" (au sens de la Relativité). La modélisation supplémentaire mentionnée fait intervenir le postulat épistémologique courant d'existence de propriétés intrinsèques pour des entités réelles isolées. On déduit de ce postulat des inégalités du même type que celle de Bell, concernant des statistiques de résultats de mesures sur des entités supposées *isolées*. On établit donc une connexion entre des tests sur des inégalités observables d'une part, et d'autre part le postulat épistémologique d'existence de propriété intrinsèques pour des objets isolés au sens de la Relativité. Sur cette base on admet (il me semble?) que la violation de l'inégalité de Bell infirmerait à elle seule la signifiance de la conceptualisation en termes d'entités isolées possédant des propriétés intrinsèques. Or j'ai montré ailleurs (10) (en termes trop techniques pour être reproduits ici) que cela n'est pas possible. Ici je ne ferai à ce sujet que quelques remarques qualitatives.

Tout d'abord, les considérations faites plus haut concernant la création d'une paire peuvent aussi se transposer d'une manière évidente au cas de l'interprétation épistémologique. Mais prolongeons encore autrement ces considérations: plaçons-nous cette fois d'emblée à l'instant $t=t_o$ où $S_I$ et $S_2$ sont créés. Pour $t>t_o$, $S_1$ et $S_2$ occupent maintenant deux domaines d'espace disjoints $\Delta s_1(t)$ et $\Delta s_2(t)$ qui s'éloignent l'un de l'autre et qui rencontrent ensuite respectivement les appareils $A_1$ et $A_2$, produisant des interactions de mesure. L'interaction de mesure de $S_1$ avec $A_1$ est elle-même un évènement qui occupe un domaine non nul d'espace-temps $\Delta s_{m1}(t_{m1})$. $\Delta t_{m1} \neq 0$ (l'indice m se I mesure) où $t_{m1} \in \Delta t_{m1}$ et le facteur de durée $\Delta t_{m1}$ *dépend* de l'extension spatiale $\Delta s_{m1}(t_{m1})$ liée à l'époque $t_{m1} \in \Delta t_{m1}$ (en supposant que cette extension spatiale reste constante au cours de l'époque $t_{m1} \in \Delta\ t_m$). Il en va de même pour l'évènement de mesure sur $A_2$ dont l'extension d'espace-temps est $\Delta s_{m2}(t_{m2}).\Delta t_{m2} \neq 0$. Comment définir maintenant la distance d'espace-temps entre ces deux évènements de mesure? Quelle que soit la distance spatiale fixée entre $A_1$ et $A_2$, comment savoir si la distance correspondante d'espace-temps entre les évènements de mesure est ou non du genre espace? Car c'est cela qui décide si oui ou non la condition cruciale "d'isolement" réciproque de ces évènements de mesure, se réalise, et c'est sur la base de cette condition que l'on s'attend à l'inégalité de Bell pour les statistiques des résultats enregistrés. Que la distance d'espace-temps entre les évènements de mesure soit ou non du genre espace, cela dépend évidemment (entre autres) des facteurs d'extension spatiale $\Delta s_{m1}(t_{m1})$ et $\Delta s_{m2}(t_{m2})$. Or, que savons-nous de la valeur de ces facteurs? $S_1$ et $S_2$ se déplacent-ils "en bloc", "mécaniquement", comme le suggèrent le modèle de Louis de Broglie et le concept récent de soliton, ou bien s'étalent-ils comme le suggère le concept quantique courant de paquet d'ondes à évolution linéaire Schrödinger?

On pourrait peut-être espérer avoir une réponse plus claire dans le cas où $S_1$ et $S_2$ seraient des photons "dont la vitesse est C". Mais la vitesse de quoi? Du front de l'onde photonique, oui, mais que penser du "reste" du photon? Comment est fait un photon, comme un microsystème de Louis de Broglie, avec une singularité et un phénomène plus étendu autour? Le comportement manifesté par des ondes radio le laisse supposer. De quelle extension alors? Dans la phase actuelle, que savons nous, exactement et individuellement sur ces entités que l'on dénomme "photons"? La Mécanique Quantique newtonienne ne les décrit pas ; l'électromagnétisme ne les décrit pas individuellement. La théorie quantique des champs a été marquée, au cours des années récentes, par des essais "semi-classiques" dont le but est d'éliminer tout simplement la notion de photon afin d'éviter les difficultés conceptuelles liées aux algorithmes de re-normalisation (11).

On peut donc conclure en toute généralité que, quelle que soit la distance spatiale fixée entre $A_1$ et $A_2$ (qu'il s'agisse de microsystèmes à masse non nulle au repos ou de photons), pour savoir si les évènements de mesure sur ces microsystèmes sont séparés ou non par une distance d'espace-temps du genre espace, il faudrait connaître (entre autres) l'extension spatiale des états de ces microsystèmes, en fonction du temps.

Sans détailler plus des enchaînements logiques non essentiels, ces seules remarques suffisent pour indiquer la base de l'affirmation suivante.

A eux seuls, les tests de l'inégalité de Bell ne permettront jamais de conclure concernant la signifiance de l'assignation de propriétés intrinsèques à des entités réelles isolées au sens de la Relativité d'Einstein. Donc pour l'instant aucun face-à-face n'est encore défini entre la Mécanique Quantique et les postulats épistémologiques de notre conceptualisation courante de la réalité. Seule une direction de pensée est tracée, qui suggère l'intérêt de recherches nouvelles sur la structure d'espace-temps de ce que l'on appelle des microsystèmes individuels. Cette direction de pensée me paraît courageuse et très importante, mais dans la mesure où elle se reconnaît et s'assume. Elle s'associe alors naturellement à des recherches récentes sur l'extension des microsystèmes à masse non nulle au repos (12), (13) et sur le concept de photon (11). Il est très remarquable de voir que toutes ces recherches se concentrent sur les phénomènes et concepts d'interférence. En effet c'est là qu'à travers le statistique peut apparaître l'individuel. C'est là que peut se trahir – si on l'y cherche – la confusion entre des interférences *mathématiques* de statistiques standard et d'autre part des statistiques d'interférences physiques d'une entité individuelle qui se superpose avec elle- même (14), (15).

A travers le problème de localité, j'ai dirigé volontairement les regards sur la couche sémantique qui se trouve sous les mots qu'on emploie. L'état de celle-ci est en quelque sorte l'objet principal de ces remarques. La boue sémantique au dessus de laquelle nous voltigeons salubrement d'algorithme en algorithme, accrochés à des cordes de mots, me paraît mériter d'être connue de plus près. Il faudra bien y plonger pour forger les concepts nouveaux qui manquent et en fixer les contours d'une manière qui permette de s'élever jusqu'à des syntaxisations.

Le concept d'objet au sens macroscopique de ce terme est cerné avec rigueur – bien que qualitativement – à l'intérieur de la logique des classes d'objets et de prédicats. Celle-ci est par essence une théorie des objets macroscopiques explicitement structurée et de généralité maximale. Mais cette théorie est *foncièrement* inapte à une description non restreinte des changements. En effet, la logique des classes d'objets et des prédicats est fondée sur la relation d'appartenance $\in$: si pour l'objet x le prédicat f est vrai, alors x appartient à la classe $C_f$ définie par f: f(x) → $x \in C_f$. Mais cette relation fondamentale d'appartenance ∈ est conçue au départ d'une manière statique, hypostasiée. Aucun aménagement ultérieur ne peut compenser les rigidités introduites ainsi au départ. La théorie des probabilités d'une part et d'autre part les différentes théories physiques (la mécanique, la thermodynamique, les théories des champs, la Mécanique Quantique, la Relativité) sont arrivées à combler cette lacune à des degrés différents. Mais chacune pour une catégories particulière de faits et par des méthodes implicites et diversifiées. Une théorie générale et spécifique des évènements et des processus, une logique des changements absolument quelconques, à méthodologie explicite et unifiées, n'a pas encore été construite*.

Considérons maintenant de nouveau la logique des classes d'objets et de prédicats. Elle transgresse foncièrement l'individuel, puisqu'elle décrit des classes. Elle semblerait donc être vouée naturellement à une quantification numérique de type statistique ou probabiliste, à l'aide d'une mesure de probabilité définie sur les classes. Pourtant, à ce jour, une telle quantification numérique de la logique n'a pas pu être accomplie. Les "quantificateurs" logiques ∃, ∀, Ø, sont restés *qualitatifs* !

Complémentairement en quelque sorte, à ce jour, la théorie des probabilités n'a pas encore développé explicitement un traitement classificateur. Le concept fondamental employé est celui d'espace de probabilité $[U,\tau, p(\tau)]$ où $p(\tau)$ désigne une mesure de probabilité posée sur une tribu d'événements $\tau$, définie sur l'univers $U=\{e_i, i=1,2,\dots\}$ d'événements élémentaires $e_i$. Cette tribu peut refléter, en particulier, une classification des événements élémentaires $e_i$ commandée par un prédicat f et en ce cas des propriétés spécifiques "logiques" s'ensuivent pour l'espace $[U,\tau, p(\tau)]$. *Via* ces propriétés classificatrices, la connexion entre logique et probabilités pourrait être amorcée. Mais ceci n'a pas été tenté, et la connexion reste pour l'instant non élaborée.

Considérons maintenant la Mécanique Quantique. Elle introduit des espaces de probabilité. Pourtant les relations entre ces espace sont telles que certains mathématiciens affirment que "la Mécanique Quantique n'est pas une théorie de probabilités". La connexion entre la théorie des probabilités et la Mécanique Quantique reste pour l'instant elle aussi très obscure.

---

* J'ai pu prendre connaissance d'une tentative originale et courageuse de formaliser la durée (I6). Jusqu'ici seules les valeurs associables à la durée ("le temps") ont fait objet de certaines formalisations.

D'autre part les relations de la Mécanique Quantique avec les divers concepts suggérés par le langage qu'elle introduit – 1 système, 1 système de 2 systèmes corrélés, etc. – restent elles aussi très obscures. La Mécanique Quantique n'indique en fait strictement rien concernant ces concepts tels que l'on pourrait vouloir les imaginer en dehors de l'observation. Même la probabilité de présence n'est qu'une probabilité de résultats d'interactions d'observation: il est permis par la Mécanique Quantique d'imaginer qu'un "système" qui fait une marque sur un écran à un moment t, se trouvait, en lui-même, aussi loin que l'on veut de cette marque, aussi peu que l'on veut avant le moment t. La Mécanique Quantique laisse parfaitement non conceptualisée en elle-même, "la réalité" dont elle codifie de manière si riche et détaillée les manifestations observables à travers les interactions de mesure.

Considérons enfin la théorie de la Relativité. Cette théorie est, à sa base, individuelle, non statistique, et continue, non quantifiée. En outre, elle décrit "ce qui est", bien que relativement à l'état d'observation. Sa connexion avec les espace de probabilité à évènements foncièrement *observationnels* et quantifiés de la Mécanique Quantique, soulève des problèmes bien connus et très résistants.

Ainsi nous sommes actuellement en possession de plusieurs structurations syntaxiques constituées, chacune très complexe, riche et rigoureuse. Mais ces structurations sont comparables à des icebergs émergeant de la mer de boue sémantique, sous le niveau de laquelle les bords et les bases disparaissent. Quand à l'ensemble des concepts liés à la propriété fondamentale de durée, les concepts de processus, d'évènement, de changement, de permanence, de succession, de TEMPS, ils n'agissent librement qu'à l'état épars, primitif et subjectif, tels que l'expérience et le langage les a diversement induits dans les esprits. Car les organisations auxquelles ces concepts ont été soumis à l'intérieur de la théorie de la Relativité, de la théorie des probabilités, ou à l'intérieur de telle ou telle autre théorie physique, sont toutes particularisantes et amputantes. La situation est encore telle que la décrivait Bergson: «La déduction est une opération réglée sur les démarches de la matière, calquée sur les articulations mobiles de la matière, implicitement donnée, enfin, avec l'espace qui sous-tend la matière. Tant qu'elle roule dans l'espace ou dans le temps spatialisé, elle n'a qu'à se laisser aller. C'est la durée qui met des bâtons dans les roues' » (17).

Je résume une fois encore par un schéma:

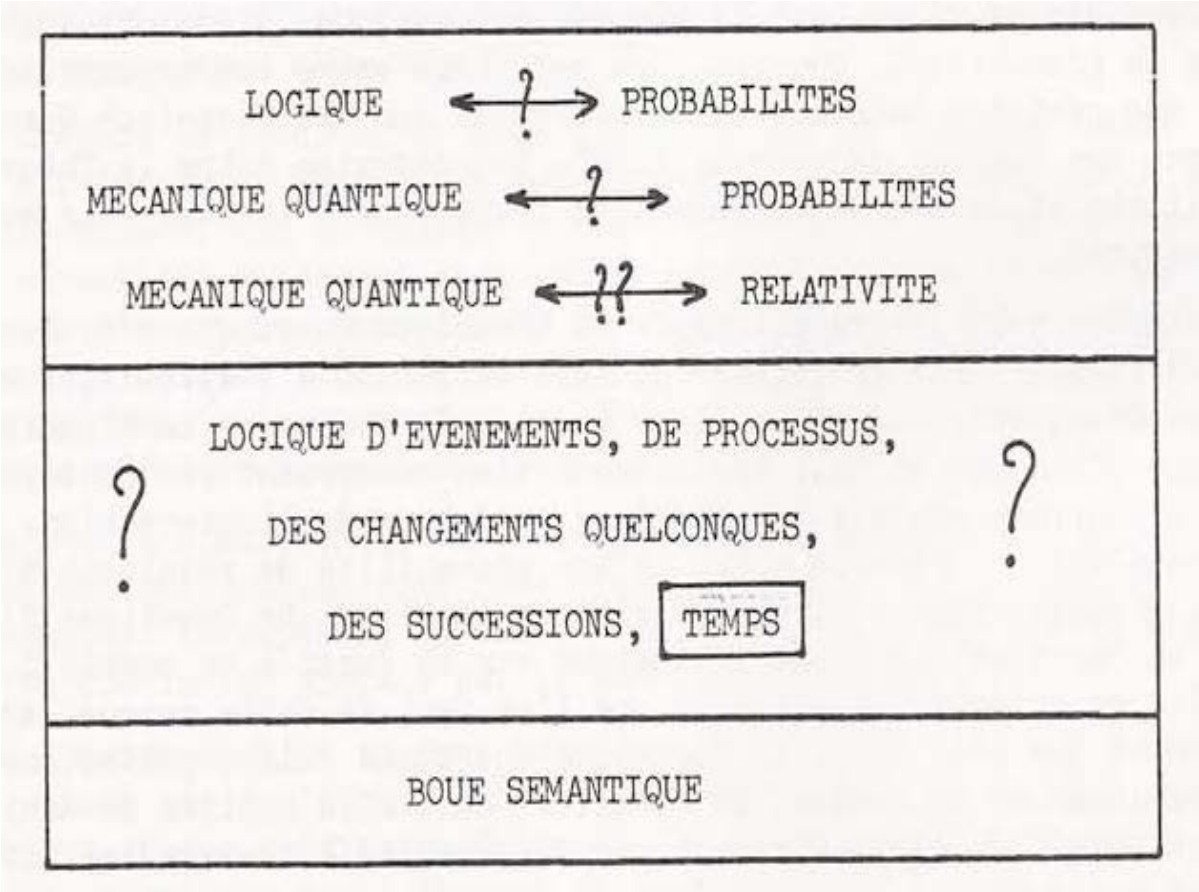

Quand il n'existe encore aucune unification entre la démarche statistique, discrète, observationnelle, orientée vers le microscopique, de la Mécanique Quantique, et d'autre part la démarche individuelle, continue, réaliste, orientée vers le cosmologie, de la Relativité, quant tout ce qui touche à la durée et au temps est encore si peu élucidé, quand tout ce qui touche à la manière d'être de ces entités que l'on appelle des microsystèmes – ou plus encore, de microétats – est encore tellement inexploré, quel sens cela peut-il bien avoir d'affirmer qu'on se trouve – sur la base de tests de "non-localité" – devant un face-à-face *contraignant*, direct ou pas, entre la Mécanique Quantique et la Relativité? Ou bien entre la Mécanique Quantique et notre conceptualisation du réel?

**Conclusion**

Je ne puis qu'écarter, pour ma part, les face-à-face que les autres physiciens pensent percevoir. Pour moi la valeur du théorème de Bell réside ailleurs: ce théorème, et l'écho qu'il soulève, illustrent d'une manière frappante la puissance d'action des modélisations *mathématisées*, lorsqu'elles sont connectables aux tests expérimentaux. Pendant des dizaines d'années, les tabous positivistes ont fait obstacle aux modèles. Le résultat est ce vide vertigineux de modèles syntaxiques, et même seulement qualitatifs, que l'on découvre maintenant sous les algorithmes quantiques. Or, la modélisation de Bell a déclenché une dynamique de conceptualisation et de syntaxisation. Cette dynamique atteindra peut-être l'attitude positiviste. Elle ébranlera peut être la Mécanique Quantique et la Relativité. Car elle attire et maintient longuement l'attention sur l'état du milieu conceptuel dans lequel les théories actuelles sont immergées. De ce contact prolongé sortiront peut-être des théorisations nouvelles, plus unifiées, plus étendues et plus profondes. Je perçois (ici comme en théorie de l'information) les premiers mouvements de formalisation de l'épistémologie, les premières ébauches, peut-être, d'une méthodologie mathématisée de la connaissance. Et cela pourrait s'avérer plus fertile que toute théorie particulière d'un domaine donné de réalité.

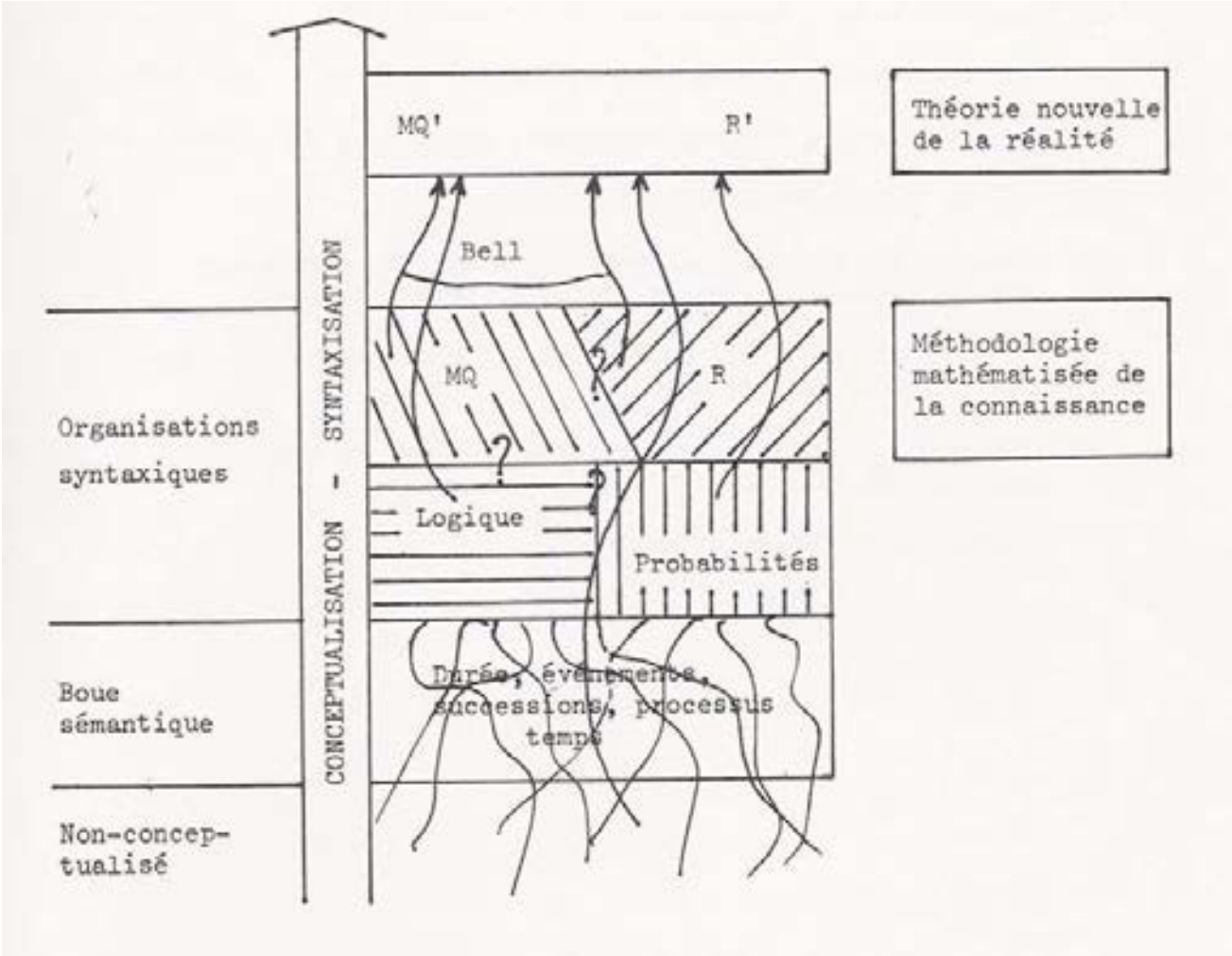

********************

The most striking in this account from 39 years ago is that meanwhile the public conceptual situation concerning microphysics did not notably change in its essence.

As for the author of this work, she believes that by precisely what is exposed below – and from her subjective viewpoint – the program delineated in the Conclusion reproduced above – that since 1979 has been kept in work without interruption – has finally reached its boundary.

# 1.I [6]

# THE FIRST GERM OF A DESCRIPTION OF A "MICROSTATE":
# * GENERATION OF A MICROSTATE
# * QUALIFICATION OF ONE SPECIMEN OF A MICROSTATE

## (1.I).1. THE OPERATION OF GENERATION *G* OF A MICRO-STATE-TO-BE-STUDIED AND A BASIC METHODOLOGICAL DECISION

In agreement with Dirac I distinguish between stable characteristics assigned to a 'micro-*system*' (mass, spin, etc.), and unstable dynamical characteristics assigned to a *mechanical state* of a micro*system*, so to a (mechanical) 'micro*state*', namely position, momentum, etc. So far however 'a micro*state*' is just a verbal sign posited to point toward a physical thing that is still entirely unknown as to all its specificities.

As already stated in the introduction, in this first part of the present work the global aim is to establish the general features that – a priori – are necessary conditions for *any* process of creation of scientific knowledge on microstates (communicable, consensual, provisional and verifiable knowledge). In order to reach this degree of generality I *start at the **extreme** 'bottom', and I proceed **bottom-up***: this will avoid missing any possible step.

I want to establish how, out of the as yet never qualified physical substance, it is in principle possible to generate scientific knowledge on microstates.

### (1.I).1.1. The basic question

In current languages and in classical grammars an object-to-be-qualified is usually supposed to pre-exist as such. It just "is" there. Its definition is realized by use of grammatical predicates ("Bring me the brown thing from that drawer", etc. And for generality, look in a dictionary). As for the predicates, they also are considered to pre-exist, but in the air of thought, platonically. They are expressed by just naming words, or by only verbal pointers of location in space and time ('there', 'now', etc.), or even by just pointing physically toward the object-to-be-qualified. In the classical logic these assumptions are sanctified. The objects-to-be-qualified are represented by a set of letters *('x', 'y', 'z',....)* and the functional expressions like $f_P(x)$ that contain such a letter (*x* in this case) and where *P* designates a 'predicate', are called propositional functions and they become true or false according to whether *x* satisfies the predicate *P* or not, which is a *fact* that is directly perceived by the human observer. So the generation of knowledge on classical 'objects' appears as a natural activity that in general raises no basic problems. As for the generation of *scientific* knowledge on classical 'objects', it involves only some natural methodological prolongations for insuring consensus, predictions, and verifiability of these.

All this is rooted into the naïvely realist postulate that the objects-to-be-qualified are perceived 'such as they really are', in an *absolute* way, via their intrinsic 'properties' represented by predicates. These beliefs are strongly favoured by the circumstance that in our current life most of the objects-to-be-qualified are directly perceived.

---

[6] To be read 'chapter 1 from Part I'.

But while the domain of scientific construction of knowledge includes more and more objects-to-be-qualified that escape direct perception, the aim to ever reach a notion of 'how things really are, intrinsically and absolutely' recedes into a thickening tissue of procedures where it seems doomed to disintegrate.

Indeed consider the case of the microstates. How can a radically non-perceivable and unknown microstate be introduced as that-what-is-to-be-studied, when in general it does not even pre-exist? In this respect the *un*-bound microstates oppose the most striking obstacles. How can an unbound, freely evolving microstate be obtained in the role of object-to-be-qualified? And moreover in a way compatible with some sort of availability for further cognitive action on it, such that it shall permit also consensual qualifications and verifiability of these? For *knowledge* of some thing – by definition – is qualification of that thing, and scientific knowledge is just communicable, consensual, predictive and verifiable qualification.

How is it possible to generate *scientific* knowledge on an unbound microstate?

Let us detail this question. Of course as soon as we presuppose an unknown microstate and we indicate it by some word or label we already have presupposed that it is tied with something that pre-exists and out of which that microstate can be brought into the role of an object-to-be-qualified. But in order to effectively bring it into this role, some definite macroscopically controllable *conceptual-physical operation of* ***generation*** *of such a 'role-situation'* should be realized. For obviously such a situation does not *pre*-exist. But an appropriate operation of generation cannot be conceived, nor realized, otherwise than on the basis of some previously established knowledge, and on some specified space-time support. If not we cannot even think of this microstate, so a fortiori we cannot study it. Furthermore the operation of generation of a microstate in the role of object-to-be-qualified has to permit some gain of information on that microstate, some qualification, some knowledge, *and* in a way that admit consensual *verification*, for if not a scientific study is out of reach. And when the way of realizing factually verifiable qualifications is analysed, it becomes obvious that the obtained qualification *also* cannot be conceived to pre-exist, that it also has to be created. So, more and more, the aim to gain knowledge on microstates reveals an essentially *constructive* character and this entails results indelibly marked by *relativities* to the constructive procedures that have brought the results into being.

The problems mentioned above and their effects do not arise with respect to the directly perceivable objects from our current life because these – admittedly – just subsist perceptibly while we cease observing them. And when we want to perceive them again we only have to manage to bring them again into our domain of perception [7]. But for a radically non-perceivable micro-entity the problems enumerated before emerge basically and dramatically. Throughout the historically realized top-down approach, from one small new step that has been realized, to another small step, these problems remained more or less hidden by an implicit and uncontrolled increasing extension of the classical models and assumptions that, surreptitiously, transgressed more and more any possibility of factual realization. But when one wants to start the approach at the extreme conceivable 'bottom' and therefrom to proceed bottom-up, the veil of classical models and assumptions is radically abandoned and these problems explode into the attention from the start with their whole content, and the unique choice left is to either solve them explicitly or to stop to start.

[7] Cessation of this possibility is usually irreversible.

### (1.I).1.2. The general concept of operation of generation *G* of a micro-entity-to-be-studied

Then let us now focus first upon the initial necessity of an operation of generation of a micro-entity in the role of a microstate-to-be-studied in the scientific sense. We denote this a priori by *G*.

Since a micro-*state* is by definition conceived as an evolving sort of entity, the *repeatability of G* is an unavoidable pre-condition for constructing consensual and verifiable, scientific knowledge on a given micro-state. But how can we know whether yes or not each reiteration of the operation denoted '*G*' yields the *same* result?

Well, we simply ***cannot*** know whether *G* comes out the same when it is repeated, or not. Nor can we insure factually that it come out the same. This is so because the operation *G* is a factual physical process that has to be inter-subjectively specified and communicated, which is possible only by some ***finite** definition*. And any finite factual definition is quite essentially unable to constrain into absolute ('total') sameness the whole factual singularity of each realized replica of the operation *G* (Umberto Ecco has said that as soon as we speak or write we conceptualize and thereby we quit and lose irreversibly the infinite singularity of any piece of factual entity).

> Here the unconceivable *in*finity of possible ways of being of any fragment of factual physical reality stays face-to-face with the *finiteness* of the human capacity to constrain and to control in predefined ways.

However giving up because of this the whole project of establishing how it is possible to create some sort of knowledge on the dynamical states of micro-entities would be an unacceptable weakness from the part of a human mind. We are in presence of a problem of strategy, of method. So we have to conceive an appropriate cognitive strategy.

### (1.I).1.3. A methodological decision ("the microstate *corresponding to G*" and "one specimen" of it)

We organize a *methodological decision denoted MD* that introduces a global strategy of speaking and thinking on the basis of which it becomes possible to start, to act, and to achieve the bottom-up construction of *IQM*.

***Formulation of MD***

- Each time that one individual operation denoted *G* of generation of a dynamical state of a micro-entity-to-be-studied, is realized in agreement with a definition expressed in terms of a finite number of parameters that are controllable factually from our macroscopic level of existence – which, for us, is the only sort of possible factual definition – this operation *G* itself is *admitted by construction* to come out the same *with respect to its factual finite definition*.
- *That* what emerges in consequence of ***one*** realization of *G* (the result) is not directly observable by our bio-psychical apparatuses, but – whatever it be – it is posited to be observable indirectly via future appropriate operations of *qualification*; *and* it is conceived as one *specimen* (or variant) denoted $\sigma(ms_G)$ of something more global than any individual specimen $\sigma(ms_G)$.
- The more global entity posited above can be labelled $ms_G$ and we call it *the factual microstate corresponding to G*.
- This amounts to denote: $ms_G \equiv \{\sigma(ms_G)\}$ ($\sigma$: specimen)
- On this basis we enter upon the bottom-up construction by positing an observable and verifiable, *law*-like one-to-one relation

$G \leftrightarrow ms_G$ (1)

This was our first step.

***Comments***

Together, the necessarily finite character of the human definition of *G*, the action of this operation on – directly – the still a-conceptual unlimited factuality, and the obvious fact that *absolute sameness is just nonsense*, ***impose*** inside *MD* a *factual and* ***multiple*** content $ms_G \equiv \{\sigma(ms_G)\}$ for the concept called 'the ***factual*** microstate corresponding to *G*' and denoted $ms_G$. This is *a new concept of microstate*. The classical concept is defined *abstractly* and its content is specified *individually*. So indeed:

*A factual microstate '$ms_G$' in the sense of MD is essentially different from a microstate in the classical sense.*

Nevertheless the word 'microstate' is kept in use because it can play the very useful role of a recurrent element of continuity, reference and comparison between the classical top-down approach specified conceptually via abstract definitions of which the possibility of effective realization has not been examined, and the factual bottom-up approach initiated here. Inside the definition of 'the factual microstate $ms_G$ corresponding to *G*' this classical word will work as a memento that the *origin* of the process of construction of knowledge has been changed and that this has consequences.

For the sake of commodity in what follows the classical word "microstate will often be used alone. But the meaning assigned to it will always be that defined inside *MD*.

So we start now via (1) from *local* ***zeros*** of previously constructed knowledge on – *specifically* – each individual specimen $\sigma(ms_G)$ of $ms_G$ brought in as an entity-to-be-studied [8]. And therefrom we want to construct bottom-up. This changes also the *order of construct**ability** of concepts* (abstract concepts or factually defined ones), as well as the place inside this order of each sort of piece of verifiable knowledge. It can be hoped that the explicit awareness of a new order of constructability entailed by a bottom-up approach, when compared with that from the classical top-down approach that started spontaneously from our everyday level of ancestral conceptualization, will bring forth many clarifications concerning the problems of interpretation of the modern microphysics.

***The principle of determinism versus "essential indeterminism"***

The posit (1) – via the definition $ms_G \equiv \{\sigma(ms_G)\}$ – introduces in a natural way the so much discussed "*essential* indeterminism" of the modern microphysics. Namely by the imperative necessity to define a new concept of factual microstate, associated with the ineluctable *finiteness* of our capacity to produce operational definitions and so, to realize "totally" identical physical results by repetition of a physical operation.

The "essential indeterminism" induced by (1) consists of a *factual* human impossibility that can be immediately sensed to entail observable effects, and in particular to mark the possible predictions. Whereas the classical postulate of determinism is an *abstract, purely conceptual assertion* that, though largely suggested by certain facts, is devoid of any rigorously realizable factual proof [9].

But it is very noteworthy indeed that:

---

[8] By its individual significance the concept of a specimen $\sigma(ms_G)$ of $ms_G$ is more than $ms_G$ alike to the classical concept of microstate, but *it emerges entirely* ***undefined in its individuality****, its specificity*.

[9] This is classically explained by the notions of "imprecision of measurement" and of unpredictable (chaotic) development of always imprecise realizations of the initial conditions, etc. The mathematical investigations on "chaos" have brought forth that the posit of 'determinism' ***has*** to be composed with factual outcomes that are ***never*** strictly precise.

*In fact the whole posit MD and in particular the one-one relation (1) $G \leftrightarrow ms_G$, which – together – found the "essential* ***indeterminism****" of the modern microphysics,* ***still express a basically deterministic view****.*

Indeed, the relation (1) $G \leftrightarrow ms_G$ amounts to the assertion of existence of a probability measure, which still is the assertion of a 'law', of a one-one causal connection "*if* this *G*, *then* that ($ms_G$)": Our human minds – such as they have been modelled by optima of adaptation of our ways of perceiving, thinking and acting – have selected and imprinted upon our minds a principle of causality. This is a mental fact. This principle works so strongly that in order to transgress it, we still *use* it, but in a way that displaces its frontier upon 'probabilistic' factual-operational-observational contours, instead of point-like individual assertions. So what acts in this circumstance is not in the least an "essential indeterminism"; it is – *in general –* ***a factual*** *impossibility to insure* ***a rigorously individual prediction***. And the Methodological Decision (1) permits to nevertheless save a global inner coherence founded upon a general deterministic postulate, by *distinguishing* explicitly between: ***(a)*** a general abstract posit of punctual causality, and on the other hand ***(b)*** the *sort* of scientific consensual, predictive-verifiable ***knowledge*** that can be generated in a cognitive situation that is *entirely* founded upon factual-physical operations, in the strict absence of any direct sensorial human perceptibility[10].

### (1.I).1.4. Mutation of the classical concept of 'definition': a split

We have noted already that in the classical conceptualization the entity-to-be-studied is conceived to pre-exist as a stably available potential support for qualifications achieved by a convenient predicate conceived to select from inside the set of all the pre-existing "objects" those that "possess" the "intrinsic" "property" expressed by the predicate. The direct perceptibility permits this confortable ellipsis that *absorbs* in it the necessity of an explicit operation *G* of generation of the considered entity as an entity-to-be-studied. But for microstates such an ellipsis is not possible. And that is why:

> *MD* splits the classical concept of definition into a sequence of two distinct operations that can be achieved only separately, namely; an initial action of generation of the object-entity-to-be-studied that is already specified inside *MD*; and a *subsequent* act that still remains to be specified, of qualification of the object-entity-to-be-studied generated before.

### (1.I).1.5. Composed operations of generation $G(G_1,G_2,...G_k)$: a principle of composition of physical operations of generation

From its start, the study of microstates has brought into evidence a class of microstates that have been called '(auto)-interference-states' and that have played a founding role in the emergence of quantum mechanics (the paradigmatic case is Young's two slits experiment). The process of generation of an interference-state permits to distinguish at least two operations of generation $G_1$ and $G_2$ that are involved, in the following very peculiar sense: Each one of these two operations *can be produced* ***separately***, and if they are, then two different corresponding microstates $ms_{G1}$ and $ms_{G2}$ do emerge. But when $G_1$ and $G_2$ are

[10] In fact the difference with the classical Physics is rather feeble. It is arbitrary to choose to call "errors of measurement" dispersions that emerge *unavoidably* in consequence of basic aspects that are systematically at work in any act of measurement, namely – just like in *MD* – an opposition between the finite *definability* of any human physical operation (the result of any act of measurement can be expressed numerically only via the choice of some non-null unity that entails unspecifiable variations inside its extension) and on the other hand an unknowable infinity of effects from the underlying flux of physical reality. Similar considerations hold concerning the "imprecision in the factual realization of initial conditions". This illustrates to what a degree the most basic conceptual 'problems' can depend on language, so on construction and method.

'composed' into only *one* operation – let us denote it $G(G_1,G_2)$ [11] – then, accordingly to (1), there emerges only *one* corresponding microstate $ms_{G(G1,G2)}$ that manifests 'auto-interference effects'.

On this factual basis tied with the just indicated way of speaking, we introduce here an only qualitative but nevertheless a general *principle of composition of operations of generation* according to which:

> In certain operations of generation of a microstate, two or more separately realizable operations of generation – either deliberately produced by human researchers or brought forth by natural processes – can 'compose' to act upon *on*e unspecified *preliminary microstate*, so as to generate together *one* microstate-to-be-studied, in the sense of *MD*.

When this happens we shall speak of one microstate $ms_{G(G1,G2,...Gn)}$ with a composed operation of generation $G(G_1,G_2,...G_k)$ [12]. When this does not happen, we can – for contrast and precision – speak of a 'simple' (non-composed) operation of generation.

Here the concept of composition of operations of generation of a microstate defined above, as well as the corresponding underlying principle of possibility to compose an operation of generation, are only very feebly defined. And deliberately we leave it in this state, for the sake of minimal assumptions with respect to our present aim. Indeed *such as it is*, in the Parts II and III of this work this feeble defined principle will suffice for entailing most essential consequences.

### (1.I).1.6. Universality of *G*

At a first sight it might seem that the concept of operation of generation of an entity-to-be-qualified constitutes a radical novelty of which the necessity is specific of microphysical entities. But a deeper analysis reveals that in fact, in this respect, there is no mutation. The case of microstates only brings into full evidence a ***universal*** phase in the human conceptualization (MMS [2002], [2006]) that acts already inside the fully classical conceptualization. Indeed any definition presupposes – more or less implicitly but quintessentially – an operation of initial specification of the entity-to-be-defined (the grammars stress this fact). Often this is a specification via a merely psycho-sensorial out-cut from the continuum of the directly perceived surrounding 'exterior reality'; or even only a reflex human gesture (turning the whole head or only the eyes toward some delimited domain of direct perceptibility); or even an exclusively mental selection via a focalization of the attention. But in many 'classical' situations the act of specifying the entity-to-be-qualified consists of a deliberate and laborious *physical* operation of separation and of supply into immediate accessibility (think of medical analyses or geological or archaeological procedures). And sometimes, exactly like in the case of microstates, a classical operation of generation consists of a deliberate radical *creation* of the entity-to-be-qualified (production of prototypes in the industry of artefacts[13], simulated test-situation in a detective research, etc., etc.). In short:

---

[11] This notation stresses that only *one* operation of generation has been effectively achieved by 'composing' other operations of generation that *could* have been achieved separately but have *not* been separately achieved.

[12] We do not try to specify the conditions that restrict the possibility of composing operations of generation (in particular, the space-time conditions) though such conditions do certainly exist. Nor do we try to specify some limit to the possible number of composed operations of generation. These are features that are still unexplored from both a factual and a conceptual point of view because inside nowadays quantum mechanics – together with the concept of operation *G* of generation of a microstate itself – they remain hidden *beneath* what is mathematically expressed, *in consequence of a basic confusion between 'superpositions' in the* ***mathematical*** *sense, and* ***factual*** *superpositions in space-time, of operations or of physical entities*. The consequences of this basic confusion will be narrowly surveyed and in the third part of this work they will play a quite essential role.

[13] Cf. H. Boulouet [2014].

Strictly always a human being, in order to acquire some knowledge on some thing, somehow singularizes this thing from inside the continuum of the physical reality, explicitly or implicitly[14].

That is so because a human being can have only finite perceptions and can perform only finite actions, whether these actions are psychical, or psychophysical, or physical. So he is obliged to somehow *delimit* that what he wants to qualify, out from the inside of the infinite whole of what we call 'reality', to parcel this out in some sense, more or less *creatively*. This inescapable necessity induced by the human imprisonment in finiteness has very basic and unexpected consequences. In particular, *this* is the source wherefrom, inside *MD*, is instilled a basic impossibility to assert an *individually* deterministic one-to-one ***factual**-observational* relation; which in the case of microphysical entities-to-be-qualified becomes obvious and systematic ***before*** considering the question of qualification (so of measurement) and *thereby* entails the non-classical posit $ms_G \equiv \{\sigma(ms_G)\}$ and the corresponding displacement of the one-to-one deterministic relation (1) upon the probabilistic level of qualification.

The character and the effects of the operations of generation considered in *MD* are indeed specific of the human cognitive situation with respect to microstates. But the presence alone of an operation *G* of generation of the entity-to-be-studied is *not* specific of the human cognitive situation with respect to microstates. On the contrary, this presence alone is a quite universal cognitive fact; a more or less hidden fact but a universal fact. This simply has not been explicitly remarked, precisely because it is universal, but also no doubt because in the current life – historically and during a very long time – inside the domain of physical reality that was accessible to direct perception it has very often been possible to "naturally" put a physical entity in the role of entity-to-be-qualified, or even to realize this in reflex unconscious ways. While inside the various global methodized approaches, the act of bringing an entity in the role of entity-to-be-studied got lost in an ocean of other norms (think of the *global* requirement of 'reproducible experimental conditions' in classical physics).

The obliteration of the operation of generation *G* is a huge lacuna.

This will become more and more obvious while the consequences of the operations *G* – and their importance – will progressively appear, together with the consequences of an obliteration of these basic operations.

It is true that the acts of measurement introduce systematically observational dispersion of the results *with respect to* the strictly causal character posited for the behaviour of the physical reality. But *the acts of measurement are **not** the unique source of the statistical character of what can be observed*. The methodological decision *MD* specifies how, in what a sense, the operation of generation that is involved, in general is *also* a source of dispersion in the results of measurements, and a source that in general comes in *before* the operation of measurement. This is so because any act of measurement f*ollows* an act of generation of the entity on which that act of measurement is performed, so that in rigorous and fully general terms one has to always consider ***pairs** [G.MesA] of*

*[(an operation of generation G of the entity-to-be-qualified),*
*(an operation MesA of qualification of this entity)]*

and so *the dispersions due to G and to MesA mix with one another*.

All this becomes clearer by the following examination of also the acts of qualification of a microstate.

[14] And probably any living being does this.

## (1.I).2. BASIC FEATURES OF THE GENERAL CONCEPT OF QUALIFICATION OF ONE SPECIMEN OF A MICROSTATE

### (1.I).1.4. Classical qualification

Inside the classical thinking an act of qualification involves more or less explicitly a *genus-differentia* structure. The *genus* can be conceived as a *semantic dimension* (an abstract ground with *spatial* character, in a generalized sense); while the *differentia* can be regarded as point-like *"values"* from a *spectrum of values* singularized inside this semantic dimension. The spectrum can be numerical or not, ordered or not, and it can be specified as to their nature by the help of material samples, or by conventional labels (words, symbols, mathematical functions). Let us denote the semantic dimension by $X$ and by $X_j$, $j=1,2,...J,$ the values from the spectrum assigned to $X$ (for instance $X$ can be 'colour' and then the spectrum of values $X_j$ consists of a finite number of freely chosen definite colours defined conventionally by words, ({red, green, blue, etc.}) and in their nature by a finite set of material samples.

As already recalled, inside classical thought with its languages, grammars, and logic, a given semantic dimension and the spectrum of values carried by it are currently imagined to somehow pre-exist in the realm of ideas, even if only potentially. But here – and even for classical acts of qualification – we conceive them as being *constructed* more or less deliberately by the human observer who conceptualizes accordingly to his local *aims* of description and under the general and permanent though ignored control of the irrepressibly restrictive human ways and possibilities of thinking and doing and of the cognitive situation that is at work: these, considered globally, act like a net of a priori *constraints*.

According to the classical conception again, there also usually just 'exists' some possibility to estimate what value $X_j$ of $X$ has been found for a given entity-to-be-qualified when this has been 'examined via $X$'. This amounts in essence to imagining more or less explicitly a sort of act of *measurement-interaction* – biological or not, spontaneous or scientific – between some *measurement apparatus (or instrument)* $\mathcal{A}(X)$ and the entity to be qualified. Let us denote by *MesX* such an act of measurement-interaction. The result $X_j$ of an act of *MesX*, when perceived by the observer, becomes a piece of *knowledge* concerning the examined entity.

> Indeed – by definition – ***human consensual, scientific knowledge of some thing is just consensual qualification of this thing by human beings; what is not consensually qualified in any way is not scientifically knowable***[15].

Now, the operation *MesX* – just like *G* – cannot be defined otherwise than by some *finite* specified set of controllable parameters. Unavoidably, features and circumstances that cannot be conceived a priori transcend the control entailed by these parameters. So again, just like in the case of *G* and (1), there is no other way than just admit that all the realizations of *MesX* are the 'same' with respect to a necessarily finite set of specified parameters. *This is not reducible to "imprecision" or "errors"; it is an essential feature* unavoidably entailed by the battle between the finiteness of the human capacity to define and realize 'identities' and the uncontrollable flow of a-conceptual factual being.

When the registration of the value $X_j$ of a semantic dimension or 'quantity' $X$ that is posited to be able to qualify an 'object' in the classical sense, is performed directly via a human biological sensorial apparatus, it generates in the observer's mind a *quale*, a strictly subjective perception of a 'quality'. As it is well known, this quality itself cannot be communicated, but often its subjective existence can be asserted by words, gestures, or other signs that label it consensually *in connection with its exterior source* that is publicly perceivable, namely the considered classical 'object' [16]. We denote globally this classical coding-process by *cod.proc*$(X_j)$ and we represent a classical *grid of qualification (gq)* by writing

$$gq[X, X_j, MesX, cod.proc(X_j)] \qquad (2)$$

---

[15] This apparent triviality is simply ignored by our spontaneous conception on what we call 'reality'. We shall come back on this very fundamental point in the last part of this work when the involved contents will have become totally clear.

### (1.I).2.2. Qualification of one specimen of a microstate

But how can be qualified a microstate $ms_G$ that cannot be directly observed? The answer, if it is thoroughly constructed in the terms imposed by our general way of thinking, appears to be a genuine saga.

Consider a qualifying quantity $A$ with 'values' $a_j$. In order to qualify by a value $a_j$ of $A$, the specimen of microstate $ms_G$ brought forth by a given realization the operation of generation $G$, this realization of $G$ must be followed *immediately* by a qualifying measurement interaction *MesA* realized inside the space-time neighbourhood of the space-time support of the operation $G$. Indeed each outcome of a specimen $\sigma(ms_G)$ of $ms_G$ is quintessentially pre-conceived as a *dynamic* of this specimen (we want to characterize any *mechanic* of the microstates). So, even though any specific knowledge of this changing entity is still lacking in our minds, nevertheless – insofar that the researched knowledge concerns the specimen $\sigma(ms_G)$ *such* as it has been generated by the considered realization of $G$ (not something that has evolved out of that) – the measurement interaction *MesA* must immediately follow the operation $G$. For this purpose a whole succession *[G.MesA*] has to be realized in order to obtain one qualification via $A$ of the considered specimen of $ms_G$. And since a measurement-interaction requires an appropriate *non*-biological apparatus, its result can only consist of some publicly observable marks registered by devices of this apparatus. Furthermore, in general the measurement-interaction destroys the involved specimen $\sigma(ms_G)$ generated by the previously accomplished operation $G$ of generation. And so on. All these questions have been already discussed very much indeed and they have suffered heavy trivialization, but without having been genuinely studied.

But much more radically, and rather curiously, a huge gap seems to have been unanimously left entirely implicit, namely *the coding problem*. Our own examination of the process of qualification of one specimen of a microstate $ms_G$, is centred on precisely this problem and, deliberately, it will be exposed in an outrageously explicit way.

#### *(1.I).2.2.1. The coding problem* versus *"theory" of the factual microstates*

What *criteria* do permit to define the procedure that deserves being called a measurement-interaction *MesA* between a specimen $\sigma(ms_G)$ of a given microstate $ms_G$ and a device for measuring the quantity $A$? What procedure can endow the publicly observable marks produced by one given act of 'measurement-interaction' *Mes(A)*, with *meaning*, and in terms of – precisely – a *given* value $a_j$ of – precisely – the quantity $A$ that one wants to measure? Whatever it be, we call such a procedure *a coding procedure in terms of a value* $a_j$ *of* $A$, we denote it *cod.proc*$(a_j)$ and, quite dramatically, we restate our question:

---

[16] We recall a classical example: Each one of us experiences the feeling of a quality that he has learned to call 'red' while referring to the *source* (say a flower) to which he connects this 'value' of the quality called 'colour'. Thereby – by learning and via the involved sort of context – that quale and its values acquire common inter-subjective *verbal* labels that inside each given mind point toward strictly subjective, non-communicable events. So in classical circumstances each very currently arising quale acquires an inter-subjective labelling and this is tied with the illusion that the quale 'exists objectively', outside there, in the object itself, as a property possessed by it.

*How can one define the coding procedure cod.proc($a_j$) for a specimen $\sigma(ms_G)$ of a factually defined microstate $ms_G$* when the physical characters of such a specimen are still entirely unknown so that not even the applicability to it of qualifications by a dynamical quantity *A* that so far has been defined only inside the classical mechanics (position, or momentum, or energy, etc.), can be asserted a priori?

This is a most fundamental problem. Nevertheless it has been left implicit. So it has been taken into account only intuitively, without generality, nor rigor. Let us stop on this problem.

The general content of a grid for mechanical qualification of a specimen $\sigma(ms_G)$ accepts the same general form (2) of a classical grid. But when a specimen $\sigma(ms_G)$ is the object of qualification, the signs *A*, $a_j$, *MesA*, *cod.proc*($a_j$) point toward contents – entities and circumstances – that with respect to the human observer involve cognitive constraints that are radically different from those that act in the case of 'mobiles' in the classical sense:

- That what is to be qualified – one specimen $\sigma(ms_G)$ of a microstate $ms_G$ for which the one-to-one relation (1) $G\leftrightarrow ms_G$ is posited – has been extracted by the operation *G* of generation directly from the as yet a-conceptual physical reality. It is still radically unknown in its physical specificities inside the class $ms_G\equiv\{\sigma(ms_G)\}$. It is only posited to exist and is labelled.

- Every individual specimen $\sigma(ms_G)$ remains constantly and entirely non-perceptible itself by the observer. Suppose that a given sort of measurement *MesA* (for instance with *A* meaning 'momentum' *P*) *does* make sense with respect to what the symbol '$\sigma(ms_G)$' represents, and that we know how to perform such a measurement. When an the act *MesA* is performed upon a specimen $\sigma(ms_G)$, exclusively groups $\{\mu\}_{kA}$ of some publicly observable marks (with $k_A=1,2,...m_A$) can be obtained on registering devices of some corresponding apparatus $\mathcal{A}_p(P)$ (a spot on a sensitive screen, a sound-registration at a time *t*, etc., some group of such marks).

- Since the registered group $\{\mu\}_{kA}$ of observable marks is the result of a measurement-*interaction MesA* between $\sigma(ms_G)$ and an apparatus $\mathcal{A}_p(A)$, its meaning cannot be *conceived* in terms of some property assignable to $\sigma(ms_G)$ ***alone***. The marks $\{\mu\}_{kA}$ characterize exclusively the achieved measurement interaction as a whole. While in the radically incipient cognitive situation that is considered here no criteria are conceivable for separating a posteriori inside $\{\mu\}_{kA}$ the contributions from the two sources $\sigma(ms_G)$ and $\mathcal{A}_p(A)$.

- A fortiori, since $\sigma(ms_G)$ itself is not directly perceivable, no qualia tied with exclusively this entity can be formed and triggered in the observer's mind via the measurement-interaction *MesA*: The observer gets no inner subjective feeling whatever tied with the nature of *A* and with the particular specimen $\sigma(ms_G)$ that has been involved in the considered interaction.

So the classical contents assigned to the designation "a measurement-interaction *MesA*", such as these are implicitly instilled by the top-down extrapolating approach *that has failed*, are fundamentally different from the contents revealed by the bottom-up approach considered here.

The characters listed above will be globally indicated as the result of one *primordial transferred qualification of a specimen $\sigma(ms_G)$ of a microstate $ms_G$*, which means: a strictly first compact whole of observable marks that are transferred on the registering devices of an apparatus, that do *not* entail any sort of qualia tied with – separately – the involved specimen of a microstate *σ(ms$_G$)* itself, and that cannot be analysed further in effects of this specimen and effects of the involved act of *MesA* [17].

We come now back to the central question: How are we to conceive an act of measurement-interaction *MesA* in order to found the assertion that the registered marks *{μ}$_{kA}$* do qualify the involved specimen *σ(ms$_G$)* of the studied microstate *ms$_G$* in terms of a given value $a_j$ of a given measured quantity *A*? In what a way can an observable group of brute marks be brought to *signify* consensually in terms of one definite value $a_j$ of *A*? ***How can the observable result of a measurement-interaction MesA be adequately endowed with a definite consensual meaning***?

I assert that in the absence of ***any*** *commonly accepted generic* ***MODEL*** *of one specimen of a factually defined microstate ms$_G$* it is not conceivable to produce an a priori meaningful consensual definition of the set of possible results of a measurement-interaction with a specimen *σ(ms$_G$)* of *ms$_G$* (an only *mathematical* representation is not necessarily also a definite such generic model); and without such a model a consensual study of a 'mechanics' of the factual microstates cannot even begin.

> For this purpose – just to be able to begin – *a* ***general MODEL of any specimen of the studied factual microstate MUST be given as a basic primary datum*** *in any theory of the microstates*.

And *this involves beforehand also a* ***global model*** where*from* the individualization of "*one* specimen *σ(ms$_G$)*" of a factually defined microstate *ms$_G$* can be realized in a ***physically operational*** way, accordingly to MD and (1), via some inter-subjectively specified and repeatable operation of generation *G*.

The orthodox "positivistic" interdiction of models was an a priori impossibility.

Like a fly in a room that hits indefinitely the transparent glass of a window, we hit again and again the unperceivable walls that imprison us inside our human ways of thinking *and* ***acting***. This sort of imprisonment is a major *fact* that cannot be transgressed. Then we must take it into account explicitly so as to let it work freely though under control. We must organize a framework where, working freely accordingly to the specific laws of our thought, we can develop ***consensual****, meaningful, communicable, predictive and verifiable, scientific* ***knowledge***. But ***not*** a knowledge of absolute truth of how a studied physical entity "is in itself"; knowledge of *how this or that studied entity appears via the human grid that is acting in order to qualify it*. For without any such grid there is no qualification, so there is no scientific knowledge: this comes out irrepressibly when everything involved is spelled out explicitly.

> A framework that permits to achieve *this* – and iff it involves also mathematical representations – is what here is called "a *theory* of the microstates that belongs to the Mathematical Physics".

But a mathematical global representation *alone* is ***not*** such a theory. It is not rich enough for generating *consensual knowledge*. It is only a mental mathematical representation, either an "interpretation" of some physical-conceptual-operational representation from an already achieved theory from the Mathematical Physics, or just a heap of signs that, via words, generates in the minds submission to calculated results arbitrarily identified to "predictions":

---

[17] *Any* very first – primordial – registration of the result of a measurement interaction is 'transferred', *even in the case of directly perceived entities like in the classical domain* (private exchanges with Henri Boulouet). What is specific indeed of the microstates is the fact that *no qualia can be formed in the observer's mind*. As soon as the studied entity accedes to some sort of direct perceptibility via some apparatuses (microscopes, etc., as it happened historically for molecules and atoms), this absence of qualia *ceases*. But this does *not* entail that the qualia that have been produced in *this way* can be confounded with 'intrinsic properties' of the studied entity.

"A theory from the Mathematical Physics" involves a coherent structure of *conceptualization, **operations**, and **models**, **and** mathematical representations*; it is necessarily *a conceptual-**operational** construct* where certain features *have* to be also ***physically*** operational; for if not then the object-entities-to-be-studied cannot be obtained in this role ***factually*** and furthermore cannot be qualified in a communicable and consensual way; so they cannot be objects-of-science [18].

In the classical physics the models, whether individual or global, emerge spontaneously from the perceptions generated by our biological sensorial apparatuses; they emerge in a hidden, implicit, even reflex way. This strongly supports the tendency to identify a self-consistent representational mathematical structure – alone – with, both, a *theory* from Mathematical Physics *and furthermore with "the reality such as it truly is in itself"*: the coalescence of two hiatuses of which the second one disguises a semantic abyss into a self-contradicting concept: into the impossible notion of knowledge of ways of "being in itself", in the absence of *any* qualification-grid.

But when no direct sensorial perception of the entity-to-be-studied can emerge any more, when no models can be formed any more spontaneously and in a reflex way, these two lumped-together illusory beliefs, dissolve. And then it appears that any action of creation of scientific, consensual, predictive and verifiable knowledge on a considered domain of physical facts is simply blocked as long as are not explicitly defined via some other means:

***(a)*** a *model of any "**one** entity-to-be-studied" that has been **deliberately and operationally** generated* via some operation of generation *G*;

***(b)*** a ***global*** *model* of the physical substance on which *G* is conceived to work in order to there*from* "delimit" the entities-to-be-studied;

***(c)*** individual acts *MesA* of measurement to be performed on a specimen of the entity-to-be-studied generated by one realization of *G*, each one of which produces publicly observable marks that are coded accordingly to specified coding-rules in terms of a definite value $a_j$ of the considered qualifying quantity *A*, *on the basis of some definite reasoning*.

Of course, such conceptual models and such reasoning can be suggested only by some connection with a definite cognitive situation where *direct perceptibility* offers an initial ground for elaborating the modelling and the reasoning; so for elaborating a connection with the *classical* conceptualization, because this, for us, is the unique domain of organized meaning that emerges spontaneously. Only there*from* can we begin a process of modelling, even if later we change the initial modelling grounds [19]. On the other hand, inside *IQM* – that is deliberately required to define with full generality the features of *any* acceptable theory of microstates – no particular conceptual model of any one microstate, nor any corresponding global model (nor any associated mathematical model) can be asserted without perpetrating vicious circularity.

---

[18] All this might seem too obvious to be worth being stated explicitly. But in fact it is very far from being currently accepted. Many proposed "theories" are just a mathematical representation disconnected from any sort of operationality.

[19] An exclusively *mathematical* model can involve a conceptual model, but *this is not always the case* (think of Heisenberg's matrix-formulations); so a strictly mathematical model can leave without any support for deciding what is to be called "an act of measurement of the momentum" (not of the "velocity", nor of the "energy" assigned to one specimen of a factual microstate, but of, specifically, "momentum"). In MMS [2013] (pp. 117-126) I have constructed a "space-time coding" procedure that *identifies* – so labels a posteriori – the results of an *arbitrarily* constructed "test-interaction" *T* between a corresponding test-apparatus and the specimens $\sigma(ms_G)$ of a factually defined microstate $ms_G$' but *without* endowing these results with any meaning that relates them to some previously achieved conceptualization. But – and this is noteworthy – such a coding-procedure, even though it cannot signify in terms that possess some meaning in terms of pre-established conceptualization and so it cannot directly connect to the classical science, nevertheless can *initiate* a process of conceptualization that, indirectly, via intuitive substrata, take profit from the already established conceptualization. Let us also note that nowadays any material entity is conceived to merge with the universal 'sub-quantic substance' so that it is devoid of delimiting contours. ***Delimitation by some G is only a human necessity in the processes of conceptualization, not a "property" assigned via some model to the "studied physical entity" posited to pre-exist as such***. *Furthermore even the absence of spatial delimitation is just a model conceived by human mind, not some sort of unconceivable* representation of a "true property" of the studied entity such as "it really is in itself". This Fata Morgana notion is doomed to vanish.

So *the coding problem cannot be* ***treated*** *inside IQM*.

But inside any mathematical theory of the microstates, coding problems do irrepressibly emerge and *there* they ***have*** to be treated, and the solutions to these problems require well-defined individual models and a corresponding global model, and specification of the reasoning that leads to this or that posit of a coding-measurement-evolution *MesA*.

The preceding point is a very important one. The conceptual situation brought into evidence above refutes the very *possibility* to obey Bohr's positivistic interdiction of any model of a microstate. Which proves that in fact this interdiction has never been genuinely taken into account. It has only enormously intimidated the physicists and pushed them into passive and abstruse acceptance of arbitrarily posited interdictions that they have incessantly violated [20].

The global content of (1.I).2 is summarized graphically in the Fig.1.

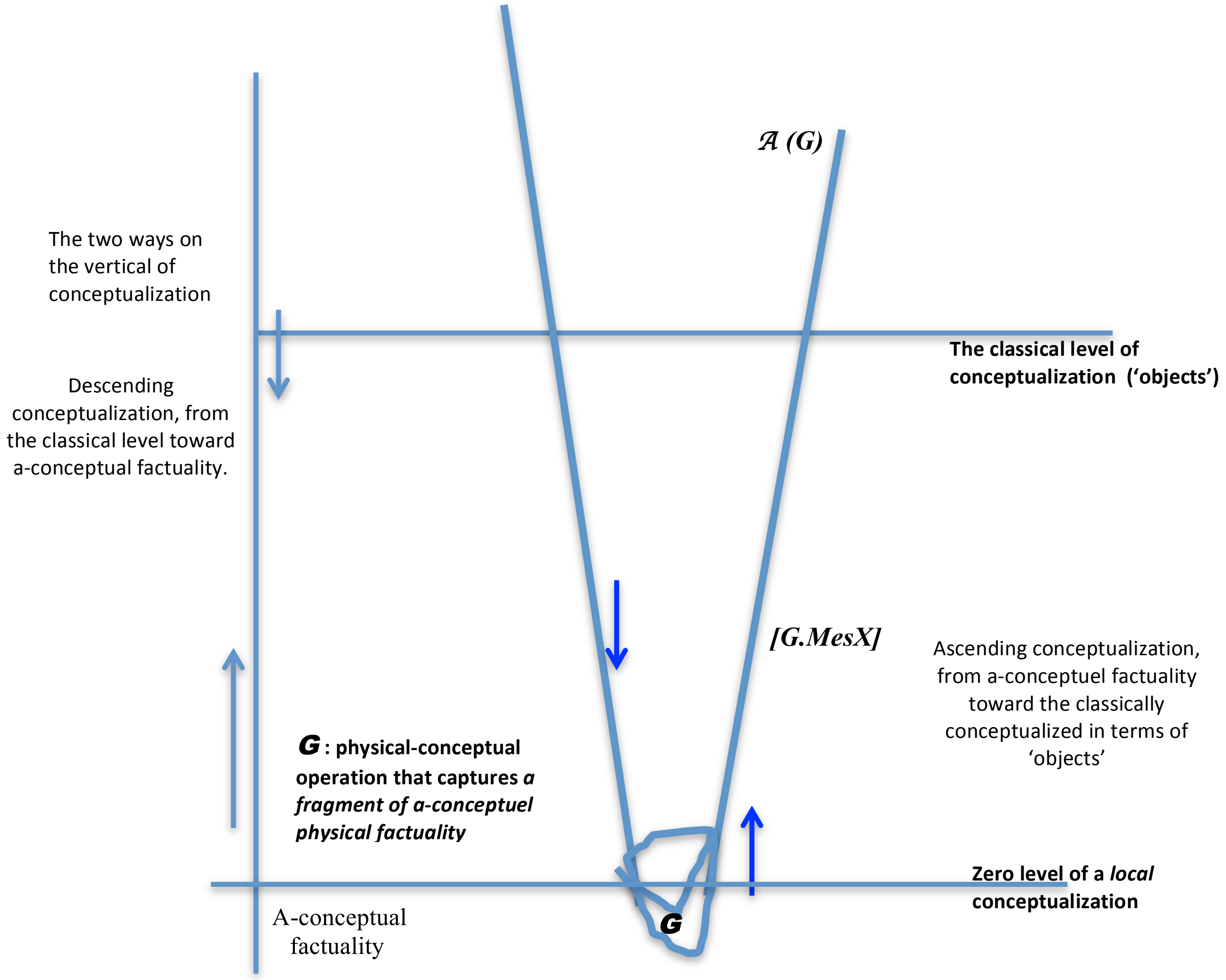

**Fig.1**. *One qualification of one* ***specimen*** *of a microstate: the germ of the structure of a primordial transferred description*

[20] This, quite certainly though implicitly, flaws the nowadays Quantum Mechanics in ways that will have to be explicated and suppressed.

The apparatus for producing the operation of generation $G$ is denoted $\mathcal{A}_{pp}(G)$; the apparatus for producing the measurement interactions for the dynamical quantity $A$ is denoted $\mathcal{A}_{pp}(MesA)$. The basic operational construct that generates the result of only *one* act of measurement-interaction performed upon *on*e outcome of one specimen of the microstate $ms_G$ defined in (1) can be represented as a chain:

$$[(G \leftrightarrow ms_G)\text{-}[G.MesA]\text{-}\{\mu\}_{kA} \textit{ coded in terms of one } a_j)], \quad k_A=1,2,...m_A, \quad j=1,2....J \qquad (3)$$

*The chain (3) that brings forth just one act of qualification of one specimen $\sigma(ms_G)$ of a factually defined microstate $ms_G$ will be called a (one) coding-measurement-succession. It constitutes the very first germ of the factual constructive representation of the process of generation of knowledge on such a microstate. This germ is already endowed with a rather complex inner structure and it already specifies in what a sense the pairs*

***[(****one operation of generation of the entity-to-be-qualified****),***
***(****one operation of qualification of this entity****)]***

*play a basic role in the construction of consensual predictive and verifiable knowledge.*

A chain (3) acts like a fragile narrow bridge over the frontier between the a-conceptual universal physical substance of which the existence is posited by our minds, and the volume of human conceptualization.

In what follows this germ will be developed into a still far more complex concept, namely a general form of a full *scientific description* of a microstate, a deliberate, consensual, predictive and verifiable piece of stable knowledge on a microstate $ms_G$: *the primordial transferred description* of a factually defined microstate $ms_G$ in the sense of *MD*.

# 2.I

# BOTTOM-UP CONSTRUCTION OF THE TRANSFERRED DESCRIPTION OF A FACTUALLY DEFINED MICROSTATE

## (2.I).1. PRELIMINARY CONSTRUCTION OF LANGUAGE: DEFINITION OF 'MICRO-*SYSTEM*', 'MICRO-*STATE* $ms_G$', '*TYPES* OF MICRO-STATES $ms_G$'

### (2.I).1.1. The general problem

In our current life we begin by embedding structures of thought in structures of some current language that emerged and evolves collectively by an anonymous and spontaneous process. But from a scientific point of view the structures of thought expressed inside a current language are most often beds of Procustes because the aim of the natural languages is to be ***contextual*** in order to maximally permit rapid, allusive, suggestive, approximating transmissions of meaning, of poetic connotations, of humour, etc. The accent falls upon local and contextual efficiency in space and time, and upon the harmonics of the central meaning. Whereas the aim of a scientific language is to induce maximally strict and stable *consensus* inside some definite group of consensus, via a priori definitions that point as precisely as possible toward a uniquely defined significance; which can be realized – nearly strictly – only via axiomatic constructions. These two just mentioned two sorts of aims are opposite to one another. And quantum mechanics, like the majority of the mathematical theories of Physics, is not axiomatic, it is a mathematized representation imbedded in the natural language where one relies on contextual communication. This blurs the significance of many basic words that occur currently in the feebly defined verbal support of the quantum mechanical mathematical representations (to *'prepare'* (the 'system', the 'state'); to *'measure'*; *'superposition'* (mathematical, or physical, or both?) etc.). Thereby much confusion is induced. In what follows we suppress beforehand the possibility of several such basic confusions.

### (2.I).1.2. The specific problem

Consider a measurement-interaction involving a specimen $\sigma(ms_G)$ generated by the operation *G* that corresponds to the studied microstate $ms_G$. This produces observable marks that have to be translatable in terms of *one* value $a_j$ of ...... of *what*, exactly? Of one value $a_j$ of only *one* measured dynamical quantity *A*, for ***any sort*** of 'involved microstate', or possibly of *several* such quantities or values of quantities permitted for *some* sorts of microstates? Shall we organize our concepts-and-language so as to require that *one* act of measurement on only *one* specimen $\sigma(ms_G)$ of the studied microstate $ms_G$ brings forth *necessarily* only *one* value $a_j$ of each measured dynamical quantity *A*? Or that it shall necessarily involve – at most – only *one **set*** of 'compatible' quantities (which is not the same thing as in the preceding question)? And, in this case, what exactly does 'compatible' mean? What restrictions are we prepared to accept?

Furthermore, according to (1) *each specimen* of the one micro-*state* tied with one operation of generation *G* can involve one or more other micro-entities (like when *G* creates a pair). How can we name these micro-entities? If we call them 'particles' – as it is very often done – we suggest a model, which we want to avoid inside *IQM*. Could we then speak of one, or two, or more micro*systems* involved by each specimen of a given micro-state? This is less specific about the nature of what is conceived – some sort of more individualized entities – does not contradict the current way of speaking inside quantum mechanics. If then we indeed do choose to call micro-*systems* the somehow more individualized component entities from one specimen of a given micro-state, how are we to count them, according to what observational criteria? What presuppositions have to be incorporated in order to stay in clear agreement with the concepts of a factually defined micro-state and of a specimen of it, in the sense of (1), as well as with the current ways of speaking and thinking that accompany the formal quantum mechanical writings?

The answers are not at all obvious. Inside the current languages the word "system" points usually toward a complex whole that contains 'components'. But inside quantum mechanics, on the contrary, the word "system" – even "*the* system" – points often just toward *what is studied*, no matter whether it is posited to involve one or several components; moreover the term 'micro-state' indicates the dynamical characters of the *whole* studied entity, and the word 'system' points toward exclusively the constant characters of 'a particle'.

All these ways of using words are not severely regulated, while in what follows we want to install rigor in order to avoid false problems. So we define a language that stay in agreement with:

*(a)* The general fact that the concept of 'dynamical state' [21] designates a variable behaviour that involves an invariant material support: violating such a fundamental slope of natural human conceptualization would uselessly waste energy.

*(b) MD*, that introduces the basic posit (1) $G \leftrightarrow ms_G$, with $ms_G=\{\sigma(ms_G)\}$, according to which one operation of generation *G* produces factually one 'specimen' $\sigma(ms_G)$ of the micro-state denoted $ms_G$; while the number of the involved 'systems' is *not* restricted by (1) because this concept is not involved by *MD*.

*(c)* The hidden consensual assumptions that can be identified inside the moving ways of speaking and writing practised inside quantum mechanics.

***Definition [(micro-state) and (micro-system).*** The concept associated to the *persistent* characters (mass, charge, etc.) assigned to any element from the set $\{\sigma(ms_G)\}$ of mutually distinct specimens of the micro-*state* $ms_G$ in the sense of *MD* is called a micro-*system* involved by $ms_G$.

***Definition [(*one *micro-system) and (*one *micro-state of* one *micro-system)].*** Consider a micro-state $ms_G$ that is such that *one* act of measurement accomplished upon *one* specimen $\sigma(ms_G)$ *of* $ms_G$ can bring forth only *one* group $\{\mu\}_{kA}$ of observable marks significant in terms of a value of the measured quantity. We shall say that this micro-state $ms_G$ brings in specimens $\sigma(ms_G)$ each one of which consists of *one* micro-*system S* and so we shall call it in short *a micro-**state** of (with) **one** micro-**system***.

***Definition [*one *micro-state of* n *micro-systems].*** Consider now $n>1$ micro-*systems* of a type of which we know that, for each one of them *separately* it is *possible* to generate a micro-state in the sense of the preceding definition; which, if done, would lead to '***n** micro-states of **one** micro-system*' in the sense of the preceding definition. But let *G(nS)* (*nS* : *n* systems) denote only *one* operation of generation that, acting upon some physical initial support that relatively to *G(nS)* is regarded as 'prime matter', has generated *one common* micro-*state* for *all* these *n* micro-*systems*; or even, out of some initial substratum, *G(nS)* has simultaneously radically created the *n* micro-*systems themselves* that are contained by each specimen of the studied common *one* micro-state[22, 23]. In both these cases we shall say that the micro-state generated by *G(nS)* is *a micro-state of (with) n micro-systems* and we shall denote it by $ms_{G(ns)}$[24].

[21] A somewhat self-contradicting expression.
[22] This is the case when *G(nS)* consists of some interaction with pre-existing elementary particles that brings forth "a pair".
[23] This way of speaking seems convenient in both fundamental quantum mechanics and the fields-theories.

***Definition [complete measurement on* one *micro-state of* n *micro-systems].*** *One* act of measurement performed on one specimen $\sigma(ms_{G(ns)})$ of a microstate $ms_{G(ns)}$ of *n* micro-systems, *can produce at most n distinct groups of observable marks* signifying *n* observable values of dynamical quantities. An act of measurement that effectively realizes this maximal possibility will be called a *complete* act of measurement on the one specimen $\sigma(ms_{G(ns)})$ of the one micro-state $ms_{G(ns)}$ of *n* micro-systems. The quantities *A* and the values $a_j$ to which these *n* distinct groups of marks are tied, are permitted to be either identical or *different*.

***Definition [*in*complete measurement on* one *micro-state of* n *micro-systems].*** One act of measurement accomplished upon one specimen $\sigma(ms_{G(ns)})$ of a microstate $ms_{G(ns)}$ of *n* micro-systems that produces *less* than *n* distinct groups of observable marks, will be called an *incomplete* act of measurement on $ms_{G(ns)}$.

Finally, for self-sufficiency of this sequence of definitions, we restate here telegraphically the definition from 1.I of a micro-state $ms_{G(G1,G2,..Gk)}$ generated by a composed operation of generation:

***Definition [*one *micro-*state *generated by a* composed *operation of generation].*** Consider – indifferently – either *one* micro-state of *one* micro-system, or *one* micro-state of *n>1* micro-systems. If the specimens of this micro-state are generated by a composed operation of generation $G(G_1,G_2,..G_k)$ in the sense defined in 1.I then we call it *a microstate with composed operation of generation*.

***Definition [*one *'bound' micro-*state *of several micro-*systems*].*** This is the usual verbal designation of the result of a *natural* operation of generation, i.e. accomplished in consequence of the physical laws of nature, before any human aim of investigation (like in the case of the natural realization of an atomic structure). But in principle it can be also thought of in terms of the result of a *composed* operation of generation (so much more so as a bound micro-state of several micro-systems manifests systematically 'interference-effects').

We hold that the preceding definitions insure, both, global coherence relatively to the implications carried by the language practised inside nowadays microphysics, and continuity with the basic principles of the classical conceptualization and language, and they realize this without asserting any definite conceptual model of a specimen of a factual microstate. If one contests the adequacy of some feature from these definitions, he should specify the reasons for the contestation and propose a better usage of words. Meanwhile the definitions from (2.I)1 are adopted throughout what follows. Their force to clarify and to solve stubborn problems will soon become obvious.

We now announce the following

***Choice***. *In this work the bound microstates will occupy a very marginal position*.

We make this choice on the basis of two reasons. The first one is that a bound state can pre-exist any desired investigation, just as it is supposed for classical 'objects'. The second reason is that furthermore, to a bound state it is possible to assign – in a certain relative sense of course – *a definite spatial delimitation*, again as in the case of a classical mobile. These two features might explain why the mathematical representation of bound microstates has constituted the natural passage from classical physics to quantum mechanics when the practised approach still was top-down. But in this work we want to explicate and stress the radical novelties imposed by a bottom-up representation of microstates. Only these novelties will permit to bring into evidence:

- To what a degree the scientific representations can become a deliberate consensual construction of which the necessary and sufficient conditions of possibility depend strongly on the involved cognitive situation (that can evolve with the evolution of the sciences and the techniques).
- To what a degree this should modify our conception on scientific representation, and stress the utmost importance of the relativities to the constraints and the aims that act.

[24] The posit (1) entails that the uniqueness of the operation *G(nS)* is to be a priori conceived as a source of certain global observational specificities of each specimen of $ms_{G(ns)}$ and so of $ms_{G(ns)}$ itself.

And these novelties are brought forth – specifically – by unbound microstates. So here we are quasi exclusively concerned with unbound microstates. The bound microstates will finally be naturally absorbed in the new representation constructed here.

On the basis of the contents from (2.I).1 we enter now upon the construction of the general concept of description of a microstate.

## (2.I).2. DEFINITION OF A *PRIMORDIAL* TRANSFERRED DESCRIPTION OF AN UNBOUND FACTUAL MICROSTATE

What follows is formulated in terms that are valid for *any* factual microstate.

### (2.I).2.1. Preliminary requirements

We start again from the remark that inside current thinking and speaking the qualifications are in general just asserted freely concerning an object-for-qualification that is conceived to pre-exist *such* as we qualify it (this tree *is* big, today the air *is* cold, etc.); whereas a scientific description is required to be endowed with explicit consensual definitions that are *communicable* with precision and without restriction to co-presence of the members of a specified group of consensus, and to be predictive and verifiable. All these requirements *subsist* when it is recognized like in (1.I) that the qualifications that have been obtained cannot be considered to be properties of the entity-to-be-described alone, isolated from the measurement-interaction. And the requirement of verifiability entails *repeatability* of the involved operations as well as the existence of some definite descriptive *invariant* brought forth by many repetitions of the action of qualification: only such invariants can permit prediction and verification. Now, in the case of microstates these implications of the condition that restrict scientific knowledge entail specific and non-trivial consequences among which the following are the most important.

#### *(2.I).2.1.1. Consequences of the requirement of repeatability*

A classical mobile is conceived as an "object" that in general pre-exists to qualifications of it; it stays available "there outside". So in general a measurement operation *MesA* on a classical "mobile" can be conceived separately from an operation of generation *G* of that mobile [25]. But an unbound microstate-to-be-studied *does **not** pre-exist in some known and attainable way*, like a macroscopic "object"; and furthermore in general it is destroyed by the act of qualification. So (as already observed in *(1.I).2.2.2*) the observer-conceptor, if he wants to create a germ of knowledge on such a microstate, has to radically generate that microstate before achieving on it an act of qualification, so to realize a whole 'measurement-succession' *[G.MesA]*: There steps in explicitly the necessity to realize repeatedly and in a physical-operational way whole pairs

***[****(one operation of generation of the entity-to-be-qualified)****,***

*(one operation of qualification of this entity)****]***

And this, for scientific descriptions, entails an arm-wrestling between *IQM* and the classical presuppositions. Indeed:

In classical mechanics the studied mobile is admitted to be publicly observable, and the registration of the result of an act of measurement does not destroy the studied mobile, nor does it necessarily perturb notably its dynamical state. So it has been possible to conceive and to formulate basic classical mechanical laws as *individual* invariants with respect to repetitions of only a given act of measurement *MesA*. Furthermore a classical law is posited to characterize exclusively the studied entity *itself*, it is regarded as the revelation of a behavioural 'property' of, exclusively, the studied mobile; a classical mechanical law is not conceived to characterize only the whole of the measurement interaction. When the results of repeated measurements on the studied mobile manifest a statistical dispersion this is posited to be due exclusively to imprecisions in the acts of measurement, which withstands the *knowledge* of the exact individual value $a_j$ of the measured quantity $A$ that is "possessed" by the entity-to-be-studied, but does not challenge the assumption of pre-existence of this value. Furthermore, according to the classical thinking this obstacle in the way of exact knowledge of the value $a_j$ of $A$ is doomed to disappear asymptotically while progress is achieved in the techniques of measurement. On the basis of these arguments it is admitted that one can advance toward knowledge of how the studied physical entities "truly are, exactly and in themselves".

---

[25] This, in fact, is confusion. Indeed – by definition – an operation of generation *G* in the sense of (1) is what entails an entity in the *role* of entity-to-be-studied. And a classical mobile that just is conceived to 'exist' is not thereby automatically in the role of entity-to-be-studied. Always some supplementary act is necessary from the part of the observer-conceptor, even if this consists of just bringing the mentioned mobile inside the domain of perceptibility by the observer-conceptor and focusing attention upon it. As already remarked, the existence of an operation of generation *G* is a universal character of any act of qualification, so of any act of creation of a piece of knowledge. This fact is far from being trivial: it is part of the hidden key that opens up access to a path toward unification of microphysics and quantum gravitation.

This sort of illusory scientific realism is quasi unanimous.

Whereas the factually defined concept of microstate $ms_G$ from *MD* is organically tied with a conceptual segregation of a radically different nature that we recall synthetically:

***(a)*** Since the unavoidably *physical* operation of generation *G* can be defined by only a finite set of parameters while the domain of physical reality from which this operation stems, as well as that on which it acts, are endowed with the unlimited singularity of the being, it would be *unconceivable* that repetitions of *G* bring always forth specimens $\sigma(ms_G)$ of the studied microstate $ms_G$ that are all mutually identical, i.e. the posit $ms_G \equiv \{\sigma(ms_G)\}$ where $\{\sigma(ms_G)\}$ is a whole *set* of mutually distinct specimens, is quintessential for factual microstates; it introduces a basic 'statisticity' that is *not* asserted as a physical scientific truth about the studied entity itself, but only as an unavoidable fact involved by the deliberate human aim to construct consensual, predictive and verifiable knowledge on factual microstates.

***(b)*** Since one act of measurement *MesA* also cannot be defined otherwise than by a finite set of macroscopically specified parameters, when it is repeated its own effects equally cannot be conceived otherwise than dispersed, in general [26].

***(c)*** The specimens $\sigma(ms_G)$ are not observable, while the observable result of one succession *[G.MesA]* (cf. figure 1) characterizes exclusively this succession *as a whole* in a way that cannot be analysed further.

***(d)*** So repetitions of the *whole* succession *[G.MesA]* are unavoidable and these lead in general to a statistical distribution of the observable results of the achieved successions that – quintessentially – cannot be removed nor analysed into the dispersing effects entailed by *G* and those entailed by *MesA* (MMS [2002B], [2006], [2017B]).

***(e)*** So: The researched *law*-like invariant – a concept that is 'deterministic' by definition – can manifest itself observably *only* in terms of *probabilistic convergence of repeated statistical distributions of results of sequences of very numerous repetitions of the whole succession [G.MesA].*

Let us remind now that according to the classical view specified before, the entity-to-be-studied is defined via individually specifying predicates and it obeys laws that, in principle, are posited to be individual invariants but that are only imperfectly observable as such because of imprecisions of measurement. In consequence of *MD* this view *cannot be regarded to be fundamental and general.* Indeed the existence of an operation *G* of generation of the entity-to-be-studied at the beginning of any process of generation of consensual knowledge has a *universal* character ((1.I).1.6) so that in fact we *always* perform measurement-successions *[G.MesA]* (not directly measurements) and so the framework *(G, MesA, [G.MesA])* is general, it only works differently in different cognitive situations. So the classical view holds – approximately – only in the cases in which the considerations from *(a)* fade out because the operation *G* introduces a dispersion that is negligible in some sense (for instance, for the majority of the macroscopic directly perceived "objects" (MMS [2002B], [2006]), or for the purely *mental* conceptual-mathematical representations of celestial entities-to-be-studied (black holes, galaxies) introduced by a purely mental operation of generation *G* and that can be confirmed or invalidated by – exclusively – verification of consensually observable predictions that have been drawn from these representations in a purely *deductive* way)[27]. But for consensual predictive and verifiable knowledge on factual microstates all the requirements *(a),(b),(c),(d),(e)* are fully active and the classical view breaks down. So let us examine the consequences imposed by these requirements upon the possible sort of descriptional invariant brought forth by repetitions of a given measurement succession *[G.MesA]* on a microstate.

[26] When a unity is defined it sets a conventional lower bound to the dispersion that is taken into account. The nanotechnologies might reduce strongly the dispersion of certain specifically targeted observable effects.

### *(2.I).2.1.2. A consensual, observable, predictive and verifiable descriptional INVARIANT concerning factual microstates*

In general when one given succession *[G.MesA]* is repeated one obtains *different* results $a_j$. So in general a whole statistic of results $\{a_j\}$, $j=1,2,...J$ emerges, notwithstanding that in each succession *[G.MesA]* each one of the two operations *'G'* and *'MesA'* is 'the same' *with respect to the two finite groups of parameters that define it.* This is a fact. We are placed on an observational ground that – factually and with respect to knowledge – has a primordially statistical character. Whereas on the other hand any 'law' that permits predictions and verification of these is an invariant with respect to repetition. So the unique possible sort of observational law consists of a primordially *probabilistic* invariant of the statistical distributions of the possible results $a_j$ of realizations of the succession *[G.MesA]*. Which involves [a big set of [*N* repetitions of the succession *[G.MesA]* with *N* very big]] *and the concept of 'probabilistic convergence' of these statistical distributions* introduced by the classical theory of probabilities. This, in *MD*, shifted us upon a postulated level of ***'deterministically*** *probabilistic' conceptualization expressed by the one-one relation (1)* $G \leftrightarrow ms_G$ *with* $ms_G=\{\sigma(ms_G)\}$.

It might seem counter-intuitive to assert that a probabilistic qualification is a deterministic qualification, but – considered globally – it *is* one, in this sense that the recurrence of the convergence and of its target-value are predictable.

So we consider now the mathematical classical concept of probabilistic convergence: This is a *purely formal and non-effective* global concept embodied in the *mathematical* weak law of large numbers; from A to Z this concept is constructed inside the formal, general theory of measures in the *mathematical* sense that posits by definition the general relations between any probability-law ($\pi(e_j), \forall j$), the corresponding universe of elementary events ($e_j, \forall j$), and an algebra of events posited on this universe.

But the weak law of large numbers says strictly nothing concerning the probability law *from a particular well defined factual situation: it does not* ***specify*** *numerically this probability distribution; it does not specify one by one the real numbers* $\pi(e_j)$ *for this or that definite value of j.*

I have called this "the aporia of Kolmogorov" (MMS [2006], [2014]) in order to just draw attention upon the fundamental difference between the purely mathematical concept of probability and *a factual concept of probability* of which the semantic content is *not* exhausted by the weak law of large numbers. I spell all this out in a so childishly explicit way because the physicists seem to believe that for physics it “suffices” to dispose of the non-effective classical mathematical theory of probabilities of Kolmogorov. But *this is a fundamentally false belief*. Kolmogorov’s mathematical framework not only is non-effective, but moreover *it is very insufficiently* ***comprehensive***. This will appear below where can be watched the progressive outline of something new that can be called "a theory of physical probabilities" belonging to the Mathematical Physics.

[27] The de Broglie-Bohm formal representation of the Universal Substance introduces a limiting conceptual situation: *both G* and *MesA* are simply absent – basically – and so there is no source of observational dispersion any more, we are in presence of just a global and mathematically expressed metaphysical model that remains to be explicitly connected to this or that local consensual, predictive and verifiable knowledge that – necessarily – involves a superposed specification of local *factual* successions *(G,MesA)* and repetitions of these.

Let us begin by defining below a factual equivalent of a mathematical probability-law. Consider the weak theorem of large numbers:

$$\forall j,\ \forall(\varepsilon,\delta),\quad (\exists N_0:\ \forall(N\geq N_0)) \Rightarrow [\boldsymbol{\Pi}\,[|n(e_j)/N-\pi(e_j)|\leq\varepsilon]]\ \geq\ (1-\delta) \qquad (4)$$

(The significance of the notations is well known). From this it is possible to extract explicitly a relativized *finite* implication that is defined below:

The probability $\pi$ and the meta-probability $\boldsymbol{\Pi}$ are ***limit***-(real numbers) toward which, at infinity, converge the corresponding distributions of relative frequencies. But consider a universe of events $U=[e_1,e_2,....e_J], j=1,2,...J$, with $J$ a finite integer. *If* the probability $\pi(e_j)$ of an event $e_j$ is postulated to exist for any $e_j$, *then* (4) asserts that for any pair of two arbitrarily small real numbers $(\varepsilon,\delta)$ there exists an integer $N_o$ such that – for any $N\geq N_0$ and with an uncertainty not bigger than $\delta$ – [the meta-probability $\boldsymbol{\Pi}$ of the event $[|n(e_j)/N-\pi(e_j)|)\leq\varepsilon]$ that the relative frequency $n(e_j)/N$ observed for the event $e_j$ inside a sequence of $N$ events from $U$ does not differ from $\pi(e_j)$ by more than $\varepsilon$] is bigger than $(1-\delta)$. This assertion alone, such as it stands, i.e. *the passage to the limit being suppressed* – with $N_0$ chosen freely and with the corresponding pair $(\varepsilon,\delta)$ – will be considered in what follows to define a general and ***factual***, *finite* numerical probability-value of the individual event $e_j$.

The set $\{(\varepsilon,\delta,N_0)$*-probabilities* $\pi(e_j),\ \forall j\}$ will be called *the factual* $(\varepsilon,\delta,N_0)$*-probability law of emergence of the* $e_j$*, with respect to the triad* $(\varepsilon,\delta,N_0)$ [28].

In our case $U$ consists of the finite spectrum of values $a_j$ assigned to $A$. And we make the – strong – ***assumption*** that the systematic repetition, for any $A$, of the corresponding succession *[G.MesA]* introduces sufficient constraints for entailing a factual $(\varepsilon,\delta,N_0)$-probability law $\pi(a_j)$ for any association between a chosen pair $(\varepsilon,\delta)$ and the relative frequencies $n(a_j)/N$ found for the values $a_j$ that are present inside the chosen qualification grid (2) $gq[A,a_j,MesA,cod.proc(a_j)]$ [29], with $j=1,2,...J$ (which amounts to an alternative formulation of the posit (1) $ms_G\Leftrightarrow G$).

So, given a factually defined microstate $ms_G$, the posit (1) introduces for any couple of pairs $((G,A),(\varepsilon,\delta))$ a corresponding ***factual*** $(\boldsymbol{\varepsilon},\boldsymbol{\delta},\boldsymbol{N_0})$***-probability law***

$$(\varepsilon,\delta,N_0)\text{-}\{\pi(a_j),\ \forall j\}_G,\quad A\ \textit{fixed} \qquad (5)$$

This involves ***a new concept of probability, a physically defined concept of probability*** not a purely mathematical concept that specifies only the *general* structure of any probability law, while the probability law that works in a particular concrete situation remains to be established by introduction into the mathematical void mould of factual data of which the genesis is not defined. ***We call this "factual probability"***.

---

[28] In (MMS [2014B]) this ***factual*** and ***numerically defined*** probability law has been constructed from an *interpretive assumption* on the concept of probability and ***it has been proved compatible with the weak theorem of large numbers*** (cf. also (Wasserstein&Lazar [2016], Leek&Penn, [2015] concerning the conceptual status of – merely – a *statistic*, with respect to the conceptual status of a *probability law*).

[29] The event $a_j$ being identified from a group of observable physical marks, via the utilized coding-procedure that inside *IQM* cannot be defined but that is supposed to have been defined inside the employed theory of microstates.

***(2.I).2.1.3. Compatibility of quantities versus specificity of the 'knowledge' on a microstate***

The initial factual and methodological definition (1) of the microstate-to-be-studied amounts to merely label the unknown and unobservable microstate '$ms_G$' by the operation $G$ that is supposed to have produced it; the final purpose is to substitute to the mere label '$G$', a 'description' of the microstate-to-be-studied in terms of predictive and verifiable knowledge tied with – specifically – this entity *itself*. Now, does a factual probability law (5) constitute such knowledge?

No, not yet, because nothing entails that only one probability law (5) established for $ms_G$ relatively to only one dynamical quantity $A$, cannot be observed also for *another* microstate different from $ms_G$, i.e. generated by another operation of generation $G'\neq G$. The law (5) alone might not be specific of $ms_G$.

It seems likely however that two probability laws (5) corresponding to two mutually different dynamical quantities $A$ and $A'\neq A$ – considered conjointly – might already constitute an observational factual specificity associable to the considered particular microstate $ms_G$ generated by $G$. While a fortiori – in as far as the language introduced in *MD1* resists to the observable facts – the *totality* of all the mutually different laws (5) that are defined for $ms_G$ is quasi-certainly specific of this microstate.

But what sort of *difference* between two dynamical quantities $A$ and $A'\neq A$ is determining in this context? Consider two distinct dynamical quantities $A$ and $A'\neq A$ and a given type of microstate $ms_G$, in the sense of the definitions from (2.I)1.

> We shall say that *A and A'≠A are mutually compatible* ***with respect to the microstate-to-be-sudied AND a given coding-measurement-evolution MesA≡MesA'*** *(we write MesAA')* iff it is possible – via ***one*** act of *MesAA'* performed upon ***one*** specimen of the considered microstate – to obtain one group of marks $\{\mu\}_{(AA'),k}$, $k=1,2,...m_{AA'}$ wherefrom, on the basis of previously established knowledge and assumptions, are calculable – ***both*** – one value $a_j$ of $A$ and one value $a'_{j'}$ of $A'$.

***This is a necessary and sufficient general condition of RELATIVE compatibility***. If it is reazlized then $A$ and $A'$ "exist simultaneously with respect to $ms_G$", if not, they do not.

Nothing entails a priori that this condition can be met factually in one and same way for any given sort of microstate (2.I)1 and any given pair *(A,A')*; nor that for a given sort of microstate (2.I)1 and a given pair *(A,A')* it can necessarily be satisfied in some way. *The answer to these questions depends upon the* ***model*** *that is made use of, upon the* ***definition of the qualifying quantity A****, upon the available technical possibilities,* etc. So it is not possible to establish a general answer inside *IQM*; answers can be established only inside a given *theory* of the microstates, relatively to its specific assumptions and requirements.

But in order to explore more thoroughly the conceptual situation entailed by the definition of relative compatibility of two qualifying quantities $A$ and $A'$ posited above, we examine below an important example that is familiar because inside the nowadays theory of the microstate it is posited in absolute terms (and with a more or less implicit physical meaning) and is asserted as an expression of a "factual truth".

Suppose that the microstate-to-be-studied $ms_G$ is a microstate of *one* micro-system. In this case each specimen of $ms_G$ consists of only one microsystem and in these conditions the requirement posited above ***is*** insured by the possibility to achieve for both $A$ and $A'$ *a unique* ***physical*** *common measurement-interaction upon* ***one*** *specimen* of $ms_G$. So the common interaction has to cover a unique *common* space-time support and to finish by the registration of a unique group $\{\mu\}_{(AA'),k}$, $k=1,2,...m_{AA'}$ of brute observable marks. In this case a 'difference' between $A$ and $A'$ can be worked out only *after* the realization of this unique common physical-operational interaction, by the adjunction of exclusively conceptual specifications and of corresponding calculi that construct two conceptually distinct *values*, $a_j$ and $a_j' \neq a_j$ to be assigned, respectively, to *A* and to $A' \neq A$ [30]. If the condition required above can be realized, here we shall say that *A* and *A'* are *mutually compatible quantities with respect to a microstate of one microsystem*; if this condition cannot be realized we shall say that *A* and *A'* are *mutually incompatible quantities with respect to a microstate of one microsystem*. In the first case the two factual probability laws (5) constructed for *A* and *A'* introduce a poorer factual constraint than in the second case, so the corresponding knowledge is less specific. Then a maximally specific knowledge on the studied microstate is obtained by establishing the probabilistic behaviour of this microstate with respect to *all* the groups of mutually in-compatible dynamical quantities that are defined for the studied microstate.

But suppose now that $ms_G$ is a microstate of two (or more) microsystems. In this case two (or more) mutually distinct measurement-interactions *can* be accomplished on different systems from a unique specimen $\sigma(ms_G)$ of the studied microstate $ms_G$ (cf. the definitions from (2.I)1 and also the future point (3.I)2). So in this case here we shall say that ***any*** *two different quantities A and to* $A' \neq A$ ***can*** *be compatible* **with respect to** *a microstate of two or more microsystems*. And the maximally specific knowledge on the studied microstate is obtained by establishing its probabilistic behaviour with respect to *all* the dynamical quantities that are defined for it [31].

This settles the question of specificity with respect to the studied microstate, of the knowledge on this microstate captured in a factual probability law (5).

Let us note that: The concept of compatibility of dynamical quantities that has been defined here in connection with the question of specificity of the knowledge created concerning the studied microstate, is essentially ***relative*** to:

- the concept of ***ONE individual*** specimen of the studied microstate;
- the *sort* of considered microstate, in the sense of the definitions from (2.I)1.
- the *coding procedure* that is involved, so also the *model* of a microstate that is presupposed in the theory that is made use of;
- the available techniques for measuring (which in general *vary* while time passes).

This conclusion is striking when it is compared to the well-known concept of compatibility of qualifying quantities defined in the nowadays Hilbert-Dirac formulation of the quantum mechanics [32].

### (2.I).2.2. The primordial transferred descriptions of a factual microstate.

The considerations from the preceding point lead us to posit by definition that – even though the laws (5) do *not* concern *exclusively* the studied microstate $ms_G$ itself, i.e. separately from the measurement interactions from the successions *[G.MesA]*, $\forall A$ that led to them – nevertheless:

The set

$$\{(\varepsilon,\delta,N_0)\text{-}\{(\pi(a_j)\},\ \forall j)\}_G\},\ \forall A \qquad (5')$$

of *all* the factual $(\varepsilon,\delta,N_0)$-statistical-probabilistic laws (5) established with respect to *one* given operation of generation *G* and *all* the dynamical quantities *A* defined for a microstate, can be regarded as a *mechanical description 'of $ms_G$'*. Indeed, it is the maximally specifying characterization that can be realized for the considered microstate $ms_G$ in the sense of *MD*, and it is a characterization that is specifying with a strong degree of certainty. So, to the initial definition (1) of the microstate $ms_G$ that only labels this microstate by the operation *G* that generates it, and then, to one chain (3) that endows us with a very first unstable dot of qualification tied with this microstate itself, (5') substitutes finally:

- a characterization of $ms_G$ in terms of a whole *stable* and dense structure of communicable, consensual, predictive and verifiable pieces of observable factually-probabilistic data,
- that exhausts the defined possibilities to qualify this microstate,
- and that are all tied with this particular microstate *itself*, with effects of ***its*** interactions with measurement procedures.

---

[30] For instance, for the classical quantities ***p*** and $p^2/2m=T$ it is possible to first determine in a physical-operational way the numerical value of the common basic quantity $|p|=m(v_x+v_y+v_z)$, and out of this basic operational determination, to work out afterward, conceptually, the two results '***p***' (a vector) and '$p^2/2m$' (a scalar) that are mutually distinct from a conceptual point of view as well as by their numerical values).

[31] This is the case in the nowadays theory of the microstates, $QM_{HD}$.

[32] So the definition posited here involves that according to *IQM* the nowadays quantum mechanics introduces concepts of mutual compatibility or incompatibility of dynamical quantities ***that are uncritically assigned AN ABSOLUTE, INTRINSIC NATURE***.

Moreover, via the coding-procedures *cod.proc($a_j$)*, *∀A*, posited to be necessarily involved by the definitions of the measurement interactions *MesA*, *∀A*, from the theory of microstates that is employed, the information contained in (5') is intelligible, in this sense that it is connected to the already previously constructed classical mechanics.

So (5') finally installs the concept of a microstate $ms_G$ as a scientific concept that is endowed with a definite, stable and specific, intelligible 'own' content[33].

Nevertheless the sort of knowledge represented in (5') violates strongly the current classical ways of thinking in terms of "objects" that – as delimited wholes – are endowed with a delimited and stable global space-time location entailing a definite inside and a corresponding outside, as well as an *inner* organization conceived in terms of *properties* that these objects would *possess*.

Moreover the genesis and the content assigned to (5') violates surreptitiously but radically the clear-cut conventional views on 'objective' facts. The set of relativities that mark (5') concerns characters of the human observer-conceptor (his ways of conceiving, thinking and acting and his technical possibilities) at least as much as it concerns the studied microstate. One is led to speak now much more cautiously, namely in terms of only inter-subjective consensus on predictions and verifications of outcomes of human methodological ways of operating. Thereby the classical notion itself of "knowledge *of some 'thing'*" recedes also.

### *(2.I).2.2.1. Notations, denominations, comments*

Let us now immediately organize and denote in detail the new sort of knowledge involved by (5'). In order to deal efficiently with all the unusual descriptional elements introduced here we shall now improve and summarize the names and notations associated with this knowledge[34].

- The grid of qualification introduced by a dynamical quantity *A* defined for microstates will be called *the aspect-view A*. The definition of each aspect-view *A* is assumed to contain the explicit specification of a coding-rule, in order to compensate the absence of direct perceptibility and of qualia assignable to the studied microstate itself. This is what insures a way to associate a meaning in terms of a definite value $a_j$ of *A,* to the group of brute observable marks $\{\mu\}_{kA}$, $k_A=1,2,...m_A$ produced by one act of measurement-interaction from a succession *[G.MesA]*.

- The whole set of all the dynamical quantities defined for a microstate will be called *the mechanical view defined for a microstate* : $\{A\}\approx V_M$ ('$_M$': mechanical)

---

[33] From now on, for the sake of simplicity, for a usual repetitive index like '*j*' in $a_j$ we shall write *∀j* instead of *j=1,2,....J*, keeping in mind that the cardinal *J* is finite.

[34] These insert *IQM* explicitly in the general Method of Relativized Conceptualization, *MRC*, both conceptually and verbally.

- A pair

$(G,A)$ (6)

that founds the operational succession *[G.MesA]* is called an *epistemic referential*; in particular, the pair $(G,V_M)$ is called *the mechanical epistemic referential.*

- A triad

$(G, ms_G, A)$ (6')

of the basic ***genetic*** *elements of (5')* will be called *the genetic triad of the factual probability law (5')*. It can be regarded like a sort of inorganic physical-conceptual string of *DNA*.

- The whole set

$\{[G.MesA]\}, \quad \forall A \in V_M$ (7)

of repeated successions of operations of the general form *[G.MesA]* achieved by use of all the genetic triads (6) realized inside the process (5') will be called the *genesis* of (5').

- The brute result of the genesis $\{[G.MesA]\}, \forall A \in V_M$ of (5') consists exclusively of the set-of-sets of observable marks

$\{\{\mu_{kA}\}, k_A=1,2,...m_A, \forall A \in V_M\}\}$ (8)

These are the factual data produced by (5').

The totality (8) of all the factual data emerges at very dispersed moments, and also very dispersed spatially, on various registering devices of possibly various apparatuses. Observationally, this totality consists of just heaps of traces of vanished interactions, transmuted into meaning by a man-made operational-conceptual-methodological machine[35]. These heaps of traces however hide inside them a very elaborate unity of *human* curiosity, project and method. In a still non-expressed way, the factual data from (8) are already marked in their inner content by all the organizing relativities that inside (5') have been endowed with an explicit, intelligible and consensual final expression via the use of some definite model of a microstate. Nevertheless the factual data from (8) and their explicitly meaningful final expression (5') are devoid of any own space-time organization, as well as of any qualia assignable to the studied microstate $ms_G$ alone. This, of course, is a striking feature of *any* probabilistic description. But here, in consequence of total non-perceptibility of the entity to be studied, it acquires a limiting degree of purity.

The definitions (5) and (5') of the primordial probabilistic predictive laws concerning $ms_G$ – separated from their geneses (7) – will be *re*-noted now as, respectively:

$(D/A)(ms_G) \equiv \{(\varepsilon,\delta,N_0)\text{-}\pi(a_j), \forall j\}_G$, $A$ fixed (9)

$D_M(ms_G) \equiv \{\{(\varepsilon,\delta,N_0)\text{-}\pi(a_j), \forall j\}_G, \quad \forall A \in V_M$ (9')

The notation $(D/A)(ms_G)$ from (9) will be called *the primordial transferred description of the microstate* $ms_G$ ***with respect to the mechanical qualification A*** (a description entirely 'transferred' on registering devices of apparatuses). It is the basic concept of transferred description.

The notation $D_M(ms_G)$ from (9') will be called *the primordial transferred* ***mechanical*** *description of the microstate* $ms_G$. The writings

$(D/A)(G,ms_G, A)$ or $D_M(G,ms_G, V_M)$ (10)

[35] Let us stop a moment to realize how simplistic it would be to assert that this knowledge pre-existed and has been 'discovered', when so obviously it has been invented and constructed.

can replace the expressions from the second members from (9) and (9'), respectively, when one wants to recall the geneses of, respectively, the factual probability law from (9) or the factual probability law from (9'): they stress that in the case of microstates the gained consensual knowledge, and the conceptual-(*physically*-operational) generation of this knowledge by the human observer-conceptor, constitute together an intimate unity wherefrom the intelligibility stems.

Considered globally, this whole point *(2.I).2.2.1* is an immediate application of the general Method of Relativized conceptualization *MRC* (MMS [2002A], [2002B], [2006]).

# 3.I

# THE PROBABILITY TREE OF A PRIMORDIAL TRANSFERRED DESCRIPTION OF AN UN-BOUND MICROSTATE

A primordial transferred description is a radically basic and new cognitive concept. In spite of all the specifications and comments from the section (2.I)2this new concept still remains too abstract for triggering an intuitive and sufficiently detailed as well as integrated perception of the whole novelty of the concept of a primordial transferred description. This is so because this novelty is quintessentially genetic. So in order to bring it into evidence we shall now construct graphic representations of the contents carried by the written representations [(1)→(10)]. We shall do this only for the two main sorts of unbound microstates defined in (2.I)1, namely a microstate of one micro-system and a microstate of two (or several) micro-systems. This will suffice for bringing forth that this concept involves a genuine revolution of classical probabilities and logic [36].

## (3.I).1. THE PROBABILITY TREE OF AN UNBOUND MICRO-*STATE* OF *ONE* MICRO-*SYSTEM* WITH NON-COMPOSED OPERATION *G* OF GENERATION

Throughout what follows we distinguish clearly between distinct levels of conceptualization.

We begin with the basic case of one unbound microstate of one micro-system. For this case we shall be able already to reveal non-classical specificities involved by (9) and (9').

### (3.I).1.1. Individual level of conceptualization

By definition the very numerous successions of operations *[G.MesA]*, $\forall A \in V_{Mec}$ involved in a genesis (7) start all with one same operational realization of a 'trunk'-operation of generation *G*. But afterward – in consequence of *individual* and *relative* compatibilities and incompatibilities between dynamical quantities in the sense defined in (2.I)2 – the set of all the individual space-time supports of these successions of operations *[G.MesA]* falls apart, in general, in distinct space-time genetic 'branches'. So in general there emerges a tree-like graphic structure. For simplicity we presuppose here only two non-compatible quantities *A* and *B*. The generalization is obvious.

The two mutually incompatible dynamical quantities *A* and *B* introduce respectively the two grids of qualification of form (2)

$$gq[A, a_k, MesA, cod.proc(a_k)],\ j=1,2,....M; \quad gq[B, b_r, MesB, cod.proc(b_r)],\ r=1,2,....M \quad (2')$$

For simplicity we have endowed them with the same number *M* of possible values $a_j$ and $b_r$, respectively, and accordingly to the note attached to (5') we shall write only $\forall j$ or $\forall r$.

[36] Inside *MRC* it appears that this revolution reaches and incorporates also Shannon's theory of information and the representations of complexity.

Let $[d_G.(t_G-t_o)]$ denote the invariant space-time support of each one realization of the operation $G$ of generation of the studied microstate $ms_G$; this plays the role of a common 'rooting' into the microphysical factuality. Let $[d_A.(t_{MesA}-t_G)]$ and $[d_B.(t_{MesB}-t_G)]$, respectively, denote the mutually distinct space-time supports of a measurement-operation *MesA* and a measurement-operation *MesB*, the time origin being re-set on zero after each time-registration (obvious significance of the notations). So each realization of one whole succession *[G.MesA]* covers a same global space-time support

$$[d_G.(t_G-t_o)+d_A.(t_{MesA}-t_G)]$$

and it produces a group of observable marks $\{\mu_{kA}\}_j$, $k_A=1,2,...m_A$, $\forall j$, that is coded in terms of a value $a_j$ accordingly to (2'); while each realization of a succession *[G.MesB]* covers another same global space-time support

$$[d_G.(t_G-t_o)+d_B.(t_{MesB}-t_G)]$$

and produces a group of observable marks $\{\mu_{kB}\}_r$, $k_B=1,2,...m_B$ that is coded in terms of a value $b_r$ of the quantity $B$.

Thereby for the considered case the genesis (7) plced on the level of the *individual* conceptualisation involved by the representation (9), is achieved. The dominant character of this individual phase of conceptualization is *physically* operational.

### (3.I).1.2. Probabilistic level of conceptualization

Let us now start from the final result of the phase of individual conceptualization: values $a_j$ of $A$. The coding values $a_j$ are stored. Mutatis mutandis, the same holds for a succession *[G.MesB]*.

Suppose now that a sequence of a very big number $N$ of realizations of a succession $[G.MesA]_n$, $n=1,2,....N$, has been realized. The relative frequencies $n(a_j)/N$, $\forall j$ (where the symbol $n(a_j)$ is to be read 'the number $n$ of values $a_j$ of $A$') have been established *and* by global repetitions of the whole process an $(\varepsilon,\delta,N_0)$-convergence in the sense of (5) has been found to emerge indeed for these relative frequencies. In these conditions the primordial transferred description (9) has been factually specified *fully*, operationally *and* numerically. Furthermore on the top of the branch we have *effectively* constructed for the pair $(G,A)$ a relativized Kolmogorov-like factual $(\varepsilon,\delta,N_0)$-probability-space. The universe of elementary events from this probability space is $U=\{a_j\}$, $\forall j$, and the probability law from this space is the primordial transferred description relatively to $A$, (9) $(D/A)(ms_G)\equiv\{(\varepsilon,\delta,N_0)\text{-}\pi(a_j),\ \forall j\}_G$, $A$ fixed, (we do not yet consider explicitly the algebra on the universe of elementary events).

Mutatis mutandis, the same holds for the quantity $B$ and its values $b_r$.

Thereby the primordial transferred description (9) relatively to $B$, $(D/B)(ms_G)\equiv\{(\varepsilon,\delta,N_0)\text{-}\ \pi(b_r)\}_G$, $\forall r$ is also effectively constructed.

So we have the transferred description (9') for the considered case: Out of the brute observable data $\{\mu_{kA}\}_j$, $k_A=1,2,...m_A$, $\forall j$, and marks $\{\mu_{kB}\}_r$, $k_B=1,2,...m_B$ we have worked out factually for the qualifying quantities $A$ and $B$ a purely *numerical* probabilistic content, via individual genetic, physical-operational actions (7). So when this second level of conceptualization is also achieved, the probability laws obtained on it – considered *separately* from their geneses [37] – possess a purely abstract *mathematical* character [38].

[37] We stress this because inside quantum mechanics the asserted probability laws *are* indeed considered separately from the corresponding probability spaces, so in particular separately from the universe of elementary events that generate these laws. Furthermore *they are **not** defined factually for the purpose of prediction, their factual (re)production serves **exclusively the purpose of verification** of the predictive statistics*. This circumstance deserves being noted immediately and kept in mind because it plays a major role in the parts II and III of this work.

### (3.I).1.3. A meta-probabilistic level of conceptualization

But the geometric representation from the Fig.2 does not permit to stop here, it pushes further. Indeed the striking awareness of the role of the *unique* operation $G$ of generation of the specimens of the studied microstate $ms_G$ from *both* branches hinders to stop because it strongly stresses that the two different effective probability laws $(D/A)(ms_G) \equiv \{(\varepsilon,\delta,N_0)\text{-}\pi(a_j), \forall j\}_G$, with $A$ fixed, and $(D/B)(ms_G) \equiv \{(\varepsilon,\delta,N_0)\text{-}\pi(b_r), \forall r\}_G$, with $B$ fixed, that crown the space-time branches from the zone of individual conceptualization stem both from one same trunk-operation of generation $G$, i.e. they concern one same microstate $ms_G$. So it seems unavoidable to *posit* that there exists some sort of meta-probabilistic correlation between these two probability laws $\{(\varepsilon,\delta,N_0)\text{-}\pi(a_j), \forall j\}_G$, and $\{(\varepsilon,\delta,N_0)\text{-}\pi(b_r), \forall r\}_G$. Such a correlation accepts an expression of the general form

$$\pi(a_j)=\mathbf{F}_{aj,B}\{\pi(b_r), \forall r\}_G, \qquad \forall A,\ \forall B \tag{11}$$

$$\mathbf{F}_{AB}(G)= \{\mathbf{F}_{aj,B}\{\pi(a_j), \forall j\}_G, \qquad \forall AB \in V_M \tag{11'}$$

where $\mathbf{F}_{aj,B}\{\pi(b_r), \forall r\}_G$ and $\mathbf{F}_{AB}(G)$ are two functionals that represent, respectively, the *individual* probability $\pi(a_j)$ in terms of the *whole* probability law $\{(\varepsilon,\delta,N_0)\text{-}\pi(b_r), \forall r\}_G$, and the global correlation between the two whole laws $\{(\varepsilon,\delta,N_0)\text{-}\pi(a_j), \forall j\}_G$, and $\{(\varepsilon,\delta,N_0)\text{-}\pi(b_r), \forall r\}_G$. Together the relations (11) and (11') will be called *the meta-probabilistic correlations involved by (1)* $G \Leftrightarrow ms_G$ with respect to $(A,B)$ and will be symbolized by $(M\pi c(G))_{AB}$ ($M\pi c$: 'meta-probabilistic correlation') [39]. So the description (9') of the studied microstate has to be explicitly completed:

$$D_M(ms_G) \equiv \{[(\varepsilon,\delta,N_0)\text{-}\ \pi(a_j), \forall j\}_G,\ (M\pi c(G))_{AB}\ ]\},\quad \forall A,\quad \forall AB \in V_M, \tag{9''}$$ [40]

(in $(M\pi c(G))_{AB}$ the indexes $j$ and $r$ remain implicit).

In order to distinguish clearly between the factual probability-laws $\{(\varepsilon,\delta,N_0)\text{-}\pi(a_j), \forall j\}_G$ from (9), (9'), with $A$ fixed, and the meta-probabilistic correlations $(M\pi c(G))_{AB}$, $\forall AB \in V_M$ defined by (11), (11'), we shall say by definition that (9), (9') contain exclusively *probabilistic qualifications of the first order* whereas $(M\pi c(G))_{AB}, \forall AB \in V_M$ from (9'') expresses also *probabilistic qualifications of the second order* [41, 42].

---

[38] Notice how, out of the qualitative and physical operational approach practised here, the factual $(\varepsilon,\delta,N_0)$-probability laws induce *spontaneously* a promontory into the realm of the mathematized because they express the results of effective counting.

[39] The two functionals $\mathbf{F}_{aj,B}\{\pi(b_r), \forall r\}_G$, $\forall A, \forall B$ and $\mathbf{F}_{AB}(G)$ can acquire a precise numerical definition only inside a theory of microstates where are specified the general model posited for a microstate and the corresponding general concept of an act of measurement *MesA*, $\forall A$, with the involved coding procedure.

[40] Mackey [1963], Suppes [1966], Gudder [1976], Beltrametti [1991], and probably quite a number of other authors also, have tried – directly by purely mathematical means – to establish a satisfactory formulation of a meta-probability law associable with a quantum mechanical state-vector. The tree-like structure constructed here explicates the qualitative and semantic foundations of such a law. This, in the future, should much facilitate the specification of a consensual mathematical expression for what is here denoted $M\pi c(ms_G)$.

[41] We note that the whole process of description (9'') has been developed inside an a priori given *cell for conceptualization*, namely the pair $(G,V_M)$, that acted like a *local 'epistemic'* referential.

[42] For the sake of brevity, from now on we cease to always write explicitly the specification '$(\varepsilon,\delta,N_0)$'; but it will be constantly presupposed: we consider exclusively factual, effective probability laws.

### (3.I).1.4. The global result of the preceding genesis

All what precedes is represented on the Fig.2:

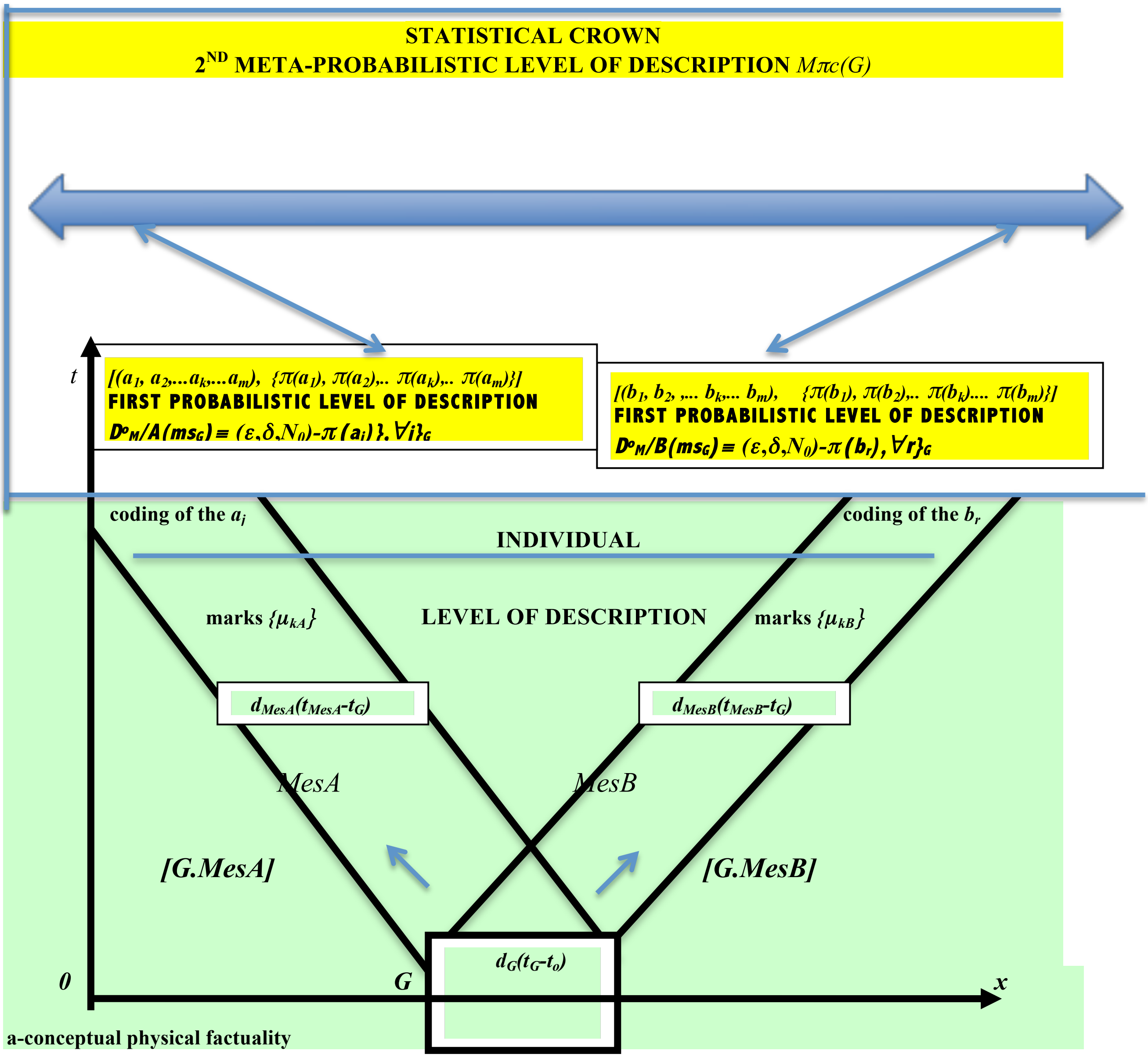

**Fig.2**. *The probability-tree $T(G,(A,B)$ of an unbound microstate $ms_G$*

We have remarked that – in contradistinction to its purely numerical results – the genesis of these results does possess a definite space-*time* structure. But the temporal character of emergence via successively registered individual results is abstracted away. It evaporates from the structure while it is progressively accomplished. So, in the exclusively spatial tree-like representation that persists, only the existence of distinct branches still just recalls the genetic temporal classifying features entailed by the mutual compatibilities or incompatibilities between the measured dynamical quantities, with respect to the considered type of microstate [43].

Let us denote by *T(G,(A,B))* (*T*: 'tree') this entirely geometrized residual structure of the genetic process of a description (9") [44]. The green zone of genetic conceptualization – purely individual and *physical*-operational – is clearly separated from the posterior superposed yellow zone of exclusively abstract conceptualization.

### (3.I).1.5. More detailed probabilistic examination of *T(G,(A,B))*

The concept of probability-tree of a microstate involves significances that are far from being trivial: they have already helped us to expand Kolmogorov's purely abstract, mathematical concept of a probability-space – where in particular the distributions of probability remain an only *general* pure concept that is *not* specified numerically – into a new and much more complex tree-like probabilistic whole where each element is defined, while the probability measure (5) $(\varepsilon,\delta,N_0)$-$\{\pi(a_j)\}, \forall j)$, emerges endowed with a finite conceptual definition *and* a factual numerical specification. Let us explicate this a little more.

***- Random phenomenon.*** The classical theory of probabilities offers no formalization of the notion of random phenomenon. It just makes use of the word 'experiment'. Whereas on the fFig.2 one literally sees how – from nothingness – a whole group of Kolmogorov probability-spaces emerges for a microstate $ms_G$, mutually connected by the corresponding operation of generation *G*, *and* by meta-probabilistic correlations between these. Thereby the basic concept of 'random phenomenon' acquires for a detailed inner structure, expressed in definite terms, namely [*G*, *MesA* or *MesB*, etc., marks $\{\mu\}_{kA}$ or marks $\{\mu_k\}_{kB}$, etc., code $\{a_j\}$ or code $\{b_r\}$, etc.], wherefrom factual finite Kolmogorov probability-spaces are then constructed. Inside these mutually connected probability spaces are lodged *numerically* specified factual $(\varepsilon,\delta,N_0)$-probability laws that are *effective and* ***relativized to all the actions and features that determine them***.

This result can be generalized to any physical entity and it can be induced in a strongly enlarged abstract theory of probabilities that accepts naturally a deep-set unification with *a relativized and extended logic* (MMS [2002A], [2002B], [2006], [2013], [2014]). Thereby these two fundamental structures of the human thought merge into a unification that has been shown to include Shannon's informational approach (MMS [1980], [1982], [2006]).

***- Probabilistic dependence.*** The factual Kolmogorov-type probability spaces that crown the two branches from the Fig. 2 admit, respectively, the denotations

$$[U(a_j),\ \tau_A,\ \{\pi(a_j,\forall j\}_G], \qquad [U(b_r),\ \tau_B,\ \{\pi(b_r,\forall r\}_G],$$

where $\tau_A$ and $\tau_B$ are the involved algebras of events. Let us consider now explicitly these algebras also. Inside the classical theory of probabilities the concept of probabilistic dependence is defined *only* for events from the algebra from *one* given space. Kolmogorov ([1950], p.9) has written:

---

[43] I perceive this as a hint that time *could* be conceived as just an artefact of Nature brought forth *inside human thought* by the biological evolution, as a feature of fitness to subsist as a species though we are *constrained to parcel in order to conceptualize* (other authors seem to reach a similar notion, for instance Carlo Rovelli [2015] and Donald D. Hoffmann&Ananda Gefner [2016]). In such a perspective – in the scientific *consensual* conceptualization of the ***physical*** 'reality' – space would be more basic than time, though according to Kant's postulate space-and-time are "equally" basic a priori forms of the *human* intuition: I deeply agree with Bergson that this is a very simplifying identification of the *epistemological* status *inside the* ***individual*** *minds* where consensus is neither required nor possible.

[44] The expression "probability tree" is already much made use of, with various significances. All these should be very carefully distinguished from the particular significance represented in the Fig.2.

> **«….one of the most important problems in the philosophy of the natural sciences is – in addition to the well known one regarding the essence of the concept of probability itself – to make precise the premises which would make it possible to regard any given real events as independent. »**

And he has posited – just posited by definition – that two events $a_1$ and $a_2$ from the algebra $\tau$ of a probability space are mutually independent from a probabilistic point of view if the numerical product $\pi(a_1).\pi(a_2)$ of the probabilities $\pi(a_1)$ and $\pi(a_2)$ of their separate occurrences is equal to the probability $\pi(a_1 \cap a_2))$ of their (set)-product-event $a_1 \cap a_2$ from $\tau$ ; whereas if this is not the case, then $a_1$ and $a_2$ are tied by a probabilistic dependence. But inside the classical theory of probabilities the concepts of probabilistic dependence or independence are ***not*** defined for *elementary* events from *one same* universe *U*. Such a sort of 'dependence' can be apprehended only indirectly, by comparison with the probability law that acts upon a universe of elementary events defined as a Cartesian product of two other universes, one of which is *U*. But this involves *another* random phenomenon, distinct from the random phenomenon that generates the space where *U* is the universe of elementary events [45] and a rigid *juxtaposition* of these two random phenomena. Whereas inside *IQM* this limitation will be circumvented in (3.I).2 via the definition (1) of an operation of generation *G* combined with the definition of 'one microstate of two or several micro-systems'.

In the case of a microstate of one micro-system, the classical definitions of probabilistic dependence are sufficient only if each one of the two probability spaces that crown the two branches from the Fig.2 is considered *separately* from the other one. But consider now an elementary event $a_j$ from the space that crowns the branch *MesA*, and an elementary event $b_r$ form the space that crowns the branch *MesB*. *Observationally* these two events *are* mutually in-dependent in the sense of Kolmogorov: Since the quantities *A* and *B* are mutually in-compatible in the sense defined in (2.I).2, the two measurement-operations *MesA* and *MesB* cannot be realized together, simultaneously, for only *one* specimen $\sigma(ms_G)$ of the studied microstate $ms_G$, so *the elementary events $a_j$ and $b_r$ cannot even coexist in an* ***actualized*** *way*. But nevertheless, the events $a_j$ and $b_r$ concern the *same* microstate $ms_G$ – in the sense of (1) – generated by one same operation of generation *G*. And even though inside our approach a microstate in the sense of (1) is distinct by definition from any specimen $\sigma(ms_G)$ of it, the considerations that led to (11)+(11') entail with a sort of necessity the assertion of a *meta*-probabilistic correlation *($M\pi c(G)$)* and of the corresponding extension (9") of (9), because both spaces that are considered stem from one same operation of generation *G*. This argument amounts to the assertion of a sort of 'probabilistic dependence' of second order that somehow knits into one whole all the distinct branch-random phenomena of which the common operation of generation *G* from the trunk of the tree introduces a priori the ***potentiality***.

The classical theory of probabilities also defines the general concept of probabilistic correlations, quite explicitly. But it does not *singularize* inside it a class of meta-probabilistic correlations that manifests specifically the fact that one same basic concept of a physical entity ($ms_G$ or *G*) is involved in different and *separately actualized* random phenomena [46]. This however is obviously an important case because it can be extremely frequent and it can entail subtle explanations for queer but observable behaviours.

---

[45] This detour could stem from the desire to stay inside the domain of the *actualized*. But let us notice that outside an algebra of events, the Kolmogorov concept of probabilistic dependence between two elementary events from the universe *U* is also a mere potentiality when these elementary events are not compatible in our sense, while in *this* case inside an algebra this dependence could be regarded as actual only because the concept of event from an algebra involves already potentiality by construction.

[46] K.J. Jung has introduced a concept of 'synchronicity' that seemed rather mysterious and has much struck Pauli, possibly because quantum mechanics – via the "principle of exclusion" – had suggested to him implicitly similarities with the behaviour of microstates, and this has been discussed in the correspondence Jung-Pauli (MMS [2002B], note pp. 279-281).

In short, considered globally *the probability-tree of a microstate constitutes* ***one whole of potential knowledge****, a closed cell of explicitly specified potentially possible fabrications of different but interconnected sub-wholes of actualized knowledge* [47].

Whereas the Kolmogorov *mathematical* conceptualization – even though it makes verbal use of the concept of 'experiment' – ***offers no formal possibility of location for the factual-conceptual successions (7) [G.MesA]*** when these are combined with the definitions from (2.I).2). The classical mathematical theory of probabilities is blind with respect to the *genetic* features of the *factual* concept of probability that works in the case of a transferred description of a physical entity.

Why is this so?

The answer, we think, is the following one, and it is striking: because Kolmogorov's concept of a probability space – conceived entirely on the classical level of conceptualization and then extrapolated top-down into micro-physics just as it has been obtained there – simply ***does not realize any contact*** *with the factual root of the factual probabilistic whole that works inside the definition (9") of a transferred description constructed bottom-up*. Whereas the bottom-up approach practised here *starts much deeper*, on the frontier between the already conceptualized and the a-conceptual physical factuality, wherefrom it proceeds upward via the methodological decision (1) $G \Leftrightarrow (ms_G \equiv \{\sigma(ms_G)\}$ that introduces the factual-operational definition of precisely this factual root of a new probabilistic whole.

This illustrates strikingly the differences of nature and of content between a top-down abstract and factually blind conceptualization, and a bottom-up factually constructive conceptualization.

The entire probabilistic output of the factual root of a primary transferred description – with respect to an arbitrary but *given* collection of mutually incompatible branch-qualifying mechanical quantities – can be represented inside *one* new sort of enriched probability space:

$$[\ U_T(e_{jb}) = \cup_b U(e_{jb}), \qquad \tau_T = \cup_b \tau_b, \qquad \{\pi_T(e_{jb}, \forall jb\}_G = \cup_b \{\pi_b(e_{jb}, \forall j\}_G\ ]$$

where the index '$T$' labels the considered probability-tree; the index '$b$' labels the considered branch from $T$; the index '$jb$' labels the elementary event $e_{jb}$ from the branch labelled by $b$, and $\tau_T$ designates the *total* algebra of events from this enriched probability space.

### (3.I).1.6. The particular case of a one-trunk probability tree

What happens if no sort of relative mutual physical-conceptual in-compatibility does act *in the considered circumstance, i.e.* ***WITH RESPECT TO the involved factual microstate $ms_G$ and ALL the considered qualifying quantities***? In this case the space-time domain covered by the involved operation of generation $G$ leads to only one 'branch' that is common to all the considered mutually compatible mechanical quantities; which amounts to saying that the common trunk-and-branch of the tree is crowned by a *set* of probability spaces – one for each quantity $A$ – that, inside (9") – *are only conceptually distinguished from one another and then conceptually meta-correlated to one another*, as indicated below.

---

[47] Human psychic "time" is strongly populated by potentialities, by the virtual; so the representation of probabilities should fully encompass also such 'states' of events. A probability tree in our sense is overtly constructed as a structure-of-potentialities that spreads out freely inside the whole domain of possibilities where develop the individual inner times of human beings, wherefrom the public time is then constructed (MMS [2006]).

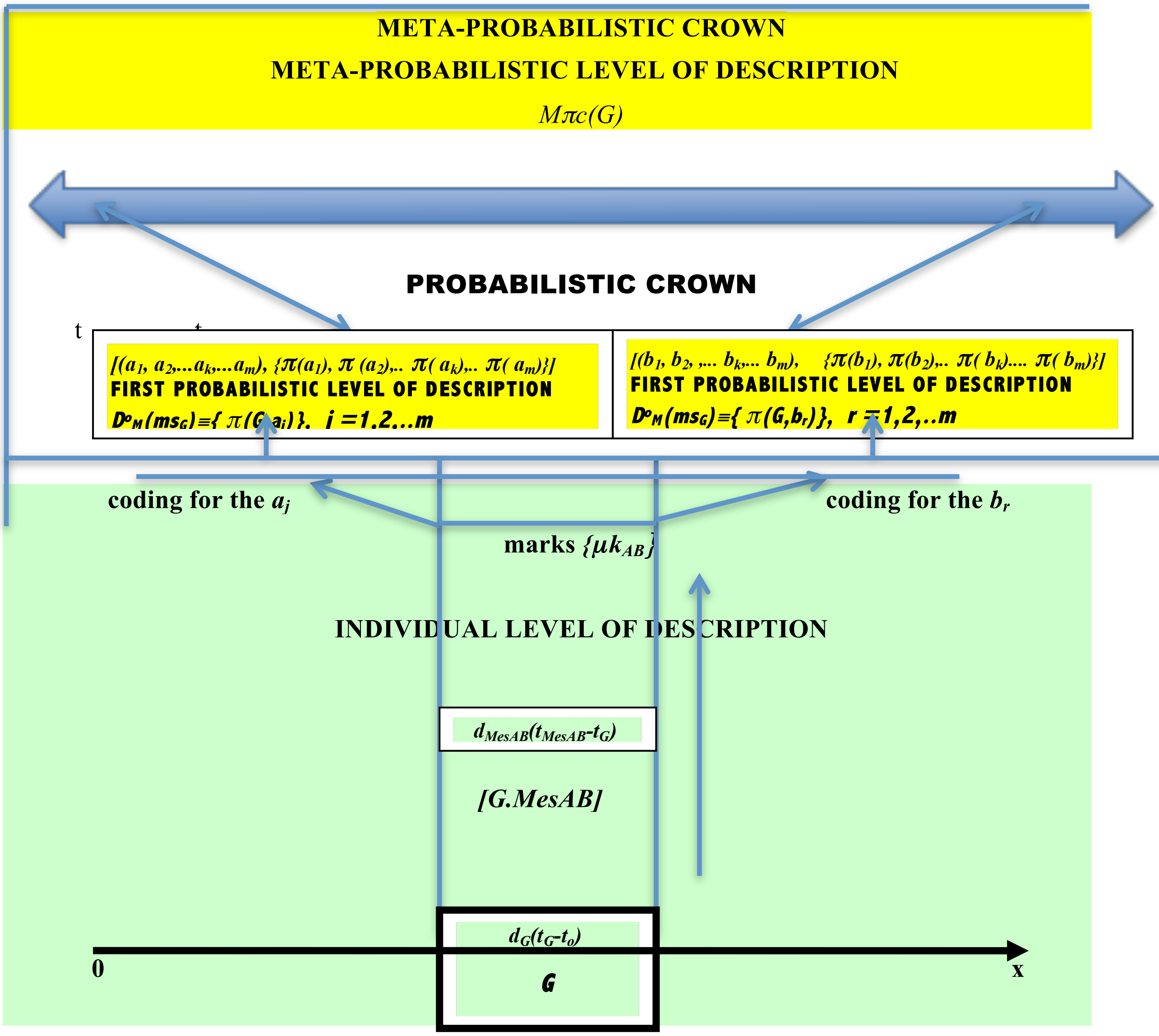

**Fig. 3**. *The probability-tree T(G,(A,B) of two mutually compatible observables*

Here the capacity of the case of an unbound microstate of *one* microsystem to reveal non-classical probabilistic contents of (9"), comes to exhaustion. But the most surprising such contents appear just below.

## (3.I).2. PROBABILITY TREE OF *ONE* UNBOUND MICRO-*STATE* OF *TWO OR MORE* MICRO-*SYSTEMS*: *A THEOREM ON BELL'S "PROBLEM" OF NON-LOCATITY*

We now enter upon the case of micro-states of two or several micro-systems. Thereby we come face-to-face with what is called *the problem of non-locality*. This case finally brings strikingly forth to what a degree the concepts-and-language introduced by the definitions from (2.I)2 and by the basic concept of a probability tree constructed bottom-up, introduce a structure of conditions of inner coherence that entails intelligibility.

***General treatment***. Consider *one* progressive micro-state $ms_{G(2S)}$ of *two* micro-systems $S_1$ and $S_2$, in the sense of the definitions from (2.I)2). How shall we construct the probability tree of $ms_{G(2S)}$ ?

According to (1) one microstate $ms_{G(2S)}$ is generated by one corresponding operation of generation *G(2S) to which it is tied in the sense of (1)* and of the identity $ms_{G(2S)} \equiv \{\sigma(ms_{G(2S)})\}$.

According to the definitions from (2.I)2) in this case *one* complete operation of measurement-interaction on *one* specimen of the factually defined microstate $ms_{G(2S)}$ involves *two* partial measurement-interactions, a partial measurement-interaction *MesA* with $S_1$, *and* a partial measurement-interaction *MesB* with $S_2$ (in particular the quantities *A* and *B* can identify, but in general they are permitted to be different). For maximal graphic clarity instead of *A*, $a_j$ and *B*, $b_r$ we shall exceptionally write in this case, respectively, *A1*, $a1_j$ and *B2*, $b2_r$. So a complete act of measurement will be denoted *Mes(A1,B2)*.

In (3.I)1, for the case of one micro-state of *one* micro-system, we have assigned by construction an *own* branch to *each* given sort of 'complete' act of measurement that is involved i.e. which involves fully one specimen of the microstate-to-be-studied (this happens always for a microstate of only one micro-system). In order to stay in agreement with all the constructive definitions from (2.I)2) and (3.I)1, here we must apply this same procedure: [one given sort of complete act of measurement involving fully one specimen of the microstate-to-be-studied] corresponds to [one branch of the tree]. So to each sort of complete act of measurement of the same form as *Mes(A1,B2)* we assign *one* branch from the probability tree of *G(2S)*. Then the *two* partial measurements *MesA1* and *MesB2* from one complete act of measurement *Mes(A1,B2)* operated respectively upon the two micro-systems $S_1$ and $S_2$ from each *one* specimen $\sigma(ms_{G(2S)})$ of the studied micro-state $ms_{G(2S)}$, are ***both*** lodged inside ***one*** *same branch* of the probability tree of *G(2S)*. So we must assign another branch of this tree to the complete measurements that involve another pair of quantities denoted for instance *(C1,D2)* with values, respectively, $c1_k$ and $d2_z$, where *at least* either *C1* is *in*-compatible with *A1* or *D2* is *in*-compatible with *B2* in the sense defined in (4), or where both these possibilities are realized; *there is no condition then concerning the compatibility of C1 and D2*.

So a two-branches-tree from the figure 4 founded upon the operation of generation *G(2S)*, can be denoted *T(G(2S),(A1,B2; C1,D2))*.

Let us now focus upon the following fact: For one micro-state of *two* micro-systems the two dynamical quantities *A1* and *B2* that are involved in *one* complete act of measurement *Mes(A1,B2)≡(MesA1 and MesB2) are **always** compatible* in the sense defined at the point 3 from (2.I)2, because the measurements *MesA1* and *MesB2* are performed, respectively, upon the *two* mutually distinct systems *S1* and *S2* that are involved in any *one* specimen of the microstate $ms_{G(2S)}$ and so no incompatibility between these space-time supports of these two partial acts of measurement comes in necessarily [48] (if in some circumstance these two space-time supports tend to overlap it should be easily possible to separate them).

Since *one* complete act of measurement *Mes(A1,B2)* contains by definition an act of measurement *MesA1* and an act of measurement *MesB2*, the corresponding pair of observable marks $(\{\mu\}_{kA1},\{\mu\}_{kB2})$ – let us denote it $\{\mu_{kA1B2}\}$ – once it has been coded in terms of a pair of values $a1_j,b2_r$, $j,r=1,2,...M$, constitutes ***one*** elementary event from the universe of elementary events $U=\{a1_j,b2_r\}$, $j,r=1,2,...M$ from the probability-space that in the Fig.4 crowns the *unique* branch of the complete measurements *Mes(A1,B2)*; while the factual probability distribution on the universe of elementary events from this probability space consists of the transferred description (9) with respect to the pair of quantities *(A1,B2)* and has to be denoted as

$$D/(A1,B2)(ms_{G(2S)}) \equiv (\varepsilon,\delta,N_0)\text{-}\{\pi(a1_j,b2_r), j, r=1,2,...M\}_{G(2S)}, \quad j, r=1,2,...M$$

[48] We recall that inside the approach developed here the compatibility or incompatibility of two dynamical quantities is defined only for *one specimen* of the studied microstate and it is *relative* to both the nature of these quantities and to the type of the microstate that is considered, in the sense of the definitions from (2.I)2.

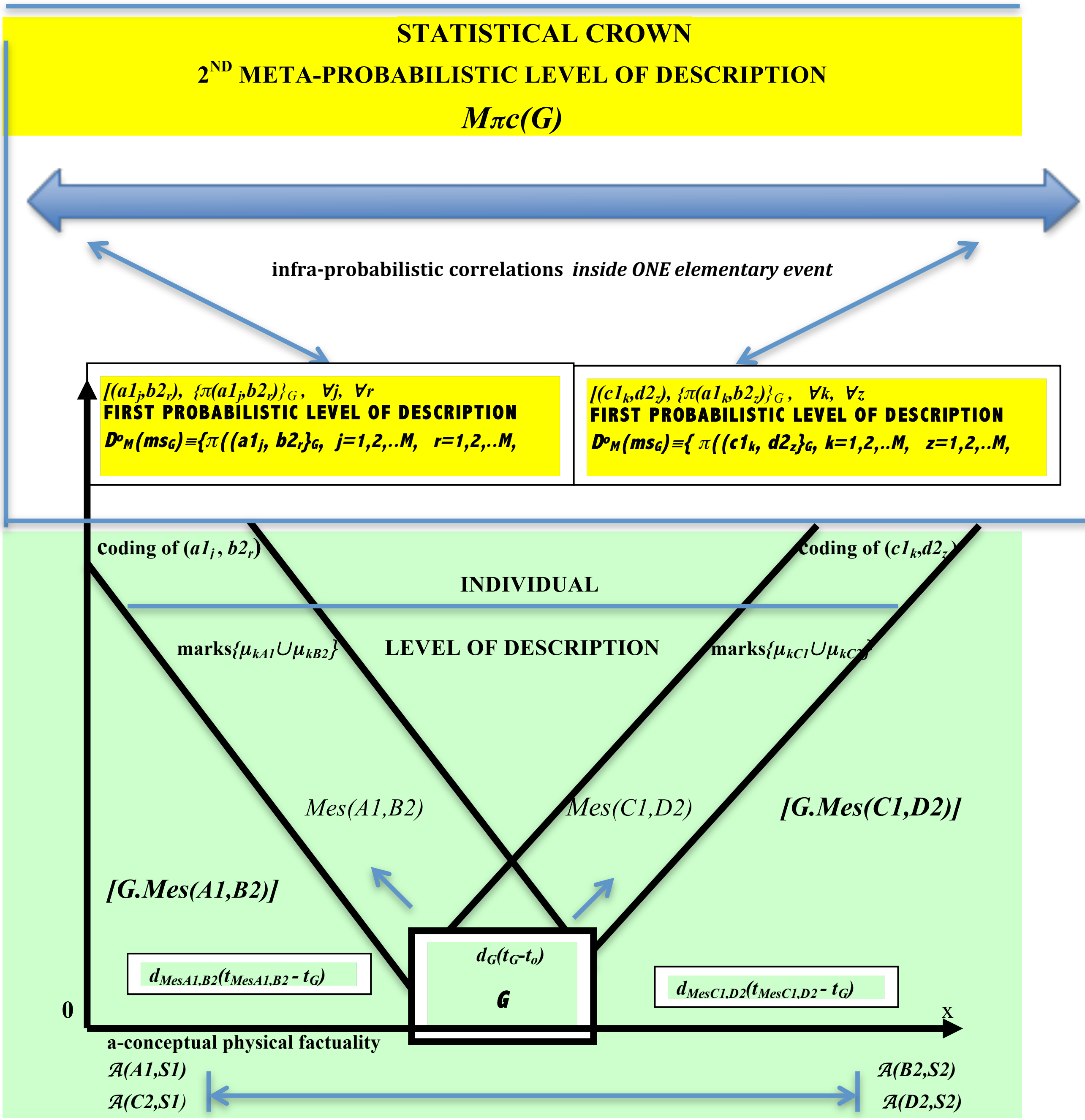

**Fig. 4**. *The probability-tree T(G(2S),(A1,B2; C1,D2)) of a microstate* $ms_{G(2S)}$*: the case of the 'problem of non-locality'*

So the *two* quantities *(A1,B2)* of which one qualifies the system *S1* and the other one the system *S2* are involved *both* in each ***one*** *– and* ***elementary*** *– 'event'* in the probabilistic sense, that concerns only ***one*** complete act of measurement *Mes(A1,B2)* on ***one*** specimen of the studied micro-*state*. And nevertheless the *here-now's* of the corresponding *two* observable and registered events – in-the-***physical***-sense this time – namely [the observation by a human observer, of marks $\{\mu\}_{kA1}$ coded by a value $a1_j$ that qualifies *S1*] and [the observation by a human observer, of marks $\{\mu\}_{kB2}$ coded by a value $b2_r$ that qualifies *S2*], *can be separated from one another by an arbitrarily big space-time distance*. While *the* *corresponding description (9) – **one** factual probability law – is **itself** devoid of a defined space-time structure.*

This is obviously connected with the "problem" of locality. The Fig.4 represents graphically a most explicit analysis of the *inner **probabilistic** structure of this "problem"*. But inside *IQM* this structure arises free of any assumption concerning "influences" tied with "non-locality" in the Einstein-Bell sense; it is brought forth exclusively by the general algorithm for conceptual-probabilistic organization involved by the construction of the probability-tree of the studied microstate of type $ms_{G(2S)}$.

Indeed *inside IQM, the basic relation (1) and the definitions (2.I)2 **suffice** for entailing the probabilistic structure brought into evidence above, as a general and **purely probabilistic** consequence, quite **independently** of any supplementary condition specifically tied with the notion of non-local influences.*

This directs attention upon the following question:

How – exactly – does emerge the connection between Bell's problem of non-locality, and on the other hand features brought forth inside *IQM* in a way so immediately tied with a so basic, general and only qualitative tree-like representation of the set of observational events tied with what here, by a definition from (2.I)2, we have called "one microstate of two or more micro-systems"?

***On a qualitative proof inside IQM concerning Bell's problem of non-locality***. Any formal definition delimits a corresponding class inside a larger given class posited as an initial ground, in the role of a "genus". The definition of "one microstate of two or several microsystems" delimits, inside the class of any microstate as defined by (1), the class of microstates denoted $ms_{G(2S)}$. And – deliberately – it achieves this inside *IQM* in absence of any definite model, in order not to arbitrarily restrict a priori the as yet un-defined fully intelligible theory of the microstates that will be later researched in this work. However we have already brought into evidence that this still non-specified theory will *have* to make use of some model of a specimen of any sort of studied microstate; for if not it would lack criteria for defining "adequate" acts of measurement *MesA* on that sort of microstate.

This being recalled, consider now what follows.

- In MMS [1987] I have shown that the de Broglie model of a specimen of one microstate of two micro-systems $ms_{G(2S)}\equiv\{\sigma(ms_{G(2S)})\}$ (cf. (6.II).2 in this work), can be found *compatible* with the $QM_{HD}$ predictions on spin-measurements on $ms_{G(2S)}$ *without any recourse to non-local "influences" in the Einstein-Bell sense*; and in Bordley [1989] it has been shown that the representation from that work is compatible with established experimental results. Furthermore in MMS [2013B] I have shown that *the formulation of Bell's conclusion of his proof does not follow from the proof.*

- In Muchowski ([2018]) a similar result is established. I reproduce the abstract below:

Abstract – It is shown that there is no remote action with polarization measurements of photons in singlet state. A model is presented introducing a hidden parameter which determines the polarizer output. This model is able to explain the polarization measurement results with entangled photons. It refutes Bell's Theorem.

- In Shanahan [2019] is constructed another model that asserts the same sort of compatibility with respect to a specified model as in the previous quotations. Again I reproduce the abstract:

*Abstract*. Effects associated in quantum mechanics with a divisible probability wave are explained as physically real consequences of the equal but opposite reaction of the apparatus as a particle is measured. Taking as illustration a Mach-Zehnder interferometer operating by refraction, it is shown that this reaction must comprise a fluctuation in the re-radiation field of complementary effect to the changes occurring in the photon as it is projected into one or other path. The evolution of this fluctuation through the experiment will explain the alternative states of the particle discerned in self-interference, while the maintenance of equilibrium in the face of such fluctuations becomes the source of the Born probabilities. In this scheme, the probability wave is a mathematical artifact, epistemic rather than ontic, and akin in this respect to the simplifying constructions of geometrical optics.

Of course, each one of the works mentioned above remains open to examination and confirmation or invalidation. But let us just *suppose* that at least one of the mentioned works has been consensually declared to be confirmed. In this case – *inside IQM* – the confirmed work constitutes a proof by construction of the assertion that:

(1) and (2.I)2 (via the concept of factual probability expressed by the probability-tree ***T(G(2S),(A1,B2; C1,D2))*** of a microstate $ms_{G(2S)}$ from the Fig.4), delimit a class of possible models of one specimen of a microstate $ms_{G(2S)}$ that includes the model involved in the confirmed example, for which the $QM_{HD}$ predictions qualified by Bell as "non-local" can be explained without any recourse to non-local "influences" in the sense of Bell.

Which means that ***what is perceived as a "problem of locality" is certainly tied with hidden particular modelling assumptions***; which entails that indeed ***Bell's conclusion is not imposed with formal necessity by the content of his proof***.

***General insufficiencies of the nowadays concept of a microstate.*** Quite generally now, let us note that Bohr's interdiction of models has imposed lethal stagnations upon the conceptualization of the microstates. This conceptualization has been kept *unachieved* from various points of view:

- Consider the *two* "micro-*systems*" *'$S_1$'* and *'$S_2$'* from *one* specimen $\sigma(ms_{G(2S)})$ of the studied microstate $ms_{G(2S)}$. The human thought cannot bear a strict absence of any model. So the entities denoted *'$S_1$'* and *'$S_2$'* have been spontaneously and implicitly imagined more or less like two spatially delimited small balls radically *exterior* to one-another, so mutually 'separated' by a ***void*** 'distance', which raises irrepressibly the question of *what* 'exists' and 'happens' ***outside and between*** *'$S_1$'* and *'$S_2$'* [49, 50]. Whereas the experimentally registered ***time***-distance between *'$S_1$'* and *'$S_2$'* seems to be quasi null, or in any case smaller than is entailed by the Einstein-velocity of a 'light-signal'. This conceptual situation acts as a strident common call for a general model of any specimen of a microstate inside any ***theory*** of the microstates, in spite of the orthodox interdiction imposed by the Copenhagen school; a theory of the microstates cannot exist without such a model, in contradistinction to *IQM* that – by deliberate construction – is only a general framework of all the acceptable theories of the microstates.
- But also other uncontrolled presuppositions are involved by the concept of microstate that works inside the nowadays microphysics. For instance, the non-locality problem emerges in a particularly striking way because it is explicitly and essentially *lodged inside the space-time frame* ***of the human observers with their apparatuses***. One complete act of measurement $Mes_{12}(A1,B2)$ involves two blocks of macroscopic apparatuses $\mathcal{A}(A1,S1)$ and $\mathcal{A}(B2,S2)$ that – themselves – ***are perceived*** with delimited volumes and are endowed with registering devices that pre-structure classes of possible space-time locations of the *observable results* of measurements coded $a1_j$ *and* $b2$, which can define perceived spatial and temporal distances between potential locations ***of these pre-constructed observable space-time locations of the perceivable marks***. This entails an inextricable mixture between: a mathematical formalism; implicit expectations induced by the classical human macroscopic conceptualization; and cognitive human actions, registering 'objects', and pre-organized observable events, that are unavoidably involved by the acts of measurement. While obviously, such heterogeneous features with their respective conceptual roles have to be strictly distinguished from one another via a well defined methodological structure imposed upon the study.

[49] The question of "separability" has been much discussed.

[50] Quite remarkably, Descartes held that *void space* ***"does not exist"***. Which I understand as the assertion that only void space ***with respect to some definite aspect*** can exist, because 'space' ***is** exclusively* the universal "bearer" posited a priori in the human minds for any given quality that 'exists' in some ***physical*** sense with respect to some given ***physical*** entity.

- *The purely conceptual* **probabilistic** *situation is also unintelligible, from a very basic probabilistic point of view.* The space-time distances, *together* with the observed correlations, emerge in relation with only *each* ***one-branch***-probability distribution

$$(\varepsilon,\delta,N_0)\text{-}\{\pi(a1_j,b2_r,\ \forall(j,r)\}_{G(2S)} \quad \text{or} \quad (\varepsilon,\delta,N_0)\text{-}\{\pi(c1_k,d2_z),\ \forall(k,z)\ \}_{G(2S)}$$

– i.e. ***not*** inside the meta-probabilistic correlations denoted by us *'Mπc(G)'* between the ***two*** branch-probability distributions that are involved – (cf. the Fig.2). In this case, the considered 'correlations' appear as tied with a sort of 'probabilistic dependence' that stems from the ***insides*** of the observable probabilistic 'events' $(a1_j,b2_r), \forall(j,r)$ or $(c1_k,d2_z), \forall(k,z)$ from ***one*** probability law, i.e. either $(\varepsilon,\delta,N_0)\text{-}\{\pi(a1_j,b2_r, \forall(j,r)\}_{G(2S)}$ or $(\varepsilon,\delta,N_0)\text{-}\{\pi(c1_k,d2_z), \forall(k,z)\}_{G(2S)})$, , respectively. While in the probabilistic sense these are ***elementary*** *events*.

> The classical concept of probabilistic dependence defines exclusively a mutual probabilistic dependence for two distinct *non*-elementary events *from the algebra* posited on the universe of elementary events from one probability space. This classical concept of probabilistic dependence *cannot deal with features of the* ***inner*** *structure of elementary events*.

Here the classical probabilistic conceptualization is literally overwhelmed.

> The bottom-up approach practised here brings forth the possibility and the *utility* of a Factual Theory of Probabilities from the Mathematical Physics that clearly exceeds the abstract framework of Kolmogorov's mathematical theory of probabilities: a new and much more complex mathematical framework can be achieved in order to lodge this factual theory.

- This, once more, focalizes the attention upon the *direction* of conceptualization, top-down or a bottom-up. In this circumstance also this direction plays an essential role, via the universal but classically hidden presence of an operation of generation *G* at the beginning of any act of conceptualization. Historically the human conceptualization has been developed top-down on the vertical that connects the macroscopic level of conceptualization, to the microscopic one, and this entailed that the notion of a *common* trunk *G* of a possible probability-tree from which stem distinct branches, had not yet been noticed, nor even conceived at the time when Kolmogorov elaborated his theory of probabilities. The general genetic concept of operation of generation *G* of the entity-to-be-studied had not yet emerged itself. So Kolmogorov has defined only probability spaces entirely separated from one another, each one of which tops only one and incompletely represented 'experiment' (or 'random phenomenon').

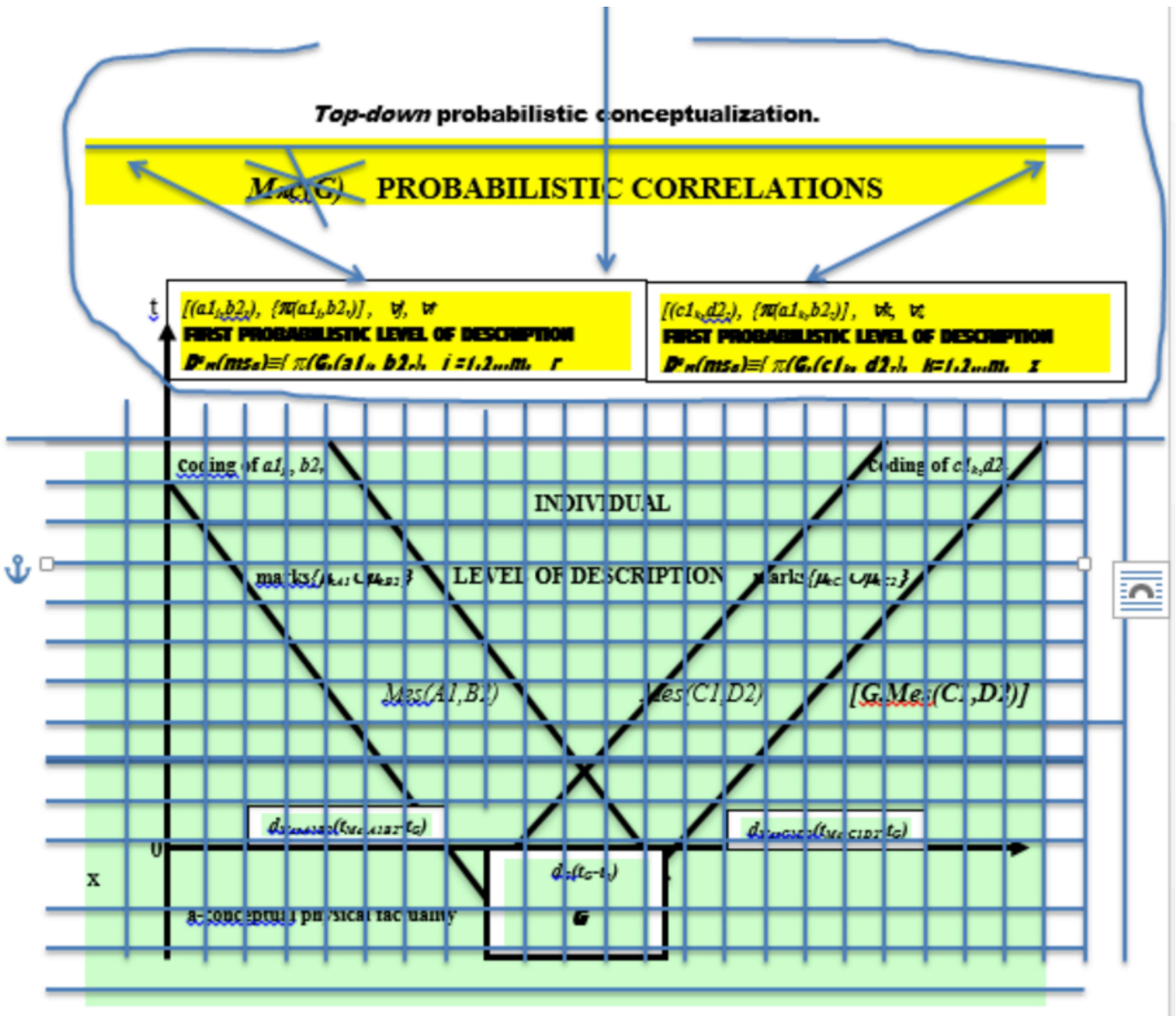

**Fig. 5**. *A probability-tree T(G(2S),(A1,B2;C1,D2)) as encountered by a top-down approach that installs Kolmogorov's spaces and then* ***stops*** *its progression downward.*

And then Kolmogorov *stopped*.

Via his concept of a whole probability-*space* he had already realized a very remarkable progress with respect to the preceding Bernoulli-von Mises concept of – directly and ***alone*** – the mathematical concept of a probability-*measure*.

But from the viewpoint of a bottom-up approach it appears that the concept of one Kolmogorov probability-space lies ***isolated*** *on the top of only* ***one*** *branch from the set of all the processes of genesis of an elementary event from a whole tree-like probabilistic structure like that from the Fig.2 and the Fig.4*. Indeed one probability-space from a probability-tree *gives no access to the common roots of all the different probability spaces from one multi-branches probability-tree*. So for human beings that confine their conceptualization on what we call 'the macroscopic level', factual common genetic roots can stay hidden a very long time, in contradistinction to what happens in a bottom-up approach where the roots of a process of conceptualization appear and are specified in the very first place. While in order "to make precise" the premises of probabilistic dependence (cf. the Kolmogorov-quotation from (3.I).1)) a sine-qua-non condition is to be aware of the existence – of the *universal* existence of an 'operation of generation *G* of the entity-to-be-studied' in any avatar of a process of conceptualization, physical or sensorial or only mental – and of its basic role[51]**.** But this existence and its role, notwithstanding their universality, become striking and are endowed with general contours only inside microphysics, and only if one is attentive to the consequences of the *physically* operational character of the operation of generation *G* and of the successions *[G.MesA]* when the entity-to-be-qualified is "a factual microstate" in the sense of (1).

---

[51] It might come out in the future that ***any*** probabilistic correlation can be assigned to a certain class of distinct branches from one huge probability-(meta-tree)-of-probability-trees). This would found in a very toned way Gustav Jung's concept of 'synchronicity' that has interested Pauli. (MMS [2002B], the note pp. 279-281). (In particular, it is not a priori absurd that certain subconscious psychical perceptions of 'synchronicity' of physical events come out to be connected with some sort of instinctive, reflex reactions to physical events from a same probability tree that are separated from one another by an arbitrarily big spatial distance *internal* to some basic sort of physical substance (like that from the de Broglie-Bohm view) relatively to which Einstein's 'limit-velocity' of, specifically, **light**-'signals', simply has no meaning).

And furthermore, *even* if the concept of operation *G* of generation *is* taken into account, *still* the Bell-case can stay probabilistically non-intelligible, because the awareness that for unbound microstates the performance of whole successions *[G.MesA]* is a condition for obtaining consensually perceptible marks, if it is active *alone*, entails only a monolithic concept of probabilistic correlation *that cannot distinguish between correlations that are* ***interior*** *to one* ***elementary*** *event in the probabilistic sense, and correlations between* ***distinct*** *probabilistic events*. This last distinction, as stressed, requires additionally also explicit recourse to *also* the definitions from (2.I)2.

So a full perception of all the sorts of probabilistic dependence that can emerge, still remains to be developed.

In short: The case of the probability tree of one microstate of two or several microsystems is paradigmatic from various points of view, and very basic ones. It is not surprising that – implicitly – it is tied with so many astonishments, researches and considerations (cf. the very interesting researches quoted above, by E. Muchowski [2018] and D. Shanahan). The *IQM*-analyses of such cases illustrate strikingly the elucidating force of the definitions from (2.I)2, and more generally of an explicitly constructed structure of reference.

> These – remarkably – have permitted inside this point (3.I).2 to generate inside *IQM* ***intelligibility*** on the inner structure of the "problem of non-locality" – namely what stems from (1) and (2.I)2 and what stems from a proposed model of the specimens of the studied microstate, *without any recourse to the formalism of the nowadays Quantum Mechanics and without any mathematical arguments*, via an independent and purely qualitative conceptual examination.

Which illustrates the close connection between *that what strikes human eyes* thus asking for conceptualization, the chosen *modalities of conceptualization*, and the "problems" that these arise:

> *A structure of conceptualization of a physical entity is never entailed exclusively by the involved "facts" and "entities"*; it never expresses truth that is intrinsic to – exclusively – that what is conceptualized; it ***always*** emerges by interactions between the human observer-conceiver and what we call "physical reality" and it ***always*** characterizes ***THIS***, alone and as a whole.

## (3.I).3. PROBABILITY TREE OF ONE MICROSTATE WITH *COMPOSED* OPERATION OF GENERATION

Consider now a *composed* operation of generation $\mathbf{G}(G_1,G_2)$ (cf. (1.I),(2.I).1) of a microstate in which only two simple operations of generation $G_1$ and $G_2$ are involved, like in the Young two-slits experiment. And consider an effectively realized microstate $ms_{G(G1,G2)}$. Let us compare its factual description (9') with the factual descriptions (9') of the two microstates $ms_{G1}$ and $ms_{G2}$ that *would* be obtained, respectively, *if* the two operations of generation $G_1$ and $G_2$ were each one *fully realized separately*. According to our present knowledge on microstates such a comparison would bring forth the physical *fact* that in general, between the probability $\pi(G(G_1,G_2),a_j)$ of realization of the value $a_j$ of a dynamical quantity *A* via acts of measurement *MesA* performed on one outcome of $ms_{G(G1,G2)}$, and the probabilities $\pi(G_1,a_j)$ and $\pi(G_2,a_j)$ of this same value $a_j$ established, respectively, via measurements *MesA* performed onthe microstates $ms_{G1}$ and $ms_{G2}$, there holds an *in*-equality

$$\pi_{12}(\mathbf{G}(G_1,G_2),a_j) \neq \pi_1(G_1,a_j) + \pi_2(G_2,a_j) \qquad (12)$$

This circumstance deserves being noticed. It suggests that a microstate tied with $\mathbf{G}(G_1,G_2)$ belongs to a domain of phenomena that is of another nature than the domain of phenomena entailed by $G_1$ and $G_2$ *when* these are realized separately. But 'different' *in what sense*, exactly? In this preliminary stage of the conceptualization of the microstates this question remains *open*. Nevertheless we can already formulate the following important remark.

The inequality (12) is usually expressed verbally in *positive* terms by saying that '$ms_{G1}$ and $ms_{G2}$ *interfere* inside $ms_{G(G1,G2)}$'. But inside the present approach – according to the one-to-one relation (1) between a given operation of generation and the corresponding microstate – this expression is misleading from a conceptual point of view. The relation (1) entails that *only* the *one* microstate $ms_{G(G1,G2)}$ is effectively generated when the operation of generation $\mathbf{G}(G_1,G_2)$ is performed. So $\mathbf{G}(G_1,G_2)$ cannot be coherently conceived to generate *also* the two microstates $ms_{G1}$ and $ms_{G2}$ when the microstate $ms_{G(G1,G2)}$ has been generated. When the microstate $ms_{G(G1,G2)}$ has been generated, the microstates $ms_{G1}$ and $ms_{G2}$ have to be conceived as somehow *non*-achieved or non-'completed' microstates that, by construction, can at most possess the status of partial effects of two a priori *possible* full operational state-individualizations via $G_1$ and $G_2$, but that in fact have *not* been fully actualized when $G_1$ and $G_2$ are composed inside $\mathbf{G}(G_1,G_2)$.

The symbol $\mathbf{G}(G_1,G_2)$ adopted here suggests that – if and when this seems useful – the mentioned effects can be *referred* two the unachieved microstates $ms_{G1}$ and $ms_{G2}$. So, in terms of probability-trees, the trees $T(G_1)$ and $T(G_2)$ are only two reference-trees, ghost-trees, only the one tree $T(\mathbf{G}(G_1,G_2))$ is factually realized.

And, since $ms_{G1}$ and $ms_{G2}$ have not been both and separately effectively realized by $\mathbf{G}(G_1,G_2)$, they do not 'exist' inside $ms_{G(G1,G2)}$, as it is implied by the assertion that 'they' *interfere* inside $ms_{G(G1,G2)}$. Such a language – *like also the mathematical writing (12)* or some equivalent one – are misleading inside the present approach: the natural language involves many shades, and these work inside the minds, so they have to be carefully dominated.

The preceding considerations can be generalized in an obvious way to the case of an operation of generation $\mathbf{G}(G_1,G_2,...G_m)$ that composes several operations of generation.

So inside the present approach the mathematical representation of the whole category of 'microstates with *composed* operation of generation' will have to be openly considered in a critical state of mind and in the third part of this work it will play the role of a discriminating test of the construction submitted there.

This point (3.I).3 closes our exploration on probability trees of progressive microstates.

## (3.I).4. ON THE EVOLUTION OF ANY UNBOUND MICROSTATE

Is it possible, inside this qualitative and general approach, to assert something concerning the evolution of a progressive microstate? The answer is yes, and again it brings into evidence the crucial role of the concept of operation $G$ of generation of a microstate.

Imagine the *final* moment $t$ assigned to an operation of generation $G$ from (1) that introduces the microstate $ms_G$ to be studied. In contradistinction to what has been assumed before, let us admit that during some time interval $\Delta t_1=t_1-t$ subsequent to $t$ the human observer does *not* act upon the microstate $ms_G$. Nevertheless during $\Delta t_1=t_1-t$ the initial microstate $ms_G$ can be *posited* to 'evolve' in the 'exterior' conditions $EC$ that it encounters (exterior known macroscopic fields or obstacles). Indeed it would seem weird to posit that '$ms_G$' remains immobilized from any conceivable point of view. Now – formally – this evolution *can be integrated in (1)*, in the following way.

Consider ***(a)*** the effect denoted $ms_{Go}$ of the initially conceived operation of generation $G_o$ realized at an initial time $t_o$, and ***(b)*** the change, the *evolution* of $ms_{Go}$ during a laps of time $\Delta_1=t_1-t_o$, $t_1>t_o$, that yields another microstate denoted $ms_{G1}$. Nothing hinders to posit in full logical coherence with the preceding development that *the association between (a) and (b) act together like another operation of generation* – let us denote it $G_1=F(G_o, EC(t_1-t_o))$ ($F$ : some functional of the exterior conditions $EC$ that act on the initial microstate during $\Delta_1$) that generates the microstate $ms_{G1}$ corresponding to $G_{t1}$ in the factual sense defined in *MD*. As stated before for any factually defined microstate $ms_G$, the microstate $ms_{G1}$ can be studied via sequences of successions $[G_1.MesA]$, $\forall A\in V_M$. The time interval $t_1-t$ can be chosen with any desired value, and the external conditions $EC$ can be freely conceived. So – given the initial operation $G_o$ – one can study successively a set of mutually 'distinct' microstates $ms_{Gk}$, $k=1,2,...K$, that correspond respectively to the set of successive operations of generation:

$$G_o,\ G_1=F(G,EC(t_1-t_o)),\dots G_k=F(G_o,EC,(t_k-t_o)),\dots.G_K=F(G_o,EC,(t_K-t_o));\quad k=1,2,...K \quad (13)$$

For each operation of generation $G_k$ from (13), can be constructed the whole corresponding probability tree $T(G_k,A)$, $\forall A\in V_M$, i.e. also the corresponding descriptions (9') and (9''). So, denoting now $k\equiv t$ and $G_o\equiv G(t_o)$, we can write in general terms:

$$(G(t_o).(t-t_o)) \equiv G_t(G(t_o), EC(t-t_o)) \quad (13')$$

The relation (13') *absorbs* into the general concept of operation of generation $G$, the phase of ***individual evolution before measurement*** of any involved specimen $\sigma(ms_G)$ of the studied microstate $ms_G$.

So – by definition – inside a succession $[G_t.MesA]$ the 'initial' state of the involved specimen of $ms_G$ is to be understood as the state of this specimen *at the time* $t\equiv t_1$ *when* ***begins*** *the act of measurement MesA*. So the act of measurement lasts an interval of time $(t_f-t_1)$ and for a measurement succession $[G_t.MesA]$ we can also write more explicitly

$$[G_{t1}.MesA(t_f-t_1)]=[(G(t_o), EC(t_1-t_o)) . MesA(t_f-t_1)]$$

All this permits to replace $G$ by $G_t$ (or by $G_{t1}$) in (9), (9'), (9''). So we get:

$$D_M(ms_{Gt}) \equiv \{ (D/A(ms_{Gt}), (M\pi c(G_t))_{XY} \},\ \forall A,\ \forall AB,\ \forall t,\ \forall j, \quad (9''')$$

Together the relations (13') and (9''') express an essential new concept, namely ***a factual*** *statistical-probabilistic law of evolution* tied with the studied microstate.

## (3.I).5. CONSTRUCTION OF THE DESCRIPTION OF A MICROSTATE *VERSUS* VERIFICATION OF THE DESCRIPTION

What follows here is very brief to be stated, but it becomes so important in the Part III of this work that it deserves this separate sub-section.

How can we *verify* a description (9) of any sort – (9') or (9'') or (9''')? The answer is obvious: *Only by reconstructing it* as many times as one wants and by modifying the parameters $(\varepsilon,\delta,N_0)$ from the definition (5) $(\varepsilon,\delta,N_0)$-$\{\pi(a_j)\}$, $\forall j$ until one observes the desired degree of stability. If this is not possible the prediction cannot be verified.

> Inside *IQM* the sequence of operations *[G.MesA]* from (7) or the generalization $[G_t.MesA]$ in the sense of (13') of this sequence, constitute the operational basis for *both* the construction and the verification of a description of a microstate. And the *IQM* factual-probability measures from any sort of primordially probabilistic description (9) – (9') or (9'') or (9''') – emerges by construction endowed with factual truth *and* verified to be such.

## (3.I).6. THE PROBABILITY TREE OF AN OPERATION OF GENERATION AS A REPRESENTATION FROM *A THEORY OF FACTUAL PROBABILITIES CONSTRUCTED BOTTOM-UP*

Inside *IQM* the "probability-tree $T_G$ of the operation of generation $G$" (or $G_t$) stems from the general method of relativized conceptualization *MRC* (MMS [2002A], [2002B], [2006]) where it emerges endowed with full generality: Inside *MRC* this concept is involved in ***any*** non-"individual" relativized description of any physical entity put in the role of object-entity-to-be-studied, denoted $œ_G$. And ***even the "individual" relative descriptions do involve it*** because according to *MRC* the *diagnosis of "individuality"* of a description $D/G,œ_G,V/$ of some physical object-entity $œ_G$, with respect to the qualifying view $V$ from that description – so with respect to the posited measurement-successions $[G.MesV_g]$ – requires by definition *many repetitions* of *all* the involved successions $[G.MesV_g]$: by definition, inside *MRC* the description $D/G,œ_G,V/$ is called "individual" iff the results of *all* the realized successions $[G.MesV_g]$, for any $V_g \in V$, are found to be individual with respect to that $V_g$, which then entails ***a posteriori*** that a one-branch probability-tree is generated.

> *So according to the general method of relativized conceptualization MRC* ***any description of a physical entity-to-be-studied is tied with a probability-tree****. Each one cell $D/G,œ_G,V/$ of relativized conceptualization can be expressed by a corresponding probability-tree.*

Consider now a probability-tree of the general form from the Fig.2:

- in its first-level statistical-probabilistic crown this probability-tree, if it has more than only one branch, includes *more than only one* finite $(\varepsilon,\delta,N_0)$-Kolmogorov-space;

- in its second-level statistical-probabilistic crown it includes *the correlations* between *the factual probability measures from these two or several spaces*;

- the probability measures mentioned above *are* ***numerically*** *specified*;

- *inside a from a probability-tree, the factual "truth" of the numerical specification of a factual probability measure is insured by construction, which means* ***only*** *that such a measure – by its very definition – emerges already "verified", it is stable with respect to repetition of its construction,* ***NOT in the least that it expresses how the studied physical entity "truly is in itself"****;*

- *so let us notice that*

> ***the factual probability measures from this probability-tree are not flawed by the principle of Laplace of a priori equipartition of the probabilities of the elementary events***;

- the trunk and the branches of the tree represent explicitly also the ***geneses*** of the whole content of each probability-space from the tree, via individual, ***physically*** realized and explicitly defined operations and processes: none among the involved descriptional elements consists of just a void conceptual mould, like in the top-down purely mathematical approach to the concept of probability that is practised nowadays. On the contrary, in a probability-tree all the involved descriptional elements emerge as conceptual moulds *already charged by construction with a* ***defined*** *sort of factual content.*

So we are in presence of a *far* more concrete and exhaustive representation of the concept of probability than that of Kolmogorov.

This is so because a probability-tree is constructed bottom-up by a creative factual process that brings forth into the focus of the constructive attention each one of the descriptional stages to be accomplished, separately, successively, in mutual connection with one-another. This excludes that these stages stay hidden like in a top-down extrapolation that does not *reach* – nor permits to only conceive – the factual roots of the physical-conceptual operational processes that have to be achieved accordingly to the constraints imposed by the involved cognitive situation.

So inside *IQM* – via the examination of the particular case of the factual microstates – have become very explicit the *necessity* as well as the main lines of *a new* ***sort*** *of theory of probabilities,* ***the "Theory of Factual Probabilities" belonging to the Mathematical PHYSICS***.

It is worthwhile stopping to consider this situation.

A theory of probabilities in the sense of a theory of ***physical*** facts simply does not exist as yet, notwithstanding that, strictly speaking, *inside Physics* ***any*** *prediction is* ***only*** *probabilistic if it is considered from a factual point of view*.

There exists exclusively a *mathematical* theory of probabilities, as a sub-domain of the theory of measures (nowadays Kolmogorov's treatment is dominant). The physicists make use of that as they make use of equations with one or two variables, or of the other mathematical structures. They introduce in it "factual data", pre-fabricated as such, they accommodate the provisional whole obtained in this way, to the desired application, and they complete it with various other mathematical techniques *and with much intuition*, like in any application to Physics of a mathematical instrument; and they obtain amazing results explained by the magic powers of Mathematics.

But in the case of the concept of probability *a much more complete* ***general FORMAL specification is possible***, so the mathematical framework appears as insufficient.

Indeed the examination of the particular case of the factual microstates has brought forth that the semantic outline of the *general* probabilistic features that are involved, are much more complex than it has been imagined in any framework available inside Mathematics. Kolmogorov's mathematical framework – the most modern one – is abundantly exceeded. And what has here acquired a rather clear contour, is a new Theory of "Factual Probabilities" *in the sense of the Mathematical* ***Physics***.

This will generate a new corresponding mathematical structure.

For the moment however it constitutes only a striking detailed illustration of the way in which Physics can generate mathematical structures. And finally, let us also notice that:

The whole *IQM* – considered globally – is synthetically encompassed by the concept of probability-tree of an operation of generation of a factual microstate.

The *way* in which the concept of factual probability expressed by probability trees of the operation of generation of a factual microstate, emerges together with *IQM* itself, deserves much attention because it brings strongly into evidence the decisive role, in the representations of scientific knowledge, of the *cognitive situation introduced by the entity-to-be-studied,* ***and of the direction of conceptualization – top-down or bottom-up*** *– which determines the order of conceptualization and the nature of the involved physical-conceptual-methodological operations*. In a top-down approach acts a strong inertial tendency to conserve indefinitely all the features and meanings from the starting-level of conceptualization. While in a bottom-up approach, on the contrary, the conceptualization is constantly under the variable and un-predictable control of the involved cognitive situation, of the global representational *aim*, as well as – step by step – of the local representational aim that commands that step. This entails a sort of conceptual Darwinism, an imperative to constantly adapt the process of conceptualization, to the achieved-and-evolving context, whichcleans away the inertial tendencies.

## (4.I)

# INFRA QUANTUM MECHANICS

We have isolated and organized the methodological framework of conceptual-*physical* cognitive operations that must be contained by any fully intelligible theory of the microstates. This has been named a priori Infra-Quantum Mechanics and denoted *IQM*.

*IQM* has been developed independently of any mathematical formalism.

In order to insure explicit control on all the levels of conceptualization we have started on the level of *zero* pre-accepted knowledge concerning the individual physical and fully singular specimens of any factual-microstate-to-be-studied[52]; and therefrom we have constructed *bottom up*. Thereby the origin and the order of progression on the vertical of conceptualization have been changed with respect to the classical science where, historically, the generation of knowledge on microphysical entities has necessarily begun, and so has been oriented top-down.

> This change of the origin of the process of conceptualization, from the start, has entailed a basic modification of the concept of 'microstate': we have been confronted with the problem of generating knowledge on "*factual* microstates", not on a mere concept of microstate.

Via an *unavoidable* methodological decision *MD*, the definition of the classical concept of microstate – that consists of a precise, *individual* and an ***abstract*** definition – has been transmuted into a *factual* definition that involves a *physical operation of generation G*. And this operation, being a human action, cannot itself be defined otherwise than by a *finite* number of controllable parameters. Whereas the fragment of as yet never conceptualized physical reality produced by such a factual operation of generation *G*, entails a priori an open unlimited set of unpredictable possible effects. And – with respect to the aim of constructing scientific, predictive and verifiable knowledge – this contrast that starts on a local zero of *specific* knowledge on the fragment of physical reality extracted by *G*, entails a "primordially" statistical character of the constructible scientific *knowledge*.

> The *cognitive situation* entailed by the choice to achieve a representation of scientific knowledge concerning micro-entities, that be constructed bottom-up, is what has unavoidably introduced a "primordially" statistical character of this knowledge.

So, starting from local zeros of specific knowledge, and according to general criteria of factual or logical necessity or of declared methodological choices, *IQM* has emerged as *a network of symbolizations of classes of conceptual moulds* of different sorts (methodological procedures, physical operations, probabilistic laws, etc.) in each one of which – later, inside a given *theory* of the microstates – will have to be lodged an element from the same class but more specified semantically via the constraints imposed by a general *model* of the entities to be studied *and* a global model wherefrom theses entities are drawn by adequate operations of generation. We have endowed this basic network of conceptual moulds with all the foundational elements that are required for achieving *any* acceptable scientific "theory" of the microstates: the whole construct has been organized in a logically coherent way; we have excluded elements that are not generally necessary as well as arbitrary a priori restrictions; etc.

[52] Unavoidably, the concept of a microstate itself must be *given* – as a receptacle where to pour future specifications – because from nothing, nothing more can be drawn.

The final result, *IQM*, can be regarded as a structural definition proposed here of *a **general** concept of "a **theory** of the **factual** microstates".*

The content of this result can be characterized as follows.

***1.** Consider first separately only the achieved probabilistic "descriptions" denoted D.*

The core of *IQM* consists of a primordially probabilistic *transferred description* constructed inside a minimal representational cell delimited by a pair *(G,A)* where *G* symbolizes a physical operation of generation of specimens of the studied factual microstate, in the sense of (1), and *A* indicates only one mechanical qualifying quantity that introduces a corresponding grid of qualification (2), in the sense specified in (1.I).2).

The corresponding most basic form of transferred description is

$$D/A(ms_G) \equiv [(\varepsilon,\delta,N_0)\{\pi(a_j,\forall j\}_G], \quad \forall A\in V_M \tag{9}$$

It is written time-*in*dependent, it involves only one qualifying quantity *A*, and it stops on the first probabilistic level.

The most comprehensive form of transferred description involves all the qualifying quantities *A* that are defined for the studied microstate inside an epistemic referential of the form $(G_t, V_M)$ and all the levels of conceptualization are included:

$$D_M(ms_{Gt}) \equiv [(\varepsilon,\delta,N_0)\{\pi(a_j),\forall j\}_{Gt}, (Mpc(G_t))_{AB}], \quad (G_t, V_M) \tag{9'''}$$

*Throughout IQM the physical operation G of generation of the individual specimens of the factual microstate to be studied reveals a basic and central role. Nevertheless this unavoidable operation has **never been noticed before**.*

The descriptional structure (9''') is marked by very remarkable peculiarities:

* It is strongly *relative* to a triad (6) $[G_t,ms_G,A]$ of genetic elements where the cell of conceptualization delimited by a given pair $(G_t,A)$ (an epistemic referential) is constituted in adequacy with a particular descriptional *aim*.

* The global basic genetic process of type (7) $\{[G_t.MesA]\}$, $\forall A$, that by repetitions of all the successions of the same general form brings forth a description (9'''), involves explicitly the fact that *each* one act of measurement performed on a specimen of the factual microstate $ms_G$ to be studied, requires in general a *previous* corresponding realization of also the operation of generation of that specimen, because in general a measurement-interaction destroys this specimen of the involved micro-*state* even if the involved micro-*systems* do persist.

* The brute observable result (8) $\{\mu\}_{kA}$, $k_A=1,2,...m_A$ of each one genetic succession $[G_t.MesA]$ from (7) – a group of publicly observable physical marks – is entirely meaningless by itself because it carries no perceivable qualia associable with, separately, the involved specimen of the studied microstate, nor with the qualifying quantity *A*.

Furthermore, in order to gain indirectly for the observable marks $\{\mu\}_{kA}$, $k_A=1,2,...m_A$, a meaning in terms of a value $a_j$ of the previously defined quantity *A*, and that be somehow tied also with the involved specimen of the studied microstate, ***the measurement-evolution MesA has to incorporate an adequate coding-procedure that depends on the considered pair (ms$_G$, A)***.

In its turn such a coding-procedure, in order to be definable in a non-arbitrary way, requires a ***general model*** *of a microstate, and a model of "a specimen of the studied microstate".*

So any acceptable "scientific *theory*" of the factual microstates *must* introduce a generic *fragmentary model* of the specimens of any sort of "one factual microstate" (accordingly to the definitions from (2.I).1.2)). And such a model unavoidably presupposes also a *global model of the physical reality* wherefrom it can be singularized in a physical-operational way via an adequate operation of generation *G* in the sense of (1).

But such a global model – even if it is mathematically expressed or if it consists exclusively of a mathematical structure – should never be confounded – itself – with *a* "scientific *theory*" of the microstates. A a full "theory" – in the sense defined here – has to prescribe explicitly the *whole* process by which one can create step by step consensual predictive and verifiable knowledge on the factual microstates, via operations of generation of specimens of the entity-to-be-studied and acts of measurement on these specimens. A representation where the pairs (*G,MesA*) of conceptual-*physical* operations are not explicitly defined and represented and made use of accordingly to *IQM*, is *not* a theory of the microstates because it does not prescribe how to be applied.

A mathematical representation ***alone*** (in particular of a global model of the physical reality [53]) that is devoid of also a conceptual-*physical* operational representation of the way in which the corresponding knowledge is concretely created, is not by itself a consensual, predictive and verifiable theory of the microstates; it is only a necessary element from such a theory (cf. *(1.I).2.2.1*).

* A description (9''') cannot be assigned to the studied microstate itself considered separately, but only – globally – to the whole successions of operations $\{[G_t.MesA]\}$, $\forall A \in V_M$ that generated the description (where $G_t$ is reducible to $G_o$ in the sense defined in (13'). So *a description (9''') is strongly and* ***indelibly*** *tied to its genetic process*. ***In a description (9''') the classical notion of "object" is not yet extracted from its genetic process***. This feature is intimately tied with the *absence* of any defined space-time structure assigned to the description (9''') *itself* [54].

The remarks listed above have pointed out new *general* characteristics of the descriptions of the factual microstates.

***2.*** *Consider now the descriptions D from (9) an (9''') inside their whole* ***probabilistic*** *context, either genetic or epigenetic*.

We shall now be very redundant with respect to (3.I).6. But the point we want to make is so important that we" choose to do this.

A description (9''') itself consists of – exclusively – a set of factual probability *measures* that have been specified numerically in *finite* terms. In contradistinction to a Kolmogorov purely mathematical, ineffective and generic concept of probability measure, a set (9''') of factual probability measures is explicitly tied inside its probability-tree to all the genetic successions of operations (7) $[G_t.MesA]$ that have been achieved by the observer-conceiver in order to construct it. So, while a Kolmogorov probability space is a-temporal, the full probabilistic description from a probability-tree involves also quite essentially specific space-time features. This is a vast *extension of the classical purely mathematical concept of probability*. Furthermore, the definitions from (2.I).1.2 entail also another sort of extensions of the classical concept of probability. Namely, in the second case there emerge probabilistic extensions that violate brutally the classical probabilistic ways of thinking because in this case the involved probabilistic significances concern also the ***interior of the "elementary events"*** in the probabilistic sense, which never happens inside Kolmogorov's approach[55], [56]. Precisely thereby, inside *IQM* the 'problem' of non-locality becomes intelligible.

---

[53] Like, for instance de Broglie's interpretation or D. Bohm's one.

[54] In MMS [2002B] and [2006] it has been shown that the construction of the concept of material "object" in the classical sense involves precisely assignation of an own space-time support.

Together, the time-features mentioned above and the consequences of the definitions (2.I).1.2 introduce a deeply modified and enriched concept of probabilistic dependence that involves an explicit distinction between what is presupposed to be only potential and what is presupposed to have been actualized, as well as incorporation of both these ways of 'existing' conceived by the human mind.

Let us insist now more upon the time-features. When one progresses mentally bottom-up along the vertical of conceptualization, inside the probability-tree from the Fig.2 one can watch step by step *how* a radical scission sets in between all the individual ***physical***-conceptual genetic *human actions* – that *do* involve space-*time* – and on the other hand the final global *result* (9''') of all these actions once they are achieved. On the graphic representation from the Fig.2, the purely numerical probabilistic content of the final description (9''') appears separately, displayed on the tops of the mutually disjoint purely spatial branch-zones of the tree. But this static, geometric tree-like disposition is only a globalized and residual purely spatial *trace* of the factual space-time emergence of the successive effects of the individual genetic operations that have generated the tree. The temporal aspects that, in their actuality, had *individually* and *successively* contributed to the globalized spatial splitting in distinct branches of the tree, have now evaporated in – literally – 'the air of time'. Only the global final probabilistic description (9''') – considered by itself – subsists *radically devoid of any* ***own*** *space-time organization* and definitively separated from its genesis entailed by human aims and ways of thinking and by the human ways and technical possibilities of acting.

> The description (9''') is a purely symbolic-and-numerical drop-off from the whole physical-operational *factual* realization in space-time, of a conceptual-methodological design for reaching a cognitive purpose. The nature of this drop-off, namely a set of *$(\varepsilon,\delta,N_0)$-convergent* '*probabilistic*-statistical' distributions that has been factually specified in numerical terms, *is –* ***itself*** *– entirely abstract*, and *its own final global structure is a-spatial and a-temporal*.

Scissions of the same kind appear of course in any human process of construction, whether of an abstract entity, or of a material one. But nowhere do they appear with a so radical character. Nowhere are they erected upon a strict absence of *any* ***perceived*** material instance of that what is studied, not even inside only memories from human minds; and this, notwithstanding that *what* is to be studied is posited to be of physical nature and to exist inside space and time. This radical character stems from the fact that inside *IQM*, for the case of factual microstates, we have been coercively led to ***entirely suppress*** any *specified* assignation to the specimens of the entity-to-be-studied, of any previously well-defined properties, and to ***replace*** such assignations by exclusively a composition of a small number of basic *classes* of only posited and named concepts of only pre-formatted ***void*** *conceptual* ***receptacles*** *where to lodge* **later** *– in a prescribed way – a* ***specified knowledge*** *that is still left to be generated and poured into these receptacles inside a* **theory** *of the microstates (cf.*

---

[55] The mentioned extensions of the classical concept of probability are intimately connected with basic extension of also the classical *logical* conceptualization. It has been shown in MMS [2002A], [2002B], [2006] that these extensions possess a general character and they admit *a unification of the logical and the probabilistic approaches* thereby dissolving obstacles that resisted since a long time.

[56] The concept of a primordial transferred description (9''') itself, in fact *founds universally the whole human conceptualization* (MMS [2002], [2006]), the macroscopic classical one as much as the conceptualization of the microstates. The only difference with microphysics is that inside the classical domain of conceptualization the direct perceptibility of the involved physical entities permits to economize an explicit *knowledge* of this foundational fact.

*for instance MD and the definitions from (1.I).2)*. This entails for the global space-time structuration of the mentioned receptacles, elements with a very purified semantic nature and very clear-cut contours, and this induce a very definite and detailed ***intelligibility***.

Such a result cannot emerge inside a directly mathematical theory of the microstates because there the general conceptual-methodological-operational imperatives get mixed from the start with the features of pre-fabricated mathematical tools, and with also the particular consequences of some definite particular model that certainly ***is*** working inside any theory – even if only implicitly – in order to specify appropriate measurement operations and coding-procedures. For in the absence of such specifications there is no possibility whatever to define adequate measurement-operations, so no possibility whatever to define predictions and to verify them.

> Via the *particular* case of the factual microstates *IQM* illustrates – only illustrates – *a new and general "Theory of Factual Probabilities" belonging to the Mathematical* ***Physics***. The content of this new theory consists of all the relativized structures each one of which is called "a probability tree of an operation of generation *G*", and the relations between these trees. And the summed up, abstract but perennial output of each such tree – considered alone, isolated from its genesis – is a set of type (9''') of primordial transferred *descriptions* $D/A(ms_G) \equiv [(\varepsilon,\delta,N_0)\{\pi(a_j,\forall j\}_{Gt}]$, $A \in V_M$ where *all the involved probability laws as well as their correlations of higher orders emerge* ***numerically*** *specified and genetically endowed with* ***verified factual truth***, *and concern definite qualifications of definite physical entities*, instead of concerning just symbols called "elementary events" or "events" like in the purely mathematical structure defined by Kolmogorov.

This new general Theory of Factual Probabilities ***incorporates*** effective and relativized realizations of the purely mathematical concept of a Kolmogorov probability-space (a non-ineffective and absolute concept where moreover the probability law is not specified numerically). But it strongly exceeds this concept from various points of view, both semantic and structural.

> This is already a most noteworthy result because ***any*** *scientific description of physical entities is* $(\varepsilon,\delta,N_0)$*-probabilistic when it is considered from a factual point of view* **[57]**.

And it is noteworthy that inside *IQM*, this new Theory of Factual Probabilities works *already* notwithstanding that inside *IQM* it is as yet devoid of a mathematical expression. It has already clarified:

* the endlessly discussed question of the 'primordial' or 'essential' statistical character of the modern microphysics;

* the general semantic essence of the "locality problem";

It has achieved these clarifications merely in consequence of the explicit structure of *IQM*, mainly in consequence of (1) and the definitions from (1.I).2) **[58]**.

---

[57] In the Part III of this work the probability measures from this Theory of Factual Probabilities will acquire also a general mathematical representation, namely a Hilbert-Dirac representation. Thereby it will become **the most general one among all the possible theories from the Mathematical Physics** (precisely because **any** scientific description of physical entities is $(\varepsilon,\delta,N_0)$-probabilistic when it is considered from a factual point of view**)**. Then QM2 as a whole will appear to be just **an application to the particular study of the factual microstates,** of this new general Theory of Factual Probabilities; a particular application but a radical one, in this sense that it can be rooted directly into the a-conceptual physical factuality and therefrom it can be entirely constructed bottom-up. In this particular but radical application the concept of probability-tree of the operation of generation from a given epistemic referential (G,V) constitutes just one **basic** confined cell of relativized conceptualization of the studied microstate, in the sense defined inside the general Method of Relativized Conceptualization (cf. the general framework in the Appendix1 and in MMS [2006] chapter 4, and a more specific exposition in MMS [2014]).

[58] While the Hilbert-Dirac formulation of the nowadays quantum mechanics $QM_{HD}$ continues since decades to nourish these two basic unsolved questionings.

*3. We now spell out two essential* ***methodological facts*** *brought into evidence by IQM and a major problem of the scientific conceptualization raised by it.*

The first *methodological fact* is the following one.

* Taking systematically into account any involved descriptional relativity, *restricts*, and thereby it specifies, thus entailing *precision*. This is directly opposed to the meaning of the word 'relativ***ism***'.

This huge confusion should be suppressed.

The descriptional relativities are organically tied with reference, and reference installs methodological specifications instead of the vagueness governed by absolutes, that quasi-systematically are *false* absolutes.

Descartes' concept of system of reference has organized our thought and our power of communication; it has enhanced to an unspeakable degree our conceptual and practical efficiency.

The second *methodological fact* is the following one.

* The genesis of a description is the vehicle of the semantic contents poured into that description. So explicit geneses are precious to be known explicitly.

The *problem of scientific conceptualization* raised by *IMQ* is the following one:

* *How, exactly, can the radically singular and potentially so complex conceptual-****physical*** *content carried by a description of type (9'''), be pertinently* ***loaded*** into a pre-fabricated mathematical construct like Schrödinger's differential equation?

More generally:

What, exactly, happens at each ***junction*** – inside a given theory of the microstates – between a factual effective realization of an output of the form (9'''), and a pre-established mathematical descriptor chosen in order to represent it?

This question, when one stops on it long enough, triggers a sort of stupefaction. I think that Wigner's famous considerations on *the* "unreasonable" power of mathematics concern very precisely this sort of questions. One senses a void of satisfactory analysis of this question, disguised in just a feeling of miracle.

This sort of void should be suppressed.

*4.* Finally let us recall once more that *IQM* is – by construction – marked by two related, big, *deliberate absences*. The absence of a general model of microstate, and the absence of specified coding rules for assigning meaning to the observable result of a measurement succession (7) *[G.MesA]*. These two deliberate absences are conditions of the full generality of this "theory of *any* theory of the factual microstates" that we have called Infra-Quantum Mechanics and have denoted *IQM*. Indeed any manner of compensating these absences can stem only from *a priori* ***restrictive*** *postulations* that can find place only inside one given particular theory of the factual microstates.

By contrast and paradoxically, these absences are precisely that what entails with full evidence a highly non-trivial assertion:

> Without a *model* of a microstate that permit to conceive 'appropriate' modalities for measuring a given quantity *A* on a given sort of factual microstate and without corresponding explicit *coding procedures* for translating the observable marks produced by any one act of measurement *MesA* into *a meaning in terms of a definite value $a_j$ of the measured quantity A*, the primordial transferred descriptions (9''') are just a heap of inert puppets.

The necessary and sufficient strings that can bring these puppets to work in a controlled way and to create effective knowledge on a factual microstate consist precisely of *a general model of a specimen of a factual microstate extracted via an operation of generation G from a corresponding global model of the physical reality*. Only such a hierarchical duo of models can permit to identify measurement-interactions of which the observable results are *intelligibly* coded in terms of a definite value of the measured quantity.

So, when the general imbedding-and-reference-structure from *IQM* will be compared with the nowadays Quantum Mechanics – that, necessarily, does somehow involve a duo of models of the sort mentioned above – the confrontation, by comparison, will reveal some deficiencies inside inside the nowadays Quantum Mechanics, and thereby it will act like a machine that produces guides for constructing a more intelligible mathematical theory of the factual microstates: A whole set of *referred criteri*a will be at work to help to reach this purpose. For reference is a very powerful instrument, and *IQM* offers an organized recourse to a structure of references. In its essence *IQM* ***is*** just a structure of references organized so as to guide the generation of – specifically – scientific knowledge on factual microstates. *IQM* is nothing more.

Now we conclude:

Out of nearly a nothingness of explicit previously available knowledge on *how* consensual, predictive and verifiable knowledge on factual microstates can emerge, we have drawn the Infra-(Quantum Mechanics) *IQM*, an explicit methodological-conceptual-operational construct where *any* acceptable mathematical theory of the factual microstates should be embeddable for multiform reference. This construct has been endowed with a *formalized* though qualitative structure that is tied step by step with specifications of a ***semantic*** nature. The whole construct consists of just *a composition of void* ***semantic*** *moulds* for lodging in each one of these a more specified content of the same basic semantic nature as itself, but *specified* via measurement-evolutions defined so as to permit to code the observable marks produces by any act of measurement, in consistency with a corresponding hierarchic duo of models, in the sense indicated above. This construct is a sort of formal system for the construction of any theory of the microstates.

***IQM is the first "infra-theory", fully relativized***.

But *IQM* has been organized inside the general Method of Relativized Conceptualization (MMS [2002A], [2002B], [2006]). Thereby it will generate a Second Quantum Mechanics *QM2* that opens up a perspective that largely exceeds microphysics, by tracing on the horizon:

- a fully specified "Theory of the Factual Probabilities" *belonging to the class of the Mathematical Theories of Physics*;

- the path along which can be realized a deep-set methodological unification of the *whole* of modern Physics; a unification founded upon the mutual characterization, inside this vast domain, of the various human *cognitive situations* that are involved.

Moreover, beneath the contours of the domain of the whole of the modern Physics, and far beyond these contours, will appear still pale lines that sketch out a synthetic perception of a genetic methodologica**l** unity of the whole of the human Sciences of the bio-physical reality *and* of the techniques drawn from these.

**SYMBOLICALLY EXPRESSED SYNTHESIS OF *'IQM'***

***Operation of generation G of one factually defined microstate $ms_G$:***

**(1)** $G \leftrightarrow ms_G, \quad ms_G \equiv \{\sigma(ms_G)$

***One qualification of a microstate $ms_G$ by a value $a_j$ of a measured qualifying quantity A:***

**(2)** $[(G \rightarrow ms_G)$. $(MesA \rightarrow$ (group of observable marks $\{\mu\}_{kA}$ coding for one value $a_j$ of $A)]$, $k_A=1,2,...m_A$, $j=1,2..J$,

In short **(3)** $[G.MesA] \rightarrow (\{\mu\}_{kA} \approx a_j)$, $k_A=1,2,...m_A$, $j=1,2....J$, $A$ given

***The definitions from* (2.I).1 *of the main types of microstates $ms_G$***

***The factual predictive $(\varepsilon,\delta,N_0)$-probability law on the statistic of outcomes of A-measurements on the microstate $ms_G$, so inside the epistemic referential (G,A):***

**(5)** $(\varepsilon,\delta,N_0)\text{-}\{(\pi(a_j)\}, \forall j)\}_G, \quad (G,A)$

**** The set of all the factual predictive $(\varepsilon,\delta,N_0)$-probability laws (5) for one given microstate $ms_G$, so 'description ofms$_G$ inside the epistemic referential $(G,V_M)$:***

**(5')** $\{(\varepsilon,\delta,N_0)\text{-}\{(\pi(a_j)\}, \forall j)\}_G\}, \quad (G,V_M)$

**** The 'genetic triad' of one factual $(\varepsilon,\delta,N_0)$-probability law (5)***

**(6)** $(G, ms_G, A), \quad (G,A)$

**** The 'genesis' of one given law* (5)**

The set of successions of operations

**(7)** $\{[G.MesA]\}), \quad (G,A)$

**** The brute observable output of the genesis {[G.MesA]} of one law (5)***

$\{\{\mu_{kA}\}, \ k_A=1,2,...m_A, \quad (G,A)$

**** The brute observable output of all the geneses {[G.MesA]} of all the laws (5'):***

The set of all the factual data produced by (5')

**(8)** $\{\{\mu_{kA}\}, \ k_A=1,2,...m_A, \quad (G,V_M)$

**** Re-notation: 'primordial transferred description of $ms_G$ with respect to the mechanical qualification A, so inside the epistemic referential (G,A)'***

**(9)** $\{(\varepsilon,\delta,N_0)\text{-}\pi(a_j)\}_G \approx D/A)(ms_G), \quad (G,A)$

**** Re-notation: the primordial transferred mechanical description of the microstate $ms_G$, so inside the epistemic referential $(G,V_M)$'***

**(9')** $\{\{(\varepsilon,\delta,N_0)\text{-}\pi(a_j)\}_G \approx D_M(ms_G), \quad (G,V_M)$

**** Genetic symbolizations of the two sorts of primordial transferred descriptions:***

**(10)** $(D/A)(G,ms_G,A), \ (G,A)$ or $D_M(G,ms_G,V_M), \ (G,V_M)$

**** The meta-probabilistic correlations $(M\pi c(G))_{AB}$ involved by (1) $G \leftrightarrow ms_G$ with respect to the pair (A,B) of qualifying quantities:***

**(11)** $\pi(a_j)=\mathbf{F}_{aj,B}\{\pi(b_r), \forall r\}_G$

**(11')** $\mathbf{F}_{AB}(G)= \{\mathbf{F}_{aj,B}\{\pi(a_j), \forall j\}_G$

where $\mathbf{F}_{aj,B}\{\pi(a_j), \forall j\}_G$ and $\mathbf{F}_{AB}(G)$ are two functionals that represent, respectively, the *individual* probability $\pi(a_j)$ in terms of the *whole* probability law $\{\pi(a_j), \forall j\}_G$ and the global correlation between the two whole laws $\{\pi(a_j), \forall j\}_G$ and $\{\pi(b_r), \forall r\}_G$.

**** The description (9') completed by (11), (11'):***

**(9'')** $D_M(ms_G) \equiv \{[(\varepsilon,\delta,N_0)\text{-}\{\pi(a_j), \forall j\}_G, (M\pi c(G))_{AB}], \ \forall A, \forall AB \in V_M, \ (G,V_M)$

**** Qualitative logical specification on the individual probabilistic predictions on one microstate with one micro-system and with composed operation of generation :***

**(12)** $\pi_{12}(a_j)_{G(G1,G2)} \neq \pi_1(a_j)_{G1} + \pi_2(a_j)_{G2}$

where all the probabilities are *individual.*

**** Absorption in the operation of generation, of the evolution of a microstate $ms_G$ :***

**(13)** $G_o, \ G_1=F(G,EC,(t_1\text{-}t_o)),..,G_k=F(G_o,EC,(t_k\text{-}t_{k-1})),..., \ G_f=F(G_o,EC,(t_f\text{-}t_{K-1})), \ k=1,2..K$

in short **(13')** $[G_o.(t\text{-}t_o)] \equiv G_t$

**** Consequence of (13') on the transferred description (9''):$D_M(ms_{Gt})$:***

**(9''')** $D_M(ms_{Gt}) \equiv \{ (D/A(ms_{Gt}), (M\pi c(G_t))_{XY}\}, \ (G_t,V_M)$.

# CONCLUSION ON THE PART I

When one watches the way in which *IQM* emerges the naïvely realistic view that scientific knowledge is 'discovery' of pre-existing 'truth' collapses into dust. And in its place one sees, one *feels* in what a sense conceptual-operational *procedures* – involving physical operations or abstract ones – can progressively be assembled into ***a METHOD*** born from the unlimited human curiosity and inventiveness, from the constraints imposed by the human ways of thinking of what we call physical reality and of acting upon this, and from explicit ***PURPOSES*** freely chosen by men. What has been obtained here, in the Part I of this work, is such a particular piece of *metho*d. It is a global, internally coherent piece of method for constructing a definite particular piece of procedural *scientific* knowledge directed by a definite project. It is not in the least a discovery of pre-existing 'intrinsic truths' about how physical reality *is*, absolutely, 'intrinsically', 'in itself'; not even is it – not in the least – a way of 'approaching' such a discovery. Such "discoveries", such "asymptotic reaching", are mere illusion; just an emanation from the self-contradicting notion of *'scientific knowledge of reality-in itself'*; a genuine Fata Morgana, the original sin of scientific thought [59].

We are trapped in a cage where "absolute intrinsic truth" is irrepressibly felt to pre-exist but constantly stays out of the reach of "science", changes of direction, negates any definitive convergence, marks new starts, unpredictably, frustratingly. This mental phantom acts from inside a non-organized and changing swarm of lures and it precipitates us blindly toward ill-defined targets. The hope for impossible final and absolute scientific truths unavoidably entails assaults by a feeling of impotence, of inefficiency, of enslavement.

I perceive only one attitude that preserves from this sort of major fail: To realize fully that – from the point of view of science – *a posit of* ***existence*** *of a physical reality, and on the other hand,* ***consensual predictive and verifiable knowledge*** *of "how this physical reality truly is", are of* ***different ESSENCES***; that an absolute bare ***EXISTENCE*** of a "physical reality" *can* be posited inside what we call "science", *but just as such*, as exclusively ***a posit of BARE existence***. While the intuitive source of this posit and its qualifying contents are definitively imprisoned in metaphysics; it is inaccessible to *consensual* knowledge, notwithstanding that in the absence of this posit, "science" would seem to be just a game.

While inside science, with a blindfold deliberately fixed on our metaphysical eye and on the basis of entirely declared posits – metaphysical or not – and *data*, to *construct* consensual, predictive and verifiable knowledge, humbly, hypothetically, *relatively*, provisionally, respecting step by step the unavoidable constraints as well as the deliberately chosen ones; and to construct *from the maximal possible depth,* ***upward***. Thereby only restricted, finite and methodized knowledge can emerge; but a fully definite and consensual knowledge endowed with *an entirely exposed genesis* where the unending inflow of relative *meaning* can be watched and is constantly left open to return and to indefinite optimization, precisely because it is only hypothetical and finite and relative.

---

[59] MMS [2006], pp. 127-136. The human desire of knowledge, by itself, is not a sin, even if in this respect the Bible is somewhat ambiguous. But the posit that the *scientific, consensual, predictive and verifiable knowledge* is "discovered", not deliberately *constructed* accordingly to methodological-factual laws, certainly is a huge sin, because it imprisons the minds and exhausts them in efforts toward impossible aims.

# PART II

# PRELIMINARY CRITICAL-CONSTRUCTIVE GLOBAL EXAMINATION OF THE HILBERT-DIRAC QUANTUM MECHANICS, BY REFFERENCE TO *IQM*

**ABSTRACT OF PART II**

The mathematical representation of the microstates accordingly to the Hilbert-Dirac Quantum Mechanics $QM_{HD}$ is compared with the Infra-Quantum Mechanics *IQM*. It is found that:

**(a)** $QM_{HD}$ is *void of any definite representation of* ***the individual level of conceptualization***.

**(b)** $QM_{HD}$ involves quite essentially, though in a hidden way, Louis de Broglie's model of a microstate, incorporated to his global model of the physical reality.

**(c)** Via the *IQM*-concept of operation of generation *G* of a factual microstate, these de Broglie models permit to specify explicitly and in defined relation with operational-observational features the – posited – result of one realization of an operation of generation of a specimen of the studied factual microstate.

**(d)** The *IQM*-concept *G* is furthermore shown to clarify the semantic contents of several basic formal writings from $QM_{HD}$.

**(e)** The "theory of measurements" asserted by $QM_{HD}$ is then examined and is found to be ***fundamentally unacceptable***.

The results indicated in **(a)**, **(b)**, **(c)**, **(d)**, **(e),** of the critical-constructive comparison of $QM_{HD}$ and *IQM* leave us with a cleansed and semantically improved conceptual-mathematical ground for entering upon the construction of a fully intelligible Second Quantum Mechanics *QM2* inside an explicitly defined new and evolving global framework for achieving this that is denoted $[IQM\text{-}QM_{HD}]$.

# INTRODUCTION TO THE PART II

The second part of this work is devoted to a global preliminary examination of the Hilbert-Dirac formulation of Quantum Mechanics $QM_{HD}$, by reference to *IQM*.

*Throughout what follows* $QM_{HD}$ *is supposed to be well known*.

The main goals of the Part II are:

* By use of *IQM* to yield – from the outside of $QM_{HD}$ – a critical perspective on the general structural features of $QM_{HD}$.

* To identify the general characteristics of the model of a specimen of a factual microstate of a given sort, extracted from the physical reality by an operation of generation *G* in the sense of (1) from *MD* (such a model certainly does somehow work inside $QM_{HD}$ since in the absence of *any* such model this theory would have been impossible).

* To establish explicitly all the *immediately* specifiable clarifications induced by *IQM* into current mathematical representations from $QM_{HD}$.

* To identify the *precise* reasons why, since soon a whole century now, the "theory of quantum measurements" asserted in $QM_{HD}$ raises so stubbornly an unending variety of "problems" and of "solutions" to these.

The goals mentioned above are pursued inside two chapters, 5.II and 6.II.

- The chapter 5.II is devoted to a brief comparison between $QM_{HD}$ and *IQM*.

- The chapter 6.II is devoted to critical and constructive clarifications, namely:

-- Identification of the general model of a specimen of a factual microstate of a given sort that works inside $QM_{HD}$.

This amounts to a conceptual-physical-operational insertion of de Broglie's seminal model, into the network *IQM*, via the concept of operation *G* of generation.

-- Introduction of an explicit and general use of the operations *G* of generation of a factual microstate in all the main mathematical writings from $QM_{HD}$. This amounts to a step toward an explicit conceptual-*physical* operationalization *of* $QM_{HD}$.

-- An explicit *refusal* of the von Neumann representation of the quantum measurements.

-- Extraction of the essence of the $QM_{HD}$ representation of the quantum-measurements and, by reference to *IQM*, a thorough critical identification of its implications and of the conclusion entailed by these, namely a radical refusal of what is called inside $QM_{HD}$ "the theory of quantum-measurements".

Throughout the Part II, *IQM* and $QM_{HD}$ are both made use of. In this sense they constitute together a new framework that will be denoted *[IMQ-*$QM_{HD}$*]*. At the start *IQM* and $QM_{HD}$ act inside *[IMQ-*$QM_{HD}$*]* by confrontation. But progressively, by integration of the consequences of each confrontation, a fusion between $QM_{HD}$ and *IQM* will develop. So in the Part II of this work we start an organic process of conceptual growth. At the end of the Part II this process will have brought us in possession of a framework *[IQM-*$QM_{HD}$*]* already endowed with a non-null degree of inner unity, and correlatively, on the very edge of a guided action of constructing a fully intelligible second quantum mechanics.

# 5.II [60]

# COMPARISON BETWEEN $QM_{HD}$ AND *IQM* AND THE PLAN OF THE PART II

As already stated, $QM_{HD}$ is considered to be known. But for self-sufficiency and commodity we recall telegraphically the core-features of both representations to be compared.

## (5.II).1. THE $QM_{HD}$-REPRESENTATION

***The basic posits***

The $QM_{HD}$-formalism is founded on:

° two basic *representational definitions*, namely a rule that defines the mathematical representation of the studied microstate and a rule (of 'quantification' [61]) that defines the mathematical quantum mechanical representation of the classical mechanical qualifying quantities;

° a *mathematical* principle (Born's principle) founded upon a basic equation;

° three measurement postulates;

° a postulate of evolution.

* ***The rule of representation of a microstate***. At any fixed time *t* the studied microstate is defined by a state-vector (a *ket*) $|\psi(\mathbf{r},t)>$ from a generalized Hilbert-space $\mathcal{H}$ that contains also tangent ket called eigenket and defined by an equation mentioned later [62].

* ***The rule of representation of* the qualifying mechanical quantities.** Any classical measurable mechanical quantity $A(\mathbf{r},\mathbf{p})$ is represented by a corresponding operator $\boldsymbol{A}$ called an *observable*, that acts on the elements from $\mathcal{H}$; this observable is constructed from $A(\mathbf{r},\mathbf{p})$ as follows:

- The classical quantity 'position' $\boldsymbol{r}(x,y,z)$ is represented by a position-observable $\boldsymbol{R(X,Y,Z)}$ where: the classical functional form that in $\boldsymbol{r}(x,y,z)$ relates the symbols $x,y,z$ is conserved; $\boldsymbol{X,Y,Z}$ represent, respectively, the classical space-coordinates entailed by the chosen Cartesian referential but each one of which is posited to act operationally by multiplying what follows it, which is expressed by re-writing it also as operators $X.$ , $Y.$ , $Z.$ .
- The classical quantity of momentum $\boldsymbol{p}(p_x,p_y,p_z)$ is represented by the momentum-observable $\boldsymbol{P}(\boldsymbol{P_x},\boldsymbol{P_y},\boldsymbol{P_z})$ where: $\boldsymbol{P_x}=i(h/2\pi)d/dx$, $\boldsymbol{P_y}=i(h/2\pi)d/dy$, $\boldsymbol{P_z}=i(h/2\pi)d/dz$; the classical functional form that in $\boldsymbol{p}(p_x,p_y,p_z)$ connects $p_x$, $p_y$ and $p_z$ being conserved for connecting also $\boldsymbol{P_x}$, $\boldsymbol{P_y}$ and $\boldsymbol{P_z}$.
- The classical mechanical quantity $A(\mathbf{r},\mathbf{p})$ is represented by the function $\boldsymbol{A(R,P)}$ of the operators $\boldsymbol{R}$ and $\boldsymbol{P}$ that is first constructed with the same functional form as in $A(\mathbf{r},\mathbf{p})$, and then is *symmetrized*.

*The equation for eigenket and eigenvalues of an operator $\boldsymbol{A}$: $\boldsymbol{A}|u(\mathbf{r},a_j)>=a_j|u(\mathbf{r},a_j)>$, $\forall j$.

---

[60] This notation is to be read: chapter 5 from Part II: The numbering of the chapters *continues* the numbering from the first part.

[61] Cohen-Tannoudji, C., Diu, B. & Laloë, F., [1973],

[62] This term has no plural.

* ***Born's mathematical principle of spectral decomposability***. Any ket from $\mathcal{H}$ – state-ket or eigenket – can be decomposed on the basis of eigenket $\{|u(\mathbf{r},a_j)>\}$ introduced in $\mathcal{H}$ by any observable $\boldsymbol{A}$, which yields the expansion of $|\psi(\mathbf{r},t)>$ with respect to $\boldsymbol{A}$: $|\psi(\mathbf{r},t)>/\boldsymbol{A}=\sum_j c(a_j,t)|u(\mathbf{r},a_j)>, \forall j$, where the $|u(\mathbf{r},a_j)>$ are the eigenket that form the basis of eigenket of $\boldsymbol{A}$ in $\mathcal{H}$ and the $a_j$ denote the corresponding eigenvalues.

* ***The three measurement postulates***

- ***Measurement postulate 1.*** The observable result of an act *MesA* of measurement of the observable $\boldsymbol{A}$ represents by definition an eigenvalue $a_j$ of $\boldsymbol{A}$.

- ***Measurement postulate 2 (Born's probability postulate)***. When the observable $\boldsymbol{A}$ is measured at the time $t$ the probability $\pi(t,a_j)$ of outcome of the eigenvalue $a_j$ of $\boldsymbol{A}$ is calculated via the expression $|<u(\mathbf{r},a_j)|\psi(\mathbf{r},t)>|^2$ where $|u(\mathbf{r},a_j)>$ is the eigenket of $a_j$ and $|<u(\mathbf{r},a_j)|\psi(\mathbf{r},t)>|^2 \equiv |c(a_j,t)|^2$ is the squared value of the coefficient $c(a_j,t)$ from the expansion of $|\psi(\mathbf{r},t)>$ with respect to $\boldsymbol{A}$, $|\psi(\mathbf{r},t)>/\boldsymbol{A}=\sum_j c(a_j,t)|u(\mathbf{r},a_j)>, \forall j$.

- ***Measurement postulate 3 (the projection postulate).*** If the act of measurement *MesA* has produced the result $a_j$ then immediately after the measurement the studied micro-*state* is represented by the re-normed projection

$$P_n|\psi(\mathbf{r},t)> / \sqrt{|<\psi(\mathbf{r},t)|\, P_n\, |\psi(\mathbf{r},t)>|}$$

of $|\psi(\mathbf{r},t)>$ on the direction in $\mathcal{H}$ of the eigenket $|u_{aj}(\mathbf{r})>$ of $a_j$.

* ***The postulate of evolution***: The evolution of the state-ket $|\psi(\mathbf{r},t)>$ is defined by the Schrödinger equation of the problem

$$i(h/2\pi)d/dt\,|\psi(\mathbf{r},t)> = \boldsymbol{H}(t)$$

where $\boldsymbol{H}(t)$ is the hamiltonian observable that represents the total energy assigned to the studied microstate.

***The main algorithms***

The essence of the way of working of the $QM_{HD}$-formalism can be regarded to consist of four *purely formal* types of problems and of the correlative algorithmic procedures for obtaining the solution, and a fifth ***factual***-formal problem with its own solution:

***Problem 1***: Determine the state-ket $|\psi(\mathbf{r},t)>$ that represents the microstate to be studied inside the Hilbert space $\mathcal{H}$ assigned to this microstate.

*Solution to problem 1*: Write the Schrödinger equation of the problem; solve it; *introduce the limiting conditions* and ***identify the initial state-ket*** $|\psi(t_o)>$. Therefrom the Schrödinger equation is asserted to determine the state-ket $|\psi(t)>$ of the studied microstate for any time $t$.

***Problem 2***. For any state-ket $|\psi(\mathbf{r},t)>$ and any observable $\boldsymbol{A}$, determine the predictive probability law concerning the possible outcomes of measurements *MesA* performed on the studied microstate.

*Solution to the problem 2*:

Write the equation $\boldsymbol{A}|u(\mathbf{r},a_j)>=a_j|u(\mathbf{r},a_j)>$, $\forall j$, and calculate from it the eigenket $\{|u(\mathbf{r},a_j)>\}$ and the corresponding eigenvalues $a_j$. According to the measurement-postulate 1, the result of any act of measurement *MesA* represents by definition an eigenvalue $a_j$.

In order to determine the probability of outcome of any *given* value $a_j$, form the set of squared absolute values $|c(a_j,t)|^2$, $\forall j$, drawn from the expansion $|\psi\ \mathbf{r},t)>/\boldsymbol{A}$ on the basis

$\{|u(\mathbf{r},a_j)>\}$ of the eigenket of $\boldsymbol{A}$, and accordingly to the measurement-postulate 2 write Born's predictive probability law $\{\pi(a_j) \equiv |c_j|^2\}, \ \forall j$.

***Problem 3***. Specify the way in which you can transform the representation of the studied microstate in $\mathcal{H}$ relatively to $\boldsymbol{A}$, into the representation in $\mathcal{H}$ of the same microstate but relatively to another observable $\boldsymbol{B}$ with eigenvalues $b_k$ and eigenvectors $|v(\mathbf{r},b_k)>, \ \forall k, \ k=1,2,....K$.

*Solution to the problem 3*: Apply Dirac's "theory of transformations":
For *any* given value of the index $k$ we have inside $QM_{HD}$ that

$$<v_k(\mathbf{r},b_k) \ |\psi_{G,\boldsymbol{H}}(\mathbf{r},t)> \ = e^{i_\gamma(\boldsymbol{B},k)} \ |d_k(b_k,t)| = \sum_j \tau_{kj}(\boldsymbol{A},\boldsymbol{B}) \ c_j(t,a_j) \ , \quad \forall j, \ \forall t$$

where $\tau_{kj}(\boldsymbol{A},\boldsymbol{B})=<v_k|u_j>, \ \forall j$. So for any complex factor of given index $k$ there is a separate condition

$$e^{i_\gamma(\boldsymbol{B},k)} = <v_k|\psi_G(t)> \ / \ |d_k(t,b_k)| \ = \ \sum_j \tau_{kj}(\boldsymbol{A},\boldsymbol{B}) \ c_j(t,a_j) \ / |d_k(t,b_k)|, \quad \forall \boldsymbol{A},\boldsymbol{B}, \quad \forall t$$

(where '$/$' is to be read: *divided by*) so that $d(b_k,t) =|v(\mathbf{r}, b_k)>$ and $|\psi(\mathbf{r},t)>/\boldsymbol{B}=\sum_j \tau_{bk,aj} \ c(a_j,t)$ where $\tau_{bk,aj}=<v(\mathbf{r}, b_k)|u(\mathbf{r},a_j) >, \ \forall j, \ \forall k$

***Problem 4***. Represent mathematically the measurement processes by which can be verified the predictive probability laws $\{\pi(t,a_j)\equiv|c(a_j,t)|^2\}, \ \forall j$, drawn from $|\psi(x,t)>/\boldsymbol{A}, \ \forall \boldsymbol{A}$.

*Solution to the problem 4*: Write "the measurement-evolution state-ket" of the problem (defined in (6.II) by the equation (19)).

***Problem 5***: Verify the statistical predictions of the formalism via use of (19).
*Solution to the problem 5*: Accordingly to the quantum theory of measurements, '*prepare the measurement-evolution state-ket*' and operate the verification-measurements.

## (5.II).2. THE *IQM* WAY OF REPRESENTING A MICROSTATE

This representation consists of the whole Part I of the present work where it stays available. It will act with all its features throughout what follows.

## (5.II).3. THE COMPARISON

When the two representations *IMQ* and $QM_{HD}$ are compared, the most striking conclusions are the following ones.

***Global critical characterization of*** $\boldsymbol{QM_{HD}}$.

Inside $QM_{HD}$ all that is explicitly ruled is defined *directly* in purely mathematical and algorithmic ways; moreover the mathematical instruments that are made use of allow continuity and infinities. The main descriptional element is the concept of a state-ket $|\psi>$ from the generalized Hilbert-space assigned to the studied microstate. This state-ket is obtained directly and *exclusively* via mathematical procedures and in their turn the probabilistic predictions from $QM_{HD}$ are generated directly and exclusively by the solution $|\psi>$ of the Schrödinger equation of the problem and by the equation for eigenket and eigenvalues of the involved observables, so in a strictly abstract *abstract* mathematical way.

Furthermore (with the unique exception of the eigenvalues of the observables introduced by the measurement-postulate 1) *not even mere* ***representations*** *or symbolizations of* ***individual and factual*** *entities or events or operations do come in*, neither mathematically expressed ones, nor only qualitatively *defined* ones; even the concept of 'micro-*state*' – that is precisely the main concept that the whole formalism represents – is left devoid of a clearly stated definition. It is even devoid of merely a specific symbolization. In the current way of *speaking* where the theory is imbedded, the most basic concept of a microstate is ambiguously indicated by two words, 'system' and 'particle', that are not defined, they are just picked up from the current language and are made use of in absence of any explicitly ruled mutual specification.

A fortiori *there is no defined concept of operation of generation of a microstate.*

> Any systematic structuration of the involved individual and physical operations lacks, the observables are defined to work on the statistical state-ket while the meaning of the eigenket remains obscure and they are made use of in purely algorithmic ways (cf. 6.II)). A model of a microstate is interdicted and correlatively *the coding problem is neither declared nor, a fortiori, treated overtly.*

In this context the representation of measurements remains strictly *abstract*, it is achieved via postulations and algorithms, *notwithstanding that an act of measurement has a quintessentially factual and* ***individual*** *nature*.

And even on the statistical level there is a representational gap between factual features and their mathematical expression. *The basic act of "giving" the whole initial statistical state-vector* $|\psi(t_o)>$ *is not clearly distinguished from giving the limiting conditions and moreover* ***in general it is not realizable***. This deserves being detailed and cried out on the roofs.

Let us very generously suppose that it is always possible to write down the Schrödinger equation for any factual situation and to calculate the corresponding general. The requirement to specify the initial *limiting* conditions – that is reflection upon the statistical level of the space-time domain on which effectively acts the involved *individual* operation of generation – should be clearly distinguished from the requirement to specify the whole initial state-ket $|\psi(t_o)>$, which involves the whole physical context where are introduced *all* the specimens of the factual microstate-to-be-studied. This physical context – that influences the evolution of the probabilistic predictions – cannot be reduced to exclusively well-defined macroscopic fields and obstacles; also unperceivable microscopic *factual* data act inside it, and in natural non-idealized conditions *these are not knowable* so a fortiori in general they cannot be specified in mathematical terms; they cannot even be verbally expressed[63]. Thereby already – and *in principle* – the domain of *rigorous* applicability of the $QM_{HD}$-formalism is restricted a priori to the domain of physical problems that do permit to produce a mathematical expression of the initial state-ket; which can be realized for bound microstates, but usually ***not** for unbound ones*. So when the considered problem concerns an unbound microstate the mathematical treatment is quasi ***systematically*** flawed by approximations (parcelling of the spatial support of the global phenomenon that is studied, more or less arbitrarily simplified mathematical translations of just imagined and posited micro-facts, etc.) of which the effects – or absence of effects – upon the statistical predictions, remain out of control.

But let us come back to the feature of major importance, namely the absence of an organized representation of the individual level of conceptualization.

Any predictive statistic, both *mentally* and *factually*, stems from *individual* entities, fragments of physical substance, operations, results of operations, individual concepts. These are abstracted away from the statistical representation of big sets of "*results*" of what is called "measurements". But – genetically – they have determined these results. And the individual determining-factors that have acted for constructing a predictive statistic will act *again* when one wants to *verify* it. *For its verification a predictive statistic has to be constructible factually from individual measurement operations, and in a consensual way.* But this is not possible when a general and systematic representation of any individual concept and any individual operation, is entirely lacking.

$QM_{HD}$ is flawed by a giant gap of representation of *modalities of factual generation of statistics of results of measurements*. This hinders the possibility of rigorous and systematic specification of the initial state-ket $|\psi(t_o)>$ as well as the possibility of verification. In such circumstances arbitraries and approximations can rush in uncontrollably. So, contrary to what is is currently asserted, we have to leave open the a priori possibility of considerable dissentions between factual truth and the $QM_{HD}$ statistical predictions concerning unbound microstates.

Historically, the Schrödinger equation has been a marvellous tool for the representation of bound microstates. But for unbound microstates, in a hidden way it has certainly been just a temporary unreliable solution imposed by the top-down initial approach toward a deep modern microphysics fully and explicitly endowed with its specificities.

### *Global characterization of IQM*

Inside *IQM* the descriptions (9), (9'), (9''), (9''') are *directly rooted into the factual microscopic a-conceptual physical reality* and are constructed out of this on the basis of an organized structure of definitions of individual concepts, of basic *physical* individual operations and of methodological posits concerning these and declared as such.

Inside *IQM* the concept of an *individual* and *physical* operation *G* of generation of a microstate manifests a quite determining role, namely via (1) it *leads* to:

- The classification of the sorts of operation of generation (simple or composed, and actual or revolved inside the past of a natural physical genesis (in the case of bound states).

- In consequence of the posit $ms_G \equiv \{\sigma(ms_G)\}$ the operation of generation *G* entails also with necessity the classification of the microstates defined in (2.I).1.

[63] In microphysics the concept of identically reproducible experimental conditions is far less effective than in the classical physics.

- Inside *IQM* the explicit specification of a model of a microstate and of coding procedures for each sort of possible microstate and of possible qualifying quantity, have appeared as sine qua non conditions, for any theory of microstates.

*The whole statistical conceptualization from IQM **follows** from the individual conceptualization*, via distinct progressive stages: the basic tree-like structures from the figures 2, 3 and 4 that summarize graphically the whole content of *IQM*, stem all from *one* operation *G* of generation; this is followed by individual acts of measurement which forms branch-successions *[G.MesA]* that lead to factually constructed probability spaces of first order; and these spaces are mutually correlated on a meta-level of probabilistic qualifications of second order.

***The comparison***

On the basis of the previously formulated characterizations the comparison follows now immediately. It is represented in the following figure Fig.6:

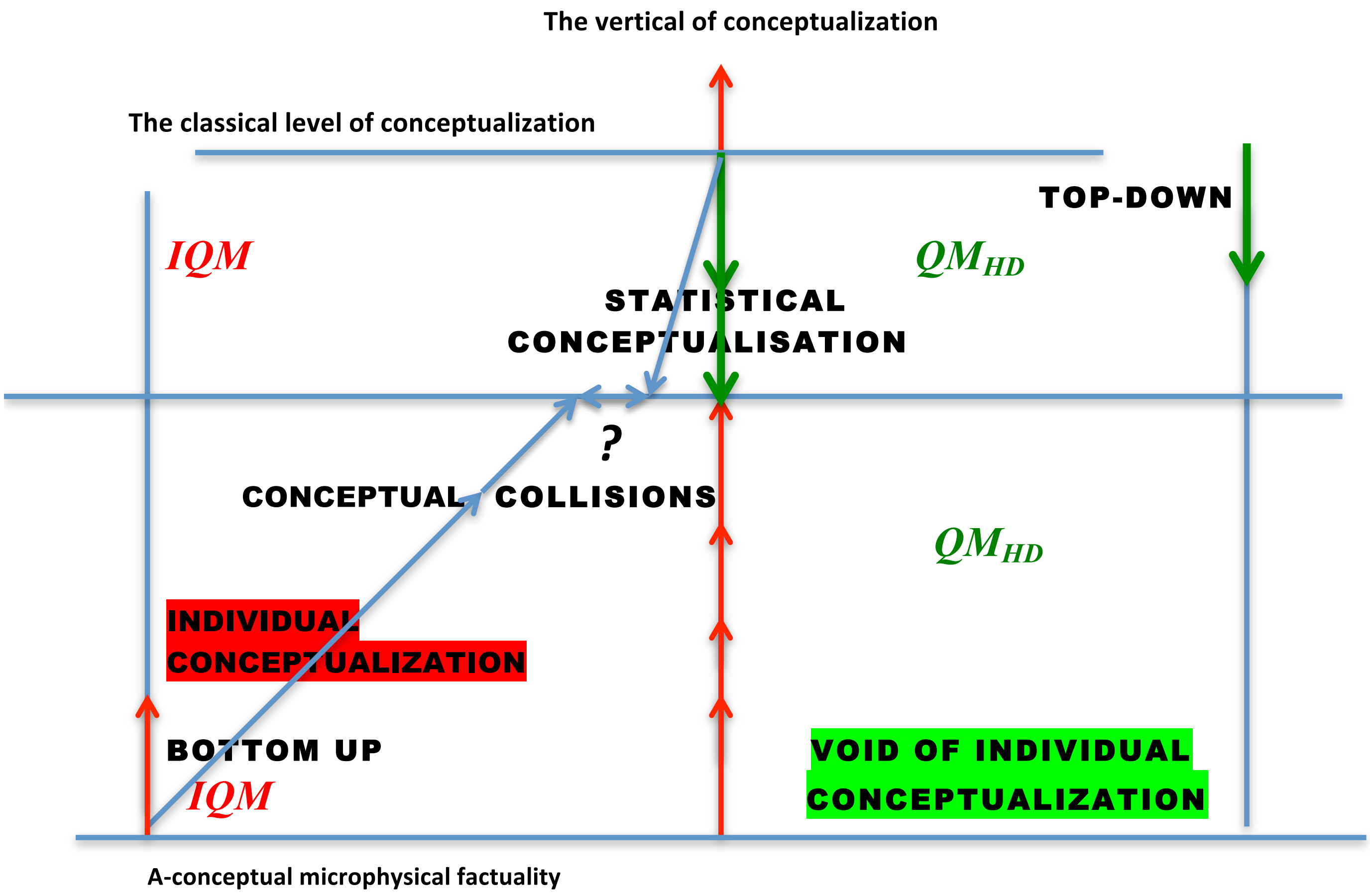

**Fig. 6.** *Comparison between IQM and* $QM_{HD}$.

We reproduce also below the Fig.5, for recalling the more global perception of the genesis of probabilistic predictions, as established inside *IQM*.

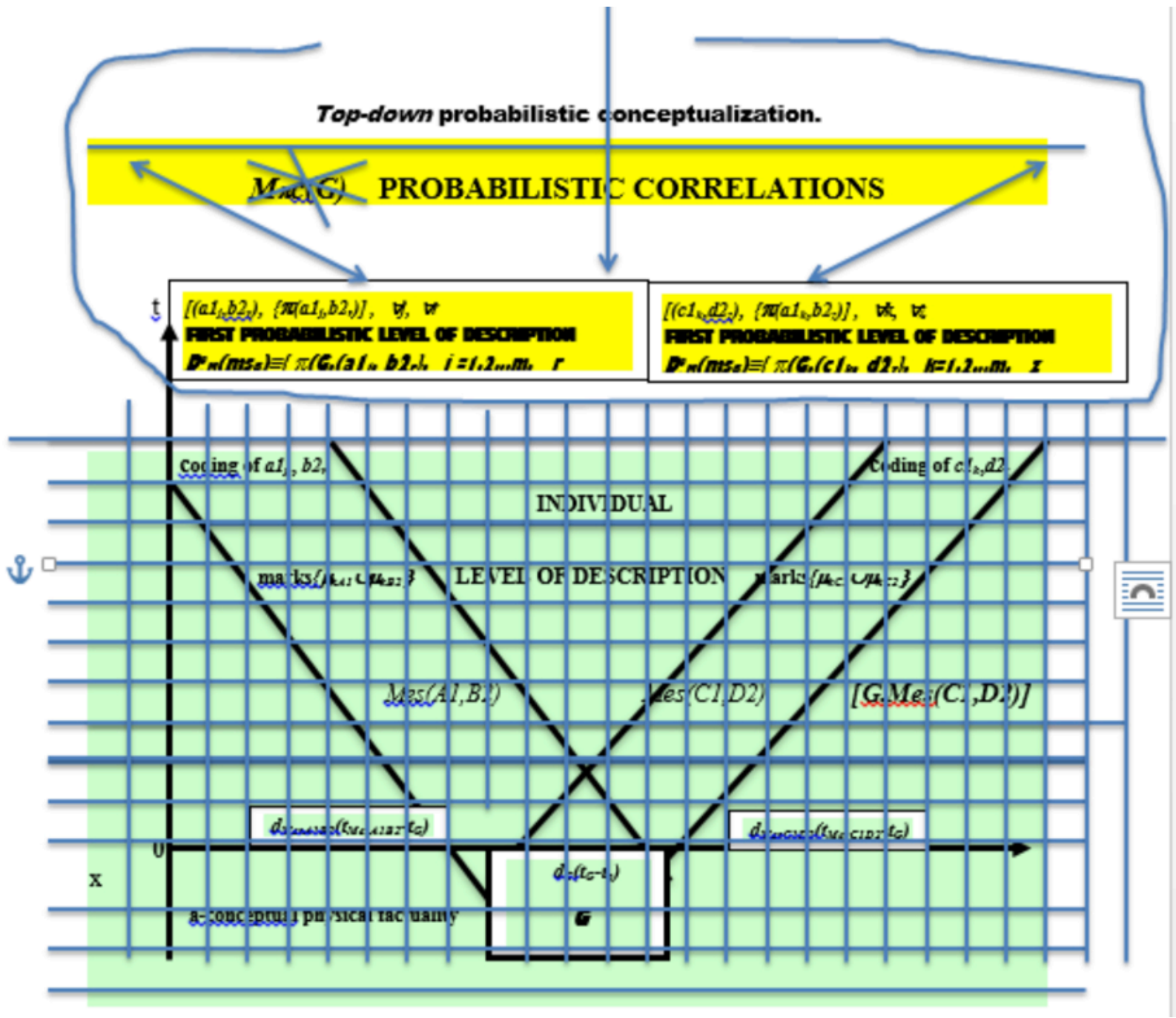

**Fig. 5**. *A probability-tree T(G(2S),(A1,B2; C1,D2)) as encountered by a top-down approach that stops at the level of the probability spaces (what is covered by green remains non-perceived).*

The hidden antagonism represented on the Fig. 6, between mathematically generated predictive statistics and factually generated verification-statistics, has settled in because in the historically achieved top-down initiation of a deep microphysics the *statistical* aspects have been reached *first*, experimentally, and so they have been represented first, such as they were registered and in directly abstract mathematical terms.

Moreover the *bound* states were at the core of all the attentions. These can be conceived as spatially delimited and pre-existing structures of several micro-entities, much alike to the classical 'objects', and they permit measuring-interactions via "effects" (Stark, Zeeman) produced by appropriate particular ways of making use of macroscopic fields, which *eliminates* major and systematic specificities from the *IQM* representation of unbound microstates.

So the main specificities of a *general* representation of the microstates, founded upon ***individually**-factual* measurement-interactions, remained hidden for a long while. Louis de Broglie's individual model of a specimen of a microstate – that brought the scientific thought down upon the very frontier between the as yet never conceptualized physical substance – and the whole conceptual volume of the scientific conceptualization involved by this model, emerged only some 25 years after Bohr's initial work. And it emerged precisely at the time when Heisenberg was developing his new top-down offensive of mathematical representation by matrixes, where the bound microstates were still taken into a quite dominant consideration. So de Broglie's individual model – together with the corresponding Schrödinger equation – had to confront the already organized and solidly installed positivistic attitude of the Copenhagen school, deliberately optimized for statistical predictions tied with bound microstates and for measurements on these. The Copenhagen top-down approach opposed de Broglie's model and occulted its bottom-up potentialities under the protection of strong socio-psychological inertial forces.

> ***And so – up to this very day – nowhere inside $QM_{HD}$ does one find a clear distinction between individual and statistical representations***.

So now, when *IQM* and $QM_{HD}$ are brought together like in the figures 5 and 6, they arrive at the statistical level of conceptualization from the two opposite directions on the vertical of construction of knowledge, and the abstract outputs claimed by the Schrödinger equation, in general fit only loosely with the factual outputs of long sequences of successions *[G.MesA]*.

In the light of the comparison achieved above the Hilbert-Dirac statistical formulation $QM_{HD}$ appears as a conceptual bas-relief of which the surface of algorithms is very finely crafted but of which the basic underlying structure of conceptual-*physical*-operations and meanings, is simply *undone*; it still lies lost in an amorphous substratum.

# 6.II

## BASIC CLARIFICATIONS:
## * A GENERAL MODEL OF A MICROSTATE,
## * USEFULNESS OF '*G*';
## * REFUSAL OF:
## - VON NEUMANN'S REPRESENTATION OF QUANTUM MEASUREMENTS;
## - THE WHOLE $QM_{HD}$ REPRESENTATION OF QUANTUM MEASUREMENTS

In what follows the local problems or insufficiencies from $QM_{HD}$ will constantly be referred to *IQM*. Therefore we introduce for our framework the symbol *[IQM-$QM_{HD}$]*. But to begin with, *IQM* and $QM_{HD}$ will act as two separate structures. The reference to *IQM*, however, will bring forth several basic clarifications that initiate a process of fusion.

### (6.II).1. THE *[IQM-$QM_{HD}$]* MEANING OF AN EIGENFUNCTION OF AN OBSERVABLE AND CONSEQUENCES

***Digging out the meaning of an eigenfunction***. Let us place ourselves inside $QM_{HD}$. Consider the equation

$A|u_j(\mathbf{r},a_j)\rangle = a_j|u_j(\mathbf{r},a_j)\rangle, j=1,2,...J\ (\forall j)$, with *J* finite [64]

that determines the eigenfunctions $\{u_j(\mathbf{r},a_j)\}$ from the basis of eigenket introduced by ***A*** in the Hilbert space $\mathcal{H}$ of the studied microstate. In general such an eigenfunction is not square integrable. This is considered to be a 'problem', in the following sense. A state-function $\psi(\mathbf{r},t)$ from a state-ket $|\psi(\mathbf{r},t)\rangle$ is *required* to be square-integrable because it represents a set of distributions of probability. But an eigenfunction – in general – is not square-integrable and furthermore *it is **not** required such!* [65] Why, exactly, is that so*?* That is the 'problem'.

Bohm ((1954) p. 210-211) writes:

> « …We obtain $\psi = e^{ipx/\hbar}$…… Strictly speaking, the above eigenfunctions cannot, in general, be normalized to unity...Let us recall, however, ...that in any real problem the wave function must take the form of a packet, since the 'particle' is known to exist somewhere within a definite region, such as in the space surrounded by the apparatus. To obtain a bound and therefore normalizable packet, we can integrate over momenta with an appropriate weighing factor.»

So Bohm adopts an exclusively mathematical-technical point of view. Not a moment does he focus upon the involved meaning. He does not even make use of a specific notation for distinguishing between eigen-function and state-function. And in order to deal with the mathematical situation he accepts approximations without any hesitation, notwithstanding that the considered question seems to be a question of principle.

---

[64] From now on any index is posited *finite* and any finite spectrum is denoted by a notation of the type $\forall j$.

[65] This – if one refuses to regard it as a mathematical miracle (a reputed author on the present status of $QM_{HD}$ asserted inside a central conclusion that *the formalism of $QM_{HD}$ is smarter than we are*) – can only be an indication that a number of physicists have sensed strongly enough that that an "eigen-*state*" is not a "state".

The same attitude is usually found in the textbooks.

Dirac (1958, p. 48), on the contrary, writes:

> « It may be that the infinite length of the ket-vectors corresponding to these eigenstates is connected with their unrealizability, and that all realizable states correspond to ket vectors that can be normalized so that they form a Hilbert space ».

(« connected with their "unrealizability" » suggests that Dirac, at least in a certain subliminal way, did perceive the possibility of some specific significance of the non-integrability, in general, of the eigenfunctions, as well as the absence – and the *utility* – of *a representation of the way of generating a microstate*).

As for the outstanding didactic exposition of $QM_{HD}$ [66], it proposes *«a "physical" solution to the "difficulties"»*, namely to replace the eigenfunction by a $\delta$-distribution centred upon the corresponding eigenvalue (the same solution has been proposed also by Bohm [1954], p. 212).

However recourse to history reveals that the 'problem' of non-integrability of an eigenfunction is a *false* problem because the concept of eigenfunction has been endowed with a specific *meaning* that is radically different from that of a state-function. So the problem is not mathematical, it is conceptual. Indeed the meaning of an eigenket stems from Louis de Broglie's Thesis (1924, 1963). Louis de Broglie has derived his famous relation $p=h/\lambda$ from his well-known model of a microstate, erroneously named the wave-*"particle"* model. The model itself stems from the use of Fourier decompositions inside the classical electromagnetism. In a Fourier decomposition of an electromagnetic wave each constant value $\lambda$ of a monochromatic wavelength is associated with a corresponding plane wave. By analogy, to each value $\boldsymbol{p}_j$ of the classical mechanical fundamental quantity of momentum $\boldsymbol{p}$ of an unbound electron, de Broglie has associated a plane wave with a *'corpuscular **phase**-function'* $\Phi(\boldsymbol{r},t)=a(\boldsymbol{r},t)exp((2\pi i/h)\beta(\boldsymbol{r},t)$ where '*a*' denotes an arbitrary and *constant* amplitude of vibration and the 'corpuscular phase' is written as $)\beta(\boldsymbol{r},t))=(Wt-\boldsymbol{p}_j.\boldsymbol{r})$ where $W$ is the energy of the *"corpuscular-(**like-aspect)**"* of *the "corpuscular wave"* while $\boldsymbol{p}_j$ denotes the constant value posited for the momentum *of this **corpuscular-like aspect***[67]

The 'corpuscular-like aspect of the corpuscular wave' remained devoid of representation inside the mathematical expressions that Louis de Broglie associated to his model. This has been a huge strategic error because in mathematical physics only what possesses a definite mathematical expression does subsist in the minds. The rest does not strike sufficiently numerous attentions and so at last it evaporates into the air of history. But verbally, de Broglie has clearly specified in his writings that he conceived the 'corpuscular aspect' to consist of a *singularity in the amplitude of the corpuscular wave*. Namely a very localized space-domain where this amplitude is so much bigger than its surrounding values that it concentrates in it practically the whole energy of vibration of the corpuscular-*like* singularity in the amplitude of the 'corpuscular ***wave***'. This singularity was posited by de Broglie to be "like a small classical mobile" that – in consequence of its strong spatial localization and its relatively very high energy – admits at any time the '*mechanical*' qualifications of position and momentum, from which in classical mechanics all the other mechanical qualifications can be constructed; whereas the rest of the wave, of course, does *not* accept mechanical qualifications. In short:

***de Broglie's model does not introduce any "particle" in the classical sense****; it introduces exclusively a corpuscular-like aspect of a wave, namely a very localized*

[66] Cohen-Tannnoudji C., Diu B. and Laloë, F., [1973].

[67] Here we introduce the notations '$\boldsymbol{\Phi}$' and '$\beta$' for de Broglie's *initial* concept, in order to *distinguish* from the start the representation of a *physical* phase of a *physical* wave as introduced by Louis de Broglie, from the phase $\varphi(\boldsymbol{r},t)$ of a *mathematical* "state-function" $\psi(r,t)=a(r,t)exp((2\pi i/h)(r,t)$ introduced later by Dirac inside a $QM_{HD}$ "state-ket" $|\psi(r,t)>$ that represents a formal tool for statistical predictions on results of measurements on a microstate (de Broglie himself, in his famous Thesis, glided rapidly into confusions between these two concepts, like also Schrödinger; but in de Broglie [1956] he went back to a very clear distinction.

> *singularity of the amplitude that admits mechanical qualifications in consequence of localization and of the fact that, while it performs displacements it carries with it the major part of the energy of the wave.*

Which required a "*wave-**mechanics***".

In the course of the construction of the relation $p=h/\lambda$ de Broglie has proved the 'theorem of concordance of the phases' [68] according to which:

> The model of a microstate of an unbound electron can be *stable* if and only if, at any given space-time location *(**r**,t)* of the up-down-up *localized* vibration of the singularity from the *amplitude* of the global wave, the phase of the involved elongation of this localized vibration is identical to the phase defined by the phase-function $\beta(\mathbf{r},t)$ for the elongation of the portion from the *extended-in-space wave* that surrounds the singularity (that constitutes the vibrational vicinity of the singularity from the amplitude).

*This theorem – notwithstanding the objections that can be opposed it – is crucial for just understanding the meaning of the $QM_{HD}$-concept of eigenket*: Louis de Broglie's *physical* wave-function $\Phi(\mathbf{r},t)=a.\exp(i(2\pi/h)\beta(\mathbf{r},t))$ satisfies the equation $\mathbf{P_x}\Phi(\mathbf{r},t)=\mathbf{p_{rj}}.\Phi(\mathbf{r},t)$ for eigenket and eigenvalues of the momentum observable from the Hilbert-Dirac formalism, just like the $QM_{HD}$ eigenfunctions of the $QM_{HD}$-observable $\mathbf{P}$. And the $QM_{HD}$-equation $\mathbf{A}|u_j(\mathbf{r},a_j)>=a_j|u_j(\mathbf{r},a_j)>$, $(\forall j, \forall A)$, ***generalizes*** this particular mathematical fact to any quantum mechanical observable and introduces it in the bra-ket expressions of the Hilbert-Dirac formalism, *without any condition on integrability*. This leads immediately to the following identification of the general meaning of the equation $\mathbf{A}|u_j(\mathbf{r},a_j)>=a_j|u_j(\mathbf{r},a_j)>$, $(\forall j, \forall A)$:

> The eigen-function $u_j(\mathbf{r},a_j)$ from the eigenket $|u_j(\mathbf{r},a_j)>$ associated with the eigenvalue $a_j$ of the observable $\mathbf{A}$, plays the role of *a mathematical functional representation of **a SAMPLE of a definite sort of wave-movement** around the spatial location of the corpuscular-like singularity in the amplitude of the involved de Broglie "corpuscular-like wave": This wave-movement must be **such** that the **mechanical** qualification of the involved corpuscular-like singularity from the amplitude of the wave, by the qualifying quantity **A** shall yield the (eigen)value $a_j$ associated with **A** when an "appropriate" measurement-operation MesA is performed (so involving an "appropriate" coding rule)*

So the eigenfunction $u_j(\mathbf{r},a_j)$ is a mathematical expression of an aspect of de Broglie's model.

As soon as this has been spelled out, it leaps to one's eyes that the form itself of the equation $\mathbf{A}|u_j(\mathbf{r},a_j)>=a_j|u_j(\mathbf{r},a_j)>$, $(\forall j, \forall A)$ simply cries it out on the roofs. So – no offense to Bohr – de Broglie's model of a microstate is quite basically present inside the whole formalism of $QM_{HD}$. It defines the physical-conceptual meaning of all the *bases* in the Hilbert-space of any microstate, as well as all the spectral decompositions of any state-ket. While furthermore *these spectral decompositions are the core of the predictive formalism from $QM_{HD}$*.

No more, no less.

The whole predictive $QM_{HD}$-algorithm is an undeclared infusion from de Broglie's model, wherefrom the physical significances are mutely drawn. Indeed in any spectral decomposition

$$|\psi(\mathbf{r},t)>/\mathbf{A}=\sum_j c(a_j,t)|u_j(\mathbf{r},a_j)>, \ \forall j$$

---

[68] In the last chapter of this work we come back critically upon this theorem in a way that might be useful for an also 'inner' unification of the modern Physics, besides an 'exterior' unification via a general methodological frame where the contents change with the involved cognitive situation.

of a state-ket $|\psi(\mathbf{r},t)>$ with respect to the basis $\{|u_j(\mathbf{r},a_j)>\}$, $\forall j$, introduced in $\mathcal{H}$ by the observable $\boldsymbol{A}$, *the eigenket* $|u_j(\mathbf{r},a_j)>$ *from the term* $c(a_j,t)|u_j(\mathbf{r},a_j)>$ *symbolizes the sample of that what is counted by the real squared modulus* $|c(a_j,t)|^2$ *of the complex coefficient* $c(a_j,t)$ (exactly as, in the expression *34m*, the symbol '*m*' means that the length that is measured is *34* times the length of the sample of a meter from the National Bureau of Standards of Weights and Measures).

And *a sample of wave-movement – by the definition of the concept of 'sample' –* ***has an arbitrary spatial extension***.

With respect to this significance for an eigenfunction, squared modulus integrability is simply senseless.

***Consequences of the identification of the meaning of an eigenfuction***. The preceding conclusion has noteworthy consequences.

- It evaporates the false 'problem' why an eigenfunction is in general [69] not required to be square-integrable: If it *were* required to always be square-integrable, *that* would be a real problem.

- In classical thinking a unique semantic dimension (for instance 'color') suffices for carrying *all* the 'values' ('red', green', etc.) that one wants to singularize on this dimension. But when a microstate has to be qualified it obviously is very useful – if not even necessary – to analyse the representation more, namely so as to compensate for the absence of any perception of a quale for assigning meaning to the brute result of each one act of measurement. The Hilbert-Dirac formalism realizes this analysis by a formal splitting: A $QM_{HD}$ observable $\boldsymbol{A}$ represents – separately – the considered semantic dimension – namely a qualifying quantity that qualifies from a mechanical point of view ('a momentum' 'a total energy', etc.); it introduces a 'grid' for qualification, in the sense of (2). And on the other hand – like in a catalogue joined to $\boldsymbol{A}$ – inside the set of pairs $\{(|u_j(\mathbf{r},a_j)>,a_j)\}$, $\forall j$, are represented separately each one of the 'values' singularized on the semantic dimension $\boldsymbol{A}$; and *each* 'value' is specified by *a pair* $(|u_j(\mathbf{r},a_j)>,a_j)$, $\forall j$, because the wave-movement of a corpuscular wave *and* a numerical mechanical qualification of the 'corpuscular aspect' of that wave via a corresponding eigenvalue $a_j$, are both involved, and are tied in a one-to-one connection [70]. This is marvellously expressive. And when it is discretized via an explicit adjunction of a unit for measuring the quantity *A* represented by the observable $\boldsymbol{A}$., it becomes also *effective*.

- This explains the high adequacy of the use of a Hilbert space $\mathcal{H}$ for representing mathematically the predictions on issues of measurements on a microstate: Each 'value' $a_j$ of $\boldsymbol{A}$, $\forall j$, can be placed on a separate axis reserved to it, on which the state-ket $|\psi(\mathbf{r},t)>$, when projected onto that axis, determines the complex number $c(a_j,t)$, so also the probability $|c(a_j,t)|^2$ postulated by Born for the emergence of what is represented by the pair $(|u_j(\mathbf{r},a_j), a_j)$ if a measurement of $\boldsymbol{A}$ is performed upon the microstate with state-ket $|\psi(\mathbf{r},t)>$ (this mimics geometrically the expansion $|\psi(\mathbf{r},t)>/\boldsymbol{A}=\sum_j c(a_j,t)|u_j(\mathbf{r},a_j)>$, $\forall j$, of $|\psi(\mathbf{r},t)>$).

Thereby the representational roles of, respectively, a state-ket or an eigenket, are radically distinguished.

- The preceding remarks indicate also that Dirac's 'theory of transformations' expresses in mathematical terms passages from a given *'semantic* space*'*, to another one: A given semantic is defined formally by the corresponding pairs $(|u_j(\mathbf{r},a_j)>,a_j)$:

---

[69] In a bound state of a microsystem the eigenket of the total energy has the same mathematical expression as the state-ket; it is confounded with the state-ket and the eigenket is required to be square-integrable and it *is* such: ***from a conceptual point of view this is a mathematically 'degenerate' situatio***n.

[70] Degenerate spectra are not considered here.

***Dirac's bra-ket calculus is potentially a GENERAL calculus with semantic specifications***[71, 72].

Let us keep this in mind. Indeed thereby the Hilbert-Dirac formalism – in itself, *independently* of microstates and $QM_{HD}$ – can be useful in many disciplines.

[71] In Dirac's mind might have worked implicit general criteria that he did not care to capture and communicate.
[72] Cf. also MMS [1993].

## (6.II).2. FROM THE HIDDEN PRESENCE IN *$QM_{HD}$* OF de BROGLIE'S MODEL TO ITS EXPLICIT PHYSICAL-*OPERATIONAL* INCORPORATION TO *[IQM-$QM_{HD}$]*

We have posited a framework denoted *[IQM-$QM_{HD}$]*. Inside this framework we shall now absorb de Broglie's ideal – i.e. *non*-(physically operational)) – model of a specimen of a studied microstate, into the general physically-*and*-operational concept *G* of factual generation of the microstate-to-be-studied defined by the methodological decision *MD* from the Part I of this work.

What follows – installing explicitly a *model of* the specimens of the studied microstate – might be perceived as shocking after the cure of positivistic purity suffered by microphysics since nearly a century, according to which models were interdicted. But let us keep in mind that just above this interdiction has been shown to have never acted. And the announced absorption will soon appear to have been an essential step forward in the process of construction of a a second quantum theory directly rooted in the physical factuality (cf. 4.I). So let us vanquish the diktats. Everywhere in the sciences and the techniques the models are unavoidable and precious and their efficiency is increased when they are specified with detail.

***Association via G of de Broglie's model of a microstate, with a factual microstate $ms_G$ in the sense of MD and (1)***. The guiding ideas are the following ones:

***(a)*** Consider the definitions from (2.I).1 of various sorts of microstates. It is clear that *what is called 'system' in these definitions has to be identified with de Broglie's 'corpuscular-**like** singularity' in the amplitude of a corpuscular wave*. So we posit that:

> An operation of generation *G* tied with 'one micro-*state* of *one* micro-*system*' introduces into the domain of what a human observer can qualify individually in a factual-operational and consensual way, a specimen $\sigma(ms_G)$ of the studied microsystem $ms_G$ that contains only one corpuscular-like singularity; whereas an operation of generation *G(ns)* (*ns*: *n* systems) tied with 'one micro-*state* of *n* micro-*systems*' introduces a specimen with *n* de Broglie-singularities.

***(b)***. This point is fundamental, so we express it with much detail, in a redundant way.

The de Broglie's corpuscular-like singularities are conceived to be organically integrated to an unlimited "corpuscular wave". It seems obvious that it would be nonsense to conceive that an operation of generation *G* defined by the use of macroscopic apparatuses and macroscopically controlled parameters, cuts off radically from this wave the "corpuscular-like" singularity(ies)[73] contained in the generated specimen $\sigma(ms_G)$. Outside this wave, by its definition, a de Broglie singularity *cannot be* ***conceived***. It *is* just an aspect of this unlimited wave. So we *posit* that – in some sense and for some time – *G* just captures into the domain of what can be operated upon by human observers and can produce observable marks, a *portion* of a corpuscular wave that carries *one* de Broglie singularity when *G* entails *one* micro-state of *one* micro-system, or carries *n* such singularities when *one* micro-state of *n* micro-systems is generated by *G*. While *observationally* the main part of the wave-like phenomenon to which this portion of corpuscular wave *continues* to be incorporated after the action of *G*, remains in the physical substratum, even though via the generated specimen of the studied microstate it has been connected to the domain of the observable by men, during the time of the subsequent act of measurement from the succession *[G.MesA]* intiated by *G*.

Such a view violates the most current significance of the classical concept of 'object', namely that it is *endowed with a delimited spatial volume*, even if this 'object' is in a liquid or even a gaseous state. Here we stay face to face with a manifestation of the modern view on the frontier between a representation of the Spinozian "Universal Substance" and a very *first* process of *individualized* human conceptualization. Indeed nowadays we are aware that any definite "entity", a chair, a living body, cannot be conceived to be cut in an absolute sense from the surrounding 'physical waves' that are posited to exist (electro-magnetic, gravitational, etc.). Any "object" in the classical sense is admitted to emit and to absorb waves of various natures and to be traversed by such waves.

[73] We use singular expressions for the sake of simplicity.

But the *global* de Broglie model of the *whole* of the 'physical reality' (de Broglie [1956]) [74] does not incorporate the concept defined in *MD* of a *physical*-operational generation of individualized entities-to-be-studied, of specimens of factual microstates. In de Broglie's global approach – like in Bohm's one, and also like in ***any*** only mathematical representation – *the physical-operational concept 'G' from IQM is replaced by purely mental, wandering focalizations of the inner attention*. But mental focalizations of the inner attention of the human conceiver-observer – alone – cannot entail consensually observable physical effects subjected to conditions of predictability and of verifiability.

> *A* ***global*** *mathematically expressed model of the whole of the 'physical reality' – alone – cannot constitute a scientific theory of the microstates in the sense defined in IQM.* As such and alone it is just a sort of mathematically expressed piece of metaphysics. And it is in consequence of ***this*** that it ***can*** be posited to be deterministic, like in the classical disciplines of Physics[75].

Whereas in order to achieve what in *IQM* is called "*a scientific theory*" of the microstates – a structure of *consensual* predictive and verifiable ***knowledge*** belonging to the Mathematical *Physics* – it is unavoidably necessary to repetitively generate one by one factual microstates via some operation of generation *G* – in the sense of *MD* and (1) – and on each one of these to then perform an act of measurement *MesA* by a succession *[G.MesA]*. And in order to achieve this it is unavoidably necessary to make use of also a global.

These considerations close the point *(b)*. We furthermore posit now that:

***(c)*** The *location* of the de Broglie singularity(ies) inside the corpuscular wave of any specimen of a studied $ms_G$, in general *varies* arbitrarily from one individual specimen $\sigma(ms_G)$ of $ms_G$ to another one (this is an essential element from de Broglie's own view (cf. de Broglie [1956])) and is intimately tied with the preceding point *(b)*.

***(d)*** We do not try to specify other characters of the portion of a corpuscular-wave that is trapped into the domain of possibility of deliberate interaction with the one specimen $\sigma(ms_G)$ of the studied microstate $ms_G$. But we admit that, whatever these characters are, they constitute similarities that justify their common designation in *MD* and (1) as elements from the class $\{\sigma(ms_G)\}\equiv ms_G$ of emergences of the '*one* microstate $ms_G$ corresponding to *G*': The language posited in this way has already shown its pertinence.

***(e)*** Finally, in agreement with de Broglie's works[76] *we also posit the famous guidance relation* according to which the phase of the corpuscular wave in the neighbourhood of the singularity 'guides' the singularity by determining its momentum.

---

[74] De Broglie has worked out the *global* De Broglie-Bohm model *after* Bohm's [1952]-work. This global model includes Bohm's work as well as the de Broglie's initial model of a microstate. As far as I know it is the unique global model of the 'physical substance'.

[75] Let us immediately and explicitly stress that in our view de Broglie's representation of the Universal Substance does not in the least offer "consensual scientific knowledge of the physical reality such-as-it-is-in-itself"; it remains just *a human* ***model*** *that is relative to the human way of perceiving and thinking from which it stems*. And the same holds concerning Bohm's model. According to our view this sort of limitation is definitively impossible to be transgressed. Louis de Broglie's "individual" model is *more specified* than Bohm's un-defined concept that underlies the word "particle" (that is rather generally in use whether for designating Bohm's approach *or de Broglie's one*). Indeed Bohm does not specify the nature of the "particles" of which he speaks; while de Broglie introduces the notion of a corpuscular-like *aspect* of the "corpuscular waves", namely a singularity in the amplitude of such a wave. Nevertheless Bohm's "particles" are "piloted" by waves exactly like de Broglie's "singularities": so mathematically the two global models are entirely compatible. We choose de Broglie's individual and global models only because later in the present work they offer a more specified support for conceiving adequate coding-measurement postulates.

[76] And with those of the nowadays physicists from Bohm's school (in particular Peter Holland 1993).

***(f)*** Consider now a *composed* operation of generation ***G****$(G_1,G_2,...G_k)$*. The methodological posit (1) requires a one-one relation between any operation of generation *G* and its result denoted $ms_G$. So – accordingly to the definitions from (2.I).2 – we must admit that:

An unbound *one* micro-state of *one* micro-*system* produced by a composed operation of generation involves only ***one*** singularity in the amplitude of the corpuscular wave of the generated microstate $ms_{\mathbf{G}(G1,G2,...Gk),cw}$.

But we also admit that the other operations of generation $(G_1,G_2,...G_k)$ mentioned in connection with ***G****$(G_1,G_2,...G_k)$* – that *can* be realized separately but have *not* been separately realized when ***G****$(G_1,G_2,...G_k)$* has been realized – produce together some *specific* effects on the dynamic of this unique singularity such as it is asserted by the guidance law quoted in *(e)*. This will appear in the Part III to be a very important point in the process of construction of a satisfactory representation of the quantum measurements on unbound interference-microstates.

On the basis of the assumptions *{(a),(b),(c),(d),(e),(f)}* we now finally take the following new step:

***The factual physical-operational modelling postulate MP(***$\{\sigma(ms_{G,cw})\}$***).*** In agreement with *MD*, (1), it is posited that any one realization of an operation of generation *G* introduces one specimen of the studied factual microstate that is specified accordingly to the assumptions *{(a),(b),(c),(d),(e),(f)}*. Such a specimen will be denoted $\sigma(ms_{G,cw})$ (cw: corpuscular wave) and it will be called *the G-(corpuscular-wave)-model of a specimen of the studied microstate* (in short, a *G-cw* model). Correlatively the corresponding microstate can be re-noted $ms_{G,cw}$.

When necessary, the relation (1) $G\leftrightarrow ms_G$ from *MD* can be maximally specified in agreement with (13'):

$$G_t \leftrightarrow (ms_{Gt,cw} \equiv \{\sigma(ms_{Gt,cw})\}) \qquad (1')$$

But in general we shall make use of the basic notations *G-cw* and *MP*($\{\sigma(ms_{G,cw})\}$).

The modelling postulate introduces a crucial connection between *IQM*, $QM_{HD}$, and furthermore de Broglie's global *interpretational* approach constructed in [1956]), tied with Bohm's *interpretation of* $QM_{HD}$. But inside [*IQM-*$QM_{HD}$], via the *G-cw* model, the concepts of microstate and of specimen of a microstate as defined in *MD cease to be imprisoned inside the category of* ***'interpretations'*** *of* $QM_{HD}$: They become basic elements of the process attempted here of construction of intelligible consensual knowledge on the microstates, predictive and verifiable and pragmatically utilizable, not only a painting of these in mental mathematical symbols and words. Let us briefly elaborate on this point.

Initially, in the relation (1) from *IQM*, the concepts $ms_G$ and $\sigma(ms_G)$ have been provisionally qualified exclusively by *the label 'G' of the way in which they are produced*; later inside *IQM*, in (8), this initial definition has been enriched by a whole *set* of brute observable 'transferred' marks $\{\mu\}_{kA}$, $k_A=1,2,...m_A$ registered via measurement-interactions. Both these definitions from *IQM* possess already a consensual scientific character because they are realized consensually and physically-operationally, they are publicly observable and reproducible, so communicable and verifiable. Nevertheless they still remain *exterior* to the concepts '$ms_G$' *and* '$\sigma(ms_G)$' in consequence of the absence of any specified model of a microstate. While now – because inside the new framework [*IQM-*$QM_{HD}$] we have introduced a model that from (1) leads to (1') $(ms_{G,cw}\equiv\{\sigma(ms_{G,cw})\})$ – the results $\sigma(ms_{G,cw})$ of one operation *G* and the result $ms_{G,cw}$ of a big set of repetitions of *G* are finally endowed with definite characters and some of these characters even concern the *insides* of the specimens $\sigma(ms_{G,cw})$ *themselves*, so they specify the *own nature* assigned to the concepts $\sigma(ms_{G,cw})$ and $ms_{G,cw}$.

Thereby the present approach draws in, not only the possibility to investigate concerning factually individualized microstates, but *also the possibility to take into consideration features assigned to the* ***interior of one studied microstate***.

> Thereby the fundamental inside-outside opposition (cf. Atmanspacher& Dalenoort [1994]) enters the mechanic of microstates that we are constructing here, and this can have many consequences because *the main specificities of the notion of a factually defined microstate stem from its inner non-classical 'quantum-fields' and these are quasi* ***omnipresent*** [77].

So the *G-cw* local model enriches the initial concept (1) of a factually and consensually defined microstate. With respect to the absolute and causal approach of de Broglie that has a purely deterministic character[78] *the price of this gain is a primordially statistical-probabilistic* ***observational*** *character*.

The results from (6II).2 constitute a noteworthy advance: They offer from now on clearly defined assumptions, concepts, and words for investigations that are placed – strictly – on the frontier between the still a-conceptual factuality and what is extracted therefrom *factually* for a radically first conceptualization able to lead to 'scientific' knowledge. They also suggest subsequent developments[79]. Moreover they stress explicitly the following major fact.

> *Absolute* and intrinsic spatial delimitating contours of *individual* 'objects' in the classical sense are just a human construct. *Individuality*, otherness, and also stable and homogeneous space-time *inner* structure are only intuitive and pragmatic human assumptions of which the utility is relative to the local cognitive purposes and actions.

In particular, any model of a microstate is, of course, just a human construct. But a model generates *coherence* if it is referred to systematically in any construction, any operation, and any reasoning – very especially if it is a mathematically expressed model – and this, for human minds, is source of intelligibility, so of efficiency[80].

Globally considered, the content of this point (6.II).2 amounts to a conceptual-physical operationalization of Louis de Broglie's global model of the universal physical substance, realized via the modelling postulate (1') $G_t \leftrightarrow (ms_{Gt,cw} \equiv \{\sigma(ms_{Gt,cw})\})$. This offers a first and detailed illustration of the radical distinction made in this work between a global and mathematically expressed model conceived for the universal substratum of physical reality posited by the human mind, and on the other hand a *scientific theory* of a given domain of physical entities and facts, subjected to a specific cognitive situation and to the corresponding conceptual, ***physically*** *operational structure of* ***cognitive actions*** that entail consensually observable effects. According to *IQM* this structure is founded upon pairs of operations *(G, MesA)* related with coding-rules[81].

---

[77] The definitions from (2.I).1 have left open this possibility, and this has permitted to understand the main features of the "problem" of locality inside *IQM* ((3.I).2) even in the absence of definite model). This illustrates the importance of "inside" specifications.

[78] In the de Broglie-Bohm (*dBB*) approach the equation of evolution that yields solutions in terms of statistical state-ket is just *superposed* via Born's postulate to the deterministic basic representation, *in* ***absence*** *of any organically incorporated explanation of the statistical-probabilistic character assigned to the solutions*.

[79] For instance: a microstate of one system (in the sense of the definitions from (2.I).1) with electric charge or magnetic moment can be drawn into the realm of the observable by use of classical macroscopic fields. But how could be *manipulated* the result of an operation *G* if this operation generates (for instance by a nuclear reaction) a microstate that is sensitive *exclusively* to a gravitational field? (Which probably means *a maximally 'simple' de Broglie singularity, a 'pure quantum of de Broglie-mass' (a 'graviton'? what exactly does this mean? Is it assigned a non-null spin?*). Such questions touch as much the most modern researches in gravitation, In relation with gravitation, teleportation, etc., in de Broglie [1956] the chapter XI, pp. 119-131 is fascinating. And *the pair* $(MP(ms_{G,cw}), (1'))$ *offers now to the dBB representation* ***access to*** *an operational* ***utilization***.

[80] Why this is so, however, remains an open problem.

[81] This recurrently asserted distinction might surprise those readers of this work that perceive it as a triviality. But I discovered with astonishment that many physicists do not make this distinction and moreover probably reject it when they find it claimed. So that it seems to me correct to declare openly that in my own view this distinction is essential and to try to bring forth why.

Though in this point 6.II).2 we have incorporated in our constructive process an element from de Broglie's interpretation of $QM_{HD}$, we consider that we continue working inside the framework *[IQM-$QM_{HD}$]*, since de Broglie's model has been shown to be basically contained in $QM_{HD}$.

## (6.II).3. CLARIFICATIONS VIA THE CONCEPT OF OPERATION *G* OF GENERATION OF A MICROSTATE

It appeared above that the use of the concept of operation *G* of generation of a microstate would have economized the false problem of why eigenket in general are not square-integrable. In this point (6.II).3 we shall bring into evidence other three fundamental current writings from $QM_{HD}$ where accordingly the framework *[IQM-$QM_{HD}$]* the concept '*G*' entails clarification of ambiguities or of latent problems that inside $QM_{HD}$ vitiate the intelligibility. The first of these follows just below.

***Expansion-writings and mathematical superposition writings, versus G***. Inside $QM_{HD}$ works a mathematical principle of spectral decomposability according to which for any state-ket and any basis of eigenket $\{|u_j(\mathbf{r},a_j)>\}$, $j=1....$, introduced by a $QM_{HD}$-observable $\boldsymbol{A}$ it is justified to assert the expansion-postulate:

$$|\psi(\mathbf{r},t)>/\boldsymbol{A}=\sum_j c_j(t)\ |u_j(\mathbf{r},a_j)>, \quad \forall j \tag{14}$$ [82]

Furthermore, the choice of a vector-space-representation *permits* to write the state-ket associated to a microstate $ms_{\mathbf{G}(G1,G2,...Gn)}$ generated by a composed operation of generation ***G****(G1,G2,...Gk)* ((1.I).1), as a *mathematical* superposition

$$|\Psi_{12...k}(\mathbf{r},t)>=\lambda_1|\Psi_1(\mathbf{r},t)>+\lambda_2|\Psi_2(\mathbf{r},t)>+......\lambda_k|\Psi_k(\mathbf{r},t)> \tag{15}$$

of the state-ket of the microstates $ms_{G1}$, $ms_{G2}$, .... $ms_{Gk}$ that *would* have been obtained *if* each one of the operations of generation *G1,G2,...Gk* that have been composed inside ***G****$(G_1,G_2,...G_k)$* *would* have been realized *separately* [83].

In $QM_{HD}$ the Hilbert space $\mathcal{H}$ of a state-ket is extended into a 'generalized-Hilbert space' $\mathcal{H}$ where the eigenket are included as a limiting sort of vectors. This entails that from a strictly mathematical point of view both writings (14) and (15) are just superpositions of vectors inside $\mathcal{H}$, permitted by the mathematical *axiom* – included in the definition of the algebraic structure called a vector-space – that any two or more elements from a given vector-space admit an additive composition; which is expressed by saying that they can be 'superposed', i.e. summed. These formal features installed a purely mathematical language that calls *indistinctly* 'superposition' *any* additive combination of ket, whether only state-ket like in (15), or state-ket *and eigenket* like in (14), or only eigenket as in Dirac's theory of transformations.

One senses immediately what confusions can stem from such a mixing in-distinction between formal elements – state-ket and eigenket – that represent concepts that in (6.II).1 have been shown to be deeply different from a semantic point of view. No physical criteria, nor conceptual ones, are made use of inside $QM_{HD}$ in order to make specifications inside the general category of additive compositions of ket. In (6.II).2 we have seen an illustration of the consequences that precisely this sort of blindness can entail inside a theory of mathematical physics concerning the *conceptual-physical meaning* of the mathematical writings. Inside mathematical physics the relation between physics and mathematics is so intimate that a blindness of this sort is a major danger for intelligibility; under the protection of this intimacy mathematics can simply chase the intelligibility out of the mathematical physics.

---

[82] We recall that an adaptation of the results from this work to our choice of effectiveness, will entail, for the sake of strict inner coherence, only *finite* mathematical representations. So, in our view, the usual infinities involved in the now usual mathematical expressions, are a provisional feature of the present work.

[83] In the Part III of this work the choice *to make use* of this formal possibility inside a physical theory of the factual microstates, will be criticized and will be replaced by another choice of mathematical representation.

In the case of the writing (14) the mathematical form induces into the minds the more or less explicit *semantic* interpretation that all the eigenket-terms $c_j(t)|u_j(\mathbf{r},a_j)>$ from the second member are of the ***same*** *semantic nature* as the state-ket $|\psi(\mathbf{r},t)>$ from the first member. Which has been shown in (6.II).1 to be false and to raise false problems.

And in (15) this same sort of semantic blindness suggests that the state-ket $|\Psi_{12...k}>$ points toward a *physical* 'superposition' of all the ***separate*** mathematical representations of the factual microstates $ms_{G1}$, $ms_{G2}$, …. $ms_{Gk}$, supposed to *coexist* inside $|\Psi_{12...k}>$. This formally induced suggestion is strongly privileged by the fact that inside Physics there exist *principles of* ***physical*** *superposition* that – *if* from a conceptual-physical point of view they are adequate indeed – *can* also be expressed mathematically by an additive composition; but only in this case, not in general.

The adequacy of an assertion of conceptual-*physical* superposition should be examined separately, before making use in the mathematical writings of a formal possibility of mathematical superposition writings. For if, in a theory from the Mathematical Physics, an assumption of physical superposition is not adequate from a conceptual-physical point of view, a representation of the involved entities and facts by a mathematical superposition in general introduces problems of global consistency. So inside the provisional framework *[IQM-QM$_{HD}$]*) we want to avoid as much as possible any ambiguity of the sort specified above. This aim should be much favoured by notational distinction, in the mathematical writings, between formal descriptors that are directly tied with a *physically* operational act *G* of generation of a factual microstate, and on the other hand formal descriptors that are not directly tied with an act *G*.

*Consider first the writing (14).* According to *[IQM-QM$_{HD}$]* in (14) only the state-ket $|\psi(\mathbf{r},t)>$ from the first member corresponds to the studied factual microstate $ms_{\mathbf{G}}$, namely on the statistical level of conceptualization; while all the terms $c_j(t)|u_j(\mathbf{r},a_j)>$ from the right-hand expansion of $|\psi(\mathbf{r},t)>$ are symbols of a product of a *number* $c_j(t)$ and a *model* $|u_j(\mathbf{r},a_j)>$ of a possible corpuscular-wave-movement. So in (14) is involved only one operation of generation, namely for the state-ket from the first member.

*Consider now the writing (15).* According to *[IQM-QM$_{HD}$]* in (15) the resulting *one* microstate $ms_{G(G1,G2,...Gk)}$ that is effectively generated by the *unique* though composed operation of generation $\mathbf{G}(G_1,G_2,...G_k)$, cannot be coherently conceived as a coexistence of all the factual micro-states $ms_{G1}$, $ms_{G2}$, …. $ms_{Gk}$ that *would* have been obtained via the *separate* realizations of $(G_1,G_2,...G_k)$. (This has been already remarked in the Part I and then in (6.II)2). So – according to *[IQM-QM$_{HD}$] – the state-ket from the second member of the QM$_{HD}$ writing (15) can be regarded* ***at most*** *as only* ***virtual representational elements***[84], *if* the representation (15) is conserved inside the new theory of the factual microstate that we construct in this work. But for this decision we do not posess as yet a sufficiently evolved basis. So we postpone it. Here we only recall the essence of the remark from (3.I).3. Inside *QM$_{HD}$* the writing (15) has been considered to be useful ***in order to introduce a mathematical representation of the experimental fact that expressed in (12)*** that is generalized by the in-equality *in*-equality

$$\pi_{12...k}(\mathbf{G}(G_1,G_2,...G_k),a_j) \neq \pi_1(G_1,a_j)+\pi_2(G_2,a_j)+.........+pk(G_k,a_j) \qquad (16)$$

Indeed, via the spectral decomposition (14') of $|(\Psi_{12...k})_G(\mathbf{r},t)>$ on the basis of eigenket of an observable $\boldsymbol{A}$ and application of Born's postulate of probability to the complex expansion coefficients $c_j(t)$, between the expansion coefficients there emerge *mathematical* 'interference'-terms that entail the inequality (16). This, at a first sight, seems very adequate for expressing the involvement of the operations of generation $(G_1,G_2,...G_k)$ in the unique effectively realized operation $\mathbf{G}(G_1,G_2,...G_k)$. However, according to (1) in the methodological decision *MD* from *IQM*, to the only ***one*** physically realized operation of generation $\mathbf{G}(G_1,G_2,...G_k)$ there corresponds only *one* factual microstate $ms_{\mathbf{G}(G1,G2,...Gk)}$ inside which the the the operations $(G_1,G_2,...G_k)$ that are involved in $\mathbf{G}(G_1,G_2,...G_k)$ must be conceived to have induced certain effects, but ***not*** also to have fully generated the factual microstates $ms_{G1}, ms_{G2},... m_{Gk}$ that they *would* have generated *if* they would have been realized separately.

[84] We do not assert here any physical facts, only consequences of a basic requirement of inner consistency inside the framework *[IQM-QM$_{HD}$]*. How this can be connected with physical facts will be examined later.

> So $QM_{HD}$ *and IQM involve distinct views concerning the semantic acceptability of writings of the form (15).*

This is possible because *IQM* is constructed precisely in order to represent the involved *meanings*, while $QM_{HD}$ *occults the involved meanings under purely mathematical requirements*.

Such a situation is *doomed* to come to some factual confrontation. We are aware of this, and we stay attentive.

> *Quite generally now*:
>
> *When inside* $QM_{HD}$ *a state-ket seems to be 'absent', inside the new framework [IQM-* $QM_{HD}$*] this – in consequence of (1) from MD – means that the microstate that corresponds to this state-ket* ***is not generated separately****. So it has not been brought into individualized factual existence.*

The most striking case of such an "absence of a state-ket" is that of *one* micro-*state* of *two or several* micro-*systems*, tied with the problem of locality. The formalism of $QM_{HD}$ – rightly – represents such a *one* factual micro-*state* (cf. the definitions from (2.I).1.2) by only *one* state-ket. But for each involved micro-*system* it introduces a distinct *representation-space* (the mathematical relation between these representation spaces is specified in a way that is indicated by the now current words 'intrication' and 'non-separability'). Certain authors speak of « 'absence' of an 'own' state-ket for each 'system' » ; other authors speak of « absence of 'information' » (in what a sense, exactly?). These ways of speaking express that a state-ket is conceived to be like a secretly placed planet or lake, something that 'is' somewhere outside, but nobody knows how to go and see where and how it 'is'. In the textbooks it is written that « often a micro-system 'is' represented by a state-ket and, if so, the state is 'pure' »; while if it 'is' not pure, then it 'is' a 'mixture', but in such a case (happily) one can nevertheless dispose of a statistical operator. But in the case of the "problem of locality" not even a true statistical operator does 'exist'; only 'a partial-trace' operator 'exists'; but this cannot change the *fact* that there is 'non-separability' because – curiously – a statistical correlation is observed even when the spatial distance between the involved systems is very big. All the mentioned ways of speaking suggest that the state-ket, the statistical operators, etc., are conceived to possess an existence quite independently of the representational *choices, decisions, constructions, realized by human beings, by physicists*. As soon as 'there *is*' a 'system', 'its' state-ket should also 'be', and nevertheless sometimes it is absent and we do not know why, nor where it is gone. The special case of one micro-state of *two* micro-systems in the sense of (2.I).1.2 is a particularly strong discloser of how the whole mathematical formalism of $QM_{HD}$ is currently conceived.

> *We are in presence of a huge reification of the mathematical* ***formalism*** *of* $QM_{HD}$ *considered to constitute the whole of* $QM_{HD}$ *by itself,* ***by itself alone****.*

The fact that the whole $QM_{HD}$ is just a human construction achieved by men in order to represent what men can name, if they coose so, 'factually definable microstates', has receded far out of the minds. How much more far, then, we still are from conceiving that not even the microstates to be studied do 'exist' out there, and sometimes not even the involved 'systems', in this sense that in general one has to generate them in order to study them!

This situation entails a sort of consternation. It even produces a sort of religious admiration for $QM_{HD}$ because the experiments on locality have 'confirmed the predictions of the formalism'; as if predictions could be made without free choices of definitions that involve words.

*We are in presence of a huge reification of the whole conceptualization.*

But inside *[IQM-$QM_{HD}$]* one understands that, and how, these attitudes stem from the following circumstance:

Notwithstanding that in general the mathematical writings from $QM_{HD}$ are in agreement with the definitions from (2.I).1.2 *these definitions are not spelled out inside $QM_{HD}$*. And furthermore, in the case of one micro-*state* of two micro-*systems*, in the current language that accompanies the use of the formalism one speaks of two or several *'systems'* – never of ***'one*** micro-***state*** of two micro-*systems'*. Therefore inside $QM_{HD}$ the *indirect* and non-explicated one-one connection

$$G \leftrightarrow |\psi_G(r,t)>$$

is simply out of perceptibility, notwithstanding that it is generally accepted that always « a state-ket represents the studied *'system'* ». But inside *[IQM-$QM_{HD}$]* this connection is logically *entailed*, via (1) $G \leftrightarrow ms_G$.

Going now to the roots, one finds that – quite generally – all the preceding examples illustrate how inside $QM_{HD}$ unintelligibility is entailed by the fact that no clear and systematic distinction is made between ***individual physical entities*** (one individual realization of an operation *G*, one specimen $\sigma(ms_G)$ of a factual microstate $ms_G$, one act of measurement *MesA*), or ***individual concepts*** (eigen-'state' $|u_j(\mathbf{r},a_j)>$, which in fact is not a state but a sample of wave-movement), and on the other hand the ***statistical*** *descriptor* $|\psi(\mathbf{r},t)>$ tied with the new physical-*statistical* concept $ms_G \equiv \{\sigma(ms_G)\}$ of a *factual* microstate defined by (1) from *MD* [85].

In these conditions, inside the minds used to $QM_{HD}$ like a New-York boy is used to Manhattan, an explanation is badly needed indeed, *why* sometimes some of the two or several state-ket that would be so 'necessary', nevertheless are stubbornly 'absent'.

We are now ready to close this point by the following convention:

***Notational convention 1***. Inside *[IQM-$QM_{HD}$]* any state-ket $|\psi(\mathbf{r},t)>$ that corresponds to a physically generated micro-*state* $ms_G$ will be re-noted as $|\psi_G(\mathbf{r},t)>$ and the *sort* of operation *G* that indexes it will be explicitly stated, and when necessary its specific structure will be distinguished graphically.

So from now on (14) will be re-written as

$$|\psi_G(\mathbf{r},t)>/\mathbf{A} = \sum_j c_j(t)\ |u_j(\mathbf{r},a_j)>, j=1\ldots\ldots \qquad (14')$$

and ***inside $QM_{HD}$*** (15) will be re-written as

$$|\Psi(\mathbf{r},t)_{\mathbf{G}(G1,G2,\ldots Gk)}> = \lambda_1|\Psi_{G1}(\mathbf{r},t)>+\lambda_2|\Psi_{G2}(\mathbf{r},t)>+\ldots\ldots\lambda_k|\Psi_{Gk}(\mathbf{r},t)> \qquad (15')$$

[85] On the other hand it is true that it *does* seem amazing to find out to what a degree the mathematical formalism – in contradistinction to the physicists – is *observant* of (is compatible with) the involvement or not, in a given state-ket-symbol, of an operation of generation *G* of a corresponding micro-state, and with the significance of the involved state-ket from the viewpoint of the definitions (2.I)1; and this, *notwithstanding that inside the formalism the concept 'G' is neither defined nor represented, and the definitions (2.I)1 are not stated, while the specific meaning of an eigenket has not been recognized either.* Indeed: ***(a)*** a spectral decomposition (14) is usually conceived to involve an infinite number of terms, the coefficients from these terms are complex numbers dependent on time, while the eigenfunctions – models of wave-movement – are correctly written as *in-dependent of time*; whereas ***(b)*** a superposition (15) of state-ket tied with a composed operation of generation is written as a finite number of terms, the coefficients are usually constant real numbers, and the ket from the superposition *are dependent on time*. Everything in the mathematical writings is fully concordant with the analyses made here inside *[IQM-$QM_{HD}$]*. This raises strongly a very interesting question concerning something that could be called "the **semantic** expressivity of the mathematical internal syntactic coherence". The unique explanation seems to be that the mathematicians, when they construct a mathematical structure, somehow **imply the existence of operations of generation !!!**

where, in the global symbol $\boldsymbol{G}(G_1,G_2,...G_k)$, the first operation of generation ***G*** from the first member is written in *bold font* in order to express that it is the *unique* operation of generation that has been fully accomplished physically; while the operations of generation $G_1,G_2,...G_k$ from the second member – *each one of which was possible a priori but none of which has been factually and separately accomplished fully* – will be written in *non-bold font* in order to stress that their role is to act as only *virtual 'microstates-**of-reference**'*.

But the writing (15') from the the ***provisional*** framework *[IQM-QM$_{HD}$]* will have to be reconsidered later. It constitutes only a basis for this announced future re-examination.

These specifications will entail much clarification [86]**.** They also are an illustration of the fuzzy character of the level where takes place the collision between a *top-down* and a bottom-up approach and of the way in which the bottom-up approach can *incorporate* progressively the preceding top-down approach in a new, more precise, improving elaboration.

Other two clarifying notational improvements founded on the concept '*G*' are introduced in the course of the following point (6.II).4.

[86] The clarification will lead in the Part III to a ***suppression***, inside *QM2*, of the whole concept-and-writing (15').

## 6.II).4. CRITICAL EXAMINATION OF THE $QM_{HD}$ THEORY OF MEASUREMENTS

Here we definitely walk into Absurd-Land, so abruptly and totally that, by fear of being considered subjective and malevolent, I do not dare to immediately make use of my own voice. So I first offer an objective look at the conceptual situation by reproducing an extract from Wikipedia that consists of a disparate collection of views.

« The **measurement problem** in quantum mechanics is the problem of how (or *whether*) wave-function collapse occurs. The inability to observe this process directly has given rise to different interpretations of quantum mechanics, and poses a key set of questions that each interpretation must answer. The wave-function in quantum mechanics evolves deterministically according to the Schrödinger equation as a linear superposition of different states, but actual measurements always find the physical system in a definite state. Any future evolution is based on the state the system was discovered to be in when the measurement was made, meaning that the measurement "did something" to the system that is not obviously a consequence of Schrödinger evolution.

To express matters differently (to paraphrase Steven Weinberg[1][2]), the Schrödinger wave equation determines the wavefunction at any later time. If observers and their measuring apparatus are themselves described by a deterministic wave function, why can we not predict precise results for measurements, but only probabilities? As a general question: How can one establish a correspondence between quantum and classical reality?[3].

**Schrödinger's cat**

The best known example is the "paradox" of the Schrödinger's cat. A mechanism is arranged to kill a cat if a quantum event, such as the decay of a radioactive atom, occurs. Thus the fate of a large scale object, the cat, is entangled with the fate of a quantum object, the atom. Prior to observation, according to the Schrödinger equation, the cat is apparently evolving into a linear combination of states that can be characterized as an "alive cat" and states that can be characterized as a "dead cat". Each of these possibilities is associated with a specific non-zero probability amplitude; the cat seems to be in some kind of "combination" state called a "quantum superposition". However, a *single, particular observation* of the cat does not measure the probabilities: it always finds either a living cat, or a dead cat. After the measurement the cat is definitively alive or dead. The question is: *How are the probabilities converted into an actual, sharply well-defined outcome?*

**Interpretations** (*Main article: Interpretations of quantum mechanics*)

**Hugh Everett's** many-worlds interpretation attempts to solve the problem by suggesting there is only one wave-function, the superposition of the entire universe, and it never collapses—so there is no measurement problem. Instead, the act of measurement is simply an interaction between quantum entities, e.g. observer, measuring instrument, electron/positron etc., which entangle to form a single larger entity, for instance *living cat/happy scientist*. Everett also attempted to demonstrate the way that in measurements the probabilistic nature of quantum mechanics would appear; work later extended by Bryce DeWitt.

**De Broglie–Bohm** theory tries to solve the measurement problem very differently: this interpretation contains not only the wavefunction, but also the information about the position of the particle(s). The role of the wave-function is to generate the velocity field for the particles. These velocities are such that the probability distribution for the particle remains consistent with the predictions of the orthodox quantum mechanics. According to de Broglie–Bohm theory, interaction with the environment during a measurement procedure separates the wave packets in configuration space which is where apparent wave-function collapse comes from even though there is no actual collapse.

**Erich Joos and Heinz-Dieter Zeh** claim that the phenomenon of quantum decoherence, which was put on firm ground in the 1980s, resolves the problem.[4] The idea is that the environment causes the classical appearance of macroscopic objects. Zeh further claims that decoherence makes it possible to identify the fuzzy boundary between the quantum microworld and the world where the classical intuition is applicable.[5][6] Quantum decoherence was proposed in the context of the many-worlds interpretation[*citation needed*], but it has also become an important part of some modern updates of the Copenhagen interpretation based on consistent histories.[7][8] Quantum decoherence does not describe the actual process of the wavefunction collapse, but it explains the conversion of the quantum probabilities (that exhibit interference effects) to the ordinary classical probabilities. See, for example, Zurek,[3] Zeh[5] and Schlosshauer.[9]

**The present situation is slowly clarifying, as described in a recent paper by Schlosshaue**r as follows:[10]

Several decoherence-unrelated proposals have been put forward in the past to elucidate the meaning of probabilities and arrive at the Born rule ... It is fair to say that no decisive conclusion appears to have been reached as to the success of these derivations. ...

As it is well known, [many papers by **Bohr** insist upon] the fundamental role of classical concepts. The experimental evidence for superpositions of macroscopically distinct states on increasingly large length scales counters such a dictum. Superpositions appear to be novel and individually existing states, often without any classical counterparts. Only the physical interactions between systems then determine a particular decomposition into classical states from the view of each particular system. Thus classical concepts are to be understood as locally emergent in a relative-state sense and should no longer claim a fundamental role in the physical theory.

**A fourth approach is given by objective collapse models**. In such models, the Schrödinger equation is modified and obtains nonlinear terms. These nonlinear modifications are of stochastic nature and lead to a behaviour which for microscopic quantum objects, e.g. electrons or atoms, is unmeasurably close to that given by the usual Schrödinger equation. For macroscopic objects, however, the nonlinear modification becomes important and induces the collapse of the wavefunction. Objective collapse models are effective theories. The stochastic modification is thought of to stem from some external non-quantum field, but the nature of this field is unknown. One possible candidate is the gravitational interaction as in the models of Diósi and Penrose. The main difference of objective collapse models compared to the other approaches is that they make falsifiable predictions that differ from standard quantum mechanics. Experiments are already getting close to the parameter regime where these predictions can be tested.[11]

**An interesting solution** to the measurement problem is also provided by the hidden-measurements interpretation of quantum mechanics. The hypothesis at the basis of this approach is that in a typical quantum measurement there is a condition of lack of knowledge about which interaction between the measured entity and the measuring apparatus is actualized at each run of the experiment. One can then show that the Born rule can be derived by considering a uniform average over all these possible measurement-interactions. [12][13] ».

### (6.II).4.1. Refusal of von Neumann's representation of quantum measurements

I now dare to continue by my own summary of the situation. In what follows immediately I place myself inside $QM_{HD}$, not inside $[IQM\text{-}QM_{HD}]$. So I just express first the current nowadays language and reasoning about quantum measurements. The representation of measurements on microsystems is that one proposed by von Neumann in 1932, namely:

***The orthodox $QM_{HD}$ assertions on quantum measurements***. The Schrödinger equation of the problem endows us with the state-ket of the problem, $|\psi(\mathbf{r},t)>$. So this state-ket is given mathematically, we dispose of it from the start in consequence of purely mathematical operations. We want now to *represent* the measurements. Therefore we have to write the state-ket for the measurement-interaction. For this we proceed as follows: Let $t=t_o$ be the initial moment given in $|\psi(x,t)>$. At a time $t_1 \geq t_o$ we want to measure the observable $\boldsymbol{A}$ on the 'system' represented by $|\psi(\mathbf{r},t)>$. We take now into account that for $t \geq t_1$ there is *interaction* between the studied system and the measurement-apparatus. "So", is it said :

> For $t \geq t_1$ the measurement-evolution ***must*** *represent also the apparatus* ***"because" the apparatus is also constituted of microsystems***.

So the measurement-evolution is to be represented by a state-ket of [(the studied system S)+(the apparatus for measuring $\boldsymbol{A}$)]. Let us then write, say, $S+app(\boldsymbol{A})$ and $|\psi_{S+app(A)}(\mathbf{r},t)$. Since we measure the observable $\boldsymbol{A}$, the expansion of $|\psi_{S+App}(\mathbf{r},t)>$ with respect to the basis of $\boldsymbol{A}$ comes in. Accordingly to the well-known quantum theory of '*a system* composed of two *systems*' we write the tensor-product expansion:

$$|\psi_{S+App}(\mathbf{r},t)>= \sum_k\sum_j c_j(t)d_k(t)\ |u_j(\mathbf{r},a_j)>|q_k(\mathbf{r},a_k>, \ \ \forall j, \ \ \forall k \qquad (17)$$

where $|q_k(\mathbf{r},a_k>$, $k=1,2...$ are the eigenket of the 'observable' called the 'needle-position of the $app(\boldsymbol{A})$', that can be denoted $\boldsymbol{\chi(A)}$, with eigenvalues, say $v(a_k)$, $\forall k$, that express, respectively, 'the needle-positions of $app(A)$ that correspond to the eigenvalues $a_k$ of $\boldsymbol{A}$'. Furthermore – by the definition of the concept of 'apparatus for measuring $\boldsymbol{A}$' – the set $\{c_j(t)d_k(t)\}$ of all the product-expansion coefficients $(c_j(t)d_k(t))$ from (17) reduces to a set $\{\alpha_{jj}(t)\}$ (with $\alpha_{jj}=c_jd_j$) of only the coefficients with non-crossed indexation, because the needle position $v(a_j)$ of the $app(\boldsymbol{A})$ *is what –* ***alone*** *– indicates the obtained eigenvalue* $a_j$ *of* $\boldsymbol{A}$ [87]. So in fact in this case we have only

$$|\psi_{S+App}(\mathbf{r},t)>= \sum_j\sum_j \alpha_{jj}(t)\ |u_j(\mathbf{r},a_j)>|v_j(\mathbf{r},a_j>, \ \ \forall j \qquad (17')$$

[87] So no coding problem arises according to this 'measurement-theory': One is protected from this problem, the apparatus will know where to settle its needle, since it is conceived for this aim.

The measurement evolution is produced accordingly to *a measurement-Schrödinger evolution where the hamiltonian operator **H(A)** commutes with **A***. And it is *posited* that this evolution finishes with a definite needle position $\chi(a_j)$ that indicates ***one*** definite result $a_j$ [88].

Now, the above-mentioned representation is considered to raise two 'problems'.

- The *reduction problem*: what happened to all the terms from (17') with index $k \neq j$ that accordingly to a linear formalism should subsist? Where have they disappeared?

- The problem of 'decoherence': how can we *prove* that after the realization of the position $\chi(a_j)$ of the apparatus-needle that announces the result $a_j$ the measurement interaction really ceases? [89]

Here finishes my own summary of the general framework accepted for the representation of measurements. In what follows I go now back inside *[IQM-$QM_{HD}$]* and I speak again for myself and by use of the language introduced up to now in this work.

Bertrand Russell has written somewhere that aims are induced by temperament while the choice of a method when an aim is given is induced by intelligence. If this is accepted then, with respect to the aim to represent the measurements on microstates, von Neumann's choice of a method is stunning. If we followed his argument, in order to measure the position of a star by use of a telescope, given that the telescope and the star are both made of microsystems, we should represent [(the telescope)+(the star)+(the measurement interaction between these two entities)]; and we should *prove* in terms of the theory thus conceived, that the star and the telescope do really separate physically once the star's position has been established. Such an argument manifests a total blindness with respect to the rather obvious fact that in science what decides the optimality of a representation is the *cognitive situation* of the observer-conceptualiser with respect to that on what he wants to obtain some knowledge, etc.; the inner constitution of that what has to be qualified, or of the instruments that are made use of, has nothing to do with the criteria for generating the desired knowledge. Quite generally the functionality of a construct is not in a one-to-one relation with its material (or abstract) structure. Those who have finally realized the aeroplanes have had to thoroughly understand this before succeeding. Moreover, in the case of microstates, most often what *can* be registered is just marks on a sensitive registering device and/or durations determined by chronometers, not 'needle positions'. From these data one has then to *construct conceptually* the researched 'value' of the measured 'quantity', and this quantity in its turn is constructed beforehand on the basis of conceptual-mathematical operations. And finally, von Neumann's representation of measurements dodges the crucial coding problem. It simply makes it disappear behind an amorphous heap of words and symbolic writings *void of definition*, so of meaning. Indeed von Neumann's representation of measurements *transgresses $QM_{HD}$*:

> '*The* observable' $\boldsymbol{\chi(A)}$ called the 'needle-position of the *app(A)*' is not a quantum mechanical observable; it cannot be constructed formally in a definite way from a definite classical mechanical quantity, and so its eigenfunctions and eigenvalues cannot be *calculated*. The eigenvalues are just *postulated* to be the eigenvalues of the observable ***A***.

All this simply is not acceptable. A measurement-apparatus that is made use of in a scientific description of something else than this apparatus itself has to be introduced as a

[88] As far as I know, this has never been proved inside $QM_{HD}$ to be generally insured by the condition imposed upon the measurement evolution.

[89] The locality-problem incites to think that it might not do this, but so what?

*primary datum* that stays outside the representation of the act of measurement; if not, one enters indefinite regression [90]**.** This is a general interdiction of logical nature. Etc.

So I declare without shades that *I quite radically reject von Neumann's framework for representing quantum measurements*.

### (6.II).4.2. Global critical-constructive considerations on "the $QM_{HD}$ theory of measurements"

We shall now concentrate on the essential features of the $QM_{HD}$ representation of the quantum measurements. The core of the unintelligibility of $QM_{HD}$ is hidden there. The developments from 6.II and the refusal of von Neumann's 'theory of quantum measurements' leave us now face to face with the deepest problems of intelligibility. These problems will now be stated independently of any superfluous representational clothing, in order to have a chance to draw into light the prime source of the vices that obstruct the tortuous channels toward intelligibility. To guide toward this source I begin by a global sketch of the conceptual situation.

#### *(6.II).4.2.1. Preliminary remarks*

One *constructs* statistics of numbers (or of other sorts of observable 'results') in order to be able to *predict statistically* on meaningful consequences of acts of examination indicated by the verbal label 'measurements'. Usually the construction of a predictive statistic is achieved by performing sets of individual measurements. When this is the case let us speak of *construction-measurements*. And then, in order to be certain that the construction of the predictive statistic has succeeded, one *verifies* the obtained statistic before announcing it as useful for further prediction. In this case let us speak of *verification-measurements*. It seems natural to assume that factually the results of repeated construction-measurements and those of repeated verification-measurements identify when the measurements are the same, notwithstanding that they are performed with different aims.

> So knowledge of the *way* of factually performing the individual acts of measurement should be insured before trying to construct and to verify the statistics that involve such measurements.

But *where* can one find inside $QM_{HD}$ a thorough definition of the way of performing *factually* an individual act of quantum-measurement? ***Nowhere****, strictly nowhere*. Indeed: In $QM_{HD}$ – via the Schrödinger equation – the representation of, directly, the predictive statistics is generated mathematically. This circumvents the necessity to define individual construction-measurements in order to predict, as is the case in *IQM*.

However *the individual verification-measurements remain unavoidably necessary*.

So the absence of any definition of the individual acts of measurement is a striking gap, even *if* indeed such a definition were necessary exclusively for verification of the statistical predictions[91].

> But inside $QM_{HD}$ the *process* of verification of the statistical predictions asserted by the state-ket of the studied microstate – considered globally – is represented mathematically by the Schrödinger "measurement-evolution-ket*" which is a* ***statistical*** *descriptor*.

---

[90] Wittgenstein has written somewhere: «There is one thing of which one can say neither that it is one metre long, nor that it is not one metre long, and that is the standard metre in Paris ». I dare to complete: At least one class of things cannot be absorbed into the quantum mechanical representation of measurements: The class of the measurement apparatuses.

[91] For bound states, collective acts of verification-measurement can often suffice (absorption or emission of radiation versus registration of the intensity of spectral lines of emission or of absorption of radiation). But for unbound microstates, individually defined verification-measurements are not avoidable.

And it is admitted that the measurement-evolution-ket somehow *involves the factual **individual** acts of measurement* that should verify the predictive statistic: a strange knot[92]. Such a common formal location of two sorts of descriptive elements that belong to two different levels of conceptualization is immediately highly suspect.

But furthermore, throughout this suspect representation, the explicit definition of an individual act of measurement is just supposed to be known and it is alluded to in mere words, without even being also *defined*, at least in mere words: the mentioned definition is just pointed to with a verbal finger. And the *end* of such an undefined act of measurement that cohabits with the statistic to the verification of which it is associated, is expressed formally via a sudden modification of the representation of the involved Schrödinger ket of measurement-evolution: a modification *that is **not** entailed by the mathematical formalism* but is just *posited separately*: the famous "collapse of the measurement-evolution-ket" [93].

Some authors have remarked that in the classical theory of probabilities "also" one considers from the outset a whole probabilistic distribution of the elementary events from a whole universe of possible such events, while every individual act from the process of verification of this distribution produces just one definite result. This seems to be conceived as a sort of legalizing jurisprudence. But in this respect we notice that in the theory of probabilities the mathematical representation of the probabilistic predictive distribution is not used *itself, and globally*, for representing the individual acts from the process of verification of the predictive probabilistic distribution. This sort of coalescence between a statistical-probabilistic descriptive element, and the representation of individual operations for a progressive verification of a pre-existing probabilistic predictive distribution, is *specific* of the quantum representation of measurements.

We finally add that – according to *IQM* – any individual act of measurement is accomplished on *one* specimen of a microstate and so it requires an *immediately* previous realization of an operation of generation *of that specimen*; and furthermore, *it also requires sine-qua-non an operational definition that involve a factually realizable coding procedure*; *which in its turn **requires a general model of a microstate***. Inside *IQM* – that has been constructed as only a reference-and-imbedding structure of any theory of microstates – these requirements have been left blank, to be specified in any given theory of microstates. But here we try precisely to draw from *IQM* and $QM_{HD}$ a fully intelligible theory of the factually defined microstates $ms_G$ from (1) $G \leftrightarrow ms_G$, $ms_G \equiv \{\sigma(ms_G)\}$. So now we have to specify what model is posited, and, for each sort of microstate, what coding rule is involved for the measurement-evolution of any given observable ***A***.

In (6II).1 it appeared that the formalism of $QM_{HD}$ involves surreptitiously but essentially de Broglie's model of a microstate and in (6II).2 we have re-defined this model in operational terms via the *modelling postulate* $MP(\{\sigma(ms_{G,cw})\})$ and (1'). So in what follows, for each one of the various types of unbound microstates distinguished in the section (2.I)2, it will have to be specified for any individual quantum measurement *MesA* of any observable ***A*** *what explicit sort of coding-rule* is entailed via (1').

The problems sketched out above can be summarized as follows: It is generally agreed that a statistical representation only mirrors the physical individual facts and operations that it involves genetically. But this does not exonerate from stipulating *of what consists that what is mirrored, when precisely that is studied*.

[92] Lacan would have interesting comments on it.

[93] Since already a century this situation seems scandalous to a big number of physicists and many solutions have been proposed, verbal or mathematical ones. The best-known is the very extravagant "interpretation" of Everett that many physicists finally *do admit* "because it does not involve any *inconsistency*" (which is to be read as "*mathematical* inconsistency", any semantic constraint being expurgated: This manifests the relations that work inside some minds of physicists concerning facts versus mathematics-and-logic inside a discipline of *Physics*.

### *(6.II).4.2.2. A fundamental distinction:*
### *Individual physical wave-function* versus *abstract statistically predictive 'state'-function*

Let us go back to the fact that the wave function initially introduced by Louis de Broglie was conceived to represent a *physical* 'corpuscular wave' $\Phi(\mathbf{r},t)= a(\mathbf{r},t)exp((2\pi i/h)\beta(\mathbf{r},t)$ assigned to any micro-system, like in Bohm's later approach. But very rapidly this initial concept has transmuted into a mathematical representation of predictive statistics of results of measurements performed on some given sort of microstate. In de Broglie's mind thereby the content of the initial descriptor $\Phi$ just became more 'complex'. The amplitude $a(\mathbf{r},t)$ of the function $\Phi$ pointed now toward a statistical-probabilistic prediction for results of repeated registrations of the position of the corpuscular-like singularity posited to be involved by the amplitude of any specimen of the considered microstate, while the phase $\beta(\mathbf{r},t)$ from $\Phi$ was conceived to continue to point toward the physical individual wave movement from the corpuscular wave of any one specimen of the studied microstate. But in fact the representational evolution mentioned above violated one of the numerous still very little known semantic rules that do work inside our minds concerning the processes of conceptualization. (For instance, when one applies arithmetic to a factual problem it is explicitly interdicted – for semantic reasons – to add prunes with apples, notwithstanding that inside arithmetic the operation '+' is defined without restriction). In particular, inside mathematical physics works implicitly an interdiction to represent *by one same mathematical descriptor* two features that qualify physical entities of different semantic natures, so two different sorts of physical entities: if this interdiction is transgressed, sooner or later this leads to confusions. This is just a fact. So de Broglie's wave-mechanics has started with a violation of a hidden semantic rule. The spontaneous public reasoning – that feels the semantic rules – has tended to compensate this violation by selecting *only* the *statistical* – mathematical – representation of a microstate. And this contributed to simply abandon in the non-spelled-out the individual physical descriptive elements.

Here however, before entering upon the examination of the $QM_{HD}$ representation of quantum measurements, we introduce a clear and stable distinction between the descriptor of an individual physical corpuscular-like de Broglie wave, and on the other hand a mathematical descriptor of statistics of results of quantum-measurements. This is done via the following second notational convention (to be added to the first one from (14') and (15')):

***Notational convention 2***. The physical individual wave-like phenomenon introduced in the domain of scientific conceptualization by one realization of the operation *G* of generation – *in the sense of (1')* – called "one specimen $\sigma(ms_{G,cw})$ of the studied microstate $ms_{G,mw}$" will be systematically denoted by an ***individual*** *wave*-function denoted $\Phi_{G,cw}(\mathbf{r},t)$[94](this mathematical form is posited here to *include* the representation of also the 'corpuscular singularity(ies)' from the amplitude of the wave-like phenomenon). While the *state*-function from the $QM_{HD}$-state-ket $|\Psi_G(\mathbf{r},t)>$ that is associated with the factual microstate $ms_{G,mw}$, will represent *exclusively* a mathematical tool for predictive statistics of results of measurements on individual specimens $\sigma(ms_{G,cw})$. Keeping present in memory all the details of this convention, we usually shall go back to the simplified notations $\Phi_G$ and $|\psi_G>$.

---

[94] We maintain the notation $\Phi$.

The distinction introduced above does not in the least interdict any possible degree of similitude between the mathematical forms of $\Phi_G$ and $|\psi_G>$ [95].

Now, on the ground enriched by the preliminary considerations from *(6.II).4.2.1* and the distinction from *(6.II).4.2.2*, we enter upon the examination of what is called "the $QM_{HD}$-theory of quantum measurements".

### (6.II).4.3. The coding rule implied by the $QM_{HD}$-formalism

Since we refuse von Neumann's representation of quantum measurements we go back to the initial representation of these where the measuring-apparatus is *not* represented. But the Hilbert-Dirac representation is conserved. The other notations as well as the global point of view remain those entailed by the association *[IQM-$QM_{HD}$]* enriched with the *new* contents already gained in the chapter 5.II and in (6.II).1 to (6.II).4.2. So we are no more inside the $QM_{HD}$ alone.

Thus equipped, the purpose in this section is to identify *how* – inside the nowadays $QM_{HD}$ – the observable result of one physical measurement-evolution of the one specimen of the studied one microstate of one microsystem that is involved, is supposed to be translatable in terms of one definite eigenvalue of the measured quantity. In other terms, to identify what coding rule is at work more or less implicitly inside $QM_{HD}$ for one act *MesA* of measurement of a given observable $\boldsymbol{A}$.

Let us admit that via the Schrödinger equation of the problem acted by some initial evolution-Hamiltonian operator $\boldsymbol{H}$, it has been possible to identify the state-ket $|\psi_{G,\boldsymbol{H}}(\boldsymbol{r},t)>$ of the microstate $ms_G$ to be studied, where $t \geq t_o$ and $t_o$ "the initial moment". If at a time $t_1 \geq t_o$ one wants to know the probability of the outcome $a_j$ if the observable $\boldsymbol{A}$ is measured on a specimen of the studied microstate $ms_G$, the $QM_{HD}$-procedure is as follows (cf. (5.II)1). Write the expansion of $|\psi_{G,\boldsymbol{H}}(\boldsymbol{r},t_1)>$ on the basis $\{|u(\boldsymbol{r},a_j,)>, \forall j\}$, of eigenket of $\boldsymbol{A}$ for the moment $t_1$ :

$$|\psi_{G,\boldsymbol{H}}(\boldsymbol{r},t_1)>/\boldsymbol{A} = \sum_j c_j(a_j,t_1)|u(\boldsymbol{r},a_j,)>, \qquad \forall j \qquad (18)$$

The *statistical* prediction concerning the outcome of an $\boldsymbol{A}$-measurement on the microstate represented by $|\psi_{G,\boldsymbol{H}}(\boldsymbol{r},t_1)>$ is given by the set of numbers $\{|c_j(a_j,t_1)|^2, \forall j\}_{\boldsymbol{A}}$.

As for the *verification* of this prediction, one proceeds accordingly to the following algorithm that brings in the famous reduction-problem: From $t_1$ on, the action of the evolution-Hamiltonian operator $\boldsymbol{H}$ that has worked during the time-interval $(t_1-t_o)$ is stopped. $|\psi_{G,\boldsymbol{H}}(\boldsymbol{r},t_1)>$ is replaced by a new $\boldsymbol{A}$-measurement-evolution-ket $|\Psi_{G,\boldsymbol{H}(A)}(\boldsymbol{r},(t_o \leq t_1 < t_f))>$ of which the evolution is performed accordingly to the Schrödinger equation acted by an $\boldsymbol{A}$-measurement-Hamiltonian $\boldsymbol{H(A)}$ that commutes with $\boldsymbol{A}$. This $\boldsymbol{A}$-measurement-evolution takes a time $(t_1 \leq t \leq t_f)$ where $t_f$ marks the end of the considered act of $\boldsymbol{A}$-measurement, when the *individual* eigenvalue $a_j$ of $\boldsymbol{A}$ is registered. ***Thereby*** this evolution exhibits an *individual* feature. Nevertheless the ket

[95] For an unbound microstate there certainly exists a strong similitude between the mathematical function $\Phi_{G,cw}(\boldsymbol{r},t)$ appropriate for representing $\sigma(ms_{G,cw})$, and the state-function $\psi_G(\boldsymbol{r},t)=a(\boldsymbol{r},t),t)e^{(i/\hbar)\varphi(r,t)}$ from the state-ket of $ms_G$. But in consequence of the predictive task assigned to $\psi_G(\boldsymbol{r},t)=a(\boldsymbol{r},t)e^{(i/\hbar)\varphi(r,t)}$ – also certainly – there is no identity between these descriptors (this is now clear by the definition of the amplitude function $a(\boldsymbol{r},t)$; but even the phase $\varphi(\boldsymbol{r},t)$ from $\psi_G(\boldsymbol{r},t)$ might indicate only some sort of mean-phase with respect to the unknown individual phase functions $\beta(\boldsymbol{r},t)$ that are involved in the set of specimens $\{\sigma(ms_{G,cw})\}$ of the studied microstate $ms_{G,cw}$ (cf.(14). Anyhow, for now the fact is that in general we do not know what equation can yield as solutions the functional representations of the *individual specimens* $\sigma(ms_{G,cw})$; and even though de Broglie has asserted a *common* equation for $\psi_G$ and $\Phi_{G,cw}$ and has characterized it in detail (de Broglie [[1956],[1957]), nothing insures a priori the general validity of a common equation. During the years 1960-1980 neither de Broglie nor anybody else seems to have been aware of the essential role – for consensual verifiable knowledge – of the operation of generation of the entity-to-be-studied, *when the considered microstate is* ***put in this role***; nor, a fortiori, of the other related mathematical representations of individual operations. Correlatively, nobody seems to be aware that *'giving' the initial state-ket with the initial wave-function constitutes a key-action for quitting the realm of purely mental mathematical representation and stepping over into the domain of factually predictive and verifiable consensual* ***knowledge***.

$|\Psi_{G,H(A)}(\boldsymbol{r},(t_o \leq t_1 < t_f))>$ of measurement-evolution is itself a ***statistical*** descriptor that is made use of via its expansion with respect to $\boldsymbol{A}$

$$|\Psi_{G,H(A)}(\boldsymbol{r},(t_1 < t_f))>/\boldsymbol{A} = (\sum_j c_j(a_j,t_1))|u(\boldsymbol{r},a_j,)>)_{G,H(A)} \quad \forall j,\ \forall \boldsymbol{A},\ t_o \leq t_1, \qquad (19)$$

where ***all*** the expansion coefficients $c_j(a_j,t_1))$ are present.

Furthermore in (19) the expansion coefficients are the same as in (18). So the new evolution (19) of the initial expansion $|\psi_{G,H}(\boldsymbol{r},t_1)>/\boldsymbol{A}$ – but with ***H(A)*** and $(t_o \leq t_1 < t_f)$ – is supposed ***not*** to change the expansion coefficients from (18); while the eigenket $|u(\boldsymbol{r},a_j,)>$ are time-independent. So that a question that might emerge naturally is: *what* ***does*** *change* – mathematically – during an "evolution" (19)?

This possible question however – to be kept in mind – concerns mathematical changes. Whereas we, in order to understand what coding rule is supposed to work, we have to first specify what is conceived to change *physically* during $(t_f - t_1)$, and how. This purpose is likely to somehow lead us from the representation with respect to $\boldsymbol{A}$, to the *space-time* representation of the measurement-evolution-ket $|\Psi_{G,H(A)}(\boldsymbol{r},(t_1 < t_f))>/\boldsymbol{A},\ \forall j,\ \forall \boldsymbol{A},\ t_o \leq t_1$.

At this point let us make a break in order to bring in guiding data.

Remember Bohm's analysis [(1951)] of the Stern-Gerlach method for spin measurements, de Broglie's analyses ([1957]), and Gottfried's presentation of quantum measurements ([1966]). But quite especially, *remember the method time-of-flight for measuring the momentum observable* ***P***. This last method, mentioned by Feynmann, has been thoroughly studied by Park and Margenau ([1968]) and when it is examined closely it appears quite clearly that what is supposed to go on from a physical viewpoint during the process involved in (19) is *such* that:

For any given index $j$ a measurement-evolution (19) represents globally the emergence via sufficiently many repetitions of one act of ***P***-measurement, of a statistical correlation between:

- On the one hand, *a more or less extended space-time domain* $(\Delta\boldsymbol{r}.\Delta t)_{j,\boldsymbol{P}}]$ *that codes for* ***one*** *eigenvalue* $p_j$ *of* ***P***.

- And on the other hand, the (finite) number of times – posited to determine factually *the probability for registering the result* $\boldsymbol{p_j}$ – in which the group of all the observable marks registered during one act of ***P***-measurement is located inside $(\Delta\boldsymbol{r}.\Delta t)_{j,\boldsymbol{P}}$; ***anywhere*** inside $(\Delta\boldsymbol{r}.\Delta t)_{j,\boldsymbol{P}}$, whereby the correlation is statistical.

***Which amounts to a coding-procedure***.

The method 'time of flight' has been formulated exhaustively, and the case of the observable ***P*** is both ***basic*** and paradigmatic because the 'presence'-observable ***R*** codes trivially for its registered value $\boldsymbol{r_k}$ so that a pair $(\boldsymbol{r_k},\boldsymbol{p_j})$ yields indirectly a coding for any eigenvalue $a_j$ of any observable $\boldsymbol{A}$ via the fact that any observable $\boldsymbol{A}$ is a function ***A(R,P)*** (cf. 5.1) (whereas the case of spin-measurements has been studied separately). Therefore below I summarize explicitly the method 'time of flight, in $[IQM\text{-}QM_{HD}]$-terms.

Let $\delta E(G)$ be the space-domain covered by one realization of the operation $G$ via an apparatus $\mathcal{A}(G)$ for generating specimens $\sigma(ms_{G,cw})_n$, $n=1,2,...N$, of the microstate $ms_G$ to be studied. Place a very extended detection-screen $S$ sufficiently far from the space-domain $\delta E(G)$ for permitting to assimilate $\delta E(G)$ to only a point denoted $O$, *relatively* to the distance $OS$ between $\delta E(G)$ and $S$ measured on an axis $Ox$ that starts at $\delta E(G) \approx O$ and is perpendicular on the plane of $S$. We act as follows:

***(a)*** We effectively carry out with $\mathcal{A}(G)$ an operation $(G)_n$ and we denote by $t_{no}$ the time when $(G)_n$ *ends*, indicated by a system of two interconnected chronometers, one of which is connected to the generation-apparatus $\mathcal{A}(G)$ and the other one is connected to the screen *S* (the index '$_n$' individualizes the considered realization $(G)_n$ of *G*). The duration $\delta t(G)$ of the operation of generation *G* does *not* come in, alone will matter the time elapsed between the moment $t_{no}$ when the operation of generation *G ends* and the time $t_n$ when an impact is recorded anywhere on *S*.

***(b)*** If between $\delta E(G)\approx O$ and the screen *S* there pre-exist macroscopic fields or material obstacles, at $t_{no}$ the fields are extinct and/or the 'obstacles' are removed by a convenient device.

***(c)*** After some time an impact is produced on a spot of the screen *S* that we indicate by $\mathcal{P}(x_n,y_n,z_n,)$ where the coordinates are written with respect to a Cartesian system of reference that includes *Ox* and *(Oy*, *Oz)* are in the plane of *S*. When the impact-point $\mathcal{P}(x_n,y_n,z_n,)$ emerges on *S* the system of chronometers indicates the time $t_n$ and *"the 'time of flight' $\Delta t_n=t_n-t_{no}$"* [96] (that has been automatically calculated in the system).

***(d)*** Let $\boldsymbol{d_n}$ designate the vector-value of the distance between $\delta E(G)\approx O$ and $\mathcal{P}(x_n,y_n,z_n,)$. The square of the absolute value of this distance is $|\boldsymbol{d_n}|^2=d_{xn}^2+d_{yn}^2+d_{zn}^2$ where $d_{nx}$ , $d_{ny}$, $d_{nz}$ are calculated with respect to the specified Cartesian referential.

*(e)* The vector-eigenvalue $\boldsymbol{p_n}$ of the quantum mechanical momentum-operator ***P*** and its absolute value $|\boldsymbol{p_n}|$, are then calculated according to the formulas $\boldsymbol{p_n}=M(\boldsymbol{d_n}/\Delta t_o)=M\boldsymbol{v}$ *and* $|\boldsymbol{p_n}|=M\surd(d_{xn}^2+d_{yn}^2+d_{zn}^2)/\Delta t_o$, where *M* is the 'mass' associated with the involved specimen $\sigma(ms_{G,cw})_n$ of $ms_G$ such as this mass is defined in classical mechanics and in atomic physics[97], and ***v*** means 'velocity'. This completes the considered act of momentum-measurement.

Now we note what follows. The global apparatus is made up of: a generation apparatus $\mathcal{A}(G)$; a system of two chronometers; the suppressor of external fields and/or obstacles; the screen *S*. The observable physical marks produced during the considered act of measurement, are: the point-like mark $\mathcal{P}$, the position of the needle of one chronometer at the beginning of the act of measurement, and the position of the needle of the other chronometer at the end of this act. The observable manifestations enumerated above are not themselves numerical values, nor do they trigger any quale of which the direct perception is necessarily associated in the observer's mind with the involved specimen of the studied microstate. They are only perceptible physical marks – say $\mu_{1n}$, $\mu_{2n}$, and $\mu_{3n}$, respectively – registered on 'recorders' of the utilized global apparatus. The *meanings* associated with the recorded marks as well as the numerical values associated with these are defined:

- With respect to permanent elements of reference.
- By the way in which is *conceived* what is called "one act of measurement of, specifically, the eigenvalue $\boldsymbol{p_n}$ of the momentum-observable ***P***". *This, in agreement with IQM, presupposes a general model of a microstate* with respect to which this way of conceiving the measurement procedure makes sense and can be conceptually integrated in the previously achieved structure of scientific knowledge. And it seems utterly clear that:
- The general concepts of 'momentum-value ***p***' and of 'mass *M*' that are involved in the time-of-flight method, such as it is *nowadays* formulated, are founded upon the ***classical*** model of a 'mobile'. This indicates that in the required conditions (in particular the extinction of all the exterior fields) the method works for a classical model even though in fact a specimen of a microstate is involved.

And the method as a whole seems 'reasonable' precisely *because, and **only** because* it is designed so as to change during one act of momentum-measurement certain features of the posited classical mobile – namely its position – but in a way that does not alter *that what has to be measured* for the involved specimen of such a mobile – namely the momentum-value $\boldsymbol{p_n}$ of the momentum observable $\boldsymbol{P}$ at the *beginning* of the considered act of measurement. Indeed the procedure would be totally arbitrary in the absence of this assumption.

---

[96] 'Flight' of what? Obviously a model is involved. Probably the model of a classical 'mobile'.

[97] So indeed here the ***classical** model of 'mobile' is involved*. For the mass *in de Broglie's sense* has a *different* definition, and this might come out one day to be very important. But probably the de Broglie-Bohm approach (Bohm [1952]), de Broglie [1956]) was not yet much known in 1968.

**-** The vector-value $\boldsymbol{d_n}$ calculated from the observable marks $\mu_{1n}$, $\mu_{2n}$, and $\mu_{3n}$, has an origin that is permitted to *vary* inside some non-negligible space-domain $\boldsymbol{\Delta r}$ (because the origin of the 'flight' of the posited 'mobile' simply cannot be defined strictly when the mobile indicates a physical entity that is a specimen $\sigma(ms_{G,cw})_n$ of a microstate $ms_G$).

The method time of flight is the only one that, accordingly to nowadays $QM_{HD}$, seems to be regarded as 'legal' for measuring the momentum observable. This is so because *it is implicitly supposed to realize the inner structure* ***of an eigenstate of the momentum observable*** *throughout the considered act of measurement***,** namely the plane-wave structure required by the expansion $|\Psi_{G,H(A)}$ $\boldsymbol{(r,\ (t_o \leq t_1 < t_f) > /P}$ for the observable $\boldsymbol{P}$.

In short, we conclude that the measurement-evolution (19) that is implied by the time-of-flight method, ***does*** *entail a coding rule*, via a statistical correlation. Indeed: Let us denote by $\mu_{pj} \equiv \{(\mu_k,\ k=1,2,3)_j\}_{pj},\ j=1,2,...J_P,$ the group of *all* the observable marks produced by any one act of $\boldsymbol{P}$-measurement. On the basis of a reasoning drawn from the classical mechanics but that can be coherently related with the modelling postulate $MP(\{\sigma(ms_{G,cw})\})$ and the relation (1') $G_t \Leftrightarrow (ms_{Gt,cw} \equiv \{\sigma(ms_{Gt,cw})\})$, it is possible to specify for any eigenvalue $\boldsymbol{p_j}$ of $\boldsymbol{P}$ a corresponding space-time domain $(\boldsymbol{\Delta r}.\Delta t)_j$ such that if the whole group of marks $\{(\mu_k,\ k=1,2,3)_j\}_P$ is registered *anywhere* inside $(\boldsymbol{\Delta r}.\Delta t)_j$) this means 'registration of the eigenvalue $\boldsymbol{p_j}$'. So that one can write:

**[(**registration of the marks $\{(\mu_k,\ k=1,2,3)_j\}_P$ anywhere inside $(\boldsymbol{\Delta r}.\Delta t)_j$ ) *means* the result *'* $\boldsymbol{p_j}$*'*] (20)

This, as announced, is a statistical correlation that acts like a coding rule for ***individual*** *acts of* ***P****-measurement*, in the sense defined in (1.I).2. So it seems likely that – on the basis of the method time-of-flight and comforted by the analyses of the Stern-Gerlach method for spin-measurements which mutatis mutandis leads to a similar conclusion – de Broglie, Bohm, Park and Margenau have admitted more or less implicitly that ***any*** act of quantum measurement involves a coding rule of the general form (20). We denote this view by *BBGPM.*

In the present context the conclusion (20) is very interesting because it immediately suggests the following new considerations.

- No general proof of the coding rule (20) has been worked out inside $QM_{HD}$. This can be understood: $QM_{HD}$ is void of any treatment of the individual entities or processes that are involved in the statistical predictions that it asserts; *the individual measurement-evolutions find no formal place inside* $QM_{HD}$. But for our re-constructive purpose this absence of proof is not a crucial circumstance. Indeed, more or less explicitly, Park and Margenau have succeeded to somehow 'show' the presence of the correlation (20) inside $QM_{HD}$ for the particular case of the momentum observable $\boldsymbol{P}$. Moreover, as already remarked, for the basic position observable $\boldsymbol{R}$ a correlation of the same form is also realized, tautologically. And the $QM_{HD}$ 'postulate of representation' of an observable $\boldsymbol{A(R,P)}$ permits to form $\boldsymbol{A}$ by a simple calculus from the pair $\boldsymbol{(R,P)}$ of the two basic observables, so that also the eigenvalues $a_j$ of $\boldsymbol{A}$ can be constructed via definite algorithms from the eigenvalues of $\boldsymbol{R}$ and $\boldsymbol{P}$. So inside $QM_{HD}$ the coding procedure (20) can be applied to any observable $\boldsymbol{A}$, in some specifiable conditions.

But:

- The method time-of-flight involves the classical 'mobile'-model of a microstate, while inside $[IQM\text{-}QM_{HD}]$ the modelling postulate $MP(ms_{G,cw})$ from (6.II).2 introduces the *G-(corpuscular-wave)-model* defined by (1') for any specimen $\sigma(ms_{G,cw})$ of the studied microstate $ms_G$. And according to this model, *when the studied microstate involves inner quantum fields* *the central condition of extinction of all the fields that act upon the corpuscular-like singularity from the wave of the involved specimen of the microstate* ***cannot*** *be realized any more.*

This means that inside *[IQM-$QM_{HD}$]* the way of *conceiving* what can adequately be called 'one act of measurement of the observable ***A***', might in general *change* when the studied microstate involves inner quantum fields.

Nevertheless it remains quite conceivable a priori that for unbound microstates without inner quantum fields the form (20) of a coding rule might subsist and be coherently inserted in also the framework *[IQM-$QM_{HD}$]*, in agreement with the *G*-(corpuscular-wave)-model of a microstate. But in order to decide whether this is so indeed, or not, it seems necessary to better *understand intuitively* – just to thoroughly understand – on what grounds deeper than merely the above-mentioned algorithmic constructability of a one-to-one coding-rule with form (20) – the *BBGPM*-approach could have been led to admit implicitly a coding rule of this form for any observable ***A***. So we concentrate now upon discerning such grounds.

This brings us back to comparison with a classical mobile. The main obstacle in the way of such a comparison is that the formalism from $QM_{HD}$ does not distinguish between statistical descriptions and individual descriptions, the last ones are not even specified. Correlatively, in (19) the measurement-evolution-hamiltonian ***H(A)*** conserves only the *mean* value of the eigenvalues of ***A***. Whereas a coding relation of the form (20) for the result of one *individual* act of measurement on a microstate of one micro-system is quite essentially required to specify just *one* definite eigenvalue $a_j$ of ***A*** (accordingly also to the general concept of coding relation introduced in (1.I).2). In relation with this remark, in order to be able to obtain the sharpest possible comparability of (19) with a measurement-evolution of a classical mobile we should focus as strongly as possible the representation from (19) upon *one* eigenvalue $a_j$. For this aim let us make use of an artifice of thought, namely let us go upon the legal *limit* of the concept "$|\Psi_{G,\mathbf{H(A)}}(\mathbf{r},(t_1<t_f))>/\mathbf{A}$, $\forall j$, $\forall \mathbf{A}$, $t_o\leq t_1$" *by considering only* ***one*** *term from the sum (19)*: We require $(c_{j'}(t_1)=0$ *for* $\forall j'\neq j$, $|c_j(t_1)|=1$, $c_j(t_1)=1.exp(i\alpha(\mathbf{r},t)$, and so we have

$$|\Psi_{G,\mathbf{H(A)}}(\mathbf{r},(t_1<t_f))>/\mathbf{A}=(1.exp(i\alpha(\mathbf{r},t_1)|u(\mathbf{r},a_j)>)/\mathbf{A} \qquad (21)$$

where $\alpha(\mathbf{r},t)$ is an arbitrary phase-function. This – strictly speaking – does not lead outside the formalism of $QM_{HD}$, nor does it still belong to $QM_{HD}$ any more, it places just *upon* its frontier. *In (21) the representation (19) becomes 'tangent' to the core-formalism of $QM_{HD}$.*

According to $QM_{HD}$ the unique eigenvalue $a_j$ that is involved in (21) coincides with the mean value of the possible eigenvalues, so it is *itself* conserved by the measurement-evolution generated by ***H(A)***.

*Thereby – in a conceptual no man's land – we have finally extorted an individual qualification from the statistical formalism of $QM_{HD}$.*

However *what* ***changes*** in time in (19) – both mathematically *and physically* – is not yet clear.

Then let us go over into the 'physical'-space ***R***–representation of (21) where the considered measurement-evolution does play out inside the framework of our natural factual perceptions. In the physical space the ket from (19) *>/A* is conceived as a 'wave-*packet*' endowed with one maximum to which it is possible to associate an 'individual' dynamic comparable with that of a classical 'mobile'. *But with the limit-form (21) of (19) the corresponding* ***R****-representation does not entail any more any mechanical-like evolution because it ceases to be a wave-packet that possesses a moving maximum.* It simply is a globally stagnant plane wave. So, while the limiting form (21) itself of the ***A***-representation (19) has permitted to obtain an individual *insight* by having started with a statistical descriptor, this same limiting form (21) cuts off quite radically any comparability between the statistical descriptions from $QM_{HD}$ via wave-packets and the individual description of a mobile from the classical mechanics. This brings into evidence that – in general, and unsurprisingly – *it is* ***impossible*** *to pinpoint individual mechanical qualifications via the mechanical movement of the peak of a statistical wave-packet*. It explodes into the attention that any quantum-mechanical state-ket (19) $|\Psi_{G,\mathbf{H(A)}}(\mathbf{r},\ (t_1<t_f)>/\mathbf{A},\ t_1\geq t_o$, even if it is represented in the 'physical' space, consists of just a set of classified *numbers* each one of which symbolizes a result – with respect to a referential – of a position-measurement on one specimen $\sigma(ms_G)$ of the studied microstate $ms_G$ from a big set of such measurements. But the *one* physical position of *one* physical entity remains radically disconnected from such a statistic of numbers. This of course is well-known but also well forgotten [98].

While according to *[IQM-QM$_{HD}$]* the ***R***-localization of that what *can* produce observable marks – namely the singularity from the amplitude of the corpuscular wave of the one specimen $\sigma(ms_{G,cw})$ of the involved studied microstate – is on the contrary very localized in the physical space – at any time – but it lies on the level of *individual* conceptualization and so it remains exterior to the formalism of $QM_{HD}$.

Between the $QM_{HD}$ representation (19) and that what acts in the coding-procedure (20) there is a silenced semantic representational rupture, a split between mathematical wave-like representations of abstract statistics of numbers, and individual features of the physical individual entities that are involved, namely specimens $\sigma(ms_{G,cw})$ of the involved studied microstate, that refuse a mathematical representation of this sort. This clash is silenced by the fact that the formalism of $QM_{HD}$ either tends to absorb any individual entity or feature into a conceptually inadequate way, by placing it on the statistical level of conceptualization (think of the case of the eigenket in (II.6).1)), or it tolerates from these merely ghost-like verbal reflections upon the statistical representations.

This is a prelude to the "reduction problem".

Anyway, here we have touched the limit of the investigation that can be made inside $QM_{HD}$ in order to found a general coding rule of the type (20). And notwithstanding the loss in the ***R***-representation of (21), of any indication on an individual mechanical significance, the hints entailed by the preceding considerations suffice already for having conveyed a direct, intuitive, analogical understanding of the *way* in which inside *[IQM-QM$_{HD}$]* it *will* be possible to connect intelligibly the $QM_{HD}$ representation (19) of the ***A***-measurement-evolution-ket $|\Psi_{G,\mathbf{H(A)}}(\mathbf{r},\ (t_1<t_f)>/\mathbf{A},\ t_1\geq t_o$, with the *G*-(corpuscular-wave)-model of a microstate; namely by going over into the *IQM*-domain of individual representations of the specimens of the studied microstate. Indeed: going back now inside $QM_{HD}$, we can add that the space-time parameters that define a wave-packet are very adjustable. They permit to quite satisfactorily *approach* – via a statistical representation – the two essential features to be researched for the coding-purpose, namely:

***(a)*** Conservation of a mean value of the eigenvalue $a_j$ peaked around *one* value $a_j$ as strongly as one wants.

***(b)*** Choice of a definite direction and a convenient degree of stability for the dynamics of the maximum of the wave-packet throughout the time interval $(t\leq t_1<t_f)$, via the control of the 'external' fields involved by ***H(A)*** [99].

---

[98] This illustrates into what conceptual-descriptional impossibilities and mixtures and inadequacies is cornered an exclusively statistical representation by wave-functions – realized ***before*** an individual representation, because the approach has been developing ***top-down*** – when this representation is confronted to the purpose to generate ***later*** the individual physical entities and features on which any statistical description is necessarily ***founded***: a war between two opposed temporal orders bursts out. (We are in the domain of applicability of the theorem of Ehrenfest that "connects" $QM_{HD}$ with classical mechanics; but the "connection" is just a split-and-separation of the type represented in the Fig.6 by the double arrow on the statistical level.

[99] 'External' in the sense that they can be manipulated (generated, suppressed) from our macroscopic level.

***(c)*** The possibilities *(a)* and *(b)* permit to identify the optimal choices for the *locations of the registering devices* (this is what permits the method 'time-of-flight'). (Inside *classical* mechanics it is obvious that a mechanical displacement of a mobile throughout which the value $a_j$ of a given mechanical quantity $A$ keeps constant, leads the mobile into a *predictable* spatial domain $\boldsymbol{\Delta r}_j$ if the displacement lasts sufficiently; and that this domain can become as distant as one wants with respect to any other domain $\boldsymbol{\Delta r}_{j'}$ that corresponds to another value $a_{j'}$ of $A$, with $j'\neq j$ (the source-domain being the same). Which permits to mutually singularise these spatial domains).

So the $QM_{HD}$-parameters of the measurement-wave-packet $|\Psi_{G,\mathbf{H(A)}}(\mathbf{r},(t_1<t_f)>/\boldsymbol{A}$, $t_1\geq t_o$, from (19) assigned to each one measurement evolution of the studied microstate, can be adapted to the goal of insuring that the initial – *individual* – value $a_j$ of $\boldsymbol{A}$ from the corresponding specimen of this microstate, stays constantly, with *controllable* approximations, on the direction of displacement that reaches a more or less spatially extended registering device *where* ***any*** *impact means '$a_j$'*.

Together, all the preceding considerations permit to understand inside *[IQM-$QM_{HD}$]* on what an extraordinary sort of mixture of concepts can have taken form the – implicit – supposition that *any* quantum measurement does admit a coding rule of the form (20); namely that the statistical $QM_{HD}$ representation of a microstate *can* be brought to ***consist*** – wholly – of exclusively an evolution of a form of the type (20) [100].

However, in spite of the extraordinary character of this path, it leads us to *admit* here a priori – in agreement with *BBGPM* – the possibility of principle to transpose in *[IQM-$QM_{HD}$]*-terms the formulation of coding-measurement-evolutions of the formal type (20).

But we know in advance that the reasons that ground this acceptance apply ***exclusively*** to *unbound* microstates of *one* microsystem and with *simple* operation of generation, i.e. to microstates that evolve ***in the absence of quantum-fields*** that can work in an uncontrollable way on the 'singularity' from the involved specimen of the studied factual microstate. *Which is a very severe restriction*. We know from the start that we shall have to face the problem of defining also another coding procedure valid for microstates that involve quantum fields.

### (6.II).4.4. The major confusions from the $QM_{HD}$-representation of a measurement-evolution

We shall now concentrate upon the significances involved by the mathematical expression of the $QM_{HD}$-representation of quantum measurements. What follows might sometimes seem much too repetitive and trivial. But finally, we hope, this feeling will fade out and leave place to an opposed feeling of full elucidation of questions that stay unsolved since a very long time precisely because nobody dares to be sufficiently repetitive and trivial for certainly covering the lacunae in the personal way of wondering and of understanding of any potential reader. Only an extravagantly explicit, exhaustive and redundant analysis can dislodge sediments of ways of thinking and of expressing that have been installed by more than a century of randomly local conceptual analyses, of formal research, and of *teaching*.

---

[100] What a prowess we have finally accomplished! The only way to find answer to *a basic* question of physics – how to know the meaning of the observable result of an act of quantum-measurement inside the fundamental 'quantum mechanics' – has been to do, ***what***? ***(a)*** To laboriously unmask individual intruders $\{|u(\mathbf{r},a_j,)>, \forall j\}$ injected into an abstract statistic of mere *numbers* $\{|\, c_j(a_j,t_1)\,|^2, \forall j\}_A$ drawn from the set of all such statistics of which a $QM_{HD}$-state-ket $\Psi_G$ – by itself – does *entirely* consist. ***(b)*** And therefrom, by use of faint reflections, by these individual intruders $|u(\mathbf{r},a_j,)>$, of aspects of physical individual entities and features, to draw – at distance – on the level of the classical representation of the dynamics of another sort of individual real physical entities called 'mobiles', *a virtual trace* that is alike to the shadow of a 'mobile'. (The convolutions from this last verbal expression only translate the meanders of the reasoning that has permitted to knock out the conclusion formulated above). *This conveys a feeling of the distance introduced between meaning and representation by the top-down constructed statistical formalism from $QM_{HD}$*. And let us recall that – via the expansion-postulate – this concept is quite essentially tied with the reference-concept of *'eigenstate' – an* ***a-temporal*** *sample of arbitrary spatial extension* – of the wave-movement in the neighbourhood of the singularity from a de Broglie corpuscular-like wave.

### *(6.II).4.4.1. Preliminary recall*

The evolving *state-ket* $|\psi_{G,H}(\mathbf{r},t)>$ of the studied microstate $ms_G$ is supposed for the moment to be always specifiable by use of the Schrödinger equation of the problem, so directly in mathematical terms, if the initial form $|\psi_{G,H}(\mathbf{r},t_o)>$ is given.

- $|\psi_{G,H}(\mathbf{r},t)>$ works as a reservoir of all the *potentially* available predictive-numbers that $QM_{HD}$ can offer via algorithms concerning the studied microstate $ms_G$; namely the probabilities of results of measurements of any observable $\boldsymbol{A}$, performed on $ms_G$ at any time $t \geq t_o$ after the moment $t_o$ when the initial form $|\psi_{G,H}(\mathbf{r},t_o)>$ of $|\psi_{G,H}(\mathbf{r},t)>$ has been specified. The descriptor $|\psi_{G,H}(\mathbf{r},t)>$ has an only *potential* content that remains to be partially worked out according to particular predictive purposes. Its outputs are purely abstract, numerical and statistical.

- The expansion (18) $|\psi_{G,\mathbf{H}}(\mathbf{r},t_1)>/\boldsymbol{A}=\sum_j c_j(a_j,t_1)|u(\mathbf{r},a_j)>$ of $|\psi_{G,H}(\mathbf{r},t)>$ on the basis of eigenket $\{|u(\mathbf{r},a_j>, \forall j\}_A$ for $\forall j$, and for any time $t_1 \geq t_o$ and any given observable $\boldsymbol{A}$, permits to explicate from $|\psi_{G,H}(\mathbf{r},t)>$ – via Born's postulate and for any moment $t_1 \geq t_o$ and any observable $\boldsymbol{A}$ – the predictive probability law denoted $(\boldsymbol{A},\{|c_j(a_j,t_1)|^2, \forall j\})$ on the emergence by measurement at the time $t_1$ of the eigenvalues $a_j$ of $\boldsymbol{A}$. This permits to *verify* this law via a subsequent sufficiently long succession of effective realizations of coding-measurement-evolutions for $\boldsymbol{A}$. So (18) permits to express the *total* predictive content of $|\psi_{G,\mathbf{H}}(\mathbf{r},t_1)>$ in the following more explicit form:

$$|\psi_{G,\mathbf{H}}(\mathbf{r},t_1)> \approx_{\text{pred.}} \{\forall \boldsymbol{A},\ \forall t_1 \geq t_o,\ |\psi_{G,\mathbf{H}}(\mathbf{r},t_1)>/\boldsymbol{A}\ \} \qquad (22)$$

where "$\approx_{\text{pred.}}$" is to be read "equivalent in prediction".

- In (18) and so in (22) the particular amount of numerical predictive content from $|\psi_{G,H}(\mathbf{r},t)>$ that is tied with any *given* observable $\boldsymbol{A}$ and any given time $t_1 \geq t_o$, ceases to be potential, it is explicated by already achieved calculi. Thereby, though they remain strictly abstract, the writings (18) and (22) literally ex-presses now the whole essence of the $QM_{HD}$ predictive algorithm, namely: The equation $\boldsymbol{A}|u_j(\mathbf{r},a_j)>=a_j|u_j(\mathbf{r},a_j)>, \forall j$ determines the eigenket in connection with the algebra of observables defined inside $QM_{HD}$; the concept of eigenstate of an observable $\boldsymbol{A}$ plays the role of a sample of wave-movement around a singularity from a corpuscular wave (6.II)1); so de Broglie's 'corpuscular-wave' model is made use of; and finally, Born's postulate asserts that $\pi(a_j)=|c_j(a_j,t_1)|^2$.

*The concept (22) is the explicit core of the $QM_{HD}$ predictive representation.*

- Consider now the second member of the coding-*measurement*-evolution ket (19) $|\Psi_{G,\mathbf{H}(A)}(\mathbf{r},(t_1<t_f))>/\boldsymbol{A}=(\sum_j c_j(t_1)|u(\mathbf{r},a_j)>)_{G,\mathbf{H}(A)}$, with $t_1 \geq t_o$. It has the same expansion-form with respect to the state-ket $|\psi_{G,\mathbf{H}}(\mathbf{r},t_1)>$ and the observable $\boldsymbol{A}$ as in (18) $|\psi_{G,\mathbf{H}}(\mathbf{r},t_1)>/\boldsymbol{A}$, but that – from the time $t_1$ when one act of measurement-evolution begins and up to the time $t_f$ when this one act of measurement-evolution finishes – *continues* the evolution from (18) *in the coding-conditions defined by a measurement-Hamiltonian* ***H(A)*** *that commutes with* $\boldsymbol{A}$. So, as stressed before:

> The $QM_{HD}$ descriptor of any ***one*** coding-measurement-evolution is a ***statistical*** descriptor.

But this statistical descriptor is posited to finish its evolution by a 'reduction' of its statistical character that reveals the *individual* result $a_j$ that the *one* considered act of an $\boldsymbol{A}$-measurement-evolution has actualized out from the whole spectrum $\{a_j\}, \forall j$ of a priori possible results $a_j$.

*Thereby, from the beginning, the descriptor (19) is flawed by an ambiguous character in what concerns the level of conceptualization on which it is placed.*

- Let us finally notice also that the three descriptors $|\psi_{G,\mathbf{H}}(\mathbf{r},t)>$ with $t \geq t_o$, $|\psi_{G,\mathbf{H}}(\mathbf{r},t_1)>/\mathbf{A}$ and $|\Psi_{G,\mathbf{H}(A)}(\mathbf{r}, (t_1<t_f))>/\mathbf{A}$ designate meanings of very different natures, in particular in what is tied with the involved *time*-parameters. So we stay attentive to the involved time parameters.

### *(6.II).4.4.2. Critical remarks and questions*

From now on we adopt the point of view of *[IQM-QM$_{HD}$]*.

Consider the descriptor (19) $|\Psi_{G,\mathbf{H}(A)}(\mathbf{r},t_1<t_f)>/\mathbf{A}=\sum_j c_j(t_1)|u(\mathbf{r},a_j)>, \forall j,\ t_1 \geq t_o$, of a coding measurement-evolution. It is often said that "at $t_1 \geq t_o$ the 'system' is 'prepared' for measurement, and correspondingly is also 'prepared' the new ket $|\Psi_{G,\mathbf{H}(A)}(\mathbf{r},t_1< t_f)>/\mathbf{A}$, $t_1 \geq t_o$, that represents the measurement-evolution". What happens *factually* during these 'preparations' is neither represented formally nor explicitly stated by consensually established ways of using words. In particular ***it is not said how the "system" is OBTAINED, physically, operationally, in order to (then? thereby?) "prepare" it***. Nor is it explicitly stated how is to be carried out the factual ***coding***-measurement-evolution *itself* for – specifically – a given observable ***A*** and a given sort of microstate. It is only specified that the Hamiltonian must commute with ***A***, but on the physical coding-process there are no other indications than the directly *postulated* assertion that each act of ***A***-measurement produces, for *any* 'system' (i.e. microstate? specimen of a microstate?), an eigenvalue $a_j$ of ***A*** that is associated with the corresponding eigenket $|u(\mathbf{r},a_j)>$ and – just like in von Neumann's unacceptable representation – is "indicated by the 'needles' of the registering devices". Everything hovers calmly exclusively inside the mathematical-verbal spheres. No particular stress whatever is placed upon the fact that, in order for $a_j$ to become ***known***, and since obviously a statistic of abstract numbers cannot itself trigger physical marks by interaction with physical registering devices, one should be informed how the physical interaction has to be organized in order that the 'needles' inform us that they indicate this or that definite $a_j$ and not another $a_{j'} \neq a_j$, or even something else. While in fact the written expression of the descriptor $|\Psi_{G,\mathbf{H}(A)}(\mathbf{r},(t_1<t_f)>/\mathbf{A}=(\sum_j c_j(t_1)|u(\mathbf{r},a_j>)_{G,\mathbf{H}(A)}$ with $t_1 \geq t_o$, places us on the statistical level of conceptualization, as just *a continuation of the expansion (18)* in new external conditions expressed by the measurement-Hamiltonian ***H(A)*** instead of the initially acting Hamiltonian ***H***. And just like in the case of von Neumann's unreasonable representation of quantum measurements, here also is totally occulted the question of the procedure to follow in order to register observable physical marks $\{\mu_{kA}\}_j$, $k= 1,2,...n$ from which it be then possible *for **us** to **construct** the eigenvalue* $a_j$ that is to be regarded as the result of the considered act of coding-measurement-evolution.

Even the concept itself of 'coding'-measurement-evolution – that in *IQM* is central – is devoid inside *QM$_{HD}$* of a defined and general equivalent, notwithstanding that *precisely the requirement of coding the registered observable marks $\{\mu_{kA}\}_j$, k= 1,2,...n in terms of one definite eigenvalue $a_j$ of the observable A determines entirely that what can be called an act of MesA, because the specification of the genesis and meaning of these marks constitute **all** that permits the coding*, as it appeared clearly in (6.II).4.3.

The method time-of-flight studied by Margenau and Cohen, and Bohm's analysis of the Stern-Gerlach procedure have not been followed by an explicit conclusion formulated in *general* terms. In the text-books it is only asserted that when $a_j$ 'is obtained' this 'fact' is accompanied by a ‘reduction’ to only one of its terms, of the measurement-ket (19) $|\Psi_{G,\mathbf{H}(A)}(\mathbf{r}(t_1<t_f))>/\mathbf{A}=(\sum_j c_j(t_1)|u(\mathbf{r},a_j>)_{G,\mathbf{H}(A)}$.

*But the above-mentioned 'fact' – the **postulated** outcome of '$a_j$' and the correlative reduction of (19) – is a **formal** fact called 'the position of the needle of the apparatus' that, as such, is ill defined and of which the physical correspondent remains simply non-specified, while the mathematical status of a $QM_{HD}$ 'operator' is abusively asserted for it.*

Let us continue. The identification between mathematical writings and physical facts has become so deep in the nowadays physicist's minds, and so perfect, that *what is formal finally banishes what is factual*. Indeed strictly nothing is specified concerning what happens physically to the necessarily involved *one **individual** specimen* of the studied microstate while the formal ***statistical*** evolution symbolized in (19) goes on. Namely that in general, at the time when $a_j$ 'is obtained' – which in *IQM*-language means when $a_j$ is *constructed* conceptually from:

- the mathematical representation of ***A*** as a function ***A(R,P)***;
- *a model posited for $\sigma(ms_G)$*;
- the *way* induced *by this model*, of registering *relevant* physical marks $\{(\mu_{kA}\}_j$ ;
- the physical marks $\{\mu_{kA}\}_j$, $k= 1,2,...n$.

At that time *the physical **state** of the involved specimen is usually destroyed*, even if the involved ***system*** still subsists accessibly [101].

All this, though in a certain sense it is well known by many physicists (mainly by experimentalists) is not treated inside $QM_{HD}$. Consequently, the necessity, in general, to generate *another* specimen of the studied microstate before entering upon a new measurement-evolution (19), does not in the least trouble the attention. So the physical operation of generation *G* of a physical and individual specimen $\sigma(ms_G)$ that somehow *must* be realized for any factual measurement-evolution, remains entirely hidden in the void of *factual* individual meaning of the verbal expressions 'preparation of the system' and 'preparation of the measurement-evolution-ket'.

***The basic concept of operation of generation G is not formed and named in $QM_{HD}$.***

In these conditions the unavoidable and obvious necessity to make use of a clearly individual representation of measurement-successions *[G.MesA]* for *verifying* the mathematically obtained statistical predictions, remains very obscure inside $QM_{HD}$, precisely because the predictions are obtained mathematically and top-down, in contradistinction to what happens inside *IQM* where they are constructed factually and bottom-up.

As for the ***possibility*** to make use of individual measurement-successions *[G.MesA]* that involve ***coding***-measurement-evolutions *MesA **also*** for *constructing **factually** any statistical **prediction***, as it is done inside *IQM*, this does not even appear on the far horizon of $QM_{HD}$[102].

In short, all the individual descriptors *(G* from (1), $G_t$ from (13'), $ms_G$ from (1), $\sigma(ms_G)$ from (1), *MesA, [G.MesA], [$G_t$.MesA], etc.)* that inside *IQM* are defined and mutually distinguished, not only are not represented in the mathematical formalism of $QM_{HD}$, but moreover, in the merely verbal expressions that accompany the mathematical $QM_{HD}$-representations, *they act without being defined*, in a conceptually lacunar and chaotic way, intermittent and uncontrolled. And so a thick semantic mud has banked up. The Fig.6 is *strictly, radically* valid.

---

[101] Which is one of the reasons that led us to speak inside *IQM* in terms of micro-*states* of micro-*systems*, *not* directly in terms of 'systems', in order to stay entirely clear. Correlatively the concept of "successive measurements" has gained much attention and it has been variously represented in more or less ideal-forms that float high above genuine physical operability because they *neglect* the destruction, in general, of the *state*, even if the system subsists.

[102] It is occulted by the fact that the basic descriptor $|\psi_{G,H}(t)\rangle$ is posited to be always available and to offer the possibility to perform any expansion (18) $|\psi_{G,H}(\mathbf{r},t_1)\rangle/A$, which generates directly the probability-law to be verified via exclusively mathematical means.

### *(6.II).4.4.3. The reduction problem*

Let us now focus upon the postulated 'reduction' of the statistical $QM_{HD}$ descriptor (19) $|\Psi_{G,\mathbf{H(A)}}(\mathbf{r},(t_1<t_f))>/\mathbf{A}=(\sum_j c_j(t_1)|u(\mathbf{r},a_j>)_{G,\mathbf{H(A)}}$, $t_1\geq t_o$, $\forall\mathbf{A}$, $\forall j$, at the final moment $t_f$.

For self-sufficiency of the present argument let us recall again that some authors have asserted that from a general conceptual viewpoint this 'reduction' does not seem unacceptable when one thinks of the general formal representation of the calculus of probabilities. In this assertion the existence of a precedent example acts as a legalizing circumstance. But the two conceptual situations are different. It is true that in the representation of probabilities also each one realization of the involved 'experiment' that generates by repetition the whole universe $U=\{(e_j, j=1,2,...J)\}$ of possible elementary outcomes $e_j$, actualizes only one outcome from $U$, which *can* indeed be indicated verbally in a loose way by saying that "each realization of this experiment 'reduces' the permanent and a priori global potentiality $U$, to only one actualized elementary event $e_j$". But inside quantum mechanics the 'reduction' of the $QM_{HD}$-descriptor $|\Psi_{G,\mathbf{H(A)}}(\mathbf{r},(t_1<t_f))>/\mathbf{A}$, $t_1\geq t_o$, is not compatible with the mathematical rules imposed by a linear formalism, whereas in the case of general probabilistic writings no incompatibility of this sort is involved.

Furthermore, a mathematical inadequacy usually reflects some semantic impossibility[103]. In the present case any physicist can *feel* that it is conceptually inadequate *to **express** one actual individual outcome of one given eigenvalue $a_j$, by the help of an irregular behaviour of the mathematical representation of the whole a priori set of potential individual outcomes*[104].

But beyond any matter of mathematical regularity or non-regularity, what remains most mysterious is this:

*Why should one **DESIRE*** that [a physical individual procedure of measurement performed on an individual physical specimen of a physical entity-to-be-studied in order to verify via repetitions of this procedure a statistical prediction concerning this sort of physical entity] *be represented by a mathematical statistical descriptor*? Why should one *want* to get entangled in such a knot?

This strange attitude has been brought into evidence so repetitively that it vanishes by trivialization. But if nevertheless one succeeds to stop a sufficiently long moment focusing genuine attention upon it, all of a sudden, like in certain optical illusions, a surprising sort of summarizing explanation leaps to one's eyes as an obviousness: Just because historically the statistical descriptor became available first, and then for a very long time it stayed the *unique **available*** conceptual resource for formal expression; and so it has been tried again and again to express *by this available descriptor* not only what has to be verified, but also the process of verification *itself.*

The $QM_{HD}$ measurement-evolution-ket $|\Psi_{G,\mathbf{H(A)}}(\mathbf{r},(t_1<t_f))>/\mathbf{A}=(\sum_j c_j(t_1)|u(\mathbf{r},a_j>)_{G,\mathbf{H(A)}}$, $\forall j$, $t_1\geq t_o$, with the 'reductions' that it requires, is just a desperate aborted attempt at crowding inside a unique available mathematical statistical descriptor all that is mentioned below:

***(a)*** On the one hand, the representation of ***all** the very numerous*, physical, actual, *individual* coding-measurement-successions *[G<sub></sub>$_t$.MesA]* defined in (3.I).4, each *one* of which ends with the individual factual, actual registration of its own result that consists of a certain group $\{\mu_{kA}\}_j$, $k=1,2,...n$, of *physical* marks.

---

[103] This circumstance is very remarkable: the definition of a formal mathematical system (axioms, well-formed expressions and transformation rules, etc.) always introduces inexplicitly a strictly defined class of semantic structures, which constrains the acceptable meanings.

[104] Nevertheless the distribution of the reactions to the mentioned formal situation pushes into satirical hypotheses: In certain particularly tolerant minds such an expression might suggest that it also is *'a fact'* that the statistic *itself* generates *physically* the individual outcomes, so that the only thing that remains to be done is to find out ***what** physical individual outcome* has been produced the mathematically irregular behaviour posited for the statistical representation? Who knows? Concerning $QM_{HD}$ anything succeeds to seem conceivable. Think of Everett's infinity of parallel universes that is asserted by some to be 'really' generated *by each 'reduction' of a mathematical writing on a sheet of paper*. In such views the in-distinction between formal descriptors written on paper or screens, and physical facts, reaches a sort of metaphysical grandeur. This ***is*** a "*fact*" indeed, a psycho-social one.

And on the other hand ***also***

***(b)*** The ***unique***, globalized, statistical distribution of the set $\{|c_j(t_1,a_j)|^2\}, \forall A, \forall j$ of all the $QM_{HD}$ predictive *pure numbers*, each one of which – i.e. with $A$ and $j$ both *fixed* – is the cardinal (and also the posited probability) of a class of outcomes of an a priori possible numerical eigenvalue $a_j$ of a quantum mechanical observable $A$ that, itself, is just the coded meaning assigned to a corresponding group of physical marks $\{\mu_{kA}\}_j$, $k=1,2,...n$.

All this has been crowded inside the statistical descriptor (19). At the time when the "theory of quantum measurements" has been formulated this descriptor had taken form in the free imagination of von Neumann, possibly because of the formal analogy with the Schrödinger state-ket assigned to the studied microstate. Anyhow, the analogy benefitted of the prestige of two famous mathematicians and it triggered an effect of familiarity. One can imagine that this permitted to the descriptor (19) to install roots into the minds of physicists that were still too astounded by the new theory that was emerging for being able of a genuine critical opposition.

Such a desperate attempt however quite certainly involves gross confusions. And from the standpoint of *IQM* the main confusion certainly concerns the *time-parameters* that come in. Below we bring this under a magnifying glass, hoping to finally pinpoint the first conceptual source of the fallacy involved in the reduction problem.

Let us begin by recalling all the descriptive elements that are involved in the $QM_{HD}$ representation of the quantum-measurements and by introducing more explicitly written time-symbols, with upper indexes that denote the level of conceptualization on which they are made use of: the upper index '*i*' means individual and '*s*' means statistical, so we write $t^{(i)}$ or $t^{(s)}$.

Consider now first the following $QM_{HD}$ chain of statistical descriptors:

$$[|\psi_{G,\mathbf{H}}(\mathbf{r},t^{(s)}_o)> \rightarrow |\psi_{G,\mathbf{H}}(\mathbf{r},t^{(s)})>,\ t^{(s)} \geq t^{(s)}_o \rightarrow |\psi_{G,\mathbf{H}}(\mathbf{r},t^{(s)}_1)>/A,\ t^{(s)} = t^{(s)}_1 \rightarrow |\Psi_{G,\mathbf{H}(A)}(\mathbf{r},\ t^{(?)}_1 < t^{(?)}_f)>/A \quad (ch)(QM_{HD})$$

that is involved by a given problem of measurement.

* The $QM_{HD}$-time-parameter $t^{(s)}$ from $|\psi_{G,\mathbf{H}}(\mathbf{r},t^{(s)})>$ is tied with a Schrödinger-evolution of the initial implicit statistical content of the state-ket $|\psi_{G,\mathbf{H}}(\mathbf{r},t^{(s)}_o)>$.

* The $QM_{HD}$-time-parameter $t^{(s)}_1$ from $|\psi_{G,\mathbf{H}}(\mathbf{r},\ t^{(s)}_1))>/A$ – $\forall A$ – indicates all the explicit predictive ***statistical distributions*** $\{|c_j(t_1,a_j)|^2\}$, $\forall j$, of the numerical results $a_j$ of $A$-measurements that are involved by $|\psi_{G,\mathbf{H}}(\mathbf{r},\ t^{(s)}_1))>$ such as these are realized at the moment $t^{(s)}_1$ *of **their** Schrödinger-evolution*, and each one of these $A$-statistics being explicitly singularized by a corresponding expansion $|\psi_{G,\mathbf{H}}(\mathbf{r},t_1))>/A$; $t^{(s)}_1$ *indicates only this, nothing else.*

* The measurement-evolution descriptor (19) $|\Psi_{G,\mathbf{H}(A)}(\mathbf{r},(t^{(?)}_1<t^{(?)}_f))>/A$, that *corresponds* to one given expansion $|\psi_{G,\mathbf{H}}(\mathbf{r},t_1)>/A$, introduces the two time-parameters $t_1$ and $t_f$ – written here as $t^{(?)}_1$ and $t^{(?)}_f$ because their implication concerning the level of conceptualization on which they act, is confuse: Verbally, the time-value $t_1$ from (ch) is indicated as "the time when a measurement-evolution begins". *But **what** begins, exactly*?

(1) Does a "*Schrödinger*-measurement-evolution" of the *whole statistic* $\{|c_j(t_1,a_j)|^2\}, \forall j$ *from* $|\psi_{G,\mathbf{H}}(r,t_1)>/A$, *of results of A-measurements* begin at a ***statistical*** time-moment $t_1=t^{(s)}_1$, and then continue inside its statistical time until this Schrödinger-measurement-evolution of the whole statistic $\{|c_j(t_1,a_j)|^2\}, \forall j$ finishes at its ***statistical*** time $t_f=t^{(s)}_f$?

(2) Or does *one **individual** act of measurement-evolution MesA* begin at an ***individual*** time $t^{(i)}_1$, and then finish at the subsequent ***individual*** time-value $t^{(i)}_f$? If this is so then $t_1 \neq t^{(s)}_1$, and $t_f \neq t^{(s)}_f$ and this entails that the global symbol $|\Psi_{G,H(A)}(\mathbf{r}, t^{(?)}_1 < t^{(?)}_f)\rangle/A$ points toward a hybrid self-contradicting unconceivable concept written in the form of a statistical Schrödinger ket but that contains individual time-parameters in it.

Though the answers seem already obvious, we nevertheless detail this point even more, via questions:

(3) Consider the chain *(ch)(QM<sub>HD</sub>)*.

* ***Can we write the identity*** $t^{(s)}_1 \equiv t^{(?)}_1$ for
[the time-parameter $t^{(s)}_1$ from the third link, "$|\psi_{G,\mathbf{H}}(r,t^{(s)}_1)\rangle/A$ with $t^{(s)}=t^{(s)}_1$"]
and
[the time-parameter $t^{(?)}_1$ from the fourth link, "$|\Psi_{G,\mathbf{H}(A)}(\mathbf{r}, t^{(?)}_1 < t^{(?)}_f)\rangle/A$"]?

* ***Can we write the inequality*** $(t^{(?)}_1 \geq t^{(s)}_o)$ for
[the time-parameter $t^{(?)}_1$ from the fourth link, $|$"$\Psi_{G,\mathbf{H}(A)}(\mathbf{r},(t^{(?)}_1 < t^{(?)}_f)\rangle/A$"?
and
[the time-parameter $t^{(s)}_o$ from the first link "$|\psi_{G,\mathbf{H}}(r,t^{(s)}_o)\rangle$"]

* Are the ***natures*** of the symbols $t^{(s)}_o$, $t^{(s)}_1$, $t^{(?)}_1$, $t^{(?)}_f$, mutually identical, or not?

* ***Does the time-parameter*** $t^{(s)}_o$ ***from the first link*** "$|\psi_{G,H}(r,t^{(s)}_o)\rangle$" ***possess some individual "equivalent"*** $t^{(i)}_o$ ***that admit the assertion of the identity*** $t^{(i)}_o \equiv t^{(s)}_o$ ***?***

These are crucial questions, for if the answers to them are 'yes' then the requirement of homogeneity of nature is satisfied, like it has to be is in any equation. But if the answer is 'no', even for only one of these questions, then a problem of semantic homogeneity concerning the $QM_{HD}$-representation of the measurements becomes perceptible inside the framework $[IQM\text{-}QM_{HD}]$. And history has taught us how important it is for the mathematics of Physics to avoid any semantic inhomogeneity.

So let us explore the questions formulated above.

This requires extending explicitly the analysis to also the individual level of conceptualization as it is specified inside *IQM*, and then to construct a comparison.

So consider now the individual level of conceptualization from *IQM*. As announced we re-write *all* the descriptive elements from this level with an upper index '*i*'. Furthermore let us write explicitly all the time parameters, *as arguments*. Re-written in this striking way the *IQM*-descriptors involved by the individual acts of measurement are:

* The operation of generation (1) $G^{(i)}(t^{(i)}_o)$.

* The generalization of (1) $G^{(i)}(t^{(i)}_o)$ defined in (13'):

$$(13')\quad G^{(i)}(t^{(i)}_o)\,.\,(t^{(i)}-t^{i)}_o) \equiv G^{i)}(t^{(i)})\Big(G^{(i)}(t^{(i)}_o),\ EC(t^{(i)}-t^{(i)}_o)\Big),\quad \text{with } t^{(i)} \geq t^{(i)}_o$$

* The corresponding generalized concept of a coding-measurement-succession defined in (3.I).4.:

$$[G^{(i)}(t^{(i)}_1)\,.\,MesA(t^{(i)}_f - t^{(i)}_1)] = \Big[\big(G^{(i)}(t^{(i)}_o), EC(t^{(i)}_1 - t^{(i)}_o)\big)\,.\,MesA(t^{(i)}_f - t^{(i)}_1)\Big]$$

where $G^{(i)}(t^{(i)}_1)$ is drawn from (13') by setting $t^{(i)} = t^{(i)}_1$ where $t^{(i)}_1$ is the time when the act of *A*-measurement *MesA* begins, and $t^{(i)}_f$ is the time when it finishes.

Inside *IQM* the notations employed above possess primary, clear and self-sufficient meanings that are induced by the bottom-up constructive approach and by the fact that each time-parameter is tied with a definite, individual, physical entity (an operation or a conceptually specified physical entity like $ms_G$ or $\sigma(ms_G)$; and *by construction* these individual time-parameters do admit the obviously semantically homogeneous order ($t^{(i)}_o \leq t^{(i)}_1 \leq t^{(i)}_f$).

We now will try to spell out fully and exactly why inside the statistical chain (ch)($QM_{HD}$) the ordering of the involved time-parameters ($t^{(s)}_o$, $(t^{(s)}_1, t^{(?)}_1)$, $t^{(?)}_f$) raises questions.

Let us begin by stressing that in the statistical chain *(ch)*($QM_{HD}$) the time-parameter $t^{(s)}_o$ is an abstract time-parameter that has a purely algorithmic role: *It qualifies the initial form* $|\psi_{G,\mathbf{H}}(\mathbf{r},t^{(s)}_o)>$ *assigned to the statistical state-ket* $|\psi_{G,\mathbf{H}}(\mathbf{r},t^{(s)})>$ *associated to the studied microstate*. This permits to just singularize the state-ket $|\psi_{G,\mathbf{H}}(\mathbf{r},t^{(s)})>$ inside the general infinite set of solutions admitted by the Schrödinger equation of the problem, *without* tying its initial form with some definite physical event. So :

> Both semantically and factually – the two time-parameters $t^{i)}_o$ and $t^{(s)}_o)$ are radically disconnected from one another. And the same holds concerning $t^{i)}_1$ and $t^{(s)}_1)$.

But nothing interdicts to connect them *conventionally*, as follows. Below on the Fig.7 we introduce a graphic representation that contains a meta-temporal-dimension-of-comparison denoted *mtdc*, and we just require that the projections on *mtdc* of $t^{i)}_o$ and $t^{(s)}_o$ shall coincide, as well as the projections of $t^{i)}_1$ and $t^{(s)}_1$. Thereby we establish two conventional correspondences $t^{i)}_o \leftrightarrow t^{(s)}_o$ and $t^{i)}_1 \leftrightarrow t^{(s)}_1)$ that facilitate comparability while respecting the non-suppressible difference of nature between time-parameters $t^{i)}$ and $t^{(s)}$.

**bottom-up**

**CONCLUSION**

**In consequence of the difference of *semantic nature***
**between the contents from the statistical level of conceptualization of the microstates**
**and the contents from their individual level of conceptualization,**
**the time parameters $t^{(i)}_o$, $t^{(i)}_1$ and $t^{(i)}_f$ tied with individual meanings**
***CANNOT* COEXIST *FACTUALLY ON THE* STATISTICAL *LEVEL***
**with time parameters like $t^{(s)}_o$, $t^{(s)}_1$ that are tied with statistical meanings:**
**$t^{(?)}_1$ and $t^{(?)}_f$ from$|\Psi_{G,H(A)}>/A$ are just *fictitious* time parameters imagined via an implicit unexpressed operation of projection on a meta-temporal-dimension of comparison, for establishing a mental framework for correspondences.**

**for the projections on *mtdc* we have BY POSITED CONVENTION :**

$$proj_{mtdc}\ t^{(s)}_o \text{ from } [|\psi_{G,H}(r,t^{(s)}_o)> \equiv proj_{mtdc}\ t^{(i)}_o \text{ from } G_o$$

$$proj_{mtdc}\ t^{(s)}_1 \text{ from } |\Psi_{G,H(A)}>/A \equiv proj_{mtdc}\ t^{(i)}_1 \text{ from } G \text{ and } MesA$$

$$proj_{mtdc}\ t^{(s)}_f \text{ from } |\Psi_{G,H(A)}>/A \equiv proj_{mtdc}\ t^{(i)}_f \text{ from } G \text{ and } MesA$$

***mtdc*: a meta-temporal dimension of comparison between the statistical $QM_{HD}$ times and the *IQM* individual times**

$[|\psi_{G,H}(r,t^{(s)}_o)> - |\psi_{G,H}(r,t^{(s)}_1)> - |\psi_{G,H}(r,t^{(s)}_1)>/A - |\Psi_{G,H(A)}(r, (t^{(s)} \leq t^{(?)}_1 < t^{(?)}_f))>/A$ **(ch)([*IQM-QM$_{HD}$*)]**

**statistical level of conceptualization**

$[G(t^{(i)}_1).\ MesA(t^{(i)}_f - t^{(i)}_1)]],\ \forall A$ **with** $(t^{(i)}_o \leq t^{(i)}_1 < t^{(i)}_f)$

**individual level of conceptualization with :**

$G_o(t^{(i)}_o);\quad G_o(t^{(i)}_o)\ .\ (t^{(i)}_1 - t^{(i)}_o) \equiv G(t^{(i)}_1)(G(t^{(i)}_o), EC, (t^{(i)}_1 - t^{(i)}_o));\quad t^{(i)}_o \leq t^{(i)}_1 < t^{(i)}_f$

Fig.7. *The source-fallacy in the "theory of quantum measurements".*
**This figure is to be browsed bottom-up.**

With the Fig.7 in view consider what follows:

According to *IQM* the verification of the predictions from any given statistical chain $(ch)(QM_{HD})$ is realized via

[many repetitions of [a very long sequence of [repetitions of the coding-measurement-succession $[(G^{(i)}(t^{(i)}_o), EC(t^{(i)}_1 - t^{(i)}_o))\ .\ MesA(t^{(i)}_f - t^{(i)}_1)]$ that corresponds to that chain.]]]. Now,

If the conventional correspondences defined above – $t^{i)}_o \leftrightarrow t^{(s)}_o$, $t^{i)}_1 \leftrightarrow t^{(s)}_1)$, $t^{i)}_f \leftrightarrow t^{(s)}_f)$ – are applied, then *each **one** realization of the coding-measurement-succession*

$$[(G^{(i)}(t^{(i)}_o), EC(t^{(i)}_1 - t^{(i)}_o)).\ MesA(t^{(i)}_f - t^{(i)}_1)]$$

*runs beneath the **whole** statistical chain $(ch)(QM_{HD})$.*

But it does this inside its *own* specific, factual and individual temporal universe. And each such succession *carries in it **one** physical specimen* $\sigma(ms_G)$ *of the studied microstate* $ms_G$, from the moment $t^{(i)}_o$ when this specimen has been *factually* generated by the realization of the operation $(G^{(i)}(t^{(i)}_o), EC(t^{(i)}_1 - t^{(i)}_o)$[105]$)$ of generation, and until the final moment $t^{(i)}_f$ when this specimen $\sigma(ms_G)$ has been *factually* qualified via a group of physical and observable coding-marks $\{\mu_{kAj}\}_j$, $k=1,2,...n$, that are then *decoded* in terms of one eigenvalue $a_j$.

Thereby this individual act of measurement triggers its *own* contribution of observable marks that codes for just ***one*** given eigenvalue $a_j$.

This restores explicitly the quite obvious and trivial fact that the one *physical* specimen $\sigma(ms_G)$ carried by each one individual *physical* coding-measurement-evolution triggers upon physical devices one group of *physical* observable marks $\{\mu_{kAj}\}_j$, $k=1,2,...n$, that are then coded in terms of one eigenvalue $a_j$, of which the realization is immediately counted as *a contribution of just one unity to the as yet unknown final number* that will verify factually – or ***contest*** factually – the statistical $QM_{HD}$-predictive abstract *total* number $|c_j(t_1)|^2$ of outcomes of the value $a_j$ of $\boldsymbol{A}$ that is asserted by the term from the mathematically obtained expansion $|\psi_{G,\boldsymbol{H}}(\boldsymbol{r},t_1)>/\boldsymbol{A}$ from (ch) that is there distinguished by the index '$j$'.

No "parallel universes" are necessary for this.

And:

***This is "the needle of the apparatus",*** *nothing else.*

The von Neumann "operator of needle-position" is just nonsense.

One *common* formal container for – both – the statistical time-parameters from $(ch)(QM_{HD})$ and the individual time-parameters involved by one coding-measurement-succession $[(G^{(i)}(t^{(i)}_o), EC(t^{(i)}_1 - t^{(i)}_o))$**.** $MesA(t^{(i)}_f - t^{(i)}_1)]$ *cannot be even only **conceived***: the operational-conceptual process exposed in *IQM* by which a statistical representation of the studied microstate emerges by repeated realizations of a very long sequence of repetitions of this factual succession, is such that the individual physical content from *one* realization of the succession is entirely eliminated as soon as its own input in the statistical representation from the chain (ch)($QM_{HD}$) is achieved and inscribed; for thereby it has been transformed in just that humble contribution of one unity to *the still emerging **constitution** of a 'verification-cardinal'* – let us denote it ***ver.***$|c_j(t_1)|^2$ – that later, at the end of the whole process of verification, will establish via the finally fully emerged ***ver.***$|c_j(t_1)|^2$, a confirmation or an invalidation of the $QM_{HD}$ predictively asserted cardinal $|c_j(t_1)|^2$ indexed by $j$ in the expansion $|\psi_{G,\boldsymbol{H}}(\boldsymbol{r},t_1)>/\boldsymbol{A}$ from $(ch)(QM_{HD})$: If at the end of a sequence of a very big number $N$ of repetitions of a succession $[(G^{(i)}(t^{(i)}_o), EC(t^{(i)}_1 - t^{(i)}_o))$**.** $MesA(t^{(i)}_f - t^{(i)}_1)]$, all the final verification-cardinals ***ver.***$|c_j(t_1)|^2$ – with any $j$ – are (sufficiently) equal to those from the element $|\psi_{G,\boldsymbol{H}}(t^{(s)}_1)>/\boldsymbol{A}$ from $(ch)(QM_{HD})$, then the verification of the $QM_{HD}$ prediction concerning $\boldsymbol{A}$ will have succeeded. But this, if projected on a meta-temporal public dimension of comparison, can factually happen only *long **after*** the individual and repeated *verification*-time-parameters $t^{(i)}_1$ and $t^{(i)}_f$ have already all worked *on the individual level* for *generating* the verification-number ***ver.***$|c_j(t_1)|^2$ for the prediction of index $j$, $|c_j(t_1)|^2$, involved by $|\psi_{G,\boldsymbol{H}}(\boldsymbol{r},t^{(s)}_1)>/\boldsymbol{A}$ and $|\Psi_{G,\boldsymbol{H}(A)}/\boldsymbol{A}$.

[105] While passes the interval of ***public*** time $(t^{(i)}_1 - t^{(i)}_1)$ before the moment $t^{(i)}_1$ when we want to perform an act of measurement upon $\sigma(ms_G)$, the specimen $\sigma(ms_G)$ evolves in 'exterior conditions' *EC* that – though they are specified by the $QM_{HD}$-Hamiltonian ***H*** that has worked to define the statistical state-ket $|\psi_{G,\boldsymbol{H}}(\boldsymbol{r},(t^{(s)} \geq t^{(s)}_o))>$ associated to the microstate $ms_G$ – nevertheless are *themselves* qualified by the same sort of time parameters time-parameters that qualify also $G_t\,G_o$ , $G_t$ and $[G_t.\ Mes\boldsymbol{A}]$.

*Factually, the individual time-parameters $t^{(i)}{}_1$ and $t^{(i)}{}_f$ from one act $MesA(t^{(i)}{}_f\text{-}t^{(i)}{}_1)$ of **A**-measurement are **disconnected** from **all** the statistical times $t^{(s)}$, in a fundamental and definitive way.*

Any *factual* temporal regulation of the physically meaningful temporal features of one coding-measurement-evolution, acts exclusively on the individual level of conceptualization of the microstates, via the individual coding-measurement-successions. The insertion *into* the descriptor $|\Psi_{G,\mathbf{H(A)}}>/\mathbf{A}$ from a statistical chain, of time-parameters like $t^{(i)}{}_1$ and $t^{(i)}{}_f$ that are imported from the individual acts of measurement, is just an impossible, a misleading conceptual in-fraction, in the literal sense. Time-parameters like $t^{(i)}{}_1$ and $t^{(i)}{}_f$ are *qualifying* physical entities; they simply *do not 'perceive' the distribution of predictive* ***numbers*** $|c_j(t_1)|^2$ *from a* $QM_{HD}$*-descriptor* $|\Psi_{G,\mathbf{H(A)}}/\mathbf{A}$. They are *blind* with respect to this sort of global, abstract, *pre*-constituted elements from an expansion of a statistical ket inside which nothing 'begins' nor 'finishes' any more, that simply stays now there for being verified by radically *individual* measurement successions

$$[(G^{(i)}(t^{(i)}{}_o), EC(t^{(i)}{}_1 - t^{(i)}{}_o)).\ MesA(t^{(i)}{}_f - t^{(i)}{}_1)]$$

that begin and finish indeed, but that can act for verification only from a total *outside* of the whole denoted $|\Psi_{G,\mathbf{H(A)}}/\mathbf{A}$, without any possibility to get *inside* this whole and squat its pre-constituted statistical inner structure via disguised individual times. For indeed, *what inside the two statistical descriptors* $|\psi_{G,\mathbf{H}}(\mathbf{r},t^{(s)}{}_1)>/A$ *and* $|\Psi_{G,\mathbf{H(A)}}(\mathbf{r},(t^{(s)}{}_1 \leq t^{(s)}{}_f))>/A$ *is denoted* $t^{(s)}{}_1$ *is not the same entity in both.* In the first one it is a statistical time-parameter and in the second writing it is an individual time-parameter. And the same holds for what is denoted $t^{(s)}{}_f$.

***The "measurement-ket"*** $|\Psi_{G,H(A)}(\mathbf{r},(t^{(s)}{}_1 \leq t^{(s)}{}_f))>/A$ ***is a fallacious descriptor. It mixes two different levels of conceptualization, and such mixtures always lead to conceptual aberrations.***

The condition of general semantic homogeneity of the content on any given level of conceptualization, that in particular inside the chain *(ch)(*$QM_{HD}$*) restricts* the usage of the time-parameters on the statistical level, permits now to genuinely understand the public mental mechanism that brought forth the 'reduction problem':

The mentioned condition of semantic homogeneity itself remained hidden to the public mind of the physicists.

But when, in the botched statistical "measurement-ket" $|\Psi_{G,\mathbf{H(A)}}(\mathbf{r},(t^{(s)}{}_1 \leq t^{(s)}{}_f))>/A$ the *last* moment $t^{(s)}{}_f$ is reached and an individual outcome $a_j$ is explicitly asserted, this, I dare think, brings forth in the individual minds a conceptual uneasiness; and, may be, also a glimmer of wonder on ***where*** *some representation of an agent of* ***physical*** *interaction with the apparatus has entered the chain (ch)(*$QM_{HD}$*) of purely abstract structures from the formalism of* $QM_{HD}$; some resurgence of the entirely forgotten common-sense *necessity* of some *material* entity able to bring into *physical being* the asserted observable effect.

And so, in order to face at least this common-sense necessity, notwithstanding that exclusively a purely statistical formalism was available, the 'reduction' of the measurement-ket $|\Psi_{G,\mathbf{H(A)}}(\mathbf{r},(t^{(s)}{}_1 \leq t^{(s)}{}_f))>/A$ has been postulated.

But ***this*** – unanimously – has been felt to be scandalous ***from a mathematical point of view !!!!***

While the physical point of view seems to have never been declaredly required – ***nor a conceptual one!*** – in spite of the fact that this whole saga happens inside a "theory of *physical* entities".

So a fortiori no conceptual-physical explanation has been identified.

That is how the *general* mixture between vague and lacunar individual concepts and statistical mathematical representations that flaws the whole $QM_{HD}$-formalism, finally mounts to an apogee and bursts its banks in the last link from the chain *(ch)(*$QM_{HD}$*)*. This happens because *there – inside one same descriptor* $|\Psi_{G,\mathbf{H(A)}}(\mathbf{r},(t^{(s)}_1 \leq t^{(s)}_f))>/\mathbf{A}$ – while the time-parameters $t^{(s)}_1$ and $t^{(s)}_f$ do possess a verbal significance that focuses attention upon meanings, in order to generate some minimal intelligibility the formal representation *has* to be dashed explicitly onto the individual level by asserting that a definite eigenvalue $a_j$ is registered. And this verbal diktat violates the rules of the utilized *mathematical* language. Which is sacrilege, consensually.

But the general conclusion that deserves being conserved as a gain has another formulation: The "reduction problem" brings into full light this gross, utterly trivial fact that conceiving as factually realizable a *common* ordered succession of time-parameters some of which qualify material and individual events while other ones qualify abstract and globalized statistical-probabilistic predictive structures, is just non-sense.

And ***Mathematics knows this***.

On the basis of this conclusion and of its violation *we refuse, not only von Neumann's representation of quantum measurements, but also the* ***essence*** *itself of the* $QM_{HD}$*-representation of quantum-measurements*.

### (6.II).4.5. Conclusion on (6.II).4

The analyses from (6.II).4 entail that:

***$QM_{HD}$ is simply DEVOID of an acceptable representation of measurements***.

What is called nowadays "the theory of quantum measurements" does not specify:

- how is obtained the object-entity-to-be-studied;
- how, upon it, this or that qualifying quantity can be "measured";
- how the result of an act of measurement can be identified, and so, nor how a prediction can be verified.
- And on the top of all these absences lies a thick veil of *confusion* between the individual level of conceptualization, and the statistical-probabilistic one.

How has it been possible for such a situation to establish itself in the most fundamental among the nowadays theories of physical domains of reality?

How has it been possible that so well known and trivial considerations as those brought forth above have stayed inactive such a very long time?

These questions, we think, deserve their own and detailed answer, focused upon social-historical and psychological forces that act upon the development of scientific actions. Here we only summarized remarks that have already emerged throughout the point (6.II).4.

The nowadays representation of quantum measurements has emerged under the pressure of the fact that the $QM_{HD}$-formalism, such as it has been formulated in its first phase by a purely mathematical top-down approach, offered first only mathematically pre-organized statistical conceptual moulds where could be lodged exclusively what was immediately useful for prediction. So the individual physical operations – that necessarily *are* involved by any representation of statistics of results of measurements on individual physical entities – remained entirely non-described by the initial representation. And then this lasted, because in the case of microstates the predictions being "essentially" statistical in any case – in the sense of *MD* from *IQM*, i.e. no matter whether the approach is top-down or bottom-up – *an individual level of conceptualization was not particularly useful for prediction*. So the question of the *conceptual-physical constructability* of the predictive statistics from $QM_{HD}$ did not even arise. Correlatively, the very distinction between the concept of a predictive statistic of abstract numbers and on the other hand its individual, physical genesis, glided out from the domain of attention: So these two fundamentally different concepts tended to identify on a fictitious *common* statistical level of conceptualization.

But the void of organization of an individual level of representation of the microstates ***confined also the thought*** on exclusively the statistical level. The conceptualization got imprisoned in a statistical fortress that floated on a morass of vague and chaotic individual notions, *as it still does nowadays*. There was no structured individual level of conceptualization on which to leap down and break free. What was lying deeper than the mathematically defined statistics was not visible in the undone beneath these.

It is very noteworthy that such a process has been possible: This draws attention upon the control that must be kept active, inside Physics, upon the connections between achieved mathematical expressions and physical-conceptual forms.

*IQM* with its bottom-up approach that starts from local zeros of knowledge and generates intelligibility *while* it constructs the representations was not even conceivable at that initial time. This, probably, was why de Broglie's *physical and individual* concept of a corpuscular-like wave has been hauled upside into the emergent *statistical* formalism, quartered and disguised in two *ill*-understood mathematical concepts, namely the concept of eigenstate of an observable, and the concept of a *wave*-function of which the meaning oscillated freely between that of a descriptor of a *physical* wave involved by this microstate, and that of a descriptor predictive of the *statistical* results of measurements on the studied microstate. But under the veil of absence of any defined conceptual control, the initial concept of *wave*-function *moved* very rapidly into the purely formal extension called a *state*-function possessing the *form* of "a *wave-**packet***", and then it moved further into the still more abstract concept of a "*state-ket*"; while the very essence of the concept of eigenstate was assimilated to that of a mathematically unintelligible limiting sort of *state*-ket. Thereby this genetic tremolo ended in the stable hegemony of a mathematic of "direct statistics", devoid of any representation of a physical genesis via individual physical operations.

This might have been favoured by the "essentially" statistical character assigned to the micro-entities *themselves*, i.e. not to our *knowledge* on these.

And so all the physical-conceptual meanings were silenced for a long time; while the individual features – that irrepressibly *do* impose themselves to the minds when one deals with measurements – have been stuffed together with statistical features into *one common statistical descriptor flawed by a fundamental inner semantic inconsistency*:

The botched measurement-evolution-state-ket $|\Psi_{G,\mathbf{H(A)}}(\mathbf{r},t^{(s)}{}_{1}\leq t^{(s)}{}_{f})>/\mathbf{A}$ stepped in.

This offered at least a way of speaking. And therefore, in spite of all the confusions and inadequacies that somehow, from the very start, *must* have acted in the minds as a resistance, the fallacious descriptor $|\Psi_{G,\mathbf{H(A)}}(\mathbf{r},t^{(s)}{}_{1}\leq t^{(s)}{}_{f})>/\mathbf{A}$ continues to be *taught* up to this very day.

On the other hand the fact that the mathematical formalism of $QM_{HD}$ ***itself*** rejects the in-distinction between an individual level of conceptualization and the statistical one, precisely by the reduction problem, maintains a persistent and various criticism:

*The inner consistency of the mathematical structures is very sensitive to the semantic contents.*

This also is noteworthy. The whole history of the reduction problem deserves being kept alive in the minds and its implications deserve being made use of [106].

---

[106] E. Morin has made the fundamental assertion that "only a new construction can ruin a previously installed construction". But in the case of *$QM_{HD}$ the new constructions that are researched are of **an entirely mathematical nature***. But the result is stagnation. And I hold that it it is very likely that mathematical changes alone will never succeed to genuinely compensate the total absence of an individual conceptualization of

Notwithstanding the situation brought forth above, *$QM_{HD}$* has worked and it continues to work. This theory has achieved remarkable successes and it still will achieve other successes even if it is left just such as it stands now. Indeed the curious omnipresent genius of human mind invents spontaneously local and more or less implicit understandings that permit to act adequately there and when one genuinely needs and wants to *act*. General methods for thinking well are massively left to the theorists. Whereas for the experimenters, it seems, it suffices to believe that a quantum theory of measurements exists for measuring adequately and for constantly making progress.

This teaches humility to those who try to construct theories.

This also proves that a fully satisfactory theory of quantum measurements is possible, since no doubt it is quite often applied without being known.

So, practically, there is no urgency.

But conceptually there is urgency. Indeed, what value of principle – as a *theory of Physics* – does a representation of *non*-perceptible microstates possess if it predicts via purely mathematically constructed predictions *and* does not state in clear, organized and generally valid terms how to conceive-and-perform measurements for *verifying* these predictions?

---

the factual microstates via the specification of the structure of the physically operational cognitive actions *G* and *MesA* that are involved and of the connection between these operational actions and the physical micro-entities to be studied.

# CONCLUSION ON THE PART II

We have first identified in 5.II the void, inside $QM_{HD}$, of an individual conceptualization of the microstates. This has led to a first clear perception of the concept of *(top-down)–(bottom-up) anachronistic collision* between the approach from $QM_{HD}$ and the approach from *IQM*.

Then, inside 6.II we have brought forth that an eigenket $|u(\mathbf{r},a_j,)>$ of a quantum mechanical observable $\boldsymbol{A}$ has the meaning of a definite mathematically expressed model of wave-movement around the singularity in the amplitude of de Broglie's "corpuscular wave" model of any specimen of the studied microstate; namely, a wave-movement that, during a measurement-evolution *MesA*, keeps constant the corresponding eigenvalue $a_j$ of a qualifying mechanical-quantity $\boldsymbol{A}$ defined for the corpuscular-like singularity from the amplitude of this corpuscular-wave.

This, inside the framework $[IQM\text{-}QM_{HD}]$, has triggered a first – important – constructive step toward the new theory of microstates that is researched here. Namely, we have defined a "*G*-corpuscular-wave model" of a microstate, denoted $ms_{G,cw}$, and to this we have associated a modelling postulate $MP(ms_{G,cw})$. Thereby de Broglie's model of one specimen of a microstate – *an ideal purely mental model incorporated in a global model of the same kind* – is transposed in physical-operational terms that permit to incorporate it to the approach developed here, subjected to the requirement of a consensually predictive and verifiable character indelibly tied with cognitive actions of a conceptual-physical and operational nature.

Correlatively, the relation (1') $G_t \leftrightarrow (ms_{Gt,cw} \equiv \{\sigma(ms_{Gt,cw})\})$ by which the mentioned model is expressed, enriches the initial relation (1) $G \leftrightarrow (ms_G)$ where $(ms_G) \equiv \{\sigma(ms_G)\}$. Inside *IQM* the relation (1) $G \leftrightarrow (ms_G)$ where $(ms_G) \equiv \{\sigma(ms_G)\}$ defines already a new concept of *factually* generated microstate $ms_G$, but it assigns to this concept an only very general semantic content entailed by methodological requirements; while the relation (1') $G_t \leftrightarrow (ms_{Gt,cw} \equiv \{\sigma(ms_{Gt,cw})\})$ *endows this initial content with features that specify the inner structure assigned to the specimens of a factual microstate*.

We have then brought into evidence the power of clarification entailed by a systematic specification of the existence – or *not* – of a connection between a ket from a mathematical $QM_{HD}$-expression, with the physical operation of generation *G* of specimens of the studied microstate, and a fortiori with the character of this operation of generation itself (simple or composed). This led to useful new denotations that distinguish between the nature of a state-ket and the nature of an eigenket.

Finally we have examined the quantum theory of measurements from $QM_{HD}$. We have refused von Neumann's representation on the basis of general conceptual reasons. Then we have identified the coding rule that is implicitly assumed inside $QM_{HD}$ and we have explicated that it is likely to be devoid of a *general* validity. We have then shown that the essence itself of the $QM_{HD}$-representation of quantum measurements is not acceptable so that in fact $QM_{HD}$ is devoid of a representation of the quantum measurements. Which is a radical conclusion.

Considered globally, the Part II of this work brings into evidence a surprising realization: the so much idolized $QM_{HD}$ is *not* a conceptual structure that can be clearly called a "scientific theory of the microstates belonging to the Mathematical Physics". $QM_{HD}$ as it now stands does not distinguish clearly between a purely mental representation – a mathematical structure of *signs*, just abstract signs – and on the other hand a representation that, structurally, incorporates also explicit definitions of the unavoidable and basic conceptual-*physical* ***operations*** "*G*" and "*MesA*" and of their actions, as specified in *IQM*. While *for microscopic entities*, in the absence of such definitions the consensual verification of the mathematically obtained predictions on the results of measurements cannot even be clearly *conceived*.

# PART III

## CONSTRUCTION ACCORDINGLY TO *IQM* OF A SECOND QUANTUM MECHANICS

**ABSTRACT OF THE PART III**

This is the constructive core of the present work. It consists of two chapters, 7.III and 8.III, that bring forth the essential lines of a Second Quantum Mechanics *QM2* possessing all the characters required here for a *scientific theory of the microstates belonging to the Mathematical Physics*, i.e. to be a consensual, predictive and verifiable mathematical representation of the microstates that obeys strictly all the physically-operational, methodological and structural conditions imposed by the Infra-Quantum Mechanics *IQM* established in the Part I of this work.

In the chapter 7.III, inside a framework *[IQM-*$QM_{HD}$*]*, is formulated a new sort of representation of the quantum measurements *on* ***unbound*** microstates, where all the specificities of quantum-measurements appear clearly. This representation is founded upon Gleason's 1957-theorem on the possibility of a Hilbert-space representation of any ***given*** probability measure. It is directly rooted into the a-conceptual physical substratum and it establishes – *both* – the process of elaboration of statistical predictions ***and*** the process of verification of these. ***The result entails that – when convenient or necessary – the recourse to Schrödinger's equation can be circumvented***.

In the case of unbound microstates with simple operation of generation the $QM_{HD}$-Hilbert-space-representation of the studied microstate conserved unchanged.

But for unbound interference-microstates with *composed* operation of generation it appears that a new Hilbert-space representation is necessary. This, via a critical analysis of the concepts of "observable" and "beable", leads to experimental recourse to the de Broglie-Bohm guiding-law. Thereby the framework *[IQM-*$QM_{HD}$*]* is extended to include the de Broglie-Bohm approach *dBB*. Inside the new framework denoted *[IQM-*$QM_{HD}$*-dBB]* is defined a new coding-measurement-evolution for the momentum-measurements and a generalization of the $QM_{HD}$ equation for eigenket and eigenvalues of the momentum operator, as well a re-definition of the Hilbert-space state-ket associated to the studied interference-microstate.

The case of bound microstates continues to require the use of the Schrödinger equation; but it involves no obstacle whatever in the way of the radical conceptual changes required by *IQM* for any acceptable theory of the microstates.

The mentioned development entails a genuine melding of *IQM,* $QM_{HD}$ and *dBB*, into one conceptual whole.

In the chapter 8.III, on the basis of the representation of the quantum measurements constructed in 7.III and of the fusion between *IQM,* $QM_{HD}$ and *dBB*, are defined the main global features of the researched Second Quantum Mechanics *QM2* itself.

# INTRODUCTION TO PART III

> « It would seem that we have followed as far as possible the path of logical development of the ideas of quantum mechanics as they are at present understood. The difficulties, being of a profound character, can be removed only by some drastic change in the foundations of the theory, probably a change as drastic as the passage from Bohr's orbit theory to the present quantum mechanics. »
>
> P.A.M. Dirac, *The Principles of Quantum Mechanics*, Oxford at the Clarendon Press, 4th edition 1958 (1st edition 1930).

The third part of this work is resolutely constructive. On the new basis offered by the clarifications and the purges from the Part II we shall now delineate the essential features – *only* these – of a new sort of Hilbert-Dirac mathematical representation of the factual microstates.

This representation is physically-operationally rooted directly into the physical factuality, and therefrom it is constructed bottom-up.

More globally, throughout the third part of this work the Infra-Quantum Mechanics *IQM* acts as a structure of reference and insertion, accordingly to the aim for which it has been constructed. The framework *[IQM-$QM_{HD}$]* that has been introduced in the chapter (6.II).2 by simply juxtaposing *IQM* and *$QM_{HD}$* is first conserved unchanged. The conjugate use of both structures from this framework has already initiated in 6.II the emergence of a common language and a common system of notations that manifest the growth of a new whole. This growth will now continue. The definition of a convenient coding-measurement-evolution, in any given case, will be founded upon the model of a specimen of the considered sort of microstate in the sense of the definitions from (I.2).1.2 and the corresponding model. These will be taken into account overtly and systematically, with detail and rigor, and so the relativities to these two basic elements of reference will guide the construction step by step and will generate guidance and intelligibility [107]. At the end of the Part III this methodological progression will have brought forth that the framework *[IQM-$QM_{HD}$]* has never possessed the nature of a mere scaffold; that from the start, inside *$QM_{HD}$* itself already – while the core of de Broglie's approach was working covertly – *the whole structure of IQM was also contained potentially*, but implicated and compressed like in an abstract embryo. Once ex-pressed and "ex-plicated" in vitro this structure will have forcefully concurred to the conceptual growth of the second quantum mechanics developed here, *QM2*.

*QM2* is not conceived as a new "interpretation" of the nowadays quantum mechanics; nor as a here *achieved* new theory of the microstates. It is submitted as a first outline of *a **fundamentally** new representation of the factual microstates*, required general, "scientific" (i.e. consensually operational in a conceptual-physical sense, predictive and verifiable) ***and fully intelligible***.

In the chapter 7.III we construct first the main lines of a new representation of the quantum-measurements for un-bound microstates (the case of bound microstates is absorbable). This representation distinguishes constantly and explicitly between the individual level of conceptualization and the statistical one, and it has a ***factual***-formal character. The process of construction involves an explicit re-soldering of *IQM* and *$QM_{HD}$*. This entails the controlled emergence of an exhaustively exposed semantic content. This content is rooted into the physical factuality via *IQM* and its probabilistic predictive features are endowed with *a **factually** re-structured Hilbert-space mathematical representation that is founded directly upon Gleason's 1957-theorem*.

[107] This will seem queer in consequence of the contrast with the positivistic doctrine that has interdicted models for more than 100 years. At the end of this work the reader will be able to forge an opinion for or against this doctrine.

This permits to circumvent the use of Schrödinger's equation of evolution for un-bound microstates, if this is necessary or convenient.

The unbound microstates with *simple* operation of generation accept conservation of the form of their $QM_{HD}$ Hilbert-space representation by one state-ket.

But the $QM_{HD}$ Hilbert-space representation of the unbound microstates with *composed* operation of generation by a superposition of two or more state-ket leads to assertion of predictive probabilities that are not verifiable. So the form of their factually re-structured Hilbert-space representation requires change. In order to fulfil this requirement we are led to make use of de Broglie's guidance law $\mathbf{p}(\mathbf{r},t)=-\boldsymbol{\nabla}.\beta(\mathbf{r},t)$ (or of Bohm's form $\mathbf{v}(\mathbf{r},t)=-\boldsymbol{\nabla}.\beta(\mathbf{r},t)$) of this law). This draws now the whole essence of the de Broglie-Bohm approach, inside an enlarged framework re-noted *[IQM-$QM_{HD}$-dBB]*.

Inside this enlarged framework the $QM_{HD}$ equation for eigenket and eigenvalues of the momentum operator is conveniently generalized, and a convenient modified representation is assigned to the state-ket of a microstate with composed operation of generation.

This finally permits to extend the new representation of the quantum measurements constructed for the unbound microstates with *simple* operation of generation, to also the unbound *interference*-microstates that have a composed operation of generation. Thereby will have been achieved for all the unbound microstates, a mathematical Hilbert-space representation that entirely frees the specification of predictive probabilities from a necessary use of the Schrödinger equation.

But the specification of probabilistic predictions for the results of measurements on *bound* microstates continues to require systematic use of the Schrödinger equation.*The bound microstates remain a bastion of the **formalism** of $QM_{HD}$.*

The category of bound microstates, however, is found *not* to involve any obstacle whatever in the way of the basic new *IQM*-requirements concerning an acceptable theory of the microstates.

*In the chapter 8.III*, around the core constituted by the new representation of the quantum measurements constructed in the chapter 7.III, are finally sketched out very succinctly the main lines of the whole "second quantum mechanics" *QM2*.

The genesis of *QM2* projects on the near horizon two other unexpected and important contours, but that draw out of the Part III.

# 7.III

# A NEW REPRESENTATION OF THE QUANTUM-MEASUREMENTS ON UNBOUND MICROSTATES

## (7.III).1. THE SEMANTIC SELF-CONSISTENCY OF THE FRAMEWORK *[IQM-$QM_{HD}$]* AND ORGANIZATION OF A NEW PERSPECTIVE ON THE REPRESENTATION OF QUANTUM-MEASUREMENTS

This following short section is devoted to the global inner consistency of the subsequent construction.

### (7.III).1.1. Apparent absence of unity inside *[IQM-$QM_{HD}$]* on the statistical predictions and their verification

Consider $QM_{HD}$.

Inside $QM_{HD}$ the predictions on results of quantum measurements are conceived to be obtained by the use of the Schrödinger equation of the considered problem. This is not disturbing for didactical idealizations. But when real physical situations are considered that are not heavily a priori restricted, it is in general much more difficult than asserted in textbooks to mathematically work out verifiable predictions (think of Schrödinger's Memoire for solving the simplest real case of the one electron from an atom of hydrogen). *In order to dispose of the state-ket $|\psi_{G,\mathbf{H}}(\mathbf{r},t)>$ of the problem – that is the source of the whole predictive $QM_{HD}$ algorithm – one has to **give** the initial conditions via the initial state-ket $|\psi_{G,H}(r,t_o)>$*; and often this is not possible without admitting gross *basic* approximations[108]. And when the acting Hamiltonian cannot be considered to be stationary, or when it simply *is entirely unknown* because unspecified quantum fields or "obstacles" are acting, then even the writing down of the Schrödinger equation of the problem itself, is impossible. While when this equation *can* be written, quasi always the mathematical production of a general solution involves already various approximations of which the factual effects cannot be imagined a priori, so they cannot be controlled mathematically.

The general formal *constructability* of the predictions is far from being generally insured inside $QM_{HD}$.

Furthermore, as it appeared in (6.II).4, inside $QM_{HD}$ when it has been possible to acceptably establish the predictions by a purely mathematical elaboration, the *verification* of the predictions is treated in a way that is dramatically deficient as much from a conceptual point of view as from a mathematical point of view. (This has been stressed in the chapters 5.II and 6.II).

Consider now *IQM*.

*Inside IQM the predictions are constructed **only** factually and then they can be verified only by repeating the factual construction* (cf. 3.I).5).

[108] As for instance the implicit application of the Laplace principle of an a priori uniform distribution of the probabilities of the elementary events, so in particular, of *an a priori **uniform initial distribution of 'presence'** over the whole space-domain of existence assigned to the studied microstate*, etc.

So when the fundamental question of prediction and of verification of the predictions is considered, it seems at a first view that inside the framework *[IQM-$QM_{HD}$]* even the slightest degree of unity between the *IQM* representation and the representation form $QM_{HD}$, is lacking from the outset.

But this is only an appearance, and a very misleading one. Just below will appear a deep unity inside *[IQM-$QM_{HD}$]* between predictive probabilities and the verification of these that – remarkably – has emerged silently and organically in a certain sense.

### (7.III).1.2. The conditions of inner semantic self-consistency of the global framework *[IQM-$QM_{HD}$]*

*IQM* has been constructed like a reference-and-immersion-structure for understanding any given theory of the microstates, for estimating its adequacy, and for improving it. As such *IQM* has been deliberately endowed with the maximal generality compatible with its required status. This entailed leaving *undefined* the model of a microstate. And in consequence of the absence of a defined model, the content of a measurement operation *MesA* remained equally unspecified inside *IQM*, as well as the 'external conditions' *EC* from the generalized definition (13') $G_t$*=F(*$G_o$*,EC,(t-*$t_o$*))* of an operation of generation of specimens of the studied microstate. This entailed that throughout the Part I and the Part II of this work the conditions of a full comparability between the *semantic* contents of $QM_{HD}$ and *IQM* have remained un-defined.

But in the chapter 6.II we have realized several basic semantic elucidations and this has changed the conceptual situation. For self-sufficiency of the present chapter 7.III we recall these conditions below:

- In (6.II).1 we have identified the meaning of the concept of eigenket of a quantum mechanical observable and its intimate connection with the individual model of a specimen of a microstate that acts quite systematically inside the mathematical formalism from $QM_{HD}$.
- In (6.II).2 – via the generalizing redefinition (1') $G_t \leftrightarrow (ms_{Gt,cw} \equiv \{\sigma(ms_{Gt,cw})\})$ of an operation of generation, and the *modelling postulate* $MP(\{\sigma(ms_{G,cw})\})$ – we have incorporated this individual model of a specimen of a microstate, to the physically-operational approach practised in this work,.
- In (6.II).3 we have stressed the clarifying role inside the $QM_{HD}$ mathematical writings, of the existence – or *not* – of a direct connection between, respectively, a state-ket or an eigenket, and the operation of generation *G* of the involved specimens of the studied microstate, and we have introduced specifying notations.
- In *(6.II).4.3* we have brought into evidence the implicit existence, inside $QM_{HD}$, of a general type (20) of concept of *coding-measurement-evolution* that seems to be valid indeed for an *unbound microstate* ***without quantum fields***.
- In *(6.II).4.4.3* we have shown that the $QM_{HD}$ representation of one act of coding-measurement-evolution, namely (19) $|\Psi_{G,\mathbf{H(A)}}(\mathbf{r},t_1)>/\mathbf{A}=\sum_j c_j(t_1)|u(\mathbf{r},a_j,)>, \forall j$, attempts to express the second factor *MesA* from an *IQM* ***individual*** coding-measurement-succession *[$G_t$.MesA]* by use of a *statistical* descriptor that *occults the originating first factor* $G_t$ (cf. the Fig.7 and the corresponding comments) and just asserts a final fall on the individual level of conceptualization, which is unacceptable both conceptually and mathematically.
- In *(6.II).4.4* we have reached the general negative conclusion that $QM_{HD}$ is devoid of an acceptable representation of quantum measurements.

The semantic progresses listed above entail also a massive positive conclusion that concerns specifically the general question of prediction and verification of results of quantum measurements. This conclusion has taken form silently. Below we explicate it.

To begin with, suppose optimistically that we are in a physical situation that has permitted to write the Schrödinger equation of the considered problem, to solve it, to write down the initial state-ket $|\psi_{G,\boldsymbol{H}}(t_o)>$, and so, to identify the state-ket of the studied microstate, $|\psi_{G,\boldsymbol{H}}(\boldsymbol{r},t)>$ for any moment $t \geq t_o$. Consider now an expansion (18) $|\psi_{G,\boldsymbol{H}}(\boldsymbol{r},t_1>/A=\sum_j c_j(a_j,t_1)|u(\boldsymbol{r},a_j,)>, \forall j$, of this state-ket $|\psi_{G,\boldsymbol{H}}(\boldsymbol{r},t_1>$ at a given moment $t_1$, and the predictive probability law $(\boldsymbol{A},\{|c_j(a_j,t_1)|^2, \forall j\})$ defined by this. The examination from *(6.II).4.4.3* of the $QM_{HD}$ representation of the quantum measurements induces the following remark:

> In general the statistical prediction $(\boldsymbol{A},\{|c_j(a_j,t_1)|^2, \forall j\})$ from an expansion (18) *can **not*** be *verified* experimentally otherwise than via a very big number of repetitions of whole individual coding-measurement-succession $[G_{t1}.MesA]$, *in the sense defined in IQM.*

And in order for the verification to be expressible inside $[IQM\text{-}QM_{HD}]$ the merely general structural definitions imposed by *IQM* and the clarifications from 6.II have to be now completed by specifying in an organized way the *conditions of operational semantic compatibility* between *IQM* and $QM_{HD}$. These are the following ones:

***(a)*** We make use of the individual model of a microstate denoted $ms_{G,cw}$, *such* as this model is expressed by the modelling postulate $MP(ms_{G,cw})$ and by (1') $G \leftrightarrow ms_{G,\mathrm{cw}}$, with $ms_{G,cw} \equiv \{\sigma(ms_{G,cw})\}$.

***(b)*** We posit that in (13') $G_t=F(G_o,EC(t-t_o))$ with $G_o \leftrightarrow ms_{Go}$ and $G_t \equiv [G_o.(t-t_o)]$, the external conditions during $(t-t_o)$ denoted $EC(t-t_o)$ have to be those expressed inside $QM_{HD}$ by the Hamiltonian operator $\boldsymbol{H}$ that, in the Schrödinger equation of the problem, acted abstractly for determining the state-ket $|\psi_{G,\boldsymbol{H}}(t)>$, $(t_o \leq t \leq t_1)$, *when this equation can be written and solved.* This permits then to write inside $[IQM\text{-}QM_{HD}]$

$$G_{t1}=F(G_o, \boldsymbol{H}, (t_1-t_o)) \qquad (13'')$$

where '*F*' means ' a functional of ' and ***H*** *includes* what is called 'obstacles' (walls, barriers, wells).

***(c)*** When inside the succession of operations $[G_{t1}(t_1-t_o).MesA(t_f-t_1)]$ the act of a coding-measurement operation $MesA(t_f-t_1)$ begins at a time $t_1$, ***H*** is replaced by a measurement-hamiltonian ***H(A)*** that commutes with ***A***; and then, *throughout the duration $(t_{f1}-t_1)$ of this act $MesA(t_f-t_1)$ the measurement Hamiltonian **H(A)** acts upon the (unknown) **individual and physical WAVE-function** $\Phi(\boldsymbol{r},t)$ of the specimen $\sigma(ms_G)$ of the studied microstate $ms_G$ that is involved, accordingly to the distinction introduced in (6.II).4.2.2 between a **state**-ket and a **wave**-function that represents an individual physical wave*[109].

***(d)*** One act of coding-measurement-evolution $MesA(t_f-t_1)$ from the succession $[G_{t1}.MesA(t_f-t_1)]$, and its formal representation, must be explicitly defined in a manner consistent with: the modelling postulate $MP(ms_{G,cw})$ accordingly to *(a)*; the type of microstate that is considered (in the sense of the definitions from (2.I).1); the mathematical language chosen for constructing the representations.

The specifications *(a),(b),(c),(d)* define the a priori conditions of inner semantic-operational self-consistency of the framework $[IQM\text{-}QM_{HD}]$. *Inside* $[IQM\text{-}QM_{HD}]$ *they apply a*

---

[109] This condition brings forth the general ambiguity inside $QM_{HD}$ on the significance of a solution of the Schrödinger equation, namely whether this solution points toward a physical wave-phenomenon or toward a statistic of results of measurements. Indeed the condition (13'') holds also for the indirect acts of measurement on a *bound* microstate from an atomic or molecular structure where it is clear that ***one** specimen* $\sigma(ms_G)$ of the studied microstate, with its ***physical*** wave, can *subsist* for an arbitrarily long time and meanwhile it can from time to time interact for measurement with test-particles or other devices (Zeeman or Stark effects, etc.); so that in this case both ***H*** and ***H(A)*** act on one 'same' physical wave, while the solution $|\psi_{G,\boldsymbol{H}}(t_1)>$ of the Schrödinger equation permits to calculate via (18) ***statistical*** predictions concerning the whole set of results of such test-interactions. While for unbound microstates each one measurement-interaction involves its own specimen $\sigma(ms_G)$, so a whole set of wave-functions comes in.

*posteriori to IQM also*. All these specifications must be added to those made in *(6.II)4.4.3* (that are illustrated in the Fig.4) in what concerns time parameters versus level of conceptualization.

On the basis of requirements obtained in this way we shall now bring into evidence the *nature* of the unity that can be realized inside *[IQM-QM$_{HD}$]* with respect to statistical predictions on results of quantum measurements, and to verification of these. As long as this unity will be realizable we shall stay inside the framework *[IQM-QM$_{HD}$]*. When this will cease to be possible – which will happen for microstates with composed operation of generation – the framework *[IQM-QM$_{HD}$]* will have to be improved explicitly by modifying the *QM$_{HD}$*-representation. In this way at the end of this chapter we shall be endowed with a new representation of the quantum measurements endowed with intelligibility that will have incorporated *IQM* and modified *QM$_{HD}$*.

### (7.III).1.3. A basic assertion on the prediction-verification unity inside *[IQM-QM$_{HD}$]*

Consider the *IQM*-descriptor (9) $(D/A)(ms_{Gt1})\equiv\{(\varepsilon,\delta,N_0)\text{-}\pi(a_j,t_1)\}_{Gt1}, \forall j$, that has been constructed factually. We formulate the following 'assertion' *Ass.1* that involves this descriptor and is supported by a corresponding 'argument' *Arg(Ass.1)*[110] :

***Ass.1***

If the *IQM* description (9) $(D/A)(ms_{Gt1})\equiv\{(\varepsilon,\delta,N_0)\text{-}\pi(a_j,t_1)\}_{Gt1}$ has been *constructed* by use of coding-measurement-successions $[G_{t1}.MesA(t_f\text{-}t_1)]$ of which the content has been specified accordingly to the conditions *(a),(b),(c),(d)* of semantic-operational inner self-consistency of *[IQM-QM$_{HD}$]*, then the statistical predictive *QM$_{HD}$*-law $(A,\{|c_j(a_j,t_1)|^2, \forall j\})$ defined by the expansion (18) can be found to be *verified* ***if and only if*** it identifies in content – inside the limits permitted by the parameters $(\varepsilon,\delta,N_0)$ – with the statistical assertions of which the factual description (9) consists.

***Arg(Ass.1).*** Obvious, because inside *IQM* the description (9) $(D/A)(ms_{Gt1})$ is constructed *factually* and so, in order to verify this description inside *IQM* one is obliged to just repeat its construction: no other way is conceivable inside *IQM*. This means that inside *IQM the experimental verification of* $(D/A)(ms_{Gt1})$ *is certain a priori, by construction* (3.I).5)[111]. So if a very big number of repetitions of the succession of operations $[G_{t1}.MesA]$ accomplished for ***verifying*** the mathematically constructed *QM$_{HD}$* prediction $(A,\{c_j(a_j,t_1)^2\}), \forall j$, are realized in the same way permitted by the conditions *(a),(b),(c),(d)* as the successions of operations $[G_{t1}.MesA]$ by which the *IQM* description (9) $(D/A)(ms_{Gt1})$ has been constructed, then these successions of operations, while they verify $(A,\{|c_j(a_j,t_1)|^2, \forall j\})$, also reconstruct $(D/A)(ms_{Gt1})$ ■

In other words, the conditions *(a),(b),(c),(d)* of semantic-operational inner self-consistency of *[IQM-QM$_{HD}$]* entail the following unifying identifications:

- Under the constraint of the conditions *(a),(b),(c),(d)* the predictive content of the *IQM* description (9) $(D/A)(ms_{Gt1})\equiv\{(\varepsilon,\delta,N_0)\text{-}\pi(a_j,t_1)\}_{Gt1}, \forall t_1, \forall j$ – a 'factual probability law' – identifies with the predictive content denoted $\{A,|c_j(a_j,t_1)|^2, \forall j\}$ of the *QM$_{HD}$* expansion (18) $|\psi_{G,\mathbf{H}}(\mathbf{r},t_1)>/A\}$; only the notations differ.

---

[110] Throughout what follows we speak in terms of 'assertions' and 'arguments' because we are not yet inside a formally closed structure where can be given 'proofs' in the strict sense. From now on, for brevity, we drop the specification $(t_f\text{-}t_1)$.

[111] In so far, of course, that $(D/A)(ms_{Gt1})$ has been considered to have been accomplished only when a convenient choice in (9) of the set of parameters $(\varepsilon,\delta,N_0)$ has stabilized the quasi-identical recurrence of $(D/A)(ms_{Gt1})$ when one reconstructs it inside the correspondingly admitted fluctuations.

- Under the constraint of the conditions *(a),(b),(c),(d)* the *IQM* 'complete' description (9'') $D_M(ms_G)\equiv\{[(\varepsilon,\delta,N_0)\text{-}\pi_{t1}(G,a_j)],(M\pi c(G))_{AB}], \forall A, \forall AB\}, \forall j$ (that includes the meta-probabilistic correlations $(M\pi c(G))_{AB}$) ***must*** be posited a priori to possess potentially the same predictive content as the representation (22) $|\psi_{G,\mathbf{H}}(\mathbf{r},t_1)\rangle \approx_{\text{pred.}} \{|\psi_{G,\mathbf{H}}(\mathbf{r},t_1)\rangle/A\}, \forall A, \forall t_1$ of the state-ket $|\psi_{G,\mathbf{H}}(t_1)\rangle$ from $QM_{HD}$, *since both (9'') and (22) are verified by one same set of results of verification-measurements* [112].

At a first sight it might seem that the pair (*Ass.1, Arg(Ass.1)*) expresses a circularity or at least a triviality. But in fact this is not the case at all; the contrary is the case, because the identifications stated above under the sole constraint of the conditions *(a),(b),(c),(d)*, constitute the specification in rigorously defined and structured terms, of ***the factual-operational SOURCE*** *– on the* ***individual*** *physical-operational level of conceptualization from IQM –* of the two basic $QM_{HD}$-descriptors $|\psi_{G,\mathbf{H}}(\mathbf{r},t_1)\rangle$ and $\{|\psi_{G,\mathbf{H}}(\mathbf{r},t_1)\rangle/A$ that inside $QM_{HD}$ have acquired an ***exclusively*** formal a-semantic basic content (some parsimonious semantic significances have been just appended by posits). So:

> Inside the enlarged and *common* framework $[IQM\text{-}QM_{HD}]$ that acts in (*Ass.1, Arg(Ass.1)*) *IQM* offers – starting from zeros of local individual knowledge and up to the probabilistic predictions constructed from these – quite non-trivial ***"vertical"*** factual-semantic "explanations" of the predictions that inside $QM_{HD}$ are extracted algorithmically from the descriptors $|\psi_{G,\mathbf{H}}(\mathbf{r},t_1)\rangle$ and $\{|\psi_{G,\mathbf{H}}(\mathbf{r},t_1)\rangle/A$ (cf.: Fig.6 and the comments; (6.II).1; *(6.II).4.4.3*).

Via the conditions *(a),(b),(c),(d)* the pair (*Ass.1, Arg(Ass.1)*) ***knits together*** *the physical, factual* ***genesis*** *of a* $QM_{HD}$ *predictive probability-law* $(\mathbf{A}, \{|c_j(a_j,t_1)|^2, \forall j\})$, *with this law itself, so with the Hilbert-Dirac mathematical formalism.* Thereby any $QM_{HD}$ mathematical law appears now inside $[IQM\text{-}QM_{HD}]$ as the result of thoroughly defined and structured ***individual physical operations*** [113]. The void of organization of the individual-factual level of conceptualization that flaws $QM_{HD}$ (Fig.6) is now already considerably filled. This is a first and massive manifestation of *possibility* to organize a full unity between *IQM* that starts from the a-conceptual factuality and attains the statistical level of conceptualization of the factual microstates in the expression (9") after having traversed the whole individual level, and on the other hand $QM_{HD}$ that starts from the classical thinking and attains directly the statistical level of conceptualization of the factual microstates in the expression (22) via a top-down approach. When this possibility will have been realized it will entail also a uniformly bottom up progression along the vertical of conceptualization.

So the assertion *Ass.1* will act as the germ of the full intelligibility of the second Quantum Mechanics developed in this work; its force and fertility will appear throughout what follows.

---

[112] This is important because it leads to assign to Dirac's "calculus of transformations" a definite ***factual*** meaning, instead of a purely algorithmic one.

[113] Inside *IQM* the statistical writings (9) $(D/A)(G_t,ms_G,A)$ and (10) $D_M(G_t,ms_G,V_M)$ stress precisely the organic *unity*, in the case of microstates, between the statistical predictive knowledge that, once established, can be considered and made use of separately, and on the other hand the conceptual-physical-operational genesis of this knowledge – respectively the individual genetic triads $(G_t,ms_G,A)$ and $(G_t,ms_G,V_M)$ – via repeated actions $[G_t.MesA]$, $\forall$**A**, of the human observer-conceptor, *wherefrom the* ***intelligibility*** *stems*. But there this is done in only general and qualitative terms, whereas inside the framework $[IQM\text{-}QM_{HD}]$ these terms are particularized via the modelling postulate $MP(\{\sigma(ms_{G,cw})\}$ and they can be worked out quantitatively.

### (7.III).1.4. An immediate consequence of the assertion *Ass.1*: Possibility of principle to circumvent the Schrödinger equation for unbound microstates or to complete its performance

As soon as the first impression of triviality is dominated and the assertion *Ass.1* is really understood, it immediately points further toward a very surprising and remarkable possibility[114]:

The $QM_{HD}$-predictive mathematical representations should be, not exclusively *verifiable* by a long set of repetition of coding-measurement-successions *[$G_t$.MesA]* that are specified accordingly to the conditions *(a),(b),(c),(d)*, but also ***constructible*** in this same way, ***in a radically factual way, just like inside IQM***.

Indeed: Since the statistical results that are obtained by a long set of repetitions of coding-measurement-successions *[$G_t$.MesA]* realized accordingly to the conditions *(a),(b),(c),(d)*, do verify the prediction $(\mathbf{A},\{|c_j(a_j,t_1)|^2,\forall j\})$ entailed by the $QM_{HD}$ expansion (18) $|\psi_{G,\mathbf{H}}(\mathbf{r},t_1)\rangle/\mathbf{A}=\sum_j c_j(a_j,t_1)|u(\mathbf{r},a_j)\rangle$ when they only *re-construct* its predictive content, these results also *construct* the content of (18) ***exactly in so far that this content is factually true***.

So whenever this is convenient, inside *[IQM-$QM_{HD}$]* it should be possible to circumvent the use of the Schrödinger equation of evolution by generating factually *the predictive contents* of the state-ket of the studied microstate, even though in (18) their *form* is quintessentially of a mathematical nature.

Moreover, in order to conserve also inside *[IQM-$QM_{HD}$]* access to Hilbert-space *calculi* – just like inside $QM_{HD}$ – one could, while producing factually, by measurements, the numerical predictive contents of (18) – to *pour* these contents into a *pre*-imposed Hilbert-Dirac mathematical *form* of a corresponding expansion (18). Nothing interdicts such a procedure. A Schrödinger genesis of a Hilbert-space state-ket is not a necessary element of a Hilbert-Dirac representation of the predictions on results of quantum measurements. And furthermore, a factual generation of a Hilbert-space representation of these predictions *would simply suppress all the restrictions of a purely mathematical nature* that the Schrödinger equation carries with it and imposes not only upon the very possibility to obtain these predictions directly by calculus, but also upon the adequacy of the predictions themselves, in consequence of the so often unavoidable idealizing approximations. So if a factual generation of (18) were realizable this would entirely free (18) of any purely syntactical restriction that does not concern specifically the physical and conceptual nature of the knowledge that is researched. The knowledge contained in (18) would then be obtained just as free of restrictions as it is inside *IQM*.

On the other hand, when it has been possible to write down the Schrödinger equation of the problem and to solve it, once the predictions have been factually generated for a time-value $t_o$ *and* expressed in the mathematical form of an expansion (18) of a state-ket, this would permit to make then use of the equation for calculating the predictions for subsequent times $t_1>t_o$. This would:

- *simplify* notably the mathematical tasks, since the problem of giving the intial state-ket would have been eliminated;

[114] After nearly 100 years of existence the Schrödinger equation has acquired in the minds a sort of a-temporal, definitive necessity. The notion that, historically, it has been unavoidable because the scientific thought – *for the **first** time* – was trying to represent physical entities that are quintessentially and radically un-perceivable, is not clear as yet, up to this very day. So, notwithstanding the various sorts of limitations imposed by the use of this equation, it might seem somehow absurd to conceive that it is possible to do without it. But it will appear that nowadays – ***for unbound microstates*** – this is possible indeed, and it would entail a genuine liberation as well as improvements.

- *clean* the results of any effect of a priori mathematical approximations;
- *extend* the domain of effective utility of the Schrödinger equation.

The new possibility just specified is a direct consequence of the change of origin on the vertical of conceptualization that is introduced by *IQM*, so of the order of conceptualization that now progresses bottom-up.

We conclude: the two distinct representations *IQM* and $QM_{HD}$, far from being unrelated inside $[IQM\text{-}QM_{HD}]$ concerning prediction and verification, quite on the contrary appear now all of a sudden to be soldered to one another in this respect, under a vast new horizon.

### (7.III).1.5. Possibility of a 'normal' relation between a predictive statistics and the individual measurements that verify it

In (6.II).4 we have analysed why and how the reduction problem has emerged, and it appeared that the main cause has been the absence inside $QM_{HD}$ of an organized level of individual conceptualization where to lodge the representations of individual physical entities and operations. But via the conditions of inner semantic self-consistency from (7.III).1.2 and the argument *Arg(Ass.1)* that installs the assertion *Ass.1* we now dispose inside $[IQM\text{-}QM_{HD}]$ of a strongly organized level of individual conceptualization as well as of its vertical bottom-up connection with the statistical level of conceptualization. So nothing withstands any more a normalized relation between a predictive statistics and the measurements that verify this statistic, or invalidate it. Of course, specifying such a normal relation more than this is already very well known by everybody since centuries can only consist of sheer trivialities from *A* to *Z*. Nevertheless, given the particular mentality generated by the so astonishingly long-lasting more or less passive acceptance of the reduction-problem, it seems necessary to state these trivialities explicitly. So:

- We require that any statistical prediction on results of measurements on microstates be globally available *before* the *verification*-measurements *begin*. We state this common-sense requirement no matter how the predictive statistic has been constructed – whether mathematically or factually – and how it is expressed – as a formal law $(A,\{|c_j(a_j,t_1)|^2, \forall j\})$, or as a factually-probabilistic description (9) $(D/A)(ms_{Gt1})\equiv\{(\varepsilon,\delta,N_0)\text{-}\pi(a_j,t_1)\}_{Gt1}$.

If the statistical-probabilistic prediction has been constructed factually as in the case of (9) $(D/A)(ms_{Gt1})\equiv\{(\varepsilon,\delta,N_0)\text{-}\pi(a_j,t_1)\}_{Gt1}$ from *IQM*, the requirement formulated above means that before any action of verification begins, a sequence of many repetitions of an *individual* coding-measurement-succession $[G_{t1}.MesA]$ *that – itself – is* ***not*** *expressed in statistical terms*, has been *globally repeated* by adjusting the parameters $(\varepsilon,\delta,N_0)$ from (9) until the global statistical distribution that is obtained exhibits the desired degree of stability of the normed cardinals of the various classes of mutually distinct registered results $a_j$ (the $\pi(a_j,t_1), \forall j$), i.e. until all *these stay constant* ***up to*** $(\varepsilon,\delta)$ *as long as* $N_0$ *is kept invariant*. The involved predictive statistical distribution is considered to be available only once this has been done[115]. Now:

- A factually constructed probabilistic prediction (9) is verified by its re-construction (*IQM* and *Ass.1*).

- An only mathematically calculated probabilistic prediction $(A,\{|c_j(a_j,t_1)|^2, \forall j\})$ is verified *iff* the corresponding stable factually generated $(\varepsilon,\delta,N_0)$-probabilistic distribution (obtained as required in *IQM* for (9) *and* accordingly to the conditions of compatibility

[115] In the top-down approach from $QM_{HD}$ this constructive phase is absorbed in the abstract construction of the state-ket $|\Psi_{G,H(A)}(\mathbf{r},t>$ via calculus, i.e. in the construction of the Schrödinger equation of the problem, of its solution, and of the determination of the initial state-ket $|\Psi_{G,H(A)}(\mathbf{r},t_o>$.

*(a),(b),(c),(d)* from *(7.III).1.2*) is *(ε,δ,N₀)*-identical to this mathematically established prediction $(\boldsymbol{A},\{|c_j(a_j,t_1)|^2,\forall j\})$.

Any experimentalist, no doubt, conceives and works precisely in this way.

While on the individual level of conceptualization and during its own *individual* time-interval $(t_f\text{-}t_1)$, *each **one*** individual coding-measurement succession creates *outside* the stable formal statistical chain *(ch)(QM$_{HD}$)*, *one* whole thread of operational evolution $([G_{t1}.MesA]\rightarrow a_j)$ that begins with an operation of generation (13') and is closed by the registration of its own result $a_j$.

In this way no mixture is made between individual time parameters and statistical time parameters: When the individual threads of one coding-measurement-succession $[G_{t1}.MesA]\rightarrow a_j$ are repeatedly performed for verifying a statistical-probabilistic predictive distribution, the expression of this distribution – the expansion $|\psi_{G,\boldsymbol{H}}(t_1)>/\boldsymbol{A}$ and the corresponding predictive law $(\boldsymbol{A},\{|c_j(a_j,t_1)|^2,\forall j\})$ or, equivalently, the description (9) $(D/A)(ms_{Gt1})\equiv\{(\varepsilon,\delta,N_0)\text{-}\pi(a_j,t_1)\}_{Gt1}$ – are already *just history*, *just an already fully **pre**-constructed **reference** that waits to be made use of later, whenever one wants that (admitting that the substrata and the "exterior conditions" have remained unchanged)* [116]. And at such a *public* time when finally – *a posteriori* – a *global* verifying-confrontation is realized, nothing withstands it conceptually any more, because at *that* public time the statistical-probabilistic prediction $(\boldsymbol{A},\{|c_j(a_j,t_1)|^2,\forall j\})$ – no matter whether established mathematically or factually – and on the other hand the *always factually constructed **verifying***-probabilistic-distribution (9") of ***all*** the ***achieved*** results $a_j$ produced by the *repeated* coding measuring successions $([G_{t1}.MesA]\rightarrow a_j)$, ***are** mutually homogeneous* from a conceptual viewpoint, they are just two sets of numbers with a same sort of genesis, their contents are fully of the same semantic nature and so they are ready for being compared quantitatively.

*No scandalous reduction is necessary any more.* Everything is plainly normal, and is intelligible. The conclusion from the Figure 7 is fully incorporated.

### (7.III).1.6. Global formulation of the purpose formed inside (7.III).1

We conclude on (7.III).1.2.

In consequence of the inclusion of the general reference-and-embedding structure *IQM*, the framework *[IQM-QM$_{HD}$]*, such as it has been specified in (7.III).1, offers the possibility of a radically bottom-up theory of the microstates.

*The probabilistic predictions from the theory can be constructed factually, by measurements of which the results are uploaded in a Hilbert-Dirac form.* For in its predictive essence the state-ket $|\psi_{G,\boldsymbol{H}}(\boldsymbol{r},t)>$ is nothing more then a collection of Hilbert-Dirac expansions (18), as these are expressed by (22).

*And such an approach, while it generates representation, generates also the corresponding intelligibility.*

We want to realize this possibility. We are at the time of big data and of a vertiginous progress of the computing tools. The limitations that stem from the nowadays constraint to make use exclusively of the Schrödinger equation for generating predictive and verifiable consensual knowledge on microstates, can be dissolved. Only the *conceptual-semantic-operational* restrictions from *IQM* must be all kept active.

---

[116] It is not any more a unique *process*, like in (19), that, ***while*** it is statistical i.e. *abstract* and carries *in* it all the formal seeds '$c_j(t_1)$' of the statistical predictive law $(\boldsymbol{A},\{|c_j(a_j,t_1)|^2,\forall j\})$, is posited like in *(19)* to ***also*** generate *actual* verifying individual and physical data.

***But such restrictions are precisely the generators of "scientific knowledge"***.

## (7.III).2. FACTUAL CONSTRUCTION OF A HILBERT-DIRAC REPRESENTATION OF THE PROBABILISTIC PREDICTIONS ON THE RESULTS OF QUANTUM-MEASUREMENTS ON UNBOUND MICROSTATES

### (7.III).2.1. Born's postulate, Hilbert-space structure, and Gleason's 1957-theorem

***The Schrödinger-de Broglie Wave-Mechanics versus Born's postulate.*** Inside the Schrödinger-de Broglie Wave-Mechanics, consider the wave-function $\psi(\mathbf{r},t)$ defined by the Schrödinger equation of a problem concerning a studied microstate $ms_G$. Let $c(a_j,t)$ denote the coefficient from the term of index '*j*' in the expansion $\psi(\mathbf{r},t)/\mathbf{A}=\sum_j c_j(a_j,t)(u(\mathbf{r},a_j,)$ with respect to the observable $\mathbf{A}$. Born's postulate – let us denote it *Bp* posits the numerical equality

$$|c(a_j,t|^2 \quad =_{Bp} \quad \pi(a_j,\psi), \quad \forall j \qquad\qquad (Bp)$$

where: the symbol '$=_{Bp}$' is to be read 'numerically equal according to Born's postulate' [117] and $\pi(a_j,\psi)$ is the numerical value of the individual predictive probability for the outcome of the eigenvalue $a_j$ if $\mathbf{A}$ is measured on the microstate $ms_G$ represented by the wave-function $\psi(\mathbf{r},t)$.

***Dirac's formulation of $QM_{HD}$* versus *Bp*.** In the first edition of his book ([1930]) Dirac has had the very fertile idea to represent the solution $\psi(\mathbf{r},t)$ of Schrödinger's equation of a problem by a *vector* in a Hilbert-space $\mathcal{H}$. He called such a vector 'the state-ket' associated to the studied microstate and he denoted it $|\psi(\mathbf{r},t)>$. The normed set of eigen-*functions* $\{u_j(\mathbf{r},a_j)), \forall j\}$ of a mechanical qualifying quantity *A* from the wave-mechanics was made use of for constructing a "basis of eigen*ket* $\{|u_j(\mathbf{r},a_j)>), \forall j\}$" each element of which being represented as a vector tangent to $\mathcal{H}$ that specifies a "direction" *in* $\mathcal{H}$. So, when the expansion (18) is re-written in "ket"-terms as $|\psi(\mathbf{r},t)>/\mathbf{A}=\sum_j c_j(a_j,t)|(u(\mathbf{r},a_j,)>, \forall\mathbf{A}$, one can construct the direction in $\mathcal{H}$ of the vector $|\psi(\mathbf{r},t)>$ such as to have $|c(a_j,t)|^2 \equiv_{constr.\mathcal{H}} |Pr._j|\psi>|^2, \forall j$. This permits to then posit that

$$(|c(a_j,t)|^2 \quad =_{Bp} \quad \pi(a_j,\psi)) \quad \equiv_{constr.\mathcal{H}} \quad |Pr._j|\psi>|^2, \quad \forall j \qquad\qquad (23)$$

where: the symbol "$\equiv_{constr.\mathcal{H}}$" is to be read "identical inside $\mathcal{H}$ by construction", and $Pr._j|\psi>$ is the projection of $|\psi>$ on the direction in $\mathcal{H}$ defined by the eigenket $|u_j>$ with index *j* from the basis $\{|u_j(\mathbf{r},a_j)>, \forall j)\}$.

It seems that in Dirac's initial reformulation in Hilbert-space terms of the de Broglie-Schrödinger wave-mechanics, the identity "$\equiv_{constr.\mathcal{H}}$ " contained in (23) was more or less implicitly considered to be a formal feature *specifically tied with the microstates*. But this is not so. The mentioned identity can be written in consequence of exclusively the geometrical features introduced by the Hilbert-space mathematical structure; it only applies *a **general** possibility* offered by the mathematical Hilbert-space structure, to the particular case of the representation of a microstate in terms of elements of this structure. This appears below.

---

[117] This assertion is not at all obvious a priori. It is only suggested by the significance of a Fourier coefficient in the classical electromagnetism (de Broglie has made use of the suggestion) and it is active in the association between the eigenket $\{|u(r,a_j>$ and the eigenvalue $a_j$ in the equation of eigenket and eigenvalues of $\mathbf{A}$.

***Gleason's 1957 theorem (Gth).*** In 1957 Gleason proved an important theorem. Gleason's own formulation is reproduced below:

> *Gth.* « Let $\mu$ be a measure on the closed subspaces of a separable (real or complex) Hilbert-space $\mathcal{H}$ of dimension at least three. *There exists* a positive semi-definite operator $T$ of the trace class such that for all closed sub-spaces $\mathcal{A}$ of $\mathcal{H}$
>
> $$\mu(\mathcal{A}) = \text{trace}\ (TP_{\mathcal{A}})$$
>
> where $P_{\mathcal{A}}$ is the orthogonal projection of $\mathcal{H}$ onto $\mathcal{A}$. »

So *Gth* establishes the possibility of *a systematic mathematical connection between the mathematical structure called a Hilbert-space and the concept of a "measure" in the abstract purely mathematical sense*. This possibility *includes* the particular case of a *probability*-measure, ***any*** probability measure, as well as the still more particular case of probability-measures of outcomes of quantum-measurements operated on a microstate.

> *But the theorem itself is strictly **independent** of the concept of "microstate"*: Gleason's theorem concerns *also* Quantum Mechanics, but only in particular and via the general concept of probability-measure.

***$QM_{HD}$ versus (Gth and Bp).*** Notwithstanding *Gth*, Born's postulate *Bp* remains necessary inside $QM_{HD}$.

Indeed the equality $|c(a_j,t)|^2 =_{Bp} \pi(a_j,\psi)$ is not entailed by the purely mathematical genesis of the numbers $|c(a_j,t)|^2$ via the use of the Schrödinger equation. This equation, by itself, involves nothing concerning probabilities tied with microstates; it does not endow the numbers $|c(a_j,t)|^2$ with a probabilistic significance, nor a fortiori does it specify the numerical value of an individual probability $\pi(a_j,\psi)$ of the possible factual outcome $a_j$ when a measurement *MesA* is performed upon the studied microstate.

*Both* features mentioned above are separately asserted by Born's postulate.

As for *Gth*, it entails (in particular) *only* that ***if*** "an individual probability measure $\pi(a_j,\psi)$" ***is GIVEN AS SUCH***, ***then*** its numerical value can be expressed in a Hilbert space $\mathcal{H}$ as the projection $|Pr._j|\psi\rangle|^2$ of a vector $|\psi\rangle$ from $\mathcal{H}$, as written in (23) accordingly to Dirac's reformulation of the wave-mechanics: this reformulation does not replace the content of Born's postulate that *performs* precisely that required "giving as such".

So in $QM_{HD}$ Born's postulate *cannot* be suppressed.

***Gth versus ([IQM-$QM_{HD}$].*** But consider now *Gth* from the point of view of the framework *[IQM-$QM_{HD}$]* that incorporates *IQM*.

> Gleason's theorem permits also to regard the second equality from (23) as just a *generally* possible translation of the mathematical representation of the concept of probability, from Kolmogorov's language into the Hilbert-space language, asserted on purely formal grounds **[118]**.

---

[118] Pitowsky (2008) has drawn this same essence in connection with 'quantum logic'. He asserts that for him "Hilbert spaces are a new theory of probabilities". For me however the mentioned translation from Kolmogorov's language into the Hilbert-space language does not suffice for generating the new mathematical framework of an exhaustive Theory of the Factual Probabilities as expressed by *the probability-trees of transferred description in the sense of IQM* (cf. (3.I).6), because trees introduce a concept of probability that is *far more complex than Kolmogorov's one*, a probabilistic structure where can be located, not only the Kolmogorov probability spaces, but also the geneses of these, inner structural features of elementary events and events from these, as well as various correlations between the measures from the Kolmogorov spaces. Whereas Gleason's theorem offers a new and very efficient alternative mathematical language for representing exclusively the involved *probability **measures***; a language that offers abstract location for also *a semantic labelling* of the numerical probability-values, via the elements from bases in the utilized Hilbert-space, but nevertheless an insufficiently rich and complex language for exhausting the new concept of Factual Probability revealed by *IQM*.

And since anyhow the first equality from (23) has to be *given*, ***nothing*** *hinders to "give" it* ***factually*** *via the probability measures (5') (ε,δ,$N_0$)-{(π($a_j$)}$_G$, ∀j, ∀A, or as in (9") from IQM*, ***just as anyhow it HAS to be done*** *for the unavoidable verification-measurements, as the Ass.1 points out*. And *if* the set *{π($a_j$,ψ), ∀j}* is given in this factual way, as in *IQM*, instead of being drawn from the Schrödinger equation of the problem, *then* each element *π($a_j$,ψ)* from the set emerges endowed by construction with – ***both*** – the significance of the individual probability of the outcome $a_j$, *and* with the specified numerical value of this individual probability.

***In this case BORN'S POSTULATE CAN BE SUPPRESSED***.

### (7.III).2.2. Factual-formal bottom-up construction of a Hilbert-space representation of quantum measurements on unbound microstates *without* quantum fields

We come back to the remark that nothing hinders to apply *Gth* in association with the *factually* constructed probability laws (5') *(ε,δ,$N_0$)-{(π($a_j$)}$_G$, ∀j, ∀A*, defined in *IQM* involved in the transferred description (9") where are *explicitly* included also the meta-correlations ***Mπc($G_t$))$_{AB}$*** between distinct probability laws that are involved in (5'). Indeed one can construct factually and progressively the expansions (22) *|ψ>/A* ***∀A***, from *$QM_{HD}$*, via individual coding-measurement-succession *[$G_{t1}$.MesA]* exactly like in *IQM*, but furthermore under a priori organized conditions of mutual consistency between *IQM* and *$QM_{HD}$* in the sense defined in (7.III).1.2 and by lodging them progressively in the formal Hilbert-space mould from (22*)*.

Any discording breach on the statistical level of conceptualization, like that indicated in the Fig.6 would be avoided a priori by such a *uniformly* bottom-up approach, from zeros of local microscopic knowledge and up toward the classical the classical level and beyond, without limitation.

The global probabilistic result would invest the *whole* new tree-like structure of the factual quantum probabilities of the types represented in Fig.2 and Fig.4 that incorporate *both* the Kolmogorov-*spaces* from the tops of the branches of a probability-tree *and* the *correlations* between the measures from these spaces. These, furthermore, *would somehow emerge with a $QM_{HD}$ mathematical expression, since they are developed inside the mathematical expression (22) of an expansion (22) of a $QM_{HD}$ state-ket.*

And ***Born's postulate would become superfluous because a factually constructed predictive probability emerges endowed by construction with its probabilistic significance and its numerical value (that furthermore is certainly factually true)***[119].

At a first sight it might seem that such a program is not "normal" or even interdicted by some implicit commandments on the correct use of mathematics. But in fact we only are on the edge of an unexpected and deep *unification* of the so basic concept of probability. This unification has been held back by the historically imposed initial top-down penetration into the microscopic physics, via extrapolation of procedures that are *tied* with a direct perceptibility and manipulation of the studied entities. But any extrapolation involves unavoidably a cortege of arbitrary assumptions and these, unavoidably, lead to false problems and stagnation. Such extrapolating procedures have been accepted only because the physicists were still entirely unaware of the organic ties between *the human cognitive situation* with respect to a given domain of physical entities and the conceptual-mathematical representation of these entities.

[119] Thereby would finally emerge a surprisingly radical answer to many very profound researches where it has been tried to explain Born's postulate or to obtain it deductively (cf. the basic bibliography in Raichman [2003] (with, in particular, Anandan [1999], Deutch [1999], Saunders [2002], etc.).

But now these ties become obvious and *IQM* makes explicit use of them. When we act bottom-up accordingly to *IQM* we discover ***first*** the conceptual-operational foundations, and so *the formal adequacies or in-adequacies of representation can take form step by step, while constructing*. This eliminates the necessity of extrapolations; it draws attention upon the specific local problems of representation – and of *meaning* – as these arise progressively, thus permitting to deal with them successively in a way as adequate as possible.

*Nothing interdicts to proceed in this way*; quite on the contrary. In the cognitive situation in which a human observer is placed when he researches an optimally adequate representation of physical entities that – quintessentially – he cannot directly perceive and manipulate, *IQM* guides toward the optimal procedure, and this is *factually* constructive.

We begin below by considering the unbound microstates *without quantum fields*.

### *(7.III).2.2.1. The coding-postulate for unbound microstates $ms(unbound,1)_{G(n-c)}$ with non-composed operation of generation*

The analyses from (7.III).1 have suggested that the sort of measurement-evolution that is presupposed inside $QM_{HD}$ by the *BBGPM* approach can – in coherence with the Hilbert-Dirac representation – be conceived to perform implicitly a coding-measurement-evolution of the general form (20). But it also has appeared there that likely the mentioned supposition has a restricted validity in this sense that it can be 'understood' only in the absence of quantum fields. (Why this is so will appear later). So in (6.II).4.3 – founded upon the analysis of the method 'time-of-flight' for measuring the basic momentum observable, and by reference also to the Stern-Gerlach procedure for spin-measurements – we have *admitted* the efficiency of a coding-measurement-evolution of type (20) for unbound microstates with non-composed operation of generation, so in the absence of quantum fields.

Such microstates will be denoted $ms(unbound,1)_{G(nc)}$.

For these we shall now *postulate* explicitly, in strict and detailed $[IQM\text{-}QM_{HD}]$-terms, the coding-measurement-evolution of the general type (20) required by the condition *(d)* of inner consistency and that is compatible with the other conditions, *(a),(b),(c)*.

***Recall.*** In (6.II).1 it has been found that inside $QM_{HD}$ works implicitly de Broglie's general 'corpuscular-wave' model. In (6.II).2 – via the modelling postulate $MP(ms_{G,cw})$ – this general model has been brought to consistency with the general *IQM* concept (1) of an operation of generation $G$ of a specimen $\sigma(ms_{Gt})$ of the studied microstate. So here the coding-measurement-evolution admitted in $[IQM\text{-}QM_{HD}]$ for the particular sort of microstate denoted $ms(unbound,1)_{G(nc)}$ will have to be specified in consistency with the modelling $MP(ms_{G,cw})$ and with the observable $\boldsymbol{A}$ to be measured [120]; *and* furthermore it must obey all the conditions of semantic consistency formulated in *(7.III).1.2.*

Consider now a coding-measurement-succession $[G_{t1}.Mes\boldsymbol{A}]$.

- According to the conditions of semantic consistency, in (13'') we posit: $[G_o.(t\text{-}t_o)]\equiv G_t$, where: $G_o \Leftrightarrow ms_{Go}$, $[G_t=F(G_o ,EC,(t\text{-}t_o))$, and the external conditions '*EC*' during $(t_1\text{-}t_o)$ are those expressed in $QM_{HD}$ by the hamiltonian operator $\boldsymbol{H}$ from the Schrödinger equation of the problem. So $G_{t1}$ has to be written as $G_{t1}=F(G_o,\boldsymbol{H},(t_1\text{-}t_o))$ ('*F*' means: 'a functional of').

- According to the modelling postulate $MP(ms_{G,cw})$ the operation $G_{t1}$ introduces via $G\equiv G_o$ a specimen $\sigma(ms_{Go})$ of the initial state of the studied micro-state $ms_{Gt1}$ that is represented by a physical individual wave-f*unction* $\Phi_{Go}(\mathbf{r},t=a_{\Phi}(\mathbf{r},t)exp((2\pi i/h)\beta(\mathbf{r},t))$ that is unknown (cf.*(6.II).4.2.2*). So until the moment $t_1$ when the act *MesA* of measurement-evolution begins, the evolution of $\sigma(ms_{Go})$ is represented by the action upon the *individual physical* wave-function $\Phi_{Go}(\mathbf{r},t)=a_{\Phi}(\mathbf{r},t)exp((2\pi i/h)\beta(\mathbf{r},t))$ from $\sigma(ms_{Go})$ of the exterior fields from the hamiltonian $\mathbf{H}$ that determines also the evolution of the statistical $QM_{HD}$ descriptor $|\psi_{G,\mathbf{H}}(t_1,t_o)>$.

[120] In the absence of any model the concept itself of coding-measurement-evolution disappears, so no theory of the microstates is possible any more, as stressed inside *IQM*.

- In *(6.II).4.3* we have admitted the *BBGPM* implication that a measurement hamiltonian operator **H(A)** that commutes with the measured observable **A** – if it works on $\sigma(ms_{Gt1})$ ***after t₁*** and *in the absence of quantum fields* – insures for the wave-function $\Phi_{Gt1}(\mathbf{r},t)= a_{\Phi}(\mathbf{r},t)exp((2\pi i/h)\beta(\mathbf{r},t)$ of $\sigma(ms_{Gt1})$ where $t\geq t_1$ a structure of wave-movement represented by an eigenket of **A**. While correlatively, for the corpuscular-like singularity in the amplitude of $\Phi_{Gt1}(\mathbf{r},t)= a_{\Phi}(\mathbf{r},t)exp((2\pi i/h)\beta(\mathbf{r},t)$ it involves a dynamic that leads it into a space-domain $\Delta\mathbf{r}_j$ (or a space-time domain $(\Delta r\Delta t)_j$) that is in a one-one relation with a given eigenvalue $a_j$ of **A**.

***Formulation***. The above recall leads to the following $[IQM\text{-}QM_{HD}]$ version of a coding-postulate of the general form (20):

> ***P(cod)***$_{G(nc)}$. A coding-measurement-evolution *MesA* from a succession $[G_{t1}.MesA]$ performed upon a microstate $ms(unbound,1)_{G(nc)}$, admits the general representation:

$$[(G_{t1}\rightarrow\sigma_{\Phi}).MesA(\sigma_{\Phi})] \;\rightarrow_{\mathbf{H(A)}}\; [\text{any group of marks registered inside } (\Delta\mathbf{r}\Delta t)_j \text{ means "}a_j\text{"}] \quad (20')$$

> ($\sigma_{\Phi}$: abbreviation for $\sigma(ms_{Gt1})$; $G_{t1}\equiv[G_o.(t_1\text{-}t_o)]$: a functional $F(G_o,\mathbf{H(A)},(t_1\text{-}t_o))$.
>
> If in particular it is supposed that the coding-measurement-evolution $MesA(\sigma_{\Phi})$ is performed by starting it at the time $t_o$ when the initial operation of generation $G_o$ finishes, then we make use of the corresponding particular form of $P(cod)_{G(nc)}$ :

$$[(G_o\rightarrow\sigma_{\Phi}).MesA(\sigma_{\Phi})] \;\rightarrow_{\mathbf{H(A)}}\; [\text{any group of marks registered inside } (\Delta\mathbf{r}\Delta t)_j \text{ means "}a_j\text{"}]$$

So, like (20), the postulate $P(cod)_{G(nc)}$ concerns an *individual specimen of the studied microstate*: Inside $[IQM\text{-}QM_{HD}]$ this postulate 'explains' the non-analysed $QM_{HD}$-postulation of ‘emergence' of an eigenvalue $a_j$ of the measured observable **A**[121]. We notice that indeed in (20’) the $QM_{HD}$ hamiltonians **H(A)** and **H** work on the level of individual conceptualization, i.e. upon individual *physical* entities. This is coherent with our critique of the reduction problem in *(6.II).4.4.3*.

The coding postulate (20') is presupposed throughout the constructive action from the following point.

---

[121] We are so deeply used to the purely mathematical and purely statistical representations from $QM_{HD}$, that the content of the whole point *(7.III).2.2.1* might seem queer inside a work of theoretical physics. But the reader is asked to remember that we want to root quantum mechanics in factuality, and in a non-perceivable and as yet a-conceptual physical factuality. This requires suppression of the inertial psychological refusals induced by feebly intelligible, purely algorithmic top-down representations, supported by philosophical interdictions that vitiate the modern microphysics since more than 100 years. *Human minds act on the basis of models*. This is a fact and has to be incorporated explicitly and fully to the processes of generation of scientific knowledge.

### *(7.III).2.2.2. Factual-formal construction of a Hilbert-Dirac representation for the results of coding-measurement-evolutions (20') on microstates $_{G(nc)}$*

We shall now proceed in a radically constructive way. We go back to the $QM_{HD}$-concept (22) $|\psi_{Gt1,\mathbf{H}}(\mathbf{r},t_1)> \approx_{\text{pred.}} \{\forall A, \forall t_1, \ |\psi_{Gt1,\mathbf{H}}(\mathbf{r},t_1)>/A\}$ that has been recognized to be the core of the $QM_{HD}$-representation of predictions. We make the following assertion *Ass.2*:

***Ass.2***

For a microstate of the type $ms(unbound,1)_{G(nc)}$ it is possible to generate inside $[IQM\text{-}QM_{HD}]$, via a conceptual-*factual* procedure, a Hilbert-space representation that:

* ***(Ass.2)a.*** In prediction of first order factual probability laws is equivalent to the $QM_{HD}$ Hilbert-Dirac expression (22) $\{\forall A, \forall t_1, \ |\psi_{Gt1,\mathbf{H}}(\mathbf{r},t_1)>/A\}$ and furthermore asserts also *defined* correlations – in the sense of (11) and (11') – between the first order probability laws.

* ***(Ass.2)b.*** Emerges *intrinsically* endowed by construction with all its *probabilistic meanings* and with ***verified*** *factual truth of all the involved numerical values,* inside the limits of the a priori chosen and arbitrarily small $(\varepsilon,\delta,N_0)$ imprecisions from the corresponding *IQM*-description (9"). *Whereby Born's postulate becomes* ***unnecessary***.

* ***(Ass.2)c.*** By construction, is entirely independent of the Schrödinger equation of the problem.

***Arg(Ass.2). Preliminaries.*** Consider the $QM_{HD}$ state-ket $|\psi_{Gt1,\mathbf{H}}(\mathbf{r},t)>$ of a microstate $ms_{Gt1}$ of the type $ms(unbound,1)_{G(nc)}$.

- The assertion *Ass.1* and the corresponding argument have established that ***if*** the probabilistic predictions of any given $QM_{HD}$ state-ket $|\psi_{Gt1,\mathbf{H}}(\mathbf{r},t)>$ are available and are *verified* by the repeated realization of coding-measurement-successions $[G_t.MesA]$, $\forall A$, $\forall t$ that satisfy the conditions of inner consistency *(a),(b),(c),(d)* defined in *(7.III).1.2*, ***then*** these predictions are necessarily $(\varepsilon,\delta,N_o)$-identical to those asserted by the factual $(\varepsilon,\delta,N_0)$-probability laws (5') from the *IQM*-description

$$(D_M(ms_G))_t \equiv [(\varepsilon,\delta,N_o)\text{-}\{\pi(G_t,a_j)\}, (M\pi c(G_t))_{AB}], \ \ \forall A, \forall AB, \ \forall j, \ \forall t \qquad (9")$$

constructed by use of the same set of coding-measurement-successions. So we select one given description of this kind in order to work with it and we denote it in short $D_M$.

- So ***if*** inside $[IQM\text{-}QM_{HD}]$ a *factually constructed* state-ket of the Hilbert-Dirac ***formal type*** denoted $|\psi_{Gt1,\mathbf{H}}(\mathbf{r},t_1)>_{fact(DM)}$ can be associated to the selected *IQM*-description (9") $D_M$ – i.e. to the *factual* probability laws (5') from $D_M$ and to the corresponding correlations $(M\pi c(G_t))_{AB}]$ – ***then*** *the Ass.1*, (23) and *Gth* entail that up to imprecisions $(\varepsilon,\delta,N_0)$ we *must* have the equalities:

$$\left(|c_j(a_j,t1)_{fact(DM)}|^2 \ = \ \pi(G_{t1},a_j)_{fact(DM)}\right)_{(\varepsilon,\delta,No),\,(Ass.1,\,(23),\,Gth)}, \quad \forall j, \ \forall A, \ \forall t, \qquad (23')$$

where:

-$|c_j(a_j,t)_{fact(DM)}|^2$ denotes the squared modulus of the expansion coefficient of index $j$ in the expansion with respect to the observable $A$ of $|\psi_{Gt1,\mathbf{H}}(r,t_1)>_{fact(DM)}$;

- $\pi(G_{t1},a_j)_{fact(DM)}$ is the $(\varepsilon,\delta,N_o)$-probability of $a_j$ according to the factually constructed description (9") $D_M$ from *IQM*.

- the global index '$_{(\varepsilon,\delta,No),\,(Ass.1,\,(23),\,Gth)}$' is to be read as: "with $(\varepsilon,\delta,N_o)$-approximation and accordingly to *Ass.1*, (23) and *Gth*.

***Arg(Ass.2)a).***

By use of (23') we shall now generate factually the state-ket $|\psi_{Gt,H}(r,t)\rangle_{fact(DM)}$ without making any use of a Schrödinger equation of the problem[122].

***1. Specification of the descriptive elements***. We need:

***(1a).*** The ***form*** of an as yet unknown state-ket $|\psi_{Gt1,\mathbf{H}}(r,t_1)\rangle_{fact(DM)}$ conceived a priori as a Hilbert-space vector of the $QM_{HD}$-type (22) $|\psi_{G,\mathbf{H}}(\mathbf{r},t_1)\rangle \approx_{\text{pred.}} \{|\psi_{G,\mathbf{H}}(\mathbf{r},t_1)\rangle/\mathbf{A}, \forall \mathbf{A}, \forall t_1\}$ where the $QM_{HD}$-observable $\mathbf{A}$ introduces a basis of eigenket $\{|u_j(\mathbf{r},a_j)\rangle, \forall j\}$. Via this form we can lodge the factually generated data in a Hilbert space $\mathcal{H}$ associated to the microstate to be studied. So for each observable $\mathbf{A}$ we write the corresponding expansion-form

$$|\psi_{Gt1,\mathbf{H}}(r,t_1)\rangle_{fact(DM)}/\mathbf{A} = \sum_j e^{i_\alpha(A,j)} \, |c_j(a_j,t)_{\mathbf{fact(DM)}}| \, |u_j(\mathbf{r},a_j)\rangle, \ \forall j, \forall t_1 \qquad (24)$$

(the expansion coefficients are written as a product in order to be able to treat separately the question of the numerical values to be assigned to the two factors $e^{i_\alpha(A,j)}$ and $|c_j(a_j,t)_{\mathbf{fact(DM)}}|$).

***2. Specification of the descriptors that generate the contents to be lodged in the form (24)*** namely the $(\varepsilon,\delta,N_o)$-probability laws $\{\pi(G_{t1},a_j)\}, \forall \mathbf{A}, \forall t$, from the description (9") $D_M$ of the microstate to be studied $(ms_G))_t$. These contents are obtained accordingly to *IQM*, by use of (13'') and of coding-measurement-successions $[G_{t1}.MesA]$ that obey the coding postulate (20') $P(cod_{\mathbf{G(nc)}})$ and so respect all the conditions *(a),(b),(c),(d)* of semantic compatibility defined in *(7.III).1.2*).

This exhausts the phase 1 of specification of the needed descriptive elements.

***3.*** Consider now the *general mathematical form* (24), still void of any specified content. For each coefficient from (24) we have:

$$c_j(a_j,t)_{\mathbf{fact(DM)}} = e^{i_\alpha(A,j)} \, |c_j(a_j,t)_{\mathbf{fact(DM)}}| \ \forall j, \ \forall t_1 \qquad (25)$$

*We concentrate first upon the factors* $|c_j(a_j,t)_{\mathbf{fact(DM)}}|$. So far in (25) the expansion coefficients $c_j(a_j,t)_{\mathbf{fact(DM)}}$ – still elements of a mere void mould – are endowed with exclusively the general geometrical meaning that is assigned to them in (23) by their purely mathematical definition, namely that of the projections onto the directions introduced by the eigenket $\{|u_j(\mathbf{r},a_j)\rangle\}$ of $\mathbf{A}$ in the Hilbert-space $\mathcal{H}$ of the factual state-ket $|\psi_{Gt1,\mathbf{H}}(r,t_1)\rangle_{fact(DM)}$ that is to be constructed. But now we *posit* by a *"constructive definition"* that the conditions of emergence of (23') *are all satisfied*. Then, accordingly to the assertion *Ass.1* and on the basis of Gleason's theorem *Gth*, up to an $(\varepsilon,\delta,N_0)$-imprecision we *do* have the equality:

$$(|c_j(a_j,t_1)_{\mathbf{fact(DM)}}|^2 = \pi(G_{t1},a_j)_{fact(DM)}, \forall j\,)_{constr.def,\,(\varepsilon,\delta,No)} \qquad (26)$$

where:

- $\pi(G_{t1},a_j)_{fact(DM)}$ is the $(\varepsilon,\delta,N_o)$-probability of $a_j$ according to the factually constructed description (9") from *IQM*;
- $c_j(a_j,t_1)_{fact(DM)}$ is the element of a formal mould defined in (25);
- the global index "$_{constr.def,(\varepsilon,\delta,No)}$" is to be read "by constructive definition and with $(\varepsilon,\delta,N_o)$-precision";
- the index "$_{fact(DM)}$" is to be read "factual, involved by $D_M$".

[122] What follows is of the same nature as constructing a differential equation of a form that is adequate for subjecting to it a given domain of facts to be described mathematically. But in our case the adequate mathematical form is that of a Hilbert-vector that satisfies (23).

So we can now re-write (24) with the following '*factual*-mathematical' form of the right-hand member where instead of the form $|c_j(a_j,t_1)_{fact(DM)}|^2$ is inserted the number $\sqrt{\pi}(G_{t1},a_j)_{fact(DM)}$ estimated while constructing the *IQM*-description $D_M$:

$$|\psi_{Gt1,\mathbf{H}}(r,t_1)>_{fact(DM)}\ /\mathbf{A} = \sum_j e^{i_\alpha(A,j)}.\ \sqrt{\pi}(G_{t1},a_j)_{fact(DM)}.\ |u_j(\mathbf{r},a_j)>),\quad \forall j,\ \forall t_1, \qquad (27)$$

The same procedure is valid for the spectral decomposition of $|\psi_{Gt,H}(r,t_1)>_{fact(DM)}$ with respect to any other dynamical observable different fro $\mathbf{A}$.

This settles for any observable the question of the absolute numerical values of the expansion coefficients from (24), (25).

*Consider now the imaginary factors* $e^{i_\alpha(A,j)}$ *from (24), (25).* In the expression $e^{i_\alpha(A,j)}$ the observable $\mathbf{A}$ is a variable. When one passes from $\mathbf{A}$ to another *given* observable $\mathbf{B}$ the $QM_{HD}$ concept of a state-ket $|\psi_{Gt1,\mathbf{H}}(\mathbf{r},t)>$ involves conditions of mutual consistency between the expansion-coefficients of $|\psi_{Gt1,\mathbf{H}}(\mathbf{r},t)>$ with respect to $\mathbf{A}$ and the expansion-coefficients of $|\psi_{Gt1,\mathbf{H}}(\mathbf{r},t)>$ with respect to $\mathbf{B}$. These conditions are taken into account in Dirac's "theory of transformations" and they can be specified via a trivial lemma *L(Ass.2)* made explicit inside $QM_{HD}$ for any state-ket $|\psi_{Gt1,\mathbf{H}}(\mathbf{r},t)>$:

***L(Ass.2).*** If in (25) an *arbitrary* set $\{e^{i_\alpha(A,j)}\}$ of complex factors is introduced for the observable $\mathbf{A}$, then Dirac's theory of transformations determines *consistently* with this initial choice, all the complex factors to be introduced in all the other expansions of $|\psi_{Gt1,\mathbf{H}}(\mathbf{r},t)>$ corresponding to any other given dynamical observable $\mathbf{B}$, that does not commute with $\mathbf{A}$, *so* $[A,B]\neq 0$.

***Proof inside*** $QM_{HD}$ ***of L(Ass.2).*** Consider the expansion

$$|\psi_{Gt}(\mathbf{r},t)>/\mathbf{B} = \sum_k e^{i_\gamma(B,k)}\,|d_k(t,b_k)|\,|v_k(\mathbf{r},b_k)>,\quad k=1,2...K,\quad \forall t \qquad (18')$$

of a state-ket $|\psi_{Gt}(\mathbf{r},t)>$ on the basis $\{|v_k(\mathbf{r},b_k)>\}$ of eigenket introduced in $\mathcal{H}$ by a given observable $\mathbf{B}$ that does not commute with $\mathbf{A}$, so $[A,B]\neq 0$, where the numbers $e^{i_\gamma(B,k)}\,|d_k(t,b_k)|$ are the expansion coefficients. For *any* given value of the index *k* we have inside $QM_{HD}$ that

$$<v_k(\mathbf{r},b_k)\,|\psi_{Gt}(\mathbf{r},t)> = e^{i_\gamma(B,k)}\,|d_k(t,b_k)| = \sum_j \tau_{kj}(\mathbf{A},\mathbf{B})\,c_j(t,a_j)\ ,\quad \forall j,\ \forall t, \qquad (28)$$

where $\tau_{kj}(\mathbf{A},\mathbf{B}) = <v_k|u_j>$, $\forall j$. So for any complex factor of given index *k* we have a separate condition

$$e^{i_\gamma(B,k)} = <v_k|\psi_{Gt}(\mathbf{r},t)>\ /\ |d_k(t,b_k)| = \sum_j \tau_{kj}(\mathbf{A},\mathbf{B})\,c_j(t,a_j)\ /|d_k(t,b_k)|,\quad \forall A,B,\ \forall t \qquad (29)$$

(where ' / ' is to be read: *divided by*).

So the lemma is proved for a $QM_{HD}$ state-ket $|\psi_{Gt}(\mathbf{r},t)>$ *if the condition (29) is fulfilled.*

We now *do impose* for also the formal-factual state-ket (27) $|\psi_{Gt,\mathbf{H}}(r,t_1)>_{fact(DM)}$ the condition (29) specified above.

Thereby the Hilbert-Dirac representation of any expansion of the factual-formal state-ket $|\psi_{Gt1,\mathbf{H}}(r,t_1)>_{fact(DM)}$ is now achieved, all the contents of the representational elements are fully specified, semantically as well as structurally and numerically. So we finally write:

$$|\psi_{Gt1,\mathbf{H}}(r,t_1)>_{fact(DM)}\ /\mathbf{A} \equiv \sum_j e^{i_\alpha(A,j)}.\sqrt{\pi}\,(G_{t1},a_j).\ |u_j(\mathbf{r},a_j)>,\quad \forall j,\ \forall t_1,\ \forall A \qquad (27')$$

And this in its turn permits to write for the *global* factual-formal construct $|\psi_{Gt,H}(r,t_1)>_{fact(DM)}$ the synthetic integrated form:

$$|\psi_{Gt1,\mathbf{H}}(r,t_1)>_{fact(DM)}\ \approx_{\mathbf{pred.}}\ \{\forall A,\forall t,\ (|\psi_{Gt1,\mathbf{H}}(r,t_1)>)_{DM,ff}\ /\mathbf{A}\} \qquad (30)$$

like in (22) for a $QM_{HD}$ state-ket.

The representation of $|\psi_{Gt1,\mathbf{H}}(r,t_1)>_{fact(DM)}$ from (27') specifies explicitly – and directly – all the *first order* probability laws involved in (30), from all the probability spaces that crown the branches of the *IQM*-probability-tree $T(G, \forall A)$ of the studied microstate (Fig.2).

But furthermore – according to theassertion *Ass.1* and ***L(Ass.2)*** – the imaginary factors $e^{i_{\alpha}(A,j)}$ from the expansion-coefficients and the lemma *L(Ass.2)* (27') determine also an *[IQM-QM$_{HD}$]* Hilbert-space representation of the general syntactic structures (11) and (11') of the meta-probabilistic correlations $(M\pi c(G_{t1}))_{\mathbf{AB}}$, $\forall \mathbf{AB}$ that have been defined in *IQM*.

***((Ass.2)a) is fully established*.**

***Arg(Ass.2)b.***

Let us come back to the expression (26) to make the following remark:

The final numerical content of the first member from (26) – $|c_j(a_j,t)_{\mathbf{fact(DM)}}|^2$ – is *essentially different in meaning and its logical status* from that from the $QM_{HD}$ descriptor $|c(a_j,\psi)|^2$ involved in (23) $\pi(a_j,|\psi>) =_{Bp} |c(a_j,\psi)|^2 \equiv_{constr.\mathcal{H}} |Pr._j|\psi>|^2, \forall j$.

This is so because the contents assigned to the expansion coefficients $c_j(a_j,t)_{\mathbf{fact(DM)}}$ from (26) have a ***genesis*** that is fundamentally different from that of the contents of the expansion coefficients $c(a_j,\psi)$ from (23) that are those from the $QM_{HD}$ expressions (18), (22). Indeed:

* Let us first place ourselves inside $QM_{HD}$ and suppose that it has been possible to write the Schrödinger equation of the problem and to solve it so that the researched state-ket $|\psi_{Gt,\mathbf{H}}(\mathbf{r},t)>$ and its expansion (18) $|\psi_{Gt1,\mathbf{H}}(\mathbf{r},t_1)>/\mathbf{A}=\sum_j e^{i_{\alpha}(A,j)}|c_j(a_j,t_1)||u_j(\mathbf{r},a_j)>$ are now known. In this case, the set of numbers $\{|c(a_j,\psi)|^2\}, \forall \mathbf{A}, \forall t$, does not *emerge* endowed with the meaning of an $(\varepsilon,\delta,N_o)$-probability law $\{\pi(G_{t1},a_j)\}, \forall \mathbf{A}, \forall t$, because of genesis of $|\psi_{Gt,\mathbf{H}}(\mathbf{r},t)>$ has been purely mathematical, it has been only calculated from the solution $|\psi>$ of the Schrödinger equation of the problem that has been only calculated itself, and then it has been assigned a posteriori the meaning of a probability law, via Born's postulate. For such a genesis – because of the quasi ubiquitous mathematical idealizations or approximations involved by the solution of a Schrödinger equation (and also in consequence of the necessity to "give" the initial state-ket where *even after the introduction of Born's postulate* the probabilities of the elementary events are a priori posited to be equal) – in general inside $QM_{HD}$ we would be unable to know in advance, i.e. *before the separate realization of a verification*, to what a degree Born's postulate is indeed ***factually*** *true*. Regarding this there would subsist doubts. So it would ***remain*** *to be verified* that the set of real numbers $\{|c(a_j,\psi)|^2, \forall j\}, \forall \mathbf{A}, \forall t$, represents indeed a corresponding set $\{(\varepsilon,\delta,N_o)\text{-}\pi(G_{t1},a_j\}, \forall j)\}, \forall \mathbf{A}, \forall t$ of individual probabilities that constitutes a probability law; and furthermore *a factually true one*, as asserted by Born's postulate. And the verification can be realized only by new sets of repeated individual coding-measurement-successions $[G_{t1}.MesA]$ that incorporate the coding procedure (20'). That is *why* inside $QM_{HD}$ the predictions *and* the verification of the predictions constitute *two* and basically *distinct* phases of the representation.

* Consider now (26). Each descriptor $c_j(a_j,t)_{\mathbf{fact(DM)}}$ has first been just *written* in the *form* of a coefficient from the expansion of a Hilbert-Dirac state-ket $|\psi_{Gt1,\mathbf{H}}(r,t_1)>_{fact(DM)}$ that *is to be* ***constructed***, and in a way such that *it is* ***known*** that the numerical value $\pi(G_{t1},a_j)_{fact(DM)}$ of $|c_j(a_j,t_1)_{\mathbf{fact(DM)}}|^2$ that will emerge will possess the nature of a 'probability' that is factually true within an 'imprecision $(\varepsilon,\delta,N_o)$' because ***it will have been constructed SUCH*** via factual individual coding-measurement successions $[G_{t1}.MesA]$ and accordingly to the definition of

$D_M$ in (9") from *IQM* (that now inside *[IQM-QM$_{HD}$]* incorporates also the coding procedure (20')). So:

*No verification is necessary any more. The nature of the semantic contents of the descriptors $|c_j(a_j,t)_{fact(DM)}|^2$ from (26) and the factual truth of their values are established by construction*[123]*:*

***Born's postulate is not necessary any more*** [124, 125]**.**

Finally let us remind that in the representations (26), (27') and (30) the expansion coefficients inherit *also* – via the coding-postulate (20)' – the *meaning* with which they have been endowed by the posited model of a specimen of the studied microstate, namely the meaning of probabilities of realization of a given sort of wave-movement around the de Broglie-singularity from a specimen of the studied microstate, that can be represented by an eigenket $|u_j(\mathbf{r},a_j)>$ tied with the eigenvalue $a_j$ of the considered observ*able* ***A****,* for $\forall j$. This meaning cannot be regarded to be factually true, like the numerical value $|c_j(a_j,t_1)_{fact(DM)}|^2$, but it acts as an ***unavoidable*** and very precious guide for conceptualization.

***(Ass.2)b is also established.***

***Arg(Ass.2)c).***

The *[IQM-QM$_{HD}$]* Hilbert-space representations (27') and (30) that correspond to the expression (22) of the *QM$_{HD}$*-state-ket $|\psi_{Gt,\mathbf{H}}(\mathbf{r},t)>$ of the studied microstate have been constructed without making any use of the Schrödinger equation of the problem.

***(Ass.2)c is also established.***

***So the whole assertion Ass.2 is now fully established.*** ■

The very remarkable fact that *the predictive content from (30) emerges already verified* will be expressed by the following re-notation:

$$|\psi_{Gt,\mathbf{H}}(r,t_1)>_{fact(DM)} \neq_{\mathbf{pred.}} |\psi_{Gt1,\mathbf{H}}(\mathbf{r},t)>,$$
$$\Big[\ |\psi_{Gt,\mathbf{H}}(r,t_1)>_{fact(DM)} \approx_{\mathbf{pred.}} \{|\psi_{Gt,\mathbf{H}}(r,t_1)>_{fact(DM)}/A\}, \forall A, \forall t_1\}\ \Big]_{\mathbf{verif}} \qquad (31)$$

In consequence of the fact that no Schrödinger equation has been made use of, the global factual-formal state-ket $|\psi_{Gt,\mathbf{H}}(r,t_1)>_{fact(DM)}$ is *not* endowed with an own explicit integrated functional expression: Such as it has emerged, it has to be regarded as just *defined* by the second member of (30). Throughout what follows the equivalence from (31) is just *posited* to entail the possibility of principle to construct for the first member $(|\psi_{Gt,H}(r,t_1)>)_{fact(DM)}$ an *integrated* mathematical functional expression (there should even exist an infinite family of such possibilities, since in the lemma *L(Ass.2)* the initial choice of imaginary phase-factors has been arbitrary). Inside *QM$_{HD}$* this integrated expression is obtained from the Schrödinger

---

123 (In so far as the presupposed physical conditions are stable and expressed in a way that is not very deteriorated by the unavoidable approximations).

124 This is nearly a procedural and semantic *opposition* with respect to what happens inside *QM$_{HD}$* concerning the descriptors $|c(a_j,\psi|^2$ from (22) and (23). This opposition reflects the distinction between the top-down abstract and directly statistical approach that marks the formalism of *QM$_{HD}$*, and on the other hand the bottom-up factual approach practised here, where the descriptors $|c_j(a_j,t)_{fact(DM)}|^2$ are constructed by factual individual coding-measurement-successions *[G$_{t1}$.MesA]*. A given direction of progression along the vertical of construction of knowledge involves a specific order of conceptualization and this order entails fundamental consequences.

125

equation of the problem (***iff*** this equation can be written and the solution can be found on the basis of acceptable approximations [126]).

But *an integrated solution is not more useful than (31) itself* because ***any*** $QM_{HD}$-prediction-verification process is founded on expansions (22) and the Hilbert-space representation (23) of these, which (31) permits *directly*.

### ***(7.III).2.2.4. Global conclusion on*** **(7.III).2.2.**

Via the assertion *Ass2* we have constructed a new representation of the quantum measurements for the particular quasi-classical case of the microstates *ms(unbound,1)*$_{G(nc)}$, no matter whether the aim of these measurements is to construct a prediction or to verify it. This new representation is rooted directly into the a-conceptual physical factuality; it is constructed bottom-up; and it *emerges* by construction both predictive and verified.

---

[126] **We quote as it stands the following extract from the *French* Wikipedia: Rareté d'une résolution analytique exacte**

La recherche des états propres de l'hamiltonien est en général complexe. Même le cas analytiquement soluble de l'atome d'hydrogène ne l'est rigoureusement sous forme simple que si l'on néglige le couplage avec le champ électromagnétique qui va permettre le passage des états excités, solutions de l'équation de Schrödinger de l'atome, vers le fondamental.
Certains modèles simples, bien que non tout à fait conformes à la réalité, peuvent être résolus analytiquement et s'avèrent très utiles :

- particule libre (potentiel nul) ;
- oscillateur harmonique (potentiel quadratique) ;
- particule se déplaçant sur un anneau ;
- particule dans un puits de potentiel rectangulaire ;
- particule dans un guide d'onde annulaire ;
- particule dans un potentiel à symétrie sphérique ;
- particule dans un réseau unidimensionnel (potentiel périodique).

Dans les autres cas, il faut faire appel aux diverses techniques d'approximation :

- la théorie des perturbations fournit des expressions analytiques sous la forme de développements asymptotiques autour d'un problème non perturbé exactement soluble.

l'analyse numérique permet d'explorer des situations inaccessibles par la théorie de perturbation.

# INTERMEZZO ON A "THEORY OF *FACTUAL* PROBABILITIES" BELONGING TO THE MATHEMATICAL PHYSICS

This intermezzo develops the assertion made at the end of (7.III).2.2 that – via the application of Gleason's theorem to *factual* probabilities established accordingly to *IQM* for outcomes of quantum-measurements on microstates – we are assisting at the process of formal specification of the new bottom-up *general* theory of probabilities announced inside *IQM* by the concept of probability tree of an operation of generation *G* of a microstate. Indeed the contents of the argument that establishes the assertion *(Ass.2)a* develop notably the general perspective on the notion of a Theory of Factual Probabilities belonging to the Mathematical Physics, formed already in *IQM*.

Inside $QM_{HD}$ the aim of Dirac's calculus of transformations is entirely ignorant of the *IQM* operational-*semantic* features of the pairs of observable events $\{(a_j,b_k)\}, \forall(\mathbf{A},\mathbf{B}), \forall t$ tied with the studied microstate *inside a tree-like probabilistic whole founded upon the operation of generation G or* $G_t$ *that corresponds to the state-ket* $|\psi_{Gt}>$ *of the studied microstate*. This is so because the individual operations *G* or $G_t$ – like also any corresponding coding-measurement-evolution – are not represented inside $QM_{HD}$; they simply are not even conceived there. So inside $QM_{HD}$ Dirac's calculus of transformation from a 'representation of $|\psi_G(t)>$' with respect to an observable $\mathbf{A}$, to its representation with respect to another observable $\mathbf{B}$, is asserted as just a mathematical algorithm devoid of any more complex semantic content.

But inside *[IQM-QM$_{HD}$]* the *IQM* meta-probabilistic correlations $(M\pi c(G_{t1}))_{\mathbf{AB}}$ from a probability-tree and from the corresponding description (9") *DM*, are ***entirely*** determined formally by Dirac's "theory of transformations", via the expansions from (22) and the imaginary factors $e^{i_\alpha(A,\,j)}$ from the coefficients (27') from these expansions. Thereby Dirac's "calculus of transformations" acquires a *semantic significance encapsulated in probabilistic terms*:

> Dirac's whole "bra-ket" formalism from $QM_{HD}$ appears as a *calculus with semantic contents* involved in probability-trees of operations of generation of the microstates. It can be regarded as a calculus with *measures of degrees of various sorts of angular semantic degrees of proximity between various sorts of qualifications of any given sort of microstate* (cf. MMS [1993])**[127, 128]**.

Thereby Dirac's calculus "of transformations" reveals a certain sort of Universality: it points toward the new ***general*** "Theory of factual $(\varepsilon,\delta,N_o)$-probabilities" already announced inside *IQM*, a Theory of Factual Probabilities *asserted for any sort of physical entity and expressed mathematically in Hilbert-Dirac terms* that specifies the entire factual genesis of all the first-order $(\varepsilon,\delta,N_o)$-probabilities laws from any considered probability-tree, as well as the

---

[127] Retroactively, this is already perceptible in the relations (11), (11') from *IQM*. Indeed consider for instance the *IQM* descriptor

$$(11)\; p(b_k)=\mathbf{F}_{\mathbf{bk,A}}\{p(G_t,a_j)\},\; \forall k,\; \forall j,\; \forall(\mathbf{A},\mathbf{B})$$

that denotes meta-probabilistic correlations $(Mpc(G_t))_{\mathbf{AB}}$ between whole probability laws that crown two distinct branches of a given probability-tree. Inside $QM_{HD}$ the Dirac transformation from the Hilbert-space representation of the state-ket $|\psi_G(t)>$ of the studied microstate with respect to the eigenvalues $a_j$ of an observable $\mathbf{A}$, to the representation of $|\psi_{Gt}(t)>$ with respect to the eigenvalues $b_k$ of another observable $\mathbf{B}$ that does not commute with $\mathbf{A}$, is defined as

$$d_k(t,b_k)=\sum_j \tau_{kj}(\mathbf{A},\mathbf{B})\; c_j(t,\mathbf{A}),\; \forall j,\; \forall k,\; \forall t$$

The isomorphism is striking. And a similar isomorphism can be discerned for (11').

[128] For instance, the scalar product of two distinct $QM_{HD}$-state-ket of two different physical entities, but expressed inside one *same* "*A*-representation" (i.e. with respect to one same dimension of qualification denoted *A*) might be used as *a measure of a concept of "common degree of A-content"*. A calculus of this sort can very naturally be lodged inside a mathematized version of the *general* Method of Relativized Conceptualization (MMS [2002A], [2002B], [2006]), namely by *directly* associating *any* descriptive cell defined by any given epistemic referential *(G,V)* (in the sense that generalizes (6)), by use of: the *general* concept of a probability tree; Gleason's theorem; mathematical Hilbert-Dirac representations of the type (31) of the involved factual probability laws; all this *outside* the particular theory of the microstates but including it.

*numerical* specification of these laws *and* the meta-correlations between these, ***quite independently of the concept of microstate***.

This ***general*** Theory of Factual Probabilities that takes on its first contours in connection with the particular case of unbound microstates, involves a mathematical framework that is far more deeply set, more complex and more precise than Kolmogorov's one.

And it is noteworthy that – *in contradistinction to the Schrödinger equation of the problem* – this mathematical framework accepts its factually introduced semantic content without any a priori restriction, just as does in (31) the mathematical state-ket-*mould* alone, of the mathematical state-ket-representation of the studied microstate. This is so because it emerges together with physical facts that to be hosted inside them. While the classical mathematical "structures" or "systems" are *first* entirely constructed separately, then are assigned a sort of Platonian existence independently of facts, and finally are made use of for hosting in them factual contents subjected a priori to all the restrictions imposed by this previously achieved mathematical structure.

This difference suggests immediately an intimate connection with the distinction between top-down approach and bottom-up approach inside Physics.

This deserves very attentive examination in view of an *optimal* methodology of generation of scientific knowledge on physical entities.

## (7.III).2.3. Construction of a Hilbert-Dirac representation of probabilistic predictions on the results of quantum-measurements on interference microstates *$ms(unbound,1)_{cG(qf)}$*

We consider now microstates of one microsystem but *with a composed operation of generation* (cf. the definitions from (2.I).1). Such a microstate incorporates a phenomenon of 'interference of corpuscular-waves', so a non-null quantum potential wherefrom quantum fields can emerge. Such a microstate is denoted *$ms(unbound,1)_{cG(qf)}$* (the index '$_{cG(qf)}$' is to be read: 'composed operation *G* of generation involving possibility of quantum fields').

The microstates *$ms(unbound,1)_{cG(qf)}$* constitute the *key*-category in the endeavour toward an intelligible consensual predictive and verifiable conceptualization of the factual microstates. Though they are still unbound states and so essentially endowed with unstable characters, *the* ***inside*** *of the specimens of a microstate $ms(unbound,1)_{cG(qf)}$ cannot any more be conceived to be transparent to the* ***exterior context****, as it is presupposed to be by the coding postulate (20').* With respect to microstates *$ms(unbound,1)_{cG(qf)}$* the concepts of 'outside' and 'inside' (cf. Atmanspacher&Dalenoort [1994]) acquire full power and this, in the processes of conceptualization, plays an outstanding role[129]. The opposition outside-inside displaces us from just *upon* the frontier between classical physics and quantum physics, into the unlimited depth of the 'quantic' stratum of physical substance. It draws us into the unlimited and a-conceptual depth of the 'Universal Physical Substance' in the sense of Spinoza (*Natura Naturans*) that has been conceptualized by de Broglie and Bohm.

Nevertheless, for the sake of maximal intelligibility, we shall try to realize a strong and explicit conceptual-formal continuity between the constructive approach achieved in (7.III).2.3 for the representation of the quantum measurements on microstates *$ms(unbound,1)_{G(nc)}$* and the representation of these for the case of the microstates *$ms(unbound,1)_{cG(qf)}$*.

Throughout what follows the model posited in (1') for the specimens of the studied unbound microstate plays a key-role.

### *(7.III).2.3.1. Basic problems tied with momentum-measurements on interference microstate*

#### *(7.III).2.3.1.1. Inadequacies of the $QM_{HD}$-state-ket for microstates $ms(unbound,1)_{cG(qf)}$.*

The dynamic of the singularity from the physical wave of the specimens of an interference microstate *$ms(unbound,1)_{cG(qf)}$* cannot be conceived any more to be influenced exclusively by the classical macroscopic fields from a *$QM_{HD}$*-Hamiltonian; this sort of Hamiltonian becomes insufficiently rich.

Correlatively the *$QM_{HD}$* Hilbert-space representation of the *microstates $ms(unbound,1)_{cG(qf)}$* becomes itself inadequate from various points of view, and *this brings forth a limit of the framework [IQM-$QM_{HD}$].*

In order to show this it suffices to consider a microstate *$ms(unbound,1)_{cG(qf)}$* of type (15') with only two components in the operation of generation, for instance a Young-interference *$ms_{G12}$* generated by an operation of generation *$\mathbf{G}(G_1,G_2)$*. With the notations from *[IQM-$QM_{HD}$]*) the *$QM_{HD}$* Hilbert space-representation of the state-ket of such a microstate is:

$$|\Psi_{G(G1,G2)}(\mathbf{r},t)> = \lambda_1|\Psi_{G1}(\mathbf{r},t)> + \lambda_2|\Psi_{G2}(\mathbf{r},t)> \qquad (15'')$$

Consider the momentum-observable ***P***. This observable plays a most basic role inside *$QM_{HD}$*. Indeed the position-observable ***R*** is degenerate from various points of view, while the eigenvalues of any other observable ***A*** can be represented as a function ***A(R,P)*** which permits to calculate the eigenvalues of ***A*** from those of ***R*** and ***P*** (cf. 5.II).1). So the "quantic" specificities stem mainly fro the momentum observable. Now, the eigenfunctions of the momentum-observable $\boldsymbol{P}$ are *plane* wave-functions $a.exp(i(const.)\boldsymbol{p_j}.\boldsymbol{r})$ where $\boldsymbol{p_j}$ designate the eigenvalue with index $j$ of $\boldsymbol{P}$ that is tied with the plane-wave eigenfunction where it is contained. The expansion $|\Psi_{G(G1,G2)}(\boldsymbol{r},t)>/\boldsymbol{P}$ where $|\Psi_{G(G1,G2)}(\boldsymbol{r},t)>$ is defined by (15") yields for the predictive probability of the outcome $\boldsymbol{p_j}$ – let us re-note it $\boldsymbol{p_{j,12}}$ – the well-known representation of form

$$\pi_{12}(\boldsymbol{p_{j,12}}) = \pi_1(\boldsymbol{p_{j,1}})+\pi_2(\boldsymbol{p_{j,2}})+F(\pi_1(\boldsymbol{p_{j,1}}),\pi_2(\boldsymbol{p_{j,2}})) \qquad (32)$$

with $\boldsymbol{p_{j,1}}$ and $\boldsymbol{p_{j,2}}$ the eigenvalues introduced by the expansions $|\Psi_{G1}(\boldsymbol{r},t)>/\boldsymbol{P}$ and $|\Psi_{G2}(\boldsymbol{r},t)>/\boldsymbol{P}$, respectively and $F$ represents a functional.

[129] Tied with the opposition 'open'-'closed' that lies on the ground of the concept of 'formal system'.

We assert that:

* inside *[IQM-QM$_{HD}$]* the prediction (32) and correlatively also the representation (15") are ***inadequate*** *from conceptual points of view*;

* the $QM_{HD}$ expression (32) raises questions concerning the applicability of the coding-postulate (20');

* raises questions of effectiveness and of inner logical consistency.

In short, we are in presence of a sort of block of questions that manifests *basic* insufficiencies.

The formulated assertions are detailed below.

*Conceptual inadequacy of (15")*

We have already remarked before that the additive representation (15") of the state-ket $|\Psi_{G(G1,G2)}(\boldsymbol{r},t)>/$ of a microstate $ms_{G12}$ – though mathematically it is *permitted* in a vector-space – involves semantic characters that are incompatible with the basic posit (1) from *[IQM-QM$_{HD}$]*. The two state-ket $|\Psi_{G1}(\boldsymbol{r},t)>$ and $|\Psi_{G2}(\boldsymbol{r},t)>$ from the second member from (15") concern the microstates $ms_{G1}$ and $ms_{G2}$, while according to *[IQM-QM$_{HD}$]* these microstates are not *physically* ***realized*** and so *they cannot 'exist' in the sense of (1)* i.e. as the result of a ***unique*** operation of generation that has been *fully achieved*: The basic methodological decision *MD* from I.1 asserts a *one-to-one* relation (1) $G\leftrightarrow ms_G$ between any given operation of generation $G$ and its result denoted $ms_G$, and in the case considered in (15") the *unique* operation of generation that is posited to have been fully realized physically is $\boldsymbol{G(G_1,G_2)}$, which is different from both $G_1$ and $G_2$. According to *IQM* – so according to *[IQM-QM$_{HD}$]* – in consequence of (1) exclusively the microstate $ms_{G12}$ represented by the state-ket $|\Psi_{G(G1,G2)}(\boldsymbol{r},t)>$ is physically realized; the microstates '$ms_{G1}$' and '$ms_{G2}$' have *not* been *individualized* by $\boldsymbol{G(G_1,G_2)}$[130] because $G_1$ and $G_2$ have not been fully realized separately, so the microstates $ms_{G1}$ and $ms_{G2}$ do not possess entirely actualized *own* "states". *Full stop*.

One could oppose various objections. For instance, one could object that nothing interdicts to suppose that in $QM_{HD}$ the mathematical content of the second member from (15") has been defined with a purely algorithmic purpose, namely for calculating the predictive probabilities of outcomes of measurements on the unique generated and studied microstate $ms_{G12}$. If this is conveniently realized, this suffices for choosing the representation (15") of $|\Psi_{G(G1,G2)}>$. We retain this view because the answer to it permits to well detail the conceptual situation as follows.

*Mutual inadequacy of (15") and the coding postulate of type (20)-(20')*

No matter whether one places oneself inside $QM_{HD}$ or inside *[IQM-QM$_{HD}$]*, a coding postulate (20') of the type (20) appears to be inadequate with respect to momentum measurements on microstates $ms(unbound,1)_{\boldsymbol{cG(qf)}}$. Indeed:

130 This is not visible inside $QM_{HD}$ where the whole notion of an individual operation of generation of the studied microstates remains hidden.

*Inadequacy with respect to $QM_{HD}$ alone*

**1.** The eigenstates of the momentum-observable are plane waves and one of the measuring-postulates of $QM_{HD}$ asserts that the result of a measurement on a microstate is always an eigenvalue of the measured observable that emerges ***with*** the corresponding eigenstate and *remains in this eigenstate* immediately after the end of the measurement operation [131].

On the other hand:

**2**. Inside a Young interference, a plane wave-structure is ***nowhere*** spontaneously realized by the evolution of the state. Even at an infinite distance from the holed screen, strictly speaking there still is superposition of contributions from both holes.

Then what about the *coding-measurement-evolution*? Can this nevertheless realize a plane wave out of a Young interference?

Inside $QM_{HD}$ it is explicitly admitted that for the momentum-measurements one has to apply the method time-of-flight. This method requires basically and explicitly the suppression of all the "external" [132] fields that are working (no doubt precisely *in order to* start – from the time-moment when the measurement-evolution begins – a coding-measurement-evolution that shall indicate an eigen-value-*and-eigenstate* of the momentum-observable at *finite* distance-and-time (cf. (6.II).4.3.)). But the quantum fields that work in a Young interference are not classical "external" fields, and the human observer cannot suppress them without suppressing the specimen itself of the microstate-to-be-studied. In these conditions the coding-postulate (20)-(20') – by its definition – *cannot be used in order to bring forth an eigenstate of the momentum-observable*. So the predicted eigenvalues – with *their* eigenstates and *their* probabilities of outcome – cannot emerge either.

The $QM_{HD}$ prediction (32) cannot be verified via a coding-measurement-evolution of the form (20').

One might then suggest that – since the structure of the superposition-wave-function from a Young interference becomes spontaneously as *near* to that of a plane wave as one wants when the distance-and-time increase toward infinity – it suffices to end each coding-measurement-succession *[G.Mes**P**]* by a position-registration on a very distant sensitive-screen, which will permit a satisfactory evaluation of the momentum eigenvalue. But it seems clear that it is not acceptable to admit a definition of a coding-measurement evolution that – in its very principle – involves either approximations, or a non-effective distance-and-time of evolution. *A measurement operation obviously has to be a priori required rigorous and effective **in its principle**.*

**3**. In short, the situation just described in the above point 2 is in direct contradiction with the $QM_{HD}$ view recalled at the point 1.

> $QM_{HD}$ itself does not specify an acceptable way of verifying the prediction (32) involved by the representation (15"). The implicitly supposed coding postulate of form (20') cannot be applied and no other coding postulate is formulated.

So the contradiction can last undisturbed; which probably it does, since some 100 years.

*Inadequacy with respect to requirements of effectivity and of inner logical consistency*

**1'**. According to *IQM* the descriptive structure of an interference microstate with state-ket $|\Psi_{G(G1,G2)}(\mathbf{r},t)\rangle$ is tied with the two operations of generation $G_1$ and $G_2$ involved in the unique realized operation $\mathbf{G(G_1,G_2)}$, in the sense of the last definition from (1.I)1; and

---

131 Cf. Cohen-Tannnoudji, C., Diu, B. & Laloë, F. [1973]; and also the measurement postulates in (5.II).1.

132 "External" to that to what a mechanical concept of "momentum" can be conceived accordingly to the accepted model of specimens of a microstate; so "external" with respect to the corpuscular-like singularity in the amplitude of the physical wave assigned to any specimen of the studied microstate.

according to de Broglie's model of a microstate as incorporated to *[IQM-QM_HD]* via the modelling-postulate *$MP(ms_{G,cw}$, that* what is considered to have a 'trajectory' along a maximum of presence-probability (the "particle"), is the singularity from the amplitude of the individual physical wave represented by a wave-function $\Phi_{G(G1,G2,cw)}(\mathbf{r},t)=a.exp(i.const.\beta(t,\mathbf{r})$ of the physical individual wave from the one specimen $\sigma_{\Phi(G(G1,G2)}$ of $ms_{G12}$ that is involved in any one coding-measurement-succession *[**G**$(G_1,G_2)$**.**Mes.**P**]* (cf. *(6.II).4.2.2*). Now, *each* one of the *two* operations of generation $G_1$ and $G_2$ involved in the unique fully realized operation of generation ***G**$(G_1,G_2)$* can be considered to induce a directional *trend* into the movement of the ***unique*** singularity from one given wave $\Phi_{G(G1,G2),cw}(\mathbf{r},t)$; and these two distinct directional trends are quasi-unanimously conceived to *combine* so as to determine for the one involved singularity a trajectory along a maximum of the presence-probability expressed by the state-ket $|\Psi_{G(G1,G2)}>$ that corresponds to $\Phi_{G(G1,G2),cw}(\mathbf{r},t)$.

Insofar that this is correct the strict neighbourhood of the singularity from $\Phi_{G(G1,G2)}(\mathbf{r},t)$ can ***never*** be conceived to be populated by a form of wave-movement representable by only ***one*** *plane wave*, not even approximately and at ideal infinity. For in the case of a microstate *ms(unbound,1)$_{cG(qf)}$ precisely this narrow neighbourhood of the corpuscular-like singularity from the amplitude is conceived to be quite essentially populated by a – **physical** – superposition of **two** wave-movements of distinct directions that, **together**, **determine the trajectory of the singularity**.*

And requiring to wait that at nearly-infinity this impossibility – nearly – disappear, amounts in fact to waiting – ***non-effectively*** – that disappear precisely what we want to study.

**2'**. Since $QM_{HD}$ is part of the framework *[IQM-QM_HD]* posited up to now, the $QM_{HD}$ measuring-postulates are still acting inside this framework. So, just like above for the points 1 and 2 and according to $QM_{HD}$ alone, here again, the points 1' and 2' are mutually inconsistent with respect to also the whole framework *[IQM-QM_HD]*.

So we conclude that:

*The prediction (32) in fact consists of an only mathematical definition of the predictive probabilities $\pi_{12}(\mathbf{p}_{j,12})$ that is just postulated to be factually true but that **cannot be verified to be such**.*

The formal superposition-representations of the type (15") of microstates *ms(unbound,1)$_{cG(qf)}$* are just an arbitrary mathematical choice favoured by the tendency to rush into the simplest expression *permitted* by the mathematical structure that is made use of. But in the considered physical situation this choice is devoid of any necessity of semantic or of logical nature. Quite on the contrary, from the standpoint of an inner logical-semantic coherence this choice appears to be entirely inadequate.

So, since a new coding postulate ***has*** to be formulated for the category of microstates *ms(unbound,1)$_{cG(qf)}$*, correlatively, new definitions will be needed for the concepts of eigenstates and eigenvalues of the momentum-observable, as well as for the Hilbert-space representation of the involved state-ket.

*Comment on the preceding critiques*

One might believe that this whole problem is a false problem that stems from *IQM*, while nothing imposes to accept *IQM*. One might try to hold that this problem can be evacuated by just refusing the concept of a 'composed operation of generation' defined in (2.I)1 and by continuing to accept the direct postulation from $QM_{HD}$ of the adequacy of additive representations of state-ket like in (15").

This however would not suffice. A rejection of the concept of composed operations of generation would not change the fact that the prediction (32) raises questions ***with respect to basic postulates from $QM_{HD}$ itself***; nor would it change the fact that *the coding-postulate (20') cannot be used for verifying the prediction (32) drawn from (15'')*. These flaws, once perceived, persist even if the whole *IQM* is dropped. So the fact that the concept of composed operation of generation has brought them forth, manifests the relevance of this concept.

> So we are left with a basic unsolved question: What sort of coding-measurement-evolution can be associated with momentum-measurement on unbound interference microstates?

If one wants to construct a fully intelligible representation of the microstates this question *has* to be dealt with and solved. For if not, quantum momentum-measurement cannot be performed on interference states, which would reduce Quantum Mechanics to only a very limited, superficial representation of the microstates.

This leads us quite naturally toward the *dBB* approach.

### *(7.III).2.3.1.2. On the question of measurability of guiding-traces: a factual analysis to be realized.*

Since in the case of an interference microstate $ms(unbound,1)_{cG(qf)}$ the stable conservation of the initial value of the momentum throughout a lasting momentum-measurement-evolution cannot be certainly insured until a final interaction that produces observable coding-marks, one is led to try to identify for this case a different sort of coding-measurement-evolution. This aim brings tentatively to the *dBB*) approach, the unique one that penetrates explicitly into the 'inside' of the specimens of interference microstates, where the concepts of quantum potential and quantum field take source and where is lodged de Broglie's concept of a corpuscular-like singularity that accepts mechanical qualifications.

According to *[IQM-$QM_{HD}$]* the microstates $ms(unbound,1)_{cG(qf)}$ are characterized by composed operations of generation $\boldsymbol{G}(G_1,G_2,...G_m,...G_M)$, $m=1,2,..M$ and these introduce a physical wave-function $\Phi(\boldsymbol{r},t)=a(\boldsymbol{r},t)exp((2\pi i/h)\beta(\boldsymbol{r},t)$ where the involved phase-function $\beta(\boldsymbol{r},t)$ defines a sort of momentum-value that inside $QM_{HD}$ is not defined (cf. *(6.II).4.2.2)*). The present context of the problem formulated in *(7.III).2.3.1* draws attention upon these physical descriptors.

Consider now de Broglie's well-known concept of 'guiding relation'

$$(\mathcal{M}_0/(\sqrt{1-\beta^2}))\boldsymbol{v}(\boldsymbol{r},t) = \boldsymbol{p}(\boldsymbol{r},t) = -\boldsymbol{\nabla}.\beta(\boldsymbol{r},t) \qquad (33)$$

where: $\mathcal{M}_0/(\sqrt{1-\beta^2})$ is de Broglie's "quantum-mass" $\mathcal{M}$; $\boldsymbol{v}(\boldsymbol{r},t)$ is the time-dependent velocity of the "guided" singularity from the amplitude of the wave of a specimen $\sigma(ms_G)$ of the studied microstate $ms_G$ (cf. (6.II).2 and (7.III).1.2); $\boldsymbol{p}(\boldsymbol{r},t)$ is the 'guided' momentum of this singularity at the time *t*; and $\Phi_G(\boldsymbol{r},t)=a(\boldsymbol{r},t)exp((2\pi i/h)\beta(\boldsymbol{r},t)$ is the de Broglie "*phase*-function" at the time *t* of the physical individual wave from the involved specimen $\sigma(ms_G)$. Louis de Broglie has obtained the guidance law (33) *deductively* [133] and with full generality, *both* for microstates with, or without quantum fields. (It is remarkable that similar approaches have been rediscovered recently by Eliade Stefanescu [2018] and by Unnikrishnan [2018]).

> *This law however, that – in its essence*[134] *– is present also in Bohm's approach, is quasi-unanimously considered to be* ***un****-observable.*

[133] We quote from de Broglie [1956] p. 103 (our translation): «This fundamental formula will be called "the guiding formula". It shows that the global movement of the singular region is obtained by simply extrapolating beyond the limits of the of the geometrical Optique the formula $\boldsymbol{p} = -\boldsymbol{grad}\, S$ from Jacobi's classical theory, that is the ancient Mechanics ».

[134] Bohm conceives "particles" guided by waves and the velocity $\boldsymbol{v}$ of a "particle" is $\boldsymbol{v}(\boldsymbol{r},t) = -\boldsymbol{\nabla}.\beta(\boldsymbol{r},t)$ where $\beta$ denotes the phase of the wave.

Even de Broglie himself adhered to this view. It is believed that as soon as one would try to register the guidance-trajectory in a specimen $\sigma(ms_G)$ of the studied microstate, the beginning of the interaction would immediately destroy the inner structure of the phase represented by the phase-function $\beta(\mathbf{r},t)$ and this would compromise any global relevance of the data drawn from the interaction. This idea however seems to have been admitted on the sole basis of a confusion between temporary technical impossibilities and facts, as well as in consequence of the powers of a priori submission to the formalism and the edicts of $QM_{HD}$, while notwithstanding these powers:

- Trace-registrations are currently used in Wilson-chambers, since a long time.
- For *photonic* interference states a guided trace has already been experimentally registered (cf. A. Steinberg [2011]), *which is a very strong indication that an experiment with* ***heavy*** *microsystems could equally succeed.*

In fact, I think that nobody as yet has genuinely analysed and tested – specifically – whether it is possible, or not, to choose the values of the parameters involved in an attempt at trace-registration on interference microstates of ***heavy*** microsystems in a way ***such*** *as to permit computation, out of the registered data, of the value of the momentum at the time when the trace registration* ***began****.*

> *This analysis remains to be done.* And here we conceive it as exclusively a matter of *facts* and of *techniques* that are available or conceivable at the present time, not as a matter of principles, concepts and mathematical representations[135]**.**

So in what follows we propose an experiment – denoted *EXP* – of which the unique aim is to realize the factual-technical analysis specified above.

*Description of the investigation "EXP"*

Let us imagine the experiment indicated in the Fig.8 that contains a very simplified representation of the operation of generation $\mathbf{G}(G_1,G_2)$ of an interference-state. The Fig.8 is *not* referred to the mathematical $QM_{HD}$ representation (15") in terms of a state-ket. The concept of "plane wave" is not involved either. We consider exclusively experimental features that concern the individual entities and events involved in the registrations of observable marks.

---

[135] There has been an interesting debate on "Surrealistic Bohmian Trajectories". But, surprisingly, all the involved works seem unaware of de Broglie's *"individual" physical model (with one corpuscular-**like** singularity in it) as well as of de Broglie's physical **wave**-function* $\Phi(\mathbf{r},t)=a(\mathbf{r},t)exp((2\pi i/h)\beta(\mathbf{r},t)$ *of one specimen of the microstate to be studied.* The whole *own* approach of de Broglie seems to be simply unknown, it seems that Louis de Broglie's most basic works [1924], [1956], [1957] *are not yet available in English !!!.*
And all the involved authors accept $QM_{HD}$ just as it now stands – with "collapsing" measurements included – and they follow the purpose to show full agreement between Bohm's approach and $QM_{HD}$.
So these works – though they are likely to cooperate toward a new microphysics – are fundamentally different in aim and approach from the construction attempted here.

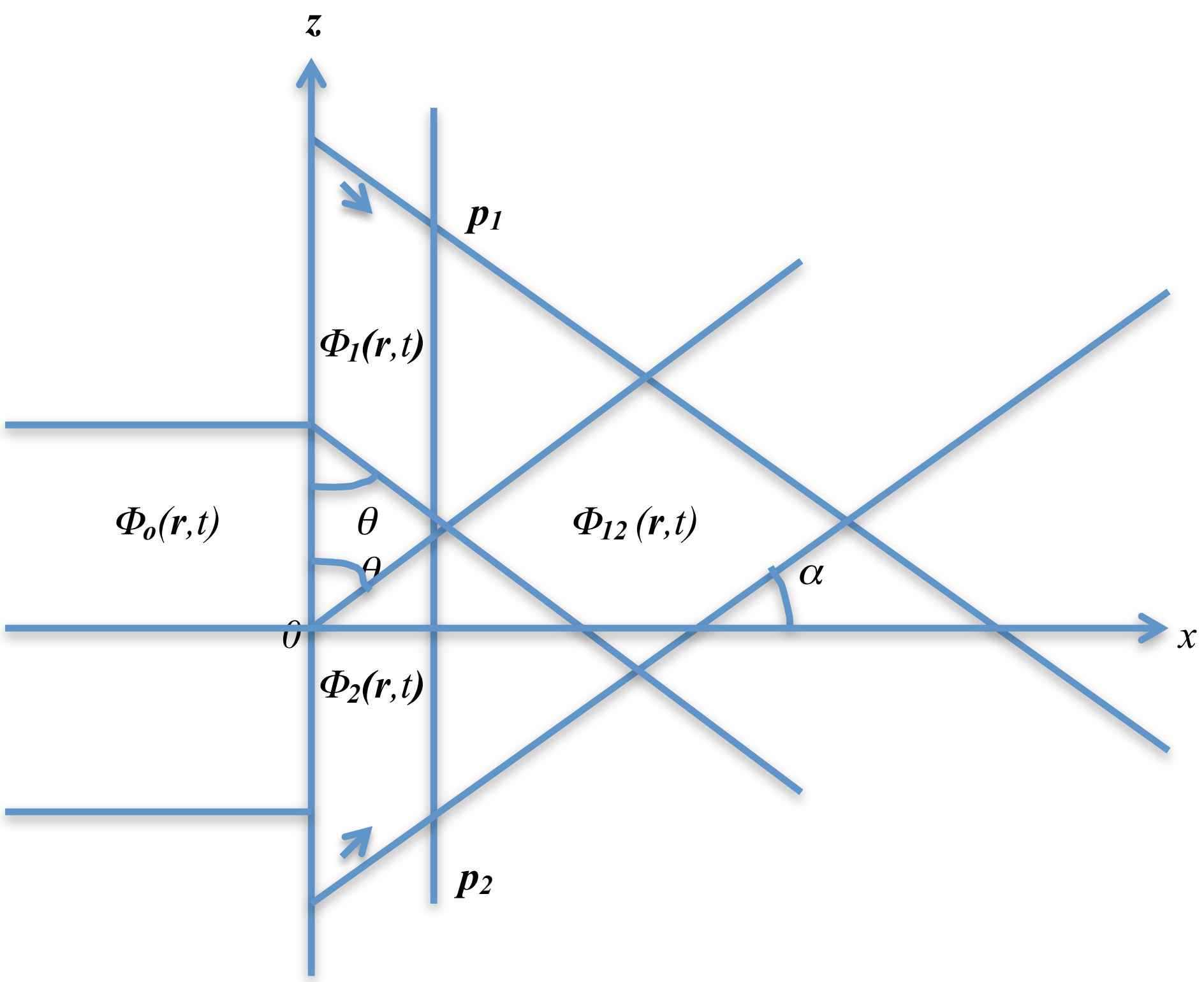

Fig.8. The operation of generation $\boldsymbol{G(G_1,G_2)}$ of a microstate $ms(unbound,1)_{cG(q\text{-}f)}$ of a heavy microsystem (very schematic graphic representation)

The symbol $\Phi_o$ denotes the wave-function from a physical individual specimen of a *preliminary* microstate $ms_{Go}$. A front-wave-divisor splits $\Phi_o$ in two parts that introduce the two new trends of wave-movement denoted, respectively, $\Phi_1$ and $\Phi_2$, that are tied in the sense defined at the end of (1.I).1.2 with the composing operations of generation $G_1$ and $G_2$. These then *combine* their effects inside the unique fully realized operation $\boldsymbol{G(G_1,G_2)}$. Accordingly to a conventional referential, the two main directions of propagation from $\Phi_1$ and $\Phi_2$ make a mutual angle $\alpha$, while with the axis $0z$ they make angles $\theta$ of the same absolute value. The considered specimens of the *interferenc*e-microstate-to-be-studied $ms_{\boldsymbol{G}(G1,G2)}$ are realized *only* on a delimited space-time domain where the physical trends of wave-movement denoted $\Phi_1$ and $\Phi_2$ superpose *physically* into one individual physical interference-wave-movement represented by the wave-function $\Phi_{12}$. Inside this delimited space-time domain the phenomenon of interference is progressive, while outside this domain it ceases to exist. This whole description is placed on the individual level of conceptualization.

For each specimen $\sigma_{\Phi,\boldsymbol{G}(G1,G2)}$ of $ms_{G(G1,G2)}$ the *dBB* representation of the guidance relation (33) asserts that – with respect to the introduced referential – the corpuscular-like singularity in the amplitude of the involved physical wave with wave-function $\Phi_{G(G1,G2)}(\mathbf{r},t)=a(\mathbf{r},t)\exp((2\pi i/h)\beta(\mathbf{r},t)$ has a velocity with components $v_x=v_o\sin\theta=const,$ $v_y=v_z=0$. So the momentum-components are

$$p_x=\mathcal{M}v_x=\mathcal{M}v_o\sin\theta, \qquad p_y=p_z=0 \qquad (34)$$

where $\mathcal{M}=\mathcal{M}_o/(\sqrt{1-\beta^2})$ *denotes the time-dependent variable de Broglie "quantum mass" introduced by the involved specimen of the studied microsystem, with rest-value*

$$\mathcal{M}_o = \sqrt{[m^2_o + (h^2/4\pi^2c^2)(\Box a/a)]}$$

(cf. de Broglie [1956] p. 205)), because the de Broglie "singularity" is not conceived to be "a particle", it is conceived as just a localized particularly energetic aspect from the amplitude of the individual physical wave represented by the wave-function $\Phi_{G(G1,G2)}(\mathbf{r},t)$.

Imagine now a factual context that is stabilized as much as possible. Imagine a big number of repetitions of the operation of generation $\mathbf{G}(G_1,G_2)$ defined above. Each one of these repetitions involves its own specimens $\sigma_{\Phi o}$ and $\sigma_{\Phi,\mathbf{G}(G1,G2)}$, so its own individual physical wave-movements described by the corresponding wave-functions $\Phi_o$, $\Phi_1$, $\Phi_2$ and $\Phi_{12}(r,t)$.

That what inside *[IQM-QM$_{HD}$]* is represented by a state-ket is the global result of all these repetitions placed on the statistical level of conceptualization and considered mentally and retroactively as a whole. We just denote this state-ket by $|\Psi_{\mathbf{G}(G1,G2)}(\mathbf{r},t)>$ but we do not know yet its Hilbert-space-representation because its $QM_{HD}$ representation (15") has been found to be deficient from the viewpoint of *[IQM-QM$_{HD}$]*). However we admit – as it is generally accepted on the basis of *factual* data – that the presence-probability inside de space-time support of $|\Psi_{G(G1,G2)}(\mathbf{r},t)>$ consists of a pattern of fringes of high presence-probability ('brilliant' fringes), all practically parallel to the *0x* axis and mutually separated by fringes of quasi-zero presence-probability ('dark fringes'). (We note that there is no way to directly *observe* these fringes, since they are only a statistical ***concept*** totalized outside any *individual* observation-time, while the individual realizations of a de Broglie guided trajectory do not leave spontaneously a perceivable and persisting trace. Only *cumulated* ***final impacts*** on a sensitive screen can be directly perceived.

We now start describing the proposed experimental analysis denoted *EXP*.
In the specified conditions the trajectory of the corpuscular-like singularity from any one physical individual *dBB* wave-function $\Phi_{12}$ is unanimously presupposed to be parallel to *0x*. Furthermore, it is also unanimously presupposed that any experimental intrusion in the inside of a given specimen $\sigma_{\Phi,\mathbf{G}(G1,G2)}$ of $ms_{G(G1,G2)}$ entails perturbations of the constant quantum-potential which generate quantum-fields that destroy the physical phase relation from the physical wave of this specimen represented in $\Phi_{G(G1,G2)}(\mathbf{r},t)=a(\mathbf{r},t)exp((2\pi i/h)\beta(\mathbf{r},t)$ by the phase-function $\beta_{12}(\mathbf{r},t)$, which in its turn destroys the possibility to register data significant for the characterization of the initial wave of $\sigma_{\Phi,\mathbf{G}(G1,G2)}$ represented by $\Phi_{G(G1,G2)}(\mathbf{r},t)$[136].

The main purpose of the following analysis is to figure out a system of choices of the involved experimental parameters that – by trial and error – shall permit to *prevent this to happen* and to control whether yes or not this purpose has been realized.

The most important parameter seems to be the kinetic energy of the specimens $\sigma_{\Phi,\mathbf{G}(G1,G2)}$ of $ms_{G(G1,G2)}$. Our hypothesis is that if the mean value of this parameter is sufficiently higher than the mean value of the energy of the ionization of a molecule from the utilized sensitive substance used for registering the researched trace, this might already suffice for insuring stability of, both, the phase relation from $\Phi_{G(G1,G2)}(\mathbf{r},t)$ and the supposed direction of displacement of the singularity from the amplitude of $\Phi_{G(G1,G2)}(\mathbf{r},t)$.

*EXP* can be structured as a sequence of distinct tests:

At a distance *$0x_1$ near the entry into the zone of interference* is placed perpendicular on *0x* a very thin layer *L1* of sensitive substance permitting with maximal probability at most *2* successive initial acts of ionization. At a second distance *$0x_2$ placed near the end of the*

[136] We just mention however that in the posited conditions the forces entailed by such fluctuations act only via the 0z component, so their possible effect should consist – at most – of a displacement of the involved singularity on another 'brilliant fringe', without suppression of the phase relation that determines the momentum-value $p_x$ from (34)).

*interference domain* is placed a thick layer *L2* of photographic emulsion with high density of molecules. When the *first* ionization occurs in *L1* at a time $t_1$ a conveniently connected chronometer registers this time. As soon as the corpuscular singularity reaches the second layer it produces there nearly certainly, and practically on the edge of its entry in this layer, a third ionization that is recorded at a time $t_2$. Then other ionizations follow until the energy of the corpuscular-like energy is consumed. *Let* $t_f$ *be the time registered for the last ionization*.

The ionizations *constitute* the desired operation of measurement of the de Broglie guided momentum-value ***p*** from (33) at the time $t_1$ when the registrations begin (***p*** is not the quantum mechanical momentum-observable ***P***). Let us denote by *Mes.**p*** this operation of measurement.

A big number of coding-measurement successions *[**G**($G_1$,$G_2$).Mes.**p**]* is realized. We keep all the cases in which *two* initial ionizations have been registered inside *L*:

* The two first ionizations permit to establish via the direction of the small segment of line that unites them whether the perturbing quantum-force has effectively displaced the corpuscular singularity on another fringe of high presence-probability, or not. This permits to be aware of the existence of perturbing quantum fields and of the strength of their effects, and to keep for use in the final calculus only the cases *without* displacement on another fringe-direction.

* The two *times* $t_1$, $t_2$ of the two ionizations registered respectively in *L1* and *L2* can furthermore define experimentally the *direction* of the momentum (34); while the difference ($t_f - t_1$) (completed with adequate considerations on the duration of one process of ionization) yields an estimation of the *value of the **velocity***: a sort of time-of-flight method ‘internal’ to the involved specimen of the studied microstate.

* The ending set of ionizations inside *L2* permits to calculate the absolute value of the momentum $\boldsymbol{p}_j$ out of the energy consumed by each one act of ionization and by the number of all the ionizations, but *in connection with the de Broglie-mass* $\mathcal{M}_o$.

* The statistic of all the positions registered at an individual time $t_2$ permits to know whether the position distribution after the first one or two ionizations is still organized in maxima and minima indicating interference fringes; so *it verifies the conservation of the initial phase-relation*.

* Since the first impact defines also the initial position ***r*** with respect to the referential, the set of registrations specified above – considered globally – *would violate Heisenberg's non-mathematical principle*[137] (not the $QM_{HD}$ mathematical "uncertainty theorem"). This would establish that the validity of Heisenberg’s principle in fact is relative to the experimental procedure. It would also establish that:

> The concept of incompatible "observables" in the sense of $QM_{HD}$ *is tied with coding-measurement-evolutions that – throughout the duration of each one act of measurement Mes.**p** – **freeze** the eigenvalue to be identified, as required in (20').*

While for microstates *ms(unbound,1)*$_{cG(qf)}$ this concept – via impossibility to insure (20') – escapes human control and in consequence of *thi*s, the domain of validity of the mathematical "uncertainty " from $QM_{HD}$ is devoid of *general* validity, in solidarity with (20').

The preceding considerations are obviously generalizable to *any* microstate *ms(unbound,1)*$_{cG(qf)}$.

---

[137] Such a violation – of which the possibility has been very explicitly asserted for heavy microstates in MMS [1964] – has been recently *proven* experimentally for photons (cf. Piacentini & altera [2015]).

We stress again that, strictly speaking, $QM_{HD}$ concerns exclusively *heavy* microsystems. So only a realization of *EXP* for systems with non-null rest mass would possess a full significance of principle. The best choice would be to work with a neutron-Young-interference (two holes) that would introduce relatively high kinetic energies even for moderate velocities and would involve exclusively quantum potentials and quantum fields, thus avoiding any electromagnetic effect during the ionizations.

A priori, nothing in the preceding sequence of conditions seems to be technically unrealizable. And if it were realized then the statistic of the registered momentum-values would *permit* now confrontation with the $QM_{HD}$-prediction (32): *it would have become possible to* ***invalidate*** *(32) experimentally* (a confirmation seems unlikely because no plane eigenwaves do realize on a finite space-time support)[138].

*(7.III).2.3.1.3. Very important final remarks on EXP*

What is directly measured in *EXP* ?

***a)*** The total energy spent for the ionizations that mark an observable trace *while they exhaust the kinetic energy of the involved specimen of the studied microstate.*

***b)*** The direction of the movement of the involved corpuscular-like singularity, so the direction of the velocity from (33),(34).

*But the values* $\mathbf{p_j}$ *of the momentum* $\mathbf{p}$ *from (33),(34) are not directly measured*, they have to be calculated from the directly observable data and the result of the calculation *depends more or less explicitly on* **Einstein's** *concept of "mass"* defined for macroscopic and even c osmic "mobiles" as $m=m_o/\sqrt{(1-(v^2/c^2)}$ and *that is involved in (33),(34)* ***if*** *de Broglie's concept of momentum* $\mathcal{M}\mathbf{v}$ *is made use of* (cf. de Broglie [1956] p.205 eq. (42)) with $\mathcal{M}=\mathcal{M}_o/\sqrt{(1-(v^2/c^2)}$ and $\mathcal{M}_o=\sqrt{[m^2{}_o+(h^2/4\pi^2c^2)(\Box a/a)]}$). *The de Broglie concept* $\mathcal{M}$ *for what is called "mass" presupposes that Einstein's concept of "rest-mass"* $m_o$ conserves – both – meaning and necessity with respect to the scientific representation of microscopic unbound entities conceived to belong to the universal physical substance at spatial dimensions of the order $10^{-12}$, $10^{-13}$ and time-dimensions of the order $10^{-33}$ (in particular also with respect to the representation of the observational interactions that generate communicable, consensual and verifiable knowledge on such entities).

> ***But this is very far from being obvious***. With respect to the concept of "mass" de Broglie's treatment does *not* seem satisfactory. It mixes in an uncritical uniformity, different levels of space-time dimensions, *and* so it implicitly asserts a posteriori a fallacious sort of conceptual simultaneity that obscures the perceptibility of the genetic order of *possibility* of conceptualization. This, in its turn, entails factitious basic conceptual assertions that entail factitious problems (cf. MMS [1994] and Appendix2).

---

[138] And I add that insofar that the technical possibility of *EXP* can indeed be distinguished and separated from any mathematical argument, any confusion between formal arguments and technical ones should be strictly eliminated. For at this point, I think, it is already clear that the semantic differences that separate from one another the approach practised inside $QM_{HD}$ and that practised in *dBB* are so numerous and fundamental that any argument on the *theoretical* possibility or not to register a 'trace' in the sense of *dBB* is simply devoid of possible conclusion if it is formulated by the use of elements of mathematical formalism, whether that of $QM_{HD}$ or that of *dBB*. On the other hand the stake of a conclusion to this problem – centred upon ***exclusively*** the well-defined question of the technical, experimental possibility of trace-registrations that would establish the vector-values of the de *dBB* 'guided momenta' (33) – appears to be as for *dBB* as it is for the the second quantum mechanics researched here. Indeed for *dBB* this answer can decide between:

- ***Either*** a *dBB* approach that conserves its present status of only a mathematically expressed meta-physics of the modern microphysics that keeps fighting for the title of "interpretation" of a $QM_{HD}$ that, itself, is un-intelligible.
- ***Or*** a *dBB* approach that might play a basic role if it were incorporated in, both, a conceptual way *and a physically-operational* way into a fully intelligible theory of the microstates that belongs to the Mathematical Physics.

While on the other hand, in (33) the second equality alone, $\boldsymbol{p}(\boldsymbol{r},t)= -\boldsymbol{\nabla}.\beta(\boldsymbol{r},t)$, seems to *suffice* for globally defining $\boldsymbol{p}$ exclusively from the phase $\beta(\boldsymbol{r},t)$ of the physical wave with wave-function $\Phi_{G(G1,G2)}(\boldsymbol{r},t)=a(\boldsymbol{r},t)exp((2\pi i/h)\beta(\boldsymbol{r},t)$ of each involved specimen $\sigma(ms_G)$ of the studied microstate $ms_G$[139].

By such considerations *EXP* leads to the root of all the "mechanical" conceptualizations. Indeed along the whole vertical of conceptualization the same few basic ***words*** [energy, mass, action, momentum] are made use of. *But* ***at different levels of space-time dimensions these words point toward different semantic contents****. So:*

*EXP involves a double challenge:*
** to measure that what for a "microstate" is* **now** *called "momentum";*
** and furthermore to re-define adequately the involved meanings when these need to be defined at the considered level of space-time dimensions.*

Word and meaning are mutually very distinct elements of the human arsenal for thought and communication of thought.

We shall come back later to the preceding remarks.

In any case, for the microstates $ms(unbound,1)_{cG(qf)}$ recourse to (33) seems to be the unique solution for measuring the basic mechanical quantity that now is called "momentum". So *EXP has* to be realized and its results have to be expressed in carefully organized terms that take into account the levels of considered space-time dimensions as well as the corresponding possibility of "mechanical" conceptualization [140].

---

[139] In Bohm's interpretation – where the basic model of the physical substance involves just "particles" of which the velocity $\boldsymbol{v}$ is determined by surrounding waves – no connection with Einstein-mass is explicitly asserted; but then the basic duality "particle"-wave might seem conceptually embarrassing.

[140] Just an additional remark on the debate Englert&alt., [1992], Dürr&alt;, [1993], Finkelstein [1995], Hiley&alt, [2000], cited already in a preceding note, versus the experiment *EXP*. At this point, I think, it is already clear that the semantic differences that separate from one another the approach practised inside $QM_{HD}$ and that practised in *dBB* are so numerous and fundamental that any argument on the *theoretical* possibility or not to register a 'trace' in the sense of *dBB* is simply devoid of significance if it is formulated by the use of a now available mathematical formalism, whether that of $QM_{HD}$ or that of *dBB*. On the other hand the stake of a valid answer to this problem – centred upon ***exclusively*** the definite question of the technical, experimental possibility of trace-registrations that would establish the vector-values of the de *dBB* 'guided *velocities*' from (33) and the involved *energies* – appears to be very high for *dBB*. Indeed this answer can decide between:
- ***Either*** a *dBB* approach that conserves its present status of only a mathematically expressed meta-physics of the modern microphysics that keeps fighting for the title of "interpretation" of a $QM_{HD}$ that, itself, is un-intelligible.
- ***Or*** a *dBB* approach that might play a basic role if it were incorporated in, both, a conceptual way *and a physically-operational* way into a fully intelligible theory of the microstates that belongs to the Mathematical Physics.

### *(7.III).2.3.2. Construction of the researched new Hilbert-Dirac representations tied with interference-microstates ms(unbound,1)$_{cG(qf)}$*

**Rule seventeen**
The proposed difficulty must be gone through be making abstraction of the fact that some of the involved terms are known and other ones are unknown, by following in a genuine walk the mutual dependences.
**Rule nineteen**
It is by this method that one must research all the dimensions (physical quantities) that are expressed in two different ways, in terms either known or unknown, in order to browse directly through the difficulty; for by these means we shall obtain just as many comparisons between equal things.
**Rule twenty.**
*It is **after** having obtained the equations that we must achieve the omitted operations, without ever making use of multiplication when division is necessary* [141].

René Descartes, The Rules for the direction of mind
(Regulae ad directionem ingenii), toward 1628 – 1629,
in Letters to Elisabeth.

Wikipedia (our translation from French)

In the present work I shall not await the verdict of *EXP* before achieving the construction started in this chapter (7.III).2**,** of a global outline of an acceptable mathematical representation of the quantum-measurements on unbound microstates. I shall just admit by hypothesis that *EXP* has been performed and has established the possibility to register experimentally instantaneous data on guided momentum-values (33). This, in our era of nanotechnologies, is very likely to happen soon, I think. And this likelihood can only be increased by an already pre-existing global outline of a view where an investigation of the type of *EXP* possesses an own and central conceptual place.

So we now continue the construction begun in (7.III).2.

*(7.III).2.3.2.1. Extension of the framework*

From now on we include the *dBB* approach in our framework, so we shall work inside the enlarged framework *[IQM-QM$_{HD}$-(dBB)]*.

*(7.III).2.3.2.2. A coding-postulate for the momentum-values of ANY unbound microstate*

We first recall that in *dBB* the guidance law (33) is asserted deductively and with *full generality*, both in the presence *and* in the absence of quantum fields. So inside *[IQM-QM$_{HD}$-(dBB)]* we can formulate the following coding-postulate for momentum-measurements on *any* sort of unbound microstate, *without* or *with* inner quantum potential, as well as with *one* or with *several* micro-*systems* (in the sense of the definitions from (2.I)1).

We posit that:

***P(cod.p$_j$)$_{unb.ms}$*.** The instantaneous momentum-value $\boldsymbol{p_j}$ of *any* unbound microstate – whether a microstate *ms(unbound,1)$_{cG(qf)}$* or *ms(unbound,1)$_{G(nc)}$* – can be determined by a coding-measurement-succession that obeys the representation

$$[G_t.Mes(\mathbf{r},\mathbf{p})] \rightarrow_{\mathbf{trace}} (\mathbf{r_k},\mathbf{p_j})_t\,, \quad k=1,2,..K; \quad j=1,2,..J; \quad \forall ms_{Gt} \qquad (35)$$

where: the sign "$_{unb.ms}$" is to be read " unbound microstate"; $G_t$ is posited to generate at the time *t – in the sense of (13")* – one unbound individual physical specimen $\sigma_{(Gt)}$ of the studied microstate $ms_{Gt}$, initially represented by an unknown individual wave-function $\Phi_{Gt}$; *Mes(r,p)* denotes one full act of measurement of the *pair* $(r,p)_t$ considered at the time *t* when the measurement ***begins***; the sign '$\rightarrow_{trace}$' is to be read ' identified via a process of trace-registration in the sense of *EXPER*; $(r_k, p_j)_t$ is the pair of *values* registered for the pair $(r,p)_t$ of qualifying quantities.

[141] My italics and bold fonts.

Like the coding postulate (20'), the coding-postulate $P(cod)_{unb.ms}$ also acts exclusively on the *individual* level of conceptualization. Furthermore, the symbol "$_{unb.ms}$" can indicate a guiding interference effect (33) even in the *absence* of any classically definable field, and in such a case *it transgresses the capacity of representation of a* $QM_{HD}$ *measurement-Hamiltonian* ***H(A)***. So in such a case (35) falls entirely outside the domain of facts that can be represented by a $QM_{HD}$-Hamiltonian. Which means that the Schrödinger equation of evolution simply ceases to be applicable: what is applied in such a case, in fact consists of additional 'techniques', i.e. ad hoc representations via "obstacles" (barriers, "walls", tunnelling "effects", criteria of continuity, approximations, etc.)

And finally, let us note explicitly that:

*The coding postulate (35) violates overtly Heisenberg's principle, the principle of complementarity, as well as the Heisenberg theorem from* $QM_{HD}$ [142].

The postulate $P(cod)_{unb.ms}$ is a new descriptive feature from the *dBB* approach that, inside the present approach, is injected into the domain of 'scientific' knowledge i.e. communicable, consensual, predictive and verifiable knowledge. In this phase of the present construction the channel of adduction into 'scientificity' of descriptional prime-matter from the purely 'interpretative' *dBB* theory is:

$$[MP(ms_{G,cw}),\ (1')\ G_{cw} \leftrightarrow ms_{G,cw},\ (14)\ ms_{G,cw} \equiv \{\sigma(ms_{G,cw})\},\ (35)\ P(cod)_{\forall msG})] \quad (36)$$

The channel (36) might entail a radical transmutation of the mutual status of the conceptualization from *dBB* and that from $QM_{HD}$; a genuine *inversion* of their relative conceptual position, in fact. Indeed, while $QM_{HD}$ is found to be devoid of verifiability for its predictions on momentum-measurements on unbound interference microstates – which is a *major* gap – the present approach might suppress this flaw by use of elements from the works of de Broglie and Bohm that, in an arbitrary and inertial way, are considered to express 'purely' interpretive approaches.

### *(7.III).2.3.2.3. Elucidation on observables and beables*

The coding postulate (35) draws attention upon the now current notions of "beable" or "observable". It has become quasi consensual to distinguish between beable qualifications and on the other hand $QM_{HD}$-observables. The position-vector observable ***R*** is considered more or less implicitly to behave like a beable, in this sense that the registration of perceivable marks produced by a specimen of the studied microstate is considered to trivially show where "the system" [143] was "really" placed when the mark emerged. But the case of ***R*** is regarded as a degenerate case. In general a $QM_{HD}$ observable – the momentum observable included – is currently conceived to manifest eigenvalues that are *created* by the measurement-interaction: What is observed and coded by an eigenvalue of the measured quantity is conceived to be in general different from the initial beable value, i.e. it is conceived to have been created by the measurement-evolution *out* of this beable value [144]. This even constitutes one of the major specificities of the modern microphysics.

---

[142] These 'principles' and this theorem – formulated in absolute terms – constitute together *a knot of* ***confusions*** *between individual temporal characters and statistical characters*. This knot deserves a detailed analysis of the type of that from (*(6.II).4.4.3*) for the $QM_{HD}$ representation of quantum-measurements (Heisenberg's 'principle' mixes present and future individual data, while the 'principle' of complementarity is in fact a consequence of the $QM_{HD}$ *Hilbert-space* ***choices*** *of representation* for ket-states and for observable-operators).

[143] In terms of the model posited in (1') this beable-value concerns the momentum of the singularity in the amplitude of the physical wave of the involved specimen of the studied microstate (cf. (2.I).1.2 and (6.II).2)).

However the analyses from (6.II).3 and (6.II).4 – and particularly the examination of the method time-of-flight for measuring momentum observable ***P*** – bring forth a view that is *directly* ***opposed*** *to that recalled above*.

> A coding-measurement-evolution of the general form (20') (that is implicitly conceived inside $QM_{HD}$ to be applicable to any sort of microstate) is ***expressly*** constructed such as to ***conserve unchanged*** the initial beable value of the measured quantity while drawing it into the realm of consensual perceptibility – so into the realm of possibility of consensual knowledge – via observable marks that permit to identify this initial value retroactively, on the basis of theoretical arguments and calculi.

*This*, in fact, is the very essence of the criterion that acts in (20) and in (20') for asserting that the procedure *can* be considered to act as a *coding*-measurement-evolution. And precisely *this* ceases to be realizable in a controlled way for momentum measurements on microstates $ms(unbound,1)_{cG(qf)}$.

But why, exactly, does it cease? This question has been already considered in *(7.III).2.2.1*. Nevertheless we re-express here the answer in a more concentrated way.

In the case of an interference microstate $ms(unbound,1)_{cG(qf)}$ – in contradistinction to what happens with microstates $ms(unbound,1)_{G(nc)}$ – the dynamics of ***that*** what, in a specimen of the studied microstate, does admit *mechanical* qualifications [145] – namely the singularity in the amplitude of the physical wave of the involved specimen of the studied microstate – can *never* be brought under the dependence of ***exclusively*** the macroscopic context of this specimen, that is controllable by the human observer. This is so in consequence of the definition of a composed operation of generation $\boldsymbol{G}(G_1,G_2,...G_m,...G_M)$ of a microstate $ms(unbound,1)_{cG(qf)}$. Such an operation of generation entails that inside *any* specimen of a microstate $ms(unbound,1)_{cG(qf)}$, the mentioned singularity from the amplitude of the involved physical wave is genetically immersed in a physical superposition of mutually distinct trends of wave-movement instilled by the mutually distinct components $(G_1,G_2,...G_m,...G_M)$ of the unique fully realized operation of generation $\boldsymbol{G}(G_1,G_2,...G_m,...G_M)$. According to the model posited in (1') this superposition acts upon the mechanical displacement of the singularity via the *dBB* guiding-law, *and* – in the same time – it shields this displacement from full control by the human observer. Indeed *the outside* ***of the singularity*** from the involved specimen of the studied microstate stays continuously exposed to an emergence, from fluctuations of the internal non-null quantum potential, of unpredictable quantum forces that can add unpredictable changes to the initial value of the momentum that characterizes the mechanical displacement of the singularity, that is what has to be qualified. No deliberate "extinction" of the possibility of emergence of quantum fields *inside* the involved specimen of the studied microstate can be conceived, like in the method time-of-flight, nor other sort of deliberate control of the internal stability. *The initial –* ***beable*** *– vector-value of the mechanical momentum to be measured cannot be experimentally preserved from unknowable internal changes throughout the measurement-evolution*. All that seems possible to be done in this respect is to maintain as stable as possible the global immediate macroscopic context of the involved specimen[146]. So what about the momentum-value brought forth a coding-measurement-evolution (35) ?

---

[144] I have myself held this view, strongly and for a very long time.

[145] In Bohmian mechanics, for instance, the spin quantity is considered *not* to be a 'mechanical' quantity; it is regarded as a qualification of the wave movement.

[146] Here, at this point of the present inquiry, one can realize to what a point the physical reality opposes obstacles to factually gained *knowledge* – not obstacles to ***mental*** conceptual or mathematical modelling, but to ***factual*** *generation of knowledge*. In the case of a specimen of a microstate such a factual generation cannot even insure a global spatial delimitation (cf. (6.II).2) of a specimen of the studied microstate, so that the very basic separation 'inside-outside' cannot work any more in the same simplified way in which it works for microstates without possible inner quantum fields. On the other hand we are constrained to construct models; and once a model is posited we are constrained to

*"Value-**conserving**" or "value-**destructive**" coding-measurement evolution.* Let us go back now to the notions of beable or observable. Consider again the microstates *ms(unbound,1)*$_{cG(qf)}$ (for simplicity we restrict to the basic case of only two components $G_1$ and $G_2$ inside the unique fully realized operation of generation **G**. So we write in short **$G(G_1,G_2)\equiv G_{12}$**, *$\Phi_{G12}(r,t)$* and $|\Psi_{12}(\mathbf{r},t)>$). Accordingly to the framework *[IQM-QM$_{HD}$]* already we have:

**(a)** *Any* act of momentum-measurement *Mes**P*** is performed on only ***one specimen*** $\boldsymbol{\sigma_{G12}}$ *of the microstate-to-be-studied* $ms_{G12}$, never on this microstate itself that is just *an abstract **concept*** (cf. (1.I).1.3 and the relations (1), (1')).

**(b)** In *each* coding-measurement-succession *[$G_{12,t}$. Mes($\mathbf{r_t}$,$\mathbf{p_t}$)]* from (35), the structure of the phase *$\beta_{12}(\mathbf{r},t)$* from the individual physical wave represented by the wave-function *$\Phi_{12}(\mathbf{r},t)=a_\Phi(\mathbf{r},t)exp((2\pi i/h)\beta(\mathbf{r},t)$* from the involved individual physical specimen $\sigma_{G12}$ plays a determining role in the determination of the registered vector-value $\boldsymbol{p_j}$.

**(c)** *For the particular case of a microstate ms(unbound,1)$_{cG(qf)}$, the structure of the wave-movement of the singularity from the wave of $\sigma_{G12}$ can **never** be conceived to rigorously be integrated into one **plane** wave.*

Inside *QM$_{HD}$* any perception of these *[IQM-QM$_{HD}$]*-features is entirely absent; only shadows of these features float there, in a basic mist tied with the fact that the concept itself of a factually generated microstate is not even defined in *QM$_{HD}$* and with the use of the cover-all word 'system'. This explains how it has been possible to formulate the *QM$_{HD}$* measurement-postulates 1 and 2 according to which the result of *any* act of momentum measurement performed on *any sort of microstate* would consist of an eigenvalue $\boldsymbol{p_j}$ of the *QM$_{HD}$* momentum-observable ***P*** that:

° Is *'created'* by the measurement process, i.e. is *different* from the 'beable-value' at the moment when the act of measurement begins.

° Always emerges necessarily tied to ***ONE*** corresponding ***plane***-wave eigenket $|a_\psi.exp((2\pi i/h)p_j.\mathbf{r}>$ of the *QM$_{HD}$* momentum observable ***P*** (the measurement postulate 1).

° Subsists even *after* the closure of the act of measurement *together* with its corresponding one-plane-wave eigenstate (the measurement-postulate 2) (cf. the concept of successive measurements).

These edicts illustrate strikingly the consequences of *beginning* by calculated representations of *statistics* of results of individual measurements that: are only *imagined* beneath these statistics of results of individual measurements; imagined via schematic extrapolations of the classical thinking; and without having constructed these statistics via effectively achieved measurements; measurements realized under all the conceptual and factual constraints entailed by the specific cognitive situation that is realized when humans want to represent microphysical entities.

Whereas inside *[IQM-QM$_{HD}$-(dBB)]* a careful analysis of this situation led to the conclusion that:

* The momentum-value $\boldsymbol{p_j}$ that is observed for a microstate is ***never** "created"* by the measurement process, ***neither*** for a microstate *ms(unbound,1)$_{G(nc)}$* subjected to coding-measurement-evolutions that obey (20'), ***nor*** for a microstate *ms(unbound,1)$_{cG(qf)}$* subjected to coding-measurement-evolutions that obey (35):

In both these cases the momentum-value at the time *t* when an act of momentum-measurement ***begins*** is ***also*** the final value $\boldsymbol{p_j}$ that is coded for by the observable marks develop a representation that be logically coherent with respect to it, for if this is not the case, in a way or another we always run into unintelligible difficulties. Such are the remarkable and strange cognitive relations between human logical coherence, and predictability on physical facts.

registered at the end of the coding-measurement-evolution. *Only the coding-processes themselves are different for these two sorts of microstates but neither of them changes the initial beable value assigned to the momentum observable.*

Indeed accordingly to *[IQM-$QM_{HD}$-(dBB)]*:

- In the case of a microstate *ms(unbound,1)$_{G(nc)}$* subjected to momentum-measurements that accept the coding-postulate (20') the posited coding-measurement-evolution acts as follows: At the time *t* when a measurement-evolution *begins* the structure of the wave-movement around the singularity from the wave of $\sigma_{G12}$ determines accordingly to (33) some beable value $\boldsymbol{p_j}$ of the momentum observable that, in general, is not that of a plane wave. But as soon as the exterior fields are all suppressed it can coherently be posited that – for a microstate *ms(unbound,1)$_{G(nc)}$* – the structure of the *whole* wave of the involved specimen $\boldsymbol{\sigma_{G12}}$ acquires immediately a plane-wave structure (just in consequence of the lack of reasons for admitting any influence on the subsequent dynamics of the singularity from $\sigma_{G12}$ that settles in when no fields of any sort act any more (the Occam-razor argument)). So according to (20') *the momentum-value* $\boldsymbol{p_j}$ *is* ***conserved*** *such as it was at the moment t when the considered act of measurement has begun*: This in fact is the very *purpose* of a coding postulate of the type (20') asserted for any *$QM_{HD}$* observable ***A***. Moreover for a *momentum* measurement the generally admitted (though only implicitly asserted) coding postulate (20') simply identifies with the method time-of-flight *itself*, with coding procedure (20) where the conservation of $\boldsymbol{p_j}$ is erxplicitly required and from which (20') stems by generalization **[147]**. So:

*For microstates ms(unbound,1)$_{G(nc)}$ the observed momentum value $p_j$ has to be conceived to be the initial beable value itself, not another value created by the measurement-evolution.*

- This same conclusion is reached also for a microstate *ms(unbound,1)$_{cG(qf)}$* that is subjected to momentum-measurements accordingly to the coding-measurement-postulate (35): At the moment *t* when the coding-measurement-evolution from (35) begins, the singularity from the wave of the involved specimen $\sigma_{G12}$ of the studied microstate $ms_{G12}$ is already genetically immersed inside the quantum-potential produced by the corresponding composed operation of generation ***G(G1,G2))*** (cf. *(7.III).2.2.3*). And even though the coding-measurement-evolution imposed by (35) upon the involved specimen $\sigma_{G12}$ has to be conceived to destroy – to "consume" – the momentum-value $\boldsymbol{p_j}$ possessed by the singularity from the wave of $\sigma_{G12}$ at the time *t* when the coding-measurement-evolution begins, nevertheless the definition of *EXP* entails that ***this*** *initial beable value* $\boldsymbol{p_j}$ *is what is translated a posteriori from all the data gathered via the coding-measurement-evolution from (35)* **[148].**

In short, we can now firmly assert that according to the approach developed here:

* In *both* cases the registered momentum-value $\boldsymbol{p_j}$ is that one that is possessed by the singularity from the involved specimen of the studied microstate at the time *t* when the considered act of measurement *begin*s. *One can say that it is an* ***"observed beable value"***.

The essential difference between the two sorts of coding-measurement-evolutions considered above simply *does not concern* the momentum-value $\boldsymbol{p_j}$ that is observed: *it concerns exclusively* ***the way in which this observed value is drawn into consensual knowledge***.

---

[147] So in the case of a microstate *ms(unbound,1)$_{G(nc)}$* it is *at most* the structure of the global wave-movement from the whole wave of a specimen $\sigma_{*(G)}$ of the studied microstate $ms_G$ – not the registered momentum-value $p_j$ – that is created by the act of measurement. But this global structure does not qualify the corpuscular-like singularity in terms of the mechanical momentum-observable, it only *specifies* the global sort of de Broglie model of a microstate *ms(unbound,1)$_{G(nc)}$* that emerges by the measurement-evolution.

[148] Let us add that in *both* cases at the end of the considered measurement-evolution the involved specimen of the studied microstate has ceased, in general, to belong (in the sense of (1)) to the studied microstate.

* And ***THIS*** difference depends on whether the coding-measurement-evolution is of the type (20') or of the type (35); which in its turn depends on the sort of microstate that is studied, *ms(unbound,1)*$_{G(nc)}$ or *ms(unbound,1)*$_{cG(qf)}$:

In the case of a microstate *ms(unbound,1)*$_{G(nc)}$ the corresponding coding-postulate (20') involves a measurement-evolution that draws a beable value $p_j$ into knowledge ***by conserving it***, while in the case of a microstate *ms(unbound,1)*$_{cG(qf)}$ the corresponding coding-postulate (35) draws a beable value $p_j$ into knowledge ***by consuming it***.

*The distinction between beables and observables evaporates*[149].

So inside the framework *[IQM-QM*$_{HD}$*-dBB)]* we introduce a new language: We shall speak of:
* *"value-conserving"* coding-measurement-evolutions (20') that bring into the domain of the observed the ***beable*** momentum-value of a microstate *ms(unbound,1)*$_{G(nc)}$;
* *"value-destructive"* coding-measurement-evolutions (35) that bring into the domain of the observed a ***beable*** momentum-value from a microstate *ms(unbound,1)*$_{G(qf)}$ where this value cannot be controllably conserved [150].

All the critiques made above on the representation (15") can be dealt with when the distinction introduced above is substituted to the illusory distinction between beables and observables[151]. In particular, the coding postulate (35) appears to be founded upon value-destructive coding-measurement-evolutions, while the coding postulate (20') is founded upon value-conserving coding-measurement-evolutions, both bringing forth results with the meaning of beable eigenvalues.

### *(7.III).2.3.2.4. An equation for the eigenket and eigenvalues of an extended concept of momentum-observable*

We are now finally ready to *extend* the *QM*$_{HD}$ equation for eigenket and eigenvalues of the momentum-observable ***P*** so as to admit also eigenket where two or more practically plane waves are ***physically superposed***, so as to be able to mime a sample of wave-movement in the neighbourhood of the corpuscular-like singularity from also an interference-microstate *ms(unbound,1)*$_{G(qf)}$.

We want to *conserve* for a generalized *[IQM-QM*$_{HD}$*-dBB)]* momentum observable the same operational definition ***P*** that it has inside *QM*$_{HD}$, in order to let it stay connected with all the other operators already defined for a "mechanics" of microscopic entities. So from now on we distinguish between the concepts of "momentum-*operator*" ***P*** that works mathematically inside the equation for eigenstates and eigenvalues of the momentum observable, and a new concept of *[IQM-QM$_{HD}$-dBB)]* "momentum *observable*" that:

- permit eigen-functions of ***P*** tied with, ***either*** "observed beable-values" brought forth by value-*conserving* coding-measurement-evolutions when such coding-measurement-evolutions are possible and conceptually adequate,
- ***or*** "observed beable-values" brought forth by value-*destructive* coding-measurement-evolutions when the studied microstate permits such – *or* ***only*** *such* – coding-measurement-evolutions.

---

[149] This distinction is likely to have been introduced by Bell in an insufficiently analyzed connection with Heisenberg's principle and with the concept of "compatibility" of two observables ***A*** and ***B*** that obey the *QM*$_{HD}$ non-commuting algebra of the observables. Explicitly expressed, Bell's idea might have been that two qualifying quantities are compatible iff their values can *both* be "created" and brought into the realm of the observable and verifiable via one same act of measurement operated upon one same specimen of the studied microstate. But the requirement of "creation" of the observable value of a *QM*$_{HD}$-"observable" introduces ***un-necessarily*** a conclusion of ***absolute*** incompatibility in the fundamental case of the pair (***P,R***) (i.e. with respect to *any* sort of coding-measurement-evolution).

[150] This is remarkable. It is a strong retroactive confirmation of *IQM* with its basic specificities with respect to *QM*$_{HD}$: All the elements from the chain ***[****The definition (1) that introduces the concept of operation of generation G; a relation between each G and one 'factually generated microstate' that introduces a whole set of distinct physical specimens; the concept of composed operation of generation; the concept of coding-measurement-evolution; etc.****]*** that concerns the individual level of bottom-up conceptualization of the microstates are radically absent in *QM*$_{HD}$. While here they permit once more basic elucidations (like those obtained in (6.II) for the significance of an eigenket, for the consequences entailed by this significance, and for the *QM*$_{HD}$ representation of quantum-measurement). *I do not speak of 'truth' in an absolute sense, **which is mere illusion**, but of elucidations in the sense of **construction of intelligibility by inner semantic-formal consistency of a structure of representation***.

[151] As for what happens once a coding-measurement-evolution is finished, whether the observed value of the measured observable and the corresponding microstate subsist or not, it does not seem very useful to try to state this in general terms. This can be examined in each case accordingly to the deliberate future operational purposes of the acting observer. Anyhow, in general the final registration of the result of an act of measurement destroys the "state" of the involved specimen of the studied microstate, and this contradicts the *QM*$_{HD}$ measurement-postulate 2; while the critique of (15") and (32) incriminates the postulate 1; so in *QM2* these postulates will have to be dropped.

This momentum-observable will be denoted $\boldsymbol{P}_{vc\text{-}vd}$ ("$_{vc}$": value conserving; "$_{vd}$": value destructive). This distinction between the mathematical definition and the involved semantic contents amounts to an explicit assertion of the relativity of the possible *ways* of observing the momentum, to the sort of studied microstate[152].

Consider now a microstate $ms_{Gt(G1,G2,...Gm,...GM)}$ of type $ms(unbound,1)_{cG(qf)}$. Via (33) a given measurement-succession $[\boldsymbol{G}_{(G1,G2,...Gm,...GM)t} .Mes\boldsymbol{P}_{vc\text{-}vd}]$ involves a measurement-evolution (35) $Mes\boldsymbol{P}_{vc\text{-}vd}$ that is directly related only with the phase-function $\beta(\boldsymbol{r},t)$ from the individual wave-function $\Phi_{Gt(G1,G2,...Gm,...GM)}(\boldsymbol{r},t)=a_{\Phi}(\boldsymbol{r},t)exp((2\pi i/h)\beta(\boldsymbol{r},t)$ of the involved specimen $\sigma_{Gt(G1,G2,...Gm,...GM)}$ of $ms_{Gt(G1,G2,...Gm,...GM)}$ (Louis de Broglie has initially called the whole function $\Phi_{Gt(G1,G2,...Gm,...GM)}(\boldsymbol{r},t)$ a "*phase*"-function). The form of $\beta(\boldsymbol{r},t)$ can be conceived to stem from an inner structure of the global wave-movement expressed by $\Phi_{Gt(G1,G2,...Gm,...GM)}(\boldsymbol{r},t)$ that somehow "involves *several* plane-waves", in the following sense: It seems necessary to conceive that (in general at least) at any space-time point $(\boldsymbol{r},t)$, each composing operation $G_m$ that contributes to $\boldsymbol{G}_t(G_1,G_2,..G_m,..G_M)$ induces its own directional "trend of wave-movement". Let us denote by $\boldsymbol{k}_m(\boldsymbol{r},t)$ the unit-wave-vector of this contribution. According to (1) the operation $\boldsymbol{G}_t(G_1,G_2,..G_m,..G_M)$ is posited to come out 'the same' each time that it is re-produced, with respect to the finite set of macroscopic parameters that defines it. So – by definition – the wave-vector $\boldsymbol{k}_m(\boldsymbol{r},t)$ tied with one component $G_m$ from $\boldsymbol{G}_t(G_1,G_2,..G_m,..G_M)$ has to be conceived to be invariant with respect to the repetitions of $\boldsymbol{G}_t(G_1,G_2,..G_m,..G_M)$. But according to (1),(1') the *global* wave of the *specimen* $\sigma_{Gt(G1,G2,...Gm,...GM)}$ generated by one realization of $\boldsymbol{G}_t(G_1,G_2,..G_m,..G_M)$, that is represented by the wave-function $a_{\Phi}(\boldsymbol{r},t)exp((2\pi i/h)\beta(\boldsymbol{r},t)$, in general *varies* with respect to passage from one realization of $\boldsymbol{G}_t(G_1,G_2,..G_m,..G_M)$, to another such realization, because the position of the singularity from the amplitude of the wave varies and so also the space-time point $(\boldsymbol{r},t)$ where the phase-function $\beta(\boldsymbol{r},t)$ is significant from a mechanical point of view.

Now, in the immediate neighbourhood of each space-time point $(\boldsymbol{r},t)$ the strictly local structure of the phase $\beta(\boldsymbol{r},t)$ from de Broglie's "phase"-function $\Phi_{Gt(G1,G2,...Gm,...GM)}(\boldsymbol{r},t)$ can *always* be represented by a superposition of contributions from plane waves. So we can write this "phase"-function as:

$$\Phi_{Gt(G1,G2,...Gm,...GM)}(\boldsymbol{r},t) = a_{\Phi}.exp\left(i(2\pi/h)\left(\textstyle\sum_m p_{jm}\,\boldsymbol{k}_m\right)_t.\boldsymbol{r}\right), \quad \forall j,\ m=1,2,..M$$

[152] The purpose outlined above is not of minor importance. For in our view the case of microstates $ms(unbound,1)_{cG(qf)}$ plays the role of the ***rule*** while the case of a simple operation of generation $G$ is a rare exception (as already remarked, even in textbook examples with potential-'barriers', 'walls', 'wells', harmonic oscillators, etc. in fact we dwell with microstates $ms(unbound,1)_{cG(qf)}$ that involve interference and non-null quantum potentials). This remains non-singularized by the $QM_{HD}$ formalism where the concept of a physical-conceptual operation of generation is *absent* and its basic physical-conceptual role is replaced by the exclusively mathematical notion of 'limiting conditions' imposed upon the Schrödinger solution 'of the problem' that endows with a statistical descriptor of results of measurement on the specimens of the studied microstate. In such circumstances one is led to wonder *whether the $QM_{HD}$ predictions on the results of momentum measurements have ever been seriously subjected to verification*; and if this has been done, *how the measurements have been realized*, given that also – or only – non-classical quantum-potentials do act that are not accessible to human manipulation, so that the method time-of flight cannot be applied; while inside $QM_{HD}$ trace-registrations, in principle, are not 'legal' procedures for measuring the momentum observable. This whole – enormous – set of questions goes lost in just language-solution ("quantum-tunneling", etc.); or, for *bound* states, in global "effects" (Zeemann, Stark", etc.) that are made use of as measurement-operations, after many approximations made for defining the state-ket.

where $p_{jm}\mathbf{k}_{m\Phi}(\mathbf{r},t)$ is the contribution at the considered space-time point $(\mathbf{r},t)$ of the trend induced by one component $G_m$ from $\mathbf{G}_t(G_1,G_2,..G_m,..G_M)$, *and the factor* $p_{jm}$ *is regarded as just a numerical coefficient* that indicates the relative importance of the contribution.

Can this representation play the role of an eigen-wave-function of the momentum-observable observable $\mathbf{P}_{vc\text{-}vd}$? The answer is positive since with the operator $\mathbf{P}$ from $\mathbf{P}_{vc\text{-}vd}$ we have:

$$\mathbf{P}_{vc\text{-}vd}\,[a_\Phi.\exp\big(i(2\pi/h)\,(\textstyle\sum_m p_{jm}\,\mathbf{k}_m)_t\,.\,\mathbf{r}\big)]=$$
$$=(\textstyle\sum_m p_{jm}.\mathbf{k}_m)_t\,[a_\Phi.\exp\big((i(2\pi/h))\,(\textstyle\sum_m p_{jm}\,\mathbf{k}_m)_t\,.\mathbf{r}\big)],\quad \forall j,\ m=1,2,..M,\quad \forall j, \qquad (37)$$

where $M$ (with $m\geq 1$) is a *small* integer.

However the sum $\sum_m p_{jm}.\mathbf{k}_m$ is deliberately left non-effectuated so far because of a notable formal fact:

> Inside the exponential, the *totalized* value would assign to the wave-function $a_\Phi(\mathbf{r},t)\exp((2\pi i/h)\beta(\mathbf{r},t)$ the physical significance of ***only ONE plane wave*** instead of expressing explicitly the structure of a ***physical*** superposition of plane waves induced by the composed operation of generation $\mathbf{G}t(G1,G2,...Gm,...GM)$.

Thereby precisely our aim to generalize the currently accepted equation $\mathbf{P}|u(\mathbf{p}_j,\mathbf{r})>=\mathbf{p}_j|u(\mathbf{p}_j,\mathbf{r})>$ in a way such that the momentum observable shall admit also eigenfunctions that represent physical superpositions of plane waves, would remain *hidden* in the mathematical form of the eigen-function, notwithstanding the composed form $\mathbf{G}t(G1,G2,...Gm,...GM)$ posited for the operation of generation. So our aim would have failed ! So, just in order to avoid this (which for the moment I cannot explain) I just do *not* perform the addition inside the exponential.

But in (37) the sum $(\sum_m p_{jm}.\mathbf{k}_m)_t$ from the phase of the eigenfunction appears also in front of the expression of the wave-function and there it is separated from the mathematical expression $a_\Phi.\exp\big(i(2\pi/h)(\sum_m p_{jm}\mathbf{k}_m)_t.\mathbf{r}\big)$ of the eigen-*wave* at $(\mathbf{r},t)$. This permits to proceed as follows.

We now ***posit*** (in agreement with de Broglie and with Bohm) that at the time $t$ when the act of measurement ***begins*** (that indexes "$\mathbf{G}_t$"), the numerical *beable*-eigen-value of the momentum-observable $\mathbf{P}_{vc\text{-}vd}$ registered by a momentum-measurement-succession $[\mathbf{G}_t.Mes\mathbf{P}_{vc\text{-}vd}]$ that is acceptable with respect to the studied microstate – whether making use of the coding postulate (20') or (35)[153] – can be defined as:

$$\textit{(the numerical value of)}\,\mathbf{p}_j \;=\; \textit{(the numerical value of)}\,(-\boldsymbol{\nabla}.\beta(\mathbf{r})_t)$$
$$\textit{and}$$
$$(-\boldsymbol{\nabla}.\beta(\mathbf{r})_t) \;=\; (\textstyle\sum_m (p_{jm}\,\mathbf{k}_m))_t,\quad m=1,2,..M$$
$$(33')$$

where:

- quite *un*usually indeed, the writing from (33') stresses explicitly that *the equality is asserted only from a* ***numerical*** *point of view*, in this sense that *it does* ***not*** *assert also the semantic feature of an identity of nature* (cf. the further note 146);
- $p_{jm}$ is the projection of $\mathbf{p}_j$ on $\mathbf{k}_m$ at the time $t$ ;

---

[153] We recall that inside *[IQM-QM$_{HD}$-(dBB)]* the guiding relation (33) is consistent with the whole *QM$_{HD}$* formalism, in consequence of the significance identified in (6.II).1 for any eigenstate of an quantum mechanical observable, and of the fact that the *dBB* guiding theorem is asserted for ***any*** sort of microstate, *with or without inner quantum potential.*

- $j=1,2,.....J$ and $J$ is an arbitrarily big *finite* integer (the discrete character assumed for the succession of the values $\boldsymbol{p_j}$ being entailed and determined by, and relative to, the choices of the involved units of measurement (we write in short $\forall j$)).

The posit (33') defines now the vector-value of the ***mechanical*** momentum *associated* by de Broglie to the 'corpuscular-like singularity' from the amplitude of the wave represented by the wave-function.

The number $\boldsymbol{p_j}$ from (33') measures *a feature assigned to a wave*, but ***that can be qualified in mechanical terms*** [154].

And once this specification is achieved, we *now* make use of "both equations" of which we dispose, as required by Descartes in the motto from the beginning of *(7.III).2.3.7*; namely the equations (37) *and* (33'). So accordingly to the twentieth rule quoted in this motto we can *effectuate* the addition from (33') and write a last equation

$$\Big[\boldsymbol{P_{vc\text{-}vd}}\,[a_\Phi.exp\,\big(i(2\pi/h)\,(\textstyle\sum_m p_{jm}\,\boldsymbol{k_m})_t.\boldsymbol{r}\big)] = (\textstyle\sum_m p_{jm}.\boldsymbol{k_m})_t.a_\Phi.exp\Big((i(2\pi/h))\,(\textstyle\sum_m p_{jm}\,\boldsymbol{k_m})_t.\boldsymbol{r}\Big)\Big] =$$

$$= \boldsymbol{p_j(r,t)}.\ a_\Phi.exp\,\Big((i(2\pi/h))(\textstyle\sum_m p_{jm}\,\boldsymbol{k_m})_t.\boldsymbol{r}\Big), \qquad \forall j,\ m=1,2,..M, \qquad (38)$$

for individual eigen-***waves*** $a_\Phi.exp\,\big(i(2\pi/h)\,(\sum_m p_{jm}\,\boldsymbol{k_m})_t.\boldsymbol{r}\big)$ and corresponding eigen-***values*** $\boldsymbol{p_j(r,t)}$ of a generalized concept of a *mechanical* momentum-observable $\boldsymbol{P_{vc\text{-}vd}}$ assigned to the mechanical movement of the singularity from the amplitude of the *wave* from one specimen of the studied microstate:

*The eigenvalue (33')* $\boldsymbol{p_j(r,t)}$ *of the* ***mechanical*** *momentum-observable* $\boldsymbol{P_{vc\text{-}vd}}$ *qualifies* ***numerically*** *in mechanical terms, a very localized* ***corpuscular-LIKE-aspect*** *of a descriptor that represents* ***a phenomenon that is entirely a wave-phenomenon]***; and in (38) this mechanical qualification is *separated* both conceptually and graphically from the qualified wave-phenomenon.

The equation (38) respects now explicitly the semantic specifications that the symbol $\boldsymbol{p_j(r,t)}$ tied with a mechanical numerical qualification, acts upon an object-entity with the nature of a wave-phenomenon represented by a wave-function where the vector-sum $(\sum_m p_{jm}\boldsymbol{k_m(r)})_t.\boldsymbol{r}$ – such as it is written – represents the *genetic structure* of the phase. The *meanings* of the two writings $\boldsymbol{p_j(r,t)}$ and $(p_{jm}\boldsymbol{k_m})_t.\boldsymbol{r}$ are finally clearly distinguished from one another, clearly separated from one another, and the mathematical expressions of these two meanings are both fully achieved, separately.

[154] The necessary and ***sufficient*** condition for applicability of "mechanical" qualifications is just narrow ***space-time localization*** and changes of ***this***.

The representation (38) respects now explicitly the model posited for a 'microstate' and it avoids the confusions entailed throughout $QM_{HD}$ by the heritage of de Broglie's initial notations that had combined his *wave*-model with *mechanical* qualifications introduced inside the wave-descriptor and *without distinguishing the semantic specificities of these two descriptional elements, from their purely numerical relations*; which entailed inappropriate languages (think of the unending debates on a Young-interference: did "the electron" pass through both holes or only through one of them? (i.e. ignorance of the definitions from (2.I).1.2; or "the wave ***and** the **particle***", etc.)[155].

*(7.III).2.3.2.5. The basis of eigenket of $\mathbf{P}_{vc\text{-}vd}$ and how it works*

The equation (38) entails the following ortho-normalized basis of generalized *wave*-eigenket of the momentum-observable $\mathbf{P}_{vc\text{-}vd}$:

$$\Big\{\Big|a.exp\Big((2\pi i/h)\ \big(\textstyle\sum_m p_{jm}\,\mathbf{k}_m(\mathbf{r})\big)_t\,.\mathbf{r}\Big)\Big>\Big\},\ \forall j,\ \ m=1,2,..M$$

(in short $\{\,|\,u(\mathbf{r},\mathbf{p}_j,m(M))>\}_t$ )[156]

(39)

Consider now the way in which the index $m=1,2,...M$ works.

For a microstate $ms(unbound,1)_{G(nc)}$ we have $m=1$. So the composed operation of generation $\mathbf{G}_t(G1,G2,....Gm,.....GM)$ reduces to a simple operation of generation $\mathbf{G}_t$, the 'mechanical' momentum vector-value $\mathbf{p}_j$ is independent on time, and the generalized *[IQM-$QM_{HD}$-dBB)]* momentum-observable $\mathbf{P}_{vc\text{-}vd}$ appears reduced to the $QM_{HD}$ momentum-operator $\mathbf{P}$ that – inside $QM_{HD}$ – is implicitly tied with the coding-postulates (20')≡(20) supposed to be universally valid. In this case the equation (38) reduces to the $QM_{HD}$ equation for eigenfunctions and eigenvalues of the momentum-operator, called there momentum-"observable".

---

[155] The lesson delivered by Descartes over centuries seems to me so surprising and noteworthy that I take the liberty to confess also the subjective path that led me to (38), notwithstanding the strong redundancies with the remarks from the current text of this work: After more than a whole year of incapacity to formulate a convenient generalization of the $QM_{HD}$-equation for eigenvalues and eigenket of the momentum observable, I had an accidental rather inexplicit insight that finally led me to proceed as exposed in the text, but without understanding why. I told of this to a very learned young friend, Carlos Lobo, and he showed me the Descartes rules. These immediately released in my mind a luminous state of obviousness and stable comfort because finally I understood very clearly why I had stagnated such a long time: *Mathematically* there is no imposed distinction between the non-effectuated form of the sum $\sum_m (p_{jm}\mathbf{k}_m)_t.\mathbf{r}$ and its effectuated form $\mathbf{p}_j.\mathbf{r}$. *In consequence of this I had an irrepressible tendency to glide immediately into the effectuated form*: But thereby the expression of a superposition of plane waves – with which I was systematically starting, ***disappeared.*** This initial expression seemed curiously volatile; while ***resisting it seemed arbitrary***. Furthermore *the effectuated form $a.exp[(i(2\pi/h))p_j.r]$ describes **one** plane wave*, whereas according to the model that I was positing, at the moment *t* when the act of measurement-evolution $Mes\mathbf{P}_{vc\text{-}vd}$ from a succession $[\mathbf{G}_t.Mes\mathbf{P}_{vc\text{-}vd}]$ begins and up to the end of this*, the wave-function of a specimen of a microstate $ms(unbound,1)_{cG(q\text{-}f)}$ **must reflect via its phase at any space point r(t), the inner structure of an evolving superposition of M plane waves, not the stable and restrictive form of one plane-wave function**: **THIS, IN THE PRESENT APPROACH, WAS A CONDITION OF INNER LOGICAL COHERENCE***. Indeed ***this** is **why** the **vector***-value $\mathbf{p}_j(t)=\sum_m (p_{jm}k_m)_t$ of the mechanical displacement of the singularity from the amplitude of the wave-function cannot be conserved during a measurement evolution, as it is required for applicability of the coding-postulate (20'); and precisely this impossibility had led me to the necessity of a recourse to "value-destructive" trace-registrations (33) and to the coding-postulate (35). So I kept struggling between the apparent arbitrariness of the interdiction to perform the sum from the phase-function, and a condition of inner consistency. I finally conserved the *non*-effectuated sum $\mathbf{p}_j(t)=\sum_m (p_{jm}k_m)_t$ inside the phase of the wave-function $\boldsymbol{\Phi}_{Gt(G1,G2,...Gm,...GM)}$, while outside it I effectuated the sum. But I did that opportunistically, unwillingly, because *I did not remember the general semantic warning that when one wants to add apples and prunes one must **first** have become able to conceive the meta-category 'fruit' and inside this meta-category one must **permit specificities*** (in our case the two distinct categories are "structure of the phase of a wave" and "mechanical momentum of the singularity in the amplitude of the wave" and the common meta-category is "*numerical value of*"). This illustrates what unpredictable features can arise inside Mathematical Physics, that *must* find location in a representation guided by semantic criteria, because *when only algorithmic or/and numerical criteria work these features entail trouble*. ***Mathematical Physics cannot be reduced to "Mathematics"***. And it is impressing to discover how explicitly Descartes, centuries ago, was aware of criteria of this sort. It is also impressing that *Mathematics itself – in fact – usually somehow "knows" the rules of semantic homogeneity that act inside the conceptualization*; while we, the physicists, do not always know these criteria and apply them. (I still do not quite understand why, *"**after** having obtained the equations......we must achieve the omitted operations, **without ever making use of multiplication when division is necessary**")* .

[156] In de Broglie ([1956], pp. 119-133 – a fascinating science-fiction-like chapter – an interference-microstate obtained by reflection on a mirror of an incident state, there are places where the corpuscular-like singularity – endowed with a "***quantum* mass**" – is at rest; or *in other circumstances it becomes imaginary*; so *one should systematically work with the concept of quantum mass and admit a priori also eigenvalues $p_j(r,t)=0$* (and also 'imaginary' eigenvalues? In some more adequate treatment these might come out to be ***infinite***). This is likely to be connectable with teleportation. But here again de Broglie's concept of quantum-mass might be found inadequate because it involves the velocity of light and insertions from Einstein's theories of Relativity, which seems unnecessary (cf. III.9)

While as soon as ***m>1*** we have a microstate $ms(unbound,1)_{cG(qf)}$ and according to (33') the 'mechanical' momentum vector-value $\boldsymbol{p_j(r,t)}$ of the singularity from each *wave*-function $\Phi_{Gt}(\sigma_{Gt})$ of the specimen $\sigma_{Gt}$ of the studied microstate $ms_{Gt}$ that is involved by one measurement-succession $[\boldsymbol{G_t}.Mes\boldsymbol{P_{vc-vd}}]$ – at the space-point $\boldsymbol{r}$ where this singularity happens to be – ***depends essentially on time*** and requires *exclusively* the coding-measurement-evolution (35) (that appears a posteriori to be applicable also when $m=1$).

So the general equation (38) tied with any possible value of $m$ and the whole basis of eigenket (39), can be split – a posteriori – in the following two *separate* equations:

$$\boldsymbol{P_{vc}}[a_{\Phi}.exp\left(i(2\pi/h)\,(p_j\boldsymbol{k.r})\right)]=p_j\boldsymbol{k}.a_{\Phi}.exp\left(i(2\pi/h)\,(p_j\boldsymbol{k}.\boldsymbol{r}\right)=\boldsymbol{p_j}\,.a_{\Phi}.exp\left((i(2\pi/h))(p_j\boldsymbol{k.r}\right),\ \forall j \tag{38'}$$

with a basis of eigenket (39') where $m=1$, so this parameter is tied exclusively with the coding-measurement evolution (20');

$$\left[\boldsymbol{P_{vd}}\,[a_{\Phi}.exp\left(i(2\pi/h)\,(\textstyle\sum_m p_{jm}\,\boldsymbol{k_m})_t.\boldsymbol{r}\right)]\ =(\textstyle\sum_m p_{jm}.\boldsymbol{k_m})_t.a_{\Phi}.exp\left((i(2\pi/h))\,(\textstyle\sum_m p_{jm}\,\boldsymbol{k_m})_t.\boldsymbol{r}\right)\right]=$$

$$=\ \boldsymbol{p_j(r,t)}.\ a_{\Phi}.exp\left((i(2\pi/h))(\textstyle\sum_m p_{jm}\,\boldsymbol{k_m})_t.\boldsymbol{r}\right),\quad \forall j,\quad m=2,..M, \tag{38''}$$

with a basis of eigenket (39") where $m>1$, so this basis is tied exclusively with the coding-measurement evolution (35).

*(7.III).2.3.2.6. The Hilbert-space representation of the state-ket of a microstate ms(unbound,1)$_{cG(qf)}$*

Now we just posit for the *[IQM-$QM_{HD}$-dBB)]* Hilbert state-ket $|\psi_{Gt}>$ of an interference microstate $ms(unbound,1)_{G(qf)}$ the ***one***-*term* form

$$|\Psi_{\boldsymbol{Gt}(G1,G2,...Gm,...GM)}(\boldsymbol{r},t)>=|a_{\Psi}(\boldsymbol{r},t)exp((2\pi i/h)\varphi(\boldsymbol{r},t)> \tag{40}$$

instead of the inadequate additive $QM_{HD}$-representation of type (15’').

*(7.III).2.3.2.7. The Hilbert-space representation of the statistical predictions on results of momentum-measurements on a microstate ms(unbound,1)$_{cG(qf)}$*

We consider the expansion (22) $|\Psi(\boldsymbol{r},t)_{\boldsymbol{Gt}(G1,G2,...Gm,...GM)}(\boldsymbol{r},t)>/\boldsymbol{P_{vd}}$ of the considered state-ket (40), with respect to, exclusively, the generalized momentum-observable $\boldsymbol{P_{vd}}$ with basis (39"):

$$\left\{|a.exp\left((2\pi i/h)\ (\textstyle\sum_m p_{jm}\,\boldsymbol{k_m(r)})_t\,.\boldsymbol{r}\right)>\right\},\ \forall j,\ m=2,..M \tag{39''}$$

Gleason’s theorem permits to write accordingly to (23)

$$\pi((\boldsymbol{p_j(r,t)})\ =\ |c_j(t)|^2=|Pr._j\,|\Psi_{\boldsymbol{Gt}(G1,G2,...Gm,...GM)}(\boldsymbol{r},t)>|^2$$

where a state-ket $|\Psi(\boldsymbol{r},t)_{\boldsymbol{Gt}(G1,G2,...Gm,...GM)}(\boldsymbol{r},t)>$ (40) *without* mathematical superposition in it is projected upon the axes of a basis of a new sort of eigenket each one of which expresses clearly a ***physical***[157] *superposition* of two or several plane-waves that describes a sample of interference-wave-movement in the neighbourhood of the *unique* corpuscular-like singularity from that interference-microstate $ms(unbound,1)_{G(qf}$. So the guided momentum-value of this unique singularity that is involved is posited to be immersed into a phenomenon of physical interference that admits the value-*destructive* coding-measurement-evolutions (35) (and *only* this sort of coding-measurement-evolution): ***the problem of the definition of an adequate coding-measurement-evolution is solved***. Correlatively, the modified form (40) of the interference-state-ket suppresses directly all the critiques formulated in *(7.III).2.3.1.1*, of "inadequacy with respect to *IQM*" and "inadequacy with respect to $QM_{HD}$" (so, in particular, we get rid of the non-verifiable character of the $QM_{HD}$-predictions (32) entailed by the representation (15")).

However via the coding-measurement-postulate (35) we clearly step ***out*** of the domain of validity of the formalism of $QM_{HD}$. We traverse into – exclusively – the more profound and unlimited domain of the physical substance that has been represented by the *dBB* approach.

But we operate this passage on the basis of the assumption that *EXP* has insured the possibility of coding-measurement-evolutions (35). And thereby the interference phenomena are now for the *first* time effectively brought in reach of a coherently founded theoretical representation that insures consensual verifiability of the asserted predictions. Indeed in consequence of the representations (38), (39), (40) the formal-factual representation (31) that emerges factually true by construction, becomes transposable to also the microstates $ms(unbound,1)_{G(qf)}$, as follows.

*(7.III).2.3.2.8. Extension of Ass.2 to the interference microstates* $ms(unbound,1)_{G(qf)}$

Consider a microstate $ms(unbound,1)_{cG(qf)}$.

- The eigenket of the corresponding observable $\boldsymbol{P_{vd}}$ are defined in (39").
- Each measurement succession of type (35) $[G_{t1}(G_1,G_2,...G_m,...G_M).Mes(\boldsymbol{r},\boldsymbol{p})_t]$ yields as result a pair of values $(\boldsymbol{r_k},\boldsymbol{p_j})_{t1}$, $k=1,2,..K, j=1,2,..J$, that is recorded.
- These records permit already to write for the unknown state-ket (40) $|\Psi_{Gt1(G1,G2,...Gm,...GM)}(\boldsymbol{r},t_1)>=|a_\Psi(\boldsymbol{r},t_1)exp((2\pi i/h)\varphi(\boldsymbol{r},t_1)>$, the *factually* constructed expansions with respect to $\boldsymbol{P_{vd}}$ and $\boldsymbol{R}$

$$|\Psi_{Gt(G1,G2,...Gm,...GM)}(\boldsymbol{r},t_1)>_{fact(DM)}/\boldsymbol{P_{vd}} \quad and \quad |\Psi_{Gt(G1,G2,...Gm,...GM)}(\boldsymbol{r},t_1)>_{fact(DM)}/\boldsymbol{R}$$

by acting exactly as in the proof of the assertion *Ass.2* for the case of a microstate $ms(unbound,1)_{G(nc)}$.

- Any mechanical quantity *A* that is different from the position and the momentum is defined *classically* as a function $A(\boldsymbol{r},\boldsymbol{p})$ that inside $QM_{HD}$ is reproduced as the same – symmetrized – function $\boldsymbol{A}(\boldsymbol{R},\boldsymbol{P})$ formed with the observables $\boldsymbol{A}$, $\boldsymbol{R}$, $\boldsymbol{P}$. It seems then natural to associate to each record via (35) of a *given* pair $(\boldsymbol{r_k},\boldsymbol{p_j})_{t1}$, a corresponding *numerical* "eigen-value" $a_q$ of the quantity $A(\boldsymbol{r},\boldsymbol{p})$ *calculated in the classical way, as* $a_q(\boldsymbol{r_k},\boldsymbol{p_j})_{t1}$. In order to entirely unify the language and the notations, to the numerical value $a_q(\boldsymbol{r_k},\boldsymbol{p_j})_{t1}$ calculated in this classical way *we associate an* "eigenket-*like* label" denoted $\lambda$-$|a_q(\boldsymbol{r_k},\boldsymbol{p_j})>_{t1}$ that is inhabited by the classical function $a_q(\boldsymbol{r_k},\boldsymbol{p_j})_{t1}$ instead of an eigen-function in the sense of $QM_{HD}$. So, for the unknown state-ket (40), one can construct "formal-factual" expansions *in the sense of (31)* denoted

$$|\Psi_{Gt(G1,G2,...Gm,...GM)}(\boldsymbol{r},t_1)>/A(\boldsymbol{r},\boldsymbol{p})$$

---

[157] This choice of words is intended to stress that we would ***not*** be in the case from (15") – inadequate with respect to *IQM* – of a superposition of two or more ***full*** realizations of distinct *micro-**states***; that we are in presence of a state generated by only one fully realized operation of generation of one interference-micro-state where other operations of generation are just involved but are not fully realized separately.

*for also any other qualifying quantity $A(\mathbf{r},\mathbf{p})$: nothing interdicts this, neither Gleason's theorem nor Dirac's Hilbert-space-bra-ket calculus*. The hosting mathematical structure is a Hilbert-space, and each representational element is introduced in this structure at the place and in the form commanded by this hosting mathematical structure, exactly as when use is made by an equation.

> Thereby is finally defined a full extension to also the microstates $ms(unbound,1)_{cG(qf)}$, of the bottom-up and factually constructed representation (31) achieved before for only the microstates $ms(unbound,1)_{G(nc)}$.

Of course many points remain to be worked out (for instance the orthogonalization of the bases (39''), the Dirac-transformations, etc.). But the essence of each specific problem that has been been encountered as well as the corresponding solution, has been clarified.

### *(7.III).2.3.4. Considerations on dBB* *versus* *the representation achieved in* (7.III).2.3 *for the interference-microstates* $ms(unbound,1)_{cG(qf)}$

The assumption (33') *completes* the modelling postulate $MP(\{\sigma(ms_{G,cw})\})$ from (1') in (6.II).2 and so also the channel (36) of adduction into consensual predictive-verifiable "scientificity", of descriptional prime-matter from the purely interpretative *dBB* approach. Indeed the channel (36) becomes now

$$\Big[MP(ms_{G,cw}),\ (1')G_{cw}\leftrightarrow ms_{G,cw},\ (14)ms_{G,cw}\equiv\{\sigma(ms_{G,cw})\},\ (35)\boldsymbol{P(cod.p_j)}_{unb.ms\bullet},$$
$$(33')\boldsymbol{p_j}(\mathbf{r},t)=\boldsymbol{\nabla}.\beta(\mathbf{r})_t=\sum_m(p_{jm}\mathbf{k_m}(\mathbf{r})_t\Big]$$
(36')

Via (35) and (33') we have rooted the present conceptual-*factual*, operational bottom-up approach, deep into the the *dBB* representation. And thereby, implicitly, we have *separated* the formalism that we are building, in two distinct parts, one of which applies exclusively to the unbound microstates while the other one concerns any sort of unbound microstate.

And let us notice a very important point: it would be ***pointless*** to struggle to realize inside $[IQM\text{-}QM_{HD}\text{-}dBB)]$, for any $\boldsymbol{QM_{HD}}$-operator $\boldsymbol{A}$ different from $\boldsymbol{R}$ and from $\boldsymbol{P}$, a generalization consistent with (38) of its $QM_{HD}$-equation for eigenket and eigenvalues *such* that the eigenvalue $a_q(\mathbf{r_k},\mathbf{p_j})$ of $\boldsymbol{A}$ that involves an eigenvalue $\boldsymbol{p_j}$ of the momentum-ovservable $\boldsymbol{P_{vd}}$ tied with a parameter $m>1$, be represented by an eigenket inhabited by a *non-classical **functional*** representation, as it is inside $QM_{HD}$ for an eigenvalue of the momentum-operator $\boldsymbol{P}$ tied with a parameter $m=1$.

> This aim would simply be an impossible aim, because *it is inappropriate with respect to the considered level of conceptualization and the corresponding cognitive situation involved by the $[IQM\text{-}QM_{HD}\text{-}dBB)]$ momentum-observable* $\boldsymbol{P_{vd}}$.

Indeed $\boldsymbol{P_{vd}}$ is tied with – *specifically* – the guiding law (33) involved by the coding-measurement-evolution (35). Thereby it concerns a level of conceptualization that, inside the physical factuality and with respect to the classical level of human conceptualization, is set ***deeper*** than the level where acts the quasi-classical coding-measurement-postulate (20') presupposed fallaciously for *any* $QM_{HD}$ operator $\boldsymbol{A}$. The $[IQM\text{-}QM_{HD}\text{-}dBB)]$ momentum-observable $\boldsymbol{P_{vc}}$ from $\boldsymbol{P_{vc\text{-}vd}}$ that in (38) is tied with a parameter $m=1$, (20') identifies with the ***classical*** coding-evolution (20) called "time of flight". The value-*conserving* coding-measurement evolutions (20) and (20') are compatible with the quasi-classical, degenerate $QM_{HD}$ concept of momentum-"observable" reduced to the momentum-operator $\boldsymbol{P}$ alone, that obeys the non-commuting algebra of all the other operational $QM_{HD}$ representations $\boldsymbol{A}$ of mechanical quantities; whereas the value-*destructive* observable $\boldsymbol{P_{vd}}$ tied with a parameter $m>1$ and with (33) and (35) bring us on the level of the *dBB* approach where the classical Newtonian concepts of "mass", "acceleration", and "field" are not yet required. *The very possibility of location of these concepts is not yet formed.* On this originating level the gradient of the phase $\beta(\mathbf{r},t)$ of the wave of a specimen $\sigma_{Gt(G1,G2,...Gm,...GM)}$ of the studied microstate defines *directly* the momentum of the singularity from the amplitude of this wave, by the guiding-law (33). But in (33) the semantic contents of the word "momentum" and of its denotation '$\boldsymbol{p_j}(r,t)$' have to be conceived as just *a time-dependent velocity* $\mathbf{v}_j(\mathbf{r},t)$ *connected numerically with the gradient of* $\beta(\mathbf{r},t)$ *via an 'energetic' coefficient that – rather arbitrarily – is named "mass".* In relation with this sort of "momentum" it is neither necessary nor even possible as yet to speak of "the mass" in the strict sense of this word. Inside a bottom-up development on the vertical of conceptualization, the necessity of the concepts of "mass", "field" and "acceleration" steps in when *lumps* of increasingly numerous *bound* states are considered, "atoms", molecules", solid or liquid lumps of "matter".

At the present time Louis de Broglie's final approach ([1956]) and David Bohm's approach are still quasi unanimously considered to offer only an interpretation of $QM_{HD}$. But inside *[IQM-*$QM_{HD}$*-dBB)]*, via the channel (36') of adduction into an operational and consensual, scientific representation, this view suffers a genuine inversion. Indeed *via (36') the preceding remarks entail noteworthy predictive consequences that can be expressed in terms of probability trees*.

A coding-measurement-evolution (35) for the *[IQM-*$QM_{HD}$*-dBB)]* momentum-observable $\boldsymbol{P_{vc\text{-}vd}}$ is posited to be valid for *any* unbound microstate of one microsystem, whether a $ms(unbound,1)_{G(nc)}$ microstate or a $ms(unbound,1)_{cG(qf)}$ microstate; and it generates simultaneously, on one same specimen of the studied microstate, *both* elements from only one pair $(\boldsymbol{r_k},\boldsymbol{p_j})_{t1}$. With respect to this feature, the $QM_{HD}$ coding-measurement-evolutions (20') and the corresponding probability-trees with mutually incompatible branches float on the *surface* of the representation of the microstates where the behaviour of the corpuscular-like de Broglie singularities is still marked by classical macroscopic features, in consequence of the absence of quantum potentials (cf. (6.II).4.3).

By the mere subsistence of their logical possibility inside the framework *[IQM-*$QM_{HD}$*-dBB)]*, the definitions posited here reveal strikingly that in $QM_{HD}$ *the concepts of mutual compatibility or incompatibility of dynamical quantities are uncritically assigned* ***an ABSOLUTE, intrinsic nature embodied in an algebra of operators that is just posited ab initio*** (cf. the note from (2.I).2.1.3). The correlative 'principle of complementarity' has instilled many considerations devoid of any clear and intelligible feature of factual or logical necessity.

But the probability-trees with mutually incompatible branches can be *all* replaced by *one*-branch probability-trees tied with a coding-measurement-evolution (35) and where, for *any* classical dynamical quantity $A(\boldsymbol{r_k},\boldsymbol{p_j})$, the value $a_q(\boldsymbol{r_k},\boldsymbol{p_j})$ that corresponds to an obtained pair $(\boldsymbol{r_k},\boldsymbol{p_j})$ can be specified by mere numerical calculus, as assumed in the *step 5* from *(7.III).2.5.2*. The figure 8 represents such a probability-tree.

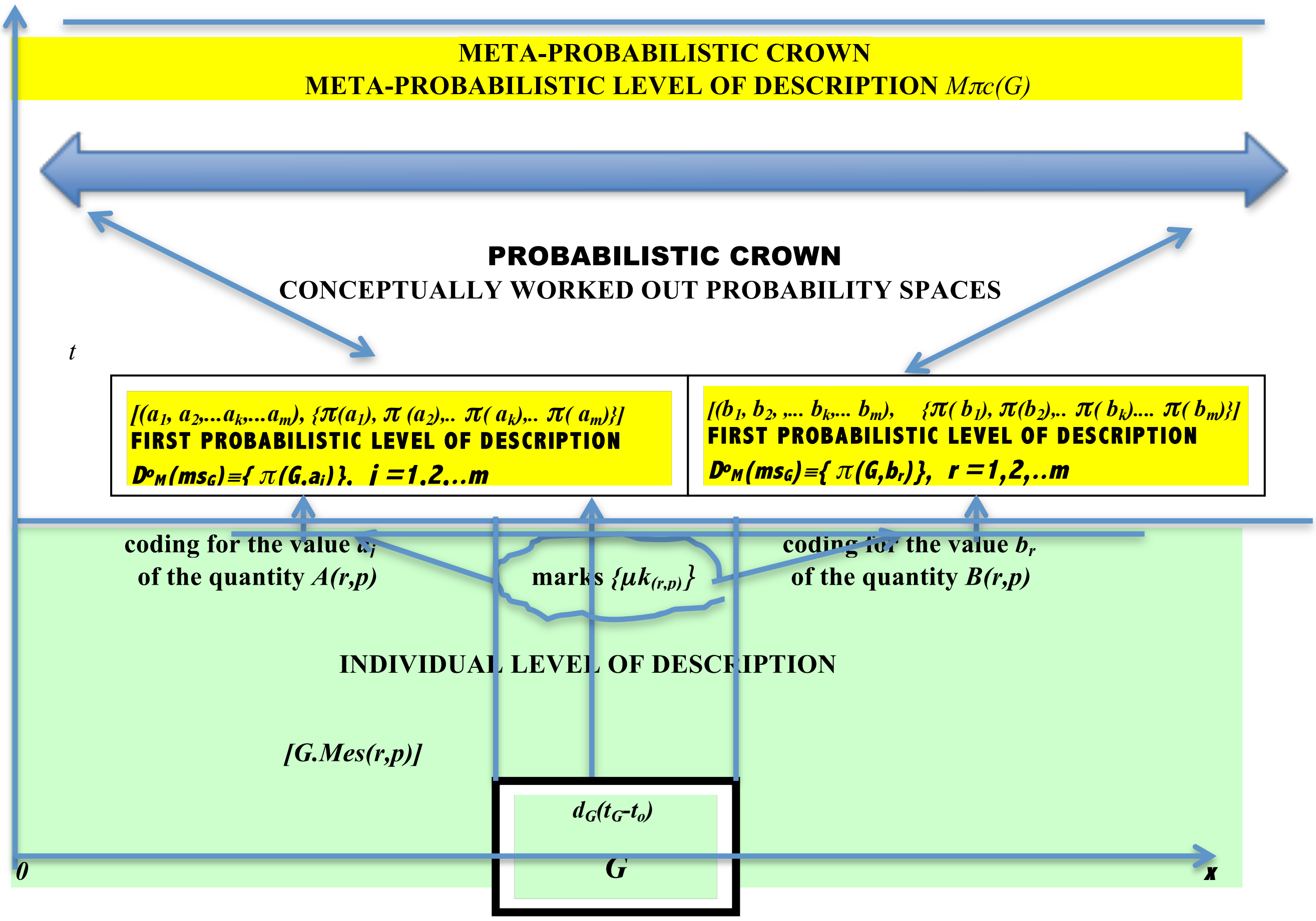

**Fig. 8.** The de Broglie-Bohm one-branch probability-tree $T_{dBB}$ [158].

We call this "a de Broglie-Bohm one-branch probability-tree" and denote it by $T_{dBB}$.

The de Broglie-Bohm one-branch trees $T_{dB}$ introduce operations of generation of a quite general, non-restricted nature and corresponding representations that can penetrate *as deep as desired* into the a-conceptual physical reality. These trees *include* the radically non-classical region of the physical factuality wherefrom are 'extracted' – in the sense of (1),(1') – the microstates $ms(unbound,1)_{cG(qf)}$ of which the corpuscular-like singularities are immersed in non-null quantum potentials that are "interior" to such microstates and so they cannot be manipulated from their respective outsides[159].

As for the coding-postulate (20') implicitly assumed in $QM_{HD}$, inside *[IQM-QM*$_{HD}$*-(dBB)]* it ***confines*** just under the frontier between classical mechanics and the mechanic of microstates. With respect to the generally possible representation from the Fig.8, the probability trees with mutually incompatible branches of the type represented in the Fig.2 appear now as *a particular case characteristic of the microstates* $ms(unbound,1)_{G(nc)}$ *that are subjected to Heisenberg's principle of uncertainty* ***iff*** *it is* ***chosen*** *to employ for them the coding-postulate (20')*. In this sense the falsely absolute $QM_{HD}$ concept of "observable" tied with a non-commuting algebra appears now as ***a huge arbitrary restriction***. In particular it appears explicitly that:

> *The non-boolean Birkhoff-von Neumann "logical" structure in fact is tied with the falsely absolute definition of mutual compatibility that characterizes just a way in which it is possible to* ***code the results of measurements of only the PARTICULAR quasi-classical category of the microstates ms(unbound,1)$_{G(nc)}$.***

[158] The tree from the Fig.8 belongs to the general category represented in the Fig.3: it is the particular case of "compatibility" of ***all*** the qualifying mechanical quantities defined in $QM_{HD}$.

[159] And this *most* non-classical region of the micro-reality can be included formally in *[IQM-*$QM_{HD}$*-(dBB)]* without any recourse to the concepts of eigenstates and eigenvalues, via a classical calculus of the values $a_q$ of $A$, when $A$ is different from ***R*** and from ***P***. It might seem paradoxical at a first sight that a case that is more complex can be represented in a less "evolved" way; but we have already noted that this is entailed by precisely the fact that this more profound region of the microphysical conceptualization accepts formalizations with a more *incipient*, a *less* specified character.

Which changes the whole conceptual status of this formal structure[160].

However here – precisely at this place in the present context – there emerges also a very fundamental question that has to be explicitly formulated and examined:

> ***Why*** should the value $a_q(\mathbf{r_k,p_j})$ to be assigned to a qualifying mechanical quantity $A$ that is posited to be determined by a pair $(\mathbf{r_k,p_j})$ obtained experimentally via a coding-measurement-evolution (35) ***of a de Broglie-singularity***, be calculated accordingly to the ***CLASSICAL*** definition $A(\mathbf{r_k,p_j})$? What insures that the classical definition can be assigned any sort of "pertinence" and methodological utility for a de Broglie-singularity? What does "pertinence" or "utility" *mean* in this context? How is this tied with the features of the model that our mind posits at the presupposed space-time dimensions?

When one re-conceptualizes bottom-up, not *all* the concepts constructed in the long historical course of a previous conceptualization, in relation with various orders of space-time dimensions, do uniformly and automatically conserve their relevance at any order of space-time dimensions. The process of construction is quintessentially creative and it involves a selective order of possibility.

This question certainly is intimately tied with the problem of unification of Physics: the epistemological considerations cannot be banished from this problem.

We now conclude as follows.

***Inside [IQM-QM$_{HD}$-(dBB)] the notion of mutually incompatible qualifying mechanical quantities could be entirely circumvented***, together with the whole $QM_{HD}$ non-commutative algebra of the observables, without thereby loosing predictive capacity on the results of quantum measurements on microstates *ms(unbound,1)$_{cG(qf)}$*.

The $QM_{HD}$-concept of "observable" could be simply disposed of.

Via [[(33), (35), (38), (39), (40), (31)] there emerges inside *[IQM-QM$_{HD}$-(dBB)]* a one-branch sort of probability tree that is topped by the two experimentally constructed statistical-probabilistic crowns for ***r*** and ***p*** that, once identified, are predictive, like in *IQM*.

Nevertheless we choose to *conserve* the branch-probability-trees inside *[IQM-QM$_{HD}$-(dBB)]*, namely for the very particular class of microstates *ms(unbound,1)$_{G(nc)}$* of the type represented in the Fig.2, that in *this* approach absorbs in it alone a whole $QM_{HD}$-like concept of an algebra of non-commuting "beable-observables with value-conserving coding-measurement-evolutions" (in the sense specified in relation with the generalized equation (38)). We make this choice in consequence of the following reasons.

- Quite generally, when in a representation the different levels of conceptualization are mixed inside some generalization, the semantic in-homogeneities that are thereby included tend to finally smash the representation. So any distinction that semantically and physically ***acts***, has to be explicitly expressed *as a memento*. Specifically now, inside *[IQM-QM$_{HD}$-(dBB)]* the quasi-classical level of conceptualization on which are defined the microstates *ms(unbound,1)$_{G(nc)}$* tied with two-or-several-branch probability-trees, has to be distinguished as clearly as possible from the deeper *dBB*-level of conceptualization on which are defined the interference microstates *ms(unbound,1)$_{cG(qf)}$* tied with one-branch-probability-trees as in the Fig.8.

[160] The whole "logic of quantum mechanics" remains to be entirely reconstructed. A ***physical*** theory cannot be reduced to the mathematical and logical aspects of the formalism that is made use of for representing the involved physical entities and facts. The semantic contents cannot be neglected, nor can they be rejected on a second plan. They are that what determine the formal structures to be chosen in order to express **them** in an adequate and most general way. ***The semantic contents are primary with respect to the formal structure***.

- Inside the nuanced, secured representation with two sorts of coding-postulates, the several-branch probability-trees tied with the microstates *ms(unbound,1)$_{G(nc)}$* introduce also *a useful element of comparison*. Furthermore they nourish a historical continuity in thought and language.

But the most basic and synthetic memento is that from now on we shall systematically distinguish between *"beable observables admitting a value-conserving or a value-destructive sort of coding-measurement-evolution"*, according to the level of microphysical conceptualization that is considered; the term of "observable" meaning exclusively what this word designates, not some algebraic properties inside a group of concepts.

Here reaches its end our search of a satisfying *[IQM-QM$_{HD}$-(dBB)]* Hilbert-space representation of the probabilistic predictions of the results of measurements on unbound microstates of only *one* micro-system. Of course many points remain to be worked out (for instance the orthogonalization of (40), the Dirac-transformations, etc.). But the specific problems that we been able to detect as well as the corresponding solutions are all clarified in their essence. So we shall now proceed to close our examination of the unbound microstates by a succinct examination of also the last case defined in (2.I).1, of one unbound microstate of two or more micro-systems.

#### *(7.III).2.3.5. Remark on spin-measurements on any sort of unbound microstate*

To close the point (7.III).2.3 we just append the following remark on spin-measurements.

For such measurements the coding postulate for both microstates *ms(unbound,1)$_{G(nc)}$* and *ms(unbound,1)$_{cG(qf)}$* reproduces in essence the Stern-Gerlach procedure that introduces a value-conserving coding-measurement-evolution of the general type (20').

## (7.III).3. THE CASE OF *ONE* UNBOUND MICRO-*STATE* OF *TWO* OR MORE MICRO-*SYSTEMS*

Consider now the case of *one* micro-state of *n* micro-systems, which we denote by $ms(unbound,n)_{cG(qf)}$. In what follows we restrict ourselves to the case $ms(unbound,2)_{cG(qf)}$ with only *two* micro-systems that is involved in Bell's theorem on non-locality (cf. (3.I).1) (the generalization of the considerations to cases $n \geq 2$ is obvious).

***On Bell's "theorem" versus relevance of relativistic invariance***. I have already examined Bell's theorem in other works, variously and thoroughly [161]**.** So here I shall add only the following very rapid considerations specifically connected with the context from this Part II of the present work.

Consider the probability tree from the part I of this work that is tied with a Bell-experiment (Fig.4, (3.I).2). Since the two involved micro-systems *S1* and *S2* belong by definition to one *common* micro-state $ms_{G(2S)}$ (in the sense from (2.I).1) there is no reason whatever to posit a priori that the spin-values registered for *S1* and *S2* by one measurement-succession $[G(2S).Mes_{12}(spin1.spin2)]$ should come out to be *non*-correlated (the sign *Mes(spin1.spin2)* just applies the general notation $Mes_{12}(A1,B2)$ from (3.I).2); quite on the contrary. And all the more so as in this case the considered sort of correlation is to be awaited to be *stronger* than the *generally* present 'meta'-correlations (11), (11') from the probability-tree of *any* sort of unbound microstate, because by definition *the correlation stems from the* ***interior*** *of each one* ***individual*** *event brought forth by each one complete measurement-succession from one branch of the corresponding probability-tree*, whereas the correlations (11), (11') refer to globally considered whole probability laws. This is a *very* unusual conceptual situation, as much with respect to the classical mechanical characterizations of 'mobiles', as with respect to the classical calculus of probabilities [162]. The *mechanics* of microstates brings in – *both* – the individual specimens of the studied microstates considered globally ***and*** *their* ***inner*** *structure*, because that what admits mechanical qualifications is included inside the wave-like 'whole' called 'one specimen of a microstate' (in the sense of (1), (1') and the definitions from (2.I).1) and is represented by *individual* physical wave-functions $\boldsymbol{\Phi_{Gt}(r,t)}$[163].

In *such* conditions, *what is the point* in hasting for changing the orientations of the spin-measurement devices just at the last moment before the registration? A predictive calculus can determine quite calmly all the possible observable spin-correlations generated by the various orientations of the registering devices. These potentialities do not depend on time. A high-pressure last-moment choice of the orientations of the devices that register spin-values is expected to do, *what*? To trick a mathematical implication of the formalism of $QM_{HD}$ that violates Einstein's principle of macroscopic locality by forcing the observational effects of micro-phenomena to dominate this mathematical violation? Such a procedure possesses meaning only with respect to:

- *An **a priori** refusal* of the possibility of existence of any correlation, founded on the presupposition that the considered events are produced by two delimited and distinct "objects" mutually separated by "void", like in the case of "mobiles" in the sense of Einstein.
- The research of a factual estimation of the value of the velocity of a 'transmission of influences' between the two involved micro-systems, that *according to the formulation of Bell's conclusion would necessarily exist*.

---

161 This "theorem" has the merit to have released a revolution inside the Bohr-orthodoxy. It has acted as a very active ferment in the scientific conceptualization of the microphysical reality. In order to fix on the time-dimension some psycho-social reference-elements, I have reproduced in the Prologue to the first part of this work an 1979 lecture on Bell's theorem; and in MMS [2013B] **v.3** and [2017] (French texts) can be found what I call a 'conceptual' invalidation of Bell's proof in which I have shown that *the conclusion – such as it is expressed verbally by Bell –* ***does not follow*** *from the mathematical proof*, though the mathematical content of this proof is *correct* ***in itself***. I have also succeeded to construct a counter-example to Bell's formulation of his only asserted conclusion (MMS [1987]) and this counter-example has been confirmed by Bordley [1989]) as "factually possible". But the most relevant *new* data concerning the problems raised by Bell consist of the content from the sub-section (3.I).2 from this work (this content is completely established only since 2012).

162 In order to bring into explicit evidence the general peculiarities of such a situation, a systematic preorganization of the involved concepts and language – like that from *IQM* – is *a sine qua non pre-condition*. In this case one gets lost for speaking and reasoning in a precise way if only concepts and words from the current languages are made use of.

163 This individual physical wave-function $\boldsymbol{\Phi_{Gt}(r,t)}$ *corresponds to* but is *distinct from* the involved state-ket $|\Psi_{Gt}>$ that – via its expansions (22) $|\Psi_{Gt}>/A$ – describes only abstract statistics of results of the individual coding-measurement-successions *[G.MesA]* where are involved the individual wave-movements described by $\boldsymbol{\Phi_{Gt}(r,t)}$. In order to genuinely understand the physical and the conceptual situation one should keep active in his mind a clear view on all this.

But *the view involved by Bell's theorem is **false***: Bell's mathematical calculus and his logical reasoning prove *exclusively* the existence of a correlation *accordingly to the $QM_{HD}$-formalism*; ***which was known***. They do *not* prove also the necessity of *'influences'* that entail these correlations (MMS [2013] [164]). The notions of 'influence' and so also of a velocity of transmission of it, *are exterior to Bell's logical-mathematical pro*of. They are just asserted – *juxtaposed directly and exclusively* – in Bell's verbal formulation of his conclusion that expresses only his own psychological beliefs. And these are founded upon a furtive supplementary and *independent* assumption, namely that: The "two systems" *must* have "separated" when they have come 'far enough' from one another [165]**.** Which amounts to wipe away the inside-outside specificities of a mechanics of a specimen of a microstate, i.e. a de Broglie ***wave-like movement*** with a singularity in its amplitude, a *wave* that – itself – cannot be delimited in space by a definite enveloping surface. A macroscopic inside-outside delimitation of the solid "objects" seems possible only because our human sensorial apparatuses do not *perceiv*e the uninterrupted intimacy of *any* material "object", with its surroundings. But does this suffice for requiring in Science to conceptualize an entirely non-perceivable microstate in a way that agrees with all the implications of our current notion of a "mobile"? [166]

The preceding remarks bring us face-to-face with the following much more crucial question:

What is the ***ground*** for imposing so dramatically a *universal* Einstein condition of locality, when this condition has been endowed with a well-defined significance and role for – ***specifically*** – the mechanics of macroscopic "mobiles" that are directly perceived with definite global volumes that individualize them mutually for [*human observers* imprisoned in differently moving inertial frames of reference wherefrom *they communicate **via light signals***]? When two or more ***such*** human observers do *all* survey ***such*** a sort of "mobile", and in ***such*** conditions, a scientific representation does indeed require some rules for insuring consensus (even concerning the mere identification of the mobile and of its dynamic); it is then necessary indeed, for instance, to insure an *instantaneous* consensual significance for even the basic assertion that the observers of this sort, in the mentioned conditions, are all perceiving the ***"same"*** mobile)[167].

But for microstates, *each human observer gathers quietly for himself, a structure of knowledge of the **kind** defined in IQM*. This is a knowledge constructed very indirectly and *in an entirely **solitary** way*; each researcher works alone inside his own Laboratory without perceiving anything else than cryptic marks on devices out of which he draws some significance ***when he wants*** [168], exclusively by use of *previously* applied rules and treatments and entirely revolved operational actions – physical or abstract – that have been consensually decided beforehand and then have been applied consensually at different, un-ruled places and times, *on the basis of a consensually assumed* ***model****. In such circumstances* ***the obtained knowledge emerges from its multiple but solitary geneses, invariant from one observer to another one, independent of time, simultaneity, etc.****;* this knowledge emerges consensual in consequence of, alone, the identical and independent of time ***ways*** in which it has been constructed by each solitary observer. So the a priori importation into fundamental *micro*-physics, of all the requirements of the *macro*-scopic and *cosmic* mechanics, is very far from being an obvious necessity. It even is a highly *arbitrary* aim that manifests the impressive blinding force of the reflex and inertial urge to assert the absolute and general validity of anything that has been strikingly efficient inside some particular context.

---

[164] The proof of this last assertion is what I have called *'conceptual'* invalidation of Bell's theorem (MMS [2013B]).

[165] In the text-books one finds systematically the assertion that "when they are still close to one another the 'systems' can interact but when they get sufficiently far from one another the 'systems' *must* have separated".

[166] This illustrates ***the major role of the model*** in a consensual and coherent predictive and verifiable description of the microstates that Bohr wanted to "free" from any model, via a philosophical diktat; when without a model one cannot define measurement-operations, nor, a fortiori, organize them coherently. In this context cf. the very well-constructed counter-example by E. Muchowski [2018], Researchgate.

[167] I remember a meeting in Geneva on non-locality where Aharonov said: "Quantum Mechanics is non-local. So what?"

[168] ***Think of "Wigner's Friend"*** who can wait as long as he wants before interpreting the marks registered on his registering devices.

Einstein's requirement of locality has a quintessentially *methodological* nature, like that of the methodological decision (MD, (1)) from *IQM*; it is ***relative*** *to the human cognitive situation that is involved in the construction of the knowledge that was researched* ***by him****.* There is no reason whatever for treating this particular methodological requirement as an absolute factual "truth". This would be just a huge confusion. ***Not any group of invariance has to be uniformly asserted in any cognitive situation****.*

> The ***cognitive situation*** of the human observers with respect to this or that domain of physical entities is what decides the relevant methodological conditions to be required for consensus on scientific knowledge concerning a particular domain of facts.

A unification of Physics cannot be realized directly by mathematical representations of assertions regarded as absolute factual truths. It can be realized only indirectly, *methodologically*, by a *general and basically relativized* ***common methodological framework****.* There – by a precise specification of each particular cognitive situation tied with a *given* domain of facts and of the corresponding consequences on the obtained relativized descriptions – it becomes possible to trace a path that leads from one domain of facts to another one tied with another cognitive situation, and thereby to define a relation between the compared representations, so *a unification* ***in the sense of a unified intelligibility****.*

As for Bell's "theorem", it seems to have been entailed by a hidden but consensual model of one microstate of two micro-"systems" where these "systems" are conceived as two solid balls that are receding from one another, while the *model* itself is regarded as an absolute ***FACT***[169].

***On the pertinence of (31) for unbound microstates of several micro-systems.*** We now quit the digression on problem of non-locality and, from the specific and humble viewpoint of the purpose of this chapter 7.III, we come back upon the case of one unbound microstate of two or more microsystems. From this point of view the significant conclusion is the following one.

Nothing withstands the extension of the assertion *Ass.2* and of the equivalence (31),(31'), to the case of one unbound micro-state of two or more micro-systems. The proofs worked out in (7.III).2.3 for microstates $ms(unbound,1)_{cG(qf)}$ can be transposed to also this case, point by point. Louis de Broglie's model of a specimen of the considered microstate suggests that inside any microstate of two or more micro-systems act *non*-null quantum potentials tied with the distinct dynamics of the two or more involved corpuscular-like de Broglie singularities. The global operation of generation *G* that has to be realized for generating the micro-state seems to involve the *necessity* to somehow mutually distinguish two or more composing operations of generation, and in this sense it is of the form $\boldsymbol{G}(G_1,G_2,...G_m,...G_M)$; it seems to entail interactions between several mutually distinct wave-movements, so quantum potentials and possible quantum fields. All this is a matter of modeling, it is not a matter of facts. The facts are the statistical results, and these show statistical correlations that have to be represented, which is not possible without modelling.

---

[169] Of course all the considerations from this work also are founded upon models. But by now, I think, it has become clear that without any model one cannot try to construct a coherent and intelligible theory of microstates; one cannot reason, prove, *conceive*; one cannot even speak and write. But the model of two balls is just not adequate with respect to the observable facts.

***From the point of view of [IQM-QM$_{HD}$-(dBB)] the unbound microstates with two or more micro-systems are as yet insufficiently modelled interference-microstates ms(unbound,1)$_{cG(qf)}$.***

## (7.III).4. THE RELATION (31) FOR UNBOUND MICROSTATES *VERSUS* THE SCHRÖDINGER EQUATION OF EVOLUTION

When $QM_{HD}$ is taught "the Schrödinger equation of the problem" holds a quasi-magic role: you write it down, you solve it, and therefrom the answer to nearly every other question follows. But when one examines the situation more closely many restrictions appear to be involved.

Consider the Schrödinger equation

$$(i2\pi/h)d/dt\,|\psi_{\mathbf{Gt,H}}(\mathbf{r},t)> = \mathbf{H}|\psi_{G,\mathbf{H}}(\mathbf{r},t)>$$ [170]

Now, the Hamiltonian from the equation contains only classical macroscopic fields. But the "external situation" can involve also "obstacles". The effects of these however *are only implicitly taken into account, via limiting conditions* that parcel the spatial domain covered by the studied microstate. In the terms of *[IQM-*$QM_{HD}$*-(dBB)]* this means that the equation simply does not involve a definite distinction between microstates $ms(unbound,1)_{G(nc)}$ and microstates $ms(unbound,1)_{cG(qf)}$: this is quite consistent with the $QM_{HD}$ in-distinction of these two categories of microstates with respect to the way of generating them and with respect to the possible coding-measurement-evolutions (only the Hilbert-space representation of the state-ket does distinguish them, but inadequately, as it has been shown in the critiques of the writing (15")).

This gaping lacuna is compensated by precisely the above-mentioned procedure of fragmenting the spatial domain in *separate zones with mutually different limiting-conditions* and by *generating a formal expression of interference effects inside a global solution, via writings of* ***mathematical*** *superpositions of the general type (15") criticized in (7.III)2.4 that – conceptually – do* ***not*** *distinguish explicitly and radically between* ***physical*** *superpositions and mathematical superpositions*. So inside $QM_{HD}$ the quasi-ubiquitous interference effects that characterize the most specifically "microscopic" behaviours of the physical matter, are introduced in the solution of the equation of evolution only a posteriori in a certain sense, via interior variations of the limiting conditions that entail "reflections" ("quantum-tunnel-effects", etc.).

> Again ***individual physical*** *concepts and suppositions are mixed inside one same writing with* ***the representation of abstract statistical features***, just like in the case of the fallacious representation of the measurements.

This, inside $QM_{HD}$, is a ubiquitous consequence of the top-down progression that has generated $QM_{HD}$. This progression has encountered *first* only the statistical features of numerical results of measurements on microstates, and – more or less – has *stopped* on the level of these features. So the basic concept of operation of generation $G$ of individual specimens of a studied microstate has not been formed, nor, a fortiori, the concept of composed operation of generation $\mathbf{G_t}(G_1,G_2,...G_m,...G_M)$. It seems very likely that a lacuna of such importance must entail upon the calculated solution some deleterious consequences that cannot be avoided by idealizations and approximations[171]. *They cannot even be **conceived**.*

---

[170] Considered from a most general point of view, this equation says: "The change in time d/dt of *that* what I study (namely $|\psi_{Gt,\mathbf{H}}(\mathbf{r},t)>$) is posited to possess the same numerical value as the effect upon that what I study (so $|\psi_{Gt,\mathbf{H}}(\mathbf{r},t)>$) of *that* what contains all the causes of the change (namely $\mathbf{H}$)". So we are in presence of a mathematically expressed application to the case of the set of the mechanical changes of [all the statistics of results of quantum measurements on a given "microstate"] ($|\psi_{Gt,\mathbf{H}}(\mathbf{r},t)>$) by all the causes of change ***of this***, such as these changes are expressed in the Hamiltonian $\mathbf{H}$ of the problem. Which is just a mathematically expressed application to the mentioned case, of the general definition of the concept of "energy", such as this is tied with the classical concept of a "Hamiltonian". And I recall that the form of *any* ***definition of a concept*** is scale-invariant, inside the domain where it is endowed with meaning.

Another fact also deserves being mentioned. When he has constructed his equation Schrödinger has imposed a priori a general wave-like form $\Psi_{Gt,\mathbf{H}}(\mathbf{r},t)=a(\mathbf{r},t).exp(2\pi i/h)\varphi(\mathbf{r},t)$ of the solutions. Therefrom stems – and stubbornly persists up to this day – a total confusion between physical *wave*-functions $\Phi_{Gt}(\mathbf{r},t)$ of individual specimens $\sigma_{\Phi(Gt)})$ of the studied microstate, and on the other hand *state*-functions from state-ket $|\psi_{G,\mathbf{H}}(\mathbf{r},t)>$ that represent a whole set of abstract statistics of numerical values of qualifying quantities obtained via acts of measurement performed on specimens $\sigma_{\Phi(Gt)})$. This ***huge*** ambiguity, tied moreover with an unintelligible dimensionality and structure of the 'propagation-space' [172], deteriorates notably the conceptual situation from $QM_{HD}$. The dimensionality and *the lack of semantic homogeneity* of the 'propagation-space' prolong and amplify the multi-faced mixture that flaws the $QM_{HD}$-representation of quantum measurements[173].

Last but far from least, the Schrödinger equation becomes useful only once the entire initial state-ket *is **"given"***. Sometimes it is said or only thought that the initial state-ket $|\psi_{G,\mathbf{H}}(\mathbf{r},t_o)>$ is determined by imposing the limiting conditions upon the general solution of the Schrödinger equation of the problem. But ***inside** the whole spatial volume that is globally cut out by the limiting conditions*, the *physical* initial situation can vary in an unlimited number of ways that escape any priori knowledge, even if the explicit data are posited to be stable. So the mathematical specification of limiting conditions might not suffice for specifying the contents of the initial state-ket to a degree that insures verification of the predictions. This, in fact, is nearly always the case in a non-idealized factual situation. *While on the other hand, on top of all that the usual calculi are likely to also imply a Laplace equipartition of the elementary events!*

In short, even when the equation exists, in general the predictions emerge flawed by ignorance, idealization, and approximation.

> *And **how** they are flawed, and **to what a degree**, can be known only a posteriori, by **verification**-measurements. Here lies the **limit** of the predictive capacities of **mathematically generated statistical-probabilistic predictions***[174].

Moreover all the preceding remarks presuppose that it has been possible to write down the Schrödinger equation of the problem. But this is not the case when the considered problem involves a non-Hamiltonian situation.

Consider now the framework *[IQM-$QM_{HD}$-(dBB)]*. Inside this framework there appears a face-à-face between the bottom-up construction and the Schrödinger predictive writings (22) constructed top-down, and:

> This – via the expression (31) – entails noteworthy new possibilities of *improvement of the predictions concerning quantum measurement*, as follows.

- The predictions from (31) emerge directly via measurements and so insured by construction to be factually true. And since these predictions are probabilistic, they require many repetitions of long sequences of reiterations of the involved coding-measurement successions, that cover a relatively non-negligible global lapse of time; which insures also a considerable *stability* of the asserted results.

---

[171] The new Portuguese School of Broca and Araujo [2010] (cf. Gatta, Rica da Silva, Araujo, Croca and Silva, Cordovil,Moreira, Magalhäes, Alves, Santos) is developing a very interesting and fundamental new representation of what here we call 'the Universal Physical Substance'. But there also, the inertial belief in the necessity to incorporate Einstein's formalism opposes illusory obstacles and generates inadequacies. We come back upon this in the chapters 8.III and 9.III.

[172] The 'propagation-space' consists partially of usual space and time, and partially on a configuration space of mechanical qualifications. This involves an inextricable *mixture* of individual physical features and statistical-abstract ones (like in classical statistical physics).

[173] By the adjunction of *IQM* to the representation of the microstates and the correlative bottom-up character of the construction, the present approach reacts to this situation also.

[174] Wigner's "unreasonable power of mathematics", in fact is less unreasonable than it is believed.

- The considered observable ***A*** being given, one can *systematically* begin by constructing factually the expansion $|\psi_{Gt,H}(r,t_1)>_{fact(DM)}/A$ from (31) for the initial time $t_o$. ***The involved operation of generation $G_o$ embodies a physical realization of the limiting conditions that act factually***, while the *factual*-formal character of the whole constructive procedure entails maximal sensitivity with respect to the specificities of *the whole factual initial ground encompassed by the limiting conditions*, *whether known and expressed,* ***or not***. Once this is done:

- *If* the situation is Hamiltonian and the Schrödinger equation of the problem exists and is solved, then one can continue by determining *mathematically, by the use of the equation,* the state-ket entailed for any moment $t_1>t_o$ by the equation ***and*** (31) at $t_o$, *if* this seems more economical then factual determinations for also the times $t_1>t_o$. But one can also generate factually the state-ket (31) for any time $t_1>t_o$, either if the use of the mathematical solution seems more difficult, or *just in order to* **compare** *the factual solutions, to the mathematical ones* and thereby improve its outputs from various points of view (suppression of idealizations and approximations, rigor, domain of applicability, control, economies of effort).

- *If* the situation is not Hamiltonian, or for some reasons the equation just cannot be written or cannot be solved acceptably, then one can radically replace the equation by (31).

However ***in any case in which the equation exists and is solved, it endows with a unifying mathematical functional expression of the state-ket of the studied microsystem***.

In short – inside *[IQM-$QM_{HD}$-(dBB)]* and for unbound microstates – the formal-factual construct (31) endows the representation of prediction-verification with a high degree of independence with respect to the Schrödinger equation, namely with various possibilities of improvement of its outputs, and with the possibility to entirely replace it.

This conclusion is natural at the present time when the computational powers have become so high and continue to increase so rapidly and the techniques for making use of big data are in such a vertiginous progress.

## (7.III).5. THE CASE OF BOUND MICROSTATES

We finally consider also the case of bound microstates. In consistent prolongation of the definitions from (I.2).1 of unbound sorts of a microstate, these can be called "*one bound stationary micro-state of several sorts of micro-systems*". In such a microstate all the involved corpuscular-like singularities from any one specimen, circulate indefinitely inside a very limited microscopic domain of the 'physical' space[175].

Thereby we come back to the general introduction to this work. Let us stabilize this loop by expressing explicitly the connexion.

Historically the representation of bound micro-states has begun inside the classical physics, via a top-down approach that assigned classical Newtonian planetary models to the basic entities posited in the molecular and atomic physics and in chemistry. Around 1900 this extrapolating top-down progression came short of logical consistency with the whole of the classical physics, and so suffered stagnation. Consequently the first 'quantum' postulates were formulated (Bohr, Plank, Einstein's interpretation of the photoelectric effect). These postulates have acted like close precursors of the full *reversion* in 1923, by Louis de Broglie ([1924], [1963]) of the *direction and the nature* of the process of construction of scientific knowledge on micro-phenomena: de Broglie's corpuscular-wave model initiated a decidedly bottom up approach along the vertical of conceptualization, that rooted this process into the as yet *never represented before* 'Universal Substance' (in the Spinozian sense). And Schrödinger's equation, by its first applications, has yielded so striking results that a new phase of construction instilled itself into the scientific thought. But this revolutionary inversion of order in the microscopic levels of the human conceptualization of the physical reality was accomplished only furtively, in an inexplicit, feebly cut out way. This happened because of Bohr's impossible "positivistic" interdiction of models in scientific constructions and because it got mixed with persisting features and results of the previous top-down approach.

> Namely, the processes of *verification* of the predictive statistical results of quantum measurements by long series of necessarily ***individual*** new acts of measurement, got stuck in the mathematical representation of the *predictive* ***statistics***; which entailed unintelligibility and long stagnation thus acting like a repulsive wall against *the continuation of de Broglie's new bottom-up start* ***by a FULL and explicit achievement of a first phase of individual conceptualization***.

Only here, after nearly a century, *IQM* defines a full explicitly structured and *general* framework of representation of also the initial phase of individual conceptualization of the microstates.

But from the point of view adopted in this work, first inside *IQM* and then inside the two successive frameworks *[IQM-QM$_{HD}$]* and *[IQM-QM$_{HD}$-(dBB)]*, the case of bound microstates appears now as placed far high, just beneath the classical roof, and there it is immersed in an accumulation of mixtures between specifically quantum features and classical ones, as well as *a mixture between formal features of mutually distinct QM$_{HD}$-concepts*.

> - For a bound microstate the human observer did not himself achieve deliberately the involved operation of generation $G_t$. This operation is conceived to have been achieved naturally, before the beginning of the human investigation. So from the point of view of its availability for being studied a bound microstate is like a classical "object", ***it just pre-exists 'there outside'.***

---

[175] According to de Broglie [1956] in certain stationary states of this sort some of these singularities are *at rest* inside the framework of the laboratory.

- Under the lighting of *IQM* one can understand what loss of specificity is entailed already by this circumstance alone. Indeed in the absence of a deliberate human operation of generation of the microstate the measurements can be conceived in the classical manner, without successions *[$G_t$.MesA]* each one of which involves first *an act of full creation* of $G_t$. So, explicitly at least, the measurement-operations *MesA* can be achieved *directly*, by use of test-microstates for which the measurement evolution can obey the coding-postulate (20') (for instance Compton collisions), or via field-effects (Stark, Zeeman). Furthermore the measurement-interaction ***itself*** is often *realized collectively* by the action from the start of a big set of simultaneous 'test-operations' upon a big set of replicas of the studied microstate (monochromatic radiation thrown upon a collection of atoms of a definite kind, etc.) so that the result is often expressed directly as a statistical *mean* drawn from many statistical distributions that – conceptually – are mutually distinguished inside $QM_{HD}$ but that are not stably singled out mutually in the declared result.

- Because the de Broglie *singularities* from the specimens of a given sort of bound microstate can be *stably* captured inside a small global space-time domain, this *same* spatial extension is assigned also to each *whole* specimen itself (the order of spatial extension assigned to "an atomic structure" is, say, of the order of $10^{-12}$). This entails that the spatially ***unlimited wave***-aspects that are devoid of an ***observably separable*** *own significance*, have been implicitly posited to be altogether devoid of any sort of significance, or even existence, just like in the case of macroscopic "objects" [176].

- For a bound microstate – like in the case of an unbound one – the *formal* distinction between the state-function from the statistical state-ket $|\psi_{Gt}>$ of a considered microstate and the corresponding de Broglie wave-function $\Phi_{Gt}$ is very feeble, and often even radically absent inside the minds. But for a bound microstate even the *conceptual-**physical*** distinction – so progressively and painfully specified in this work – is subjected to a new sort of confusion. Namely, for a bound structure of micro-systems *the **physical** specimens of the involved microstates are all **STABLY** co-present* inside a very small domain of physical space and the singularities – ***"particles"*** – involved by them concentrate upon *them* the attention and the imagination, while the abstract statistical features represented by the corresponding state-ket are usually *perceived as such*, as abstract and separated from the involved "objects", just like in the classical physics. (Whereas for an unbound microstate the strict reverse happened, the existence and the role of the physical specimens have been occulted and the abstract representation by a state-ket has been reified because the formalism works exclusively with this). Moreover:

> *The involved **state**-ket is from the start conceived to be **also** a **physically** realized **eigen**-ket of the total-energy-observable-**H***; in this sense there emerges a conceptual degeneration that confounds into one same concept two ***very*** different sorts of entities.

- Finally, and *most* importantly, accordingly to *[IQM-$QM_{HD}$-(dBB)]* an *eigen*-ket is a functionally expressed model drawn from ***ONE wave***-function $\Phi_{Gt}$ that represents "one" ***physical*** wave. But on the other hand in each one *expansion* of the form (22) of the $QM_{HD}$-***state***-ket of a studied ***ONE*** bound microstate (say, the one electron from an atom of hydrogen)

---

[176] This brings back to the configuration spaces introduced via the Schrödinger equation: only this artificially delimited space-time support is what is associated with an *abstract* representation space where are lodged the values of all the ***mechanical*** semantic dimensions of qualification of the involved micro-systems. In a classical configuration space this happens currently and nobody wonders why *more* than only four dimensions of 'physical' representation are considered because any body understands implicitly and immediately why. But in the case of a bound microstate one begins to wonder about this because the absence of intelligibility of the whole formalism entails a general attitude of suspicion and because often the physical *effects* of the individual physical waves that are involved ***do act** upon the observed results and their representation*, even if – *in consequence of lack of comparison, so of **reference*** – such actions are perceivable only in a fluctuating and a-consensual way.

***many*** mutually distinct such models are involved; so the considered *"one"* ***state***-ket – that identifies in nature with any eigenket from the expansion – is in fact conceived to represent ***a whole set*** *of potential microstates* where specimens of the studied microstate make Bohr-jumps and *actualize* for a while one or another among the potential state-ket. So, conceptually, *the considered "one" "state"-ket in fact is a sort of* ***meta****-state-ket*.

So a *huge* multiform confusion steps in between the various sorts of concepts that are involved: the general concept of a physical wave-movement in the immediate neighbourhood of an involved corpuscular-like singularity in the amplitude of the corresponding whole wave (the "particle"); the set of all the possible mutually distinct wave-movements recalled above; the set-of-sets of probabilistic predictive counts – abstract numbers – on results of measurement on the ***one*** involved sort of studied physical entity (the "state"-ket).

Under the very obscuring cover of this creased texture, the strongest conceptual and formal specificities of the conceptualization of the microstates loose any controlled contour. The mixtures enumerated above generate a mob of conceptual ghost-problems. While here we want to concentrate upon the essence of a new approach. Since moreover, for an improved representation of the *specificities* of quantum measurements the case of the bound microstates has a very marginal conceptual role, we relegate the *[IQM-$QM_{HD}$-(dBB)]* treatment of this case into other works, by the present author or by others.

> But *Gleason's theorem and the character of the argument that leads to the assertion Ass.2 permit to assert that a representation (31) is in principle possible for the bound microstates also*.

Conversely, in the case of a bound state the Schrödinger equation – when it exists and can be solved without excessive idealizations and/or approximations – endows with a very synthetic mathematical expression of the global probabilistic content of the consensual knowledge generated concerning the studied microstate.

However the preceding remarks and the fact that a bound state involves interference, so a quantum potential and quantum fields, call forth the necessity of a thorough conceptual re-examinations of this most fundamental case.

## (7.III).6. SUMMARIZING CONCLUSION ON THE CHAPTER 7.III

The global results from the chapter 7.III can be summarized as follows.

First inside the framework *[IQM-QM$_{HD}$]*, and then inside the completed framework *[IQM-QM$_{HD}$-(dBB)]*, we have constructed a new representation of the quantum measurements on unbound microstates. This representation has been explicitly founded upon:

- The necessarily individual and physical-operational character of coding-measurement-successions *[G.MesA]* that ***unavoidably*** must be performed for the *verification* of statistical-probabilistic predictions, *and that* ***can*** *be used also for the generation of these.*

- The deliberate requirement of a general and fully intelligible Hilbert-space representation (23) of the *factual* $(\varepsilon,\delta,N_o)$-probability laws constructed for the outcomes of many long series of coding-measurement-successions *[G.MesA]*, because in consequence of Gleason's theorem – ***quite independently of the concept of a microstate*** – a Hilbert-space representation of *any* probability law is indeed a particularly expressive and efficient mathematical representation of such a law, clearly superior to Kolmogorov's representation.

The mentioned results have been reached as follows.

In the section (7.III).1 – via a fundamental first assertion *Ass.1* and the argument *Arg(Ass.1)* that founds it – we have brought into evidence a "vertical" conceptual constraint that permits to organize a basic *operational and logical coherence between IQM and the formalism from QM$_{HD}$ with respect to the requirement of statistical-probabilistic predictions and of the verifiability of these.* At a first sight the assertion *Ass.1* seemed to be trivial. But a closer examination has brought forth a character tied with major consequences, namely:

*As soon as a bottom-up approach is practised the operations of quantum measurement can play a major constructive role *for also the* ***elaboration*** *of predictions* (i.e. not exclusively a role of verification of mathematically constructed predictions, as it is done in *QM$_{HD}$*).

* By a strict use of a small number of definite conditions of compatibility, *IQM* can strongly guide the elaboration of a new, fully intelligible Second Quantum Mechanics that ***incorporates the whole stratum of individual conceptualization of the microstates***, thus compensating the major lacuna that flaws and distorts *QM$_{HD}$*.

In the sub-sections *(7.III).2.2.1-(7.III).2.2.2*, by systematic reference to *IQM*, we have then proved a second assertion *Ass.2* for the *particular* case of microstates *ms(unbound,1)$_{G(nc)}$* and a corresponding coding-postulate (20'). This has produced the *formal-**factual*** equivalence

$$(31)\ \left[|\psi_{Gt,H}(r,t_1)>_{fact(DM)}\ \approx_{\mathbf{pred.}}\ \left\{|\psi_{Gt,H}(r,t_1)>_{fact(DM)}/A\right\},\forall A,\forall t_1\right\}\right]_{\mathbf{verif}}$$

where [the state-ket $|\psi_{Gt,H}(r,t_1)>_{fact(DM)}$ and all its expansions $\{|\psi_{Gt,H}(r,t_1)>_{fact(DM)}/A\}$] ***emerge already (ε,δ,N$_o$)-verified, so endowed a priori with (ε,δ,N$_o$)-factual and relativized truth. Which dissolves the necessity of Born's postulate.*** The factual state-ket (31) assert probabilistic predictions that – in general – are *different* from the corresponding probabilistic predictions drawn from the corresponding *QM$_{HD}$*-state-ket calculated by use of the Schrödinger equation:

$$|\psi_{Gt,H}(r,t_1)>_{fact(DM)}\ \neq_{\text{pred.}}\ |\psi_{Gt1,\mathbf{H}}(r,t)>$$

In consequence of Gleason's theorem the factually defined expansions from (31) $\{|\psi_{Gt,H}(r,t_1)>_{fact(DM)}/A\},\forall A,\forall t_1\}$ insure a *direct* access of the asserted $(\varepsilon,\delta,N_o)$-probabilistic predictions on the results of measurements, to a Hilbert-space representation (23), just like for the expansions (22) from *QM$_{HD}$*. But they do this *without* the use of the Schrödinger equation of the problem, directly via the *factual*-formal procedure that generates these state-vectors accordingly to the assertion *Ass.2*.

This feature is fully consonant with the very new and specific computing possibilities of our present time.

In the section (7.III).2.3 we have extended the result (31) to also the microstates *ms(unbound,1)$_{cG(qf)}$* that involve interference phenomena and a corresponding quantum potential, and are tied with the coding-postulate (35).

This extension has appeared to involve various and quite fundamental obstacles tied with the verifiability of the predictions on the outcomes of measurements of the basic $QM_{HD}$ momentum-operator ***P***. We have identified the sources of these obstacles. This led us to require a modified Hilbert-space representation (40) of the state-ket of a microstate *ms(unbound,1)$_{cG(qf)}$*, tied with the de Broglie-Bohm guiding relation (33) and with a coding-postulate (35) that is different from the postulate (20') implicitly assumed inside $QM_{HD}$ for *any* sort of unbound microstates.

The representation (40) and the coding-postulate (35) have required a generalized concept $\boldsymbol{P}_{vc\text{-}vd}$ of "momentum-*observable*" that has been distinguished from the $QM_{HD}$ momentum-*operator* ***P*** from the non-commutative algebra of the operational $QM_{HD}$ representation of the classical mechanical qualifying quantities *A(r,p)*. This distinction has been founded upon the concept of an "observable beable" that is either "value-conserving" or "value-destructive".

The momentum-observable $\boldsymbol{P}_{vc\text{-}vd}$ is tied with a generalized equation (38) for its eigenfunctions and eigenvalues that defines a generalized sort (39) of bases of eigenket: these eigenlet are modulated by a parameter "*m*" that leads to a $QM_{HD}$-basis of plane waves when *m=1*, but when *m>1* the corresponding basis is formed by a ***physical*** superposition of two or more mutually distinct plane-waves.

We have furthermore posited explicitly a rather simplistic but clarifying specification of the relation between a physical wave-function $\Phi$ and an abstract statistical state-ket $|\psi_{Gt}>$).

In this way, inside the enlarged framework *[IQM-$QM_{HD}$-(dBB)]* has emerged the new Hilbert-space representation

[(35)-(38)-(39)-(40)]

of factually established $(\varepsilon,\delta,N_o)$-probabilistic predictions on the results of momentum-measurements on microstates *ms(unbound,1)$_{cG(qf)}$* with a *non*-null quantum potential. This has permitted to extend the assertion *Ass.2* and the equivalence (31) to also the microstates *ms(unbound,1)$_{cG(qf)}$*). Furthermore *this has installed the possibility to make use of the coding-postulate (35) for* ***any*** *unbound microstate and any classical mechanical qualifying-quantity A(**r**,**p**); which in its turn entails the general possibility of one-brunch probability-trees (Fig.8*).

In (7.III).3 it appeared that nothing hinders to apply the assertion *Ass.2* and (31) to also a *one* micro-state of *two* or several micro-systems, for any observable.

In this way, inside (7.III).2-(7.III).3 have emerged the contours of a fully intelligible *factually* rooted representation of the quantum measurements on unbound microstates. These contours are defined by the set of assertions and mathematical expressions

*[Ass.1, Ass.2,* (21'), (22), (23), (31), (35), (38), (39), (40)*]*

that constitutes an explicitly constructed and clear ***semantic***-syntactic configuration.

Then we have continued as follows:

In (7.III).4 we have shown that the factually rooted representation of the quantum measurements constructed in (7.III).2-(7.III).3 permits to improve the use of the Schrödinger equation when it is available and to replace it when it cannot be solved or cannot be even written.

In (7.III).5 we have briefly indicated the specificities of the bound microstates inside the framework *[IQM-$QM_{HD}$-(dBB)]*.

So the chapter 7.III as a whole leads us from the general framework *IQM* for *any* acceptable theory of microstates and the critique of $QM_{HD}$ founded upon this framework, to a fully intelligible representation of the quantum measurements reconstructed inside the frameworks *[IQM-*$QM_{HD}$*]* and *[IQM-*$QM_{HD}$*-(dBB)]* that introduce a *physically*-and conceptually operationally defined *model* of a microstate and corresponding coding-measurement successions.

Now the task to outline *QM2* itself – a definite intelligible *theory* of the microstates among all the non-specified possible ones encompassed by *IQM* – can finally be entered upon.

# 8.III

# INTEGRATED OUTLINE OF AN INTELLIGIBLE SECOND QUANTUM MECHANICS *QM2*

For introducing this chapter I take a reviled liberty. Inside this scientific writing I shall glide into a metaphor, because it stresses well the prime importance of the representation of the measurements in a scientific theory of physical entities that are radically unperceivable.

The representation of the quantum measurements constructed in the chapter 7.III acts as the heart of the principles of the fully intelligible Second Quantum Mechanics *QM2*. This heart quivers already in the summarizing conclusion from (7.III).6. Via an act of abstract surgery I shall now integrate it into the conceptual-and-mathematical representational organism that has been baptized a priori "the principles of a Second Quantum Mechanics *QM2*".

## (8.III).1 THE THREE SOURCE-DOMAINS AND THEIR RESPECTIVE ROLES IN THE ORGANIZATION OF *QM2*

The purpose in this work is to outline now all the principles – but the principles *alone* – of a fully intelligible Second Quantum Mechanics *QM2* where all the critiques and questions that persist since more than a century, vaporize away. The prime matter for achieving this has been drawn out of *IQM*, $QM_{HD}$ and the approach *dBB*. We first recall the basic features offered for *QM2* by each one of these three source-domains.

### (8.III).1.1. The input from *IQM*.

*IQM* has been constructed with the overt purpose to obtain a structured infra-discipline that constrain a priori ***any*** theory of the microstates by acting as a pre-formalized context of immersion and reference. *IQM* has been conceived such as to induce from the unavoidable cognitive constraints that act in the involved cognitive situation, the *general* conceptual-operational-methodological features of *QM2*, in a way similar to that in which the genetic code that acts during the development of one given living being, induces into this being features that stem from the constraints imposed by the multi-personnel history of the closest bio-physical contexts that led to it.

### (8.III).1.2. The nowadays Hilbert-Dirac formulation of Quantum Mechanics, $QM_{HD}$

The $QM_{HD}$ representation of the microstates raises questions and critiques since a very long time. But on the other hand it contains very efficient Hilbert-space mathematical algorithms that deserve being preciously conserved. From the point of view of the requirements introduced by *IQM*, the mathematical representations from $QM_{HD}$ appeared as a mixture of useful algorithmic aspects and of unacceptable semantic voids or inadequacies.

### (8.III).1.3. The de Broglie-Bohm approach *dBB*

This approach started in Louis de Broglie's Thesis with the Jacobi formulation of classical mechanics-and-optics where the conditions that restrict to the 'geometric approximation' have been suppressed. Thereby there emerged a prolongation into arbitrarily small space-time dimensions (de Broglie [1924], [1956], [1963], [1987]). In continuity with the classical physics, de Broglie has proceeded toward *a sort of mathematically expressed metaphysical representation of the Universal Substance in Spinoza's sense*. Since [1956] this representation has included features developed by David Bohm ([1952]). And nowadays – remarkably – other quite similar approaches *continue to emerge* (Stefanescu,E.,[2018]; Unnikrishnan,C.S., [2018]). But historically de Broglie's marks remain far *primary*, of course. We have denoted by *dBB* the essence of the views of Louis de Broglie and David Bohm.

Both de Broglie and Bohm have asserted *dBB* as exclusively an interpretation of $QM_{HD}$. Furthermore the levels of conceptualization are not clearly distinguished from one another. The representation is quintessentially *mental and global*. The concept of 'one' microstate is not individualized, neither in a factually operational way nor only mentally. However de Broglie speaks of "microsystems"; but Bohm does not even that, I think. Louis de Broglie has elaborated on the theory of quantum measurements ([1956)], [1957)], but this elaboration remains just an explanatory substratum on the basis of the a priori and fully accepted $QM_{HD}$ theory of quantum measurements, such as it now stands. Etc.

Any trace of explicit methodology for insuring operationally scientific consensus is absent in *dBB*. In particular, the *dBB* representation does not define operations of generation.

On the other hand:

In contradistinction to Bohm's approach, *Louis de Broglie's last representations ([1956], [1963]) are subjected to all the features of Einstein's Relativities*, notwithstanding that these have been explicitly constructed on the basis of *observational* criteria for consensus induced by a cognitive situation that is fundamentally different from that presupposed in Einstein's Relativity, so that they become devoid of any necessity; moreover this induces complex and inextricable inner semantic *inconsistencies* with respect to an exclusively mental, global and purely explanatory global model that – as Hilary Putman would put it – accepts the view "from God's Eye" *where the concept of scientific* ***observationally founded consensus*** *simply is absent*[177].

Nevertheless the whole *dBB* approach is the genetic background of de Broglie's seminal model of a microstate as well as of the de Broglie-Bohm guiding relation, which – both – play quintessential roles in this work.

---

[177] This entails that for the outline of *QM2* Bohm's approach is more immediately useful precisely because it is less specified and detailed and so certain essences come out more strikingly. But de Broglie's equations might contain useful suggestions if they are regarded as observational descriptions somehow *superposed* to the background of the global a-observational model, in relation with a given level of conceptualization, and if they are associated with arguments for either their acceptance or their refusal at that level. Anyhow, it is noteworthy that the number of physicists and of thinkers that claim the importance of the models proposed by de Broglie and by Bohm is constantly increasing (cf. Gondran M.,et A., [2014]).

## (8.III).2. OUTLINE OF AN INTEGRATED ORGANIZATION OF *QM2*

This outline will be realized in a way that will stay synthetic in the extreme.

As stated from the start and often repeated, the purpose of this work is *not* to offer a fully achieved new theory of microstates, it is to only identify inside $QM_{HD}$ the conceptual loci wherefrom spouts of un-intelligibility burst out, to suppress these, and to realize a well-defined new *basic* structure that insure full intelligibility and inner semantic-formal consistency for subsequent developments.

We begin by a very brief statement of the role played by each one of the three source-domains characterized above. Then we shall enumerate of the postulates of *QM2* ; sometimes the formulation of a postulate is followed by a brief verbal specification on the semantic carried by it. Finally the whole that obeys the formulated postulates will be characterized in words in order to better perceive the intuitive substratum of the conceptualization.

### (8.III).2.1. On the roles played by the source-domain in the construction of *QM2*

**A. The constructive role of *IQM***

✚ *IQM has revealed*

+ The unacceptable character of the orthodox interdiction, in $QM_{HD}$, of a general model of a microstate, which entails absence of criteria for specifying what procedure can be considered to be adequate for measuring upon a given sort of "microstate" a given qualifying quantity.

+ *The total void inside* $QM_{HD}$ *of an explicit and organized level of individual conceptualization.*

In more detailed terms:

- the unawareness inside $QM_{HD}$ of the ***unavoidable*** actions of *a **factual, physical** operation G of generation of the microstate-to-be-studied*;
- the unavoidable *consequence* of the unavoidable actions of *"G"*, namely an "*essential indeterminism of **KNOWLEDGE***";
- the absence inside $QM_{HD}$ of explicit definitions of "***coding**-measurement **successions** [G.MesA]*" founded on a specified ***MODEL*** of a 'microstate', which simply *leaves undefined the central concept of an act of quantum measurement (**any** such act)*;
- the absence inside $QM_{HD}$ of explicit definitions of the different sorts of 'microstates', so an artificially uniform treatment unable to discern essential mutual specificities among categories of microstates that can be *observationally* distinguished from one another.

✚ By requiring explicitly a model of a microstate, *IQM* has permitted to then identify the model that – covertly – is present and works inside the whole formalism of $QM_{HD}$, namely de Broglie's model of a "corpuscular-like-wave" that endows with a *specific and basic meaning* the fundamental $QM_{HD}$-concept of "eigenket".

✚ *IQM* has permitted to identify the concept of a coding-measurement-succession (20') that inside $QM_{HD}$ is supposed – *implicitly and falsely* – to work *generally*, uniformly, for any sort of act of measurement, on any sort of microstate.

✚ *IQM* has guided a process of identification of the gross conceptual fallacies that vitiate the nowadays representation of quantum measurements – namely a radical coalescence between ***individual** coding-measurement successions [G.MesA] performed on specimens of the studied factual microstates*, and on the other hand ***statistics** of the numerical results of such successions* – thereby founding a radical rejection of this representation.

✚ Via the basic assertions *Ass.1*, *IQM* has instilled into *QM2* a remarkable *fusion* and mutual consistency between *IQM* and *QM2*.

✚ Via the basic assertion *Ass.2*, *IQM* has instilled into *QM2 a **bottom-up** "vertical" fusion* expressed by the formula (31), between:

+ on the one hand, the central *IQM*-concept (9") of a "*primordial transferred and probabilistic description*" of the studied microstate;

+ and on the other hand, the Hilbert-space mathematical formalism from *QM2* such as it is perceived in the light of Gleason's theorem on the *general* capacity of the Hilbert-space mathematics to represent *any* probability law.

✚ Thereby the assertion *Ass.2* has induced from *IQM* into *QM2* a basic illustration of the general concept of bottom-up constructed "factual $(\varepsilon,\delta,N_o)$-probabilities" – essentially different from Kolmogorov's top-down purely mathematical theory of probabilities, and *far more rich than it.*

The physical theory of factual $(\varepsilon,\delta,N_o)$-probabilities can be rooted in the a-conceptual physical factually. It is expressed bottom-up, genetically, by the concept of "factual $(\varepsilon,\delta,N_o)$-probability-trees" where can be represented all the mutually distinct levels of a probabilistic conceptualization, an "infra-individual level", the individual level, the probabilistic one, and the meta-probabilistic levels of correlations of different orders.

Which, I think, is a break-through by itself.

✚ Via the *general* concept of operation of generation, *IQM* has drawn de Broglie's model into the consensual operational and predictive approach from *QM2*.

✚ Via the class of "composed operation of generation" *IQM* has strongly guided toward the refusal inside *QM2* of the current $QM_{HD}$ Hilbert-space representation of the interference-microstates.

✚ Via the explicit requirement of adequate *coding*-measurement-evolutions for the momentum-quantity *IQM* has guided for endowing the interference microstates with a satisfactory Hilbert-space representation.

In short, throughout the chapters 6.II and 7.III of this work *IQM* has constantly played:
- a *structuring* role, by injection of its own structure into *QM2*;
- a *critical* role, by dictating *local* refusals or specifications founded upon comparisons.
- a *revealing* role (the specific meaning of the $QM_{HD}$ eigenket, the clarifying and constructive roles of the operations of generation *G*, the assumptions *Ass.1* and *Ass.2* that have prompted new and surprising conceptual organizations *whereby IQM got **incorporated** to the emerging QM2 and founded a bottom-up approach that endows by construction with factual truth and frees from the **necessity** to always make use of the Schrödinger equation).*

Considered globally, the roles played by *IQM* have been of semantic and structural nature, and they have led to radical and essential improvements.

**B. The constructive role of $QM_{HD}$**

$QM_{HD}$ has been criticized throughout the chapters II.5 and II.6 of this work. In particular the $QM_{HD}$-theory of quantum measurements has been entirely rejected. And in the chapter III.7 the additive representation (15") of the state-ket of a microstate $ms(unbound,1)_{cG(qf)}$ has been suppressed because it has been found to lead to predictions that cannot be verified while they also are tied with logical-mathematical inconsistencies.

But on the other hand it appeared that Dirac's choice of a Hilbert-space representation of the probabilistic predictions on the results of quantum measurements has been a highly and surprisingly fertile choice.

It appeared that Gleason's theorem, when combined with the *IQM*-concept of "probability-tree of an operation of generation", contributes in an essential way to the revolution of the *general* concept of probability started inside *IQM* by the concept of "factual $(\varepsilon,\delta,N_o)$-probability-trees" of an operation of generation.

Namely, a massive and intimate alliance emerges between this *IQM*-concept of "factual $(\varepsilon,\delta,N_o)$-probability-trees", and on the other hand, mathematical representations and algorithms from $QM_{HD}$. In particular, the Dirac-Hilbert bra-ket calculus reveals surprising powers for expressing mathematically *general* and fundamental ***semantic*** contents.

That is why in 7.III – notwithstanding the conceptual inadequacies of the $QM_{HD}$ representation (15") of the state-ket of an interference-microstate – we have so decidedly researched another ***Hilbert-space*** representation.

Indeed thereby the whole algorithmic essence of the Hilbert-Dirac mathematical bra-ket formalism *has been conserved inside QM2*; and so, via the chaining *IQM*-(Hilbert-Dirac-representation) we endow *QM2* with a "universal" sort of structure-and-significance, in a most basic sense that will clearer appear in what follows.

**C. On the constructive role of *dBB***

The supply from the part of the approach *dBB* is also considerable:

✙ *dBB* connects *QM2* with the Hamilton-Jacobi formulation of the classical Mechanics *and* also with the classical Electromagnetism and Optic; so it insures continuity with major essences from the classical Physics. On the other hand, via de Broglie's model of a microstate, *dBB* roots the Second Quantum Mechanics *QM2* into the microphysical factuality. So *dBB* endows *QM2* with a sort of explicitly drawn out multi-dimensional map of conceptual features that can guide the integration into one big whole of the various nowadays theories of domains of physical reality. From a conceptual point of view this is precious.

✙ Louis de Broglie's "(corpuscular-like-wave)-model" endows with intelligibility the basic equations for eigenket and eigenvalues of the dynamical observables from *QM2*.

✙ Via the channel of adduction

$$\Big[MP(ms_{G,cw}),\ (1')G_{cw}\leftrightarrow ms_{G,cw},\ (14)ms_{G,cw}\equiv\{\sigma(ms_{G,cw})\},\ (35)P(cod)_{\forall msG}),$$
$$(33')\mathbf{p}_j(\mathbf{r},t)=\boldsymbol{\nabla}.\beta(\mathbf{r})_t=\textstyle\sum_m(p_{jm}\mathbf{k}_m(\mathbf{r})_t\Big]$$

*(36')*

into the factual-operational representations from *QM2*, the metaphysical but mathematically expressed *dBB* approach is transmuted into a reservoir of concepts and representations available for conceiving inside *QM2* consensual and verifiable future developments.

✙ The guiding relation (33) has led to the proposition of a crucial experiment "EXPER" that, if it succeeds, founds the generally applicable coding-measurement-postulate (35) that: circumvents Heisenberg's principle; can penetrate arbitrarily deep into the specifically "quantic" physical reality; and permits to summarize the essence of the *QM2*-descriptions of unbound microstates in terms of *one*-branch probability trees valid for any sort of unbound microstate (Fig. 3').

✙ The guiding relation (33) and the coding-measurement postulate (35) have opened the way toward the mathematical representation [(38), (39), (40)] that permits to extend to also the microstates with non-null quantum potentials, the direct applicability of Gleason's theorem, and so to insure for any unbound microstate the validity of the factually constructed representations (31).

Together,

- the modelling-postulate $MP(ms_{G,cw})$ that led to (1') $G_{cw} \leftrightarrow ms_{G,cw}$ and to (14) $ms_{G,cw} \equiv \{\sigma(ms_{G,cw})\}$,
- the guiding relation (33) and the corresponding coding-postulate (35),
- the one-branch-probability trees from the Fig.8,

constitute a contribution to *QM2* stemming from *dBB* that is fundamental in a most deep-set sense of this term[178].

This illustrates in what a sense *dBB* instils into *QM2* new possibilities of developing the consensual, predictive and verifiable representation of the microstates[179].

✙ Paradoxically, the *critical* remarks on *dBB* formulated in (8.III).1.3 will be very useful in 9.III, in the discussion of the problem of the unification between the modern microphysics and the Einstein Relativities.

---

[178] And let us also notice a curious sort of process that concerns the future of *QM2*: Inside the framework *[IQM-QM$_{HD}$-(dBB)]* the degree of 'communication' between the purely mental representations from *dBB* and the predictive representations from *QM2 tends to* ***evolve***, which reveals potentialities instilled into *QM2* by *dBB*. For instance, consider the *[IQM-QM$_{HD}$-(dBB)]* probability $\pi(\mathbf{p_j})$ that the result of a momentum-measurement performed at a time *t* on a microstate with state-ket $|\Psi_G(\mathbf{r},t)\rangle$ be found to be $\mathbf{p_j}$. In *dBB* terms this probability can be symbolized as

$$\pi(\mathbf{p_j}) = \int [\pi(\mathbf{r}).\pi(\nabla\beta(\mathbf{r},t)=\mathbf{p_j})]d\mathbf{r} \;=\; \int [|\psi_G(\mathbf{r},t)|^2.\pi(\nabla\beta(\mathbf{r},t)=\mathbf{p_j})]d\mathbf{r}$$

(where the notation $\pi[\nabla\beta(\mathbf{r},t)=\mathbf{p_j}]$ is to be read: the probability that the guided momentum $\nabla\beta(\mathbf{r},t)$ possess the value $\mathbf{p_j}$). The purely mental (i.e. not physically-operational) approach *dBB* does not define ways of verifying the factual truth of this symbolization. But conceptually it permits to consider this symbolization as a *dBB* representation of Born's postulate:

$$|c(\mathbf{p_j},t)|^2 = \pi(\mathbf{p_j}) = \int [\pi(\mathbf{r}).\pi(\nabla\beta(\mathbf{r},t)=\mathbf{p_j})]d\mathbf{r} \;=\; \int [|\psi_G(\mathbf{r},t)|^2.\pi(\nabla\beta(\mathbf{r},t)=\mathbf{p_j})]d\mathbf{r}$$

Which acts as a specification of the *significance* of Born's postulate expressed in *dBB* terms. And furthermore inside the framework *[IQM-QM$_{HD}$-(dBB)]* the pertinence of this significance can be verified factually by use of the equivalence (31) constructed for a microstate $ms(unbound,1)_{G(nc)}$ and, alternatively, accordingly to the coding-postulate (35) or accordingly to the coding-postulate (20'), taking into account the posit P2 from *(7.III).2.3.8*.

[179] For instance, Louis de Broglie's analyses of various cases of interference states where he makes use of the concept of "variable quantum-mass" (cf. dB [1987] ***might guide inside QM2 toward a consensual and verifiable theory of teleportation***. This remark will be completed and specified in the subsequent and last chapter of this work.

### (8.III).2.2. The Principles of the Second Quantum Mechanics *QM2*

We now proceed to integrate the main lines of the global structure of the Second Quantum Mechanics *QM2* itself.

The representational features posited by definition, or reached by construction, or established by proofs, are those exposed in the body of the work. The postulates that have imposed themselves during the construction are now listed below separately.

Together,
[(the representational features posited by definition, or reached by construction, or established by proofs) *and* (the postulates listed below)]
are what we call "the principles of *QM2*".

#### *(8.III).2.2.1. The postulates of QM2*

***1. The postulate of global inclusion of IQM in QM2***

*IQM* as a whole is postulated inside *QM2* with the roles that it has effectively played throughout the chapters 5.II, 6.II and 7.III of this work; namely with the role of an explicit, pre-organized, methodological (instructional) and epistemological-operational structure that constrains a priori *all* the representations from *QM2*, their very existence as well as their meaning and their form. By this massive postulation *IQM* acquires inside *QM2* the status of a vast, incorporated, pre-*structured* ***datum*** that, mainly via comparisons and verdicts of acceptance or rejection, and *accordingly to its own rules*, permits inside *QM2*:
- to populate the ***TOTAL VOID*** left inside $QM_{HD}$ of an individual level of representation of the microstates;
- to complete or to modify as radically as necessary, in every semantic or representational respect.

The very possibility of existence of a structure like *IQM* and of its massive absorption into the theory that it contributes to construct, is *by itself* a procedural novelty.

***2. The modeling postulate***

The general *QM2*-model of a specimen $\sigma(ms_G)$ of the studied factual microstate $ms_G$ is defined by *the modelling-postulate [MP(*$ms_{G,cw}$*)* from *QM2*. It introduces de Broglie's model into *QM2* in a physically operational way.

***3. The guidance-postulate***

The definition (33) of de Broglie's guiding-law for the momentm-value $\boldsymbol{p_j(r,t)}$ is posited inside *QM2* to be valid for any sort of microstate.

***4. The postulate of choice of a Hilbert-Dirac mathematical representation of the results of quantum measurements***

The whole Hilbert-Dirac formalism from $QM_{HD}$ is massively imported in *QM2* via Gleason's theorem and Dirac's basic representation (23); its specific aim consists of the mathematical representation of the probabilistic results of the quantum-measurements.

***5. The expansion postulate (22)***

The expansion postulate (22) is posited as the foundation of Dirac's basic representation (23).

***6. The postulate of a generalized equation of the eigenket and eigenvalues of the momentum-observable***

The eigenvalues and eigenket of a generalized momentum-observable $\boldsymbol{P}_{vc\text{-}vd}$ are defined by the equations (38) and (39), respectively.

- In the case of an unbound microstate *without interference* the generalized momentum-observable $\boldsymbol{P}_{vc\text{-}vd}$ and its eigenket reduce to those from $QM_{HD}$ and they obey the non-commuting $QM_{HD}$-algebra.

- In the case of an unbound *interference*-microstate the generalized *QM2*-momentum-observable $\boldsymbol{P}_{vc\text{-}vd}$ does not belong to the non-commuting algebra of the $QM_{HD}$-observables and the eigenket of $\boldsymbol{P}_{vc\text{-}vd}$ consist of ***physical*** *superpositions of two or more plane-wav*es.

***7. The postulate of "beable observables"***

- Any observable value of a mechanical qualifying quantity $A(\boldsymbol{r},\boldsymbol{p})$ is a *"beable-observable value"* in the sense of defined in *(7.III).2.3.2.3.*

***8. The postulate of "value-conserving" or "value-destructive" coding-measurement-evolutions***

Any coding-measurement-evolution *MesA* considered inside *QM2* brings into consensual knowledge the *beable* value of the measured mechanical quantity *A* realized for the involved specimen $\sigma(ms_G)$ of the studied factual microstate $ms_G$ at the moment when that coding-measurement-evolution *MesA* has *begun*. The measurement-evolution *MesA* can conserve this initial beable value, or destroy it. In the first case it is a "*value-conserving*" measurement-evolution and it defines the measured beable value via the space-time positions of the observable parks. In the second case it is a "*value-destructive*" measurement-evolution and it defines the measured beable value via retroactive calculi involving the data registered in the course of the measurement-evolution.

***9. The coding-postulates***

- An unbound microstate ***without*** *interference* admits ***both*** the value-conserving coding-postulates (20') and the value-destructive coding-postulate (35), for ***any*** qualifying quantity $A(\boldsymbol{r},\boldsymbol{p})$ imported from the classical mechanics.

- An unbound microstate ***with*** *interference* admits ***exclusively*** the value-destructive coding-postulate (35) for ***any*** qualifying quantity $A(\boldsymbol{r},\boldsymbol{p})$ imported from the classical mechanics.

- Spin-measurements admit only a value-conserving coding-postulate of type (20').

***10. The postulate of the QM2-state-ket of an unbound interference microstate***

An unbound *interference*-microstate $ms(unbound,1)_{cG(qf)}$ is represented by a Hilbert-vector *of the form (40).*

***11. The postulate of the Schrödinger equation of evolution***

*- For unbound microstates the Schrödinger equation is replaced by the relation (31)* established by the argument that installs the assertion *Ass.2*. However the equation is conserved *as an element of comparison and a complement for calculi.*

- For *bound* microstates, *Schrödinger's equation* continues to play the role that it plays inside $QM_{HD}$ **[180]**.

---

[180] The state-ket and the eigen-ket assigned to a *bound* microstate stem from $QM_{HD}$ – via the Schrödinger equation of the problem – marked by highly degenerated formal and conceptual characters that inside *QM2* await future specifying improvements.

### *(8.III).2.2.2. Synthetic graphic representation of the contents of the postulates of QM2*

The whole probabilistic and semantic content of the state-ket of the studied microstate is incorporated in a corresponding probability-tree.

When the coding-postulate (20') is made use of the tree is a *several*-branches-tree of the basic type from the Fig. 2.

When the coding-postulate (35) is made use of the tree is a one-branch-tree of the type from the Fig.8[181].

In both cases the tree represents the *probability-laws* obeyed by results of measurements on the studied microstates; but not exclusively these laws themselves. In each considered case a probability-tree also specifies and characterizes *numerically* ***and*** *semantically* all the involved correlations as well as the whole operational genesis of these numerical and probabilistic outputs. Thereby the concept of probability-tree calls into explicit perception *a very far more defined and rich physically-operational and semantic structure than that fixed in a mathematical Kolmogorov probability-space alone*.

We go back to the contents from *(7.III).2.3.4* and we make use of the concept of probability-tree of an operation of generation for stabilizing a global and intuitive perception of the specificities of *QM2* with respect to $QM_{HD}$.

[181] The meta-probabilistic level of description can be explicitly expressed mathematically by use of Dirac's calculus of transformations: it possess its formal place inside the corresponding probability-tree. And in consequence of the *factual* genesis of the all the involved probability laws, the correlations also can be specified numerically.

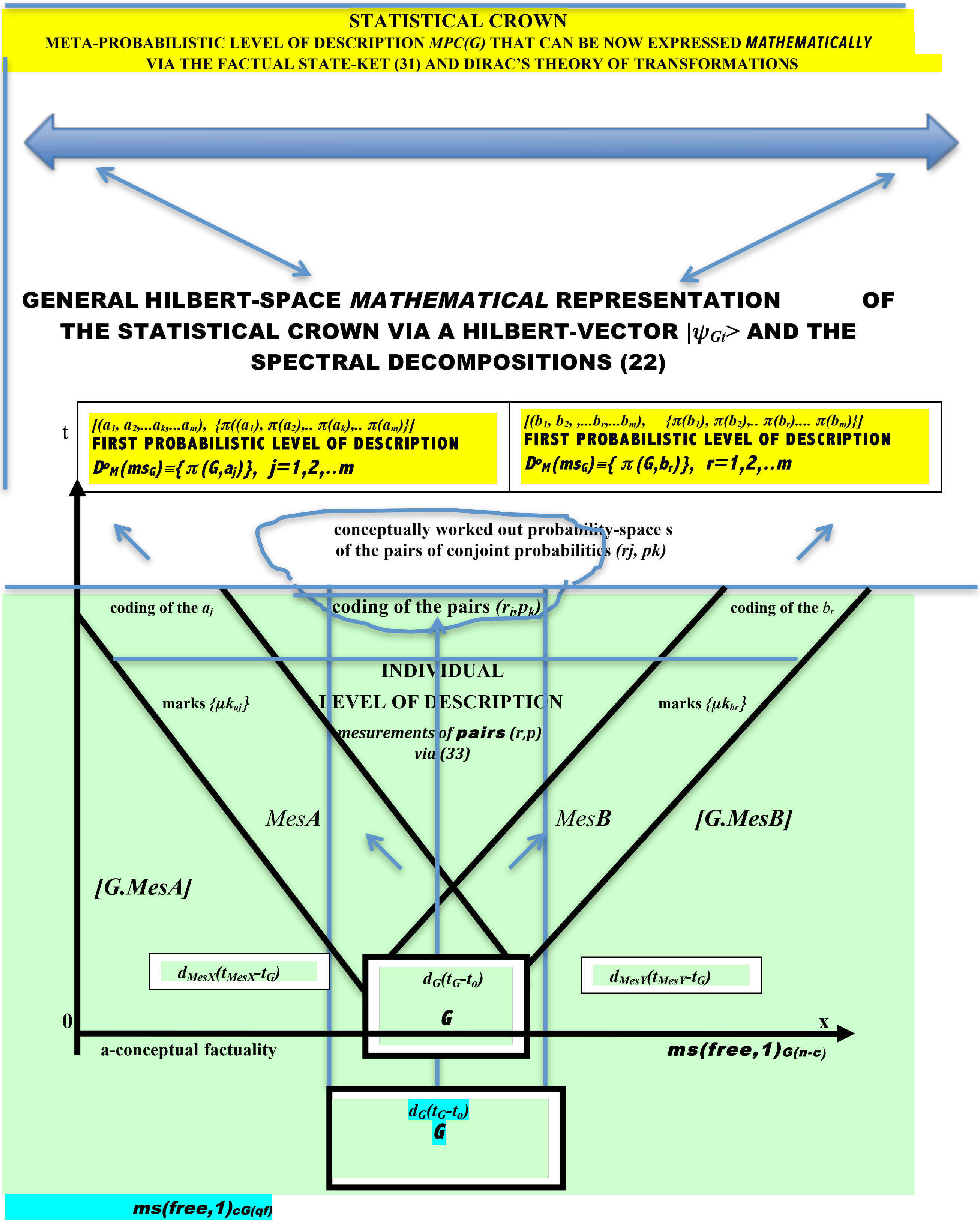

**Fig.9.** *The graphically superposed two basic sorts of QM2 probability-trees tied with the postulate 8 of two sorts of coding-measurement-evolutions.*

The *one-branch de Broglie-probability-tree* from the Fig.9 involves simultaneous coding-measurement-evolutions for ***r*** and ***p*** as posited in the coding-postulate (35); this unique branch is surmounted by the two experimentally found probabilistic crowns tied with the *QM2*-observables $\boldsymbol{P_{vd}}$ and ***R***. The content of these permits to then *calculate* also classically the predictive probability laws for any other qualifying quantity *A(**r,p**)* defined in *QM2*. According to *QM2* this one-branch factual probability-tree is constructible ***generally***, for any unbound microstate – whether a microstate $ms(free,1)_{G(nc))}$ or a microstate $ms(free,1)_{cG(qf)}$.

In the Fig.9 the one-branch-tree is placed *beneath* the multi-branches-probability-tree of a microstate $ms(free,1)_{G(nc))}$ that manifests a quasi-classical behaviour and is tied with the coding-postulate (20'). This is meant to recall that a de Broglie probability-tree can be rooted *deeper* inside the factual a-conceptual physical factuality than a branch-tree, so as to reach and incorporate also the most specifically "quantic" interference-microstates.

The formalism of the nowadays Quantum Mechanics $QM_{HD}$ takes correctly into consideration exclusively the superficial stratum of microstates with a quasi-classical behaviour; its concept of "observable" is ***falsely*** posited as a general absolute tied with a non-commuting algebra. For in fact any possibility of observation of a physical entity is relative to the characters of that entity *and* to the considered qualifying quantity (as well as, of course, to the available technical possibilities). And all these features have to be taken into account in the corresponding coding procedure that is asserted.

Inside *QM2* the quasi-classical branch-probability-trees tied with $QM_{HD}$ are conserved notwithstanding that ***they could be suppressed***. This choice permits to keep available a well-specified *formal place* where can be lodged the results of all the important pioneering works on quantum probabilities[182]; it thereby facilitates an explicit historical continuity and it acts as a reference for perceiving the changes involved by the passage from $QM_{HD}$ to *QM2*.

*Concerning QM2 the "interpretation problems" simply vanish together with their more or less fantastic solutions*[183].

The operational-formal-mathematical expression of *QM2* is subtended by a structure of meanings where the formal elements have grown in the same organic bottom-up succession as their operational and semantic contents. So all the involved features are tightly fit into one-another and have all adhered strictly to their only symbolic or also mathematical expression, as well as to the verbal one. Together, these meanings and their united forms of expression can be regarded as an application to the case of microstates, of a general "theory of the factual probabilities" expressed in Hilbert-Dirac terms on the basis of – directly – Gleason's theorem, that is valid for *any* given physical entity and *that summarizes the general Method of Relativized Conceptualization* (cf. the chapter 9.III).

So the view that has emerged via the construction of *QM2* is, both, very synthetic and very complex, and it exceeds *QM2*.

For its individual level of conceptualization via operations "*G*" and "*MesA*", this general view that encompasses the whole Method of Relativized Conceptualization accepts a mathematical expression accordingly to the theory of categories, as it has been sketched out in MMS [2002B].

As for the insertion of *QM2* into its host domains, it seems satisfactory. Via the *dBB* approach *QM2* is explicitly connected with the classical mechanics, electro-magnetism, optics;

---

182 This is in agreement with the important works by Mackey [1963], Gudder [1976], Suppes [1966], Beltrametti [1991] and many others no doubt.

183 Sometimes these solutions are even cosmically fantastic: According to the "parallel Universes"-interpretation each one humble act of measurement spells into physical existence a variant of the whole Universe; the initial phase of a small living-and-dead-cat is largely exceeded.

so practically with the whole classical Physics[184]. However in the present state of Physics these connections have not yet been explicitly re-considered and re-constructed in terms of clear specifications of the ***emergence, variation and disappearance of meanings when the order of magnitude of the space-time dimensions is changed*** (what – exactly – beyond its mathematical formula, does the concept of "mass" *mean* for a corpuscular-like de Broglie singularity? what are the differences with the case of an Einstein-mass? *In what a **sense** does the Einstein-mass depend on the velocity of the light ?!*).

This brings back to the "framed" question from the point *(7.III).2.3.4*.

In the present stage of Physics the above-mentioned connections are only potential places for later lodging in them clearly constructed assertions.

However here it is acquired already that:

> ***Via IQM and Louis de Broglie's model****, the roots of QM2 are* ***physically-operationally*** *implanted into the* ***a-conceptual*** *physical factuality, and they also are connectable – both conceptually and in a physically-operational way – to the mechanical theories and waves theories formulated before:* ***THIS IS A SPECIFICITY OF QM2****.*

For indeed, the preceding approaches, those of de Broglie and Bohm included, permit an ***only conceptual*** connection of $QM_{HD}$ with classical theories and with the a-conceptual physical factuality.

[184] The *dBB* approach is *not* the unique one that can insure the mentioned connections. Personally I know at least two other approaches, namely those by Eliade Stefanescu [2018] and by Unnikrishnan [2018]. In their essence these connections are the same as those perceived by de Broglie in his Thesis, of course, but each author perceives them with specific personal nuances that might lead to new consequences.

# CONCLUSION ON THE PART III

It is still very unusual nowadays to make explicit use of epistemological-conceptual-methodological constructions and arguments for generating scientific knowledge on physical entities. So many readers might still perceive this as a noxious intrusion of "philosophy" into Physics. But since now the essence of the elaboration of *QM2* is exposed, everyone who has read can weigh accordingly to his own criteria and standards whether the use of such constructions and arguments has flawed or enhanced the now current representation of the microstates.

In its present phase the representation of the microstates proposed here has only reached the status of a conceptual skeleton. But – as such – it might subsist and trigger the development of an acting theory. In the realm of conceptualizations this sort of evolution is current. A basic global view, the descriptive *aims*, and the method, can instil life into descripting matter, by interaction with other minds.

# INSTEAD OF A CENERAL CONCLUSION: FINAL CONSIDERATIONS

## SCHEMATIC INDICATION OF THE GENERAL METHOD OF RELATIVIZED CONCEPTUALIZATION

We have only very recently reached a stage when we try to create scientific knowledge *in the radical absence of the ancestral basis of possibility of* ***direct*** *human perception*. Thereby, surreptitiously, we have glided into a new sort of "cognitive situation". In order to acquire the capacity to deal efficiently with this basic change it has become imperative to clearly understand the notion itself of "cognitive situation", to define in what consists the passage from a *a given* cognitive situation to any *other* one when "scientific knowledge" on physical entities is researched, and to become aware of the general features of the effects of a passage from a given cognitive situation, to another one.

Below we give a very short indication of the content and structure of a general Method of Relativized Conceptualization where these requirements are fulfilled [185].

A fragment of physical reality that can play the role of *object-entity-to-be-described* has to first be *put* in this role, in this sense that it has to be endowed with some sort of *own* existence, *separately* from the whole of the physical reality. Some *operation of generation G* of this object-entity must act in the first place; for if not it is not possible to think or to speak of "*it*"; nor, a fortiori, to deal with "it" scientifically, i.e. in a way that is communicable, consensual, and can lead to predictions that can be verified.

In the current life our biological sensorial apparatuses achieve this individuation in a quasi reflex way and this conveys a deeply set ***false*** intuition that the object-entities-to-be-described pre-exist around us such as we perceive them, waiting to be observed and described by us, i.e. to be *"discovered"* by us *such as they truly are in themselves*. But in fact these object-entities-to-be-described are cut out by our sensorial apparatuses, either physically or at least only mentally; they are delimited from inside the whole of the physical reality by our ways of thinking and our reflex biological behaviours, accordingly to the own structure and functioning of *these* such as they are encrypted in our bodies. And as soon as the output of these does emerges with a blurred contour – as it happens in the case of limiting concepts like "the Universe", "infinite", etc. – the possibility of generating by their use scientific knowledge, consensual, predictive and verifiable, encounters "problems" that manifest that the frontier between science and metaphysics has been touched and stops us.

In short, the operation of generation *G* is what initially achieves for an entity-to-be-studied a separated existence that permits to then qualify this entity scientifically thus bringing it into consensual knowledge. The operation *G* can be purely factual, or factual-conceptual, or purely mental; it can consist of only a selection, or it can be entirely creative, or it can consist of some combination of these possibilities; it can be achieved exclusively by the biological apparatuses from the observer's body, or by the use of also prolongations of these by "instruments". But it ***always*** *exists*, it always ***is*** performed by the considered human observer, even if only in a degenerate and non-conscious way that realizes it in the same time with a qualification. And nevertheless, up to now, it has not been assigned systematically an explicit and ***specific*** location inside the methodology of the sciences of matter.

[185] An ***extended summary*** of only the kernel of *MRC* is reproduced in the Appendix1. The method itself has been published in three successive works of increasing degree of achievement, namely MMS [2002A] (English), [2002B] (English), and [2006] (French); a ***2019-version English version*** will very soon be available on ResearchGate.

By itself, the general concept of operation $G$ of generation, whether factual *or only mental*, is likely to remain as the ***main*** contribution of *MRC* to our awareness of the *genesis* of any human consensual knowledge.

The object-entity-to-be-described generated by $G$ can be denoted by $œ_G$.

Consider now the qualifying-structure. Let us call it *a view*. We denote this by $V$ and we posit that it possesses the formal structure of a union of *aspect-views*, $V=\cup V_g$ with $g=1, 2,...h...H$, where the index $g$ designates a qualifying *aspect* (mental or conceptual-physical-operational). We can conceive each qualifying aspect $V_g$ as a *semantic dimension of g-qualification* that carries on it a finite number $K(g)$ of freely chosen *values gk* of the aspect $g$: $g.1(g), g.2(g),....g.k(g)...g.K(g)$. So $V_g$ acts as a ground, a terrain for placing upon it the values $g.k(g)$ that, a priori, we decide to take into consideration via some well defined operation $Mes.g$ of $g$-measurement, each realization of which produces *a group of marks* $\{\mu\}_g$ *that codes in a* ***definite*** *way for one and only one value* $k(g)$.

Notice that this takes us very far away from the classical notion of "predicate" in the sense of the grammars or in the sense of logic.

So any description involves an operation of generation $G$ that generates an object-entity-to-be-described $œ_G$, and a qualifying view $V_g$ that, acting on $œ_G$ via its aspect-views $V_g$ qualifies $œ_G$ in terms of qualifying values $k(g)$, whereby $œ_G$ becomes perceptible by the observers, which means that now – ***not before*** – some knowledge on $œ_G$ can emerge in the observers' minds, that might be developed and organized in consensual, predictive and verifiable terms:

> ***Scientific knowledge is qualification. And any qualification emerges unavoidably relative to WHAT has been qualified, and HOW***.

Nothing else than consensual, predictive and and verifiable qualification can generate scientific knowledge: scientific knowledge just ***is this***, by definition. A qualification that is not marked by characters relative to the object-entity $œ_G$ that has been qualified and so to also the operation of generation $G$ that has produced $œ_G$ (space-time extension, sorts of uniformity or non-uniformity, etc.) as well as to the view $V$ that has qualified $œ_G$, *cannot be* ***conceived***. That is why the structure of words "the science leads us toward knowledge of how the physical reality truly is ***in itself***" (i.e. in absence of any qualification), is just meaningless.

At this crucial point we stop for a moment in order to formulate the following *postulate of "**minimal** realism"* that lies at the foundation of the Method of Relativized Conceptualization:

> We postulate the ***EXISTENCE*** of a physical reality independently of any scientific knowledge of it; but ***merely*** *the existence*, separated of any knowledge on how this existence is.

***This is the "postulate of minimal realism"***.

The distinction introduced by this postulate might seem trivial, but in fact its consequences mark the scientific thought in all its fibres. The physicists presuppose indeed quite currently that a *physical reality* exists independently of our *knowledge* of it (if they did not they would be solipsists, and a physicist's mind is immune to solipsism). But they tend to *also* posit that the scientific knowledge *is knowledge of how the reality truly is "in itself"*. And this is meaningless and generates endless confusion because any knowledge is relative to the ways of qualifying, while the assertion of mere existence does not. So the assertion of *existence* of a physical reality should be radically distinguished from the concept of *scientific* *knowledge* of this physical reality, in order to stay free to *fully* perceive all the unavoidable and detailed dependences of knowledge, on the cognitive situation [186].

Now we continue.

So any *description* involves a pair *(G,V)*: this is called the *epistemic referential* of the considered description. It brings in a trio ($G,œ_G,V$) where *(G,V)* works on $œ_G$ and generates *relative descriptions* of it.

The complete description itself of $œ_G$ via $V=\cup V_g$ that is realized via the epistemic referential *(G,V)* is symbolized by $D/G,œ_G,V/$ where "*D*" denotes the whole structure of *g.k(g)*-values of *H* sorts, obtained in the representation space of the view $V=\cup V_g$ with *g=1, 2,...h...H* via long sequences of many repetitions of each qualifying-succession $[G.MesV_g]$ where "$MesV_g$" is called an act of measurement on $œ_G$ of the aspect *g*. Each aspect-view $V_g$ from $V=\cup V_g$ produces a partial description of $œ_G$, $D_g/G,œ_G,V_g/$ that is relative to *G* and to $V_g$ alone.

$D/G,œ_G,V/$ is called *a relative description of* $œ_G$ *with respect to the qualifying view V* and it consists of the union $D=\cup D_g$ of the partial descriptions $D_g$. The descriptions $D_g/G,œ_G,V_g/$ and $D/G,œ_G,V/$ are placed on a *meta*-level of representation with respect to the trio of symbols ($G,œ_G,V$)).

The sort of "ground", of pre-existing "reality" on which *G* acts in order to generate the object-entity-to-be-described $œ_G$, can consist of physical factuality, or of a domain of conceptual constructs achieved before. Inside *MRC*, by definition, what is called "reality" and is denoted "*R*" consists at any given time of the reservoir of "grounds" $R_G$ that, at that time, are available to human observer-conceivers for acting on such a "ground" by a corresponding operation of generation *G* in order to construct a relative description $D/G,œ_G,V/$.

In the particular cases in which *G* works directly upon a ground $R_G$ that consists of still *a-conceptual physical factuality* (i.e. is not itself some previously accomplished description $D/G,œ_G,V/$), the notations that are made use of can be marked by an index "$_o$" (we can write $R_{Go}$, $G^o$, $(G_o,V)$, $D^o/G^o,(œ_{Go}),V/$, and $D^o/G^o,(œ_{Go}),V/$ is called a "primordial transferred description". When *G* is *not* "a $G^o$", the corresponding "ground" is already conceptualized, which means that it consists of previously established scientific knowledge, so of previously achieved scientific relative descriptions. The category of primordial transferred description has been characterized in detail inside *IQM*, for the particular case of the factual microstates, and there it appeared clearly to what a degree it possesses non-classical features.

> This category of the primordial transferred descriptions $D^o/G^o,(œ_{Go}),V/$, that is ***a universal and fundamental category of the human conceptualization***, has been brought forth inside *MRC* for the very first time.

It constitutes the universal first layer of the human conceptualization.

The passage from a primordial transferred description $D^o/G^o,(œ_{Go}),V^o/$, to a description $D/G,œ_G,V/$ in the general sense defined above, is regulated inside *MRC* by a procedure called "intrinsic meta-conceptualization" [187]. Thereby it gets involved in an unlimited number of unlimited chains of relativized descriptions that form a second and unlimited layer of conceptualization.

The intrinsic meta-conceptualizations constitute the *MRC*-expression of what is called "the quantum-classic cut".

---

[186] Already since 1788 Kant has isolated this posit from knowledge and has confined it inside metaphysics. And it is rare nowadays to find a philosopher who does not agree with Kant that "the thing in itself" cannot be known (in the sense of Science). Nevertheless many physicists – if not practically all – still believe that Science brings us asymptotically toward knowledge of "how the physical reality truly is".

[187] This – in the terms made use of inside *IQM* and during the construction of *QM2* – amounts to the definition of a model and the insertion of this model into the operation of generation *G*. Whereby the results of the operation *G* acquire also a conceptual meaning associated to the physically-operational one.

*All the chains of consensual, predictive and verifiable relativized descriptions of physical entities are rooted in the a-conceptual physical factuality via primordial transferred descriptions*.

They are developed accordingly to detailed and strict rules subjected to the current syllogistic. But the descriptional ***aims*** that define the directions of conceptualization are left fully free, and they are expressed by the choices of the epistemic referentials *(G,V)* that are made use of.

Each epistemic referential *(G,V)* introduced a descriptional cell that is finite by construction; once completed, it is *closed*, separated from any other descriptional cell accordingly to a *postulate of separation PS.* But it can be continued in other descriptional cells via meta-descriptions – with respect to it – of several distinct and strictly defined sorts.

Thereby are generated *descriptionbal chains* that meet inside "descriptional *knots*" mutually characterized according to whether they involve only operations or generation *G*, or only views *V*, or only aspects $V_g$, or some combination of such descriptional elements. From any knot can stem other separate chains. So, progressively, is generated an evolutive complex network of radically relativized scientific knowledge. This network englobes *systems of descriptions*, formal systems – logical or logical-mathematical –, or probabilistic descriptional structures, or informational-probabilistic structures, or Physical Theories.

This network constitutes at any time the volume of the conceptualized up to that time and such as it stands at that time.

Globally and such as it stands at the considered time, the conceptualized up to that time is absorbed in the *MRC* concept of "*evolving reality*" denoted *R* (by definition, *R* includes also the social reality and the artistic reality, in agreement with Karl Popper's view, but *these do not belong to MRC*). And *R* includes also quintessentially the human "consciousness-functionings" *CF* that interact with the other contents of *R*.

The scientific knowledge constructed accordingly to *MRC* ***excludes by construction any false problems and paradoxes***.

It equally excludes by construction any false "impossibility" (like the unification between the modern microphysics and the Einstein-Relativity) as well as any false "possibility" (like knowledge of how this or that physical entity truly is in itself).

# EMERGENCE INSIDE *MRC* OF A GENERAL THEORY OF FACTUAL PROBABILITIES BELONGING TO THE MATHEMATICAL PHYSICS

*MRC* has been constructed in order to generate an explicit set of necessary and sufficient representational-operational procedures for generating scientific knowledge on physical entities. This aim simply imposes itself at a time when the humans want to generate scientific knowledge on entities that can *entirely and quintessentially escape **direct** human observation* (which can happen in many different senses, for instance, in consequence of the distinction "inside-outside" the studied entity, or in consequence of the order of magnitude of the involved space-time dimensions, or by a space-time location of the studied entity that is different from that of the involved human observer-conceiver, etc.).

But for the present author the main aim for which *MRC* has been constructed has been to reach full intelligibility of the modern theory of the microstates.

Inside *MRC* there emerged naturally – and unavoidably in a certain sense – a basic concept of "probability-tree of an operation of generation *G* of ***any** physical entity* $œ_G$".

Inside the Infra-[Quantum-Mechanics] *IQM*, the ***general*** *MRC*-concept of probability-tree of an operation of generation *G* of any physical entity $œ_G$ is applied to the particular case in which $œ_G$ is a "factual microstate" $ms_G$. And for this particular case the process of elaboration of *QM2* revealed – via the assertion *Ass.2* and its constructive proof – the possibility to express mathematically the probabilistic and the semantic features from a probability-tree, via the Hilbert-Dirac formalism from $QM_{HD}$ applied ***directly** to results of measurements upon specimens of the studied microstate* (7.III).2, pp. 165-205). Thereby the necessity of Schrödinger's equation of evolution is dissolved: this necessity had been imposed by the historical top-down progression from the classical physics toward the micro-physics because in this progression the statistical level of conceptualization of the microscopic entities was encountered first. But in a bottom-up approach like that practised in this work this necessity not only vanishes, but it transmutes into a handicap, in consequence of the difficulties to specify the initial state-ket. So inside *QM2*, on the basis of Gleason's 1957-theorem – that is established for ***any*** "probability law" no matter whether its assertion is obtained by mathematical means or by a conceptual-factual method – the results of quantum-measurements *MesA* (expressed as values $a_j$ of the qualifying quantity *A* via an appropriate coding-postulate), are lodged directly into the mathematical ***form*** of an expansion with respect to *A* of the Hilbert-Dirac state-ket of the studied microstate.

*And this brings into evidence that the assertion Ass.2 can immediately be (retro)-extended to the general MRC concept of probability-tree of an operation of generation G of **any** physical entity* $œ_G$.

Thereby, in one swift and massive blow of feed-back, the probability-trees *from the whole Method of Relativized Conceptualization* – that are all constructed factually and bottom-up as it has been done for the elaboration of *QM*2 – are invested by a Hilbert-Dirac type of mathematical expression[188]. So:

*MRC becomes a **mathematized** general scientific Method of Relativized Conceptualization of the physical entities.*

This particular receptivity of *MRC* – that concerns physical entities of any sort – with respect to a procedure from the construction of *QM2* that concerns only microstates, stems from the fact that ***the deep semantic nature of MRC is probabilistic***. *MRC* as a whole is a probabilistic approach, because it is rooted in the unlimited and uncontrollable flux of the physical factuality out of which an observational stability can *never* be realized in an absolute and definitive sense, for "all" the possible aspects (here "all" is an illusory concept because the content of an open set cannot be entirely known, nor defined in the strict sense). When some sort of stability is reached, this – in all the experimental methods *of the sciences* as well as throughout the techniques – happens only with respect to particular features, by abstractions, approximately, for limited space-time domains, and via deliberate procedures invented for utilitarian aims (MMS [2006], pp. 61-86). In consequence of this the basic *MRC*-concept of relative description $D/G,œ_G,V_g/$, it is *explicitly* probabilistic by construction. The case of an "individual" relative description is defined as a particular possibility that can be perceived only a posteriori and only relatively to particular aspect-views; namely, this case is realized iff – *with respect to* a given aspect-view $V_g$ – a big but pre-established *finite* number *N* of repetitions of a very long *but finite* sequence of *N'* repetitions of the involved measurement-succession $[G.V_g]$, is found to have stably produced a ***statistical distribution*** *without dispersion*, i.e. that consists of exclusively one "same" value *k(g)* of the involved aspect *g*, *with respect to* the a priori specified references of "precision" of the involved qualifications (via tests or measurements).

---

[188] The equations for eigenket and eigenvalues can be dispensed with. A mere label of the value $a_j$ of *A* suffices for making use of Gleason's theorem *and* for benefiting of Dirac's calculus of transformations as a calculus with semantic contents.

It is in this genetically receptive context that the approach developed in the process of construction of *QM2* – from *IQM that itself is constructed inside MRC* and from Gleason's theorem – retro-induces into *MRC* a new sort of theory of probabilities that is essentially *different* from the mathematical theory of Kolmogorov[189].

And that what emerges inside *MRC* is *a general Theory of Factual Probabilities*, *constructed bottom-up*, *that belongs to the Mathematical Physics*, where:

***- Each cell $D/G,œ_G,V/$, of achieved conceptualization defined by a given epistemic referential (G,V) with $V=\cup V_g$ and g=1, 2,…h...H consists of the whole probability-tree of a physically well-specified operation of generation G.***

*- Each elementary event as well as each event is* ***a relative description $D/G,œ_G,V_g/$ of the studied physical entity $œ_G$ generated by the operation of generation G tied with $œ_G$***; and *all* these descriptions are qualifications relative to one or more aspect-views $V_g$ from *V*, **of the one physical entity $œ_G$ generated by the operation of generation *G***.

> ***This*** *is what inside one probability-tree triggers via the involved G an a priori posit of multi-faced* ***semantic*** *unity between all the involved elementary-events and events,* ***even before*** *the meaning of a given individual elementary event or event is entirely specified by the postulation of a model of the studied physical entity $œ_G$* [190].

- The probability measures from such a tree-like cell are expressed by a Hilbert-ket.

- The probabilistic correlations that connect elementary events and events from different branches of an *MRC*-probability-tree are expressed accordingly to the general functional relations (11), (11') from (3.I).1.3, in terms of applications of Dirac's calculus of transformations of eigenket, such as this is imported in *QM2* from $QM_{HD}$ (cf. this work, Intermezzo just after *(7.III).2.2.4*). Thereby the whole tree-like structure generated by the considered epistemic referential *(G,V)* is immersed in a "semantically nuanced medium" where the inner figures are defined by the general model posited for $œ_G$ and the semantic nuances and intensities are defined by Dirac's calculus of transformations.

---

[189] Entirely contained in "probability spaces" where a general concept of probability measure *in the mathematical sense* is asserted upon the sets of events from an algebra on a set of elementary events, *all these events and elementary events being just* ***symbols*** *of physical events that are not factually realized*.

[190] Inside *MRC* it is explicitly shown that the tree-like representation of the factual probability laws tied with a given operation of generation is endowed with universality and constitutes a feature of deep unity between *relativized and extended versions of the* ***three*** *most basic sorts of human conceptualization, the logical conceptualization, the probabilistic one,* ***and*** *Shannon's informational conceptualization*. The most complete account of this unity can be found in (MMS [2006].) (Less exhaustive accounts can be found in MMS [2002A], [2002B], [2014]).

- As for the *connections between* ***different*** *probability-trees*, in principle inside *MRC* they can be defined in detail on the basis of the rules for the construction of the various sorts of possible meta-descriptions. These entail *chain*s of tree-like probabilistic cells connected in various descriptional knots; and the roots of each chain consist of fragments of physical reality extracted from the whole of the a-conceptual physical reality by "basic" operations of generation denoted $G^o$, tied with *transferred* descriptions $D^o/G^o,œ^o_G, V^o_g/$, like in the case of the factual microstates. *So each one among all the elementary events and events that are involved carries a* ***meaning*** *defined in* ***physically-operational-and-conceptual*** *terms.*

> From the construction inside it, of the Second Quantum Mechanics *QM2*, the general Method of Relativized Conceptualization *MRC* re-emerges as a Formal System for the construction of scientific knowledge on physical entities, that *contains* a general Theory of the Factual Probabilities of the results of measurements on any physical entity, expressed mathematically in Hilbert-Dirac terms, and that includes factual realizations of Kolmogorov probability-spaces.

Any Physical Theory proposes acts of measurement *MesA* on the entity-to-be-studied $œ_G$, of the qualifying quantities *A* that are introduced by it. This is a general aspect of any Physical Theory. *MRC* endows this general aspect with an explicit and uniform *MRC*-definition in terms of tree-like probability-cells endowed with a Hilbert-Dirac mathematical representation.

The basic *MRC*-set of moulds that defines any tree-like probabilistic cell is indicated below with notations that generalize those introduced in *IQM* for the particular case of the microstates:

$$G \leftrightarrow œ_G, \qquad (G,V), \qquad [G.Mes.g] \rightarrow \{\{\mu_{k(g)}\}_g,\ \forall V_g \in V\}\},$$

$$D_{\mathbf{g}}/G,œ_G,V_g/ \equiv \{(\varepsilon,\delta,N_0)\text{-}\pi(k(g)), \forall k(g)\}_G,\ g \text{ fixed}, \quad D_V/G,œ_G,V_g/ \equiv \{\{(\varepsilon,\delta,N_0)\text{-}\pi(k(g)\}), \forall\, k(g)\}\}_G, \forall g \in V,$$

$$\pi(\,k(g)) = \mathbf{F}_{\mathbf{k(g),g'}}\{\pi(_{k'(g')}),\ \forall k'(g')\}_G,\ \forall g, \forall g' \qquad \mathbf{F}_{\mathbf{gg'}}(G) = \{\mathbf{F}_{\mathbf{k(g),g'}}\{\pi(k(g)), \forall k(g)\}_G, \forall gg' \in V,$$

This basic set of conceptual moulds is always present – whether hidden or expressed – and it always *works*.

As a whole now, a mathematical physical theory is never a closed formal system that permits to predict and to verify with systematic certainty. It does make use of closed formal systems (the current syllogistic and mathematical structures), but only for inner deductive transportations of delimited length of the consequences of this or that particular assertion; whereas the moulds for the inputs in a physical theory are always ***open*** *containers*; they accept – they require – to be charged with a priori ill-known fragments drawn from the changing flux of the a-conceptual purely physical, or the pre-conceptualized, reality [191]**.** These fragments are ill-known, since the aim is precisely to study them. And what this can reveal is a priori unlimited. While the operations of generation *G* that perform the charging can be conceived in only partial and finite terms, so that their repetitions cannot a priori exclude any fluctuation in the result. And so – strictly – any scientific prediction can be inserted only in a basically probabilistic approach (just as it is done inside *IQM* for the particular case of the microstates as well as inside *MRC* itself).

[191] The process of construction of *QM2* has permitted to already become aware of the importance of the *way* in which contents are poured into a mathematical instrument intended to work upon these. If this way is not fully adequate, the mathematical tools, and the semantic contents, that are supposed to "fit" with one another, in fact remain largely disconnected. They simply do not.......'*make sense*' together. This is so because a mathematical system is a formal system, so a ***closed delimited*** system. The delimitation is necessary for rigorous deduction of strictly defined results. And *the way of delimiting via axioms, definitions, symbols, well-formed expressions, laws of transformation of a well-formed expression – involves already an* ***own*** *semantic* (This is often occulted, if not even explicitly negated). This own abstract semantic of pre-constructed mathematical tools that is revealed deductively in terms of possibility of strictly defined conclusions, has to be *compatible* with the somehow *mathematically expressed* factual semantic carried by the physical facts, because otherwise the mathematical system cannot work on the data, they cannot "see" and "catch" them; the way in which the mathematical tools can be made use of should permit to *express* the whole physical semantic that one wants to represent by the help of these tools. But the factual semantic is ***open and unlimited***. It is unavoidably *relative* to the indefinitely numerous, un-named, unknown and *un-**conceived*** presences from the moreover evolving material ground on which we work operationally for generating scientific knowledge on physical entities. So, in order to be controllable, the required mutual compatibility between the closed and abstract semantic of the mathematical tools, and the *relevant* open and factual semantic that has to be poured into the mathematical tools, must somehow be methodologically restricted to a finite and small number of elements filtered out from the unspeakable bulk of potentialities and expressed in a sort of common language. As soon as this is not succeeded – or even not tried – there can emerge factual failures and paradoxes, unintelligibility, and "interpretation problems" that attract more or less fantastic "solutions".

Now, *MRC* itself establishes exclusively the general (meta)rules for consensual construction of *any* sort of piece of predictive and verifiable knowledge on physical entities. But beside its general aspects, any Physical Theory from the Mathematical Physics necessarily contains also ***models*** of the entities-to-be-studied that it introduces, and ***coding****-measurement postulates* that – quintessentially – depend on the considered entity, on its sort of state, and on the considered qualifying quantity. These models and coding-measurement-postulates are defined inside a mathematical formalism that is specific of the considered theory; obviously, a unique formalism for describing microscopic entities and cosmic ones, transports of heat or global mechanical displacements of "mobiles", etc., is simply not conceivable. So each theory from the Mathematical Physics introduces its own specific mathematical formalism that varies freely by passage from a given class of epistemic referentials, to another one.

So of what could consist a unification of Physics?

---

And when this can happen, then it does happen indeed. And if a solution to the problems that emerge – no matter how fantastic – *is itself endowed with inner syntactic coherence (is, for instance, a sort of coherent fairy-tale)*, it tends to be accepted and to last because of its own inner coherence, that focalises upon it the critical attention.

On the other hand, when inside a mathematical theory of a domain of physical entities there subsists some semantic *in*-adequacy between the considered domain of facts and their mathematical representation, *the physicists always **feel** this*. But they do not identify the source. And when this happens the physicist's minds – nowadays at least – secrete a propensity to consider the mathematical formalism as if it were itself a physical reality of some superior essence, out of reach and immutable like a galaxy: this absolves from trying to change it. *The formalism is confounded with a factual datum and thereby it slips out of conceptual control.*

Such reactions are strongly favoured by the Platonic conception of mathematics according to which mathematics is not a human construct; it "is" there, *outside*, in the air of "ideas" and inside the things "themselves" (a sort of mathematical realism). And so inside many minds the formalism undergoes a sort of divinisation. From a tool it is transmuted into an idol. This process, moreover, secretes a superposed penchant to *generalize to **everything*** the ill understood and divinised mathematical representation. This is what happens now with Einstein's Relativities; they become an obvious ***GENERAL** must*. Which entails an artificial problem that leads to arbitrary, long and very difficult elaborations that in fact are devoid of any sort of necessity.

Nowadays the forces of inertial ways of thinking are still huge.

## ON A METHODOLOGICAL UNIFICATION OF PHYSICS

In order to illustrate this general question we consider the particular case of *QM2* face-to-face with *Einstein Relativity*.

Via *IQM* the second quantum mechanics *QM2* can be connected explicitly with any domain of Physics. The fact that *IQM* has drawn its methodological structure from *MRC* and has instilled it into *QM2* induces paths for such connections. And in its turn the existence of such paths, once identified, induces possibility of *ruled comparisons* throughout the whole domain of the physical sciences. Thereby emerges the possibility of *a "methodological" unification* of all the various approaches for generating consensual, predictive and verifiable knowledge on physical entities, along paths determined by *MRC* between any two distinct cognitive situations each one of which is associated with mathematical representational features implied by the corresponding epistemic referential *(G,V)*.

When the assumed order of magnitude of the assumed space-time dimensions changes, the whole cognitive situation changes, so also sort of epistemic referential required for a given cognitive aim; while the chosen epistemic referential determines what is described, and with respect to which aspect-views, and these elements mark the whole content of the elaborated description in ways that are relative to them [192]. The cellular, invariant descriptional structure $D/G,œ_G,V/$ is, both, deeply contextual and strictly ruled, and this entails unlimited strategic adaptability.

**1.** *Consider QM2.*

Let $D/G,ms_G,V/$ be a description of an unbound factual microstate-to-be-studied. So we have $œ_G \equiv$ *"a factual unbound microstate* $ms_G$". In this case:

- The human observer ***cannot avoid*** *the necessity to first generate* ***factually*** specimens $\sigma(ms_G)$ of $ms_G$, by realizations of a repeatable operation of generation *G* that makes such specimens available in the role of object-entity-to-be-qualified as many times as this is desired (cf. *MD* and (1) in (1.I).1.3).

- Each time that this has been achieved by one given realization of *G*, the generated specimen $\sigma(ms_G)$ can be qualified by some given act of measurement *Mes.A* of a qualifying quantity *A*, where $A\equiv V_g$ consists of a whole qualifying structure in the sense of (2) and *(1.I).2.2.1* for which – according to previously established conceptualization – it is possible to specify *an "appropriate"* ***coding-measurement-evolution Mes.A***. This is a very basic and very restrictive sine qua non condition that lies in the very heart of the process of conceptualization and requires a ***model*** of "$\sigma(ms_G)$". So a whole coding-measurement ***successio****n [G.Mes.g]* must be achieved, not only a coding-***measurement***-evolution *Mes.A* alone.

- But in general the registration of publicly observable marks that closes one measurement-evolution *Mes.A* destroys the state of the involved specimen $\sigma(ms_G)$ of the microstate $ms_G$ to be-studied. So in order to realize a factual "verification" of any sort it is necessary to ***repeat*** whole successions *[G.Mes.A]*; while *repetitions of G – as well as repetitions of Mes.A –* involve the a-conceptual factual reality *R* (in the sense of *MRC*) in ways that cannot be controlled so as to be able to assert "a strict and certain absence of any change in *R*", which obviously can neither be defined nor even conceived. ***To assume the contrary has no definable meaning and it leads to an infinite regression***.

Which entails the essential conclusion that:

[192] Cf. Appendix1.

**-** *There is no basis whatever for assuming that when the succession [G.MesA] is repeated the results will be identical for all the realizations of G and of $R_G$ and for any qualifying structure $V_g$. This also is neither* ***conceivable*** *nor – a fortiori – factually realizable.*

So – unavoidably – the operations *G* and *MesA* induce into the qualifications of the microstate-to-be-studied $ms_G$ an *un*-certain statistical-probabilistic character that *cannot* be eliminated on the basis of considerations of "imprecisions in the acts of generation". Indeed, *whatever* be the precision with respect to the defined parameters that determine "*G*" and "*MesA*", *anyhow* a statistical character stems from the unavoidable and deep conceptual opposition between the finiteness of the possible conceptual and factual definability of "*G*" and "*MesA*", and on the other hand the unlimited and technically uncontrollable character of the contents of the a-conceptual flux of the physical reality *R* that is always involved by these operations at the space-time places where they are realized: one is brutally and massively faced with *"essential indeterminism"* ***of operational control, and so, of constructible consensual knowledge.***

**-** What is called "the observed values" consists exclusively of registered observable marks that can wait an arbitrarily long time to be read by the human observer (Wigner's "friend") and that are devoid of any own qualia directly associable with the studied microstate.

Given

- the cognitive situation indicated above,
- the global aim of the description,
- the way in which this aim is realized,

the ***unique*** *condition of consensus* concerning a predictive and verifiable invariant that can be required is statistical, namely the following one. The *probabilistic distributions* of the groups of observable marks each one of which codes for one observed value $a_j \equiv k_j(g)$ of the measured quantity *A* – that are obtained separately in the proper systems of reference of various Laboratories, via the same procedure applied in each one and to specimens of the same sort of microstate – must be all "identical" *inside pre-specified limits of variability*: they can consist ***only*** of *factual* $(\varepsilon,\delta,N_o)$*-probability laws*. No other sort of concept of numerical invariants is defined.

For this condition to be fulfilled, *no light signals have to be made use of.* So a fortiori the velocity of such signals is *irrelevant*, like also the state of movement of the various observers and the Lorentz transformations of the space-time coordinates of these by passage from one of the observers to another one. *All these aspects simply play no role whatever in the process.*

**2.** *Consider now RTG.*

- This is a theory of macroscopic and/or cosmic object-entities that in general *cannot be generated* ***factually*** *by the human actors*. These object-entities, no matter whether they consist of "objects" or "events", are posited to be *independent* of the cognitive actions realized for generating knowledge on them.

Inside the nowadays Physics the concept of operation of generation of the object-entity-to-be-studied is not even *conceived* with respect to such object-entities. We are in presence of a totally classical conceptualization that works in terms of "subjects" and "predicates" that all just "are there" in the Platonian sense, like in the current languages, in grammars, in the classical logic.

- The operations of generation *G* involved in *RTG* – when inside *MRC* this is researched for methodological reasons – are found to be in general *mental-perceptive*, so passive with respect to the studied entity. They consist of focusing the attention upon a certain sensorial representation that is posited to symbolize a conceptual one, that, together, isolate a corresponding physical object-entity-to-be-studied $œ_G$ from inside the whole of the physical reality, with respect to some definite qualifying aspects, *so with respect to a "View"* $V=\cup V_g$ *in the sense of MCR* (again just like in the classical languages, grammars, and logic).

The *basic* operations of generation *G* from *RTG* are realized via *classical definitions* that introduce the object-entity-to-be-studied *via conceptual-verbal qualifications*, just like in the current dictionaries.

Thereby they belong to a particular sort of operations of generation, *a degenerate sort of operations of generation that inside MRC are called "(generators G)-(of-a-view V)"* and are denoted *G(V)*.

- The object-entities-to-be-studied $œ_G$ that are produced by this sort of operations of generation absorb inside themselves their whole *MRC*-description $D/G,œ_G,V/$; they emerge classically, qualified by their generation itself that is their verbal-conceptual definition**.** The separate relativities of the descriptions $D/G,œ_G,V/$ of these object-entities-to-be-studied $œ_G$, to the two elements *G* and *V* from their epistemic referential *(G,V)*, become mutually ***in***-distinct inside *G(V)*, the separate relativities from $D/G,œ_G,V/$ coalesce via *G(V)* and then they disappear together from the classical domain of mental perceptibility[193]. Finally *RTG* writes down only trajectories of "mobiles", with respect to a representation of space-time as a whole that – by construction – is *such* that the geodesic character of each trajectory be a descriptional invariant with respect to passage from any one among the space-time systems of reference of the considered observers, to any another one.

- From the descriptional structure established in *RTG*, are then drawn by *deduction* certain events translated in terms of publicly observable signs and marks that in general cannot be produced deliberately, so a fortiori they cannot be repeated deliberately: ***coding-measurement-evolutions cannot be freely realized and repeated***. They can only be predicted to happen in *a publicly observable way* in certain definite utterly singular conditions characterized in terms of values *g.k(g)* of some aspect-views $V_g$ from a view $V=\cup V_g$ from an appropriate epistemic referential.

- The global *aim* of the considered theory is to verify that *at least* ***one*** such deductively asserted and future event that is publicly observable but cannot be repeated deliberately, does happen indeed, and such as it has been predicted.

- The measurements that are involved *for this aim* must establish values *g.k(g)* of aspect-views $V_g$ from $V=\cup V_g$. Usually these measurements concern *very big mobiles*. These are conceived to be mutually *totally separated from one another* – "truly", "in themselves", in an absolute and impossible sense. This conception stems from the fact that, *via the human biological capacities of perceiving and comparing* they can appear to "be" very distant form one another and from the human observers (the *continuity* of the physical substance into which these mobiles are immersed and inextricably webbed, escapes the direct human perceptibility such as this is organized "scientifically", and this entails inadequate reductive ways of reasoning in terms of absolute truths (like in the case of the "locality problem").

- Moreover the human observers *also* can be very distant from one another, and they can move with respect to one-another with variable relative velocities, constant or accelerated.

- So *one same very distant mobile-to-be-studied, or a non-repeatable event, can be observed from different frames of reference that can have themselves various relative movements characterized by big velocities and sometimes also by accelerations*.

---

[193] *This* is the general way in which is born and has taken firm roots the naïve realism from the classical physics according to which the Science produces "knowledge of how the reality truly is in itself". (By contrast with precisely this sort of process, the universal epistemological importance of Quantum Mechanics leaps to the eyes).

In ***such*** conditions the results of the measurements registered by different observers concerning the "same" non-repeatable event, or the "same" mobile, do indeed involve light signals, and these "same" entities appear as ***quintessentially*** different in the various space-time reference-systems that are involved. So Lorentz-Einstein transformation laws are basically necessary, for consensus on even the "sameness" of what is observed. They do indeed play a quite essential role.

And when the necessary transformations are performed, the physical operations of registration that have been performed involve differences that – in some definite sense – *can* indeed be interpreted in terms of "imprecisions" that are negligible on the basis of considerations of orders of magnitude and of an admitted principle of determinism.

*As for the involved operation of generation G, it remains hidden by the fact that it is performed in an only degenerate and implicit way, by merely taking into perceptual – or* ***even only conceptual*** *– consideration, object-entities-to-be-studied that are conceived to exist independently of the human actors and, if they are perceived, are posited to be perceived as they "truly are in themselves".*

In consequence of this sort of cognitive situations, *in Einstein's theories of relativity there is no trace of consciousness of the fact that, while scientific* ***KNOWLEDGE*** *entirely consists of consensual qualification,* ***any qualification is unavoidably and indelibly relative to WHAT is qualified and to the way in which THAT emerges into perceptibility together with the considered qualification***.

All these characteristics ***nourish*** the illusion that scientific knowledge is knowledge of how the physical reality "truly ***is*** *in itself*". And indeed ***the basic conception in RTG is "naïvely realistic"*** **[194]**. It was such in Einstein's mind (!!!) and it stays such quite generally.

The *cognitive situations* considered in *RTG* are radically different from those considered in *QM2*. And also the descriptional *aims* and the procedures employed in order to realize them are radically different. And this – via descriptional relativities of which the reflection in the formal expressions cannot be suppressed – leads in *RTG* and *QM2* to two radically different sorts of mathematical representations.

In the case of $QM_{HD}$ (we cease to consider *QM2* below because it is not made use of so far) the whole mathematical formalism consists ***exclusively*** of a Hilbert-Dirac representation of the probabilistic predictions **[195].** So *for the nowadays representation of the microstates there simply* ***IS*** *no mathematical representation of the considered entities and physical operations on these, themselves*; there is *only* a mathematical representation of the abstract distributions of numerical results of measurements that have involved these entities and operations (which is a *general MRC* feature falsely considered to be specific of the microstates)**[196]**. The whole mathematical formalism of $QM_{HD}$ is quasi ***reduced*** *to the general statistical-probabilistic predictive veil from any mathematical theory of physical entities*. Inside $QM_{HD}$ the physical entities and physical processes that are involved specifically by $QM_{HD}$, are themselves devoid of *any* specific mathematical representation.

While for *RTG* precisely the opposite happens. The probabilistic Hilbert-Dirac representation of the results of measurements that *MRC* imposes upon *any* theory from the Mathematical Physics, is reduced for *RTG* to a tight Gauss distribution of (true) "imprecisions of measurement" around the only *one* theoretically possible outcome usually predicted by this theory to *certainly* be realized in a given observational circumstance, so with probability *1*. For any other result the theoretically possible probability is a priori posited to have a null probability. In these conditions the probabilistic *MRC*-representation of the results of *RTG*-measurements is reduced to only one term *that is assimilated to a certain prediction*, and so the *MRC*-distribution of probabilities, as a whole, vanishes in invisibility. In these conditions the ***specifically*** *RTG*-mathematical description of the studied physical entities and facts themselves remains in front of our eyes *alone and naked*.

---

[194] Einstein has written to de Broglie (concerning his Ph.D. Thesis: "Vous avez soulevé un coin du Grand Voile" (d'Espagnat's title "Le Réel Voilé" refers to that, no doubt). The naïve realism is still ***very*** strongly rooted even in most outstanding minds.

[195] In *QM2*, some modelling elements have been imported from the *dBB* approach (the eigenket and the guiding law), but in $QM_{HD}$ these lie lost with non-identified meaning in the algebra of observables and in the expansions of the state-ket (the eigenket), when they are not entirely ignored (as the guiding law is).

[196] Precisely this is the huge lacuna that we have reproached to $QM_{HD}$ and that we have much compensated in *QM2*.

So there seems to be a total opposition between the formalisms of $QM_{HD}$ and of *RTG*. But in fact *there is no* ***contradiction****, there is only an* ***absence of laid out comparability***, in consequence of the fact that history has led to occult in *RTG* the mathematical statistical-probabilistic predictive feature, and to occult in $QM_{HD}$ the representation of the involved physical entities and operations.

These two theories involve cognitive situations that are so radically different that they cannot even be compared *via their specific mathematical representations*; these indeed have nothing in common, and on the other hand *only* these are what is available. While comparability presupposes common features (MMS [2006])).

But when – by method – these theories are both immersed in *MRC* where each one brings in its own epistemic referential *(G,V)*, the *existenc*e of epistemic referentials is a common feature, while the way each one of the two involved sorts of epistemic referentials can be brought to work introduces speakable differences. And this *suffices* for comparability to step in. And any comparison achieved in this way unifies because it distinguishes after having connected via a posited common substratum.

In this example it leaps to one's eyes that *all* the differences inside *MRC* stem from:

- the ***cognitive situations*** that are involved;
- the ***models*** that work;
- and correlatively, from the posited ***coding postulates*** (in *RTG* these involve the relativistic transformation laws).

In particular, it appears clearly that *the essential indeterminism of QM2 does not in the least hinder to posit a common general principle of determinism, on* ***methodological*** *grounds*, as it has been stressed from the start inside *IQM*, via *MD* and *(1)* from (1.I).1.3.

All this installs and organizes the comparability-and-unification.

Let us finally also stress the very important remark already introduced by previous considerations from *(7.III).2.3.1.2* and *(7.III).2.3.4*:

Also an ***inner*** methodological unification between all the "mechanical" theories is possible – namely *QM2*, the classical mechanics, and the Einstein Theories of Relativity – not only a global unification as considered above. Indeed, consider the words that designate the essential mechanical concepts: velocity, mass, momentum, acceleration, field, energy, kinetic energy, potential energy.

The ***meaning*** assigned to each one of these invariant words ***changes*** *when the considered space-time orders of magnitude are radically changed*. In *MRC* terms this means that in fact the aspect-views $V_g$ involved by these invariant words change.

It obviously would be important to explicate the *MRC* methodological connections that correspond to these hidden changes of meaning. The stably utilized *words* enumerated above, endowed with variable meanings, are like abstract vertebras of a fixed exterior aspect along the mechanical conceptualization of physical entities, but inside which the sort of semantic marrow changes when the order of magnitude of the considered space-time dimensions is changed. An explicit knowledge of these changes would entail a relativized inner unification between the various mechanical theories of physical entities (MMS [1994]).

A similar remark holds concerning the *definitions* of concepts, whether verbal or mathematical definitions. For instance, the Schrödinger equation

$$d/dt\ [(i2\pi/h)\ |\psi_{Gt,\boldsymbol{H}}(\boldsymbol{r},t)>] = \boldsymbol{H}|\psi_{G,\boldsymbol{H}}(\boldsymbol{r},t)>$$

is just a mathematical application of the general definition of the concept of "mechanical energy":

> The change in time of *the representation of that what I study* (in this case "$[(i2\pi/h)\ |\psi_{Gt,\boldsymbol{H}}(\boldsymbol{r},t)>]$)" is equal to the *effect upon the representation of that what I study*, of the action of the *representation of what I call "the mechanical energy" at work* (here "***H***").

This definition itself (as well as that of the involved ***representational concepts*** – here the "state-ket $[(i2\pi/h)\ |\psi_{Gt,\boldsymbol{H}}(\boldsymbol{r},t)>]$)" and the "Hamiltonian ***H***") – are invariant with respect to the changes of the order of magnitude of the considered space-time dimensions, *by construction*. But what I study, and the content assigned to the "mechanical energy" that works, *change* with this order of magnitude, together with the cognitive operations that have to be performed. So, as a whole, the mathematical representation *also* should change in its detailed inner structure such as it is defined and mathematically represented at a given order of space-time dimensions. It follows that the requirement of a common mathematical representation – in spite of radically variable cognitive situations – seems to be a radically impossible requirement that becomes conceptually paralyzing when it is imposed in an absolute way.

The considerations from 9.III are obviously generalizable to any theory of material entities, whether belonging to Physics, or not.

## UNIFICATION BETWEEN THE SCIENCES OF MATTER AND THE TECHNIQUES?

The following question arises naturally:

Would it be possible to *extend* the applicability of *MRC* to also the *"technical" aims and knowledge*? Could *MCR* be utilized as a methodology for the production of physical entities that do not "exist" but of which the existence is *desired* for utilitarian aims? i.e. as a methodology for the creation of artefacts? The process of creation of artefacts involves a first phase of mental conception, to be afterward transposed into material realization. But in a certain sense this situation is very similar to that one encountered in microphysics. There also one has to generate a sort of "*artefacts*-for-being-studied", even though these are not "artefacts-*to-be-made-**use**-of"* for ***non***-scientific aims. Inside *MRC* any consensual aim of description should find an adequate expression in terms of a network of chains of relativized descriptions; so the methodological unification of Physics indicated above should be generalizable to the technical aims also, aims of construction of artefacts instead of aims of mere knowledge.

This, if it succeeded, would lead to a relativized "systemic" and to a relativized representations of the various domains of engineering. Which would be a huge realization.

This has been the aim of the Ph.D.-Thesis of Henri Boulouet [2014].

The subject deserves utmost attention, effort and support for being developed.

# APPENDIX 1

## EXCERPT FROM MUGUR-SCHÄCHTER [2002], KLUWER ACADEMIC PUBLISHERS

### SUMMARY OF THE KERNEL OF THE METHOD OF RELATIVIZED CONCEPTUALIZATION (MRC)[197]

**IV.1. Preliminaries**

Since 1982 I never ceased developing the method of relativized conceptualization – let us denote it MRC – founded on the generalization of the descriptional scheme which I explicated from the quantum mechanical descriptions of microstates. This method can be regarded as an attempt at a certain "normation" of the processes of description of any sort, or in other terms, *a normation of the processes of communicable conceptualization*. Because of the descriptional relativizations that are explicitly built into it at each descriptional step, MRC withstands by construction the insertion of false absolutes, thus warding off false problems or paradoxes. And because it roots its constructions in physical factuality, at the lowest descriptional level that can be reached, MRC furthermore withstands any gliding into relativism:

> *MRC stands in polar opposition to what is called relativism*. It means confined, delimited, but strict precision of each descriptional step, associated with free though guided choices of the way of connecting the descriptional steps accordingly to the evolution of the descriptional aim. This insures controlled rigor throughout a progressive construction of freely decided trajectories and nets of conceptualization, always indefinitely open.

The main difficulty has been to find a way of escaping the imprisonment inside the forms which current language, surreptitiously, imposes upon thought. In all the preceding publications concerning MRC, in order to achieve this liberation I made use from the start on of certain ideographic symbolizations, but I never tried to achieve a mathematical formalization. The ideographic symbolizations, however, have been felt by many to stay in the way of a natural and full access to meaning. Therefore in this work I adopt a different strategy. In a first stage I expose the nucleus of MRC in usual language, trying to get through the stubborn implicit forms of thought induced by the current usage of words, with the help of exclusively the resources of the associations of words themselves (and of abbreviating literal notations of words). In a second stage I give a summary of the ideographic symbolization utilized in all the previous expositions of MRC, because it permits a more suggestive and economic expression of certain basic concepts and assertions. Finally, in a third stage I sketch out a *mathematical* formalization of the nucleus of MRC in terms of the theory of categories [198]. The second and third stage can be omitted without in the least hindering the understanding of the subsequent chapters.

This chapter VI is devoted exclusively to the nucleus of MRC. The way in which the nucleus works will be illustrated in the subsequent chapter V, by showing how it generates a deep and fully relativized unification between the logical conceptualization and the probabilistic one.

**IV.2. First Stage: a Presentation of MRC in Usual Language**

In what follows I formulate definitions (D), a postulate (P), principles (P), conventions (C), and assertions which are called propositions ($\pi$) because they are justified by "natural deductions" (indicated by the word "proof" written between quotation marks in order to distinguish from deductions inside a formal system). Each step is labelled by the symbol of its nature – D, P, P, C, or $\pi$ - followed by the ordinal of the step. There are 19 steps, namely 15 definitions, 1 postulate and 3 principles. When a step is splitted in sub-steps a sub-ordinal is added for each sub-step. A step is often followed by comments.

I proceed by enumeration of the steps and sub-steps. The sequence is interrupted by several intermediary titles which break the progression in small groups each one of which concentrates upon a given purpose.

**IV.2.1. Preparation of the concept of relative description**

**D1**. ***Consciousness functioning***. The activity of an observer-conceiver's mind – called here *consciousness functioning* and noted CF – is conceived to play a central generative role, acting on the exterior universe *and* on the interior universe where it belongs, and there, in particular, also on itself. This activity is regarded as the quintessence of the epistemic *actor*, irrepressibly anterior and exterior to any specified epistemic action. It is an (the ?) invariant among all the epistemic actions the observer-conceptor is aware of, it is the tissue of his continuity, and each one of its products becomes exterior to it as soon as it has been produced. It marks a mobile, permanent and non removable cut – a ultimate cut – between itself and the rest.

**Comment**. The Cartesian cut between *res cogitans* and *res extensa* is second with respect to this mobile cut.

---

[197] ***Inside this appendix all the "comments" – though precious for a thorough understanding – are suppressed.***

[198] The possibility of also another sort of mathematical formalization, more fit for calculations permitting numerical estimations – namely in terms of Hilbert-Dirac "individual" vectors (i.e. not belonging to a vector-space) – will be found in the exposition of meta-[quantum mechanics] (note 2). While in the chapter V it will become clear that the probably most natural vocation of MRC is to yield a non-mathematical formal system comparable to Russel and Whitehead's Principia Matematica, but concerning conceptualization in general instead of only logic.

Throughout what follows CF is explicitly incorporated in the representation. Thereby, from the start on, this approach breaks openly and radically with the classical concept of objectivity. It introduces basically, in a declared and systematic way, the supplementary representational volume that is necessary for a non-amputated expression of the new concept of objectivity in the sense of inter-subjective consensus, such as this concept emerged from modern physics, from quantum mechanics and Einsteinian relativity. That is, inter-subjective consensus founded on systematically extracted fragments of *pure factuality* (quantum mechanics) and qualified by qualifying quantities that are explicitly constructed in order to express definite classes of relative *observational* invariance (Einsteinian relativity). Indeed both these constraints, that are the core of modern physics, involve CF in a quite essential way.

**D2**. ***Reality***. What is called reality is posited here to designate the *evolving* pool – always considered such as it is available at the considered time – out of which any given consciousness functioning either radically creates, or delimits, or only selects, object-entities of *any* kind whatever, physical or psychical or of a mixed kind. This pool will be indicated by the letter R.

**Comment**. This non restricted definition of "reality" refuses the disputes on "existents" (do unicorns exist ? does the number 3 exist ? does a class exist ? etc.). It will appear that inside the present approach the in-distinction entailed by this absence of restrictions generates no difficulties.

**P3**. ***The realist postulate***. Throughout what follows is explicitly postulated the *existence* – but independently of any mind and of any act of observation – of also a *physical* reality.

**Comment**. In the formulation of P3, as also in D1 and D2, the specific designatum of the expression "physical reality" (that implies that a sub-realm of what is called reality is considered), is assigned the status of a primary datum. This however is *only a starting point*. In what follows the general reflexive character of MRC will manifest itself, in particular, by the fact that, progressively, a more constructed distinction between "physical" reality and reality in general will constitute itself inside MRC [199].

The posit P3 of existence of a physical reality might seem to be entailed by D2, so redundant, but in fact it is not. Indeed, though everybody agrees that what is called physical reality does contribute to the pool out of which the consciousness functioning extract object-entities to be studied, nevertheless the various disputes concerning "existents" of this or that sort of object-entity (does Jupiter exist ?) continue steadily. The association [D2+P3] is intended as ***(a)*** a *memento* of the fact stressed most by Descartes and recognized by the majority of the philosophers, that, in the order of the emergence of knowledge, the assertion of the existence of physical reality *cannot* be considered to be *primary* with respect to the assertion of the existence of subjective psychical universes (as classical physics might seem to suggest): the word «also» in the formulation of P3 is intended to provocatively remind of this; ***(b)*** an explicit refusal of solipsism, on the other hand; ***(c)*** an inclusion in what is called reality, of the concepts and systems of concepts, of the behaviours, beliefs, social and economical facts, etc. (the third world of Popper).

**D4**. ***Generator of object-entity and object-entity***. The epistemic operation by which a consciousness functioning introduces an object-entity will be regarded as an action upon R achieved by CF by the use of a *generator of object-entity* denoted G. The spot (or zone, or the sort of domain) from R where a given generator G acts upon R, is considered to be an essential element from the definition of that generator, and which has to be explicitly specified; it will be denoted $R_G$. The object-entity introduced by a given generator G will be denoted $œ_G$. For methodological reasons, a one-to-one relation is posited between a given definition of a generator G and the corresponding object-entity $œ_G$: *that* which emerges as the product of a given G-operation, *whatever it be*, is called "the object-entity produced by G" and is labelled $œ_G$.

**Comment**. Any description involves an object-entity. Usually it is considered that it suffices to name or to label this object-entity thus just directing the attention upon it before it is more thoroughly examined. This "linguistic" attitude is restrictive, since not any conceivable object-entity pre-exists available for examination. Therefore throughout what follows it is required that the basic epistemic action accomplished upon R which brings into play the considered object-entity – as such –, no matter whether this action is trivial or not, be *always* indicated explicitly and fully.

A generator G of object-entity can consist of any psycho-physical way of producing out of R an object for future examinations. Such a way involves systematically some psychical-conceptual component, but which can combine with concrete operations. A generator G can just *select* a pre-existing object or on the contrary it can radically create a new object. If I point my finger toward a stone I select a physical entity by a psycho-physical selective gesture that acts in a non creative way on a physical zone from R ($R_G$ is the volume where the stone is located). If I extract from a dictionary the definition of a chair I select by a non creative psycho-physical act, an abstract conceptual entity materialized by symbols in a physical zone from R consisting of the dictionary (so here $R_G$≡dictionary). If I construct a program for a Turing machine in order to examine the sequences produced by this program, I bring into play a creative, instructional conceptual generator of object-entity that acts on a zone from R containing subjective and inter-subjective knowledge as well as material supports of these. If, in order to study a given state of an electron, I generate it by using some macroscopic device that acts on a place from the physical space of which I suppose that it contains what I call electrons, I delimit a physical object-entity, by a psycho-physical creative action. If now I apply the *same* operation upon a mathematical theory, or upon a place from the physical space where the vibrations of a symphony can be heard but the presence of electrons is improbable, I am making use – by the definition D4 – of *another* generator, since it involves *another zone* $R_G$, and, in consequence of the one-one relation posited between G and $œ_G$, I delimit *another* object-entity (interesting, or not, probably not, in this case). When I define by words a new concept, as I am doing now, in order to later specify its behaviour, I produce a conceptual object-entity, by working, with the help of a psycho-conceptual-physical creative generator, upon the spot from R consisting of the reader's mind.

[199] This specification takes into account concurrent remarks by Jean-Louis Le Moigne, Michel Bitbol, Jean-Blaise Grize, and Gérard Cohen-Solal who – independently of one another – argued that the concept of "physical reality" seemed to them neither clear *nor necessary* in a context of the nature of MRC; that inside such a context this concept should *emerge*. Furthermore, on H. Barreau's opinion, speaking of "physical" reality might erroneously suggest some confusing necessary connection with Physics, which the word "empirical" would avoid. It will however appear that the crucial definition D14.3.1 of a basic transferred description, as well as the preparatory points 8 to13, are endowed with significance *exclusively* with respect to what is usually called physical reality, while with respect to reality in the general sense of D2 – which includes, for instance, *empirical* economic or cultural data, empirical aspects or components of what is called art, etc. – the formulations from the points 8 to 14 are *meaningless*. So I simply do not know how to avoid the assertion *ab initio* of P3 such as it is expressed above: such is the force of language. On the other hand, throughout the points 8 to 14 the concept of physical reality keeps acquiring constructed specificity. In this sense, a progressive specification of P3 *does* emerge from the evolving MRC-context, as desired by the above-mentioned colleagues, but it emerges on the basis, also, of P3 itself. So my final option is to conserve [D2+P3]. For the moment it is sufficient to understand the qualification "physical" as pointing toward anything involving an in principle definable amount of mass-energy. Then certain non-physical entities, like "art", etc., can involve physical *aspects*, while others, like the concept of the number 3, do not.

The inclusion, in the definition of G, of the "zone" $R_G$ from R where G is supposed to act, requires two important specifications.

***(a)*** $R_G$ is *not* a qualification *of the produced object-entity* $œ_G$, obtained by examining *this* object-entity in order to learn about *it*. It is a condition imposed upon the operation of generation G in order to insure the location of *all* the products of G, inside a pre-decided conceptual volume indicated by some verbal label, "microstate", "chair", "program", etc. (In the particular case of a selective generation like for instance pointing toward a stone, this pre-posited conceptual volume where G has to act, might degenerate in the conceiver's mind into an identification with the physical location of the object-entity $œ_G$: this has to be avoided). The methodological necessity of such a pre-decided conceptual location will be fully understood later, in the comment of the definition D14.3.1.

***(b)*** *The "zone"* $R_G$ *from R where G is supposed to act permits of uncontrollable fluctuations concerning what is labelled* $œ_G$. The physical region from R where I act in order to generate a given microstate of an electron, can contain non perceptible and uncontrollably variable fields, etc.; the reader of these lines can happen to be a 16 years old boy, or a mature intellectual. These fluctuations entail an unavoidable non-predictability concerning the effect labelled $œ_G$ of an operation of generation of an object-entity. However one should clearly realize that it simply is *inconceivable* to "entirely" immobilize *a priori* the effect of G denoted $œ_G$: this would require to specify "completely" $R_G$. But such a requirement is both impossible (circular) and unnecessary. One simply cannot *start* a process of representation of the way in which descriptions, i.e. qualifications of any object-entities, emerge out of R, by specifying, so qualifying R itself everywhere and for any time, and also from any point of view. Such a huge and fundamental circularity is not acceptable, and on the other hand *it cannot be realized*. While the *a priori* non-determination concerning the effect of the individual operations of generation of an object-entity, is by no means an insuperable problem or a difficulty. It simply is an unavoidable constraint that MRC is obliged to recognize, include and control. The recognition of this constraint plays an essential and very original role in the dynamics of conceptualization from MRC. It brings into evidence one of the roots of human conceptualization and it comes out to be intimately tied with a reflexive character of MRC, of maximal *a priori* freedom, followed by *a posteriori* controls and restrictions. It opens up the way toward a constructive incorporation (*via* the sequence D14 of definitions of relative descriptions) of the fundamental fact called "non-determination of reference" established by the deep analyzes of Quine (ref. 13) and Putnam (ref. 14), which marks the breaking line between factuality and mere language.

***(c)*** Consider now the one-one relation posited between a given definition of an operation G of object-entity generation and what is labelled $œ_G$. This relation is intimately tied with the above mentioned *a priori* non-determination involved by $R_G$, so also with the non-determination of reference. It is important to realize that *no other relation could be uphold ab initio*. Indeed the object-entity labelled $œ_G$ emerges still non qualified *from the standpoint of the subsequently intended examinations*, if not, in general its generation would be unnecessary for this aim. It can even emerge still entirely inaccessible to direct knowledge of any sort, if G is a radically creative and physical operation of generation (as in the case of the microstate generated by most quantum mechanical operations of state-generation). In these conditions what we called a one-one relation between a given definition of an operation G of object-entity generation, and the mere *label* $œ_G$, obviously *cannot* mean that the still unqualified replicas of $œ_G$ are all "identical" in some inconceivable absolute sense.

The one-one relation posited between G and $œ_G$ amounts to just a ***methodological*** *pre-organization of the language-and-concepts*, unavoidable in order to be able to form and express a beginning of the desired representation of a human conceptualization.
Such a methodological pre-organization is, by its nature, of a *FORMALIZING* step, like an algebraic one.

Indeed if from the start on we imagined that G might produce sometimes this and sometimes something else, how would we *speak* of what it produces, or think of it? We would have to re-label in only *one* way the product entailed by a given definition of G, whatever it be, and thus we would come back to precisely our initial choice of language and notation. On the other hand, if we asserted *a priori* a "real" one-one relation between G and what is labelled $œ_G$, we would thereby assert the sort of view that is sometimes called metaphysical realism (a God's Eye view, as Putnam puts it), which would directly contradict the very philosophical essence of the present approach. In the sequel, each time that some definite consequence of this *a priori* choice of language will appear, we shall deal with it for that definite case.

The explicitly methodological character of this constructive strategy adopted in the definition D4, is a quite crucial step. It saves premature, void, illusory questions and paradoxes that simply cannot be solved *a priori*. Instead, as it will appear, it brings forth *a posteriori* a clear, fully *relativized* operational concept of "identity" that emerges progressively in π12, π13 and D14.1 and eventually is specifically defined in π18.1; which suppresses inside MRC one of the most noxious false absolutes induced by current language. And the relativization of the qualification of identity permits then immediately to show by π18.2 and π18.3 that MRC, inside its *soma* structured from the progressively posited definitions, postulate and principles, eventually entails a well-defined sort of *minimality* of the realist postulate P3, initially posited without any further qualification. By this minimality the "metaphysical realism" will appear to by organically rejected by MRC.

**D5.** ***"Qualificators"***.

**D5.1.** ***Aspect-view***. Consider a point of view for qualification (colour, coherence, etc.). Call it an *aspect* and label it by some letter or sign, say g. Consider a finite – so discrete – set of n mutually distinct g-qualifications. Call each such qualification *a value of the aspect g* and label it gk, which reads *the value k of the aspect g*, in short *a gk-value*.

The aspect g is considered to be "defined" *if and only if*:

***(a)*** The specification of the values gk of the aspect g is associated with the explicit specification of an effectively realizable sort of *g-examination* (physical, or conceptual (in particular formal), or mixed).

***(b)*** Any result of a physical g-examination is directly perceptible by the observer-conceiver's biological senses and mind; any result of an abstract g-examination is directly perceptible by the observer-conceiver's mind.

***(c)*** Any specification of a g-examination contains a procedure for deciding what result is to be announced in terms of what gk-value, which amounts to the specification of certain *coding-rules*.

If the conditions ***(a)***,***(b)*** and ***(c)*** are all satisfied, the pair [g, {gk, k=1,2…n}] constitutes *a grid for qualification by the aspect g* that will be called *the aspect-view* $V_g$.

**Comment.** In contradistinction to the grammatical or logical predicates, an aspect-view $V_g$ in the sense of D5.1 is strongly restricted by conditions of efficiency and is endowed with a *structur*e and with coding-rules which fix a finite "gk-language" consisting of operations, signs, names, referents, and the stipulation of the relations between these. (Wittgenstein's analyses of the role of the contexts in the learning of a current language permit to apprehend the distance between a qualification in the usual sense and a qualification in the sense of D5.1).

Let us note that an order between the values gk of an aspect g is *not* required but is permitted.

The distinction between an aspect g and the set of all the gk values contained inside that aspect, takes into account the remarkable psychological fact that any set of gk-values, even only *one* such value, as soon as it is "conceptualized" (i.e. as soon as it ceases to be a mere "prime-ity" in the sense of Peirce), generates in the consciousness a whole semantic *dimension* g (a genus) that exceeds this set and constitutes a ground on which to place its abstract feet: every gk-value determines a location (a specific difference) on this semantic domain g that grows spontaneously beneath it (for instance, if gk labels the interior event toward which the word "red" points, this event, when conceptualized, generates the carrying semantic dimension toward which the word "colour" points). We are in presence of a fundamental law of human conceptualization that moulds logic, language, and even metaphysics (the concept of "substance" is the semantic ground on which are located the ways of existing of material systems, etc.). The definition D5.1 reflects this law, on which it tries to draw the attention of the cognitive approaches (what are the corresponding bio-functional substrata?).

Finally let us also note that, by definition, an aspect-view $V_g$ acts like a qualifying *filter*: it cannot yield qualifications different from any corresponding gk-value.

**D5.2**. ***View***. A grid for examination that consists of a finite but arbitrarily large set of aspect-views, is called a view and is denoted V.

**Comment**. The complexity and the degree of organization of a given view V are determined by the number of aspect-views $V_g$ from V and by the structures of the various sets of gk-values introduced by the various involved aspect-views from V (number of gk-values, "position" (central, extreme) of each set of aspect-values on the corresponding semantic dimension g, existence or not of an order among the gk-values of a fixed aspect g, a reference-gk-value (a gk-zero), etc.). In particular a view can reduce to only one aspect-view or even, at the limit, to one aspect-view containing only one gk-value on its semantic dimension g. There is nothing absolute in the distinction between an aspect-view and a view: an aspect-view can be transformed in a view by analysis of its aspect in two or more sub-aspects, and *vice-versa* the set of distinct aspects from a view can be synthesized into a unique aspect. This stresses that a view, like also a generator of object-entity, is just a construct freely achieved by the acting consciousness-functioning CF, in order to attain a definite epistemic aim.

**D5.3**. ***Physical aspect-view and view***. Consider an aspect-view $V_g$ where the aspect g is physical and requires physical operations of examination of which the results consist of some observable physical effects. Such an aspect-view will be called a *physical aspect-view*. A view containing only physical aspect-views will be called a *physical view* (concerning this language cf. note 25).

**Comment**. This definition can be best understood *per a contrario*. A mathematical or a logical view is not a physical view, though the involved examinations do involve certain physical actions (writing, drawing, etc.), because what is called the results of the examinations (not their material expression) consists of concepts, not just of physical entities (marks on a measuring device, for instance). (And of course, a physical view does not in the least necessarily involve Physics).

**D5.4**. ***Space-time aspect-views***. One can in particular form a *space-time aspect-view* $V_{ET}$. Accordingly to Einsteinian relativity the double index ET can be considered as *one* aspect-index g=ET where E reminds of the current Euclidian representations and T stands for time. However the partial aspect-indexes E and T can also be considered separately from one another, setting g=E or g=T. The space-aspect E is associated with space-values or "positions" that can be denoted E**r** (setting a position vector **r** in the role of the index k introduced in D5.1) and the time-values can be denoted Tt (setting a time parameter t in the role of k). Indeed though in general the *numerical* estimations indicated by **r** and t are not mutually independent, nothing interdicts to symbolize separately the spatial position-value and the time-value.

Infinitely many space-time views can be constructed (by varying, in the representations, the choice of the origins of space and time, of the units for measuring intervals, the form and direction of the involved reference-axes). Any space-time aspect-view introduces an *ordered* grating of space-time values. This is a specificity with highly important epistemic consequences (refs. 15 and the chapter V2 in this work) because it endows the space-time views with the power to *strictly singularize the representation of an object-entity*.

**D6**. ***Epistemic referential and observer-conceptor***. A pairing (G,V) consisting of a generator G of object-entity and a view V, is called an *epistemic referential*.

A consciousness functioning CF that endows itself with a given epistemic referential is called an *observer-conceptor* and can be denoted [CF,(G,V)].

**Comment**. A pairing (G,V) is permitted to be entirely arbitrary *a priori*. This is a methodological reaction to an unavoidable constraint: the capacity of a pairing (G,V) to generate meaning, can be examined only after having considered that pairing. This particular methodological reaction is a new manifestation of an already mentioned general reflexive strategy practised in MRC, of a tentative *a priori* approach that is entirely non restricted, but is systematically followed by *a posteriori* corrective restrictions.

An observer-conceiver [CF,(G,V)] is the minimal epistemic *whole* able to achieve epistemic actions in the sense of MRC: by itself an epistemic referential (G,V) is not yet a closed concept, nor does it designate an active entity. This concept becomes closed and activated only when it is associated with the consciousness functioning CF that generated and adopted it.

**D7.** ***Relative existence and inexistence***. Consider an *a priori* pairing $(G,V_g)$. If an examination by the aspect-view $V_g$ of the object entity $œ_G$ generated by G, never reveals to the involved observer-conceptor some value gk of the aspect g, we say that *the object-entity $œ_G$ does not exist (is not pertinent) with respect to the aspect-view $V_g$* (or equivalently, that $V_g$ does not exist with respect to $œ_G$, or that $œ_G$ and $V_g$ do not mutually exist) [200].

Suppose now, on the contrary, an act of examination by the aspect-view $V_g$ of the object entity $œ_G$ generated by G, that *does* reveal to the involved observer-conceptor one or more values gk. In this case we say that *the object-entity $œ_G$ exists with respect to the aspect-view $V_g$* (or that $V_g$ exists with respect to $œ_G$, or that $V_g$ and $œ_G$ do mutually exist).

**Comment**. The definitions of relative inexistence or existence can be transposed in an obvious way to one single value gk of an aspect g, or to a whole view V.

The concepts of mutual inexistence or existence concern, respectively, the general *impossibility* or *possibility* of the emergence of meaning, as well as the intimate connection between meaning and descriptional aims, which are induced by a tentative pairing $(G,V_g)$ or (G,V). These concepts are essentially semantic.

They express the general fact – *previous to any qualification* – that a given object-entity can be qualified only *via* the views to the genesis of which it can contribute by yielding matter for abstraction. Furthermore, the concepts of relative inexistence and existence permit to cancel *a posteriori*, among all the initially only tentative pairings $(G,V_g)$ or (G,V) that an observer-conceiver tor has introduced, those which appear to be non-significant; while the other pairings can be kept and put to systematic descriptional work. The possibility of such a selection illustrates again the general reflexive strategy of MRC: maximal *a priori* freedom followed by *a posteriori* controls and restrictions.

The concepts of relative inexistence and existence have quite fundamental consequences, but with respect to which the classical conceptualizations are more or less blind. This generates various sorts of false problems and paradoxes. Formal logic for instance, because it is posited to concern exclusively the qualifications of mutual consistency (confusingly called sometimes formal truth), decidability concerning consistency, and formal completeness, banishes the semantic concepts of relative existence. But, surreptitiously, *via* the fact that often the axioms from a formalized representation of some domain of reality are *considered as empirically true assertions, as "propositions"*, empirical truth goes back into the formal systems, and factual truth, in order to be defined, requires mutual existence as a preliminary condition. This, according to MRC, is intimately tied with the non-decidability paradoxes, and leads to certain reservations even with respect to Gödel's proof of non-decidability, though non-decidability itself, as defined for formal systems, *follows* inside MRC, by a specific chain of arguments (cf. V.2).

**P8**. ***The Frame-Principle***. I posit the following principle, called *frame-principle* and denoted FP.

---

[200] If one examined with the help of a voltmeter, a symphony by Beethoven, the operation might never produce an estimation of a difference of electrical potential (accidents being neglected). Of course during a more realistic sort of tentative research a mutual non-pertinence can be much less apparent *a priori* than in this caricatured example.

Consider a *physical* object-entity $œ_G$ that can be (or is conceived to have been) generated by some definite physical generator of object-entity, G. This entity $œ_G$ does exist in the sense of D7 with respect to at least one physical aspect-view $V_g$ (D5.3) (if not the assertion of a physical nature of $œ_G$ would be devoid of foundation (content)).

The frame-principle FP asserts the following.

- If the physical object-entity $œ_G$ does exist in the sense of D7 with respect to the physical aspect-view $V_g$, then *ipso facto* $œ_G$ exists in the sense of D7 with respect to also at least one view V formed by *associating* $V_g$ with a convenient *space-time view* $V_{ET}$ (it cannot exist with respect to *any* such association, if only because the values gk of a given aspect g can appear or disappear with respect to a given space-time view when the space-time units are changed). But the object-entity $œ_G$ is non-existent in the sense of D7 with respect to *any* space-time view that acts *isolated* from any other physical aspect-view $V_g$ where g≠ET: *the space-time views are* ***frame****-views which, alone, are blind, they cannot "see" nothing.*

- According to what precedes what is called "physical space-time" *cannot* be regarded as a physical object-entity $œ_G$. Indeed the assertion posited in the first part of this principle does *not* apply to what is called "physical space-time": the designatum of this expression *itself*, considered strictly *alone*, is non-existent in the sense of D7 with respect to *any* physical aspect-view $V_g$where g≠ET, and it is equally non-existent with respect to any association of such a physical aspect-view, with a space-time aspect-view. In this sense:

What is called "physical space-time" is – itself – only the locus of all the possible space-time frame-views (referentials), the genus of these. It is the conceptual volume where physical entities, facts or aspects, can be assigned space-time specifications which, if this is desired, can be numerically defined by the use of space-time referentials.

**Comment**. The frame principle FP adopts, transposes in terms of MRC, and specifies, the Kantian conception according to which man is unable to conceive of physical entities outside physical space-time, that he introduces as *a priori* "forms of the intuition" inside which he casts all his representations of physical entities. FP isolates and stresses certain particular implications of this Kantian conception which so far seem to have remained insufficiently noticed by physicists. Namely that any mature and normal human being, by the nature of his consciousness functioning, as soon as he perceives or even only imagines a phenomenal appearance which he connects with what he conceives to be a physical entity $œ_G$, *ipso facto* introduces more or less explicitly:

**(a)** a space-time frame-aspect-view $V_{ET}$ (the observer-conceiver's body tends to yield – vaguely – the intuitive origin, the units, and – variable – directions of the axes, whereas in the technical or scientific approaches these are explicitly and freely specified, in a precise and stable way, in a mathematical language, an integral or a differential mathematical language);

and furthermore

**(b)** at least one aspect-view $V_g$ where g is a physical aspect *different* from $V_{ET}$, relatively to which the considered physical entity $œ_G$ does exist in the sense of D7, and the values gk of which he combines with the value-indexes E**r** and Tt of the space-time aspect-view $V_{ET}$ (in mathematical terms, with the space-time coordinates yielded by $V_{ET}$). J. Petitot (ref. 5A, p. 216) writes concerning Kant's conception on space and matter:

"As quality (not as quantity any more), matter is *filling* of space. This filling is very different from a mere "occupation" (anti-Cartesianism). It is a dynamical and energetic process characteristic of the substantial "interiority" of matter."

In P8 the necessity of the presence of at least one physical aspect g different of the space or time aspects, is a way of expressing *the presence of the matter which fills the space-time, and of asserting that any phenomenal manifestation to human minds stems from this matter, not from space-time itself*; of asserting hat

**(c)** by the help of a space-time frame-view *alone*, in the strict absence of any other sort of physical aspect-view $V_g$ (colour, texture, whatever) man is unable to perceive or even to imagine a physical entity. He simply is unable to extract it from the background of exclusively space-time frame-qualifications which, by themselves, act exclusively as elements of *a grid of reference* inserted in an abstract, void *container* labelled by the words "physical space-time". By themselves these elements from a grid of reference act exclusively as potential land-marks that can be "activated" only by the values of some other aspect g≠ET.

The assertion that the designatum of the words "physical space-time" cannot be treated itself as a physical (object-)*entity* – probably obvious for most physicists – is introduced here explicitly mainly in order to emphatically block certain very confusing ways of thinking induced in the minds of non-physicists by the verbal expressions by which the physicists use to accompany their relativistic formalizations: these verbal expressions suggest that what is currently called space-time would itself ***possess*** this or that *metric*; while in fact any space-time metric is just ***assigned*** by construction to this or that space-time frame-aspect-view, on the integral level *or on the infinitesimal differential level*, on the basis of some definite (even if implicit) descriptional aim (this is discussed in the last chapter of this work).

**C9. *Conventions***. In order to take explicitly into account the frame principle FP we introduce the following conventions.

- Any view V considered in order to examine a *physical* object-entity will contain a space-time aspect view $V_{ET}$ and one or more physical aspect-views $V_g$.

- The aspects denoted g are always *different* from the space-time aspect ET.

**P10**. ***The principle of individual space-time mutual exclusion***. Consider a *physical* object-entity $œ_G$ corresponding to a physical generator G. Let V be a *physical* view with respect to which $œ_G$ does exist in the sense of D7, involving two distinct physical aspect-views $V_{g1}$ and $V_{g2}$ a well as a space-time view $V_{ET}$ (accordingly to C.9). The *principle of individual space-time mutual exclusion* posits the following.

- Any physical examination involved by V quite systematically *changes* the state of the examined physical object-entity $œ_G$, even if only to a degree which in this or that context can be neglected: the state of a physical object-entity is not a stable datum with respect to an act of physical examination (in informatics one would say that it is a "consumable" datum).

- If, when performed separately on different replicas of $œ_G$, the examinations involved by $V_{g1}$ and $V_{g2}$ can be shown to cover different space-time domains - the referential and the origins for space-time qualifications being kept the same – which involves that they change differently the state of $œ_G$ – then it is not possible to perform both these two sorts of examinations simultaneously upon a ***unique*** replica of $œ_G$ produced by only ***one*** realization of G (the word «individual» from the denomination of P10 refers to this crucial uniqueness of the involved replica of $œ_G$).

If the type of impossibility specified above manifests itself, the two physical aspect-views $V_{g1}$ and $V_{g2}≠V_{g1}$ are said to be mutually *incompatible*. In the alternative case $V_{g2}$ and $V_{g1}$ are said to be mutually *compatible*.

**Comment**. It is probably possible to show that P10 is entailed by (reducible to) the assertion of *other* more basic space-time mutual exclusions (or from some ultimately basic space-time mutual exclusions, non-reducible to still more basic ones) (an attempt has been made in ref. 22 B, p. 290). But here, for simplicity, we start from the formulation P10 because it is more immediately related with the consequences pointed out in the sequel.

The quantum mechanical principle of "complementarity" can be regarded as the realization of P10 for the particular category of physical object-entities consisting of states of microsystems. This brings into clear evidence the often only obscurely perceived fact that complementarity in the sense of quantum mechanics has an – exclusively – *individual* significance: indeed two mutually incompatible quantum mechanical measurements ***can*** be simultaneously realized on two distinct replicas of a given microstate (object-entity), and if this is done two distinct and useful pieces of information are obtained in a quite compatible way (ref. 16). But this brings already up on a statistical level, and there what is called the mutual incompatibility of two physical aspect-views *is not manifest any more*. What *is* impossible indeed is only the simultaneous realization upon *one* given replica of the considered microstate, of two mutually incompatible quantum mechanical measurements.

*The concept of incompatibility of two physical aspect-views is defined only with respect to* ***one individual replica of some given physical object-entity****: it is not intrinsic to these physical aspect-views*. This is of crucial importance from a logical point of view (cf. V.1.2)

**π11.** ***Proposition***. Consider a physical object-entity $œ_G$ corresponding to a generator G and a *physical* view V with respect to which $œ_G$ does exist in the sense of D7. In general, in order to perform upon $œ_G$ *all* the operations of examination corresponding to *all* the different aspect-views $V_g$ from V, it is necessary to realize a whole set of *successions* **[**(one operation of G-generation of $œ_G$), (one operation of $V_g$-examination of that replica of $œ_G$)**]** (in short [G**.**$V_g$]) containing (at least) one such pair for each physical aspect-view $V_g$ from V.

***"Proof"***. In order to achieve examinations of $œ_G$ *via* mutually incompatible physical aspect-views $V_g$ from V, the operation G of generation of $œ_G$ has to be *repeated* (the time parameter being re-set to its initial value $t_0$ (like in sport-measurements, in the repetitions of chemical or physical experiments, etc.)) and paired *successively* with these incompatible aspect-views.

**Comment**. This, though an obvious consequence of P10, is highly *non trivial* by itself. It is important to know explicitly that the achievement of complex examinations of an object-entity involving "consumable" characters, entails in general the condition of reproducibility of all the involved pairs [G.$V_g$] (either in succession or in simultaneity), thus involving a whole set of replicas of the involved sort of object-entity $œ_G$. (The proposition π11 and its "proof" admit of generalization to also certain *conceptual* referentials (G,V)).

**π12.** ***Proposition***. Consider a physical object-entity $œ_G$ corresponding to a given generator G, and one given *physical* aspect-view $V_g$ with respect to which $œ_G$ exists in the sense of D7. When a succession [G**.**$V_g$] is repeated a big number N of times (the time parameter being re-set for each pair to its initial value $t_o$) or when it is simultaneously realized on a big number of replicas of the object-entity $œ_G$, it is not impossible that the same observable gk-space-time-values be found in each instance; in such a case one can say that an *individual qualificational N-stability* has been obtained. But in general this does not happen: in general the N obtained gk-space-time-values are *not* all identical, notwithstanding that in each realization of a pair [G**.**$V_g$] the operations G and $V_g$ obey strictly the same defining conditions.

***"Proof"***. This follows *per a contrario*: to posit *a priori* that the results produced by repeated realizations of a given succession [G**.**$V_g$] are all identical "because" in each pair both G and $V_g$ obey the same specifications, neither follows with necessity from the previously introduced definitions and principles, nor could it be found *a posteriori* to be always factually *true*. To show this last point it is sufficient to produce a counter-example. Consider an object-entity generator G which acts by definition on a zone $R_G$ from R consisting of a piece of land, and that delimits there the object-entity $œ_G$ consisting of a definite area of one square kilometre. Let $V_g$ be an aspect-view (structured accordingly to D5.1 and C9) that permits to establish the aspect g ≡ [association of mean-colour-value-and-space-position over a surface (any one) of only one square *meter*]: inside the epistemic referential $(G,V_g)$, two distinct realizations of the succession $[G.V_g]$ in general yield two different results, even though both G and $V_g$ satisfy each time to the same operational commands.

**Comment**. Notice that if an individual qualificational N-stability *is* found for a given succession $[G.V_g]$, this does by no means exclude the possibility that in another series of N' repetitions (with N' bigger or smaller than N) no individual stability be found any more.

Furthermore, and this is more important, if for a given object-entity $œ_G$ corresponding to a given generator G, an individual N-stability with respect to the examinations by a given aspect-view $V_g$ *is* found, this does by no means involve that for the same object-entity $œ_G$ but another aspect-view $V_{g'}$ with g'≠g one will find again some individual stability for some big number

The individual stability of the qualifications of an object-entity $œ_G$ or the statistical character of these, are relative to the qualifying aspect-view $V_g$.

It is utmost important to realize that – quite generally – a generator G of a physical object-entity being fixed by some operational definition of it, it would even be *inconceivable* that for *any* association of G with some aspect-view $V_g$, the results of repetitions of the corresponding sequence $[G.V_g]$ shall all be identical: that would be a *miracle* in so far that absolute identity – independent of the considered aspect-view $V_g$, i.e. for *any* tried aspect-view $V_g$ – has never been observed concerning a *physical* object-entity which – factually – is always endowed with strict singularity (this probably holds even for a conceptual object-entity, like, say, the number 5, if its mental correspondent in a given mind is considered). As for "identity" in *absence* of *any* view – which, as many do in fact surreptitiously and vaguely imagine, would mean identity of $œ_G$ with *itself* from one realization of G to another one, not of the qualification of $œ_G$ *via* $V_g$ when the succession $[G.V_g]$ is repeated –, it is but an illusory concept tied with the quest for an impossible absolute objectivity of the thing-in-itself. (The psychological difficulty encountered for realizing this stems from the physical, "exterior" nature supposed for $œ_G$, which surreptitiously inclines to posit that – like $œ_G$ itself – the *qualifications* of $œ_G$ also exist independently of any observer-conceiver, as "properties" of $œ_G$).

The above considerations bring back to the only methodological meaning which can be *a priori* assigned to the one-one relation posited between G and $œ_G$, and, correlatively, they bring back to also the roots of the non-determination of reference.

Notice that all the preceding assertions acquire inside MRC a deductive character, in the sense of the sort of natural logical construction practised here (i.e. outside any formal system). Which is a quite non-trivial feature of MRC, manifesting already the (qualitative) formalized features with which we are progressively endowing it.

**π13. *Proposition***. Given an epistemic referential $(G,V_g)$ where both G and $V_g$ involve *physical* operations, in general no stability at all is insured for the gk-space-time values obtained by repeated or multiple realizations of the succession $[G.V_g]$, neither on the individual level of observation, nor on the statistical one.

**"Proof"**. If only a maximal, an individual N-stability is considered, i.e. identity of all the N groups of observable gk-space-time values corresponding to N realizations of a succession $[G.V_g]$, then π13 becomes a mere repetition of π12, hence the "proof" of π12 still works. But suppose that no individual N-stability has been found, i.e. that a whole statistical distribution of dispersed triads of gk-space-time-values has been found. Then it still remains *a priori* possible that a big number N' of repetitions of a series of a big number N of repetitions of the succession $[G.V_g]$ (N'≠N in general), shall bring forth, when N' is increased toward infinity, a convergence in the sense of the theorem of big numbers, of the relative frequencies of occurrence, in the mentioned statistical distribution, of the dispersed triads of gk-space-time-values. In this case one can speak of a *probabilistic (N,N')-stability*. However, up to some given arbitrary pair (N,N') of big numbers, it might appear by experiment that in fact this second possibility does *not* realize either, even though G and $V_g$ have been previously found to mutually *exist* in the sense of D7. Nothing excludes the possibility of such a situation, neither some previous MRC-assumptions, nor the empirical experience. If this negative situation does realize indeed, then only two solutions are left: either one continues the search with pairs of increasingly bigger numbers N, N', or one stops at some given pair (N,N') and announces *a posteriori* that, even though G and $V_g$ do mutually exist in the sense of D7, their pairing $(G,V_g)$ has nevertheless to be (N,N')-cancelled from the subsequent conceptualization, because, while no individual N-stability has been observed, this pairing does not generate a probabilistic (N,N')-stability either; *tertium non datur* because apart from an individual or a probabilistic stability, no other sort of still weaker stability has been defined so far (in V2 this question is treated more thoroughly). Anyhow, for *any* given pair of big numbers (N,N'), it is quite possible that no stability at all be found for the results of repeated successions $[G.V_g]$. Which establishes π13.

**Comment**. The "proof" of π13 does by no means exclude the possibility that, if the succession $[G.V_g]$ does produce a probabilistic (N,N')-stability, another succession $[G.V_{g'}]$ with G the *same* but with $V_{g'}\neq V_g$, shall produce qualifications that are endowed with some *individual* N-stability, or with no stability at all, neither probabilistic nor individual:

*The* ***existence*** *of a probabilistic stability of the qualifications of a given object-entity $œ_G$ is relative to the qualifying aspect-view $V_g$ just like the existence of an individual stability. The* ***nature*** *– individual or probabilistic – of the stable qualifications of a given object-entity $œ_G$, is relative to the qualifying aspect-view $V_g$ just like the existence of stable qualifications*.

### IV.2.2. The normed concept of relative description

**D14.** ***Relative description***.

**D14.1.** ***Relative description of a physical object-entity***. Consider an epistemic referential (G,V) where: G is a *physical* generator that generates a corresponding *physical* object-entity $œ_G$; V is a *physical* view with m aspect-views $V_g$ with respect to each one of which $œ_G$ does exist in the sense of D7; and, as required by P8 and C9, V contains also a space-time view $V_{ET}$ introducing an *ordered* space-time grating (D5.4). Furthermore consider, for *each* $V_g$ from V, a big number N of realizations of the corresponding sequence $[G.V_g]$, in simultaneity or in succession, the time parameter being re-set at the same initial value $t_o$ for each realization of a sequence $[G.V_g]$.

Suppose first that, when the succession $[G.V_g]$ is realized N times, for *each* aspect-view $V_g$ from V, *identical* outcomes of the corresponding configuration of gk-space-time-values are obtained, i.e. only *one* same "individual" result appeared N times. We shall then say that an *N-individual* outcome has been obtained (the reference to N is necessary because nothing excludes that for another sequence of successions $[G.V_g]$ some dispersion be found). The set of N-individual configurations of gk-E**r**-Tt-values corresponding to *all* the m distinct aspect-views $V_g$ from V, constitutes in the abstract representation space of V ordered by the space-time grating introduced by $V_{ET}$, a definite "form" of gk-E**r**-Tt-values. This "form" will be called an *N-individual relative description, with respect to V, of the physical object-entity $œ_G$,* (in short an individual relative description) and it will be indicated by the notation $^N D/G,œ_G,V/$ to be read «the description relative to the triad $G,œ_G,V$ and to N» (in current usage the index N, supposed to be big, will be dropped). The individual relative description $D/G,œ_G,V/$ defined above can also be regarded as the set of all the individual *relative-aspect-descriptions* $D/G,œ_G,V_g/$ with $V_g\in V$.

Suppose now that, when the various successions $[G.V_g]$ with $V_g\in V$ are realized N times, *not* all the successions $[G.V_g]$ are found to reproduce identically one same configuration of gk-E**r**-Tt-values; that at least for one $V_g\in V$ (not necessarily for all) the corresponding succession $[G.V_g]$ produces a whole set $S_{gi}=\{c_{gi}\}$ of mutually distinct, dispersed configurations $c_{gi}$ of gk-E**r**-Tt-values, (with $i\in I$ and I a finite index-set, to preserve the finitistic character of this approach); but that, for *any* succession $[G.V_g]$ which produces dispersed results, when N is increased toward infinity, the relative frequency $n(c_{gi})/N$ of occurrence of *each* configuration $c_{gi}\in S_{gi}$ converges toward a corresponding probability $p_{gi}$. In these conditions each configuration $c_{gi}\in S_{gi}$ will be called an *elementary-event-description* corresponding to the succession $[G.V_g]$ with $V_g\in V$ and it will be denoted $D_{p(gi)}/G,œ_G,V_g/$. The epistemic referential (G,V) will be said to produce *a probabilistic relative description* of the physical object-entity $œ_G$ which will be denoted $D_p/G,œ_G,V/$ [201].

**Comment**. The definition D14.1 is the core of MRC. It finally assigns a *significance* to what has been called a physical object-entity $œ_G$. A significance which, though it is relative to a view V and in certain "basic" conditions that will be specified in D14.3.1 is far from being fully "satisfactory", nevertheless is now quite definite *and* endowed with communicability; whereas G alone cannot systematically insure for "$œ_G$" a significance distinct from just the conventional label «effect of a realization of G», because the results of G might emerge still entirely non perceptible.

**D14.1.1.** ***Reference and relative meaning***. In any case of qualificational stability, individual or probabilistic, we shall say that *$œ_G$ is the reference of $D/G,œ_G,V/$* while *$D/G,œ_G,V/$ – as perceived by the acting observer – is, for that observer, the meaning of $œ_G$ relatively to V.*

**Comment**. It thus appears that the initial methodological assertion of a one-one relation between a given definition of an operation G and its result labelled $œ_G$, does not hinder the subsequent construction of all the necessary specifications. On the contrary, it founds them.

---

[201] This definition of a probabilistic description is incomplete and simplifying. It will be thoroughly reconstructed and completed in V2. A more ancient but full treatment can be found in the reference 23. In this stage of the development of MRC we are obliged to introduce it in this unachieved form, as a provisional support for essential distinctions that cannot be postponed.

Furthermore, while the "description-problem" (given (G, $œ_G$, V) find the corresponding description D/G,$œ_G$,V/) possesses a quite definite solution, the inverse "problem of reference" (given a description D/G,$œ_G$,V/, determine the referent $œ_G$ and the view V which generated it) is in general insoluble (Wittgenstein, concerning the available descriptions of the dressing that Edward II wore during the ceremony of his crowning, remarks precisely this, though of course in other terms).

The following is also worth being noted.

*The condition of existence of individual or probabilistic stability of the outcomes of the successions [G.$V_g$], with respect to repetitions of these,* ***presupposes*** *the possibility to achieve arbitrarily many successions [G.$V_g$], for all the $V_g \in V$.*

This is a strong restriction. But when it is insured it extracts out of temporality the concept of "description" founded upon it and it puts it directly on highways of communicability where reference, meaning, and objectivity in the sense of intersubjective consensus, can most immediately be attained. Furthermore, it sets a standard with respect to which relaxing generalizations can be now defined.

**D14.2. Two generalizations of D14.1.**

**D14.2.1.** ***Relative description of a non-physical public object-entity***. Let us suppress in the definition D14.1 the restriction to physical generators, while *excluding* generators that act on only one individual inner universe (there, in general at least, the sequences [G.$V_g$] cannot be repeated (in succession or in simultaneity) and so the condition of stability of their results cannot be insured). Thus relaxed, the definition D14.1 enlarges to object-entities from the non physical but public, exterior reality (economical, social) for which the repeatability of sequences [G.$V_g$] and the condition of stability of their results still do make sense. The new sort of description obtained in this way will be called a *relative description of a non physical and* ***public*** *object-entity* and it will be indicated by the notation (NPP).D/G,$œ_G$,V/, in short (NPP).D.

**Comment**. The generalization D14.2.1 holds in particular concerning any already accomplished *description* in the sense of D14.1, selected as a new, always *conceptual* object-entity, to be examined in a subsequent description *via* some new view. Thereby:

The definition D14.2.1 opens up specifically and explicitly the whole crucially important sub-realm of R consisting of a stabilized communicable *conceptua*l reality.

In the case of non-physical object-entities that admit of a description in the sense of D14.2.1, any reference to the frame-aspect of ("physical") *space* can obviously be dropped, and so the obtained relative description amounts to a "form" of only gk-time values. If moreover it appears that the considered description can be regarded to be independent also of time values, (as for instance in the study of a fixed formal system), then the reference to the frame-aspect of time can be equally dropped. (For instance, the dependence on time cannot be dropped for the relative description pointed toward by the verbal expression «this theory is true»: the truth-value yielded by the examination of the object-entity consisting of a theory, *via* the aspect-view $V_g$ where g=truth, *does* depend on the structure of knowledge (information's, understanding, modalities of verification, etc.) available to the acting observer-conceiver at the considered time; on the contrary, for the relative description indicated by the verbal expression «the sum of the angles of a Euclidean triangle is 180°», the time dependence can be dropped). Consider then a relative description where both the space qualifications and the time-qualifications can be dropped. If no one among the involved aspects g introduces by its own definition an order (cf. D5.1), this description consists of one or several *non-ordered* but *stable* configurations of gk-values. What does this mean ? It means that the involved non-ordered configurations are characterized by some *correlations*, which are stable with respect to repetitions of the sequences [G.$V_g$] permitted by the view V, i.e. a given gk-value is found to be associated with this or that other g'k'-value (g'≠g or k'≠k or both), always, or *never* (which is as strong a correlation as always), or with this or that probability.

**D14.2.2.** ***Relative testimony***. Take again as a starting point the strong definition D14.1, and suppress now in it both the restriction to only a physical generator of object-entity *and* the condition of repeatability of the sequences [G.$V_g$] for the $V_g$ from V. What becomes of D14.1 ? It reduces to a mere set of "qualifications" generated by a definite epistemic referential. Indeed as soon as an epistemic referential (G,V) is given and the condition D7 of mutual existence is satisfied for the pair (G,V), qualifications *via* V can arise for the object-entity $œ_G$ produced by the generator G. From now on any structure of such qualifications will be called a *relative testimony* and will be denoted θ/G,$œ_G$,V/, in short θ.

**Comment**. The generalization D14.2.2 of D14.1 gives a definite status inside the MRC-language to all the qualifications of *unique* object-entities of *any* nature. In the case of physical object-entities, uniqueness is often intimately connected with space-time singularity in particular with the principle P10 of individualizing space-time mutual *exclusion*. This will come out to have a surprising importance in the identification of the characteristics of the deepest stratum of an MRC-logic (V.1.2).

Furthermore D14.2.2 introduces in the MRC-language all the qualifications of psychical events from the inner universe of a conceiver-observer.

This is a huge inclusion that lays down a foundation for the future research of a clear connection *in MRC-terms*, between introspective reports and neurological facts. Which might lead to comparability of the MRC requirements on this sort of connection, with important new views on body *versus* mind, like those of Edelman (ref. 7), Changeux (ref.6), Damasio (ref.8), and more generally with the whole avalanche of results continually produced in the cognitive sciences. Thereby the problems of reference and truth that haunt this vast recent domain might find the conceptual framework for a guided approach.

Finally, the relative testimonies in the sense of D14.2.2 permit to take into consideration the historical descriptions, the poetical ones, etc. For these the fundamental concepts of reference and truth still remain wide open for discussion and for methodological organization.

**D14.3.** ***Basic transferred relative descriptions***. In what follows we finally shall touch and transpose in quite explicit and generalized terms, the fundamental epistemological innovation specifically implied by quantum mechanics.

**D14.3.1.** ***Basic transferred relative descriptions of a physical object-entity***. Consider a relative description *in the sense of D14.1* where:

- The generator consists of a physical operation and it produces a physical object-entity that cannot be perceived directly by man. Such a generator will be called a *basic generator* and will be denoted $G^{(o)}$.
- The object-entity produced by a basic generator $G^{(o)}$ will be called a *basic object-entity* and will be denoted $œ^{(o)}$ (a simplified notation standing for $(œ_{G}(o))^{(o)}$ ).
- The view able to draw phenomenal manifestations out of a basic object-entity is necessarily such that the phenomenal content of each gk-value of each involved aspect g consists of features of a material device for gk-registrations, biological or not, but which always is *different from the studied object-entity*, these features emerging as "marks" produced by the interactions between the registering-device and replicas of the considered basic object-entity. These marks acquire *significance* by their coding in terms of values gk of the aspects from the acting view. A view of the just specified kind will be called a *basic transfer-view* (in short a basic view) and will be denoted $V^{(o)}$. The aspect-views from $V^{(o)}$ will be called basic aspect-views and will denoted $V_g^{(o)}$.
- The epistemic referential $(G^{(o)},V^{(o)})$ will be called a *basic epistemic referential.*
- A relative description in the sense of D14.1, individual or probabilistic, achieved with a basic generator and *one* basic transfer-aspect-view $V_g^{(o)}$, will be called a *basic transferred relative* ***aspect****-description* and it will be denoted $D^{(o)}/G^{(o)},œ^{(o)},V_g^{(o)}/$.
- A relative description in the sense of D14.1, individual or probabilistic, achieved with a basic generator $G^{(o)}$ and a basic transfer-view $V^{(o)}$ involving at least two *mutually incompatible* basic aspect-views $V_{g1}^{(o)}$ and $V_{g2}^{(o)}$, will be called a *basic transferred relative description* (also, in short, a basic description or a transferred description) and it will be denoted $D^{(o)}/G^{(o)},œ^{(o)},V^{(o)}/$ (in short $D^{(o)}$).
- A basic transferred description $D^{(o)}/G^{(o)},œ^{(o)},V^{(o)}/$ is posited to *characterize* observationally the involved object-entity $œ^{(o)}$, which means by definition that it is posited that no other operation of generation $(G^{(o)})'\neq G^{(o)}$ can be found which, associated with the same basic view $V^{(o)}$, produces the same basic transferred description.

**Comment**. It is difficult to fully grasp the meaning and the importance of the concept of basic transferred relative description. But it is crucial to grasp it fully. Indeed it is by *this* concept that MRC penetrates beneath natural language and the forms of thought involved by it, establishing a definite relation between conceptualization and physical factuality. Therefore I shall comment on it in detail, even redundantly.

To begin with, let us stress that a basic physical object-entity produced by a basic physical operation $G^{(o)}$, *if* furthermore this sort of object-entity has *never* before been qualified *via* any transfer-view $V^{(o)}$ whatever, emerges still entirely unknown *in terms of the knowledge researched concerning it specifically*, notwithstanding that the operation of generation $G^{(o)}$ *does* singularize it out of the whole of reality. Indeed – factually – the result labelled $œ^{(o)}$ *is* entirely "specified" by $G^{(o)}$, it is "defined", since it can be held available for any possible subsequent examination and, accordingly to the posited one-one relation between the operation $G^{(o)}$ and its result $œ^{(o)}$, it can be deliberately reproduced. More. Factually, each such result emerges from the operation $G^{(o)}$ that produced it, *fully* individualized, it lies on a level of zero-abstraction, still filled with its whole untouched concrete singularity. Which no language whatever could ever realize because we generalize as soon as we speak: full singularity is unspeakable. But – consequently in fact – this result produced by $G^{(o)}$ alone, not yet followed by an operation of examination, is individualized in *another* manner than that in which knowledge concerning it specifically, is researched; namely in only a factual physical sense, not an already conceptualized, qualifying sense. It is true that the specification of the generation operation $G^{(o)}$ involves necessarily some position of a pre-decided *conceptual* space of qualification (tied with the "zone" $R_G$ from R where G is supposed to act (cf. D4 and comment on it). By its definition $G^{(o)}$ drops its products inside this pre-decided conceptual volume. That what is labeled $œ^{(o)}$ is pre-constrained to emerge inside this or that space-time domain where $G^{(o)}$ acts, it is produced so as to correspond to some definite verbal designation ("a manifestation of stellar life", or "a state of a microsystem", etc.). In this sense $G^{(o)}$ and its result labelled $œ^{(o)}$ might be considered to never be "purely" factual. But:

The preliminarily *posited* conceptual volume where the operation $G^{(o)}$ drops all its products, cannot be equated to the *new* knowledge that is *researched* concerning these products. The elaboration of this new researched knowledge is the task left by construction for examinations achieved *subsequently* upon the already produced $œ^{(o)}$, *via* this or that basic aspect-view $V_g^{(o)}$ that exists in the sense of D7 with respect to – non specifically – anything lying inside the pre-decided conceptual volume where $G^{(o)}$ drops all its products.

It is important to realize that the specification of the operation $G^{(o)}$ of generation of an object-entity *must* contain a conceptual receptacle attached to the physical action involved by $G^{(o)}$; a conceptual receptacle to be lowered with this action into the depths of pure as yet non-conceptualized physical factuality, in order to receive inside it the results of the operation $G^{(o)}$ so as to be able to hoist them up into the stratum of the concepts-and-language. This is an unavoidable condition because *only* a receptacle made of concepts-and-language can hoist up into the thinkable and speakable a lump of pure factuality. A macroscopic operation $G^{(o)}$ can be itself shown, taught, repeated, and also said. But if nothing thinkable and speakable were posited concerning what $G^{(o)}$ produces, which by hypothesis is not perceivable, then this, the product, even if factually it has been produced, would simply stay out of conceptualization. While human mind, in order to be able to *think* about a non perceivable thing, needs, not only to have labelled it by a repeatable operation of generation and by a notation, but furthermore to have endowed it with some initializing conceptual status, with at least some approximate preliminary speakable location inside the unending and infinite-dimensional space of concepts [202].

But of course a basic description $D^{(o)}$ does not *indefinitely* produce an object-entity $œ^{(o)}$ that is still unknown, specifically and precisely in the desired terms. Knowledge about $œ^{(o)}$ is a subjective and moving character. Think of a basic description that is repeated by the observer-conceiver X after having produced for him the desired knowledge concerning $œ^{(o)}$: then, even though $œ^{(o)}$ is generated by the same generator $G^{(o)}$ and emerges beneath the level of the directly observable by man, it is nevertheless already known by X (while for another observer-conceptor it can be strictly unknown, even if the knowledge acquired by X has been made socially available in public registration devices (apparatuses, catalogues, books, etc.).

The only specific and perennial features of a "basic" description $D^{(o)}$ and of what is here called a "basic" object-entity $œ^{(o)}$ stem from the constant character of the involved referential, a "basic" referential $(G^{(o)},V^{(o)})$ where $G^{(o)}$ works on the physical factuality and $V^{(o)}$ is a transfer-view as specified in the definition D14.3.1: it resides in the fact that what is called a basic description $D^{(o)}$ consists by definition of *exclusively* features imprinted upon registering devices that are all different from the studied object-entity $œ^{(o)}$ itself.

Consider now the following question which is fundamental for the MRC treatment of reference: does indeed the definition D14.3.1 of a basic description open up a way toward a communicable characterization of – specifically – the basic object-entity $œ^{(o)}$ ? The final posit from D14.3.1 concerns this question. Consider a basic *aspect*-description $D^{(o)}/G^{(o)},œ^{(o)},V_g^{(o)}/$ (the basic view consists of only one basic aspect $V_g^{(o)}$). In this case it seems clear that $D^{(o)}$ does *not* yield a characterization – individual or probabilistic, no matter, but specifically and in an isolated way – of what is labelled $œ^{(o)}$, since it points toward observable manifestations brought forth by interactions between $œ^{(o)}$ and a material device for gk-registrations. Which changes what was labelled $œ^{(o)}$ (P10) and produces perceivable results that depend on the device for gk-registrations as much as of $œ^{(o)}$. But what about a "binocular" basic description $D^{(o)}$ where the basic view $V^{(o)}$ consists of two mutually incompatible basic aspect-views $V_{g1}^{(o)}$ and $V_{g2}^{(o)} \neq V_{g1}^{(o)}$ ? In quantum mechanics, for the particular case of a basic object-entity that is a state of a microsystem, it is (implicitly) admitted that, together, two quantum mechanical descriptions of a same microstate *via* two mutually incompatible quantum mechanical views, characterize that microstate. Which means only that no other operation $(G^{(o)})' \neq G^{(o)}$ of generation of a microstate can be assumed to yield both these same two quantum mechanical descriptions. The final posit from D14.3.1 generalizes inside MRC the above-mentioned quantum mechanical implication. It would be satisfactory of course to found this posit upon a constructed argument (for instance a *reductio ad absurdum*). But so far I did not succeed to find one. So I introduce the condition as just a supplementary security for the solidity of MRC). This completes now on the observational level the methodological posit from D4 according to which a given operation of generation of an object-entity is assumed to always produce the same object-entity. The necessity of a complement of this type can be best understood *per a contrario*. In the absence of any phenomenal, specific, normed, communicable set of qualifications associated specifically with what has been labelled $œ^{(o)}$, one would have to regard "$œ^{(o)}$" as just a label that labels nothing distinct from this label itself. Then speaking and thinking of "what has been labelled $œ^{(o)}$" would be only a void sophistic trick, amounting to arbitrary implicit postulations [203]. We would be obliged to admit that pure factuality and human communicable knowledge stay for ever apart from one another. But this just does *not* happen. Quite on the contrary, our capacity to adapt to the environment and the technical powers that we are able to acquire manifest continually the astonishing, even miraculous agreement between human knowledge and factual being, attesting intimate couplings and transmissions which somehow manage to emerge between them.

[202] It was Evelyne Andreewsky who, by repeated questions and remarks, incited me to specify how, exactly, the *pre-existing* conceptualization and the descriptional *aims* act upon the extraction of new knowledge out of as yet un-conceptualized physical factuality.

[203] Putnam's thought experiments concerning the non-determination of reference (ref. 14) are very suggestive in this respect.

The posit from D14.3.1 incorporates into the MRC-representation the assertion of a definite way in which a basic object-entity produced by a basic generator $G^{(o)}$ inside pure physical factuality, *can* be conceived to be captured there and then hoisted up into the conceptual net of inter-subjective knowledge: it is that what produces a pair of sets of mutually incompatible observable manifestations which – accordingly to the final posit from D14.3.1 – cannot be obtained by the use of any other operation $G^{(o)'}\neq G^{(o)}$.

At a first sight the concept of a basic transferred description might seem very particular, and too radical. But in fact *it possesses absolute priority and non restricted generality inside the order of cognitive elaborations*. Quite universally, any object-entity corresponding to any generator, if it *did* reach the consciousness of an observer-conceiver, then it reached it *first* by some transferred descriptions. We remain unaware of this because usually the phenomenal appearance of the gk-values involved in these transferred descriptions stems from marks imprinted directly upon the biological domains of sensitivity of the observer's body which act at the same time as generators of object-entity and as views in the sense of MRC. So the involved epistemic referentials are of a nature which, with respect to the general MRC-descriptional mould, is particular and degenerate (cf. the global comments on D14, the comments o D19.4, V.1.1 and V.1.2). This entails the following effects which occur all at the same time and beyond any control of logical consistency:

(a) It hides the transferred character of the marks.

(b) It inclines toward assigning systematically a passive role to the mind, in its interactions with physical factuality. The mind is supposed to just *receive* marks irrepressibly imprinted upon the sensitive apparatuses of the body by incessant streams from the physical factuality. (How far one is thus kept from realizing the possibility and the universal methodological value of two radically distinct epistemic stages which, in general, have to be both active during a deliberate achievement of "unnatural" transferred descriptions, like those on which quantum mechanics throws light!).

(c) It pushes surreptitiously toward ontological absolutizations. Indeed one encounters severe difficulties to realize that the (various) transferred descriptions of this chair, which my consciousness functioning achieved spontaneously by the help of my biological views (involving the eyes, the nervous system, the ears and fingers, etc.), cannot, without contradiction, be identified with "the-way-in-which-the-chair-in-itself-really-is"; that nothing, never, will be able to prove that this or that model of a chair "exists" independently of any perception, of any view. More, that such an instinctive hope contradicts both philosophy and logic, since in the absence of any view the very concept of description, and even that of merely an isolated qualification, vanishes (cf. π18, D19.1, D19.2). It is really hard to withstand the irrepressible trend toward identification of our spontaneous modelling that stems from descriptions transferred on the human biological registering devices, with ontological credos that float on self-contradicting assemblages of words, alike to Magritt's tree that floats with its roots in the air. Kant, Poincaré, Einstein, Husserl, Quine, Wittgenstein, Putnam, have founded famous analyses on the explicit recognition of this fact.

But, and this is noteworthy, as soon as the transfer-view from a considered basic transferred description $D^{(o)}$ does not directly involve the biological human terminals – the nearest and which *in fine* cannot be eliminated –, as soon as the transfer-view $V^{(o)}$ from $D^{(o)}$ involves marks registered on devices that are exterior to the observer's body (as it happens indeed for microstates), it suddenly becomes quite clear that a basic description $D^{(o)}$ itself constitutes a constructed intermediary object-entity which relays the access of the basic a-conceptual object-entity $œ^{(o)}$, to the observer-conceptor's consciousness-functioning; that phenomena are not always independent of aimed volition, that they are not always just psycho-physical facts which emerge spontaneously, but might have to be planned and produced by method. Then, like in quantum mechanics, the two distinct and mutually independent stages involved in a transferred description – the stage of generation of an object-entity $œ^{(o)}$, and the subsequent stage of creation of observable manifestations drawn from $œ^{(o)}$ by interaction with gk-registering devices – appear as obvious. Their active and deliberate character strikes the mind, and the invaluable normative value of the concept of basic transferred description can be fully understood.

The basic object-entity $œ_G{}^{(o)}$ from a transferred description $D^{(o)}$ roots this description directly into the physical factuality. Correlatively the transferred description $D^{(o)}$ achieves for the involved basic object-entity $œ_G{}^{(o)}$ a very first passage from pure physical factuality, into the domain of communicable knowledge. It yields for it a first communicable form, a first observable expression that points communicably toward the involved object-entity. So the basic transferred descriptions are the *local zero-points* of the chains of conceptualization, in the following sense. Each basic transferred description $D^{(o)}$ starts from a conceptual situation where, though a conceptual environment of the basic object-entity $œ_G{}^{(o)}$ (genus, etc.) always is more or less explicitly posited *a priori* (at least *via* the definition D4 of $G^{(o)}$), nevertheless nothing is known concerning $œ_G{}^{(o)}$ *specifically*.

The very first stratum of communicable knowledge available at any given time consists of the basic transferred descriptions achieved up to that time, not of just phenomenal appearances in the Kantian sense.

*The transferred descriptions are the channels through which as yet non semantized but semantizable factual matter, is adduced into the domain of the inter-subjectively semantized. The "scientific legalization of phenomenal appearances" in Kant's sense (II.3) begins by the construction of transferred descriptions, of which $D^{(o)}$ yields a form that is* ***normed****. Which amounts to a* ***formalization*** *of the structure of the connections between knowledge and Being.*

This is a quite fundamental contribution of MRC to epistemology. It separates the volume of the known, in two essentially different strata. Indeed the whole rest of the available knowledge consists only of subsequent developments of this first – evolving – stratum of transferred descriptions which operate the very first connections between between Being and knowledge: namely, it consists of space-time modelizations which endow the basic transferred descriptions with the features required by the frame-postulate P8, thus insuring for them an "intelligibility" of which initially they are devoid. A *non limited* succession of descriptional complexifications can then indefinitely improve these space-time modelizations (cf. D16, D.19, and all the involved discussions).

I add a last remark concerning the concept of basic transferred description. From the viewpoint of MRC the quantum mechanical descriptions of micro-states appear as just particular instances of transferred descriptions of physical entities: the strategy of quantum mechanics, once identified explicitly, brings into evidence an example of the universal way in which the conceptualizations are rooted into pure physical factuality, and, for this example, it displays all the stages of the rooting. MRC recognizes the universality of this rooting and extends it to any sort of physical factuality, re-expressing it in general and *normalized* terms.

**D14.3.2.** ***Basic description of a psychical object-entity?*** Notwithstanding important difficulties (the non pertinence of the repeatability of the successions $[G^{(o)}.V^{(o)}]$ and of the stability of their results), it might turn out to be possible to forge a useful concept of basic description of "psychical basic object-entities $œ^{(o)}$", by some combination of testimonial descriptions θ in the sense of D14.2.2, with "biological basic transferred descriptions". Thereby I mean a *conscious* but *not* yet conceptualized psychical object-entity, a *primeity* in the sense of Peirce that emerges in the acting observer-conceptor's *interior* universe, and, though perceived, is still entirely *un*known, non-qualified (A. Damasio (ref. 8) has elaborated a very subtle structure of concepts-and-facts concerning events of this sort). Think for instance of all the feelings of mere *existence* of an inner fact of which one becomes suddenly aware strictly without knowing as yet explicitly what and how they are, so *a fortiori* without understanding them; think of the genuine *research* conducted by Proust in order to identify the subjective *meaning* of such feelings; think also of the psychoanalytic methods which deal with features as if transferred upon behavioural "devices" (reactions, ways of acting, feelings) by interactions between a hypothetical entirely unknown inner configuration, and various accidental or systematically arising exterior circumstances; this hypothetical inner configuration is precisely what the therapies try to first somehow delimit "operationally" (by analyses of dreams, associations, etc.) – even if by *creating* it– and then to interpret, qualify, and control or suppress. The obtained description *is* then in a certain sense precisely what seems to deserve being called a basic relative description of a psychical object-entity.

It is however clear that for the moment these are just conjectures. The central concept of basic transferred description has an indisputable pertinence only with respect to physical object-entities.

**Global comment on the definitions D14**. Finally, let us now consider globally the whole set of definitions D14 and make some comments on the general concept of relative description.

The general notation $D/G,œ_G,V/$ stresses that any description that is normed in the sense of MRC brings into play a triad $G,œ_G,V$ to which it is essentially relative: this is the general descriptional mould induced from quantum mechanics and required now for any description, whether it is basic, transferred, or not. The first location from this triad is the place reserved for an epistemic action, the generation of an object-entity, which up to now has quasi systematically been ignored, because the canonical basic transferred descriptions where the generation of an object-entity plays a separate and active key role, were ignored. Indeed for a description that is not transferred, the operation of generation of the desired object-entity is often accomplished without any difficulty, in a spontaneous or even implicit way (think of descriptions of conceptual entities, like a definition, etc.). While when the transfers occur on – directly – the biological sensorial apparatuses (views, in the sense of MRC), the involved view V *acts also like a generator G* which just selects out of R an object-entity, namely the field of perceptibility of V, and – simultaneously – also qualifies this object-entity: we can symbolize by G(V) such a *generator of a view* and by (G(V),V)) the corresponding epistemic referential. In this case the action of a generator of object-entity is still deeper hidden than in the preceding case. This highly degenerate and so wide-spread natural situation contributed strongly to the lasting occultation of the fundamental role of principle of the operations of object-entity generation. Quantum mechanics, for the first time and only implicitly, made a separate use of the operations of generation of object-entity, which permitted to this author to become aware of their general and fundamental epistemological importance.

*The generator of object-entity remained the big omission of the grammars, the logic, and of all the approaches that involve the processes of conceptualization*.

This is why the question of reference has raised insuperable problems: the basic object-entities are only *surreptitiously* drawn into the natural basic descriptions (the *degenerate* ones produced in a reflex way *via* the biological sensorial apparatuses), with the status of a present but non specified reference. The problem of identifying *a posteriori* of what this reference consists, starting from the already achieved description, has stubbornly resisted solution.

But accordingly to MRC, an operation of generation of object-entity is *always* involved, even if in a non separated and implicit or reflex way.

By construction, any relative description $D/G,œ_G,V/$ is, itself, distinct from the generator, the object-entity and the view involved by it, to all of which it is conceptually posterior; it qualifies only the object-entity which it concerns, not also the generator and the view of which it makes use, nor itself, globally. As for the generator and the view, these are by definition distinct from one another, often by their content, but in *any* case by the *role* held during the process of description.

*In the definition of a relative description the three notations $G,œ_G,V$ designate three descriptional* ***roles****, three descriptional functions,* ***not*** *the nature of the entities to which these roles are assigned in the case of this or that particular relative description*.

And all these three roles are systematically played in any relative description, even if an actor cumulates distinct roles, or plays a role superficially, or both. For instance, if I say «"red" is a too poor expression, better say "colour of blood"», the first proposition expresses verbally a relative description $D/G,œ_G,V/$ where "red", though grammatically it is an attribute, holds the role of the object-entity $œ_G$ (generated by use of a generator G which is a selector acting upon the spot $R_G$ from R indicated by the word "colour"), while "poor" is placed in the role of the view V. But if I say «my cheeks are red», "red" plays the role of the view. So the structure required by the definition D5.1 of an aspect-view, is only a necessary condition for acting *as a view*, but this condition does not hinder a view in the sense of D5.1 to act also in the *role* of an object-entity (like in the first above example) or in the role of a generator G(V) of object-entity that generates its field of perceptibility by interaction with R.

*According to MRC no operation or concept possesses intrinsically a fixed descriptional role*.

In each descriptional act, the descriptional roles are assigned by the acting consciousness functioning, and in general this roles change from one description to another one. When a natural description is examined in order to compare it to the MRC norms, the first step is to examine what plays the role of object-entity, what the role of generator, and what that of view. A description $D/G,œ_G,V/$ is a piece of constructed normed meaning which, essentially and explicitly, is relative to the epistemic actions that achieved the semantization asserted by it. *Any* asserted meaning bears inside it the genetic structure designated by the sign $D/G,œ_G,V/$, but it can include this structure in a more or less implicit, truncated, malformed way. Whereas in the normed form $D/G,œ_G,V/$ all the three involved roles $G,œ_G,V$ are explicitly indicated, each one at its own location and following the genetic order of the corresponding epistemic actions. They are to be treated as void, available, labelled rooms that have to be filled up in a reference-questionnaire to which any achieved or envisaged description must be subjected.

The distinction, inside a relative description $D/G,œ_G,V/$, between the relativity to the operation G of object-entity generation of which the role is to produce an object-entity, and the relativity to this object-entity $œ_G$ itself of which the role is to bear subsequent qualifying examinations, is one of the most subtle and important features of MRC. In particular it preserves from the very strong inertial tendency induced by classical thinking, to forget that as soon as an entity is regarded as playing in a description the role of object-entity, *ipso facto* a corresponding epistemic action of generation of object-entity has produced it *as such*, implicitly or explicitly, even if this entity somehow pre-existed and so has only had to be selected as object-entity, not to be radically created as such. The importance of a normed *memento* of this fact will fully appear in V.1 and V.2.

The association, in any relative description $D/G,œ_G,V/$, between a one-one relation $G–œ_G$ and the requirement for D of, indifferently, either a strong individual stability or an only probabilistic one, is intimately related with the impossibility, for mere language as well as for mere notations, to grasp and capture the factual individualities, neither in an absolute sense *nor* in only a relativized sense (cf. π12, its "proof" and the comments). Umberto Eco remarks: «The tragedy comes from this that man speaks always in a general manner about things which always are singular. Language names, thus covering the non transcendable evidence of individual existence» [204]. Indeed each predicate (view) is general, and no conjunction of a finite number of predicates can ever exhaust the open infinity of the possible qualifications of a physical object-entity.

In this context, let us note that *full*, non-verbalized factual singularity can be associated with the one-one relation posited between an operation of generation G and its result labelled $œ_G$, in the following sense. According to this posited relation, $G'\neq G$ entails $œ_{G'}\neq œ_G$. Which can be translated in observational language as follows: if two object-entities are introduced by two different generators G and $G'\neq G$, then it exists at least one view V that yields different descriptions of $œ_G$ and $œ_{G'}$. This assumption is what founds the belief in "experts", for instance experts able to discern an original painting from a copy, no matter how perfect.

The concept of relative description is selective. It does not admit inside the class delimited by it, illusory descriptions where one of the three roles G, $œ_G$, V is not played at all. Consider for instance the famous illusory description «this is a lie» (or «I am a lie»)» where the word "this" (or "I") masks the absence of specification of the operation G of generation of object-entity, so also *the absence of specification of the object-entity $œ_G$ itself.* This blocks any further conceptual development. Indeed, previously to any research of a truth-qualification of the description, one finds oneself in a situation of impossibility to decide concerning the mutual existence in the sense of D7 between the involved object-entity $œ_G$ – non specified – and the involved view V. If this primary non-decidability concerning the *a priori* possibility of meaning, were permitted to enter the concept of relative description, it would manifest itself later in the form, also, of a paralysis of any attempt at a meta-qualification of the *relative proposition* founded on this illusory description *via* the values gk=true or gk=false of a meta-aspect-view .g=empirical truth (cf. DL.2 and DL.3 in V.1.2).

When descriptions that violate the MRC norms, are reconstructed in a normalized way, the paradoxes stemming from them disappear. There is no need for this to introduce levelled languages of logical types, the illness is cured *locally* by the normed reconstruction of only the considered description.

---

[204] Eco, U., *Kant et l'Ornithorynque*, Grasset 1999, p. 29. My translation from the French edition.

But nothing hinders to generate (select) as an object-entity any natural description excluded by MRC, and to characterize its incapacities or specificities by reference to the MRC-norms. In this sense the methodological selectivity of the concept D/G,$œ_G$,V/ by no means constitutes an *a priori* pauperisation of the ensemble of descriptions that can be studied inside MRC.

Finally, the general concept of relative description, by its various realizations, permits to discern definite categories inside the realm of the problem of reference and of meaning, and a *dégradé* of proposed solutions: the definitions D14.1, D14.2.1 and D14.3.1 introduce, for the corresponding circumstances, what might stand as a solution or be completed to become one; the definition D14.2.2 suggests a possible approach concerning some of the circumstances to which it applies, while others are isolated as the most problematic; finally, the non achieved definition D14.3.2 concentrates in it definite questions and suggestions.

Like the one-one relation between a given generator of object-entity and the corresponding object-entity, like the definition of relative existence and then the frame-principle P8, the concept of relative description with the three roles involved by it, is an act of (qualitative) *formalization*, involving a methodological essence.

#### 4.2.3. Cells of relative description. Chains of descriptional cells. Non-reducible complexification of the conceptualization.

**P15.** ***The Principle of Separation***. Since any one relative description D/G,$œ_G$,V/, whatever its complexity, involves by construction one generator of object-entity, one object-entity, and one view, all well defined, as soon as some change is introduced in the actor designated for holding one of the roles from the triad G,$œ_G$,V, another description is considered.

By a methodological principle called the *principle of separation* and denoted PS, this other description must be treated ***separately***.

**Comment**. Any human observer-conceiver, in presence of reality, is condemned to parcelling examinations. The successivity inherent in human mind, the spatial confinements imposed by the bodily senses – whatever prolongation is adjusted to them – and the absence of limitation of what is called reality, compose together a configuration which imposes the fragmentation of the epistemic quest. MRC reflects this situation in the relativity of any one description, to one triad G,$œ_G$,V. Indeed the relativity to one triad G,$œ_G$,V specifies, but also limits the capacity of one given relative description to generate information possessed.

Relativization, limitation, and precision, are *tied* to one another in an un-separable way. *They constitute together an indivisible whole that withstands relativism*.

On the other hand any fragment generated out of reality in order to play the role of an object-entity, admits of an infinity of kinds of examinations. Moreover any examination achieved on this object-entity, raises the question of the appearance of its result *via* this or that view with respect to which this result exists in the sense of D7, or the question of the relations of this result, to other object-entities, etc., thus multiplying the conceivable subsequent object-entities and examinations. These confinements and these endless and changing vistas call forth haste and panics of the mind which entangle in knots of "paradoxes" and block the understanding. So they also block the further development of the started conceptualization. The limitations imposed by each specified description are flooded by the implicit fluxes of the rush toward more conceptualization. Without being aware of this, mind yields to whirls of implicit interrogations which generate a subliminal tendency to fluctuate between different operations of generation of an object-entity and different views; a tendency to work out simultaneously several different descriptions. But as soon as the elaboration of several different relative descriptions is simultaneously tried, the various involved generators of object-entity, object-entities and views, are offered a ground for oscillation. And then the oscillations actually happen, because it is very difficult to perceive them, so *a fortiori* to hinder them. So the different descriptions that are simultaneously entered upon, get mixed, and in general none of them can be achieved. Their interaction coagulates nonsense that stops the conceptualization.

The principle of separation hinders such coagulations. It requires the conceptualization, by method, to be achieved by explicit separation in mutually distinct, successive, closed, cellular descriptional steps.

In particular the principle of separation PS surveys the saturation of a description. It rings the bell as soon as the descriptional capacities of a started description must be considered to have been exhausted, because all the qualifications *via* the view chosen for acting in that description, of the object-entity corresponding to the generator chosen for acting in that description, have been already realized by performing a big number of repetitions of all the successions [G.$V_g$] available in that description. PS announces that once this has been done, the descriptional cell potentially delimited by the chosen epistemic referential (G,V) has been saturated with actualized qualifications; that from now on any attempt at obtaining new information inside this same epistemic referential, either is useless or it manifests the surreptitious intrusion of another generator of object-entity, or of another view, or both; that – to avoid stagnation, paradoxes or infinite regressions – one has to stop this intrusion or mixture, by identifying the new epistemic referential that weighs with subliminal pressure upon the consciousness functioning, and by putting it explicitly to work in its own turn, separately.

The systematic application of the principle of separation plays, in the development required by MRC for a process of conceptualization, a role similar to that hold by the sign "." or the word "stop" in the transmission or writing down of a message; or else, a role similar to that played in algebra by the closure of a previously opened parenthesis.

The principle of separation PS is a *formalizing* requirement of the nature of a rule of calculus.

Thereby any process of conceptualization that is normed accordingly to MRC, is clearly divided in a sequence of localized descriptional cells, and thus it develops by systematically renewed *local* frameworks, under systematically renewed *local* control.

While the tests of mutual existence (D7) detect the *a priori* impossibilities to construct meaning, the principle of separation permits to avoid any stagnation – illusory paradoxes, infinite regressions – throughout the processes of development of meaning.

The concepts of mutual inexistence, and the principle of separation, co-operate for the task of preventing sources of unintelligibility, and also of detecting and suppressing them.

The principle of separation possesses a remarkable capacity of organization of the conceptualization. This assertion will find many illustrations in the sequel of this work.

**D16.** ***Relative metadescription***. The principle of separation requires descriptional closures and new starts. These entail the necessity of an explicitly and fully relativized concept of metadescription prescribing how to transcend "legally" an already saturated description.

Consider a precedingly achieved relative description to which the *order* 1 is assigned conventionally: $D^{(1)}/G^{(1)},œ^{(1)},V^{(1)}/$ (in short $D^{(1)}$; and instead of $œ_G$ we write œ, to simplify the graphism). Consider a generator that selects $D^{(1)}$ as a new object-entity $œ^{(2)}$, denote it $G^{(2)}$ and call it a *metagenerator (or a generator of order 2) relative to $D^{(1)}$*. So we have $œ^{(2)} \equiv D^{(1)}$. Consider also a view involving *aspects of order 2* with respect to which $D^{(1)}$ does exist in the sense of D7 (for instance the aspect of factual truth of $D^{(1)}$, or else some aspect of *relation* inside $D^{(1)}/G^{(1)},œ^{(1)}{}_G,V^{(1)}/$, between the various gk-space-time qualifications produced by the examinations of $œ^{(1)}$ by the initial view $V^{(1)}$, etc.; call it a *metaview (or a view of second order) relative to $D^{(1)}$* and denote it $V^{(2)}$. The description which is relative to the triad $G^{(2)},œ^{(2)},V^{(2)}$ will be called a *metadescription (or a description of order 2) relatively to $D^{(1)}$* and it will be denoted $D^{(2)}/G^{(2)},œ^{(2)},V^{(2)}/$ (in short $D^{(2)}/ D^{(1)}$, or $D^{(2)}$).

The same denomination and notation are conserved if ***(a)*** $G^{(2)}$ selects as a new object-entity $œ^{(2)}$ not only $D^{(1)}$ considered globally, but furthermore it includes in $œ^{(2)}$ also separate elements from $D^{(1)}/G^{(1)},œ^{(1)}{}_G,V^{(1)}/$ specified explicitly ($G^{(1)}$, or $œ^{(1)}{}_G$, or $V^{(1)}$, or two or all three of them) which permits then to introduce in $V^{(2)}$ aspects of relation between such an element, and the global result $D^{(1)}$ to which it has contributed. Or if ***(b)*** $G^{(2)}$ selects a whole *set* $\{D^{(1)}{}_1, D^{(1)}{}_2,...D_m{}^{(1)}\}$ of previously achieved relative descriptions (with an explicit reconsideration, or not, of elements from these descriptions), in which case $D^{(2)}$ is relative to all these descriptions. In this way a very free and rich concept of normed relative metadescription is introduced [205].

**Comment**. The definition D.16 can also be applied to $D^{(2)}$ thus leading to a metadescription $D^{(3)}$ of order 3 relatively to $D^{(1)}$ and of order 2 relatively to $D^{(2)}$, etc. In this way it is possible for any consciousness-functioning CF to develop unlimited descriptional chains $D^{(1)},D^{(2)},...D^{(j)}...$ $D^{(n-1)}$, $D^{(n)}$ of hierarchically connected relative descriptions of successive orders j=1,2,....n – with an arbitrary origin denoted $D^{(1)}$ – in each one of which the involved metaview can contain all the desired pertinent new meta-aspects of order n.

So *in general the order of a description is not an absolute*, it labels the place where this description emerges inside the considered chain of conceptualization, while a chain can be started *conventionally* by these or those previously achieved descriptions to which the order 1 is assigned.

*But a basic transferred description can only have the* ***minimal*** *conceivable order, no matter in which chain it is involved.* Therefore this non-conventional minimal order will be denoted by 0, to distinguish it from any conventional initial order 1.

And any chain, if it has first been conventionally started with already previously achieved descriptions to which the order 1 has been assigned, can always be later completed downward until a basic transferred descriptions is identified which roots the chain into pure factuality. Thereby the chain hits an absolute end (or equivalently, it finds its absolute beginning), which entails a corresponding re-notation upwards of all the successive orders of the involved descriptional cells. But a given relative description can belong to different chains that meet in it (it can be a node of the web of chains of conceptualization). So, regarded as a cell from distinct chains, a same description can have different orders.

---

[205] Here we can go back to the important distinction from the note 20 between "objectual" qualifications – call them "objectities" – and "state"-qualifications (note 20). The objectities are (relatively) stable qualifications that apply in an invariant way to a whole class of evolving states, thereby definig the "object", in the current sense, that assumes this or that state. So according to this language *the term object-**entity** labels only* ***a descriptional role*** *in the sense of the general comment of D14, while "object" in the current sense means «endowed with some objectities»*: inside MRC these two words should not be confounded. For instance, the state-qualifications called position, momentum, energy, etc., can vary or evolve from one state to another one, thereby introducing an infinite class of states of a definite sort of "object" labelled, say, "electron", that is characterized by the *meta*-qualifications consisting of the numerical values obtained (with some given system of unities) for objectities like mass, charge, spin, that are the same inside the whole class of what is called "states of electrons". These objectities however can themselves change by creation or annihilation of the corresponding object, and when the conditions for such changes are realized they can be regarded as states of some more general object (at the limit, of what is called field or energetic substance). In this way the language introduced here can organize conveniently various hierarchies of degrees of abstraction.

*But the feature of being a metadescription (or not), is an absolute **if** transferred descriptions constitute the origin used as reference,* since the zero order of a transferred description is an absolute.

This amounts to the remark (rather obvious *a posteriori* ) that:

The (open) set of all the possible relativized descriptions falls apart in just *two* (evolving) layers: ***(a)*** the layer of transferred descriptions of physical basic object-entities which, by definition, are not themselves previously achieved descriptions, and ***(b)*** the layer of metadescriptions in the absolute sense, i.e. of descriptions of object-entities consisting of previously achieved descriptions [206]. Both layers have an evolving content.

Through the first layer, the prime matter for the elaboration of meaning is drawn into conceptualization, and inside the second layer the basic meaning produced in the first layer undergoes abstract transformations which progressively elaborate indefinitely complexified meanings.

It is essential to note that in any chain, for each passage from a descriptional level n to the following level n+1, the new epistemic referential to be used ($G^{(n+1)}$,$V^{(n+1)}$) is freely decided by the acting consciousness-functioning CF, as an expression of his own (evolving) descriptional curiosities-and-aims, such as these emerge at any given time from his own biological, temperamental, and social-cultural background: it is the consciousness-functioning CF who, step by step, chooses the "direction" of the descriptional trajectory drawn by the succession of the cellular but connected descriptional closures $D^{(n-1)}$, $D^{(n)}$, $D^{(n+1)}$,.... which, accordingly to [P15+D16], produce the indefinite progression of a hierarchical chain started by conventionally initial conceptual descriptions $D^{(1)}$ or by absolutely initial basic descriptions $D^{(o)}$.

So – as long as no method or algorithm is found for determining as a function of some definite parameters – a new epistemic referential each time that a passage from a description to a metadescription (with respect to it) takes place, a descriptional chain remains a concept that cannot be absorbed in the concept of computation. And even if such an algorithm were specified, furthermore also the determination of the parameters on which the new referential depends should emerge automatically: accordingly to what criteria ? Etc. The subjective successive descriptional aims play a decisive role in the representation of the processes of conceptualization offered by MRC. But on the other hand, the representational structure assigned by MRC to the processes of conceptualization, namely the structure of a web of chains of increasingly complex relative descriptions, is a (qualitatively) formalized structure, involving definite methodological rules and conventions.

*This brings clearly into evidence that "a formalized epistemology" in the sense of MRC is quite fundamentally distinct from a reduction to computation.*

Once this established, let us furthermore examine below the question of reductions of another sort.

**Π17.** ***Anti-reductionist proposition***. Inside MRC the "reduction" of a metadescription of order n (D.16) to the descriptions and elements of descriptions of order n-k, k=1,2,...n-1 involved in it, is in general impossible.

***"Proof"***. Consider the metaobject-entity $œ^{(n)}$) from a metadescription which, inside the considered chain, is of order n, $D^{(n)}/G^{(n)},œ^{(n)},V^{(n)}/$. An isolated element from $œ^{(n)}$ (a description $D_j^{(n-1)}$ of order n-1, or some other descriptional element of order n-1 from such a description (generator, object-entity, view)) in general simply *does not exist in the sense of D7 with respect to the new meta-aspects of order n from $V^{(n)}$*. For instance, a metaview $V^{(2)}$ of order 2 from the metadescription $D^{(2)}/G^{(2)},œ^{(2)},V^{(2)}/$ relatively to $D^{(1)}/G^{(1)},œ^{(1)},V^{(1)}/$, can contain the aspect of *distance* between two space-gk-qualifications of order 1 involved by $D^{(1)}/G^{(1)},œ^{(1)},V^{(1)}/$, with respect to which these qualifications *themselves* do not exist in the sense of D7. Or else, $œ^{(2)}$ can contain two previously achieved descriptions of physical object-entities, $D_A^{(1)}$ and $D_B^{(1)}$ involving both a same view $V^{(1)}$ (so qualifications of a same nature) while $V^{(2)}$ contains a meta-aspect of order 2 of *comparison* of these qualifications, whereas neither $D_A^{(1)}$ alone nor $D_B^{(1)}$ alone, nor descriptional elements from these, do exist in the sense of D7 with respect to this meta-aspect of comparison. In general terms now, the new qualifications of order n that can be involved in a metadescription $D^{(n)}$ while they cannot be involved in the descriptions of order n-1 contained in $D^{(n)}$, consist of *global* or *connective* metaqualifications of order n concerning two or more descriptional entities of order n-1 from the object-entity $œ^{(n)}$ from $D^{(n)}$ (consisting of whole descriptions of order n-1, or generators of object-entities, or object entities or views, of order n-1). These, when considered separately inside the descriptions of order n-1, *do not exist in the sense of D7 with respect to any of such new metaqualification of order n involved by $D^{(n)}$*.

So in general $D^{(n)}$ is not reducible to the descriptions or descriptional elements of orders n-k from the same chain.

**Comment**. On each descriptional level of a given order n from a descriptional chain (D.16), the descriptional cell $D^{(n)}$ placed on this level introduces, *via* the condition of relative existence D7, the possibility of new qualifications, of

---

[206] However it is curious to note that there are various sorts of rooting of a basic object-entity, into pure factuality: the *objectual* manifestations of a basic object-entity, in the sense of the note 31 can be *conceived* (not known, just imagined) to be tied with pre-existing "own" features of this basic object-entity (cf. D19) which, though unknown, are always the *same*. In this sense, a basic object-entity which is *a priori* researched as located inside the genus labelled micro-*object* (i.e. is researched exclusively *via* objectual manifestations) is thereby *a priori* endowed with a rooting into pure factuality which is less hidden than that of a basic object-entity researched *a priori* as located inside the genus labelled micro*state*, because it is posited to reach the level of observability by just a time-invariant coding transposition, not by the coding of the effects of a (measurement) *evolution* produced by the processes of examination. These remarks amount to the assertion of various possible deliberately chosen *depths* of the rooting of a transferred description, into physical factuality.

which the very definibility and meaningness are conditioned by the previous achievement of the descriptions from all the previous levels n-1, n-2, ....n-n:

Throughout the development of a process of conceptualization normed accordingly to MRC one can literally watch the creative complexifying work of cognitive time: *one can literally see what "emergence" means.*

It is remarkable that inside MRC this conclusion follows from the system of basic definitions, postulate and principles, in a way that permits a clear perception of the nature of each contribution to the conclusion. One can distinguish between contributions of a factual nature as for instance those brought in by a basic description $D^{(o)}$, and on the other hand contributions of psychological nature like the choices of epistemic referentials for the successive descriptional cells, or of methodological nature like the condition D7 of mutual existence and the principle of separation P15:

There is no need any more for pleading, arguments, etc., in order to draw attention upon the specific character, the mechanisms and the features of what is labelled by the words "complexity", "complexification", "emergence".

So, by normed complexification, the transferred descriptions that start from the inside of pure factuality and by which phenomena acquire a first communicable form, are then developed in unlimited chains of hierarchically connected metadescriptions of increasing order. These chains can meet and interact variously at various levels and thus they weave indefinitely compexifying and non predictable forms of communicable significance.

The consequences of the association between the principle of separation and the concept of relativized metadescription, are innumerable and always important. But in the absence of a normed descriptional structure to which any description be referable, they cannot be systematically identified and controlled.

**4.2.4. Reference, and minimality of the MRC-realism**

In this stage of the elaboration of MRC it is already possible to entirely elucidate *a posteriori* the *a priori* somewhat obscure features introduced by the definition D4 of a generator of object-entity (the posited one-one relation $G–œ_G$) and by the realist postulate P3 (cf. note 25). We shall now achieve this by a succession of three propositions. Thereby also the reflexive character of MRC will gain new illustrations, while the formalized character of MRC will become clearer.

**π18**. ***Propositions on reference and minimal realism***.

**π18.1.** ***(On comparability, identity, and the relation*** $G–œ_G$***)***. A basic object-entity is inexistent in the sense of D7 with respect to any "comparison-view": such a view is a metaview with respect to which only descriptions exist in the sense of D7, never basic object-entities.

***"Proof"***. What is not already pre-qualified cannot be compared. Only two (or more) previously achieved descriptions $D_1$ and $D_2$ can be compared, and only concerning some definite aspect-view or view with respect to which these descriptions do both exist in the sense of D7. One can for instance ask: are $D_1$ and $D_2$ identical or different with respect to this or that gk-value of the aspect-view $V_g$? If $V_g$ is absent in one or in both considered descriptions, the question is *meaningless* because $D_1$ and $D_2$ constitute together a meta-object-entity $(D_1,D_2)^{(2)}$ that does not exist in sense of D7 relatively to a metaview of g-comparison, say $V^{(2)}{}_{gc}$, so *a fortiori* a gk-identity can be neither established nor refuted. If on the contrary both $D_1$ and $D_2$ do make use of $V_g$, then $(D_1,D_2)^{(2)}$ and $V^{(2)}{}_{gc}$ do satisfy D7 and so one can research whether yes or not they do possess some gk-identities. In this example I have brought into play a most simple comparison-view, with respect to only one aspect g. Nevertheless this view is already, quite essentially, a *meta*view. One can form much richer metaviews of comparison. But all are metaviews relative to definite views with respect to which only previously achieved descriptions can exist in the sense of D7.

A basic object-entity – a bulk of pure a-conceptual factuality – is not a previously achieved description. Therefore it cannot be compared, neither to "itself" nor to something else.

**Comment**. So the whole stratum constituted by the very first products of the epistemic actions – the stratum of basic object-entities introduced by basic generators – is not reachable by the concept of comparison and by the qualifications derived from it, identity, difference, degree of similitude. For basic object-entities these qualifications cannot be established by investigation, they can only by posited by method (like in the definition D4 of a generator of object-entity). When a given basic operation $G^{(o)}$ of generation of object-entity is repeated, it simply is meaningless to ask whether yes or not the object-entities $œ^{(o)}$ produced by this operation are all identical: this finally *founds "deductively" inside MRC the* ***impossibility*** *to assign a general meaning to the question whether yes or not the repetition of a given operation G of generation of an object-entity* $œ_G$*, produces identical results* $œ_G$. So the posit of a one-one relation $G–œ_G$ appears *a posteriori* to be necessary indeed in order to be always able to speak and think fluently concerning the products of G; while the significance of this posit, already specified to a certain degree in the comment on π12, becomes now fully clear.

The one-one relation $G\text{-}œ_G$ founds a methodological strategy according to which the reference $œ_G$ – *defined from the start on* and posited to be unique – associates coherently with, both, the *a priori* condition of possibility in the sense of D7 of an as yet non-defined meaning of $œ_G$ with respect to a given view V, and with a subsequently constructed specified meaning of $œ_G$ with respect to V (while for another view $V' \neq V$, the relative existence D7, or a meaning of $œ_G$, or both, might fail to exist).

Thus *the question of reference obtains a self-consistent and effective solution.*

**π18.2**. ***"Local" proposition on the realist postulate***. Consider a *physical* object-entity $œ_G$. This is a fragment of physical reality generated by a given physical operation of generation G. The fact that any communicable knowledge is description, and the relativity of any basic description to a basic view, *entail* that the sequence of words "knowledge of how $œ_G$ is in itself" is void of significance.

***"Proof"***. Consider a physical object-entity $œ_G$. Any communicable knowledge concerning $œ_G$ amounts to some relative description D/G,$œ_G$,V/. Any relative description D/G,$œ_G$,V/ belongs to some net of descriptional chains that is rooted in pure factuality *via* a (finite) number of *basic* transferred descriptions $D^{(o)}/G^{(o)},œ_G^{(o)},V^{(o)}/$ the basic object-entity $œ_G^{(o)}$ from which somehow contributed to $œ_G$, has hereditarily transmitted into $œ_G$ some of its own semantic substance. Now, in each one of these basic transferred descriptions, the transfer-view $V^{(o)}$ acting there yields for the involved basic object-entity $œ_G^{(o)}$ a very first access to observability. But the principle P10, the propositions π11, π12, π13, and the definition D14.3.1 of a basic description, show that, and how, the basic transfer-view $V^{(o)}$, while it yields this first access, also *inserts a non removable opaque screen between the acting consciousness-functioning CF and «*$œ_G^{(o)}$*-in-itself»*, it bars the way of human knowledge toward «$œ_G^{(o)}$-in-itself». So the unavoidable and non removable descriptional relativities explicated inside MRC, and the fact that any communicable knowledge is description, entail inside MRC that [knowledge-of-the-physical-reality-as-it-is-in-itself] is nothing more than a meaningless combination of words, devoid of any designatum.

**Comment**. Since Kant the impossibility to know how a physical entity "is-in-itself", is accepted as an obvious postulate inside philosophy. But many physicists still are reluctant to fully realize this definitive limit of human rational knowledge. So is seems worth mentioning explicitly that inside MRC this limit follows from the posited assumptions without being one of these. So that there is no need to assert it as a logically independent assumption. Then those who contest this limit should specify which posited assumption(s) they contest.

**π18.3**. ***"Global" proposition on the realist postulate: minimality***. Inside MRC the realist postulate P3 can only be given a *minimal* significance: it can only be understood to assert *exclusively* the credo of the ***existence***, apart from the interior reality from my own mind, of also a physical reality independent of any act of observation; but an existence which is strictly *non-qualifiable* "in-itself", beyond the mere trivial and non-informative, idempotent assertion of its relativized qualifi*ability*, if acts of observation of it do take place in the conditions D4-D7 (in the absence of which P3 would be aimless).

***"Proof"***. According to the definition D2, "the physical reality", globally considered, is just a *posited* substratum wherefrom all the *basic* object-entities $œ_G^{(o)}$ considered in π18.1 and in the proof of π18.2, are conceived to be extracted. Only this and nothing more. It would then be an arbitrary conceptual discontinuity, a leap, a kind of spontaneous generation, of *Deus ex Machina*, and even an inner inconsistency, to assign to this substratum posited by *us*, properties that transcend the very descriptional essence of *all* the fragments $œ_G^{(o)}$ that we extract from it, namely the impossibility shown by [π18.1+π18.2], to know any qualification whatever concerning a basic object-entity $œ_G^{(o)}$ in-itself.

**Comment**. It is quite non-trivial that inside MRC this minimality of the realist postulate P3 is a feature that emerges as a consequence – in the weak sense that marks all the "proofs" – of the non removable descriptional relativities. So much more so that the forces which withstand the distinction between mere existence of something, and *knowledge* of *how* this something *is*, are huge.

**Final global comment on the realist postulate P3 (cf. note 25).** By now, I think, the specificity of the concept of "physical reality" with respect to the general concept of reality introduced by D2, has come out with satisfactory definiteness, mainly *via* the frame principle P8, the principle P10 of individual mutual exclusion, the propositions π11, π12, π13, the concept D14.3.1 of basic transferred description, and the propositions from this point 18. Thereby, retroactively, the necessity of the postulate P3 as well as its significance should have become clear. This necessity lies in the fact that *the formulations mentioned above would not have been possible without P3*. As for the significance of P3 inside MRC, it can be best grasped *per a contrario*: it is that which inside MRC makes no sense, or no clear sense, when one considers elements of reality consisting of *concepts*, social facts, etc.

As for the minimality of the realism asserted here, I suppose that notwithstanding the proposition π18.3 many will tend to continue to nurture in their minds a non-minimal realism. But reconsider in full light the quasi irrepressible hope that, in spite of all, some model or "only some invariants", might some day transpierce the obstacle generated by the descriptional relativities and inform us definitively, even if only in a coded way, on *how* the physical reality is-in-itself, independently of any perception. And on the other hand, consider the necessarily fragmenting character of the knowledge that human mind can construct, the indefinite and evolving multiplicity of the possible basic object-entities $œ^{(o)}$ as well as of the basic transfer-views $V^{(o)}$ which – now or in the future – could be found to exist in the sense of D7 with respect to a given basic object-entity $œ^{(o)}$: these stress even more, if this is still possible, the illusory character of such a hope for non-minimality. Indeed, given the non removable dependence of thought on perception, given the non removable dependence of perception on fragmenting descriptional relativities, given the unpredictable and incessant complexifications brought forth by the so various, and unbounded, hierarchical chains of metadescriptions that are growing everywhere, given the unpredictable changes of "viewpoint" (of epistemic referential) which these complexifications might bring forth – certainly radical from time to time – on *what* a rational basis could one uphold the postulation of some convergence toward a definite, definitive, terminal, absolute descriptional structure (supposing that this succession of words were endowed with some meaning) ? What a sort of invariants, magically stabilized against all the changes brought forth by the growth of thought, and magically freed of any descriptional relativity, could, thus stripped, nevertheless carry *knowledge* of the way of being of physical reality in-itself, beyond the posit of its mere existence ? When knowledge *is* nothing else than qualifications *via* some view, of a somehow delimited object-entity, so qualifications relative to some view and some generator of object-entity ? Obviously one ends here up in a whirl of circularity.

#### 4.2.5. Relative models versus minimal realism

But if any knowledge-of-how-physical-reality-is-in-itself, is indeed an illusory self-contradicting concept, *why do our minds so stubbornly keep to this concept ?* This is a question which deserves being examined.

So I close now this exposition of the nucleus of MRC as follows. First I shall show why the illusory belief in the possibility to reach knowledge of how physical object-entities are in-themselves, is quasi irrepressibly generated by human mind, in consequence of the frame-principle P8. And then I shall show how, once identified, the fallacy vanishes and leaves place to dimensions of conceptual liberty.

I proceed by defining a last group of four concepts which specify entirely the philosophical status of the minimal realism asserted here.

***On the insufficiency of the basic transferred descriptions***. Consider first an *individual* transferred description $D^{(o)}/G^{(o)},œ^{(o)},V^{(o)}/$ of a physical basic object-entity $œ^{(o)}$ (i.e. for any aspect-view $V_g^{(o)} \in V^{(o)}$, when the succession $[G^{(o)}.V_g^{(o)}]$ is repeated, always the *same* value gk is obtained). In this case, by hypothesis, the epistemic referential $(G^{(o)},V^{(o)}$ insures for the transferred results the *strongest* possible sort of qualificational stability (π12, π13, D14.1). While furthermore, according to D14.3.1 the basic transferred description $D^{(o)}$ *characterizes* observationally the involved basic object-entity $œ^{(o)}$. So one finds oneself already in possession of an observational invariant that associates a quite definite *meaning* to what has been labelled *a priori* "$œ^{(o)}$" (cf. the comments on the final posit from D14.1.3). It might then be argued that this "suffices", that in such conditions there is no reason for researching further specifications concerning what has been labelled $œ^{(o)}$. But the fact is that in general such a "sufficiency" simply is not experienced by the observer-conceptors: in presence of even an *individual* transferred description $D^{(o)}$ that produces a most immediately manifest observational stability, many thinkers (if not most) – quite modern thinkers, and even physicists – experience an irrepressible tendency toward a subsequent epistemic elaboration that shall produce a better, a clearer meaning assignable to what has labelled $œ^{(o)}$. But a "better, clearer meaning of $œ^{(o)}$", in what a sense, exactly?

When one tries to answer this question it appears that what is researched is a representation of $œ^{(o)}$ that shall endow it with an *own* form of *space-time-gk-values*, *separated* from any process of observation and any registering device; and moreover a form of space-time-gk-values possessing "unity", i.e. covering a connected space-domain obeying some definite dynamical law.

Furthermore a global and explicit space-time representation is (vaguely) desired for also the processes that have led from the basic object-entity $œ^{(o)}$ with its own space-time location, to its basic transferred description. *The frame-principle P8 is here at work*.

The requirements of the frame-principle cannot be violated definitively. One can at most postpone dealing explicitly with them. The frame-principle expresses a psychical fact which is as irrepressible as the physical fact that masses are tied with gravitation. If a basic transferred description of a basic object-entity is asserted, then one should be able to imagine some possible own form of space-time-gk-values of this object-entity, as well as some possible own structure of space-time-gk-values of the process that has generated the description. If not, the frame principle will keep active and upset us.

A basic transferred description $D^{(o)}$, though, yields no hint for satisfying these requirements. It is expressed *exclusively* in terms of observable features *of registering devices* which are all *distinct* from what is labelled $œ^{(o)}$. It yields no representation whatever concerning the space-time location of the basic object-entity $œ^{(o)}$ itself. Inside a basic description $D^{(o)}$ the involved basic object-entity $œ^{(o)}$ is not represented as an autonomous individuality endowed with an own form, it still floats behind as a mere labelled nebula suggested by the words "basic object-entity" and their notation $œ^{(o)}$. And even if, for a moment, we suspend any question concerning specifically $œ^{(o)}$, and we consider $D^{(o)}$ as a whole, again we find ourselves in presence of an absence of space-time intelligibility. Indeed, given that each registered mark gk involved by $D^{(o)}$ is found on a g-apparatus and that the transfer-view $V^{(o)}$ must involve at least two different g-apparatuses for measuring two mutually incompatible basic aspect-views, the "form" of space-time-gk-values involved by the basic transferred description $D^{(o)}$ *itself* is found to cover a scattered domain of space, tied with different registering devices *that can lie arbitrarily far from one another*. And given that the time-origin $t_o$ has to be re-established after each realization of a succession $[G^{(o)}.V_g^{(o)}]$, it is not even clear whether it is possible to somehow associate this form with some continuous evolution (or persistence) ordered by a unique increasing time-parameter.

In short, by $D^{(o)}$ alone one cannot "understand" intuitively, neither how the basic object-entity can be conceived to "be", nor in what a sense, exactly, $D^{(o)}$ is a "description" of this basic object-entity. This situation is tiring for the mind. Therefore an individual basic transferred description $D^{(o)}$ is not perceived as an *achieved* descriptional action. It is not felt to have reached a conceptual stage of epistemological equilibrium. It is obscurely felt as if loosely fixed on a steep conceptual slope where a conceptual force draws it toward a separated representation of $œ^{(o)}$ in terms of own gk-*space-time* aspect-values.

This sort of need might be regarded as a methodological instinct tied with the frame-principle, induced by the adaptive biological evolution of our minds.

All the preceding remarks hold also concerning a probabilistic transferred description. The now seventy years old debate on the interpretation of quantum mechanics proves this enough.

So one is led to consider the following question: is it possible to elaborate, out of a previously achieved basic transferred description $D^{(o)}$, a separated description of the basic object-entity $œ^{(o)}$ involved in $D^{(o)}$ ? Not a description of «how $œ^{(o)}$ really is» – by now such naïve epistemic quests can be supposed to have been entirely transcended inside MRC –, but a specification of just a possible modus of *thinking* of $œ^{(o)}$ in a self-consistent, transparent, intellectually operational way that be naturally insertable into the current language-and-conceptualization. The answer to this question is positive and it is brought forth by the following three new definitions.

**D19. Intrinsic *metaconceptualization. Intrinsic model.***

**D19.1. *Intrinsic metaconceptualization of a basic transferred description*.** Consider a basic transferred description $D^{(o)}$ of a physical object-entity $œ^{(o)}$, individual or probabilistic.

- Let $G^{(1)}$ be a metagenerator of object-entity consisting of a conceptual selector (D4) that selects for examination the meta-object-entity consisting of $œ^{(1)} \equiv [D^{(o)} + œ^{(o)}]$.

- Let $V_I^{(1)}$/indicate an *intrinsizing metaview* (I: *intrinsizing*) which, starting from the initial, purely observational, transferred description $D^{(o)}$, works out *intrinsic* qualifications of the basic object-entity $œ^{(o)}$ involved in $D^{(o)}$ (intrinsic: word used in order to distinguish from the philosophical term "in itself"). This, inside the new epistemic referential $(G^{(1)},V_I^{(1)})$, is achieved as follows.

* Let $V_{Ig}^{(1)}$ (I fixed, g=1,2,...m, Ig functioning as *one* compact index) be a set of m intrinsizing meta-*aspect*-views which, together, constitute the intrinsizing metaview $V_I^{(1)}$.

* Each intrinsizing meta-aspect-view $V_{Ig}^{(1)}$ involves an *abstract*, conceptual $V_{Ig}^{(1)}$-operation of examination *of the metaobject-entity*$[D^{(o)}+œ^{(o)}]$, namely an examination constructed in a way such that its possible results – necessarily values (Ig)k of $V_{Ig}^{(1)}$, accordingly to the definition D.5.1 – are all *conceivable* as separate intrinsic qualifications (Ig)k of the basic object-entity $œ^{(o)}$ that are compatible with $D^{(o)}$.

* The values (Ig)k of the intrinsizing metaview $V_{Ig}^{(1)}$ are furthermore constructed as: *(a)* intrinsic qualifications of $œ^{(o)}$ *at the time* $t_o$ which is the time-origin re-established at the beginning of each succession $[G.V_g]$ having contributed to the elaboration of $D^{(o)}$; *(b)* qualifications located inside a connected space-volume $\partial r$ which $œ^{(o)}$ is posited to occupy at the time $t_o$.

The relative metadescription $D^{(1)}/G^{(1)},œ^{(1)},V_I^{(1)}/$ constructed as specified above will be called an *intrinsic metaconceptualization of the basic (individual or probabilistic) transferred description* $D^{(o)}/G^{(o)},œ^{(o)},V^{(o)}/$ and it will be also assigned the alternative more specific symbol $D_I^{(1)}/[D^{(o)},V_I^{(1)}]$.

**Comment.** We speak of "an" (not "the") intrinsic metaconceptualization of $D^{(o)}$, because in general many different intrinsizing metaviews can be constructed, and each one of these yields a corresponding and possibly specific intrinsic metaconceptualization.

An intrinsic metaconceptualization of a basic transferred description $D^{(o)}$ realizes a *retro-active* localizing projection of the scattered form of $D^{(o)}$, onto a connected and instantaneous space-time domain $[\partial r.t_o]$. The uniqueness of the temporal qualification $t_o$, even though it is retro-active, suffices now for permitting to posit, starting from it, an intrinsic time-order that is hidden to observation. This permits now to assign a law of intrinsic evolution to what has been labelled $œ^{(o)}$, underlying any evolution of the observable transferred description $D^{(o)}$. As for the transferred description $D^{(o)}$, it can now finally be *explained*. The basic object-entity $œ^{(o)}$ can now be conceived to have "possessed" at the time $t_o$ – on the connected spatial domain $\partial r$ – the features assigned to it by the intrinsic metaconceptualization $D_I^{(1)}/[D^{(o)},V_I^{(1)}]$. These, one can now think, were *own* features of $œ^{(o)}$, separated from those of any measurement device, independent of them, but features which $D^{(o)}$ has been able to transpose into observable manifestations, only by disorganizing the form of intrinsic gk-space-time aspect-values constituted by them. The scattered form of space-time-gk-values involved by $D^{(o)}$ can now be thought of as the result of a bursting and change of the initially integrated intrinsic features of $œ^{(o)}$ itself. A bursting produced by the mutual incompatibility of certain aspect-views $V_g^{(o)}$ from the transfer-view $V^{(o)}$ which has obliged us to perform a set of *different* successions $[G^{(o)}.V_g^{(o)}]$, $V_g^{(o)} \in V^{(o)}$ in order to obtain the global transferred description $D^{(o)}$ (according to D19.1 at least two such incompatible aspect-views $V_g^{(o)}$ are necessary in order to characterize $œ^{(o)}$).

In short, by the assumptions from D.19.1 the basic object-entity $œ^{(o)}$ has acquired the specification of an *own* form of gk-space-time aspect- values, and the process of emergence of the basic, transferred description $D^{(o)}$ has been *causalized*: the categories of space, time and form have been restored for $D^{(o)}$ and $œ^{(o)}$, so $D^{(o)}$ has now become *intelligible*.

**D19.2.** ***Intrinsic model of a physical basic object-entity***. So the intrinsic metaconceptualization $D_I^{(1)}/[D^{(o)},V_I^{(1)}]$ constructs "explanatory" *relations* between its global meta-object-entity $œ^{(1)} \equiv [D^{(o)}+ œ^{(o)}]$ and the basic object-entity $œ^{(o)}$ involved by $D^{(o)}$, as well as an own space-time representation of this basic object-entity $œ^{(o)}$. Once this construction has been achieved it is possible to extract from it exclusively the representation of the basic object-entity $œ^{(o)}$, in the following way.

The set of intrinsic qualifications of the basic object-entity $œ^{(o)}$ produced by the intrinsic metaconceptualization $D_I^{(1)}/[D^{(o)},V_I^{(1)}]$, when considered *alone*, severed from all the other elements with which it is tied inside the intrinsizing metadescription $[D_I^{(1)}/D^{(o)},V_I^{(1)}]$, will be called an (intrinsic) *model of* $œ^{(o)}$ and will be symbolized by $M(œ^{(o)})/[V^{(o)},V_I^{(1)}]$ in order to remind explicitly of the non-removable relativity of this model to the pair of views $[V^{(o)},V_I^{(1)}]$ which determined its genesis and its characters.

**Comment**. It is important to realize clearly that an intrinsic model $M(œ^{(o)})/[V^{(o)},V_I^{(1)}]$ is *not* a relative description ***of*** $œ^{(o)}$ in the sense of the definitions D14.

The intrinsizing meta-aspect-views from $V_I^{(1)}$ that produced the qualifications assigned to $œ^{(o)}$ by the intrinsic model $M(œ^{(o)})/[V^{(o)},V_I^{(1)}]$, have examined *the meta-object-entity* $œ^{(1)} \equiv [D^{(o)}+ œ^{(o)}]$, ***not*** *the basic object-entity* $œ^{(o)}$.

The model $M(œ^{(o)})/[V^{(o)},V_I^{(1)}]$ occupies finally a position of full epistemological saturation and equilibrium of the meaning assigned to what had been initially labelled $œ^{(o)}$. Its genetic compatibility with the transferred description $D^{(o)}$, as represented by the intrinsizing metaconceptualization $[D_I^{(1)}/D^{(o)},V_I^{(1)}]$, detached it from $D^{(o)}$ like a mature fruit that has been plucked from its tree. The model $M(œ^{(o)})/[V^{(o)},V_I^{(1)}]$ superposes now to the initial purely observational basic description $D^{(o)}$, a pragmatic, economic and stable conceptual *closure*. Namely a closure consisting of *an invariant with respect to the group of transformations from one succession* $[G^{(o)}.V_g^{(o)}]$, $V_g^{(o)} \in V^{(o)}$ *that contributed to the elaboration of* $D^{(o)}$, *to any other such succession with a different aspect-view in it,* $G^{(o)}$ *being* ***fixed***: the observable effects of all these different successions $[G^{(o)}.V_g^{(o)}]$, $V_g^{(o)} \in V^{(o)}$, are now all assigned one common and definite "causal" ancestor $M(œ^{(o)})/[V^{(o)},V_I^{(1)}]$ which produces various perceptible manifestations, in a "normal" way i.e. in a way that is understandable accordingly to the frame-principle P8.

When the basic transferred description $D^{(o)}$ on which the model $M(œ^{(o)})/[V^{(o)},V_I^{(1)}]$ is founded involves exclusively the human biological sensorial apparatuses, this sort of closure emerges in an unconscious, non-mediated, genetically wired way: it *is* precisely what we believe to perceive, and *this* we automatically assign to, exclusively, the involved object-entity....*in-itself*. The stage of a transferred description $D^{(o)}$ remains unknown. And even when fabricated apparatuses are connected to the biological ones, if the whole apparatus thus obtained still offers a directly intelligible form of space-time-gk-values, this form, again, is irrepressibly felt to reveal how the perceived object-entity is *in-itself* (think of perceptions via a microscope or a telescope) More: when, like in quantum mechanics, the observable basic transferred data do not themselves offer a directly intelligible form of space-time-gk-values, so if an intrinsic model $M(œ^{(o)})/[V^{(o)},V_I^{(1)}]$ has to be explicitly constructed from these data treated as mere coding signs, still, once a model has been constructed, it is usually felt to be satisfactory and *necessary* to such a degree that its only *hypothetical, retro-active, and relative* character tends to be skipped. Implicitly and fallaciously the intrinsic models $M(œ^{(o)})/[V^{(o)},V_I^{(1)}]$ conquer inside our minds a primary and absolute status.

This is the fallacy that instates the irrepressible belief that physical object-entities can be *known* “such as they are in themselves”.

The unavoidable dependence of any intrinsic model of $œ^{(o)}$, on *both* an initial transferred description $D^{(o)}$ that has had to be achieved first and has involved some particular transfer-view $V^{(o)}$, and a subsequent process of intrinsic metaconceptualization $D_I^{(1)}$ involving a particular intrinsizing metaview $V_I^{(1)}$, tends to be overlooked. In particular, it tends to remain unnoticed that ***another*** *pair* $(V^{(o)},V_I^{(1)})$ *would have led to a* ***different*** *model of* $œ^{(o)}$.

These occultations mark all the classical descriptions, in physics, in mathematics, etc., as well as in the current thinking expressed by the current language: they are the opaque fictitious platform that floats above the physical factuality and on which is erected the classical concept of objectivity. The roots which insert the conceptualizations into physical factuality, with the relativities involved by them, are hidden beneath this fictitious platform.

Starting from the transferred data that are available for it and on which it takes support without trying to express them, human mind always rushes as rapidly and as directly as it can toward a representation of the involved object-entity by an intrinsic model. As soon as such a representation has been attained, it is spontaneously felt to be "true" in a primary, certain and absolute way, without reference to the initial transferred data on which it is founded and forgetting that it is just an economic, hypothetical, retro-actively imagined *construct*. While the initial transferred data, even though they are the sole certainties, because of their dispersed unintelligible phenomenal appearance, are implicitly and irrepressibly perceived as nothing more than "subjective" tools for finding access to the "objective truth": *a fallacious, illusive inversion*. We systematically commit what Firth [207] called «the fallacy of conceptual retrojection». Simplicity, invariance, and what we tend to call "truth" and "objectivity", have coalesced in a knot imprinted upon our minds by ancestral processes which, by implicit pragmatic causalisations, optimizes the efficiency of our behaviour, but blocks and botches the reflexive knowledge of our fundamental epistemological functioning. The interpretation as ontological assignments, of the results of our instinctive human adaptive constructs involving the frame-principle, is one of the worst and most stubborn pathologies of thought.

But in quantum mechanics this process has hit an obstacle. Up to this very day a type of intrinsic model $M(œ^{(o)})/[V^{(o)},V_I^{(1)}]$ fitting satisfactorily the quantum mechanical transferred descriptions of what is called a microstate, has not yet been found. So it has been necessary to stop the attention upon these transferred descriptions themselves such as they have emerged, and to embody these transferred descriptions in mathematical expressions able to yield, if not understanding, at least numerical predictions. And then, like a tireless insect when its instinctive constructive actions are hindered, human mind came back again and again upon these quantum mechanical transferred descriptions that resist modelling. And so it has become possible to discern more and more explicitly their specificity, which inside MRC has been redefined in quite general terms and has been called a "basic transferred" character. In this way we finally become aware of the unavoidable necessity of a quite universal first phase of conceptualization in terms of basic transferred descriptions.

Inside MRC the distinction between illusory ontological assertions concerning an absolute way in which $œ^{(o)}$ «really-is-in-itself», and relative methodological intrinsic models of $œ^{(o)}$, is quite radical, elaborate and clear cut. And the genetic order of the descriptional steps is re-constructed correctly and is fully displayed.

In these conditions the irreplaceable pragmatic and heuristic power of intrinsic models can be put to work without triggering any more insoluble philosophical pseudo-problems. Correlatively, the vain and exhausting battle between positivists and defenders of modelling, evaporates. The transferred descriptions are the unavoidable first stage of our processes of conceptualization, while the intrinsic metaconceptualizations of the initial transferred descriptions and the relative models extracted from these are a stabilising subsequent stage which, if realized, brings us down onto a (local and provisional) minimum of our potential of conceptualization.

There is no choice to be made. There is just an unavoidable *order of elaboration* to be observed, in a normed way, or to be recognized when it occurs implicitly.

**D19.3.** ***Minimal intrinsic metaconceptualization. Minimal intrinsic model.*** Consider a basic transferred description $D^{(o)}$ of a physical basic object-entity. The effect labelled $œ^{(o)}$ of the basic operation $G^{(o)}$ of generation of an object-entity can *always* be trivially metaconstructed accordingly to D19.1 so as to be conceivable as:

A bulk of potentialities of future observable manifestations, determined by $G^{(o)}$ on a finite space-domain $\partial^3 r$, at the time $t_o$ when $G^{(o)}$ comes to an end, each one of these potentialities being relative to an aspect-view $V_g^{(o)}$ from the basic view $V^{(o)}$ operating in $D^{(o)}$.

For this it suffices to posit in D19.1 the *minimal intrinsizing view corresponding to* $V^{(o)}$ – let us denote it $[min.V_I^{(1)}/V^{(o)}]$ – defined as follows. For each basic aspect-view $V_g^{(o)}$ from the basic view $V^{(o)}$, $[min.V_I^{(1)}]$ contains a corresponding intrinsizing minimal meta-aspect-view $[min.V_{Ig}^{(1)}]$ possessing a unique *minimal meta-aspect-value* denoted $Ig_{min}$ that consists of the intrinsic *potentiality*, assigned to what has been labelled $œ^{(o)}$, to produce at a time $t_g>t_o$, any one among the transferred observable aspect-values gk of the basic aspect-view $V_g^{(o)}$, iff $œ^{(o)}$ is subjected at $t_o$ to an $V_g^{(o)}$-examination ($t_g$-$t_o$: the duration of a $V_g^{(o)}$-examination, characteristic of the considered aspect g) (I recall that "intrinsic" means here assigned to $œ^{(o)}$ itself as an own feature, the word having been chosen in order to distinguish from the meaning of the philosophical term "in itself").

---

[207] Firth, R., Reply to Sellars, (1981) Monist vol.64 pp. 91-101 (the quotation is from p.100).

The trivial realization of the definition D19.1 specified above will be called *the minimal intrinsic metaconceptualization of the basic transferred description* $D^{(o)}/G^{(o)},œ^{(o)},V^{(o)}/$ and it will be denoted [min.$D_I^{(1)}/D^{(o)}$] (the relativity to the acting intrinsizing view $V_I^{(1)}$ is now included in the definition of the minimal intrinsizing view [min.$V_I^{(1)}/V^{(o)}$]). The intrinsic model of $œ^{(o)}$ extracted from [min.$D_I^{(1)}/D^{(o)}$] will be called *the minimal intrinsic model of* $œ^{(o)}$ and will be denoted [min.M($œ^{(o)}/V^{(o)}$].

**Comment.** The following consequence of the final posit from D14.3.1 is quite worth being noticed. Any basic view $V^{(o)}$ that involves two mutually incompatible basic aspect-views $V_{g1}^{(o)}$ and $V_{g2}^{(o)} \neq V_{g1}^{(o)}$ entails a minimal intrinsic model [min.M($œ^{(o)}/V^{(o)}$] which now characterizes $œ^{(o)}$ *conceptually* (by predication). It yields a conceptual definition of $œ^{(o)}$ that can now be added to the purely factual definition of $œ^{(o)}$ insured initially by the operation $G^{(o)}$ alone (whereby $œ^{(o)}$ still remained outside knowledge) and to the subsequent purely observational description of $œ^{(o)}$ offered by the basic description $D^{(o)}$ (whereby $œ^{(o)}$, though characterized observationally, nevertheless was still devoid of an own conceptual representation). MRC brings forth *degrees* of characterization of a basic object-entity $œ^{(o)}$, which compose the complexifying sequence [purely factual→purely observational→conceptual]. From that stage on, chains of *non* minimal intrinsic metaconceptualizations can indefinitely increase the degree of conceptual characterization of $œ^{(o)}$. This illustrates the reflexivity of the method and its unlimited character.

As any intrinsic metaconceptualization and any intrinsic model, the trivial minimal models also may be perceived as "opportunistic" constructs where what is actually observed is posited to stem from an *a posteriori* imagined *ad hoc* explanatory potentiality. This however does not in the least diminish the pragmatic importance of the fact that a minimal model of what is labelled $œ^{(o)}$ is a representation that permits a most natural, easy insertion of $œ^{(o)}$ into the conceptualization. Moreover it is always and automatically realizable. It is however useful to remember again and again that inside MRC this sort of representation is accepted as just an unavoidable strategic step that must be carefully distinguished from an ontological credo: nothing whatever is naïvely asserted concerning the impossible question of how the basic object-entity $œ^{(o)}$ «really-is-in-itself». It is only stated how this object-entity can be most simply conceived in order for us to become able to speak and think of it in structured, consistent, fluent terms.

#### 4.2.6. Final comment on the realism involved in MRC

«…Thus the aim of the book is to draw a limit to thought – not to thought, but to the expression of thoughts: for in order to be able to draw a limit to thought, we should have to find both sides of the limit thinkable (i.e. we should have to be able to think what cannot be thought »

Ludwig Wittgenstein, in the Preface to the *Tractatu*s

The concept of *minimal* realism possesses, I think, an essential philosophical importance. Imagine an abstract surface on which are displayed all the grammatically correct structures of words that human mind can compose about the physical reality. On this surface, the concept of minimal realism is delimited by a boundary which coincides strictly with the boundary that separates the domain of communicable *knowledges*, from the domain inside which can be found only expressions which are grammatically correct but are *devoid of reference*: this boundary defines the extreme limit which expressions of communicable knowledge can reach. The communicable knowledges cannot transcend this frontier. They can just advance toward it and eventually hit it by this or that basic transferred description which acts like a small squad carrying a local net of pre-conceptualization inside which it captures a small load of as yet unknown physical factuality which it hoists up on the very first level of speakable, communicable knowledge. But thereby the progression of the squad from inside the zone of knowledge, toward the physical reality, is stopped. The squad is reflected *back* like an elastic ball toward the inside of the realm of relative *descriptions*, where it delivers its load which, from that moment on, can indefinitely be elaborated along innumerable branches of complexification by intrinsic metaconceptualizations and/or by extraction of corresponding intrinsic models. But each one of these complexifying elaborations introduces new descriptional relativities which ***thicken*** the screen between physical reality in-itself and our mind's representations of it, they thicken this screen so as to *improve intelligibility* and thereby the capacity to think and to act. Such is the paradoxical relation between physical reality and mind.

It is crucial to become aware, intensely, of the surreptitious advent of this *inversion* in our direction of conceptualization, of these unavoidable rebounds in the opposite direction each time that the extreme frontier of the domain of communicable knowledge is hit by a basic description. If not, we remain imprisoned in the inertial illusion that by modelizing more and more we *approach* more and more the knowledge of how the physical reality "is-in-itself". The grammatically correct associations of words which express this illusion are founded upon a self-contradicting concept of reality-in-itself, namely the concept of a *qualifiable* reality-in-itself. Whereas reality-in-itself – by definition – is precisely what cannot be qualified more than by its mere qualifi-ability. By these words, "in-itself", what is pointed toward deliberately is *nothing* more than a posited existence, *posited* also to be qualifi-*able* but to be devoid of any other more specifying qualific-*ation*. Any further qualification, even the most feeble one, the most vague, is either idempotent, or generates contradiction.

*This is not a matter of* ***fact****, it is a matter of organization of language-and-concepts*.

The words “description” and “physical reality in itself” *had* to be somehow endowed with a definition (even if only implicitly). And, inside the system of language, this definition happens to be such that what is called description has been opposed (implicitly) *by construction* to what is called “physical reality in itself”.

One might perhaps believe, for instance, that it is possible to gain one more inch by specifying that the reality-in-itself is “such” that the qualifications which it admits from our part are precisely those which are elaborated by our senses and our investigations. But when we focus attention on this supposedly supplementary specification, trying to capture an element of positive novelty added by it to the minimal realist postulate, we find only nothingness. We find ourselves placed on exactly the same content of information as before. Any attempt to superpose some nuance expressible in terms of approximations or of asymptotic apprehension of *how* the physical reality is in itself, would only manifest a misunderstanding of the nature of what is here involved, namely an optimized organization of concepts-and-words. One can reasonably try to fight against a physical circumstance, even if it is a “physical law”, trying to master it in order to realize some technical aim. But trying to fight against the limitations entailed by a conceptual-linguistic organization, manifests a confusion concerning essence: what meaning would that have, for instance, to fight against the limitations imposed by, say, arithmetic, which one does not criticize and inside which one has placed oneself in order to work accordingly to its rules? “The-way-of-existing-of-reality-in-itself” is a self-contradicting notion stemming from a confusion between empirical circumstances and conceptual organizations of which on the other hand one makes current use.

In his Conference on Ethics Wittgenstein said (concerning the more or less similar confusion between value and truth): « it is perfectly, absolutely hopeless to thus bump our forehead against the walls of our cage » (my own retro-translation in English, from the French translation). One can apply the same assertion to the confusion between an impossible ontological quest, and an organization of language-and-concepts constructed by man. This confusion entails chimerical aims and fictitious problems. Or, like in the quantum mechanical orthodoxy, an arbitrary positivistic interdiction of intrinsic metaconceptualizations and intrinsic models *because these are confounded with impossible qualifications of reality-in-itself*. This mythic fauna that spouts from the bursting of an inertially oriented impetus to understand more, against the barrier placed by thought between all that is speakable, and a posited and denominated rest, must be exorcised.

So the minimal realism involved by MRC has a composite logical status. While the feature of minimality follows “deductively” inside the method (π18), the main term, realism itself, is just a posit, the postulate P3. It is a declaration of metaphysical belief, wholly subjective. Any question of truth or objectivity is meaningless concerning it. But this metaphysical belief plays a fundamental role for MRC: it seats the method on a unifying ground. It asserts that beneath the endless proliferation of branching relativities which mark the contents of descriptions, there exists a substratum of non referred absolute, wherefrom the relativities emerge together with the conceptualizations. I say “beneath” in order to stress that the thesis of realism draws *out* of the domain of language and descriptions. By the mysterious powers of self-transcendence of language, this thesis acts like a verbal directional indicator, pointed from inside the volume of the expressible, but which points toward an existence from outside this volume. It grasps the attention, displaces it, and installs it at the very core of the non expressible. There, inside this background of unconceptualisable which it succeeds to designate, the realist thesis fixes the ends of the threads with the help of which the basic transferred descriptions *web to one another* – operationally, beyond words – the two regions that stretch out on the two sides of the ghostly but insuperable wall between what is by construction devoid of communicable expression, and the formulated and communicable. In spite of the fact that we cannot « find both sides of the limit thinkable ». This is the fundamental, the huge epistemological innovation hidden inside the quantum mechanical formalism, which inside MRC is explicated, generalized, and organized in detail. Whereby *all* the false absolutes are suppressed, not only those which vitiate esthetics, ethics and metaphysics; for *everywhere* thought is invested by hosts of false absolutes that generate pathological tissues of illusory problems and paradoxes that blur out the sound limit between the thinkable, and mere non sense.

It might seem that this background of non referred, because it is absolute, is incompatible with the method of relativized conceptualization. But, and it is important to stress this, MRC by no means banishes *any* absolute. It banishes exclusively the false absolutes, those which hide descriptional relativities of which the presence can be identified, and which, if ignored, can generate illusory problems. But it is clear that when one constructs, it is unavoidable to posit certain absolutes. All the definitions from MRC, principles, etc., as *such*, have nothing relative about them. They are absolutes of the method, by the help of which the descriptional relativities are defined. And the existence of a physical reality posited in P3 is one of the legal absolutes of the method. This severely restricted concept is introduced as the final, absolute reference without which thought would get lost in an unexplained profusion of diversity; an absolute reference which unifies in one coherent whole all the indefinitely evolving descriptional relativities defined by the method.

I confess that the beauty which, to my eyes, emanates from this unification, appears to me irrepressibly as a sign of pertinence. Man and “reality” form a whole, and the feeling of beauty that can emerge in a human mind, intimately tied with coherence, has for me the significance of an announcement that certain slopes of the real have been embodied without having been violated. Whatever the unimaginable designatum of the succession of words which I just aligned, I want to align them, for we must somehow speak in order to communicate, in spite of all, concerning the unspeakable.

### 4.2.7. Global remarks on the nucleus of MRC

***MRC is:***

***Explicitly founded upon the functioning of human mind, with its cognitive aims***. The choices of the epistemic referentials that generate the relativized descriptions, stem from the consciousness functioning of the acting observer-conceptor. Each such choice expresses a curiosity, a descriptional aim of this consciousness functioning. The descriptional aims expressed by the successive choices of an epistemic referential, inside a chain of conceptualization, express the evolution of the descriptional aims of the acting consciousness functioning, and thereby they determine the "direction of conceptualization", step by step. Inside MRC, in its present stage at least, the descriptional aims do not follow from methodological prescriptions. This means the following.

No AI-machine could, by applying MRC, work like a human being, without being directed by a human being. But an AI-machine endowed with an "MRC-program" (if this is possible) and drawn by a man, would work exactly like that man.

This specifies the difference between AI and MRC as well as the particularity of an "MRC-program".

***Explicitly rooted in pure factuality***, which entails the possibility of a systematic and constructed distinction between potentiality of an infinity of processes of actualization of relative observable manifestations, and this or that actualized observable manifestation (cf. V.2.2). *Thereby it brings in the modal dimension potential-actualization-actualized.*

***Radically relativizing***. The whole approach bears the seal of the mutual existence of object-entities and views (or, equivalently, of generators of object-entity, and views) and of the relativities of descriptions to the triads $G,œ_G,V$.

***Methodological, normative, legalizing***. MRC is not an attempt at describing the natural processes of conceptualization. Though data (introspective, linguistic, etc.) concerning these natural processes are strongly taken into account, nevertheless MRC recognizes the *impossibility* of a "purely" descriptive account on the processes of description. So, deliberately, it takes distance with respect to such an aim, by constructing definitions and principles conceived in order to optimize the processes of conceptualization in compatibility with definite goals, namely the *a priori* elimination by systematic relativizations of any false absolutization, reflexivity, construction of a conceptual structure with respect to which it be possible to "localize" any other descriptional structure, natural or not, etc. *Thereby MRC is formalized.* Not mathematically and quantitatively formalized, like a modern physical theory, but already formalized, qualitatively formalized.

***Finitistic, cellular, local***. The fact that the construction of knowledge requires parcelling, steps, is taken into account quite fundamentally throughout MRC, *via* the principle of separation P15 and the concept D16 of relative metadescription.

***Globally unlimited***. Though everywhere there are strict *local* delimitations of the descriptional quest that withstand any gliding into relativism, globally nowhere a boundary is pre-imposed: *the finalized finitism of MRC generates infinities*.

***Hierarchical***. MRC generates *hierarchical* ***trajectories*** of conceptualization, in contradistinction to the theory of logical types, or that of levels of language, which introduce extended hierarchical *strata*.

***Directional and reflexive***, endowed with a capacity for an *a priori-a posteriori* double way progression. Before starting a given descriptional cell, a free choice of the direction of conceptualization desired by the observer-conceiver is expressed in a corresponding choice of an epistemic referential. Later the results of this choice can be rejected or kept and developed, on the basis of explicitly defined criteria.

The various features enumerated above are not exhaustive. Nor, by no means, are they mutually independent. Quite on the contrary, they all stem from one core-structure that induces an innumerable host of connections between these features. This core-structure is dominated by the systematically recurrent role of the consciousness-functioning which introduces the epistemic referentials. Along the whole hierarchy of distinct descriptional cells of increasing order from each chain of conceptualization from the web of such chains, the same fundamental MRC-requirements for a relativized normed conceptualization manifest themselves with a sort of fractality: each time that an epistemic referential has been chosen – no matter on what level of conceptualization – the generator of object-entity, the object-entity and the view from it – no matter of what they consist –entail non removable descriptional relativities to them.

### 4.2.8. On the conceptual status of MRC

To what class of conceptual beings does MRC belong ?

*Any* representation of "natural facts" is more or less normative, never purely descriptional as the classical myth of objectivity involves. But:

In the case of MRC the explicitly and resolutely ***methodological*** character is a major feature of the approach.

Any confusion between ontological assertions or implications, and methodological constructs, is most carefully avoided.

Nevertheless MRC can also be regarded as:

An attempt at a finitistic representation of the natural processes of generation of meaning where both relativi*sm* and false absolutizations are excluded *ab initio* by explicit rooting into pure factuality and by deliberate systematic relativizations.

The fact that throughout the process of constructing MRC one acts "logically", is neither a circularity, nor does it involve that MRC is reducible to a logic. It only illustrates the general reflexive, (*a priori*)-(*a posteriori*) character of any approach and in particular of this one: *a priori* the logical criteria are supposed to be fulfilled and they are utilized implicitly [208], but later, at a convenient level of development of the approach, the logical criteria – as it will be shown in V.1 – become ***a posteriori*** explicitly expressible in MRC-terms. (This sort of inner evolution partakes of the general reflexive character of MRC that has permitted to admit *a priori* the possibility of any pairing (G,V) and to introduce only *a posteriori* criteria concerning the relevance of a given pairing (G,V): first became expressible the criterion of mutual existence D7, and then the subsequent criterion of stability involved in the definition D14.1).

So probably the best characterization is as follows.

MRC is a strongly normative representation of the processes of conceptualization, of which the major specificities are: the place explicitly reserved to the consciousness functioning; the radical descriptional relativizations; and the fact that it explicates the structure of the very first step in the construction of objectivity, in the course of which intakes of ***a***-[conceptual-linguistic] fragments of *pure* factuality adduce into language and thought the hard semantic core of scientific objectivity.

**IV.3. The Second Stage: an Ideographical Symbolization of MRC**

In all the expositions of MRC that preceded the present one I included in a presentation made in usual language, an ideographic symbolization which - without being neither a formalization *stricto sensu* nor a mathematical representation - permits certain suggestive and economic expressions. In this work I present it simplified and separately. In this way the symbolizations are made available while the drawbacks as well as the advantages appear clearly.

- A *consciousness functioning* CF is represented by the sign 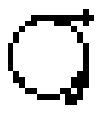 suggesting the whirling place from D1 that acts on both the Exterior Universe and the Interior Universe where it belongs, and in particular also on itself.
- *Reality* is again symbolized by the letter R.
- A *generator G of object-entity* will be represented by the sign Δ and will be *re-named* a *delimitator*[209] of object-entity, in order to stress that, whatever the nature of G, the final result is a delimitation, out of R, of a corresponding object-entity. Thereby however one *looses* the accent placed by the term “generator” upon a (possibly) of a *radically creative* character of an operation of object-entity generation. Then:
- The "place" from R where Δ works will then be denoted $R_\Delta$.
- The object-entity produced by Δ will be denoted by $œ_\Delta$.
- The *process of delimitation* by Δ, of an object-entity $œ_\Delta$, will be represented indifferently by

$\Delta R_\Delta \Rightarrow œ_\Delta$ or $œ_\Delta \Rightarrow \Delta R_\Delta$

where the two arrows do *not* have a logical meaning and cannot be considered separately, they are cemented into the global symbolizations which read respectively: "the delimitator Δ, acting on R at the place $R_\Delta$, produces the object-entity $œ_\Delta$", and "the object-entity $œ_\Delta$ produced by the delimitator Δ that acts on R at the place $R_\Delta$". Notice that the introduction of these symbolization permits to *distinguish* between:
* Δ: an epistemic operat*or* (in the sense of usual language, *not of mathematics*);
* $\Delta R_\Delta \Rightarrow œ_\Delta$: *a process*, that mentions its beginning and its result;
* $œ_\Delta \Leftarrow R_\Delta$: an explicit specification of an object-entity *via the process that produced I,* which permits to specify an unobservable object-entity, by the way of producing it.

Thereby the expressivity concerning this zone from MRC is considerably increased.

- An *aspect-view* will be symbolized by the same sign $V_g$ as before;
- The *operation of examination of* $œ_\Delta$ *by* $V_g$ will be represented by

$V_g œ_\Delta$

Notice that the introduction of these symbols permits to *distinguish* between:
* the epistemic operat*or* $V_g$ (in the sense of usual language, not of mathematics)
* the operat*ion* of examination $V_g œ_\Delta$.

Which, again is an increase of expressivity.

- A *view* will be symbolized as before by V.
- The global *operation of examination of* $œ_\Delta$*, by V* (achieved accordingly to π11), will be represented by

---

[208] Grize, J.B., (1993) *Pensée logico-mathématique et sémiologie du langage*, in Pensée logico-mathématique...... Nouveaux objets interdisciplinaires, Olivier Houdé et Denis Melville, P.U.F. The "natural logic" developed by J.B. Grize is the sort of logic that seems the nearest to that which acts throughout the elaboration of the nucleus of MRC.

[209] Later I have renounced to this denomination of "delimitator" because it suggests falsely passivity of the involved operation, a certain degree of pre-existence inside R of the result of the operation. So I began to use the words "operation of generation" and the sign *G*.

$V\oe_{\Delta}$

The remarks concerning $V_g$ hold also concerning V.

- An *epistemic* referential continues to be represented as before by $(\Delta,V)$.

- The representation of an observer-conceptor [CF,(G, V)] becomes [ 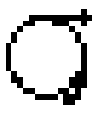 , $(\Delta,V)$].

- The *mutual inexistence* between an object-entity $\oe_{\Delta}$ and a view V will be symbolized by

$$\nexists\, \oe_{\Delta}/V \quad \text{or} \quad \nexists\, V/\oe_{\Delta}$$

which reads, respectively, "the object-entity $\oe_{\Delta}$ does not exist with respect to the view V", "the view V does not exist with respect to the object-entity $\oe_{\Delta}$".

- The *mutual existence* between an object-entity $\oe_{\Delta}$ and a view V will be represented by

$$\exists\, \oe_{\Delta}/V \quad \text{or} \quad \exists\, V/\oe_{\Delta}$$

which reads "the object-entity $\oe_{\Delta}$ does exist with respect to the view V", "the view V does exist with respect to the object-entity $\oe_{\Delta}$". (All these symbolizations can also be used, in particular, with the symbol of an aspect-view $V_g$ instead of V, which changes the meaning correspondingly).

- A *space-time view* is represented as before by $V_{ET}$.

- The *frame-principle* can be symbolized in the following way:

$$[\exists\, \oe_{\Delta}/ V_g] \rightarrow \exists V_{ET}: \exists\, \oe_{\Delta}/ (V_{ET} \cup V_g)]$$
$$[\nexists\, \oe_{\Delta}/ V_{ET}], \quad \forall V_{ET}, \forall \oe_{\Delta}$$

(where: the arrow, quite independently of any connotation suggesting *formal* logic, reads "entails that" (in the sense of natural logic) $\exists$ and $\nexists$ - outside any *formal* system, just in the sense of usual language or of "natural logic" - read, respectively, "there exists" and "there does not exist"; $(V_{ET} \cup V_g)$ considered as a one-block symbol, reads "the view formed with a space-time view $V_{ET}$ and another *physical* aspect-view $V_g$". The global reading of this symbolic picture is the verbal formulation of P8.

- The symbol of a *relative description* $D/G,\oe_G,V/$ becomes $D/\Delta,\oe_{\Delta},V/$, and the symbol for a *basic relative description* $D^{(o)}/G^{(o)},\oe_G{}^{(o)},V^{(o)}/$ becomes $D^{(o)}/\Delta^{(o)},\oe_{\Delta}{}^{(o)},V^{(o)}/$, and a relative metadescription of order n, $D^{(n)}/G^{(n)},\oe_G{}^{(n)},V^{(n)}/$, n=0,1,2,...., is symbolized by $D^{(n)}/\Delta^{(n)},\oe_{\Delta}{}^{(n)},V^{(n)}/$.

Together, these symbolizations constitute *the ideographic representation* of MRC denoted in short by { 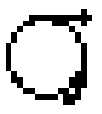 , $\Delta$, $\oe_{\Delta}$, V, ($D^{(n)}$, n=0,1,... }.

The essence of the whole representation can be reduced to 3 graphic symbolizations:

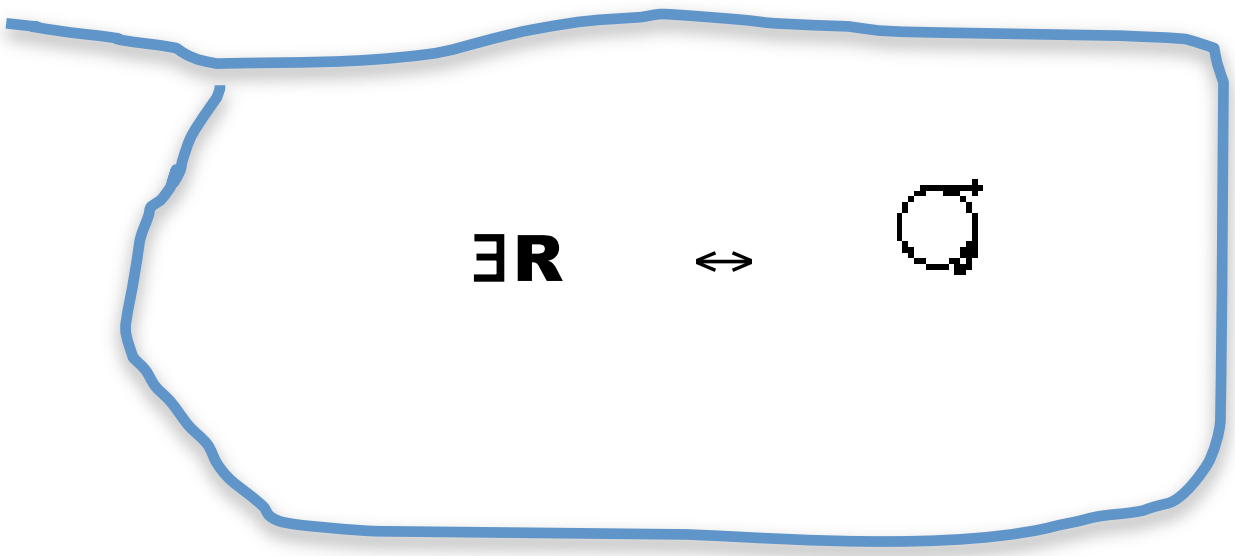

**MINIMAL REALISM**

---

**Generating triangle**

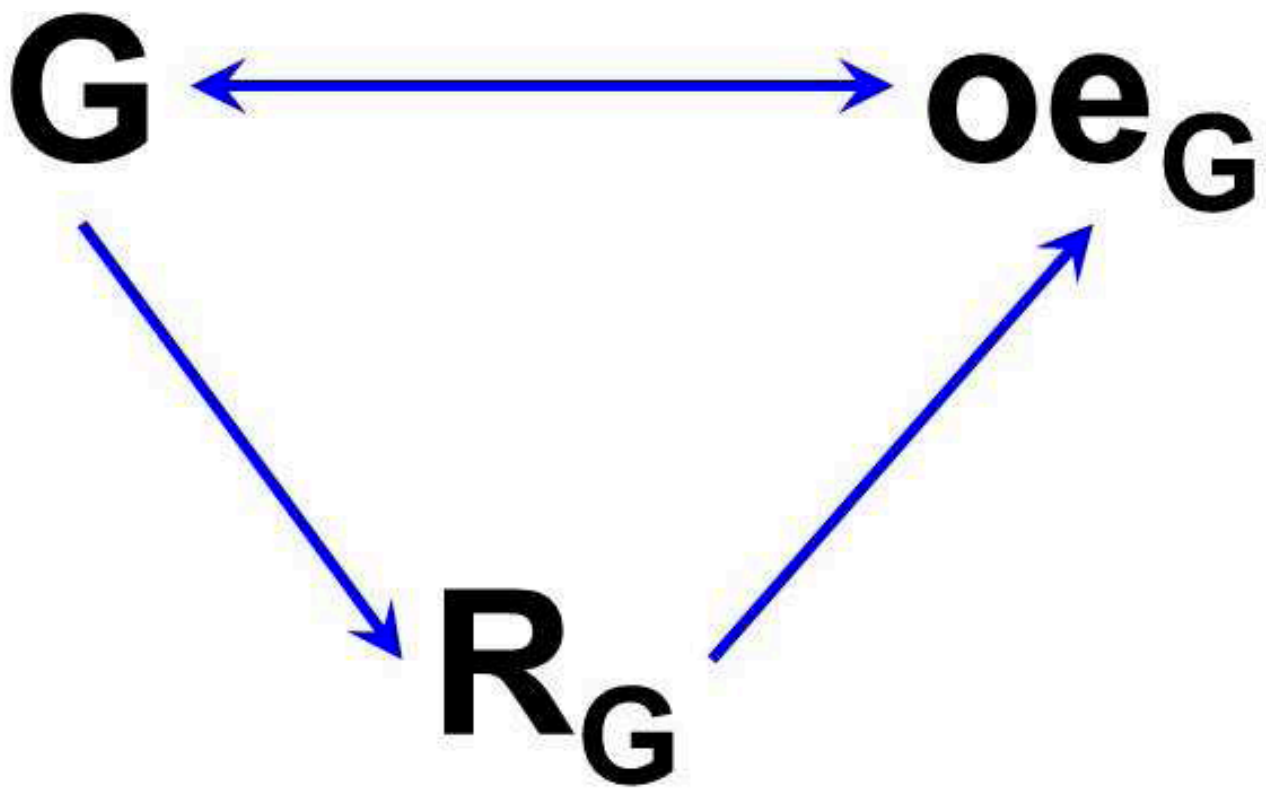

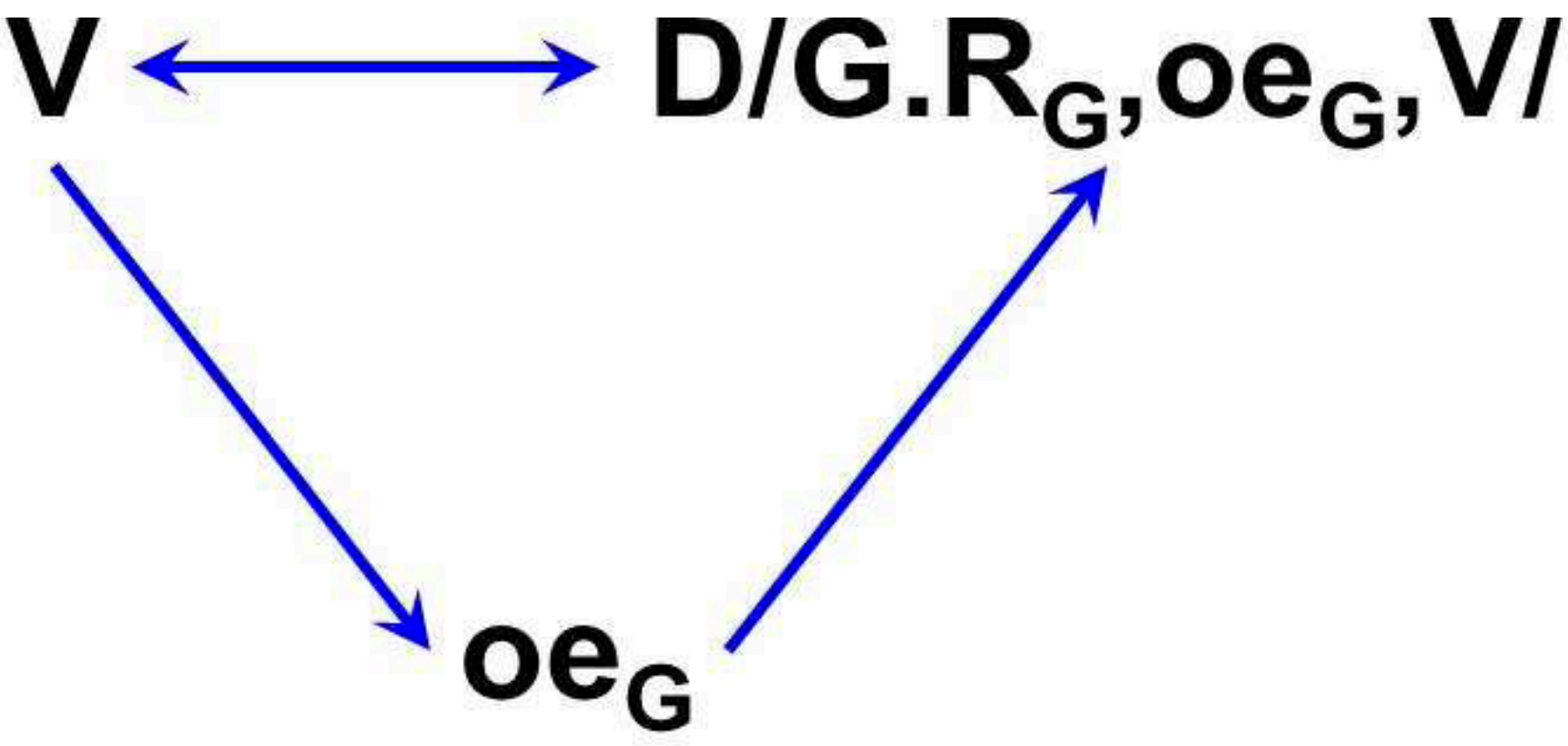

**Qualifying triangle**

---

***Descriptional AIM*** → ***(G,V) : pair of deliberate operational constructs***

→ ~~***"intrinsic truth"***~~

**$R_G$, $œ_G$ : data established *at the time when the operation is realized***

**(G,V) → D/G,$œ_G$,V/ : the result of the accomplished pair of operations**

↓

**(G.$R_G$ → $œ_G$ ) and (V.$œ_G$ → D/G,$œ_G$,V/)**

**A complete descriptional action**

**G,$œ_G$,V: "DESCRIPTIONAL *ROLES*", MODULAR, *FRACTAL***